\documentclass[twoside,openright]{cernyrep}

\usepackage[bookmarks, colorlinks=true,   linkcolor=blue, citecolor=blue, urlcolor=blue]{hyperref}

\usepackage[detect-all]{siunitx}

\usepackage[]{chapterbib} 

\usepackage[inline]{enumitem}
\setlist{noitemsep} 

\usepackage{wrapfig}

\usepackage{wasysym}
\usepackage{amsmath,amssymb}
\usepackage{graphicx}
\usepackage{verbatim}
\usepackage{placeins}
\usepackage{colortbl}
\usepackage[T1]{fontenc}
\usepackage[utf8]{inputenc}
\usepackage{etoolbox}
\AtBeginEnvironment{thebibliography}{\interlinepenalty=10000}

\usepackage[toc,page]{appendix}

\usepackage{subfigure}

\newcommand{\DOm}{\Delta \Omega}
\newcommand{\DOmvec}{\Delta \vec \Omega}
\newcommand{\dd}{\text{d}}
\usepackage{tikz}
\newcommand*\circled[1]{\tikz[baseline=(char.base)]{
    \node[shape=circle, draw, inner sep=1pt,
        minimum height=12pt] (char) {#1};}}

\newcommand{\BLKP}{
\ifthenelse{\isodd{\value{page}}}{\relax}{\mbox{}\thispagestyle{empty}\newpage}}

\usepackage{custom}

\usepackage{varwidth}
\usepackage{xcolor}

\begin{document}

\thispagestyle{empty}
\setlength{\unitlength}{1mm}


\pagenumbering{roman}
\setcounter{page}{1}
\thispagestyle{empty}
\setlength{\unitlength}{1mm}
\begin{picture}(0.001,0.001)
\put(-16,8){CERN Yellow Reports: Monographs}
\put(110,8){CERN-2021-003}

\put(-16,-80){\Huge \bfseries		
Storage ring to search for electric }
\put(-16,-90){\Huge \bfseries		
dipole moments of charged particles}
\put(-16,-100){\Large \bfseries		
Feasibility study}

\put(-5,-120){\Large CPEDM Collaboration}

\put(54,-250){\includegraphics{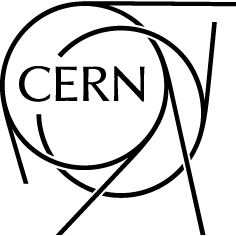}}

\end{picture}
\newpage

\thispagestyle{empty}
\mbox{}
\vfill

\begin{flushleft}
CERN Yellow Reports: Monographs\\
Published by CERN, CH-1211 Geneva 23, Switzerland\\[3mm]

\begin{tabular}{@{}l@{~}l}
 ISBN & 978-92-9083-606-3 (paperback) \\
 ISBN & 978-92-9083-607-0 (PDF) \\
 ISSN & 2519-8068 (Print)\\ 
 ISSN & 2519-8076 (Online)\\ 
 DOI & \url{https://doi.org/10.23731/CYRM-2021-003}\\
\end{tabular}\\[5mm]

Copyright \copyright{} CERN, 2021\\[1mm]
\raisebox{-1mm}{\includegraphics[height=12pt]{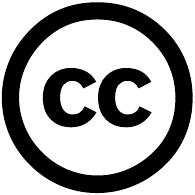}}
Creative Commons Attribution 4.0\\[5mm]

This volume should be cited as:\\[1mm]
Storage ring to search for electric dipole moments of charged particles: Feasibility study,\\ 
CPEDM Collaboration\\
CERN Yellow Reports: Monographs, CERN-2021-003 (CERN, Geneva, 2021)\\
\url{https://doi.org/10.23731/CYRM-2021-003}.\\[5mm]

Corresponding editors: \href{mailto:christian.carli@cern.ch}{Christian Carli}, \href{mailto:f.rathmann@fz-juelich.de}{Frank Rathmann} and \href{mailto:a.wirzba@fz-juelich.de}{Andreas Wirzba}.\\[1mm]
Accepted in June 2021, by the \href{http://library.cern/about_us/editorial_board}{CERN Reports Editorial Board}
(contact \href{mailto:Carlos.Lourenco@cern.ch}{Carlos.Lourenco@cern.ch}).\\[1mm]
Published by the CERN Scientific Information Service (contact \href{mailto:Jens.Vigen@cern.ch}{Jens.Vigen@cern.ch}).\\[1mm]
Indexed in the \href{https://cds.cern.ch/collection/CERN\%20Yellow\%20Reports?ln=en}{CERN Document Server} and in \href{https://inspirehep.net/}{INSPIRE}.\\[1mm]
Published Open Access to permit its wide dissemination, as knowledge transfer is an integral part of the mission of CERN.

\end{flushleft}
\clearpage

\begin{abstract} 
The proposed method exploits charged particles confined as a storage ring beam (proton, deuteron, possibly \Isotope[3]{He}) to search for an intrinsic electric dipole moment (EDM) aligned along the particle spin axis. Statistical sensitivities could approach \SI{e-29}{$e$.cm}. The challenge will be to reduce systematic errors to similar levels. The ring will be adjusted to preserve the spin polarization, initially parallel to the particle velocity, for times in excess of 15\,min. Large radial electric fields, acting through the EDM, will rotate the polarization from the longitudinal to the vertical direction. The slow increase in the vertical polarization component, detected through scattering from a target, signals the EDM.

The project strategy is outlined. A stepwise plan is foreseen, starting with ongoing COSY activities that demonstrate technical feasibility. Achievements to date include reduced polarization measurement errors, long horizontal
plane polarization lifetimes, and control of the polarization direction through feedback from  scattering measurements. The project continues with a proof-of-capability measurement (precursor experiment; first direct deuteron EDM measurement), an intermediate prototype ring (proof-of-principle; demonstrator for key technologies), and finally a high-precision electric-field storage ring.


\end{abstract} 
\newpage 
\thispagestyle{empty}
\ \newpage

\vskip 10cm
\begin{center}
{\huge\bfseries
Storage ring to search \\
 for electric dipole moments
  of charged particles \\}

  \vspace{5mm}

  {\LARGE\bfseries Feasibility study \\}
  
  \vspace{3mm}
  
  {\bfseries CPEDM Collaboration }


\end{center}



{\large
\noindent
F.~Abusaif,$^{1}$
A.~Aggarwal,$^{2}$
A.~Aksentev,$^{3}$
B.~Alberdi-Esuain,$^{4}$
A.~Andres,$^{1}$
A.~Atanasov,$^{5}$
L.~Barion,$^{6}$
S.~Basile,$^{6}$
M.~Berz,$^{7}$
C.~B\"ohme,$^{1}$
% ...
}

\vspace{3mm}




\renewenvironment{abstract}
 {\normalsize
  \begin{center}
  \bfseries \abstractname\vspace{-.5em}\vspace{0pt}
  \end{center}
  \list{}{
    \setlength{\leftmargin}{1.5cm}%
    \setlength{\rightmargin}{\leftmargin}%
  }%
  \item\relax}
 {\endlist}

\hspace{2mm}

\begin{flushleft}

{\em\footnotesize
${}^{1}$   Institut f\"ur Kernphysik, Forschungszentrum J\"ulich,  52425 J\"ulich, Germany\\
${}^{2}$   Institute of Physics, Jagiellonian University, 30348 Cracow, Poland\\
${}^{3}$   Institute for Nuclear Research, Russian Academy of Sciences, 117312 Moscow, Russia\\
${}^{4}$   III. Physikalisches Institut B, RWTH Aachen University, 52056 Aachen, Germany\\
${}^{5}$   European Organization for Nuclear Research, CERN CH-1211 Gen\'eve 23, Switzerland \\
${}^{6}$   University of Ferrara and INFN, 44100 Ferrara, Italy\\
${}^{7}$   Department of Physics and Astronomy, Michigan State University, East Lansing, MI 48824, USA\\
${}^{8}$   Institute of Nuclear Physics PAS, 31-342 Cracow, Poland\\
${}^{9}$   CNRS/IN2P3 - UGA, LPSC , Grenoble, France\\
${}^{10}$  Laboratory of Nuclear Problems, Joint Institute for Nuclear Research, 141980 Dubna, Russia\\
${}^{11}$  Department of Physics, University of Liverpool, Liverpool L69 7ZE, UK\\
${}^{12}$  Institut f\"ur Hochfrequenztechnik, RWTH Aachen University, 52056 Aachen, Germany\\
${}^{13}$  Department of Electrical and Computer Engineering, Agricultural University of Georgia, 0159 Tbilisi, Georgia\\
${}^{14}$  Budker Institute of Nuclear Physics, 630090 Novosibirsk, Russia\\
${}^{15}$  High Energy Physics Institute, Tbilisi State University, 0186 Tbilisi, Georgia\\
${}^{16}$  Institute for Advanced Simulation, Forschungszentrum J\"ulich, 52425 J\"ulich, Germany \\
${}^{17}$  Bethe Center for Theoretical Physics, Universit\"at Bonn, 53115 Bonn, Germany \\
${}^{18}$  Helmholtz-Institut f\"ur Strahlen und Kernphysik, Universit\"at Bonn, 53115 Bonn, Germany \\
${}^{19}$  Zentralinstitut f\"ur Engineering, Elektronik und Analytik, Forschungszentrum J\"ulich, 52425 J\"ulich, Germany\\
${}^{20}$  L.D. Landau Institute for Theoretical Physics, 142432 Chernogolovka, Russia\\
${}^{21}$  Engineering Physics Department, Georgian Technical University, 0160 Tbilisi, Georgia\\
${}^{22}$  Institute of Physics, Johannes Gutenberg University Mainz, 55128 Mainz, Germany\\
${}^{23}$  Research Institute for Nuclear Problems, Belarusian State University, 220030 Minsk, Belarus\\
${}^{24}$  Indiana University Center for Spacetime Symmetries, Bloomington,  IN 47405, USA\\
${}^{25}$  INFN, 70125 Bari, Italy\\
${}^{26}$  Cornell University, Ithaca,  NY 14850, USA\\
${}^{27}$  Andrzej Soltan Institute for Nuclear Studies, Warsaw, Poland \\
${}^{28}$  Lawrence Berkeley National Laboratory, Berkeley, CA 94720, USA\\
${}^{*}$  deceased

}


\end{flushleft} 

\pagestyle{plain}


\csname @openrightfalse\endcsname
\chapter*{Acknowledgements\markboth{Acknowledgements}{Acknowledgements}}

The authors would like to acknowledge important discussions with, and significant contributions  to this study by, Yannis K. Semertzidis and his colleagues from the Center for Axion and Precision Physics (CAPP) of the Korean Advanced Institute of Science and Technology (KAIST, South Korea);  a prior version of Appendix\,\ref{Chap:MagneticFields} and the present version of Appendix\,\ref{Chap:hybrid}  were authored by them.

This report is supported by an ERC Advanced Grant of the European Commission (srEDM, grant number 694340, principal investigator Hans Str\"oher,
Forschungszentrum J\"ulich, Germany). We also acknowledge the support for Transnational-Access-to-COSY within HORIZON2020 (STRONG-2020, grant number
824093) by the European Union.

This research was supported in part by the DFG and the NSFC, through funds provided to the Sino-German CRC 110  `Symmetries and the Emergence of Structure in QCD' (NSFC grant number 11621131001, DFG grant number TRR110), and by the National Science Foundation (grant number NSF PHY-1125915).

The work of N.N. Nikolaev was partly supported by the Russian MOS programme (grant number 0033-2019-0005).
The Georgian collaborators acknowledge support from the Shota Rustaveli National Science Foundation of the Republic of Georgia, SRNSFG (grant number  217854, `A first-ever measurement of the electric dipole moment (EDM) of the deuteron at COSY').

\begingroup\baselineskip.99\baselineskip
\setcounter{tocdepth}{1}
\tableofcontents
\endgroup


\BLKP

\newpage 
\thispagestyle{empty}
\ \newpage

\pagenumbering{arabic}
\setcounter{page}{1}
\renewcommand{\floatpagefraction}{0.9}
\renewcommand*\thesection{\thechapter.\arabic{section}}


\addcontentsline{toc}{chapter}{Executive summary}
\markboth{Executive summary}{Executive summary}

\begin{cbunit}

\csname @openrighttrue\endcsname 
\chapter*{Executive summary\footnote{This is an update of the version submitted to the European Strategy for Particle Physics \cite{Abusaif:2018oly}.}}
\label{Chap:ExecSum}





\section*{Science context and objectives}

Symmetry considerations and symmetry-breaking patterns have played an important role in the development of physics in the last 100\,years. Experimental tests of discrete symmetries (\eg  parity $P$, charge conjugation $C$, their product $CP$, time-reversal invariance $T$, the product $CPT$, baryon or lepton number) have been essential for the development of the Standard Model (SM) of particle physics.

Subatomic particles with non-zero spin (whether of elementary or composite nature) can only support a non-zero permanent electric dipole moment (EDM) if both  time-reversal ($T$) and parity ($P$) symmetries  are violated explicitly, while the charge symmetry ($C$) can be maintained (see \eg Ref.~\cite{Engel:2013lsa}).  Assuming conservation of the combined $CPT$ symmetry, $T$ violation also implies $CP$ violation. The  $CP$ violation generated by the Kobayashi--Maskawa  (KM) mechanism of weak interactions contributes a very small EDM that is several orders of magnitude below current experimental limits. However, many models beyond the Standard Model predict EDM values near the current experimental limits.
Finding a non-zero EDM value of any subatomic particle would be a signal that there exists a new source of $CP$ violation, either  induced by the strong $CP$ violation via the $\bar\theta$ angle of quantum chromodynamics (QCD) or by genuine physics beyond the SM (BSM). In fact, the best upper limit on $\bar\theta$ follows from the experimental bound on the EDM of the neutron. Moreover, $CP$ violation beyond the SM is also essential for explaining the mystery of the observed baryon--antibaryon asymmetry of our Universe, one of the outstanding problems in contemporary elementary particle physics and cosmology\footnote{Note that $CP$ violation in combination with $C$ violation is one of the three Sakharov conditions \cite{Sakharov:1967dj}.}. Measurement of a single EDM will not be sufficient to establish the sources of any new $CP$ violation.  Complementary observations of EDMs in a number
of systems will thus prove essential. Up to now, measurements have mainly focused on neutral systems (neutrons, atoms, molecules). We propose to use a storage ring to measure the EDM of charged subatomic particles.

The storage ring method would provide a direct measurement of the EDM of a charged particle comparable to or better than present investigations on ultracold neutrons. In the neutron investigations,   precession frequency jumps  in traps containing magnetic and electric fields are measured as the sign of the electric field is changed. These experiments are now approaching sensitivities of \SI{e-26}{$e$.cm} \cite{Afach:2015sja} and promise improvements of another order of magnitude within the next decade. Because proton beams trap significantly more particles, statistical sensitivities may reach the order of \SI{e-29}{$e$.cm}
\cite{Anas} with a new, all-electric, high-precision storage ring. Indirect determinations for the proton using $^{199}$Hg  produce model-dependent EDM limits near \SI{2e-25}{$e$.cm} \cite{Graner:2016ses}. Thus, storage rings could take the lead as the most sensitive method for the discovery of an EDM.

It should be noted that the rotating spin-polarized beam used in the EDM search is also sensitive to the presence of an oscillating EDM resulting from axions or axion-like fields, which correspond to the dark-matter candidates of a pseudoscalar nature. These may be detected through a time series analysis of EDM search data or by scanning the beam's spin-rotation frequency in search of a resonance with an axion-like mass in the
range from microelectronvolts down to \SI{e-24}{\electronvolt}\cite{Chang:2017ruk,Abel:2017rtm,Chang:2019poy}.

\section*{Methodology}

The electric dipole must be aligned with the particle spin, since it provides the only (quantization) axis in its rest frame. The EDM signal is based on the rotation of the electric dipole in the presence of an external electric field that is perpendicular to the particle spin. The particles are formed into a spin-polarized beam. Measurements are made on the beam as it circulates in the ring, confined by the ring electromagnetic fields, which always generate an electric field in the particle frame, pointing to the centre of the ring.

For a particle propagating in generic magnetic $\vec B$ and electric $\vec E$ fields, the spin motion is described by the Thomas--Bargmann--Michel--Telegdi (Thomas--BMT) equation and its extension for the EDM, see Ref.~\cite{Fukuyama:2013ioa}\footnote{More details on the application of the Thomas--BMT equation for circular accelerators are discussed in Chapter \ref{Chap:ExpMethod}, while the inclusion of gravity effects can be found in Appendix \ref{app:gravity}.}:
\begin{align}
\dfrac{\dd \vec{S}}{\dd t} & =  \left(\vec{\Omega}_{\rm MDM} + \vec{\Omega}_{\rm EDM}\right) \times \vec{S}\,, \notag \\
\vec{\Omega}_{\rm MDM} & = -\dfrac{q}{m} ~ \left[\left(G+\dfrac{1}{\gamma}\right) \vec{B}
-\dfrac{\gamma G}{\gamma+1} \vec{\beta} \left(\vec{\beta} \cdot \vec{B} \right) - \left(G+\dfrac{1}{\gamma+1} \right) \dfrac{\vec{\beta} \times \vec{E}}{c}\right] \,, \notag \\
\vec{\Omega}_{\rm EDM} & = -\dfrac{\eta q}{2 m c} \left[\vec{E} - \dfrac{\gamma}{\gamma+1} \vec{\beta} \left(\vec{\beta} \cdot \vec{E}\right)+ c \vec{\beta} \times \vec{B} \right].
\label{eq2}
\end{align}
The angular velocities, $\vec\Omega_{\rm MDM}$ and $\vec\Omega_{\rm EDM}$, act through the magnetic dipole moment (MDM) and electric dipole moment (EDM), respectively. In this equation, $\vec{S}$  denotes the spin vector in the particle rest frame, $t$ the time in the laboratory system, $\vec \beta$ is the velocity vector divided by the velocity of light and $\gamma$ the Lorentz factor. The magnetic anomaly $G$  and the electric dipole factor $\eta$ of the particle are dimensionless and introduced via the magnetic dipole moment $\vec{\mu}$ and electric dipole moment $\vec{d}$, which   point in the same direction and are proportional to the particle's spin $\vec{S}$:
\begin{equation}
\begin{array}{c}
\vec{\mu} = g \dfrac{q \hbar}{2 m} \vec{S} = (1+G) \dfrac{q \hbar}{m} \vec{S},
\qquad
\vec{d} = \eta \dfrac{q \hbar}{2 m c} \vec{S}\,,
\label{eq1}
\end{array}
\end{equation}
with $S = 1/2$ for protons and helions and $S = 1$ for deuterons,  where $q$ and $m$ are the charge and the mass of the particle, respectively.

The angular velocities ($\vec\Omega$) in \Eref{eq2} describe the rotation of the spin vector of the particle as it travels around the ring.  Because the magnetic moments of all particles carry an anomalous part, the polarization will, in general, rotate in the plane of the storage ring relative to the beam path. This rotation must be suppressed by
matching $\vec{\Omega}_{\rm MDM}$ to the cyclotron frequency
\begin{equation}
\vec\Omega_{\rm cycl} = -\dfrac{q}{\gamma m} \left( \vec B_y - \dfrac{\vec\beta \times \vec E_r}{\beta^2c} \right),
\end{equation}
\ie $\vec{\Omega}_{\rm MDM} = \vec{\Omega}_{\rm cycl}$, a condition called `frozen spin', where $\beta$ is the absolute value of $\vec \beta$. Under this condition, and with the spins pointing along the momentum, the vertical polarization can build up. In a magnetic ring, this condition requires (since $\vec\beta\cdot\vec B =0$) that a radial electric field is added to the ring bending elements, with
\begin{equation}
   E_r = \frac{GB_yc\beta\gamma^2}{G \beta^2\gamma^2-1}\, .
\end{equation}

For particles such as the proton, where $G>0$, it is also possible to build an all-electric ring ($\vec B=0$), provided that one can choose $\gamma =\sqrt{1+1/G}$. For the proton, this gives a beam momentum of $p = 0.701\UGeVc$. The corresponding kinetic energy of $T = \SI{232.8}{MeV}$ fortuitously comes at a point where the spin sensitivity of the polarimeter is near its maximum (\eg carbon target), creating an advantageous experimental situation.

The effect of the torque for a positively charged particle  is illustrated in \Fref{Fig: closed orbit}.
In this example, the magnetic and electric fields are purely vertical and purely radial, respectively. A particle is confined in an ideal planar closed orbit in the ring. Its velocity $\vec{v}=c\vec{\beta}$   is along the orbit.

\begin{figure}
\centering
\includegraphics[scale=0.7]{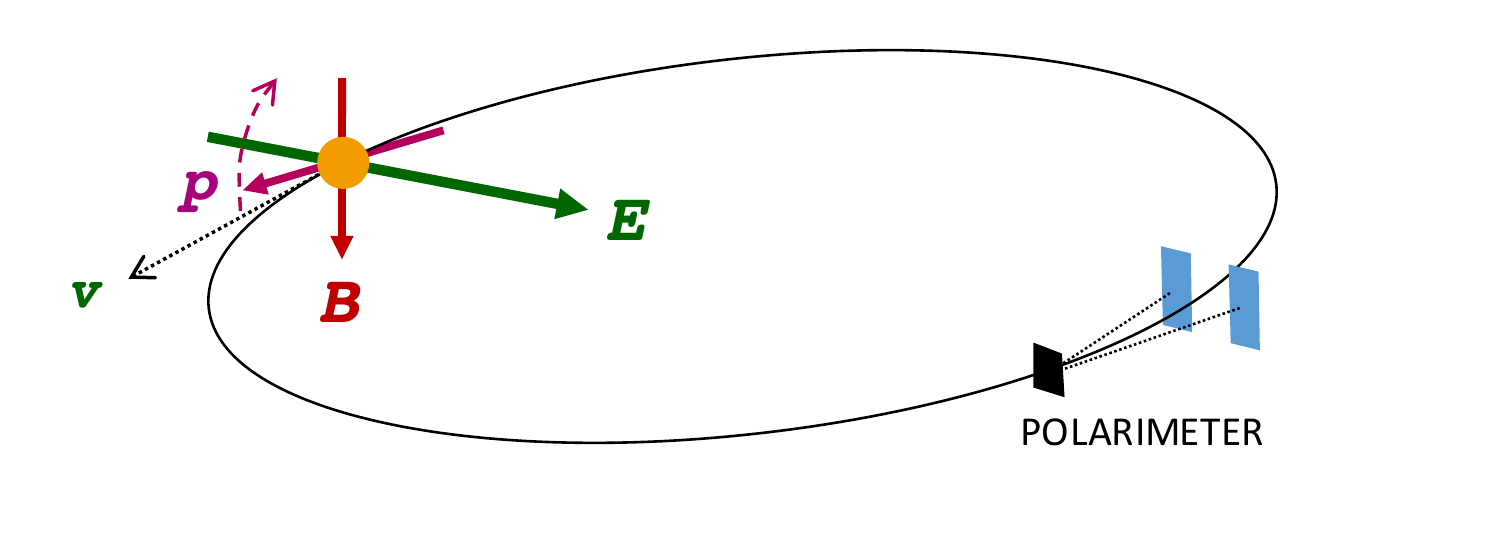}
\caption{Particle travelling around a storage ring, confined by purely vertical magnetic and purely radial electric fields. The `frozen spin' polarization, initially along the velocity, precesses slowly upwards in response to the radial electric field acting on the EDM. The vertical component of this polarization is observed through scattering in the polarimeter.}
\label{Fig: closed orbit}
\end{figure}

The spin axis is shown in \Fref{Fig: closed orbit} by the purple arrow, which rotates in a plane perpendicular to $\vec{E}$. If the initial condition begins with the spin parallel to the velocity, then the rotation caused by the EDM will make the vertical component of the beam polarization change. This rotation receives a contribution from both the external field $\vec{E}$ and the motional electric field $c\vec{\beta}\times \vec{B}$, and becomes the signal observed by a polarimeter located in the ring. This device allows beam particles to scatter from nuclei in a fixed bulk material target (black). The difference in the scattering rate between the left and right directions (into the blue detectors) is sensitive to the vertical polarization component of the beam. Continuous monitoring will show a change in the relative left--right rate difference during the time of the beam storage if a measurable EDM is present.

The statistical error for one single machine cycle is given by\cite{Anas}
\begin{equation}\label{eq:staterr}
  \sigma_\text{stat} \approx \frac{2\hbar}{\sqrt{N f} \tau P A E} \, ,
\end{equation}
(see also \Eref{sigma-EDM-final}).
Assuming the parameters given in \Tref{tab:exs:staterr}, the statistical error for 1\,year of running (\ie \num{10000} cycles of \SI{1000}{s} length) is\footnote{For further details, see Section \ref{Chap:statistics-general} and Table \ref{tab:sys:staterr} therein.}
\begin{equation}
    \sigma_\text{stat}(1\,\text{year}) = \text{\SI{4.6e-29}{$e$.cm}} \, .
\end{equation}
The challenge is to suppress the systematic error to the same level.

\begin{table}
\centering
\caption{Parameters relevant for the statistical error in the proton experiment
\label{tab:exs:staterr}}
\begin{tabular}{l l l }
\hline
\hline
Beam intensity                  &       $N=4 \times 10^{10}$ per fill \\
Polarization                    &       $P=0.8$   \\
Spin coherence time             &       $\tau = 1000\,$s  \\
Electric fields                 &       $E=8\,$MV/m \\
Polarimeter analysing power     &       $A = 0.6$ \\
Polarimeter efficiency          &       $f= 0.005$ \\
\hline
\hline
\end{tabular}
\end{table}

Many of the systematic errors in the EDM search may be eliminated by looking at the difference between two experiments run with clockwise (CW) and counterclockwise (CCW) beams in the ring.  For an ideal machine, one beam represents the time-reverse of the other, and the difference will show only time-odd effects, such as the EDM.  For the proton, the choice of an all-electric ring allows the two beams to be present in the ring at the same time, an advantage when suppressing systematic effects. \Figure[b] \ref{fig:ring_princ} illustrates two features of the all-electric proton experiment, the counter-rotating beams and the alternating direction of the polarization (along or against the velocity) in separate beam bunches, which is important for geometric error cancellation in the polarimeter.

\begin{figure}
\centering{
\includegraphics[width=0.45\textwidth]{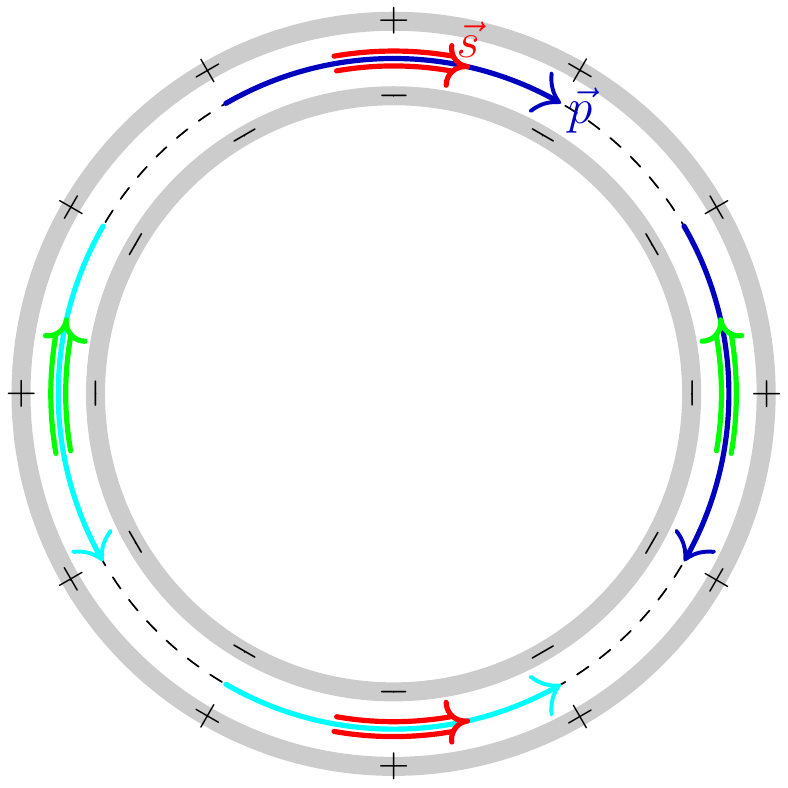}
\caption{Electric storage ring with simultaneously clockwise and counterclockwise circulating beams (dark and light blue arrows), each with two helicity states (green and red arrows for each beam). The grey circles represent electric field plates.
\label{fig:ring_princ}}}
\end{figure}

In general, any phenomenon other than an EDM generating a vertical component of the spin limits the sensitivity (\ie the smallest detectable EDM) of the proposed experiment. Such systematic effects may be caused by unwanted electric fields caused by imperfections of the focusing structure (such as the misalignment of components) or by magnetic fields penetrating the magnetic shielding or produced inside the shield by the beam itself, the RF cavity, or gravity. A combination of several such phenomena, or a combination of an average horizontal spin and one of these phenomena, may also lead to such systematic effects.

In many cases, as, for example, effects due to gravity, the resulting rotations of the spin into the vertical plane do not mimic an EDM because the observations for the two counter-rotating beams are not compatible with a time-odd effect. In this case, the contributions from the two counter-rotating beams tend to cancel out, provided the forward and reverse polarimeters can be calibrated with sufficient precision. In some cases, for example,  magnetic fields from the RF cavity, the resulting spin rotations into the vertical plane can be large.

The most important mechanism dominating systematic effects is an average static radial  magnetic field that mimics an EDM signal. For a \SI{500}{m} circumference frozen-spin EDM ring, an average magnetic field of about \SI{e-17}{T} generates the same vertical spin precession as the EDM of \SI{e-29}{$e$.cm}
that the final experiment aims at being able to identify. To mitigate systematic effects, the proposed ring will be installed in state-of-the-art magnetic shielding that reduces residual fields to the nanotesla level\cite{Anas}. The vertical position difference between the two counter-rotating beams that is caused by the remaining radial field will be measured with special pick-ups that must be installed at very regular locations around the circumference to measure the varying radial magnetic field component created by the bunched beam separation. A complete  thorough study of systematic errors in the EDM experiment is very delicate and not yet available. Studies of systematic effects have been carried out, and are underway, by several teams in the CPEDM collaboration to further improve the understanding of basic phenomena to be taken into account and to estimate the achievable sensitivity. The preliminary conclusion is that the intended sensitivity is a big challenge. Meeting this challenge requires that we proceed in a series of stages (see \Fref{fig:exs:stages}),  each of which depends on the knowledge gained from the preceding stage's experience.

\begin{figure}
\centering
\includegraphics[scale=0.45]{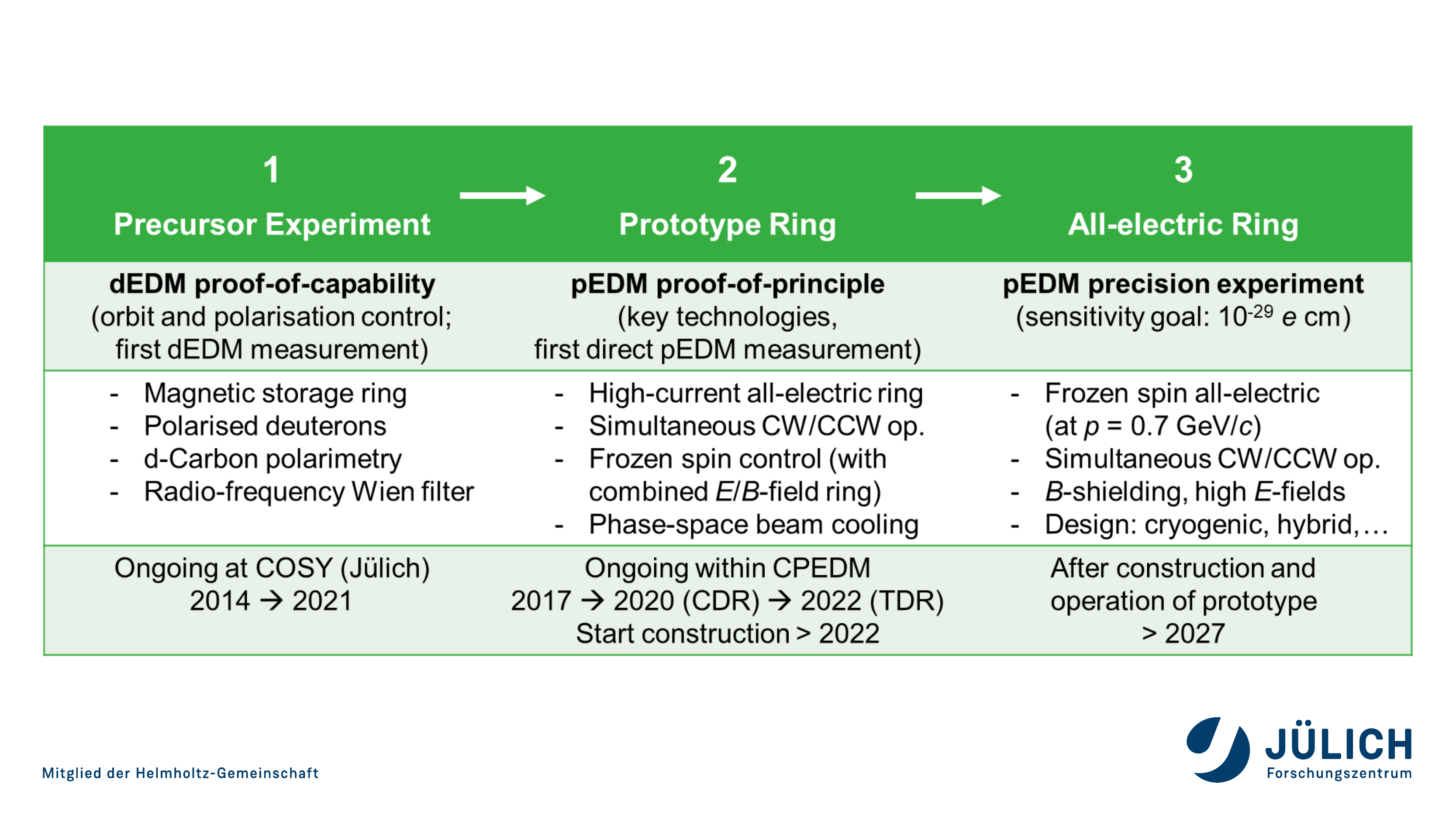}
\caption{\label{fig:exs:stages}Important features of the proposed stages in the storage ring EDM strategy}
\end{figure}

\section*{Readiness and expected challenges}

The JEDI (J\"ulich Electric Dipole Moment Investigations) collaboration has worked at COSY (Cooler Synchrotron, Forschungszentrum J\"ulich, Germany) for the last decade to demonstrate the feasibility of critical EDM technologies for the storage ring. Historically, these studies were begun with deuterons; the switch to protons has not yet been made, in order to preserve and build on the deuteron experience. Briefly, these studies are as follows.
\begin{itemize}
  \item The beam may be slowly brought to thick ($\sim$\,$\SI{2}{cm}$) target blocks, to scatter particles most efficiently into the polarimeter detectors. Favouring elastic scattering events yields the best polarimeter performance\cite{Brantjes:2012zz}.

  \item After suitable calibration, a comparison of the left--right asymmetries for oppositely polarized beam bunches may be used to reduce the polarization systematic error to less than one part per million\cite{Brantjes:2012zz}.

  \item Time marking polarimeter events\cite{Bagdasarian:2014ega} leads to an unfolding of the precession of the in-plane polarization and  measurement of the spin tune $\nu_\text{s} = G\gamma$, denoting the number of spin precessions per revolution, to one part in \num{e10} in a single cycle of \SI{100}{s} length\cite{Eversmann:2015jnk}. The polarimeter signals permit feedback stabilization of the phase\cite{Hempelmann:2017zgg} of the in-plane precession to better than one part per billion (\num{e9}) over the time of the machine store. This is necessary to maintain frozen spin.

  \item By using bunched beams, electron cooling, and trimming  the ring fields to sextupole order, the polarization decoherence with time may be reduced, yielding a lifetime in excess of \SI{1000}{s}\cite{Guidoboni:2016bdn}. Observation of the spin tune variations allows for the measurement of the direction of the invariant polarization axis with a precision of about \SI{1}{mrad}
\cite{Saleev:2017ecu}.
\end{itemize}

With deuterons in the COSY ring at \SI{970}{MeV/$c$} with non-frozen spin, the polarization precesses in the horizontal plane at \SI{121}{kHz} relative to the velocity. The EDM is associated with a tilt of the invariant spin axis away from the vertical direction. This tilt of the spin rotation axis generates an oscillation of the vertical spin component. This effect is too small to measure with  reasonable sensitivity. However, using an RF Wien filter in the ring, with fields oscillating synchronously with the spin precession in the horizontal plane, a vertical spin component builds up over the entire duration of the fill\cite{Rathmann:2013rqa,PhysRevSTAB.16.114001}. This has become the basis for the precursor experiment (see stage 1 in \Fref{fig:exs:stages}). A first precursor run at COSY showed that EDM-like signals are also caused by systematic disturbances of the deuteron spin while the particles circulate in the ring; these may be machine errors or longitudinal magnetic fields. These effects also lead to a tilting of the invariant spin axis. Signals arising from a tilt of the Wien filter axis and from longitudinal magnetic fields in the ring may actually be used to determine the orientation of the invariant spin axis at the Wien filter (see Chapter\,\ref{Chap:Precursor}).

In parallel, studies and preparations for the proton EDM measurement in a fully electric frozen-spin ring are ongoing.  Some of the key technologies are currently under development for the final ring. These include the following:
\begin{itemize}
  \item The electrostatic deflector design  requires full scale prototypes with beams to be tested to levels of at least \SI{8}{MV/m}.

  \item Beam position monitors are needed to operate at a precision of at least \SI{10}{nm} for a measurement time of \SI{1000}{s}. SQUID-based position pick-ups may be used to measure the small relative difference of the orbits of the CW and CCW circulating beams, as  discussed in \Sref{sec:tests-with-SQUID} of Appendix\,\ref{Chap:MagneticFields}.

  \item The ring must be shielded to provide isolation from systematic radial magnetic fields to the nanotesla level\cite{Anas}.
\end{itemize}

Spin tracking calculations are necessary to verify the level of precision needed in the ring construction and the handling of systematic errors. For a detailed study during beam storage and build-up of the EDM signal, one needs to track a large sample of particles for billions of turns. The COSY-Infinity\cite{BERZ1990473} and Bmad\cite{Sagan:Bmad2006,Sagan:bmad:refmanual} simulation programs are utilized for this purpose. Given the complexity of the tasks, particle and spin tracking programs  have been benchmarked and simulation results compared with beam and spin experiments at COSY to ensure the required accuracy of the results.

Finally, a strategy will be needed to verify any signal produced by the experiment after the CW--CCW subtraction, through a series of critical tests and independent analyses.

When constructed, the proton EDM experiment will be the largest electrostatic ring ever built. It will have unique features, such as counter-rotating beams and strenuous alignment and stability requirements. It may also require stochastic cooling and weak vertical focusing  by electric fields in order not to lose the beam quickly, consistent with dual-beam operation\footnote{Vertical focusing should be as weak as possible  to enhance the sensitivity of the measurement of the separation of the CW and CCW beam orbits, owing to radial magnetic fields.}.  Intense discussions within the CPEDM collaboration have led to the conclusion that the final ring cannot be designed and built in one step; instead, a smaller-scale prototype ring (PTR, see stage 2 of  \Fref{fig:exs:stages}) must be constructed to confirm and refine the critical features, as follows:
\begin{itemize}
  \item The ring stores high beam intensities for a sufficiently long time.


  \item The ring must circulate CW and CCW beams simultaneously, both horizontally polarized.

  \item The ring must support frozen spin using additional vertical magnetic field to bend the beam, albeit not at the `magic energy'.


  \item Polarimeter measurements must be made for both CW and CCW beams using the same target. Thus, a second polarimeter detector is needed.

\item Beam cooling (electron cooling before injection, or stochastic cooling) is required to reduce the beam phase space.
\end{itemize}

\section*{CPEDM strategy}

As emphasized, this challenging project needs to proceed in stages,  as outlined in \Fref{fig:exs:stages}.
\begin{enumerate}
     \item COSY will continue to be used for as long as possible for the continuation of critical R\&D associated with the final experiment design. An important requirement is to test as many of the results as possible with protons, where the larger anomalous magnetic moment leads to more rapid spin manipulation speeds.

     \item The precursor experiment will be completed and analysed. Some data will be collected using an improved version of the Wien filter, with better electric and magnetic field matching.

     \item The next stage is to design, fund, and build a prototype ring (discussed in detail in Chapter\,\ref{chap:ptr}) to address critical questions concerning the features of the EDM ring design. At 30\UMeV, the ring with only an electric field can store counter-rotating beams, but without frozen spins. At 45\UMeV, with an additional magnetic field, the frozen spin condition can be met. However, the magnetic fields also prevent the CW and CCW beams from being stored at the same time. Even so, an EDM experiment may be conducted with these two beams used in alternating fills.

     \item Following step 3, the focus will be to create the final ring design, then fund and construct it.

     \item Once the ring is ready, the longer-term activity will be to commission and operate the final ring, improving it with new versions as the systematic errors and other experimental issues are understood and improved.
\end{enumerate}

Future scientific goals may include conversion of the ring to crossed electric and magnetic field operation so that other species besides the proton could be examined for the presence of an EDM. The data may be analysed for signs of axions using a frequency decomposition and investigation of counter-rotating beams, with different species used in novel EDM comparisons.

Stages 2 and 3 of the CPEDM, \ie the prototype ring and the final electrostatic  ring, are considered host-independent. If the prototype is built at J\"ulich, it would take advantage of the existing facility for the production of polarized proton (and deuteron) beams. It may also be built at another site (\eg CERN) provided that a comparable beam preparation infrastructure is made available. In either case, the lattice design will mimic that of the high-precision ring, in order to test as many features as possible on a smaller scale.

\section*{Details of the prototype EDM ring}

The prototype ring (PTR) will be small, with a circumference of about \SI{100}{m,} and operate in two modes (see stage 2 in \Fref{fig:exs:stages} and \Tref{tab:ptr:parameters}). The ring will be as inexpensive as possible, consistent with being capable of achieving its goals. The first mode would work with fully electric deflection of the beam (at \SI{30}{MeV}) and should show that such a concept works and can, among other things, demonstrate the feasibility of the ring with simultaneous counter-rotating beams. The second mode would extend the operating range to \SI{45}{MeV} with the addition of magnetic bending using air core magnets. With this combination, frozen spins could be demonstrated for a proton beam at 30 and \SI{45}{MeV}, other spin manipulation tools could be developed, and a reduced-precision proton EDM value could be measured. Alternating fills in counter-rotating directions would allow cancellation of the average radial magnetic field $\langle B_r\rangle$  that is the leading cause of systematic error (though at the expense of a large systematic error associated with the required magnetic field reversal).

\begin{table}
\centering
\caption{Basic beam parameters for the prototype ring (PTR); further details are presented in Chapter\,\ref{chap:ptr}}
\label{tab:ptr:parameters}
\begin{tabular}{l llll}
\hline \hline
                                  &    $E$ only  & \multicolumn{2}{l}{$E$ \& $B$,} &   Unit \\
                                  &              & \multicolumn{2}{l}{frozen spin} & \\  \hline
Bending radius                    &   8.86       & {8.86} & {8.86} & m \\
Kinetic energy                    &      30      &   30    &     45      &   MeV        \\ 
$\beta =v/c$                      &    0.247     &  0.247  &    0.299    &              \\
$\gamma$ (kinetic)                &    1.032     &  1.032  &    1.048    &              \\
Momentum                          &     239      &   239   &      294    &   MeV/$c$      \\ 
Electric field $E$      &    6.67 &   4.56  &      7.00   &   MV/m        \\
Magnetic field $B$   &            &  0.0285 &    0.0327   &    T          \\
\hline \hline
\end{tabular}
\end{table}
This section describes a starting-point lattice in terms of geometry, type and strength of the elements. The ring is square with 8\Um\ long straight sections. The basic beam parameters are given in Table\,\ref{tab:ptr:parameters}.

\subsection*{Prototype ring requirements and goals}

The foremost goal of the prototype ring is to demonstrate the ability to store enough protons (\num{\sim e10}) to be able to perform proton EDM measurements in an electric storage ring. Since ultimate EDM precision will require simultaneously countercirculating beams, the PTR has to demonstrate the ability to store and control  two such beams simultaneously.

Cost-saving measures in the prototype, such as room-temperature operation, minimal magnetic shielding, and the avoidance of excessively tight manufacturing and field-shape matching tolerances, are expected to limit the precision of any prototype ring EDM measurement. Nonetheless, data for reliable cost estimation and extrapolation of the systematic error evaluation to the full scale ring must be obtained.

\subsection*{Prototype ring  design}

The lattice has fourfold symmetry, as shown in \Fref{fig:exs:ptr}. The basic parameters for the prototype ring are given in Tables\,\ref{tbl:ptr:geom} and \ref{tbl:ptr:elements}. The bending, for example for \SI{45}{MeV} protons, is achieved using eight \SI{45}{\degree} electric or magnetic bending elements. The acceptance of the ring is to be $\SI{10}{\pi. \milli \meter . \milli \radian}$ for \num{e10} particles. The lattice is designed to allow variable tuning, between 1.0 and 2.0 in the radial plane and between 1.6 and 0.1 in the vertical
plane.

\begin{figure}
\centering
\includegraphics[scale=0.48
,viewport=100 350 600 775,clip]{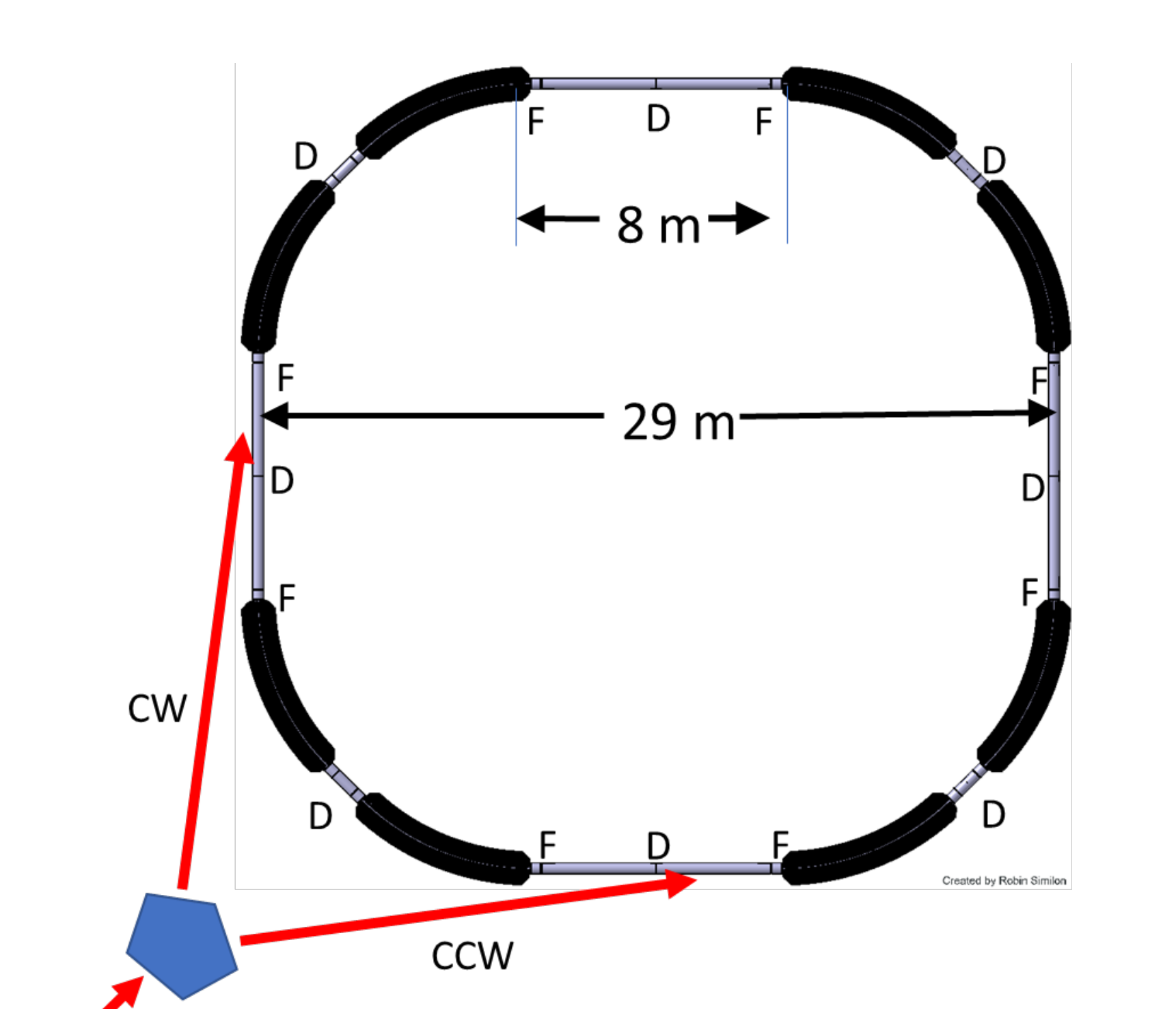}
\caption{\label{fig:exs:ptr} Basic layout of the prototype ring, consisting of eight dual superimposed electric and magnetic bends, three families of quadrupoles (focusing, defocusing, and straight-section), and four \SI{8}{m} long straight sections. The total circumference is about \SI{100}{m}. Injection lines for injecting countercirculating beams are represented  as stubs. Costs given in \Table\,\ref{tbl:costs} are restricted to just the prototype ring, which is truly site-independent. The possibly greater infrastructure costs associated with producing appropriately polarized beams are neither given nor site-independent.}
\end{figure}

\begin{table}
\begin{minipage}[t]{.5\textwidth}
\centering
    \caption{\label{tbl:ptr:geom}Geometry of the prototype ring}
    \begin{tabular}[t]{l l l}
    \hline \hline
    & & Unit \\
    \hline
    No of $B$--$E$ deflectors                   & 8       &     \\
    No of arc D quads                      & 4       &     \\
    No of arc F quads                      & 8       &     \\
   Quad length                         & 0.400   &  m  \\
    Straight length                     & 8.000   &  m  \\
    Bending radius                      & 8.861   &  m  \\
    Electric plate length               & 6.959   &  m  \\
    Arc length ($45^{\circ})$               & 15.7    &  m  \\
    Circumference total                     & 102     &  m  \\ 
    r.m.s. emittance $\epsilon_x =\epsilon_y$  & 1.0     & $\si{\pi . \milli \meter . \milli \radian}$ \\
    Acceptance $a_x=a_y$                    & 10.0    & $\si{\pi . \milli \meter . \milli \radian}$ \\
    \hline \hline
    \end{tabular}
\end{minipage}
\begin{minipage}[t]{.4\textwidth}
\centering
    \caption{\label{tbl:ptr:elements}Bending elements at \SI{45}{MeV}}
    \begin{tabular}[t]{l l l}
    \hline \hline
    & & Unit \\ \hline
      {Electric}              &              &          \\
    \hline  
                Electric field            & 7.00         & MV/m     \\
    Gap between plates        & 60           & mm       \\
    Plate length              & 6.959        & m        \\
    Total bending length        & 55.673       & m        \\
    Total straight length    & 44.800       & m        \\
    Bend angle per unit       & ($45^{\circ}$)  &  m       \\
        \hline
      {Magnetic}                &              &          \\
    \hline 
       Magnetic field            & 0.0327      &  T       \\
    Current density           & 5.000        & A/mm$^2$ \\
    Windings/element          & 60           &          \\
    \hline \hline
    \end{tabular}
\end{minipage}
\end{table}

\subsubsection*{The injector} Injection into the prototype ring will closely resemble injection into a nominal all-electric ring. In particular, there will be an even number of bunches in each beam, with alternating sign polarizations, whether in single beam or countercirculating beam operation. The beams injected will be protons in the 30--\SI{45}{MeV} range, in a cooled phase space of $\SI{1}{\pi.\milli\meter.\milli\radian}$, with the beams bunched into two bunches, to be fed into countercirculating beams in the prototype ring.

Injection into the prototype ring will be achieved using switching magnets, distributing the beams into clockwise (CW) or counterclockwise (CCW) directions, as sketched in Fig.\,\ref{fig:exs:ptr}. All beam bunches are transferred with vertical polarization, either up or down.

\subsubsection*{Electric bends}
 Each electrostatic deflector consists of two cylindrical parallel metal plates with equal potential and opposite sign. With the zero voltage contour of electric potential defined to be the centre line of the deflector, the ideal orbit of the design particle stays on the centre line. The electrical  potential vanishes on the centre line of the bends, as well as in drift sections well outside the bends. Therefore,  the electric potential vanishes everywhere on the ideal particle orbit. With the electric potential seen by the ideal particle continuous at the entrance and exit of the deflector, its total momentum is constant everywhere (even through the RF  cavity).

The designed ring lattice requires electric gradients in the range 5--\SI{10}{MV/m}. This exceeds the standard values for most accelerator deflectors, which
are separated by a few centimetres. Assuming a distance of \SI{60}{mm}  between the plates, to achieve such high electric fields, high-voltage power supplies must be employed. At present, two fully equipped \SI{200}{kV} power converters are available for testing deflector prototypes at IKP. The field emission, field breakdown, dark current, electrode surface, and conditioning will be studied using two flat electrostatic deflector plates, mounted on the movable support with the possibility of changing the separation from 20 to \SI{120}{mm}. The residual ripple of the power converters is expected to be of the order of $\pm \num{e-5}$ at a maximum of \SI{200}{kV}. A smaller ripple or stability control of the system may be required to avoid beam emittance growth and particle loss; this may constitute a dedicated task for investigations planned at the prototype ring.

\subsubsection*{Magnetic bends}
 The experiments require periodic reversals of the magnetic bending field to use symmetry to suppress systematic deviations. The magnetic field should be reversed with the best possible reproducibility. This is why the magnetic fields will be generated only by current windings, avoiding  magnetic yokes
altogether, as indicated in \Fref{fig:exs:coils}.

\begin{figure}
\centering
\includegraphics[scale=0.25]{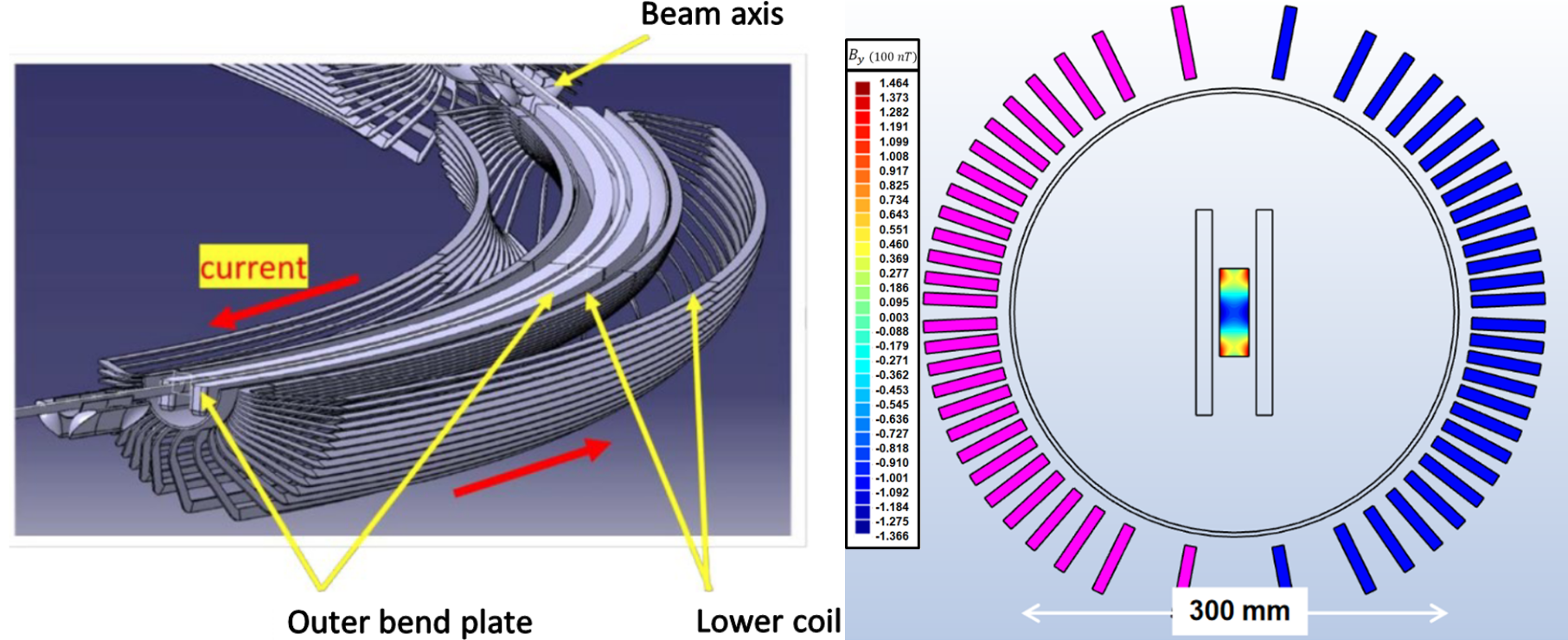}
\caption{Left: Cutaway  of  prototype ring in the $\vec{E}\times\vec{B}$ version. A side view of the lower half of a $45^{\circ}$
bend element is shown. The electrodes have a gap of 60\Umm. The magnetic coil conductors (single, $4\Umm\times 4\Umm$ copper bars) produce a highly uniform `cosine-theta' dipole field.  Right: Transverse
section, showing end view of  (inner legs of) magnet coil, as well as a field map of the good magnetic field region.
\label{fig:exs:coils}}
\end{figure}

\subsubsection*{Other components}
 All quadrupoles will be electrostatic. Their design will follow the principles of the Heidelberg CSR ring\cite{vonHahn:2016wjl}. Both DC and AC Wien filters and solenoids will be required for spin control. The RF cavity design is under study.

The requirement for the vacuum is mainly given by the minimum beam lifetime requirement of about \SI{1000}{s}. The beam emittance growth in the ring caused by multiple scattering from the residual gas is $\SI{0.005}{\pi.mm.mrad/s}$. At a ring vacuum of about \SI{e-12}{mbar}, the beam emittance right after filling is assumed to be $\SI{1}{\pi.mm.mrad}$.  Assuming a nitrogen (N$_2$) partial pressure, after \SI{1000}{s}, the emittance will have increased to $\SI{5}{\pi.mm.mrad}$. This is approximately the cooling rate expected for stochastic cooling. (One notes in passing that stochastic cooling becomes impractical for very low tunes.) For such an ultrahigh vacuum, only cryogenic or non-evaporable getter (NEG) pumping systems can be used. Bake-out must be foreseen for either cryogenic or NEG systems.

The choice of NEG requires a beam pipe with a diameter of \SI{300}{mm} over the full circumference of \SI{100}{m}. This can easily be plated with the NEG material, providing an active area of $\approx$\,$\SI{120}{\square \meter}$ for the whole ring. The required pumping speed will be about $\SI{5000}{\ell\per\second}$ per metre of length of vacuum pipe.

Beam position monitors (BPMs) are located around the ring. A BPM is placed at the entrance and the exit of each bending unit. Additionally, one BPM will be placed  in close connection to the quadrupoles in the straight sections. A new type of BPM, of Rogowski coil design\cite{Trinkel:2017jtf},  has been developed at the IKP of  FZJ. These pick-ups are in a development stage at present. The position resolution is measured to be \SI{10}{\micro\meter} over an area with a diameter of about \SI{90}{mm}. These BPMs require only a short beam insertion length of \SI{60}{mm} and an offset-bias free response to countercirculating beams.

\section*{All-electric storage ring}

This document describes the vision of the CPEDM collaboration, culminating in the design, construction, and operation of a dedicated, high-precision storage ring for protons. Operating at an all-electric, frozen-spin momentum of \SI{0.7}{GeV/$c$}, the signals from counter-rotating beams will be used to measure the proton electric dipole moment with a sensitivity of \SI{e-29}{$e$.cm}.  The major challenge is the handling of all systematic errors to obtain an overall sensitivity of a similar size. The main source of systematic uncertainties will be due to any unknown or unidentified radial magnetic field acting through the much larger magnetic dipole moment and leading to a false EDM signal. The level at which this can be mitigated remains to be determined.

Invaluable results and experiences are expected from the intermediate step,
 the construction of a smaller prototype ring. The attempts to examine the control of counter-rotating beams and study the conditions for frozen spin directly will have a huge impact on the detailed outline of the high-precision ring design.

The concept of an all-electric storage ring with extremely well-fabricated and aligned elements running two longitudinally polarized proton beams in opposite directions in the absence of significant magnetic fields serves as the current starting point. New ideas under development  offer the prospect of further mitigation of the systematic issues.
\begin{itemize}
\item A hybrid electric--magnetic ring, as discussed in Ref.~\cite{Haciomeroglu:2018nre} and in Appendix\,\ref{Chap:hybrid},  with magnetic focusing (in addition to electric deflector contributions) will change the electromagnetic environment in significant ways. The quadrupoles must be aligned very precisely to limit spin rotations resulting from magnetic fields deviating from the perfect machine. Thus, beam-based quadrupole alignment techniques must be applied, as described, \eg in Ref.\cite{Wagner:2020akw}. This substantially relaxes the requirement that radial magnetic fields be made to nearly vanish. The magnetic focusing, however, does not produce
counter-rotating beams with the same phase-space profile. Therefore, a periodic reversal of the magnetic focusing would be required to provide a set of signals that must be averaged to obtain an EDM value.
\item It is possible to find pairs of unlike polarized beams for which the same superimposed electric and magnetic bending yields a frozen spin condition for both (\eg protons and \Isotope[3]He)\cite{Talman:2018dgp}. Since the two beams would not have the same revolution frequency, to circulate simultaneously they would run with appropriately different RF harmonic numbers. Though not yielding either EDM value directly, the resulting EDM difference would be independent of the (otherwise dominant) radial magnetic field systematic error. Any EDM signal differences would be interpreted as the presence of an EDM on at least one of the two beams.
\end{itemize}
Work on these concepts can proceed using the prototype ring, with the possibility of yielding new physics results.

\section*{Prototype ring costs}
Preliminary prototype ring cost estimates are given in Table~\ref{tbl:costs}. Many items are currently receiving R\&D funding. The high-voltage supplies
for the  bend elements  are presently under development. Neither building nor injection line costs are included.  The accuracy of this cost estimation is preliminary. The magnetic bend equipment for the frozen spin experiments in a second stage will require additional costs for the magnets and a Wien filter of about 7 million euros.

\begin{table}
\caption{\label{tbl:costs} Preliminary cost estimates for the prototype ring first stage}
\centering
\begin{tabular}{ll}
\hline \hline
Component           & Cost      \\
& (k\texteuro) \\
\hline
Bends               & \phantom{1}9200      \\
Electric quads      & \phantom{1}1700     \\
Vacuum              & \phantom{1}1800    \\
Pick-ups            & \phantom{16}900    \\
Control             & \phantom{1}1500     \\
Polarimeter         & \phantom{1}1200     \\
RF equipment        &  \phantom{16}300      \\
\hline
Total         &  16600   \\
\hline \hline
\end{tabular}
\end{table}

\section*{Roadmap and timeline}

A staged approach to the CPEDM project (outlined in \Fref{fig:exs:stages} and expanded in detail in \Fref{fig:exs:timeline}) is currently ongoing, with work on the precursor experiment and feasibility studies. This is partially funded by an ERC Advanced Grant that runs until September 2021 (event 3). Since 2017, preparation has been under way on the design for a prototype electric and mixed field storage ring to verify CPEDM concepts that will appear in a conceptual or technical
design report, available for funding consideration by the middle of 2021. With approval, construction and commissioning of the prototype ring will begin. In parallel, experimental work at COSY would refocus on feasibility studies for proton beams. By the beginning of the subsequent  funding period, the first prototype results should show the best techniques for the all-electric full-energy ring. These will be the subject of another conceptual or technical
design report. If approved, efforts will switch to the construction and running of the new ring.

\begin{figure}
\centering
\includegraphics[width = \textwidth]{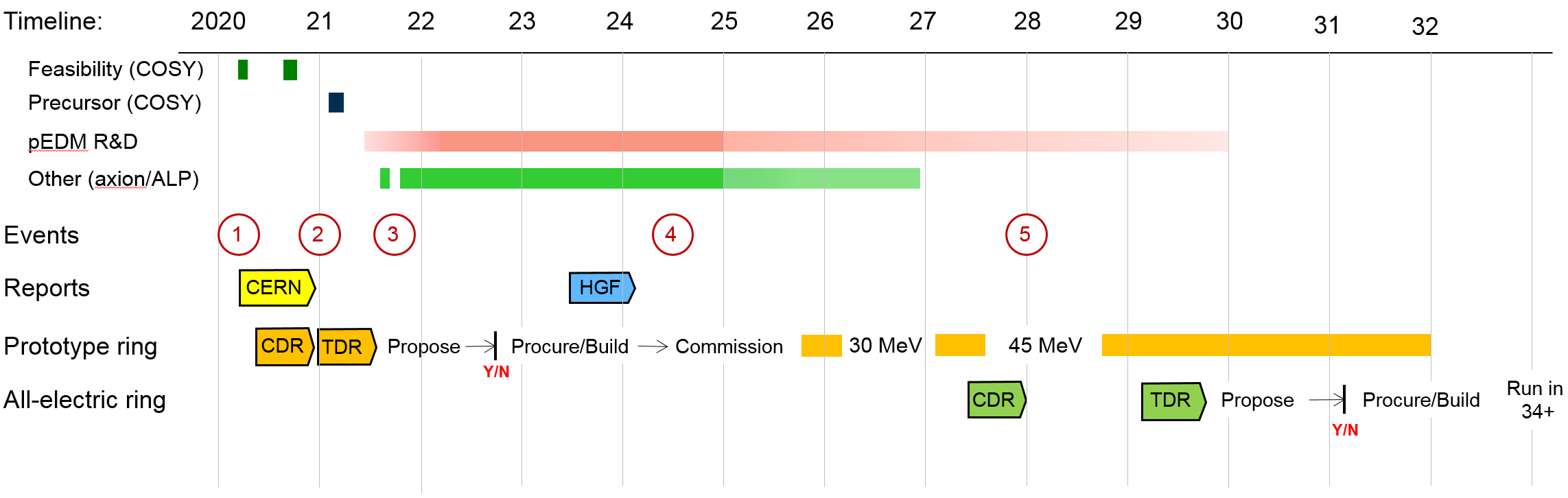}
\caption{\label{fig:exs:timeline}Timeline for  anticipated evolution of the  EDM project  storage ring. Event key: \textcolor{purple}{\protect\circled{1}} strategic programme evaluation by Helmholtz Association (HGF); \textcolor{purple}{\protect\circled{2}} start of HGF funding period; \textcolor{purple}{\protect\circled{3}} end of srEDM advanced grant from European Research Council (2016--2021)\cite{ERC-694340}; \textcolor{purple}{\protect\circled{4}} HGF midterm review; \textcolor{purple}{\protect\circled{5}} start of next HGF funding period.}
\end{figure}

The storage ring EDM feasibility studies made so far show encouraging results. Handling systematic errors is the main challenge. The path to addressing this lies in the construction and operation of a small-scale prototype ring,  which will lead to the design of a high-precision ring with the best sensitivity to new physics.



\begin{flushleft}

\end{flushleft}
\end{cbunit}

\begin{cbunit}

\csname @openrighttrue\endcsname 
\chapter{Introduction}
\label{Chap:Introduction}

\section{Project scope}

An experiment is described to detect a permanent electric dipole moment (EDM) of the proton with a sensitivity of \SI{e-29}{$e$.cm} by using counter-rotating polarized proton beams at the `magic' momentum of \SI{0.701}{GeV/$c$} in an all-electric precision storage ring.

The scientific case for such a project is based on the fact that measurements of EDMs of fundamental particles provide ``a unique, extraordinarily sensitive way to probe for a physical phenomenon of profound significance, [the] violation of microscopic time-reversal invariance'' (F. Wilczek). Assuming $CPT$-symmetry conservation, $T$-violation implies violation of the combined $CP$-symmetry, one of the ingredients required for explaining the matter--antimatter asymmetry of our Universe. Electric dipole moment searches are sensitive to physics beyond the Standard Model of elementary particle physics at a scale of the order of \SI{1000}{TeV}. Moreover, the storage ring technology will also allow a search for oscillating EDMs, which may be connected with axions or axion-like particles. The physics motivation is thus evident and well supported by the community.

The storage ring concept has been well developed over the years, with a detailed examination of the experimental method, required technologies, and involved systematics. Research and development have progressed in parallel on essential storage ring components, such as electrostatic deflectors, beam instrumentation, magnetic shielding, and polarimetry.

A good understanding of the key systematic errors has been achieved, and their potential constraints on the ultimate sensitivity of the storage ring approach have been quantified. The leading systematic uncertainty is due to a residual radial magnetic field interacting with the magnetic moment to mimic the EDM signal. A radial magnetic field will also lead to a vertical separation of the counter-rotating beams. Therefore, measurement of this separation may provide a handle to mitigate this systematic error.

The ultimate goal is to design, build, and operate an all-electric storage ring for protons at their magic momentum with clockwise (CW) and counterclockwise (CCW) longitudinally polarized beams to achieve a sensitivity of the order of \SI{e-29}{$e$.cm}. To this end, a number of ring lattice options have been developed. These options make reasonable assumptions about the achievable electric field, deflector size, instrumentation requirements, \etc, and have led to the adoption of a baseline ring design of some \SI{500}{m} in circumference.

To fully confirm the validity of the approach, a small all-electric prototype ring (PTR, see Chapter~\ref{chap:ptr}) is proposed. This would allow us:
\begin{itemize}
\item to deploy and test key hardware components of the all-electric ring;
\item to verify that an intense proton beam can be stored for at least \SI{1000}{s};
\item to deploy and use beam instrumentation, such as polarimeters;
\item to demonstrate the ability to master key systematics via the use of counterclockwise beams.
\end{itemize}
The prototype is seen as a key step in demonstrating the credibility of the full ring proposal. A baseline proposal for this prototype has been developed and foresees two phases (see Chapter\,\ref{chap:ptr}):
\begin{description}
    \item[\textnormal{phase 1:}] all-electric ring for \SI{30}{MeV} proton beams (CW and CCW);
    \item[\textnormal{phase 2:}] combined $E$ and $B$ fields for 30 and \SI{45}{MeV} proton beams (frozen spin) to allow for a first pEDM measurement.
\end{description}

It is expected that the PTR will provide invaluable information for the outline and design of the final ring.

\section{Key accomplishments}

Significant insight into the storage ring EDM physics and accompanying technological advances have been achieved in the past few years. These include early findings of the srEDM collaboration at BNL (USA) and contributions from KAIST (South Korea). Recently, a new level of measurement technology achievements has been reached by the JEDI collaboration working at  COSY, at the Forschungszentrum J\"ulich (Germany). These achievements are as follows:
\begin{enumerate}
    \item The production of an in-plane deuteron beam polarization lifetime near \SI{1000}{s} in the horizontal plane of the magnetic storage ring COSY has been achieved\cite{Guidoboni:2016bdn}. This long `spin coherence time' (SCT) was obtained through a combination of beam bunching, electron cooling, sextupole field corrections, and the suppression of collective effects through beam current limits. This record lifetime is required for a storage ring search for an intrinsic electric dipole moment of the deuteron and paves the way for a similarly large SCT for protons.

    \item A new method to determine the spin tune was established and tested\cite{Eversmann:2015jnk}. In an ideal planar magnetic storage ring, the `spin tune'---defined as the number of spin precessions per turn---is given by $\nu_\text{s} =\gamma G$ (where $\gamma$ is the Lorentz factor and $G$ the gyromagnetic anomaly). At \SI{0.97}{GeV/$c$}, the deuteron spins coherently precess at a frequency of about \SI{121}{kHz} in COSY. The spin tune was deduced from the up--down asymmetry of deuteron--carbon scattering. In a time interval of \SI{2.6}{s}, the spin tune was determined with a precision of the order of \num{e-8}, and to \num{e-10} for a continuous \SI{100}{s} accelerator cycle. This renders the new method a precision tool for accelerator physics; observing and controlling the spin motion of particles to high precision is again mandatory for the measurement of electric dipole moments of charged particles in a storage ring.

    \item Feedback from a spin polarization measurement of the revolution frequency of a \SI{0.97}{GeV/$c$} bunched and polarized deuteron beam in COSY has successfully been used to control both the precession rate ($\approx$\,$\SI{121}{kHz}$) and the phase of the horizontal polarization component\cite{Hempelmann:2017zgg}. Real-time synchronization with a radio-frequency (RF) solenoid made possible the rotation of the polarization out of the horizontal plane, yielding a demonstration of the feedback method to manipulate the polarization. In particular, the rotation rate shows a sinusoidal function of the horizontal polarization phase (relative to the RF solenoid), controlled with an error of $\sigma = \SI{0.21}{\radian}$. The minimum possible adjustment was \SI{3.7}{mHz,} for a revolution frequency of \SI{751}{kHz}, which changes the precession rate by \SI{26}{mrad/s}. Such a capability meets the requirement for the use of storage rings to look for an intrinsic electric dipole moment of charged particles.

    \item Procedures have been developed and tested that allow for systematic errors in the measurement of the vertical polarization component (that carries the EDM signal) to be corrected to a level below one part in \num{e5}\cite{Brantjes:2012zz}. This requires a prior calibration of the polarimeter for rate and geometric error effects and the use of two opposite polarization states in the measurement. The extra polarization state allows for an independent estimate of the size of the systematic error. Such corrections may be made in real time. This meets the sensitivity requirement for measuring the small vertical component of the polarization expected in the EDM search.
\end{enumerate}

An overview of the contents of this report is given  in Section\,\ref{sec:contents-of-report-by-chapter}.

\section{European and global context}

Permanent electric dipole moments are sought in various elementary and complex systems; the most recent experimental limits are given in  Chapter\,\ref{Chap:physics-case}. A  list of the international EDM efforts can be found in Ref.~\cite{edms-worldwide}.

Neutron EDM searches are being conducted or under development, or have been
proposed, at nuclear fission reactor facilities (ILL Grenoble, FRM-2 Munich, PNPI Gatchina) and spallation neutron sources (PSI Villigen, ESS Lund) in Europe, as well as in the USA and Canada (SNS Oak Ridge, LANL Los Alamos, TRIUMF, Vancouver). Molecular and atomic EDMs are sought by numerous groups worldwide, including projects at radioactive beam facilities, such as ISOLDE (CERN); molecular ions are also used as probes\cite{Cairncross:2017fip}. The muon EDM is measured as a by-product of $(g-2)_{\mu}$ experiments at FNAL (Batavia, USA) and  J-PARC (Tokai, Japan).

A new experiment, called CeNTREX, has recently been launched at Yale University (USA) to search for a deformation in the shape (nuclear Schiff moment) of the $^{205}$Tl atomic nucleus  inside a thallium fluoride (TlF) molecule\cite{centrex}. The experiment will be complementary to experiments on  $^{199}$Hg and primarily sensitive to the proton EDM and $\theta_{\rm QCD}$. It is expected that it will improve the indirect proton EDM limit of \SI{2e-25}{$e$.cm} by more than one order of magnitude in the coming years.

Storage ring EDM searches for the proton and other light nuclei (deuteron, $^3$He) have been discussed for a few years (see Chapter\,\ref{Chap:Background}). It is our strong belief that a result for the pure proton system will  eventually
be required to complement the free neutron EDM, and to shed light on the $CP$-violating sources. This is currently pursued by the JEDI and CPEDM collaborations and constitutes the motivation for the present document.

\section{Contents of the report by chapter}
\label{sec:contents-of-report-by-chapter}
\begin{itemize}
    \item The \textit{executive summary} was prepared and submitted in December 2018 for the European Strategy for Particle Physics (ESPP) update, for consideration in their review of the storage ring EDM search, along with other experimental programmes in nuclear and high-energy physics research. The summary describes the concept of the experiment, the strategic path forwards,  including plans
for a prototype ring for further feasibility testing, and an outline of plans for the final EDM ring.

    \item[{\ref{Chap:physics-case}.}] The \textit{physics case} begins with a summary of the status of other major searches for an EDM on various systems and discusses both measured and derived upper bounds. These are compared with what one would expect for an intrinsic EDM on the basis of a naive dimensional analysis. Furthermore, the predictions of non-perturbative methods (mainly lattice QCD) relating the underlying $CP$-violating parameters to hadronic matrix elements are analysed, and the results for nuclear matrix elements in the case of light nuclei are reported. The option to search for axion-like dark-matter particles through the investigation of oscillating EDMs, where the frequency is given by the axion mass, is briefly reviewed.

    \item[{\ref{Chap:Background}.}] The \textit{background} chapter gives a summary of the history of the storage ring EDM search from the original ideas developed at the Brookhaven National Laboratory in the USA. The story is followed as the search moved on to feasibility testing at the COSY storage ring, located at the Forschungszentrum J\"ulich in Germany. This led to the first direct measurement of an EDM upper limit for the deuteron, using an RF Wien filter located inside the ring. The section also explores the experience of the collaboration members and work being done on supporting technologies for the final EDM ring.

    \item[{\ref{Chap:ExpMethod}.}] The \textit{experimental method} chapter describes the storage ring EDM search, beginning with the most basic concept of the experiment and developing all of the essential ideas needed for the experiment. Various experimental possibilities are presented and considered. Essential formalism is shown for both the ring and the polarization measurement. There is also a discussion of systematic effects and the challenge of managing all aspects of the experiment. This section is intended for the novice to EDM searches.

    \item[{\ref{Chap:Strategy}.}] The \textit{strategy} chapter briefly describes  the idea of building a prototype ring (PTR) featuring both electric and mixed field designs so that more of the critical technologies for the EDM search may be demonstrated. This will lead to a design for the final ring.

    \item[{\ref{Chap:Precursor}.}] The \textit{precursor experiment} at COSY adds an RF Wien filter to the existing COSY storage ring so that the cancellation of the EDM signal due to the in-plane rotation of the polarization is broken. In principle, this allows  for a sensitivity to determine the deuteron EDM. The absence of depolarizing resonances of stored deuteron beams simplifies these EDM feasibility studies considerably. A series of first measurements were completed in the autumn of 2018 and the preliminary results are presented here.

    \item[{\ref{chap:ptr}.}] The \textit{prototype ring (PTR)} chapter presents  detailed design considerations for the small ring to be built as a site for the continuation of feasibility studies for the EDM experiment in the final ring. The prototype will operate in two modes: (i) an all-electric set-up, which allows for the simultaneous storage of both clockwise and counterclockwise travelling beams, and (ii) a combined electric and magnetic ring, which creates the conditions for frozen-spin operation. In both cases, systematic errors may be studied in an environment that uses the \textit{same} electric field structures that are proposed for the final EDM ring.

    \item[{\ref{Chap:allelectricring}.}] The \textit{all-electric proton EDM ring} is described more fully in this chapter, based on the design for the lattice for the prototype ring. Ring specifications are provided. A table is included that shows the status and preparedness of various aspects of the projects.

    \item[{\ref{Chap:Efields}.}] The \textit{electric fields} chapter reviews the status of various accelerator systems that will be needed for the EDM ring. Electric fields pose a particular challenge, since the best experiment is associated with a large field strength. This then generates requirements on the voltage holding capability and the ability to suppress dark currents. Focusing and beam injection elements will also be needed, and the PTR offers a chance to develop and  study various designs.

    \item[{\ref{Chap:polarimetry}.}] The \textit{polarimetry} chapter begins with the items needed during the beam preparation phase of the experiment to verify the polarization of the beam. For the main EDM ring, these are the polarimeter target (most probably carbon) and the detectors needed for measuring online polarization  and cancelling the systematic errors associated with this process. The requirements of the polarimeter are explained and details are  given for the choice of detector acceptance, in order to maximize the figure of merit of the device. Examples are provided for both the prototype and the final EDM ring designs. This section also details the work accomplished so far  in fashioning the calorimeter detectors and other event-tracking hardware that will be needed. Details are provided on the use of the polarimeter as a device for maintaining the frozen-spin operating condition in the EDM ring.

    \item[{\ref{Chap:SensSys}.}]  The \textit{sensitivity and systematics} chapter  gathers the derivations for the contribution of event collection statistics to the final EDM result. Connections to critical ring and experimental requirements are made. Furthermore, the chapter describes in detail the considerations that lead to an expected statistical sensitivity reach of \SI{e-29}{$e$.cm} for charged particle EDMs in a storage ring. The main part of this section is devoted to an assessment of the size and nature of systematic effects that may mimic an EDM signal and means, such as simultaneous clockwise and counterclockwise measurements, that may be used to cancel these unwanted systematic effects.

    \item[{\ref{Chap:SpinTracking}.}] \textit{Spin tracking} consists of those calculations needed to describe the history of the polarized beam as it circulates in the EDM ring. The ring is also a testing laboratory, where we can explore various sources of systematic errors (\eg magnet misalignment) and ways to mitigate them. This calls for reliably calibrated programs using well-understood techniques for treating electric and magnetic field effects.

    \item[{\ref{Chap:RoadMap}.}] The last formal chapter of the report gives a \textit{roadmap and timeline}.
\end{itemize}

\section{Special appendices}

\begin{itemize}
       \item[{\ref{app:cosy}.}] The appendix of \textit{results and achievements at Forschungszentrum J\"ulich} covers  polarimetry,  high-precision tune measurements,  long horizontal polarization lifetimes,  feedback control of polarization,  invariant spin axis measurements,  RF Wien filter construction, reference databases for deuteron- and proton-induced reactions on carbon, progress in orbit measurement and control,  electrostatic deflector development, EDM and axion theory, and  spin tracking simulations. The appendix also includes a brief summary of the results and achievements from the Jülich--Bonn theory group.

       \item[{\ref{Chap:MagneticFields}.}] The appendix on \textit{mitigation of background magnetic fields}  describes the influence of magnetic fields on the EDM experiment and why they need to be small. Stray magnetic fields need to be managed with shielding and perhaps some active elements; in addition, the effects of any residual field need to be well understood. For injection, and perhaps for spin manipulation, time-varying magnetic fields may be needed, so their effects on the experiment need to be explored.

       \item[{\ref{app:stat-err-freq}.}] In this appendix, the  \textit{statistical error in a frequency measurement} from the time dependence of counting rates is discussed.

       \item[{\ref{app:gravity}.}] The appendix on  \textit{gravity and general relativity as a `standard candle'} contains the inclusion of gravity as an explicit item in the Thomas--BMT equation. From this, the level of the signal from gravity acting on the beam may be estimated, opening the possibility of using it as a marker of sensitivity.

       \item[{\ref{app:BeamPrep}.}] The appendix on  \textit{beam preparation} contains the design principles for the polarization patterns of the bunches.

       \item[{\ref{Chap:Axions}.}] The   \textit{axion search} appendix contains preliminary plans to use the rotating polarization of the COSY  beam to search for an oscillating EDM that is a possible signature of an axion-like particle in Nature. The first experiment to develop this techniques took data in April 2019.
\end{itemize}

\section{Appendices describing new ideas}

\begin{itemize}
       \item[{\ref{Chap:hybrid}.}] The  \textit{hybrid scheme} addresses the problems of minimizing the residual horizontal magnetic field in an all-electric storage ring by imposing a magnetic focusing system. This system, along with beam-based alignment techniques, draws the beam towards the point in each quadrupole where the field vanishes. This reduces the requirements on the elimination of the residual background field by orders of magnitude. This does break the symmetry between the CW and CCW rotating beams. Symmetry may be restored by operating with both focusing field polarities and averaging the results. Independent confirmation of this scheme is under way.


       \item[{\ref{Chap:spin-tune-mapping}.}] The  \textit{spin tune mapping for EDM searches} appendix explores a more generalized method of making EDM searches by replacing the requirement of `frozen spin' with corrections applied by an RF Wien filter mounted in the storage ring. This method may be generalized to allow for a comparison of different particle species.

       \item[{\ref{Chap:FDM:Full}.}] The   \textit{frequency domain}  appendix introduces the notion of utilizing `spin wheel' rotations of the polarization about the horizontal transverse axis to obtain sensitivity to the magnetic and electric dipole contributions together. By measuring the frequency of the resulting rotation, precise subtraction to obtain the EDM contribution becomes possible.

       \item[{\ref{app:fourier}.}] The  \textit{EDM from Fourier analysis} appendix explores the idea of separating EDM effects from systematics resulting
from machine errors using  Fourier analysis of the experimental signals.

       \item[{\ref{app:extpol}.}] The \textit{external polarimetry} appendix addresses the problem that a block target located at the edge of the beam does not necessarily sample the polarization profile across the full beam. This allows the effects of a polarization distribution across the beam to become a systematic error in the results. The scheme presented in this appendix uses pellets dropped through the beam to extract a fraction of the beam into a channel branching from the main beam line, where it strikes a large and thick polarimeter target that spans the entire beam profile. The efficiency for this scheme is expected to be comparable to the block target scheme used at COSY.

\end{itemize}

\section{Final comments}

We would like to emphasize that this write-up is a status report of what has been achieved and what is known at the time of the editorial deadline---work is ongoing at COSY, CERN, and  other places towards the realization of the storage ring EDM project.

\begin{flushleft}

\end{flushleft}


\end{cbunit}

\begin{cbunit}

\chapter{Physics case for CPEDM}
\label{Chap:physics-case}

\section{Introduction}
Both continuous and discrete symmetries,  combined with possible breaking patterns,
have been decisive in the development of physics in the last 100\,years.
This was, for example, demonstrated by the construction
of the Standard Model (SM) of particle physics.  Measurements of sizes or limits with which discrete
symmetries (such as, \eg parity ${P}$, charge conjugation ${C}$, their product ${CP}$, time-reversal invariance ${T}$, the product ${CPT}$,
baryon or lepton number) are, respectively, broken or conserved  have been essential for this task in the
second part of the last century. These tests
currently play, and  will continue to play, an essential role in constructing and identifying physics beyond the SM (BSM).

As is the case for all  stationary states of finite and
parity non-degenerate quantum systems,   the
ground state of any  of the known non-selfconjugate subatomic particles  with non-zero spin\footnote{For example, the
$\rho^0$ and $\omega$ vector mesons are particles with  non-zero spin but as they are self-conjugate, \ie their own antiparticles, they cannot possess an electric dipole moment, while the $\rho^\pm$ or the $K^{\ast}$,  as non-selfconjugate vector mesons, have this possibility.} (regardless of elementary or
composite nature)
can only support a non-zero permanent electric dipole moment (EDM),
if both time-reversal (${T}$) and parity (${P}$) symmetries
are violated explicitly, while
the charge symmetry (${C}$) can be maintained.  Assuming  conservation of
the combined ${CPT}$ symmetry, ${T}$ violation also implies ${CP}$ violation.

The  ${CP}$ violation generated by the Kobayashi--Maskawa  (KM) mechanism of weak interactions
induces a very small EDM that is several orders of magnitude
below current experimental limits. However, many models beyond the standard model
predict EDM values near these limits.
Hence, there is a   window  in which the search for non-zero
EDMs corresponds to  a search for  ${CP}$ violation beyond the weak interaction
$CP$ violation.
In fact, finding a non-zero EDM value
for any subatomic particle
(above the KM limit of the SM, which, experimentally, is out
of reach for the foreseeable future)
would be a signal that there exists
a new source of ${CP}$ violation, either  induced by the
strong ${CP}$ violation via the  $\bar\theta$ angle of quantum chromodynamics\footnote{Actually, the best upper limit on this parameter of quantum chromodynamics follows from the experimental bound on the EDM of the neutron (\Eref{nEDM-exp-bound}).} or by genuine physics beyond the SM.
The latter is essential for explaining---within the framework of the Big Bang and
inflation---the mystery of the observed
baryon--antibaryon asymmetry of our Universe,
one of the outstanding problems in contemporary elementary particle physics and cosmology. Note that $CP$ violation in combination with $C$ violation is one of the three Sakharov conditions\cite{Sakharov:1967dj}.

\subsection{Current experimental bounds} \label{Chap:physics-case:bounds}
Over the years, the quest to improve
the bounds of the permanent EDM of the neutron, $d_{\sf n}$, pioneered more than 60\,years ago
by the work of Smith, Purcell, and Ramsey \cite{Smith:1957ht}, has served to rule out or, at least severely constrain, many models of ${CP}$ violation, demonstrating
the power of sensitive null results.
The current bound of  the neutron EDM resulting from these efforts is
\begin{equation}
 |d_{\sf n}|  <
     1.8 \times \SI{e-26}{\text{$e$}.cm} \ \mbox{\small (90\% C.L.)  \cite{Baker:2006ts,Afach:2015sja,Abel:2020gbr}} \,,
 \label{nEDM-exp-bound}
\end{equation}
which corresponds to $|d_{\sf n}|  <  2.2 \times \SI{e-26}{\text{$e$}.cm}$ at  a 95\% confidence
upper limit. As reported next, the prediction of the CKM matrix is at least four orders of magnitude smaller: $|d_{\sf n}^{\rm SM}| \lesssim \SI{e-30}{\text{$e$}.cm}$ (see  \Sref{Chap:physics-case:naive} for more details).

There are complementary constraints from atomic and  molecular physics experiments.
In particular,
the EDM bounds on paramagnetic atoms, \eg
\begin{align*}
   \bigl |d_{{\sf Cs}}\bigr | &< 
   1.4 \times \SI{e-23}{\text{$e$}.cm} \ \ \mbox{\small (95\% C.L.) \cite{Murthy:1989zz}},\\
  \bigl |d_{^{205}{\sf Tl}}\bigr| &< 
        1.1 \times \SI{e-24}{\text{$e$}.cm} \ \ \mbox{\small (95\%\ C.L.)  \cite{Regan:2002ta,Chupp:2017rkp}},
\end{align*}
and the constraints from dipolar molecules and molecular ions  indirectly lead to the following
upper limits on the electron EDM:
\begin{align*}
 \left |d_{e}^{ \downarrow {\sf YbF}} \right | &< 
               1.1 \times \SI{e-27}{\text{$e$}.cm} \ \ \mbox{\small (90\% C.L.) \cite{Hudson:2011zz}},\\
    \left |d_{e}^{\downarrow {\sf ThO}} \right| &< 
       1.1 \times \SI{e-29}{\text{$e$}.cm} \ \ \mbox{\small (90\% C.L.)
       \cite{Baron:2013eja,Baron:2016obh,Andreev:2018ayy}},\\
\left |d_{e}^{\downarrow {\sf HfF}^+}\right| &< 
   1.3 \times \SI{e-28}{\text{$e$}.cm} \ \ \mbox{\small (90\% C.L.) \cite{Cairncross:2017fip}}.
\end{align*}
These bounds should be put into perspective, since they are quite large
compared with the prediction of the CKM  mechanism in  the SM:  $|d_{e}^{\rm SM}| \sim \SI{e-44}{\text{$e$}.cm}$,
see, \eg\ Ref.~\cite{Pospelov:2013sca}.

Note that the EDMs of paramagnetic atoms and the ${P}$- and ${T}$-violating
observables in polar molecules are dominated by system-dependent linear combinations of the electron EDM and the
nuclear spin-independent electron--nucleon interaction, which couples to the scalar  current components of  the
pertinent nuclei. An electron EDM value $d_{\sf e}$ cannot   be independently extracted from  the extraction of this semi-leptonic four-fermion
interaction $C_{\rm S}$, while the quoted bounds of $|d_{\sf e}|$   assume that the measured
paramagnetic systems are saturated by
the electron EDM alone. For more details on this issue, on  further EDM bounds, and also on the analogous extractions of  the $|d_{\sf n}|$ and $|d_{\sf p}|$  bounds of valence and core nucleons  for diamagnetic atoms see, \eg
Refs. \cite{Chupp:2017rkp,Jungmann:2013sga} and references quoted  therein.

By contrast with the paramagnetic cases, which are sensitive to their electron clouds,
in diamagnetic atoms, the EDM-defining spin is carried by the pertinent nucleus.
 Corresponding
upper limits on  the EDMs of diamagnetic atoms are, \eg
\begin{align*}
  \bigl |d_{^{129}{\sf Xe}} \bigr| &<  6.6 \times \SI{e-27}{\text{$e$}.cm}
   \ \ \mbox{\small (95\% C.L.) \cite{Rosenberry:2001}},\\
  \bigl |d_{^{225}{\sf Ra}} \bigr| &<  1.4 \times \SI{e-23}{\text{$e$}.cm}
   \ \ \mbox{\small (95\% C.L.) \cite{Bishof:2016uqx}},\\
    \bigl |d_{^{199}{\sf Hg}}\bigr | &<  7.4 \times \SI{e-30}{\text{$e$}.cm}
   \ \ \mbox{\small (95\% C.L.) \cite{Graner:2016ses}}.
\end{align*}

Because of Schiff screening,  the indirect bounds
on the  neutron and proton EDM  obtained by applying nuclear physics
methods~\cite{Dmitriev:2003sc}
 are much weaker than their parent atom bounds. Using the current best case, $^{199}{\sf Hg}$,
 the following {\em indirect} bounds on the  neutron and proton EDM were reported~\cite{Graner:2016ses}\footnote{Note that in Ref.~\cite{Sahoo:2016zvr} the values
 $ | d_{\sf n}^{ \downarrow ^{199}{\sf Hg}} | < 2.2 \times \SI{e-26}{\text{$e$}.cm}$
and $ | d_{\sf p}^{ \downarrow ^{199}{\sf Hg}} | < 2.1 \times \SI{e-25}{\text{$e$}.cm} $ were derived as the indirect bounds on the EDMs of neutron and proton, respectively---see also Ref.~\cite{Tanabashi:2018oca}.}:
\begin{align*}
   | d_{\sf n}^{ \downarrow ^{199}{\sf Hg}} | &<  1.6 \times \SI{e-26}{\text{$e$}.cm}
      \ \ \mbox{\small (95\% C.L.)},\\
   | d_{\sf p}^{ \downarrow ^{199}{\sf Hg}} | &<  2.0 \times \SI{e-25}{\text{$e$}.cm}
   \ \ \mbox{\small (95\% C.L.)}.
\end{align*}
The indirect bound on $|d_{\sf p}|$ is about an order of magnitude weaker than the indirect or direct
$|d_{\sf n}|$ counterparts and therefore not really competitive as a bound on
the nucleon EDM.

The current status of the already excluded EDM regions derived from the experimental upper limits  of the various particles mentioned here are summarized
in Fig.\,\ref{fig:physics-case:edm-bounds}.

\begin{figure}
\centering
\includegraphics[width=1\textwidth]{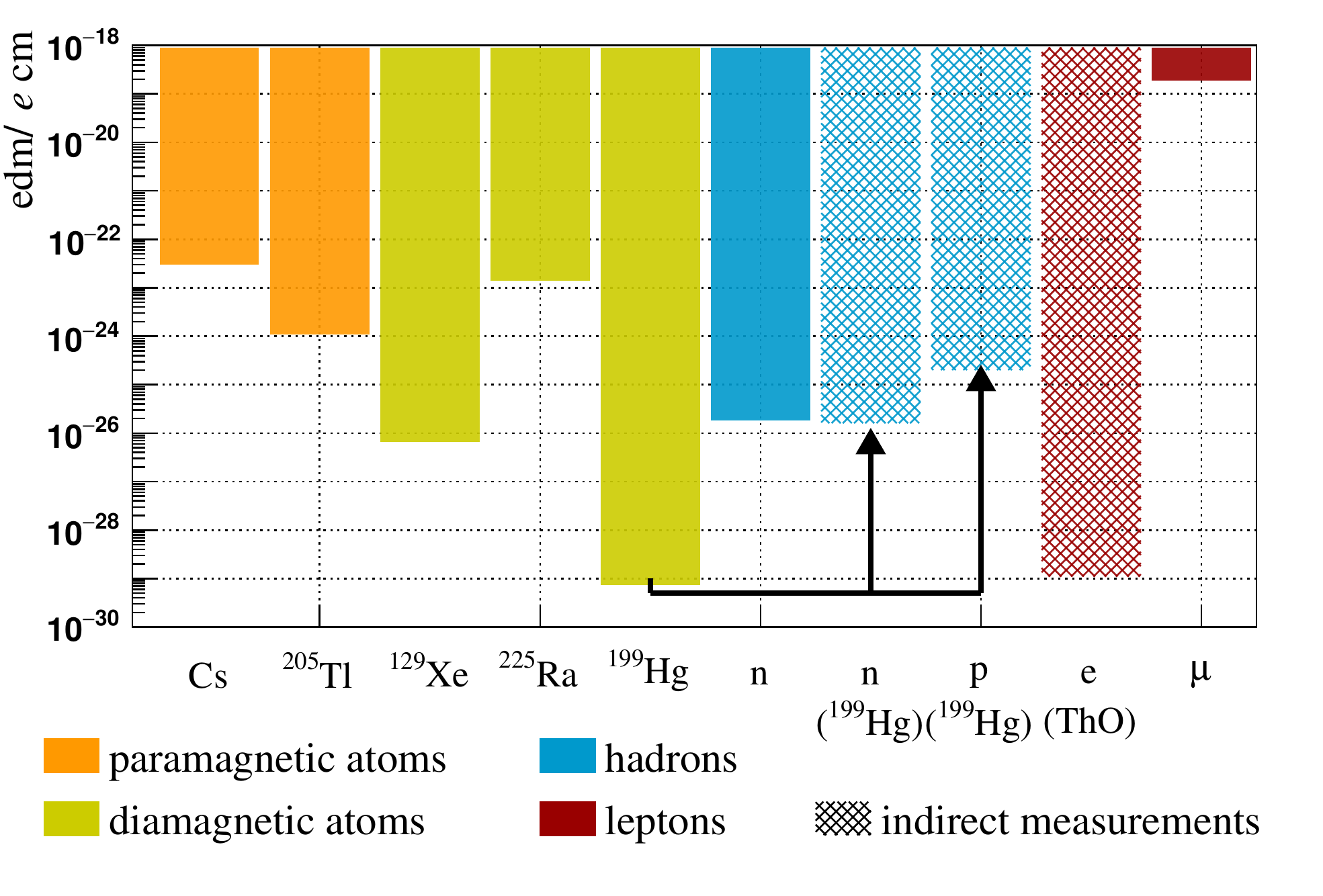}
\caption{Current status of excluded regions of  electric dipole moments. Shown are direct or derived EDM bounds of the particles and a selection of  atoms discussed
in \Sref{Chap:physics-case:bounds},  and, additionally,   the upper limit on the EDM of the muon,
$|d_\mu| < 1.8 \times \SI{e-19}{\text{$e$}.cm}$ (95\% C.L.), by the  muon $(g-2)$ collaboration \cite{Bennett:2008dy}.
\label{fig:physics-case:edm-bounds}}
\end{figure}

\subsection{Scientific potential of a proton EDM measurement}
In this proposal,
we discuss an experimental opportunity, provided by the storage ring technology, to push a
direct measurement of the proton EDM to $\SI{e-29}{\text{$e$}.cm}$ sensitivity, corresponding to an improvement by
nearly five orders of magnitude. Such a dramatic improvement can be made possible by the new ideas
and techniques described in this document.
Several new neutron EDM experiments involving ultracold  neutrons (UCNs)
have already been started worldwide, with the
aim  of eventually approaching $|d_{\sf n}| \sim\SI{e-28} {\text{$e$}.cm}$ sensitivity.
Compared with that, the storage ring studies target  a $|d_{\sf p}|$ sensitivity more
than an order of magnitude beyond $|d_{\sf n}|$ expectations, which are primarily limited by the
achievable
number of trapped UCNs.
Such an improved sensitivity
might be crucial in reaching the forefront of  the underlying mechanisms behind baryogenesis and
BSM-induced ${CP}$ violation.
In view of the entirely unknown isospin properties of the latter,
 even at the lower  $(10^{-27}\,\text{to}\,10^{-28} \,\si{\text{$e$}.cm})$ sensitivity,
the proton EDM studies are complementary to the neutron ones and will be essential in
discriminating  between---or at least constraining---various mechanisms for baryogenesis or competing models of ${CP}$ violation, \eg variants of
supersymmetric (SUSY) models, multi-Higgs models,
 left--right symmetric models, or the strong ${CP}$ violation from the  $\bar\theta$ angle of quantum chromodynamics. Note that,
 a priori, the results for $d_{\sf n}$ and $d_{\sf p}$ are independent
and could have significantly different values. Only when they are interpreted within the context of a
specific theoretical framework do their values become related and a comparison
meaningful. Even if  $d_{ \rm n}$ were found to differ from zero, the measurement of
$d_{\sf p}$ (and perhaps additional measured
EDMs of
light nuclei, \eg deuterons or helions, which might also be studied in storage ring experiments)
would prove crucial in unfolding
the new source of ${CP}$ violation.

\textbf{\textit{}}

\section{Dimensional analysis}
\subsection{Naive EDM estimate based on known physics}
\label{Chap:physics-case:naive}
Because of its inherent ${P}$ and ${CP}$ violation, the upper limit  on the  permanent EDM  $(d_{\sf{N}})$
of the nucleon  (\ie neutron or proton)  can be estimated~\cite{Khriplovich:1997ga}
from the product of the ${P}$- and ${CP}$-conserving nuclear magnetic moment (approximated by the nuclear magneton $\mu_{\sf N} = {e} / {2 m_{\sf N}} \sim \SI{e-14}{\text{$e$}.cm}$) multiplied by a  suppression scale counting
the ${P}$ violation ($\sim$\,$G_{\rm F} f_\pi^2 \sim 10^{-7}$, in terms of the Fermi constant $G_{\rm F}$ and the axial decay constant of the pion $f_\pi$) multiplied by the
{\em additional} ${CP}$-violating scale ($\sim$\,$10^{-3}$, derived from the absolute ratio of
the ${CP}$-forbidden and ${CP}$-allowed amplitudes ${\cal A}(K_{\sf L} \to \pi\pi)$ and ${\cal A}(K_{\sf S} \to \pi\pi)$, respectively).
Thus, the absolute value of the nucleon EDM  cannot be much larger than the natural scale,
 \begin{equation}
   | d_{\sf N} | \lesssim \mu_{\sf  N} \times 10^{-7} \times 10^{-3} \sim \SI{e-24}{\text{$e$}.cm} \, ,
    \label{nEDM-estimate}
\end{equation}
without coming into conflict with known physics constraints---on top of the experimental neutron EDM bound, which
nowadays is even more restrictive, as shown previously.

 In the absence of the QCD $\bar\theta$ term, the SM only possesses a non-zero ${CP}$-violating phase if
 the CKM-matrix involves at least three generations, such that  in this case the  estimate in \Eref{nEDM-estimate}
 inherently implies a flavour change. The EDM, however, is flavour-neutral. Therefore, the upper bound
 for the nucleon EDM in the SM with a zero $\bar\theta$ term  necessarily involves a further
suppression factor of $G_{\rm F} f_\pi^2 \sim 10^{-7}$  to undo the flavour change:
 \begin{equation}
  | d_{\sf N}^{\rm SM}| \lesssim 10^{-7} \times  \SI{e-24}{\text{$e$}.cm} \sim \SI{e-31}{\text{$e$}.cm}\,.
  \label{nEDM-estimate-SM}
 \end{equation}

 This simple estimate agrees in magnitude with the three-loop calculations of
 Refs.\,\cite{Khriplovich:1985jr,Czarnecki:1997bu} and also with the two-loop calculations of
 Refs.\,\cite{Gavela:1981sk,Khriplovich:1981ca} that include a strong penguin diagram and the long-distance effect of a pion loop.
 It is even  consistent  with a recent calculation on the long-distance meson loop contribution based on heavy-baryon chiral perturbation theory, see Ref.\,\cite{Seng:2014lea}, and
 with modern loopless
 calculations involving charm-quark propagators~\cite{Mannel:2012qk,Mannel:2012hb}.
  From Eqs. (\ref{nEDM-estimate}) and (\ref{nEDM-estimate-SM}),
 one can infer that an EDM of the nucleon measured in the window
\begin{equation}
  \SI{e-24}{\text{$e$}.cm} \gtrsim |d_{\sf  N}|  \gtrsim \SI{e-30}{\text{$e$}.cm}
\end{equation}
will be a clear sign for new physics beyond the KM mechanism of the SM: either
strong ${CP}$ violation by a sufficiently large QCD  $\bar\theta$ term or ${CP}$ violation by BSM
physics, as, \eg supersymmetric models, multi-Higgs models or left--right symmetric models.

\subsection{BSM scale estimate}
A rough estimate of the scale of BSM physics  probed by  EDM experiments  can be derived
from an expression of a subatomic EDM $d_i$ that is  based solely on dimensional considerations and
 that holds for dimension-six
extensions of the SM, since the SM symmetries and the pertinent
chirality constraints preclude any contribution from
dimension-five operators:
\begin{equation}
  d_{i} \approx \frac{1}{16 \pi^2} \frac{m_i}{\Lambda^2_{\rm BSM}} e_i \,\sin \phi\,.
\end{equation}
Here, $e_i$ and $m_i$ are the charge and  mass, respectively, of the relevant quark\footnote{Strictly speaking, the quark masses are scale- and scheme-dependent.} or lepton, $\sin\phi$ results from the ${CP}$-violating BSM phases,
and $\Lambda_{\rm BSM}$ is the mass scale of the underlying BSM physics. In general, the coupling of BSM physics to  subatomic particles induces at least
one quantum loop; therefore, a suppression factor
 $g^2/(16\pi^2) \sim 10^{-2}$   (assuming $g\sim 1$) is  also included.

For current quark masses of the order $m_{\rm q} \sim \SI{5}{MeV}$,
we might  therefore expect
\begin{equation}
    |d_{\sf N}| \sim 10^{-24} \left(\frac{\SI{1}{TeV}}{\Lambda_{\rm BSM}}\right)^2|\sin \phi|\ \si{\text{$e$}.cm}\,.
\end{equation}
 If $\Lambda_{\rm BSM} \gtrsim \SI{1}{TeV}$ and $\sin\phi\sim 1$, this result is compatible with
the upper limit (\Eref{nEDM-estimate}) derived from
the naive estimate, \ie it is compatible with  all the known physics, except the constraints from
direct or indirect EDM measurements. The projected sensitivity for
$|d_{\sf p}| \sim \SI{e-29} {\text{$e$}.cm}$ would, in turn, allow one to test  the ${CP}$-violating phase $\phi$ of
a theory of mass scale $M \sim  1\,\si{TeV}$ down to values of $\phi \gtrsim 10^{-5}$, while
for natural values of  the ${CP}$-violating phases, $\phi\sim 1$,
a mass scale range up to $M\sim \SI{300}{TeV}$
can be probed. (These numbers refer to one-loop processes, such as, \eg supersymmetric extensions.
They are suppressed by about two orders of magnitude  for two-loop  (so-called Barr--Zee~\cite{Barr:1990vd}) processes, as \eg in
multi-Higgs scenarios, while they are enhanced by the same factor for loop-free particle exchanges, as, \eg for leptoquarks.)

\subsection{Estimate of the strong ${CP}$-violating QCD $\bar\theta$ parameter} \label{Chap:physics-case-strong}
Even if  a natural-sized $\bar\theta$ parameter (which is given
by the sum of the original $\theta$  that couples to the  product of  the gluon and dual gluon
field strength tensors
and the phase of the determinant of the quark mass matrix) is removed,
by the Peccei--Quinn mechanism~\cite{Peccei:1977hh}, the possibility  cannot be excluded that a fine-tuned $\bar\theta$,
compatible
 with the $|d_{\sf n}|$ bound~\cite{Baker:2006ts,Afach:2015sja,Abel:2020gbr},
will re-emerge from Planck-scale physics on UV completion.

The scale of the nucleon EDM induced by the $\bar\theta$ parameter can
be estimated, in a similar way to the expression given in \Eref{nEDM-estimate}, by \cite{Baluni:1978rf,Crewther:1979pi,Bigi:2000yz} \begin{equation}
  |d_{\sf N}^{\bar\theta} | \sim |\bar\theta| \cdot \frac{m_{\rm q}^\ast}{\Lambda_{\rm QCD}} \cdot \frac{e}{2 m_{\sf N}} \sim  |\bar\theta| \times \SI{e-16} {\text{$e$}.cm}\,,
   \label{nEDM-theta}
\end{equation}
where $m_{\rm q}^\ast = m_{\rm u} m_{\rm d} m_{\rm s}/( m_{\rm u} m_{\rm d} + m_{\rm s} m_{\rm u}
+ m_{\rm s} m_{\rm d}) \sim  m_{\rm u} m_{\rm d}/(m_{\rm u} +m_{\rm d})$ is the reduced quark mass.
The additional suppression factor given by the
ratio of the reduced quark mass to
the QCD scale parameter $\Lambda_{\rm QCD} \sim \SI{200}{MeV}$ takes into account that
the $\bar\theta$-induced EDM would have to vanish if any  quark mass were vanishing, since
in that (chiral) limit the complete $\bar\theta$ term  could
be rotated away by an axial $U(1)$ transformation acting
on the quark with zero mass~\cite{Bigi:2000yz}. Applying this estimate (\Eref{nEDM-theta}) and  utilizing
the empirical
bound on the neutron EDM~\cite{Baker:2006ts,Afach:2015sja,Abel:2020gbr}, one finds the following upper
limit for $\bar\theta$:
\[
    |\bar\theta | \lesssim 10^{-10} \,.
\]
Taking into account the  limit,  \Eref{nEDM-estimate-SM}, of the Kobayashi--Maskawa  induced
nucleon EDM,
the accessible window for determining $\bar\theta$  by nucleon EDM measurements
is therefore
 \[
   10^{-10}  \gtrsim |\bar \theta| \gtrsim 10^{-14} \,,
 \]
 while the projected sensitivity for $|d_{\sf p}| \sim \SI{e-29} {\text{$e$}.cm}$ would allow a measurement of the value of, or a bound on, the parameter $\bar\theta$, down to the order  $10^{-13}$.

\section{Analysis of EDM  based on non-perturbative methods}
Measurements of EDM are of low energy in nature; therefore, all predictions of EDM values of
subatomic particles, especially nucleons, belong to the realm of non-perturbative  QCD.

\subsection{Determination of the $\bar\theta$-induced nucleon EDM}

The QCD $\bar\theta$-term is manifestly a flavour-neutral isoscalar source of ${CP}$ violation. It is instructive that the
underlying non-perturbative physics nonetheless entails $d_{\sf p} \neq d_{\sf n}$.

The best way to predict the ratios
$d_{\sf p}/\bar\theta$ or $d_{\sf n}/\bar\theta$ in the $\bar\theta$-term scenario
would be the application of lattice QCD methods.
Unfortunately, all current
high-precision lattice calculations dedicated to these predictions
have been based on the computation of the ${T}$- and ${P}$-violating $F_3$ form factors of the neutron or proton
and have not taken
 into account the fact that, in a finite volume,
 the Dirac states of the nucleon acquire an axial rotation in the mass term, such
that there is a sizeable admixture of the large $F_2$ (Pauli) in the small $F_3$ (EDM) form factor,
as first pointed out in Ref.~\cite{Abramczyk:2017oxr}. When the reported numbers
of $d_{\sf n}/\bar\theta$
or $d_{\sf p}/\bar\theta$ were reanalysed, the corrected values turned out to be
compatible with zero within one standard deviation~\cite{Abramczyk:2017oxr,Yoon:2017tag}.
Note that no general consensus has been reached about the need to perform this correction to the form
factors. Guo \textit{et al.}\,\cite{Guo:2015tla} assert that, in their method,  which
differs from other attempts by  a purely imaginary value of the vacuum angle,
the expansion is about a topologically {\em non-trivial} vacuum and that  the
mixing-angle dependence has  therefore been included.

 It is expected that lattice estimates with better accuracy will become achievable in the next one or two years, though.
A step in that direction has been made in  Ref.\,\cite{Dragos:2019oxn}.
Using the gradient-flow method with proper axial rotation, and by extrapolating from
dynamic quark masses corresponding to  admittedly large pion masses of 700, 570, and 410~MeV,   the
following results were predicted: $d_{\sf n}/\bar\theta =-1.52(0.71) \times \SI{e-16}{\text{$e$}.cm}$ and
$d_{\sf p}/\bar\theta = 1.1 (1.0) \times \SI{e-16} {\text{$e$}.cm}$. These, in turn, imply that
$|\bar\theta| < 1.98 \times 10^{-10}$ (90~\%~C.L.) as an upper bound on the QCD $\bar\theta$ term
 from
the experimental bound~\cite{Afach:2015sja} on the EDM of the neutron.\label{foot:LatticeNew}

In the meantime,   chiral perturbation theory (ChPT) can be applied to estimate
the contribution  of the pion one-loop terms to the $\bar\theta$-induced
neutron and proton EDMs \cite{Ottnad:2009jw}---note that  the leading
term, which involves a ${CP}$-violating but isospin-conserving pion--nucleon vertex,
was already estimated  nearly 40\,years ago \cite{Crewther:1979pi}, while
the loop diagram with the isospin-breaking counterpart is subleading. Both diagrams
are divergent and have logarithmic scale dependence, which, in principle, can be
cured by the addition of two independent ${CP}$-violating photon--nucleon
contact terms~\cite{Guo:2012vf,Akan:2014yha}.
The signs and sizes of the latter, however, cannot
be determined in ChPT and need external input, either from experiment
or from lattice QCD, which currently, as mentioned before, produces inconclusive results.
The leading pion loop term predicts a contribution (at the mass scale of the nucleon) of\footnote{Here, and in
the following, the signs of the EDMs always refer to the convention $e>0$.}
 \begin{equation}
  \Delta d_{\sf p}/\bar\theta = - \Delta d_{\sf n}/\bar\theta = (1.8\pm 0.3) \times \SI{e-16} {\text{$e$}.cm} \, ,
 \label{loopEDM-theta}
\end{equation}
that is, of isovector nature---see Ref.~\cite{Ottnad:2009jw}, with input
parameters from Ref.~\cite{Bsaisou:2014zwa}.
 Note, however, that  the subleading isoscalar loop-term is neglected
here and  that the sizes and signs of the two missing contact terms are not known, such that ChPT itself cannot
predict the ratio of the proton EDM to the neutron EDM.

For a real test or falsification of the $\bar\theta$ hypothesis as the leading (\ie dimension-four)
${CP}$-violating mechanism if $d_{\sf n}$ and $d_{\sf p}$ are measured,
one needs the anticipated results of lattice QCD. However,
even without lattice QCD calculations, additional measurement of the deuteron or helion, or both
EDMs would enable independent tests, since ChPT and chiral effective field theory
methods can be used to obtain an estimate of the genuine nuclear contributions of these light nuclei (including triton)~\cite{Bsaisou:2014zwa}, \ie
\begin{align}
  (d_{^2{\sf H}} - 0.94 d_{\sf p} - 0.94 d_{\sf n}) /\bar\theta &=    \phantom{-}(0.89 \pm 0.30) \times \SI{e-16} {\text{$e$}.cm}\,,\\
  (d_{^3{\sf He}} - 0.90 d_{\sf n} + 0.03 d_{\sf p})/\bar\theta &=  -(1.01 \pm 0.42)   \times \SI{e-16}{\text{$e$}.cm}\,,\\
   (d_{^3{\sf H}} - 0.92 d_{\sf p} + 0.03 d_{\sf n})/\bar\theta &=  \phantom{-}(2.37 \pm 0.42) \times \SI{e-16}{\text{$e$}.cm}\,.
\end{align}
These numbers are comparable with the predictions for  the single-nucleon EDM case---cf.  \Eref{loopEDM-theta}
and Ref.~\cite{Dragos:2019oxn}. Therefore, they can equally well
be used  to test or constrain the value of the $\bar\theta$ term to the $\sim 10^{-13}$ level, assuming that these  EDMs can be measured
to $\SI{e-29}{\text{$e$}.cm}$ sensitivity.

\subsection{Estimates of the nucleon EDM terms in the BSM scenario}

Again, lattice QCD is the first choice for an estimate of the EDM contributions of the
dimension-six ${CP}$-violating operators,
which can be grouped~\cite{Pospelov:2005pr,Ng:2011ui,deVries:2012ab} into
 quark operator terms (${CP}$-violating  photon--quark vertex terms),
 quark--chromo operator terms (${CP}$-violating gluon--quark vertex terms), the
isoscalar Weinberg
three-gluon term~\cite{Weinberg:1989dx},  isospin-conserving  ${CP}$-violating four-quark terms, and isospin-breaking
four quark terms, which can be traced back to left--right symmetric models.
While there do not exist lattice QCD calculations for any of the four-quark operators,
exploratory studies have just begun in the Weinberg three-gluon case.
In the quark--chromo
scenario, there already exist promising signals for the connected contributions, but
results with (quark-)disconnected diagrams, non-perturbative mixing, and renormalization
are still missing---see Refs. \cite{Yoon:2017tag,Gupta:2019fex} and references therein for further details.
In the quark EDM case, however, lattice QCD has delivered because the
pertinent weight factors of  the u-, d-, and s-quark EDMs  follow via a chiral rotation from the corresponding flavour-diagonal
quark tensor charges, $g_{\rm T}^{\rm u,\,d,\,s}$, \ie\footnote{Note that the flavour assignments of the
tensor charges refer to the proton case, while for the neutron, the roles of the u and d quarks  must be interchanged; the cited values refer to the $\overline{\rm MS}$ scheme at 2\,GeV~\cite{Aoki:2019cca}.}
\begin{align}
  d_{\sf n} &= d_{\rm u}^\gamma g_{\rm T}^{\rm d}+ d_{\rm d}^\gamma g_{\rm T}^{\rm u}
  + d_{\rm s}^\gamma g_{\rm T}^{\rm s}\,,\\
   d_{\sf p} &= d_{\rm u}^\gamma g_{\rm T}^{\rm u}+ d_{\rm d}^\gamma g_{\rm T}^{\rm d}
  + d_{\rm s}^\gamma g_{\rm T}^{\rm s}\,,
 \end{align}
where the predictions of the tensor charges improved  considerably from 2015 to 2018:
\begin{align}
 g_{\rm T}^{\rm u} & = 0.774 (66), & g_{\rm T}^{\rm d} & = -0.233 (28), & g_{\rm T}^{\rm s} & = -0.008(9) & \mbox{\cite{Bhattacharya:2015wna};}\\
 g_{\rm T}^{\rm u} & = 0.782(21), & g_{\rm T}^{\rm d} & = -0.219 (17), & g_{\rm T}^{\rm s} & = -0.00319(72) & \mbox{\cite{Alexandrou:2017qyt};}  \label{gT-alexandrou}\\
 g_{\rm T}^{\rm u} & = 0.784(30), & g_{\rm T}^{\rm d} & = -0.204 (15), & g_{\rm T}^{\rm s} & = -0.0027(16) & \mbox{\cite{Gupta:2018lvp}.}  \label{gT-new}
\end{align}
While the ratio $g_{\rm T}^{\rm u} / g_{\rm T}^{\rm d} \approx -4$ is compatible with the estimate
from the naive non-relativistic quark model and the model from  QCD sum rules~\cite{Pospelov:2005pr}, the
absolute values of $g_{\rm T}^{\rm u}$
and $g_{\rm T}^{\rm d}$ are smaller, approximately reduced by a factor $3/5$ relative to
the naive quark model estimate and
the central values in  the QCD sum rule case (see also the following).

These predictions of the tensor quark charges   allow for stringent tests
of the split SUSY scenario with gaugino mass
unification~\cite{ArkaniHamed:2004fb,Giudice:2004tc,ArkaniHamed:2004yi},
since in this case there is a strong correlation between
the electron and neutron (or proton) EDMs~\cite{Giudice:2005rz}, the latter governed by
the quark EDM operators, while  all other ${CP}$-violating operators are highly suppressed.
In particular,  \Eref{gT-new}
and
the indirect experimental bound $|d_{e}| < 1.1  \times \SI{e-29} {\text{$e$}.cm}$~\cite{Andreev:2018ayy}
imply $|d_{\sf n}|< 4.1 \times \SI{e-29} {\text{$e$}.cm}$
as an upper bound in the split-SUSY scenario~\cite{Gupta:2018lvp}.  This limit
is still in the  range of sensitivity of
a dedicated proton EDM storage ring experiment.

With the exception of the quark EDM case mentioned already,   there do  not currently exist any predictions
of lattice QCD or ChPT for any of the other ${CP}$-violating BSM operators. In the latter cases, only QCD sum rule
estimates of the quark and quark--chromo contributions to the nucleon EDMs
are available \cite{Pospelov:2005pr},
\begin{align}
 d_{\sf n} &\simeq  (1\pm 0.5) \times \left\{ 1.4  \left(d_{\rm d}^\gamma -0.25d_{\rm u}^\gamma\right)+
 3.2\,\left( e_\mathrm{d} d_{\rm d}^{\,\rm c} -0.25 e_\mathrm{u} d_{\rm u}^{\,\rm c}\right) \right\} \pm  (0.02\,{\rm GeV})\,e \,d^{\rm W}\,,\\
 d_{\sf p} &\simeq  (1\pm 0.5)\times \left\{1.4 \left(d_{\rm u}^\gamma - 0.25 d_{\rm d}^\gamma\right)+
 3.2 \,  \left(e_\mathrm{u} d_{\rm u}^{\,\rm c} -0.25  e_\mathrm{d} d_{\rm d}^{\,\rm c}\right) \right\}\pm  (0.02\,{\rm GeV})\,e \,d^{\rm W}\,,
 \end{align}
 where $d_{\rm u,\,d}^\gamma$ and $d_{\rm u,\,d}^{\,\rm c}$ denote the u- and d-flavour quark  and
 quark--chromo  EDMs, respectively, with $e_\mathrm{u,d}$ the corresponding quark charges, while
 $d^{\rm W}$  (of dimension mass$^{-2}$) stands for the prefactor of the Weinberg  term.
Taking these large uncertainties into account, we currently have no reliable prediction of the ratio of the
 proton to neutron EDM for any of the BSM extensions (SUSY, multi-Higgs models, left--right symmetric models),
 with the
 notable exception of the previously discussed split-SUSY case (assuming  that quark EDM ratios follow the quark mass (multiplied by  quark charge) ratios).

\subsection{Estimates of the nuclear  EDM matrix elements for light nuclei}
If, however,   storage ring experiments  are planned to measure the deuteron or helion EDMs,
these results  would determine the  genuine nuclear EDM contributions.
The relevant, \ie leading,  ${CP}$-violating nuclear matrix
elements are governed by tree-level operators and
are predicted  in the framework of chiral effective field theory (chEFT)  with reasonable
uncertainties~\cite{Bsaisou:2014zwa,Wirzba:2016saz,Yamanaka:2016umw}:
\begin{align}
d_{^2{\sf H}} - 0.94(1)(d_{\sf p} +d_{\sf n})\qquad\ &=   \left\{ 0.18(2)  g_1 - 0.75(14) \Delta_{3\pi} \right\}
\text{\si{$e$.fm}},\\
d_{^3{\sf He}} - 0.90(1)d_{\sf n}+0.03(1)d_{\sf p} &=  \left\{ 0.11(1) g_0 + 0.14(2) g_1 +0.61(14)\Delta_{3\pi} \right.
\nonumber \\
&\qquad\mbox{}-\left.0.04(2)C_1\si{fm^{-3}} + 0.09(2)C_2\si{fm^{-3}}\right\} \text{\si{$e$.fm}} \,,\\
d_{^3{\sf H}}  -0.92(1)d_{\sf p} + 0.03(1)d_{\sf n} &=  \left\{ -0.11(1) g_0 + 0.14(2) g_1 -0.60(14)\Delta_{3\pi} \right.
\nonumber \\
&\qquad\mbox{}+\left.0.04(2)C_1\si{fm^{-3}} - 0.09(2)C_2\si{fm^{-3}} \right\} \text{\si{$e$.fm}} \,.
\end{align}
Here,
$g_0$ and $g_1$ are the dimensionless low-energy constants of the isospin-conserving and isospin-breaking
${CP}$-violating pion--nucleon vertices, respectively, while $\Delta_{3\pi} \cdot m_{\sf N}$  is the prefactor of the ${CP}$-violating three-pion
term
and $C_1$ and $C_2$ are the coefficients of the two leading ${CP}$-violating four-nucleon terms.
The values of the three hadronic low-energy constants $g_0$, $g_1$, and $\Delta_{3\pi}$ can be predicted from
the coefficients of the ${CP}$-violating terms of the underlying theory at the quark--gluon level,
\eg from  $\bar\theta$ in the case of QCD~\cite{Bsaisou:2014zwa,deVries:2015una} or from
the prefactors of the quark--chromo~\cite{Seng:2016pfd} or the left--right-model-induced four-quark terms---see Ref.~\cite{Mereghetti:2018oxv} and
references therein.
While the $\bar\theta$ mechanism assigns a dominant role to $g_0$, the quark--chromo mechanism predicts that
$g_0$ and $g_1$ are of about equal magnitude, whereas $g_1$ dominates in the left--right scenario.
There are no analogous
predictions for the hadronic coefficients $C_1$ and $C_2$. The order of their contributions
can, so far, only be estimated by naive dimensional analysis and thus has to be  included in the theoretical uncertainties.
Note  that the role and magnitude of the $CP$-violating four-nucleon and
three-pion terms have not  been investigated
for $A>3$ nuclei---see Refs. \cite{Chupp:2017rkp,Jungmann:2013sga} for more
information on EDM calculations for heavy nuclei.

\section{Option for oscillating  EDM searches at storage rings}

 The storage ring technology also allows us to search
for time-varying (oscillating) components of the EDM,  in addition to the
static (permanent) one~\cite{Chang:2017ruk,Chang:2019poy}, and therefore  to test the hypothesis that the dark matter content in our galaxy is (at least partially) saturated by a classical  oscillating field\footnote{The mode occupation numbers of dark matter bosons of mass <$1\,{\rm meV}$ suffice
for the formation of a classical field.} of axions or axion-like particles (ALPs), even if the axion or ALP mass $m_\mathrm{a}$ were in the range
$\SI{e-7}{eV}$ to  $\SI{e-22}{eV}$~\cite{Graham:2011qk,Graham:2013gfa}\footnote{This assumes that the initial misalignment angle of the axion or ALP field  in this light-mass scenario is tuned so small that the
resulting `dark matter' particles  do not overclose the Universe---see, \eg Ref.~\cite{Tanabashi:2018oca} for more details.}.
This  mass range is very challenging for any
other technique to reach, since, \eg the resonance cavities of the microwave (haloscope)
method\footnote{This means that
a resonance in an RF cavity in a strong magnetic field is excited by the inverse
Primakoff effect.} would have to be unwieldy  large~\cite{Graham:2015ouw}. There are, though,
some astrophysical  constraints from  the bounds of supernova energy losses, Big Bang nucleosynthesis,
and the spatial extent of dwarf galaxies~\cite{Abel:2017rtm}. For instance, the latter give an upper bound on the de Broglie wavelength, and therefore the lower bound of  $\SI{e-22}{eV}$ (quoted just now) on the mass of a non-relativistic bosonic particle trapped in the  halo of a dwarf galaxy.

All interactions of the axions or ALPs are either suppressed by the very large axion or ALP decay constant $f_\mathrm{a}$ or are just of a gravitational nature. Thus, in the so-called {\em pre-inflationary Peccei--Quinn symmetry breaking
scenario}~\cite{Tanabashi:2018oca},  the initial displacement (misalignment) of the axion or ALP field $a$  from the minimum of its potential energy density, given by $m_\mathrm{a}^2 a^2/2$,\footnote{Starting
with the QCD epoch ($\sim$\,$\SI{e-4}{s}$ after the Big Bang),
the axion mass $m_\mathrm{a}$  is constrained as
$m_\mathrm{a} \approx 0.5 m_\pi f_\pi/f_\mathrm{a}$, where $m_\pi$ and $f_\pi$ are
the pion mass and decay constant,
respectively---see, \eg Ref.~\cite{Tanabashi:2018oca} for more details.} leads
to a coherent oscillation of the  classical
axion or ALP field at a Compton frequency $\omega_\mathrm{a} = m_\mathrm{a} c^2/\hbar$. The idea is to equate the
energy density in these oscillations with the mass-energy associated with dark matter~\cite{Graham:2011qk,Graham:2013gfa}. The axions or ALPs trapped in the
halo of our galaxy and to be observed in terrestrial experiments acquire in addition a velocity $v$ of the size of
the virial velocity of our solar system relative to the centre of our galaxy $\sim$\,$10^{-3} c$. Thus, their frequency is second-order Doppler-shifted,
$\omega' \simeq \omega_\mathrm{a} \left( 1+v^2/2 c^2\right)$. This implies a finite
coherence time of order $\tau_\mathrm{a} \approx \hbar/ m_\mathrm{a} v^2$, and thus a $Q$ value of the size $(c/v)^2 \sim 10^6$ and
a coherence length of order $\hbar /(m_\mathrm{a} v)$. For any terrestrial experiment smaller than this coherence length, which is at least
$\SI{0.5}{km}$ for $m_\mathrm{a} c^2 \lesssim   \SI{0.1}{\micro eV}$, the oscillating axion field corresponds to~\cite{Chang:2017ruk,Chang:2019poy}
\begin{equation}
 a (t) = a_0 \cos\bigl(\omega' (t-t_0) + \phi_0\bigr) \approx  a_0 \cos\Bigl({\textstyle\frac{1}{\hbar}}m_\mathrm{a} c^2 (t-t_0) + \phi_0\Bigr)\,,
\end{equation}
 where the
undetermined {\em local} phase $\phi_0$, which is approximately
constant, as long as
the measurement period $|t-t_0|$ is smaller than the coherence time $\tau_\mathrm{a}$,
is correlated with the choice of the
start point $t_0$ of the measurement cycle\footnote{The to-be-measured value of the phase $\phi_0$ ensures that, at the beginning of a measurement period, $t=t_0$,
all spectral $\omega'$ components  of the
axion field, irrespective of their velocity $|\vec v| <v_{\rm esc}$  (= the escape velocity from our galaxy), start coherently with the common phase  $\cos(\phi_0)$ and stay approximately coherent as long as $|t-t_0| < \tau_\mathrm{a}$.}.
The amplitude $a_0$ of this classical field oscillating at the frequency $\omega'\approx \omega_\mathrm{a}$ can
be estimated by saturating the local dark matter density in our galaxy,
$\rho_{\rm LDM} \approx \SI{0.4}{GeV/cm^3}$~\cite{Tanabashi:2018oca},
with the total energy density of the oscillating axion or ALP field, \ie\
$
   \rho_{\rm LDM} \approx m_\mathrm{a}^2 a_0^2/2
$.
Assuming QCD--axion coupling to the gluons   and therefore an effective angle  $\theta$,
\begin{equation}
  \theta_\mathrm{a} = \frac{a_0}{f_\mathrm{a}} \approx \frac{\sqrt{2\rho_{\rm LDM} } }{m_\mathrm{a} f_\mathrm{a}} \approx \frac{\sqrt{2\rho_{\rm LDM}}} {0.5m_\pi f_\pi}
  \sim 3 \times 10^{-19}\,,
   \label{eff_theta_a}
 \end{equation}
we  would obtain  from the naive formula (\Eref{nEDM-theta}) for the $\bar\theta$-induced  nucleon EDM the following estimate of the axion-induced oscillating component of the nucleon EDM:
\begin{equation}
  d_{\sf N}^{\rm osc}(t) \sim 10^{-16}\cdot \frac{a(t)}{f_\mathrm{a}} \sim 5 \times 10^{-35}
  \cos\Bigl({\textstyle\frac{1}{\hbar}}m_\mathrm{a} c^2(t-t_0)+\phi_0\Bigl) \,\si{\text{$e$}.cm} \,. \label{d_N_osc}
\end{equation}

The detection of  an oscillating EDM of such an amplitude  would be very demanding. In the case of an ALP, however,
there is no strict relation between its mass $m_\mathrm{a}$ and its decay constant $f_\mathrm{a}$, such that  mass regions with
$m_\mathrm{a} < 0.5 m_\pi f_\pi/f_\mathrm{a}$ and therefore {\em effective} ALP angles
with $\theta_\mathrm{a} > 3 \times 10^{-19}$ become
accessible\footnote{Only in the axion case, the estimate (\Eref{nEDM-theta})
constrains the value of the oscillating
EDM, while in
the ALP scenario the coupling strength of $a(t)$ to the nucleon is undetermined and does not
depend on the ALP mass~\cite{Graham:2013gfa}; this arbitrariness
can be taken into account by defining an effective ALP angle  as
$\theta_\mathrm{a} = c_\mathrm{a} a_0/f_\mathrm{a}$, where $c_\mathrm{a}$ is an unknown dimensionless parameter that in turn rescales the right-hand side of \Eref{d_N_osc}.}. In fact,
first exclusion bounds in the domain of  axion or ALP mass (frequency) versus axion-- or ALP--gluon coupling strength have already been extracted
from  neutron EDM measurements~\cite{Abel:2017rtm}
and   molecular-ion ({\sf HfF}$^+$) measurements~\cite{Roussy:2020ily},
and dedicated experiments applying nuclear magnetic
resonance techniques or superconducting
toroidal magnets
are currently being conducted~\cite{Budker:2013hfa,Graham:2015ouw,JacksonKimball:2017elr,Kahn:2016aff,Ouellet:2018beu}.

In complete analogy to the neutron EDM experiment,   the measurement or  bounds of the proton (or deuteron) EDM obtained by
the frozen spin method in storage ring experiments can,  of course, be analysed for slow oscillations, such
that the neutron and ALP bounds  can potentially be improved by the  ratio of the projected sensitivity of the proton EDM
measurement to the current  neutron EDM limit (\Eref{nEDM-exp-bound}). However, the advantage of the storage
ring technique is actually the search or scan for an oscillating EDM at the resonance conditions between the axion or ALP frequency
and the $(g-2)$ precession frequency of the storage ring---for further details see Appendix~\ref{Chap:Axions}. Such a  resonance enhancement would allow us to investigate
an axion or ALP frequency range of $\sim$\,$\SI{1}{mHz}$ to  $\sim$\,$\SI{1}{MHz}$, where the lower limit is simply due to the current
bound on the  spin coherence time~\cite{Guidoboni:2016bdn}, while the upper bound corresponds
to the spin-rotation frequency, as seen in the laboratory frame.
Furthermore, the resonance
method should, by fiat,
be less affected by systematical uncertainties than the frozen spin one. Moreover, in a combined electric
and magnetic storage ring (which is needed in the case of the deuteron or helion scenario), effective radial electric fields
in the centre-of-mass frame of the rotating particle can be achieved that are  one or even two orders of magnitude larger
than the presently achievable $E$ fields in the laboratory. In this way, the projected sensitivity for oscillating EDM
measurements by the resonance method may even reach  the
$\SI{e-30} {\text{$e$}.cm}$ level.

\section*{Synopsis}
Finally, let us emphasize that the physical reach of permanent proton EDM measurements
of sensitivity $\sim$\,$\SI{e-28} {\text{$e$}.cm} $
is competitive with or better than  any other  EDM
measurement, while at a $\SI{e-29} {\text{$e$}.cm}$ level
the proton EDM measurements become our best hope for finding new sources of
${CP}$ violation.

\begin{flushleft}

\end{flushleft}
\end{cbunit}

\begin{cbunit}

\csname @openrighttrue\endcsname 
\chapter{Historical background}
\label{Chap:Background}

\section{Beginnings at Brookhaven National Laboratory}

The idea of using a storage ring to confine a charged-particle beam while testing it for the presence of an EDM grew out of the  $(g-2)$ experimental effort at Brookhaven National Laboratory (BNL, New York, NY, USA). Even at low sensitivity, the data from this experiment may be checked for effects that arise from an EDM. The results from BNL\,\cite{Bennett:2008dy} and an even earlier CERN experiment\,\cite{Bailey:1977sw} reported upper limits for the muon EDM in the $\SI{e-19}{\text{$e$}.cm}$  range. Discussions in the late 1990s centred mostly on the muon experiment~\cite{Farley:2003wt}, but also considered the deuteron, which has a similar magnetic anomaly to mass ratio.

A regular pattern of BNL meetings for discussion and planning developed. In 2004, a proposal for a storage ring search on the deuteron at the $\SI{e-27}{\text{$e$}.cm}$  level was submitted to the BNL Program Advisory Committee (PAC) as Experiment 970. In light of the discrepancy between theory and experiment for the muon value of $g-2$\,\cite{Bennett:2006fi}, it was considered possible that meson-exchange contributions in the deuteron would lead to an enhancement in the EDM of the deuteron and more favourable prospects for a search. However, the BNL PAC did not find the proposal sufficiently competitive with other smaller-scale EDM searches to warrant the cost of constructing a new storage ring.

For a while, ring designs shifted to the development of resonant techniques to amplify and thus identify systematic errors~\cite{Orlov:2006su}. But eventually these schemes were discarded as unworkable at the greater sensitivities needed, and attention returned to a more standard storage ring design.

Beginning in 2005, feasibility experiments were carried out at the KVI cyclotron facility in Groningen to measure broad range spin sensitivities for deuteron scattering on carbon near $100\UMeV$. These showed large analysing powers but also sensitivities to beam alignment errors that could not be cancelled using standard first-order analysis techniques~\cite{Brantjes:2012zz}. In 2007, more definitive experiments were proposed for the COSY\footnote{Cooler Synchrotron, Forschungszentrum J\"ulich, Germany.} storage ring (Experiment 170) and approved for running. Tests began in 2008, leading to a final confirmation run in 2009 to demonstrate that, with a calibration of the sensitivity of the polarimeter to systematic errors, errors could be corrected to levels below one part in $10^5$~\cite{Brantjes:2012zz}. This was the first of what would become a series of beam studies to develop techniques needed for the EDM search in storage rings.

In 2008, a second deuteron proposal was submitted to the BNL PAC by the Storage Ring Electric Dipole Moment Collaboration~\cite{bnl-sredm}. This time, several improvements led to an anticipated sensitivity of $\SI{e-29}{\text{$e$}.cm}$,  with up to a year of data collection~\cite{bnl-sredm-deuteron-proposal}. This led to a technical review that was held in 2009  \cite{bnl-edm-review-2009}. In the meantime, it was realized that a first experiment on the proton offered some technical advantages, including the ability to have counter-rotating beams travelling along the same path in the same ring. This would optimize the cancellation of a large class of time-reversal conserving systematic errors. From this point on, proposals featured the proton rather than the deuteron. For a number of reasons, work at COSY, however, continued with deuteron beams. The primary argument was that substantial experience with beams of polarized deuterons was already available. In addition, deuterons would allow one to study mechanisms that are weighted differently than in the case of the proton\,\cite{Flambaum:1984fb}, and not even an indirect bound on the deuteron EDM is currently available. The expertise with deuteron operation at COSY is understood in the sense that any conclusions would apply to either proton or deuteron beams. Development continued at BNL and a second technical review was held in 2011, again with encouraging results \cite{bnl-edm-review-2011}. In October 2011, a full proposal was forwarded to the US DOE, but no formal evaluation was ever initiated. In collaboration with the US National Science
Foundation, the two funding agencies decided to terminate all further work along these lines.

\section{Continuation at the Forschungszentrum J\"ulich}

First contributions to the storage ring EDM effort were made at Forschungszentrum J\"ulich (FZJ, Germany) in 2008 and 2009, when members of the BNL srEDM collaboration and scientists from the Groningen KVI started experiments together with scientists from the Institut f\"ur Kernphysik (IKP) to investigate polarimetry issues at  COSY. It was soon  realized that COSY~\cite{Maier:1997zj}, with its polarized proton and deuteron beams in the energy range required for storage ring EDM experiments,  is a unique and ideal facility to perform the required R\&D.

COSY is a worldwide unique facility for polarized and phase-space cooled hadron beams, which was utilized for hadron physics experiments until the end of 2014. Since then, it has been used as a test and exploration facility for accelerator and detector development, as well as for the preparation and execution of precision experiments (J\"ulich Electric Dipole Moment Investigations (JEDI)\,\cite{jedi-collaboration},  Time Reversal Invariance at COSY (TRIC)~\cite{Aksentyev:2017dnk,Eversheim:2017zxl}). The COSY facility comprises sources for polarized and polarized protons and deuterons, an injector cyclotron JULIC (J\"ulich Light Ion Cyclotron), a synchrotron to accelerate, store, and cool beams, and internal and external target stations for experimental set-ups.

The H$^-$ (or D$^-$) ions are pre-accelerated up to 0.3 (0.55)~GeV/$c$ in JULIC, injected into COSY via stripping injection and subsequently accelerated to the desired momentum below 3.7~GeV/$c$. Three installations for phase-space cooling can be used, where the following refers to protons: (i) a low-energy electron cooler (between 0.3 and 0.6~GeV/$c$), installed in one of the straight sections, (ii) stochastic cooling above 1.5~GeV/$c$, and (iii) a new high-energy electron cooler in the opposite straight section, which can be operated between 0.3 and 3.7~GeV/$c$.

Well-established methods are used to preserve polarization during acceleration. A fast tune jumping system, consisting of one air-core quadrupole, has been developed to overcome depolarizing resonances. Preservation of polarization across imperfection resonances is achieved through the excitation of the vertical orbit, using correcting dipoles to induce total spin flips. The polarization had been continuously monitored by the internal polarimeter EDDA (which was decommissioned in 2019); an additional polarimeter, making use of the WASA forward detectors, was used for the JEDI experiments. Recently, a new polarimeter, based on LYSO scintillators, has been developed, and has been operation at COSY since  2019 ~\cite{Keshelashvili:2019ggh}. For protons, a beam polarization of 75\% up to the highest momentum has been achieved. Vector and tensor polarized deuterons are also routinely accelerated, with a degree of polarization of up to 60\%. Dedicated tools have been developed to manipulate the stored polarized beam and precisely determine the beam energy.

In 2011, the JEDI collaboration\cite{jedi-collaboration} was created, aiming to exploit COSY, not only for the development of the key technologies for storage ring EDM experiments, but also for performing a first direct EDM measurement for deuterons (as a precursor experiment). Since COSY is a conventional storage ring with magnetic bending, a dedicated insertion (a radio-frequency (RF) Wien filter) must be used to be sensitive to an EDM. This latter project towards a proof of principle for srEDM is supported by an advanced grant of the European Research Council (2016--2021)~\cite{ERC-694340}.

Meanwhile, significant experimental progress has been made at COSY and elsewhere (see Appendix \ref{app:cosy}). However, it has also become clear that between now (the precursor experiment at COSY) and then (the final clockwise, counterclockwise all-electric EDM ring), an intermediate step (a prototype or  demonstrator
step) is required to test and to demonstrate key issues, such as:
\begin{itemize}
\item   beam storage time (stochastic cooling);
\item   spin coherence time;
\item   polarimetry;
\item   clockwise (CW) and counterclockwise (CCW) operation;
\item   effects of the magnetic moment.
\end{itemize}
For further details, see Chapter \ref{Chap:RoadMap}.

The prototype may, if equipped with magnetic elements in addition to electric deflectors, be used in the frozen-spin condition to determine an EDM limit for the proton (see Chapter \ref{chap:ptr}).

\section{Ongoing activity: the precursor experiment at COSY}

During autumn 2018, the JEDI collaboration performed a first measurement of the deuteron EDM at COSY, with  analysis of the results in progress.
In a pure magnetic storage ring, such as COSY, an EDM will generate an oscillation of the vertical polarization component.
For a $970\UMeVc$ deuteron beam with a spin precession frequency of $120\UkHz$, a tiny amplitude is expected, \eg
$3\times 10^{-10}$ for an EDM of $d = \SI{e-24}{\text{$e$}.cm}$. In a magnetic storage ring like COSY, the spins cannot be frozen, thus, to allow for the accumulation of a vertical polarization proportional to the EDM, an RF Wien filter must be operated~\cite{Rathmann:2013rqa,Morse:2013hoa}. A prototype Wien filter was successfully installed and operated in COSY in 2014. A new device, providing a larger magnetic field integral (\SI{0.05}{T.mm}), was developed and constructed in collaboration with the Institut f\"ur Hochfrequenztechnik (IHF) at RWTH Aachen University and ZEA-1 in  J\"ulich~\cite{Slim:2016pim,Slim:2016dct}. This new RF Wien filter was installed in COSY in May 2017 and a first commissioning run was successfully conducted in June 2017.

\section{Charged-particle EDM initiative and experience of the collaboration}

In connection with the Physics Beyond Colliders (PBC) initiative of CERN and the European Strategy for Particle Physics (ESPP) update, a cooperation under the name `Charged Particle Electric Dipole Moment' (CPEDM) was formed in early 2017, comprising members of the srEDM and JEDI collaborations, as well as other interested scientists from CERN, in order to prepare the science case for a storage ring EDM search for the proton (deuteron, and $^3$He) and the technical design study---in other words, this document.

The JEDI members of the IKP of the Forschungszentrum J\"ulich have a decade's experience in designing,  building, and operating, as well as in further developing, accelerators: foremost, JULIC and COSY, but also  polarized and unpolarized ion sources for protons and deuterons. The IKP has also contributed significantly to the various versions of linear accelerators for spallation neutron sources and  has designed a superconducting linac, which was planned to replace JULIC as the injector for COSY. Recently, it has delivered the proton source for commissioning  the ELENA antiproton ring at CERN.

Unique experience is available in the production and acceleration of polarized beams without polarization loss and in manipulating them in COSY,  selecting polarization states and  determining the degree of polarization through the use of nuclear reactions with polarimeters, based on scintillator detectors. A huge amount of expertise has been accumulated over the years in cooling and storing beams,  accelerating and decelerating them, and using them during energy ramping or at a fixed energy at internal target stations with thin solid, gas, or pellet targets. It is also possible to provide slow (resonant and stochastic) or fast extracted beams to external target stations---this option was previously used for the TOF spectrometer and is now exploited for all kinds of detector test.

Electron cooling at low momenta (up to $600\UMeVc$) has been used in COSY early on; more recently, a high-energy electron cooler ($E_\mathrm{e} < 2\UMeV$) has been installed and commissioned in the ring.  Stochastic cooling is also used routinely in COSY (momentum range, 1.5--$3.3\UGeVc$); here, new pick-up and kicker devices have been developed at and implemented in COSY.

A group working at KAIST\footnote{Before submission of the final version of this report, KAIST dropped out of the CPEDM collaboration; the contributions by members of KAIST in Appendices \ref{Chap:MagneticFields} and \ref{Chap:hybrid} have been marked accordingly.} in South Korea (IBS Center for Axion and Precision Physics Research, CAPP) has developed a large amount of expertise in the use of SQUID magnetometers. A prototype EDM ring section has been constructed to investigate the cryogenic environment and magnetic sensitivity. This effort is in conjunction with the building of a magnetically shielded chamber to simulate conditions in an EDM beam line.

A group of scientists from CERN with extensive expertise in accelerator design  joined the CPEDM project from the start. They have already  made essential contributions to the study of electric deflection and of various kinds of systematic effect. Limiting the effects of systematic errors is the central issue in the success of the EDM storage ring project.

\section{Further developments}

Work is underway at COSY to develop electrostatic plates for use in a final EDM ring. An initial series of tests with hemispheres demonstrated fields of \SI{17}{MV/m} for stainless steel separated by \SI{1}{mm} and \SI{30}{MV/m} for aluminium separated by \SI{0.1}{mm}~\cite{doi:10.1063/1.5086862}. The next phase of the project will involve testing a prototype electric field in an $\approx$\,$\SI{1}{m}$ long section located in an existing dipole magnet with a large gap (ANKE, dipole 2) outside the COSY ring.

\section{Conclusion}

In summary, it must be emphasized that, in contrast to other EDM projects, \eg for the neutron,  electron,  muon, \etc, which are pursued in many different places worldwide, for CPEDM, Europe will be in a unique position to design, construct, and host such a project.



\begin{flushleft}

\end{flushleft}

\end{cbunit}

\begin{cbunit}

\csname @openrighttrue\endcsname 
\chapter{Experimental method}
\label{Chap:ExpMethod}

\section{Introduction}
The existence of a permanent electric dipole moment (EDM) for fundamental particles or subatomic systems is still an open question in physics, since such a quantity has never been detected. The EDM is a vector-like intrinsic property, which measures the asymmetric charge distribution along its spin axis\footnote{The dipole moment, evaluated in a quantum mechanical matrix element, must be aligned with the particle spin, the only available vector quantity in the rest system of the particle, as a consequence of the Wigner--Eckart theorem.}. Hence, experiments to measure the latter often rely on the spin
precession rate in an electric field. However, for charged particles, such a measurement cannot be made while maintaining the particle at rest, since any applied electric field leads to acceleration. Instead, those fields can be provided as a part of a particle trap. For the experiment considered here, the trap is a storage ring with crossed vertical $\vec{B}_y$ and radial $\vec{E}_r$ fields that confine a beam of spin-polarized particles (protons, deuterons, etc.) into a design orbit (see \Fref{fig:coordsystem}). The EDM ($\vec d$) couples to the electric fields, while the magnetic dipole moment (MDM, $\vec \mu$) couples to the magnetic fields, so that, for a particle at rest, a precession of its spin $\vec S$ occurs, which is given by
\begin{equation}\label{emeq0}
\frac{{\rm d}\vec S}{{\rm d}t}= \vec d \times \vec E + \vec \mu \times \vec B \,.
\end{equation}

\begin{figure}
    \centering
    \includegraphics[width=0.7\textwidth]{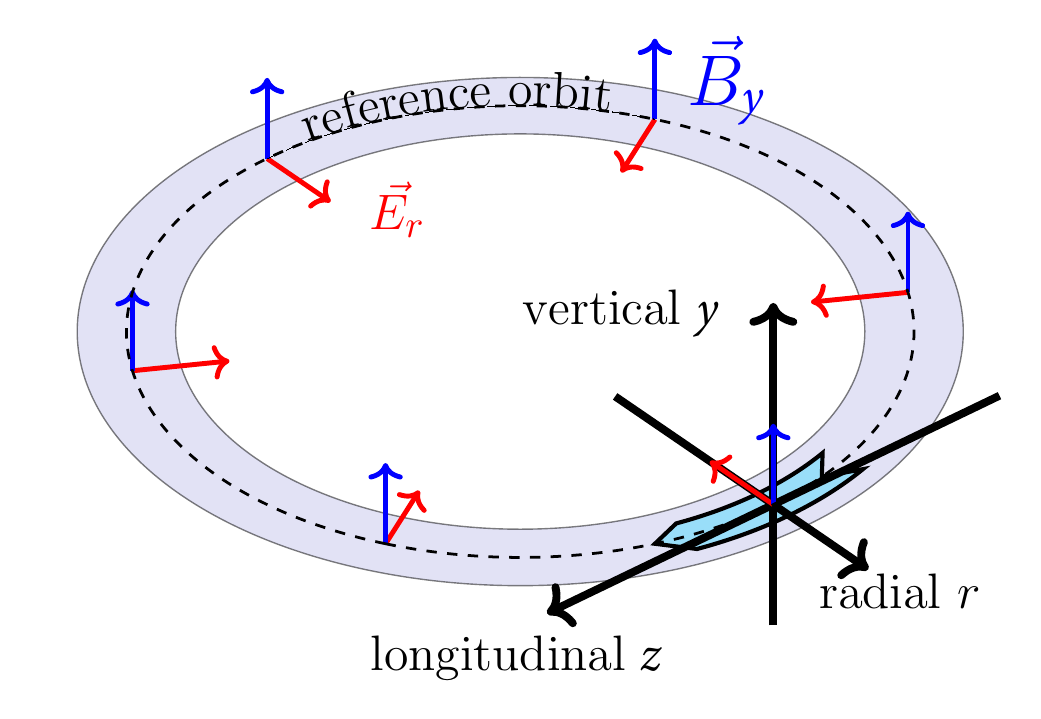}
    \caption{Coordinate system used in Eqs.\,(\ref{emeq2}) and (\ref{emeq3}) for a beam running clockwise (CW). Note that the electromagnetic fields refer to the laboratory frame.}
    \label{fig:coordsystem}
\end{figure}

In general, the MDM of subatomic particles is known to high precision; the aim of the proposed experiment is to determine the EDM part, which leads only to much smaller spin rotations. Nevertheless, since the charged particle is subject to combined electromagnetic fields and is therefore not at rest, it is necessary to account for the kinematic effect that may alter its spin precession. For this reason, we shall invoke the Thomas--BMT equation\,\cite{BMT}, which gives the precession rate of the angle between the spin and momentum vectors in the inertial rest frame of the particle.

\section{Spin evolution in electric and magnetic fields}
\label{sec:spin-evolution-electric-magnetic-fields}

In \Eref{eq2} of the executive summary, the Thomas--BMT equation is introduced for generic $\vec B$ and $\vec E$ fields. The latter are defined in the laboratory frame, while the spin is defined in the inertial rest frame of the particle. In a storage ring, where the particle is continuously deflected by the guiding electromagnetic fields to perform a closed orbit trajectory, it is convenient to rewrite the equation of motion in the non-inertial frame rotating with the velocity vector of the particle. A natural way to describe the rotation of the coordinate system is to use the Frenet--Serret frame attached to the reference orbit\,\cite{Courant:1980cf, Silenko:2015jqa}, which is therefore lying in the mid-plane of the accelerator, as illustrated in \Fref{fig:coordsystem}. In that case, the angular velocity describing the rotation of the coordinate system is given by
\begin{equation}
   \vec \Omega_{\rm cycl} = - \frac{\beta c}{\rho} \vec u_y\,,
\label{eq:omega_cycl}
\end{equation}
where $\rho$ is the bending radius, $\beta$ is the Lorentz factor, and $\vec u_y$ is the unit vector perpendicular to the midplane of the ring. (Only the field components acting on a particle on the reference orbit in a perfect machine are taken into account to explain the basic idea of the measurement method: $\vec E_r = E_r \vec u_r$ and $\vec B_y = B_y \vec u_y$, where $\vec u_r$ is the unit vector pointing radially outwards, $\vec u_z$ is the unit vector  co-linear with the velocity vector of the particle, and $\vec u_y$ is the unit vector defined such that $\vec u_y = \vec u_z \times \vec u_r$. Note that for the electric field to point inwards, $E_r < 0$.) Writing the relativistic form of Newton's second law for the reference particle in a perfect machine without any imperfections, and projecting it into the horizontal plane, it can be easily  shown  that
\begin{equation}
    \frac{1}{\rho} = - \frac{q}{m \gamma \beta^2 c^2} E_r + \frac{q}{m \gamma \beta c} B_y\,.
\label{eq:rho}
\end{equation}
Now, making use of the Thomas--BMT equation of \Eref{eq2} from the executive summary and Eqs.\,(\ref{eq:omega_cycl}) and (\ref{eq:rho}), the spin motion of the reference particle is given by the subtracted Thomas--BMT equation, extended to cover
spin precessions due to EDMs\,\cite{BMT,FandS}:
\begin{equation}\label{emeq1}
\frac{{\rm d}\vec S}{{\rm d}t}=\left[\left(\vec\Omega_{\rm MDM}
+\vec\Omega_{\rm EDM}\right) -\vec \Omega_{\rm cycl} \right] \times{\vec S} \,,
\end{equation}
where
\begin{align}
\vec\Omega_{\rm MDM} - \vec \Omega_{\rm cycl} & =  -\frac{q}{m}\left[ G \vec B_y
 - \left( G - \frac{1}{\gamma^2-1}\right)\frac{\vec{\beta} \times \vec E_r}{c} \right] \label{emeq2} \, , \\
\vec\Omega_{\rm EDM}
& =   -\frac{\eta q}{2mc}
\left[\vec E_r + c \vec{\beta}\times \vec B_y \right]\,. \label{emeq3}
\end{align}
In \Eref{emeq1}, $\vec{S}$ is the spin vector in units of $\hbar$, defined in the Frenet--Serret frame of the reference particle,
and $t$ is the time in the laboratory frame of reference. The dimensionless parameter $\eta$ is related to the EDM $\vec d$ through
\begin{equation}
\label{emeq5}
\vec d =\eta\frac{q\hbar}{2mc}\vec S\,,
\end{equation}
with $S=1/2$ for protons and helions, and $S=1$ for deuterons;
$m$ and $q$ are the rest mass  and charge of the considered particle, respectively.

In addition, it is important to note that the form of the Thomas--BMT equation shown in Eqs.\,(\ref{emeq1}) to (\ref{emeq3}) does not include the effects of gravity. The effects of gravity, which have been studied by several authors\,\cite{Orlov:2019gtt, Obukhov:2016vvk, Silenko:2006er, Nikolaev_Spin2018},
are described in  Appendix\,\ref{app:gravity}.

\section{The storage ring EDM search}
The search for a  signal of an EDM using the storage ring method relies on the direct observation of the rotation of the electric dipole $\vec d$, and thus the spin in the presence of an external electric field that is perpendicular to the axis of the particle spin\,\cite{Farley}. The particles being studied are formed into a beam that is spin polarized, and the changes in the polarization components are measured on the beam as a whole, while it is confined in the ring. However, the MDM can also contribute to the polarization build-up in the same way that the EDM does. Thus, the main idea of the storage ring EDM search (in a perfect machine) is to maintain the spin frozen along the momentum direction in order to nullify the MDM contribution and maximize the EDM signal build-up; hence, the frozen spin concept, which we discuss  next.

\subsection{Frozen spin concept}
To simplify the discussion, we shall assume that the particle is moving on the reference orbit in a perfect machine, such that the only fields acting on it are the bending fields, $\vec B_y$ and $\vec E_r$, as illustrated in \Fref{fig:coordsystem}. Then, from \Eref{emeq2}, a general relationship between the fields can be established that sets the spin precession frequency due to the MDM (or $g-2$ precession) to zero in the Frenet--Serret frame of the particle,
\begin{equation}
    G \vec B_y - \left(G - \frac{1}{\gamma^2-1} \right) \frac{\vec{\beta} \times \vec E_r}{c} = 0\,,
\label{eq:BE_relat}
\end{equation}
and the radial $E$ field that is sensed by the EDM is given by
\begin{equation}
     B_y = E_r \cdot \frac{\beta^2 \gamma^2 G - 1}{c \beta \gamma^2  G } \,.
    \label{eq:Er-as-fct-By}
\end{equation}
In other words, for each energy, there exist ($B_y, E_r$) combinations such that the spin precession frequency due to the MDM equals the particle angular velocity. Thus, if the EDM contribution is disregarded and the initial condition begins with the spin parallel to the velocity, the spin will remain frozen in the horizontal plane along the momentum direction. However, in the presence of an EDM, the spin will precess around the radial axis, leading to a vertical spin component, as  sketched in \Fref{fig:EDsEDMring}.

\begin{figure} [hb!]
\centering
\includegraphics[scale=0.75]{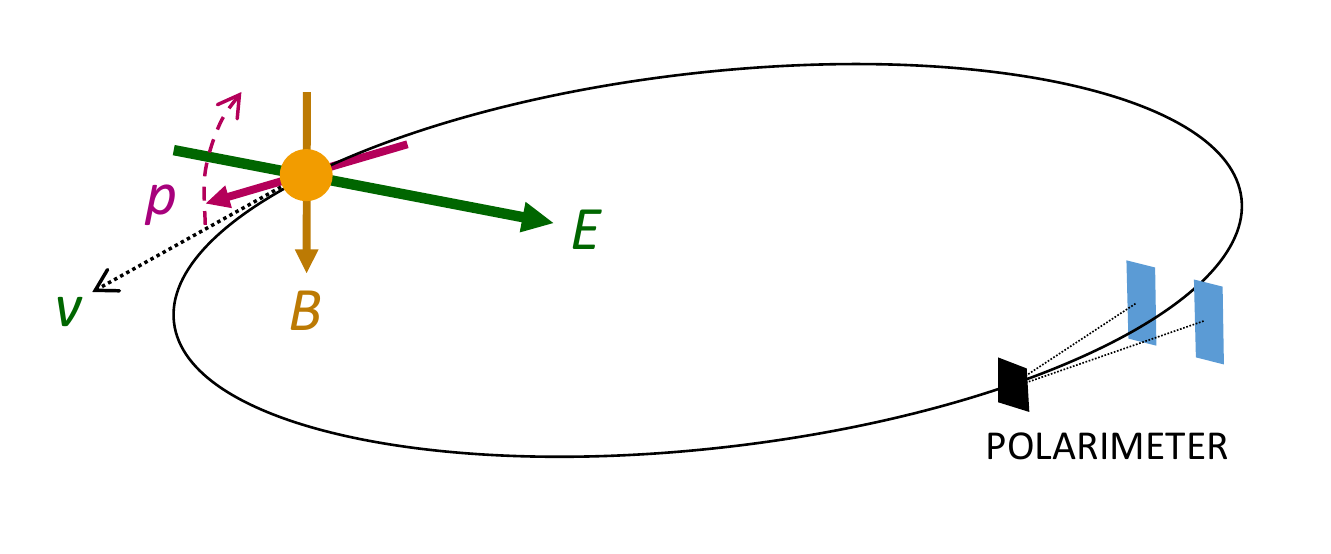}
\caption{\label{fig:EDsEDMring}
Frozen spin EDM ring with the beam circulating in the counterclockwise (CCW) direction. The particle is kept in the orbit by a  vertical magnetic field
and a radial electric field, created using field plates, satisfying \Eref{eq:Er-as-fct-By}, to maintain longitudinal polarization for particles with an MDM only. The indicated field directions correspond to a particle with $G > 0$, below the magic momentum (\Eref{eq:magic-momentum}). The electric field in the particle rest frame points towards the centre of the ring and, in the case of a finite EDM, rotates the spin out of the horizontal plane into the vertical direction, as indicated by the dashed arrow. The corresponding vertical polarization component can be observed in the left--right asymmetry of scattering from a target (black) into the detectors (blue).}
\end{figure}

Furthermore, if the anomalous magnetic moment $G$ of the particle is positive then, from \Eref{eq:BE_relat}, the frozen spin condition can be satisfied for an all-electric ring and for one specific momentum, which is generally referred to as the magic momentum $p_\mathrm{m}$,
\begin{equation}
    p_\mathrm{m} = \frac{mc}{\sqrt{G}}\,.
    \label{eq:magic-momentum}
\end{equation}
For the proton, this corresponds to a momentum $p_\mathrm{m} = 700.740\UMeVc$, \ie to a particle kinetic energy of $232.8\UMeV$.

For particles with a negative anomalous magnetic moment, such as deuterons or helions, there is no magic momentum and a combination of radial electric and vertical magnetic fields is necessary to achieve the frozen spin condition. In the  case that
$-1 < G <0$ (as for the deuteron), the electric field must be pointing away from the centre of the ring ($E_r >0$), thus reducing the bending of the beam from magnetic fields alone (see \Eref{eq:rho}). This yields an increase in the ring circumference.

\Figure[b]~\ref{fig:EDsEDMring} illustrates the frozen spin concept, where $\vec v$ is the particle velocity along the orbit, $\vec B$ and $\vec E$ are possible external fields (acting on a positively charged particle), and the spin axis is given by the purple arrow, which rotates in a plane perpendicular to $\vec E$. If the initial condition begins with the spin parallel to the velocity,  the rotation caused by the EDM will cause the vertical component of the beam polarization to change. This becomes the signal observed by a polarimeter located in the ring. This device allows beam particles to scatter from nuclei in a fixed target. The difference in the scattering rate towards the left and right sides of the beam is sensitive to the vertical polarization component of the beam. Continuous monitoring by a pair of detectors, illustrated in blue in \Fref{fig:EDsEDMring}, will show a change in the relative left--right rate difference during the beam storage time if a measurable vertical spin component due to an EDM (or perturbations, as described in the next paragraph) is generated. A practical consideration is the need for a sufficiently high polarimeter efficiency, which is the case for magic energy protons (see Chapter\,\ref{Chap:polarimetry}).

Under realistic conditions, beam particles will execute transverse `betatron' and longitudinal `synchrotron' oscillations in an imperfect machine constructed with finite mechanical tolerances, positioning errors of elements and stray fields from surrounding structures. Various effects can rotate the spin from the longitudinal into the vertical direction, even without an EDM, and may lead to systematic errors in the measurement. An example for the case of the proton EDM in the `frozen spin' scenario with an electric field only is a residual magnetic field. To mitigate this effect, the proposal includes
 the installation of state-of-the-art magnetic shielding around the ring, reducing the residual field to about \SI{1}{nT}. Even with such shielding, the residual radial magnetic field couples to the MDM and is expected to limit the sensitivity of the experiment to values well above $\SI{e-29}{\text{$e$
cm}}$.  Measures to further mitigate the effect due to the average radial magnetic field are described in \Sref{section:dual}. A more thorough analysis of systematic effects is given in Chapter \ref{Chap:SensSys}.

The kinematic diagrams in Figs.\,\ref{fig:ProtonKin} and \ref{fig:DeuteronKin} show the momentum and ring radius, respectively, as a function of the electric and magnetic fields available to fulfil the frozen spin condition for protons and deuteron beams. For the case of deuterons, no purely electric `frozen spin' solution exists; this is consistent with the observation in \Fref{fig:DeuteronKin} that none of the curves crosses the horizontal axis. The red dots in \Fref{fig:ProtonKin} labelled `pure electric ring' are for a realistic electric field of \SI{8}{MV/m,}  corresponding to a bending radius of about \SI{52}{m}. The red stars and
dots labelled `prototype ring' in Figs.\,\ref{fig:ProtonKin} and \ref{fig:DeuteronKin},
respectively,  are motivated by the prototype ring described in Chapter\,\ref{chap:ptr}, with a bending radius of \SI{8.9}{m}, as given in \Tref{tbl:BasicBeamParams}. The energy is limited by the electric field around \SI{7}{MV/m}; for protons, the `frozen' spin condition is fulfilled with a kinetic energy of \SI{45.2}{MeV}  and a magnetic field of \SI{0.0326}{T} (see \Fref{fig:ProtonKin}; both electric and magnetic fields deflect in the same direction).

\begin{figure}
\centering
\includegraphics[scale=0.31]{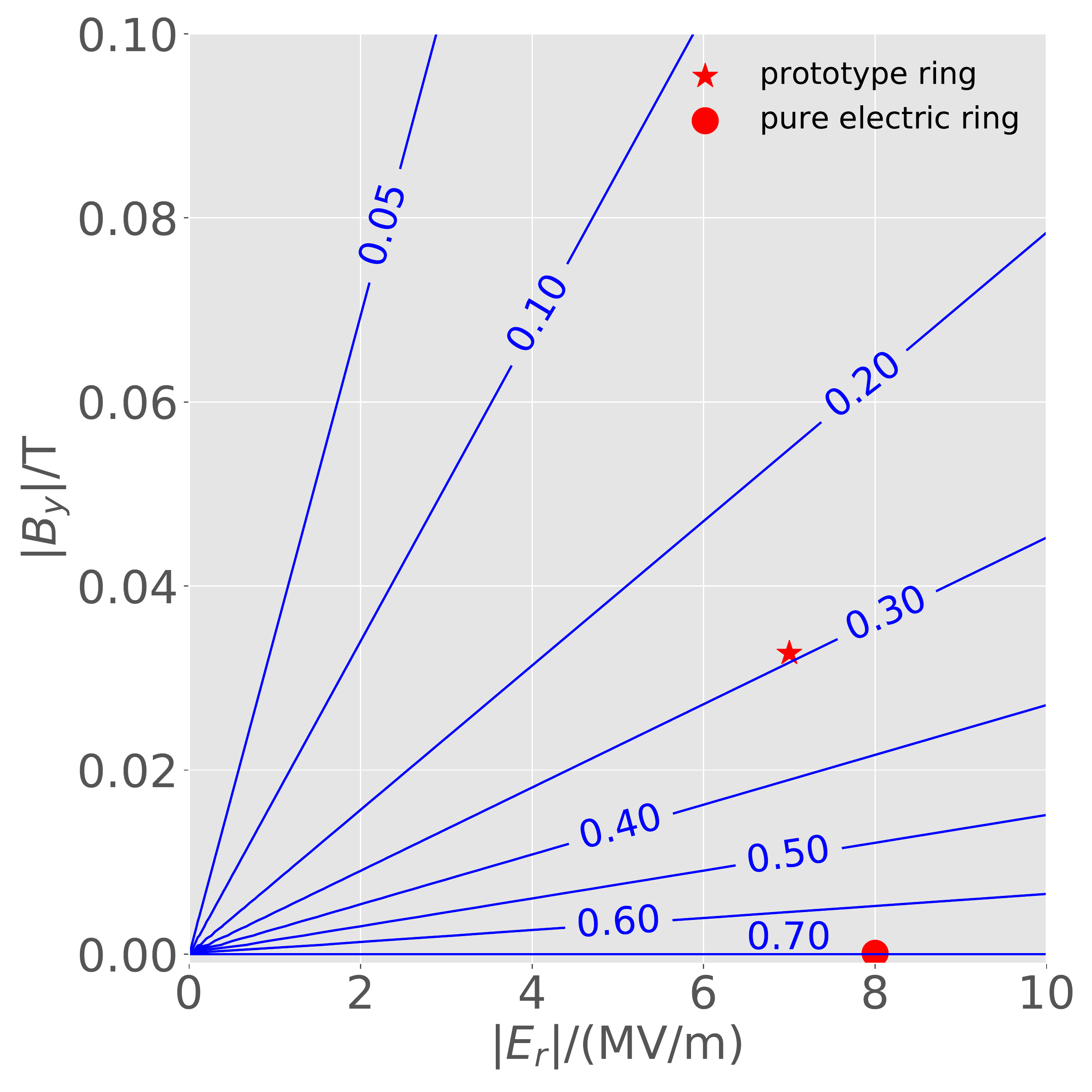}
\includegraphics[scale=0.31]{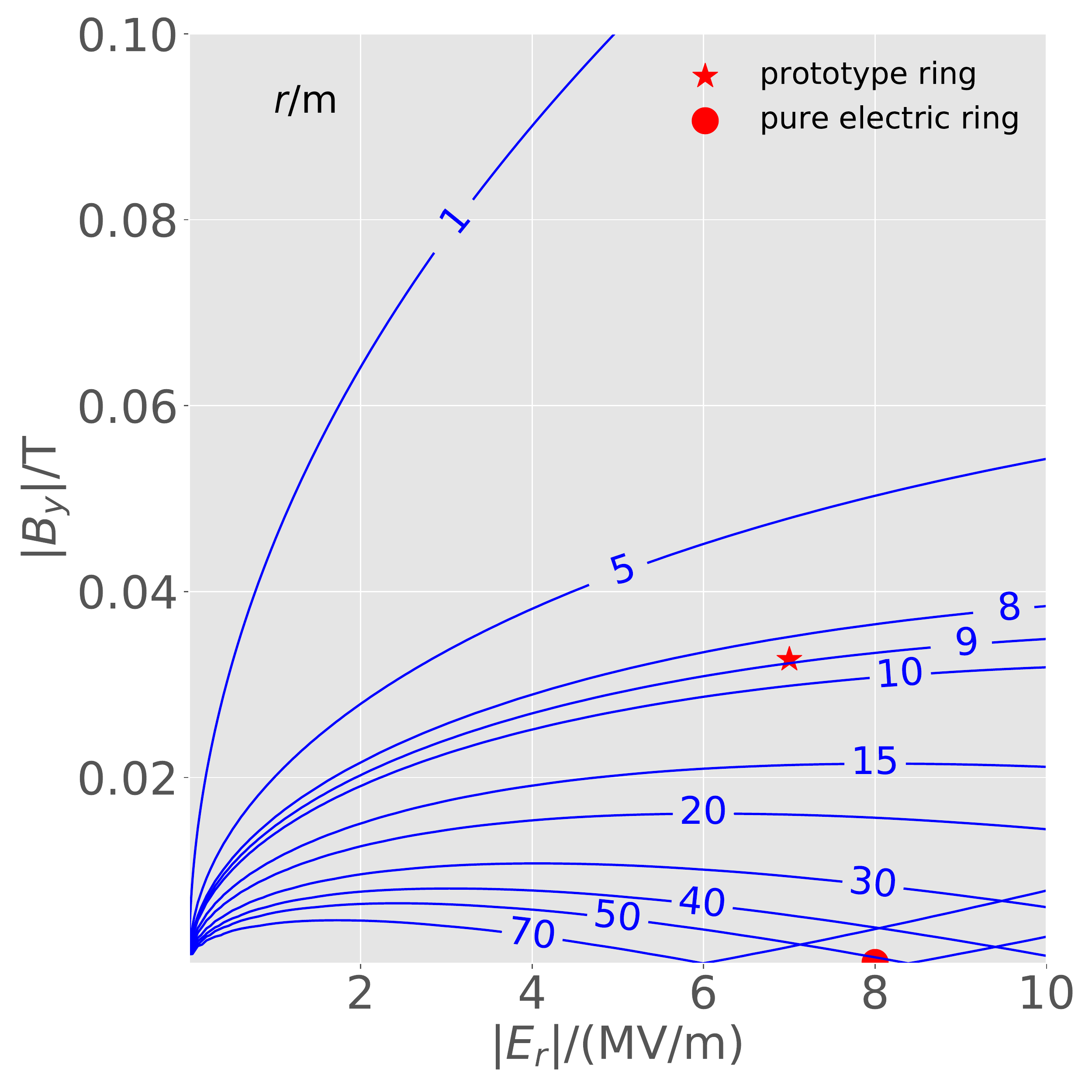}
\caption{\label{fig:ProtonKin} Proton momentum $p$ (left) and (right) storage ring bending radius $r$, for different frozen spin combinations of electric and magnetic fields (absolute field values are shown). For the pure electric ring, the momentum is fixed at \SI{0.7007}{GeV/$c$}.}
\end{figure}

\begin{figure}
\centering
\includegraphics[scale=0.30]{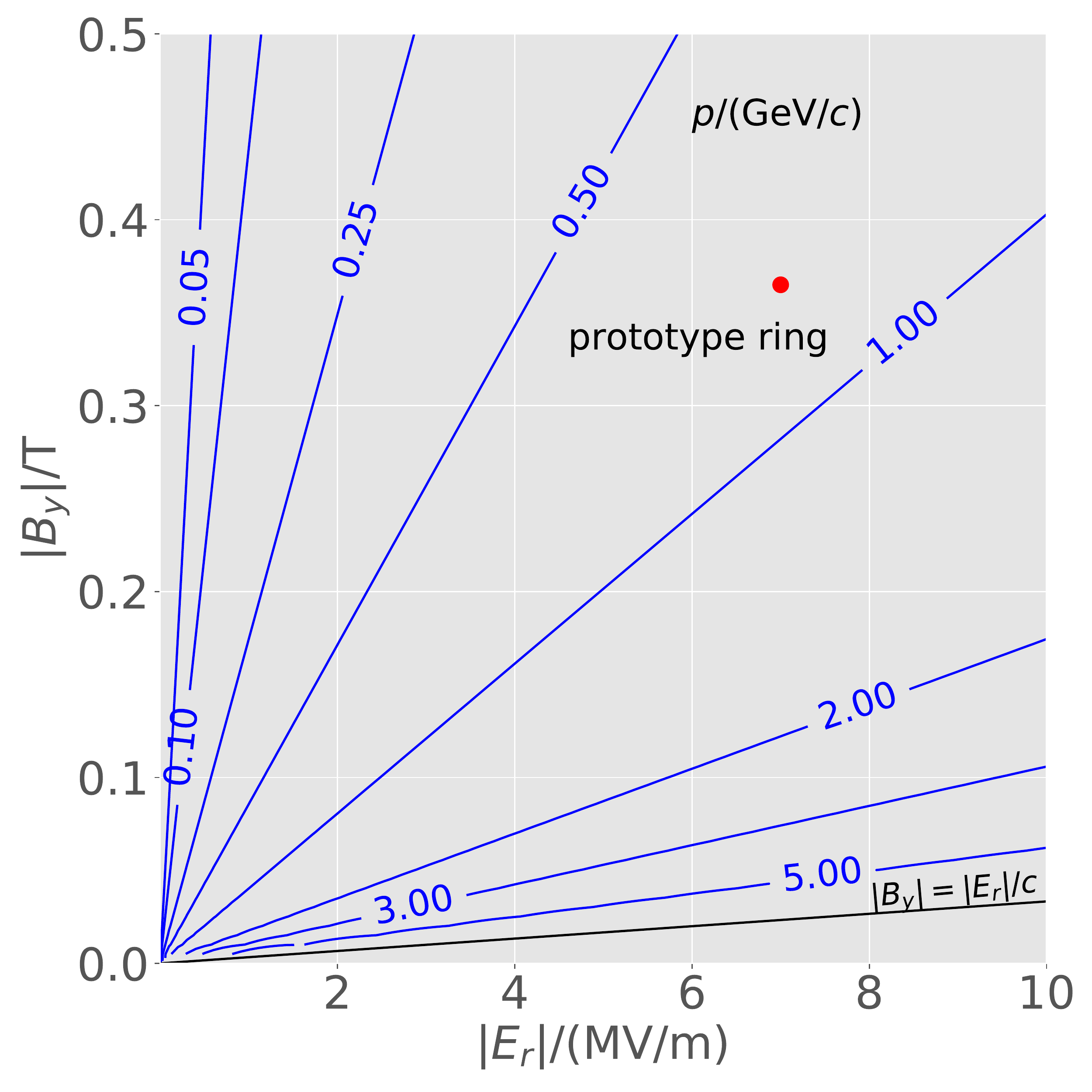}
\includegraphics[scale=0.30]{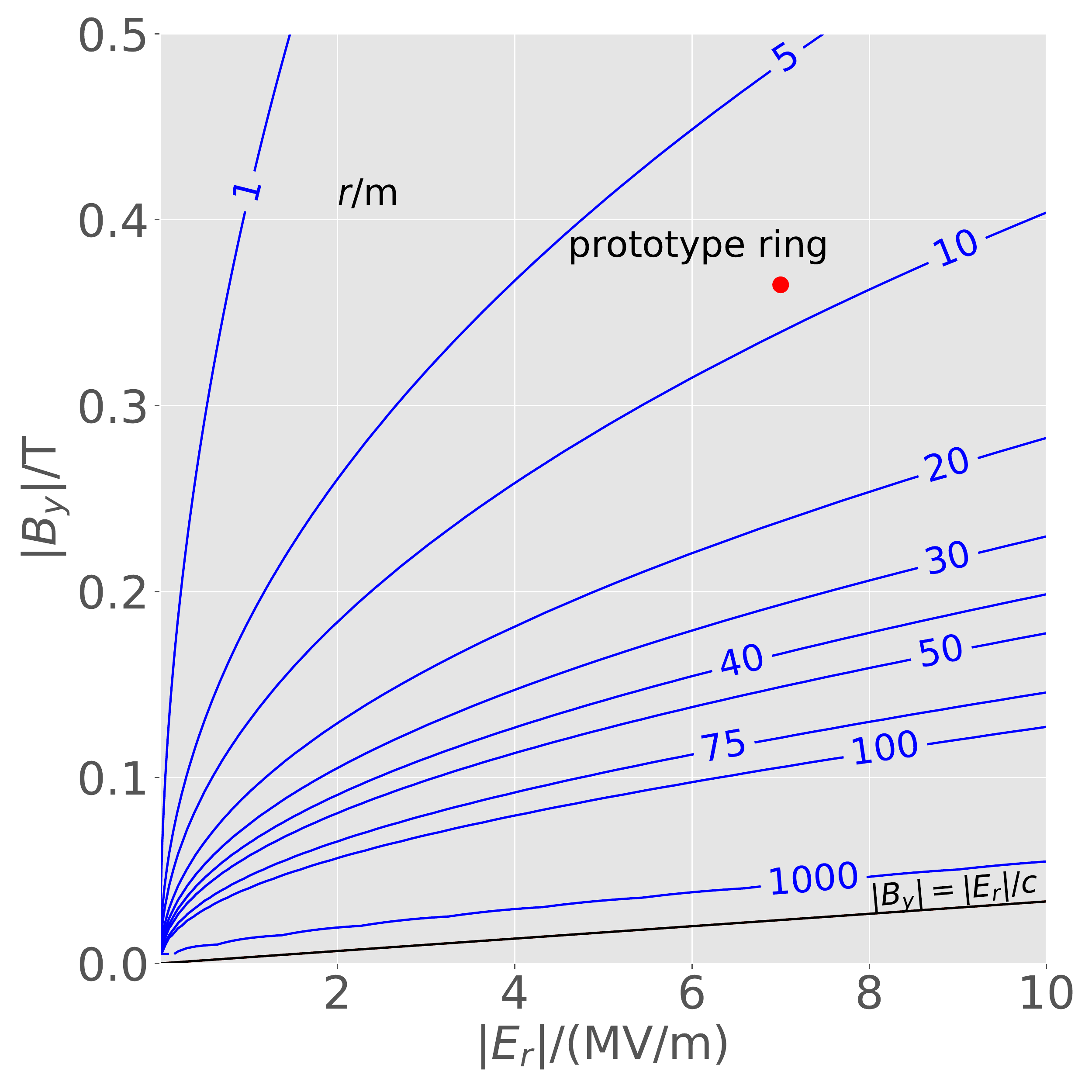}
\caption{\label{fig:DeuteronKin}Deuteron momentum $p$ (left) and storage ring bending radius $r$ (right), for different frozen spin combinations of electric and magnetic fields (absolute field values are shown).}
\end{figure}

For completeness, for deuterons, the `frozen spin' condition would be fulfilled
 for a kinetic energy of \SI{164.4}{MeV} by reversing the electric field, adding a magnetic field of \SI{0.360}{T}, indicated as red dots in \Fref{fig:DeuteronKin}. Thus, the latter figure includes only the mixed-field prototype ring operating point for the deuteron at a much higher magnetic field than is required for the proton.

\subsection{Dual beam operation}
\label{section:dual}
The large size of MDM effects compared with EDM effects also means that any storage ring experiment is sensitive to problems that might arise from issues---fringe fields, component alignment, stray electromagnetic interference, etc.---with the design and construction of the physical machine. One strategy for dealing with these problems in general is based on the realization that the EDM is time-reversal violating, while the majority of the problems are time-reversal conserving. The experiment could be changed to a time-reverse of itself by inverting the directions of all velocities, reversing all spins, and reversing all magnetic fields, while maintaining the electric fields as it is. In this case, where the time-reversed beam travels inside the same machine as the initial experiment and is subject to all of the same imperfections as the original experiment, the results could be compared directly. In other words, addition of the measured rotations of the two counter-rotating beams will cancel all machine-related systematic imperfections, such that the remaining part will correspond to the EDM signal (twice).

Nevertheless, if a residual radial magnetic field does not reverse between the two counter-rotating beams, this will yield a signal mimicking the EDM one. For the all-electric proton storage ring concept with a ring circumference $C = \SI{500}{m}$, an average radial magnetic field as low as $B_r = \SI{9.3}{aT}$ will generate the same vertical spin component as an EDM signal of $\SI{e-29}{\text{$e$}.cm}$\footnote{The sensitivity of the measurement must be evaluated carefully and may well not reach  $\SI{e-29}{\text{$e$}.cm}$.}.
This is probably the most serious systematic imperfection that needs to be corrected to reach the high-sensitivity goal of the experiment. The first line of defence against such magnetic fields is shielding. State-of-the-art multilayer shielding with degaussing procedures can reduce the ambient field to the \SI{1}{nT} level. Noting, in addition, that such a residual radial magnetic field does separate the orbits of the counter-rotating beams vertically,  the idea to remediate such an imperfection is to operate the machine with a low vertical tune, \ie with weak vertical focusing, to maximize the separation between the two beams. The latter will be measured with ultrasensitive SQUID magnetometers. For instance, with a vertical tune $Q_y = 0.1$, the same radial magnetic field of \SI{9.3}{aT} leads to an average orbit separation of \SI{5}{pm}. The measured vertical separation of the two counter-rotating beams will be reduced by an additional radial magnetic field to compensate. This method and, in particular, its limitations, are discussed further discussed in Chapter \ref{Chap:SensSys}.

\subsection{General possibilities}

Various categories of EDM storage ring are shown in \Tref{tab:exp:table1}. Of these, only the proton cases are seriously analysed in this report. The deuteron and electron cases have been mentioned earlier in the report, but are not described in any more detail. The all-magnetic case is exploited to the extent possible in the COSY precursor experiments (see Chapter\,\ref{Chap:Precursor}). However, frozen spin is not possible with only magnetic bending and an EDM measurement is possible only because an RF Wien filter synchronized to the polarization precession rate breaks the cancellation that prevents an EDM signal accumulation\,\cite{PhysRevAccelBeams.23.024601}.

\begin{table} [h]
\centering
\caption{General possibilities according to BMT equation}
\label{tab:exp:table1}
\resizebox{\columnwidth}{!}{%
\begin{tabular}{l l l l l l}\hline \hline
Field          & Particle             & $G$ factor   & Kinetic      &   Beams      &    Comment                         \\
configuration  &  type                &              & energy &  CW, CCW      &                                    \\
& &  & (MeV)  \\ \hline
 All-electric  & p                  & $+1.792847$   &     \phantom{2}232.8    & Concurrent   &  Final ring, prototype required    \\
               & e                  & $+0.001160$   &     \phantom{29}14.5     & Concurrent   &  Challenging polarimetry           \\
               & $\mu$                & $+0.001165$   &    2991      & Concurrent   & Impractically short lifetime       \\ 
 $E$ and $B$ combined  & p                  & $+1.792847$     &      Variable     & Consecutive  &  Compromised EDM precision         \\
               & d                  & $-0.142987$     &    Variable  & Consecutive  &  $E$ and $B$ technological challenge       \\
               & $^3${\sf He}$^{++}$  & $-4.183963$     &      Variable     & Consecutive  &  Must develop polarimetry          \\ 
 All-magnetic  &                      &              &              &              &  Used for precursor, no            \\
               &                      &              &              &              &  frozen spin possibility           \\ \hline \hline
\end{tabular}
}
\end{table}

Details of the ring design may be found in other chapters of this report: Chapter\,\ref{Chap:Precursor} describes the COSY precursor experiment,  based on deuterons; Chapter\,\ref{chap:ptr} describes the EDM prototype ring; and Chapter\,\ref{Chap:allelectricring} describes the all-electric proton EDM ring. The route towards the final ring, \ie the all-electric proton EDM ring will be explained in Chapter\,\ref{Chap:Strategy}. In what follows, we discuss the experimental observable and the basic measurement sequence.

\subsection{Experimental observable: beam bunch polarizations} \label{Chap:polarisation-observables}
As described in Ref.~\cite{Anas}, the \SI{232.8}{MeV} proton ring has a \SI{500}{m} circumference and a confining electric field of \SI{8}{MV/m} in the bending sections.
The accumulation rate for a signal corresponds to a rotation of the polarization according to
\begin{equation}
\Omega_\text{EDM} =\frac{2 E_r\,d}{\hbar}\; ,
\label{emeq9}
\end{equation}
where  the value \SI{5.27}{MV/m} should be applied
for the  average field around the circumference.
For an EDM of $d = \SI{e-29}{\text{$e$}.cm}$, the rate would be about $1.6 \times \SI{e-9}{rad/s}$.

The plan for an EDM-sensitive polarization measurement is to record the horizontal asymmetry in the scattering of sampled protons from a carbon target at forward angles. At the energies where the EDM search would be made, the interaction  between the polarized protons and the carbon nucleus contains a large spin-orbit term. This gives rise, in elastic scattering, to an asymmetry between left- and right-going particles when  a vertical polarization component is present. For a complete description of polarization observables and effects, see, \eg Ref.~\cite{polarizeref}.

For  spin-{\textonehalf}  particles, this effect is described by the differential scattering cross-section given in \Eref{emeq10}, with the angles defined in \Fref{fig:polcoord}. The polarization along any given axis is given in terms of the fraction of the particles in the ensemble whose spins are shown, through some experiment,  to lie either parallel or antiparallel to that axis. If these fractions are $f_+$ and $f_-$ with $f_+ +f_- =1$ for the two projections of the proton's spin-{\textonehalf}, the polarization becomes $p=f_+ -f_-$ and ranges between 1 and $-1$. The scattering cross-section $\sigma_{\rm POL}$ may be written in terms of the unpolarized cross-section $\sigma_{\rm {UNP}}$ as
\begin{equation} \label{emeq10}
\sigma_{\rm POL}(\theta )=\sigma_{\rm UNP}(\theta )\left( 1+pA_y(\theta )\cos\phi\sin\beta\right)
\, ,
\end{equation}
with the vertical component given by
\begin{equation}
p_y =p\cos\phi\sin\beta\, .
\label{emeq11}
\end{equation}
The angles are defined with respect to the coordinate system shown in \Fref{fig:polcoord}, in which a particle from the beam, travelling in the $+z$ direction, is scattered by an absorber into the $+x$ or `left' side of the $xz$ plane. The scattering angle is $\theta$. The polarization direction, shown as the red arrow, is defined by the two polar coordinate angles $\beta$ and $\phi$. The polarization effect reverses if the particles are detected at the same $\theta$ on the $-x$ or `right' side of the beam, owing to the $\cos\phi$ dependence in \Eref{emeq10}. Thus, this left--right asymmetry measures the vertical polarization component $p_y$. The size of the signal is governed by the strength of the spin-orbit interaction, which gives rise to the asymmetry scaling coefficient $A_y(\theta )$, otherwise known as the analysing power. The complete differential spin-dependent cross-section for polarized protons impinging on an unpolarized target, expressed in terms of the Cartesian components of the polarization vector $\vec p =(p_x,p_y,p_z)$, can be written as
\begin{equation}
\frac{\sigma_\text{POL}}{\sigma_\text{UNP}}(\theta) = 1 +   A_y(\theta ) \left[ p_y \cos(\Phi) - p_x \sin (\Phi) \right]\,,
\label{eq:pC-sigma-px-py}
\end{equation}
where the angle $\Phi$ denotes the azimuthal angle of the scattered proton, and the parity-violating dependence on the longitudinal polarization component $p_z$ has been omitted because of its smallness. (For more details on the formalism used for the scattering of polarized protons, see, \eg Ref.~\cite{Rathmann:1998zz}.)

\begin{figure}
\centering
\includegraphics[scale=0.45]{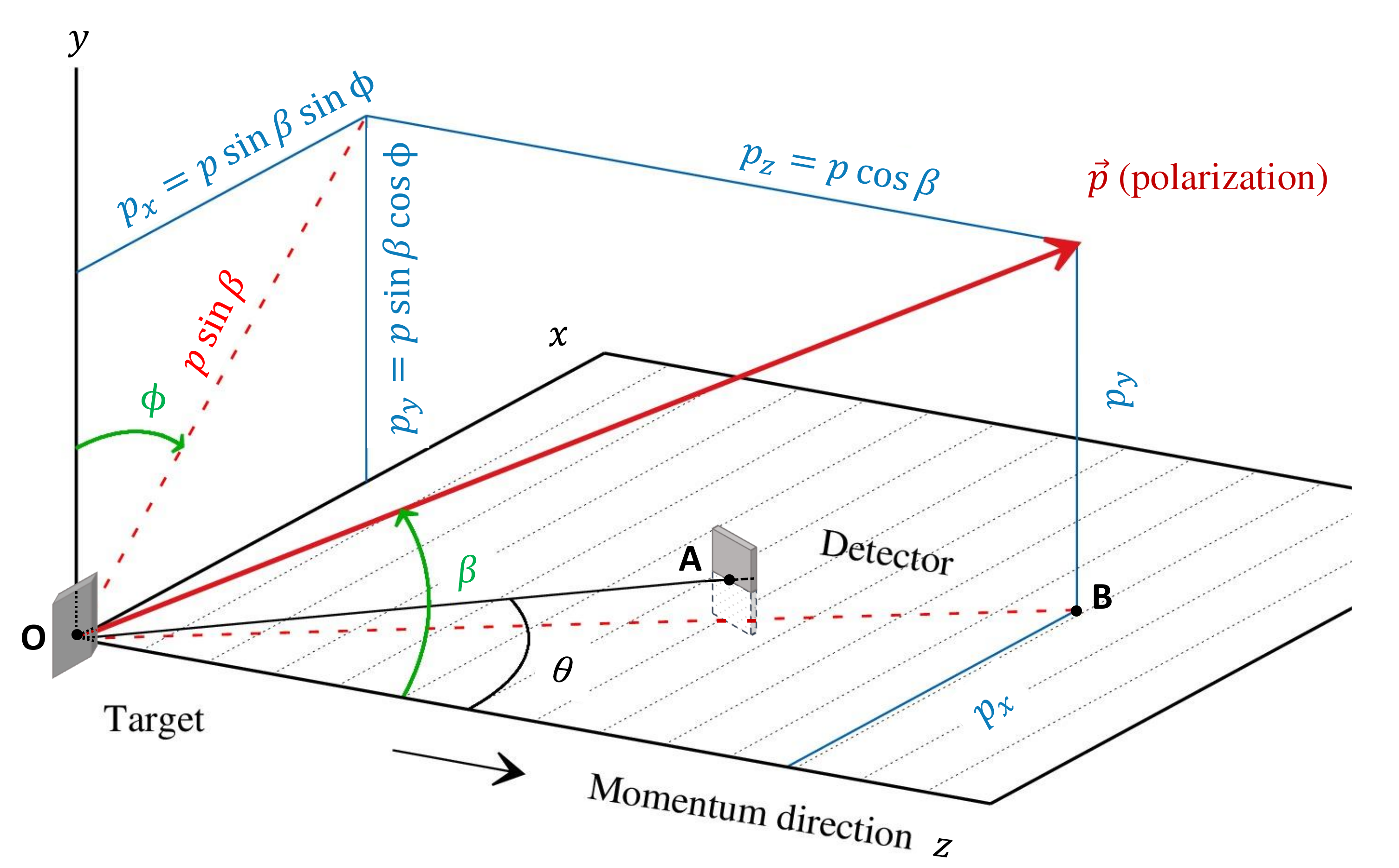}
\caption{\label{fig:polcoord}
Coordinate system for polarization experiments, where the incoming beam defines the $z$ axis, and  the particle is scattered in the $xz$ plane (see, \eg Fig. 6.1 of Ref.~\cite{schieck2014nuclear}). The points $\mathbf O$, $\mathbf A$, and $\mathbf B$ lie in the $xz$ plane, as does  the detector located at the scattering angle $\theta$, which is used to determine the spin-dependent cross-section (see \Eref{emeq10}). The angles defining the orientation $\vec p$ are indicated (see \Eref{emeq11}). Note that $|\vec p\,| = 1$.}
\end{figure}

In the case of the deuteron, which is spin-1, there are three fractions that describe the magnetic substate population, $f_+$, $f_0$, and $f_-$, where $f_+ + f_0 + f_- = 1$. The two polarizations are vector ($\mathrm{V}$), $p_\mathrm{V} = f_+ - f_-$, and tensor ($\mathrm{T}$) $p_\mathrm{T} = 1 - 3 f_0$, which can range from $+1$ to $-2$. If we are interested only in the EDM, the vector polarization suffices as a marker and the deuteron polarized cross-section (Cartesian coordinates, following the Madison  Convention~\cite{polarizeref}) becomes
\begin{equation}
\sigma_{\rm POL}(\theta ) = \sigma_{\rm UNP}(\theta) \left( 1 + \frac{3}{2} p_\mathrm{V} A_y(\theta) \cos\phi\sin\beta\right)\, .
 \label{emeq10-spinone}
\end{equation}
Tensor polarization is usually present to a small degree in polarized deuteron beams. There are three independent tensor analysing powers, which each add another `$p_\mathrm{T} A$' term to this equation. Their effects may prove useful in polarization monitoring or checking for systematic effects. Because this report explores the possibility of a proton storage ring, the deuteron spin dependence will not be further elaborated here. The corresponding description of the complete spin-dependent differential cross-section for deuterons impinging on an unpolarized target in terms of the Cartesian components of the vector polarization  $\vec p_\mathrm{V} = \left(p_{\mathrm{V}_x}, p_{\mathrm{V}_y}, p_{\mathrm{V}_z} \right)$  reads
\begin{equation}
\frac{\sigma_\text{POL}}{\sigma_\text{UNP}}(\theta) = 1 + \frac{3}{2} A_y(\theta ) \left[ p_{\mathrm{V}_y} \cos(\Phi) - p_{\mathrm{V}_x} \sin(\Phi) \right]\,.
\label{eq:dC-sigma-px-py}
\end{equation}
Here, $\Phi$ denotes the azimuthal angle of the scattered deuteron; the parity-violating dependence on the longitudinal polarization component $p_{V_z}$ has been omitted because of its smallness. (For more details on the formalism used for the scattering of polarized deuterons, see \eg Ref.\cite{vonPrzewoski:2003ig}.)

In the energy range where we would like to run the EDM search, it happens that the spin-orbit interaction between light particles, such as the proton (and deuteron), with the carbon nucleus provides a large analysing power $A_y$ ($\sim$\,$60\%$) for beam particles that scatter elastically in the forward direction from nuclei in the target. A target a few centimetres thick is positioned at the edge of the beam in such a way that beam particles can be extracted onto its front face. In this way, all of the beam particles stored in the machine may be consumed, and up to 1\% of the particles scattered from a nucleus in the target make their way into one of the forward detectors. This constitutes a very high efficiency for using the beam particles to search for any sign of an increasing vertical polarization component.

\subsection{Basic measurement sequence}
A typical single measurement sequence is outlined next, with the aim of giving some notions of the overall approach. There are still many open questions, and it is clear that experience of operating, first, a prototype ring (see Chapter\,\ref{chap:ptr}) and, subsequently, the full ring (see Chapter\,\ref{Chap:allelectricring}) will be required to firmly establish the procedures. Details of the beam preparation process and data taking may be found in Chapter~\ref{Chap:polarimetry}.
\begin{itemize}
   \item Several bunches with vertically polarized protons are injected CW and CCW into the storage ring.
   \item Beams must be injected into the ring in both directions in reasonably rapid succession. The polarization begins perpendicularly to the ring plane.
   \item Using an RF solenoid, the spins of the particles are rotated into the horizontal plane.
   \item Subsequently, the beams are continuously extracted onto the target for $\approx$\,$\SI{1000}{s}$.
   \item The increase of the vertical polarization is proportional to the EDM, and is measured via the left--right counting rate asymmetry in the detector (see  Chapter \ref{Chap:polarimetry}).
   \item Averaging the polarization measurements from the CW and CCW rotating beams cancels some of the systematic effects (\eg some  geometrical phase effects).  Other effects (\eg residual radial magnetic fields) are determined from a spatial separation of the two beams (see Chapter  \ref{Chap:SensSys}).
\end{itemize}

Although this sequence does not explicitly include spin manipulation in the injector, we do not  generally want to exclude this option. This sequence may be repeated approximately $10^4$ times per year of operation. Note that, for a single store, the statistical effects will be more than two orders of magnitude larger than any EDM effect at the expected level of sensitivity.

\begin{flushleft}

\end{flushleft}
\end{cbunit}

\begin{cbunit}

\chapter{Strategy}
\label{Chap:Strategy}

\section{Introduction}
The project to search for charged-particle electric dipole moments (EDMs) in storage rings has a strong science case. It is crucial to improve the understanding of the limitations of the new technique and to provide an informed estimate of the achievable sensitivity, \ie the smallest EDMs that can be identified. At the same time, the project  faces demanding technological and metrological challenges. Moreover, it is obvious that high-precision measurements will require commitments for a long period of time. To justify the significant expenditures for the ring(s), it will be inevitable to outline a clear plan (see Chapter\,\ref{Chap:RoadMap}) for moving towards the ultimate goal of an all-electric polarized proton EDM facility with clockwise and counterclockwise beams concurrently operating at the magic momentum: this must include not only the verification of all key technologies, but also a demonstration that the aimed-for sensitivities are feasible. This has already begun, with several polarized beam techniques meeting the EDM experimental requirements. It is now clear that the only viable way to continue this is to pursue a staged approach with a prototype ring as the essential demonstration milestone (see \Fref{fig:strat:figure3}).

\begin{figure} [hb!]
\centering
\includegraphics[scale=0.45]{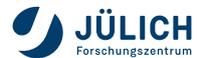}
\caption{\label{fig:strat:figure3}Important features of the proposed stages in the storage ring EDM strategy}
\end{figure}

\section{Starting point of the staged approach}
The charged-particle EDM project is in an excellent position to be pursued further, as there is a conventional (\ie using magnetic deflection) storage ring facility that provides all the required elements for R\&D and even allows a `proof-of-capability' measurement (see Chapter\,\ref{Chap:Precursor}). COSY, the cooler synchrotron at the Institute for Nuclear Physics (IKP) of Forschungszentrum J\"ulich (FZJ), Germany, is a storage ring for polarized proton and deuteron beams between 0.3 (p) or 0.55 (d) and \SI{3.7}{GeV/$c$}. Besides phase-space cooling (electron, stochastic), well-established methods are used to provide, manipulate, and investigate stored polarized beams.

Over the past decade, the  J\"ulich Electric Dipole Moment Investigations
(JEDI) collaboration has made significant progress using COSY as an EDM test facility (see Appendix\,\ref{app:cosy}). Currently, JEDI is conducting a precursor experiment (see the following and Chapter\,\ref{Chap:Precursor}) to obtain a first directly measured EDM limit for the deuteron by exploiting a radio-frequency (RF) Wien filter in the ring. The experiment is sensitive to the EDM through its effect on the direction of the invariant spin axis of the ring.  First measurements have been made and are currently being analysed. Additional measurements are planned for the first half of 2021.



\section{Route towards the final ring}
The proposed prototype ring (see Chapter\,\ref{chap:ptr}) offers new possibilities and is intended to address, on a smaller scale, issues that cannot be investigated by other means. These include electric field beam transport, with the possibility of storing two countercirculating beams using a ring lattice suitable for the final EDM experiment. With the addition of air-core magnetic bending, it becomes possible to `freeze' the beam polarization of stored protons along the direction of the beam velocity, thus allowing a more sensitive search for an EDM, compared with the precursor experiment. Tests will demonstrate the limits on beam storage and the precision of beam monitoring and control. Most importantly, systematic effects that limit the sensitivity in storage ring EDM experiments may be studied directly, along with efforts to mitigate them, in particular, the scheme for estimating and compensating the mean radial magnetic field from measurements of the vertical separation of the two counter-rotating beams.

It is the large number of uncertainties, the most fundamental of which, as mentioned previously, is the handling of the unavoidable residual magnetic fields, that currently prevents a realistic, fully fledged ring design beyond the previously published report\,\cite{doi:10.1063/1.4967465} (see Chapter\,\ref{Chap:allelectricring}). The final full-scale design will be an (essentially) all-electric \SI{233}{MeV} ring with simultaneously counter-rotating frozen-spin proton beams.

\subsection{Preparedness for the full-scale ring}
As part of the preparation of this report, the level of preparedness for construction of the full-scale ring was studied in considerable detail, with the results distilled in  Table\,\ref{tbl:Preparedness-Chap5}, in which  `lacks of preparedness' are sorted by perceived (colour-coded) `degrees of severity'. The column headed `References' is intended to guide the reader to more detailed discussions of the various aspects that are available elsewhere in this document.

\begin{table}
\caption{\label{tbl:Preparedness-Chap5}Status of  preparedness levels for the full-scale all-electric ring: green, `ready to break ground'; yellow, `promising'; red, `critical challenge'. Plus ($+$) and minus  ($-$) signs
are to be interpreted as for college course grades. Thus, the ranking, with most prepared first, is \colorbox{green}{$+$G$-$} \colorbox{yellow}{$+$Y$-$} \colorbox{red}{$+$R}. Success in meeting prototype ring goals could amount, for example, to upgrading all scores to Y($+$) or better. }
\centering
\begin{tabular}{l ll}   \hline \hline
Operations                       &  Rank    
& Reference             \\ \hline
Spin control feedback           & \cellcolor{green}G  
& Section\,\ref{appa:spincontrol} \\
Spin coherence time              & \cellcolor{green}G($-$)    
& Section\,\ref{appa:longsct} \\
Polarimetry                      & \cellcolor{yellow}Y     
& Chapter\,\ref{Chap:polarimetry} \\
Beam current limit               &  \cellcolor{red}R        
& Section\,\ref{ptr:30MeVgoals} \\
CW or CCW operation                 &   \cellcolor{red}R         
& Ref.\cite{doi:10.1063/1.4967465} \\
\hline
Theory                           &          
&
            \\ \hline
GR gravity effect                &   \cellcolor{green}G($+$)    
& Appendix\,\ref{app:gravity} \\
Intrabeam scattering             &    \cellcolor{yellow}Y     
& Ref.\cite{Lebedev:2015} \\
Geometric or Berry phase theory     &   \cellcolor{yellow}Y     
& Ref.\cite{Tahar:2019hzv}   \\ \hline
Components                       &          
&                       \\ \hline
Quads                            &    \cellcolor{green}G     
& Chapter\,\ref{Chap:Efields}  \\ 
Polarimeter                      &    \cellcolor{green}G     
& Chapter\,\ref{Chap:polarimetry}       \\
Waveguide Wien filter            &    \cellcolor{green}G     
& Section\,\ref{appa:rfwf}   \\
Electric bends                   &    \cellcolor{red}R($+$)   
& Section \,\ref{appa:deflectors}     \\ \hline
Physics and engineering           &          
&                       \\ \hline
Cryogenic vacuum                 &    \cellcolor{yellow}Y     
& Ref.\cite{vonHahn:2016wjl}   \\
Stochastic cooling               &    \cellcolor{yellow}Y     
& Ref.\cite{moehl2013stochastic}  \\
Power supply stability           &   \cellcolor{yellow}Y($-$)   
& Chapter\,\ref{chap:ptr}  \\
Regenerative breakdown           &  \cellcolor{red}R($+$)       
                                &  \\
\hline
EDM systematics                  &          
&                       \\ \hline
Polarimetry                      &   \cellcolor{green}G($-$)    
& Chapter\,\ref{Chap:polarimetry}  \\
CW or CCW beam shape matching       &    \cellcolor{yellow}Y     
& Chapter\,\ref{Chap:SensSys}  \\
Beam sample extraction           &    \cellcolor{yellow}Y     
& Chapter\,\ref{Chap:polarimetry},  Appendix\,\ref{app:extpol}  \\
Control current resettability    &    \cellcolor{yellow}Y     
& Ref.\cite{Fernqvist:2003} \\
BPM precision                    &   \cellcolor{yellow}Y($-$)   
& Chapters\,\ref{chap:ptr}, \ref{Chap:SensSys} \\
Element positioning and rigidity  &   \cellcolor{yellow}Y($-$)   
& Ref.\cite{Decker:2005mu}  \\ \hline
Theoretical analysis             &          
& Chapter\,\ref{Chap:SensSys} and refs. therein            \\
Radial $B$ field, $B_r$             &           
& Ref.\cite{doi:10.1063/1.4967465}       \\
\hline \hline
\end{tabular}
\end{table}

\subsection{Conclusion}
With only  the experience gained from the prototype ring, the information needed for a detailed design of the all-electric proton EDM ring (see Chapter\,\ref{Chap:allelectricring}) should be available. From prototype test results, we expect to be able to justify the technology that shall be employed and the sensitivity level to be achieved. Finally, detailed and realistic cost estimates will become available.

\section{Science case beyond EDM}
The rotation of the polarization (precession of the spin vector) involved in an EDM search may also couple to any oscillating EDM associated with a surrounding axion field\cite{Graham:2011qk,Chang:2019poy,PhysRevD.100.111301}. Data from the EDM search may be scanned, as has been proven possible in neutron and atomic EDM searches,  for evidence of an axion (see, \eg Ref.~\cite{Abel:2017rtm}). In addition, moving the EDM ring parameters away from the frozen spin condition enables a broader  search to be conducted. A first exploratory study to search for oscillating EDMs using a storage ring is discussed in Appendix \ref{Chap:Axions}. A first test was  carried out at COSY in the spring
of 2019 (see Section \ref{Chap:Axions-inital-test-with-beam}).

It may also be possible to find conditions where the counter-rotating beams obey frozen-spin requirements for different particle species,  allowing a class of high-precision comparisons of relative magnetic moments and EDMs, if they are observable. Thus, the EDM ring will become a facility for different experimental programmes with the  potential for discoveries at the frontier of new science.

\begin{flushleft}

\end{flushleft}
\end{cbunit}

\begin{cbunit}

\chapter{Precursor experiment}
\label{Chap:Precursor}

\section{Introduction}
The first stage of the storage ring EDM strategy, shown in \Fref{fig:strat:figure3}, is a set of `proof-of-capability' tests, referred to as the `precursor experiment'. It is performed at the Cooler Synchrotron (COSY) at the Forschungszentrum J\"ulich, Germany, which is a magnetic storage ring providing polarized protons and deuterons in the momentum range $0.3$ to \SI{3.7}{GeV/$c$} (see \Fref{fig:cosy}). The aim of the precursor experiment is to measure the EDM of deuterons using the magnetic COSY ring.

\begin{figure} [hb!]
\centering
\includegraphics[width=0.5\textwidth]{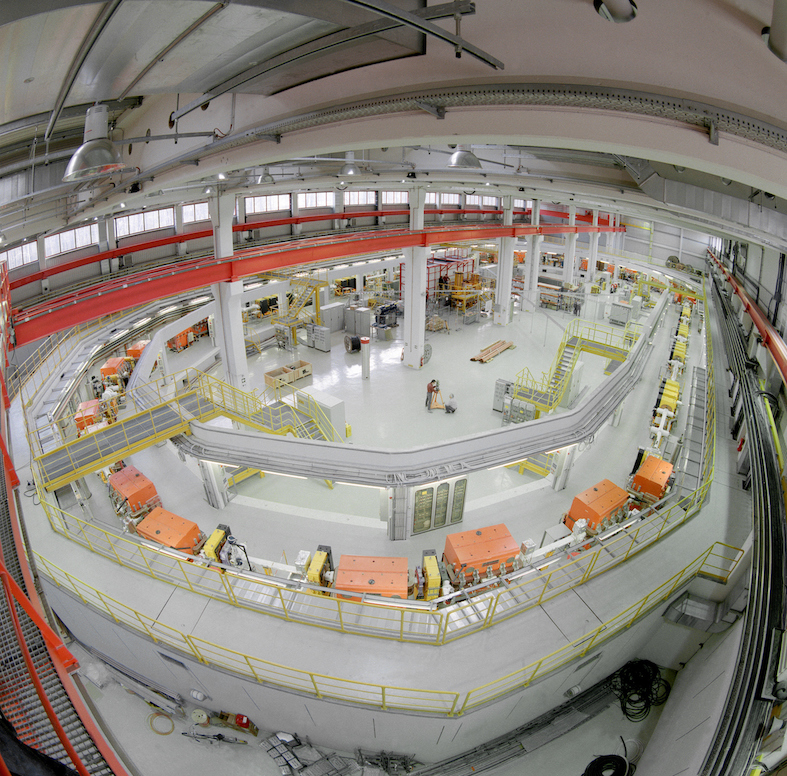}
\caption{Cooler Synchrotron (COSY), Forschungszentrum J\"ulich, Germany\,\cite{Maier:1997zj,PhysRevSTAB.18.020101}}
\label{fig:cosy}
\end{figure}

\section{Principle of the measurement}
\label{Precursor-Principle-of-Measurement}

In a purely magnetic storage ring,  operation with frozen spin is not possible. The spin precession of the polarization vector in the horizontal plane prevents a build-up of a vertical polarization due to the EDM. The EDM just causes an oscillation of the vertical polarization component with an amplitude $\xi = \arctan(\beta \eta/(2G))$\footnote{The dimensionless factor $\eta$ is related to the EDM $d$ via the relation $d = \eta ({e \hbar} / {2 m c}) S$ (see \Eref{emeq5}).}. This signature has been used in the muon $g-2$ experiment to  measure the muon EDM\,\cite{Chislett:2016jau}. For hadrons, this method is less sensitive because $|G_{\mathrm{hadron}}| \gg G_{\mu}$. The precursor experiment is performed at a deuteron momentum of $p=970\,\si{MeV}/c$. In this case, the vertical component of the oscillating polarization due to a hypothetical deuteron EDM value of $d = \SI{e-20}{\text{$e$}.cm}$ amounts to about $\xi = 3 \times 10^{-6}$. The spin dynamics of the precursor experiment are  discussed extensively in Ref.~\cite{PhysRevAccelBeams.23.024601}.

A method that involves using an RF Wien filter in the ring allows one to \textit{accumulate} the EDM-induced spin rotations when the spins are not `frozen'\,\cite{Rathmann:2013rqa,PhysRevSTAB.16.114001,PhysRevAccelBeams.23.024601}.
The RF Wien filter used for this purpose generates a vertical magnetic and a radial electric field that oscillate with a frequency $f_\text{WF}$\,\cite{Slim:2016pim}. The fields must be adjusted such that the device is operated as a Wien filter, \ie the beam particles are not deflected, because the Lorentz force vanishes. The particle spins are rotated around the vertical axis through an angle with amplitude $\psi_\text{WF}$ (corresponding to the maximum electric and magnetic field of the Wien filter). When the spin rotations induced by the RF Wien filter are \textit{in resonance} with the in-plane spin precessions of the stored particles, the EDM-induced spin rotations can be \textit{accumulated} as a function of time.

For the precursor experiment, a suitable RF Wien filter has been developed and constructed in the framework of the JEDI collaboration \cite{Slim:2016dct,Slim:2016pim,PhysRevSTAB.18.020101}, and was installed at COSY in May 2017 (see Appendix\,\ref{app:cosy} and 
\Fref{fig:cosy:wf} therein). To generate a polarization build-up, the RF Wien filter must be operated in resonance with the spin precession. The resonance condition is given by
\begin{equation}
  f_{\mathrm{WF}}  =   K\cdot f_\text{rev} + f_\text{s} = (K + \nu_\text{s})f_\text{rev} \,, \qquad \text{where} \qquad K \in \mathbb{Z}\,,  \\
  \label{eq:wien-filter-frequency}
\end{equation}
and the spin tune $\nu_\text{s}$, defined as the number of spin revolutions per turn, is given by
\begin{equation}
 \nu_\text{s}  = \frac{f_\text{s}}{f_\text{rev}} = \frac{G\gamma}{\cos \xi}\,.
 \label{eq:nus-with-EDM}
\end{equation}
Here, the  $\cos \xi$ term in the denominator reflects   the way in which
the EDM modifies the spin tune.

During the experiments at COSY, at a deuteron momentum of about $\SI{970}{MeV/c}$, the RF Wien filter may be operated at a frequency\footnote{The complete list of frequencies available using the RF Wien filter is given in Table\,1 of
Ref.~\cite{Slim:2016pim}; in this chapter, numerical values for $f_\text{WF}$ obtained from \Eref{eq:wien-filter-frequency} are treated as positive numbers.} of $f_{\mathrm{WF}} \approx \SI{871}{kHz}$, which corresponds to $K = -1$, but the device also allows  for  operation at the harmonics $K = +1$ and $K = \pm 2$\,\cite{Slim:2016pim}. The revolution frequency is $f_{\mathrm{rev}} \approx \SI{751}{kHz}$. The integral magnetic field of the RF Wien filter amounted to $\SI{0.019}{Tmm}$ and the corresponding  electric field integral to $\SI{2.7}{kV}$. The build-up depends on the relative phase $\Phi$ between the RF field and the horizontal spin precession, which is defined by the angle between the polarization vector and the momentum vector in the RF Wien filter, when the $E$ and $B$ fields are at their maximum (see Eq.\,(35) of
Ref.~\cite{PhysRevAccelBeams.23.024601}).

When, as outlined previously, the relative phase $\Phi$ is locked and constant, the parametric resonance induced by the RF Wien filter generates rotations of the spin from the horizontal into the vertical plane (and vice versa), with a constant angular velocity $\omega = 2\pi f_\text{rev}\,\varepsilon = \omega_\text{rev}\, \varepsilon  $, where the spin resonance tune $\varepsilon$ (also called the  resonance strength), which is independent of the phase $\Phi$, is given by\,\cite{Saleev:2017ecu}
\begin{equation}
\varepsilon = \frac{1}{4\pi} \left| \vec n \times \vec n_\text{WF} \right| \cdot \psi_\text{WF}\,.
\label{eq:spin-resonance-strength}
\end{equation}
As introduced earlier,  $\psi_\text{WF}$ denotes the spin rotation in the RF Wien filter, $\vec n$ is the orientation of the (invariant) stable spin axis at the RF Wien filter when the device is off, and $\vec n_\text{WF}$ denotes the direction of the magnetic field in the Wien filter around which the spins precess.

The precursor experiment requires several additional prerequisites:
\begin{enumerate}
\item a long spin-coherence time\,\cite{Guidoboni:2016bdn};
\item  precise monitoring of the $f_\text{s} = \SI{120}{kHz}$ precession in the horizontal plane\,\cite{Eversmann:2015jnk};
\item a feedback system controlling the relative phase of the polarization vector and the RF Wien filter fields\,\cite{Hempelmann:2017zgg};
\item a good understanding of effects other than an EDM that may lead to  a tilt of the invariant spin axis.
\end{enumerate}
Prerequisites 1 to 3 have all been achieved, and more details are given in Appendix\,\ref{app:cosy}. With respect to item 4, it should be noted that imperfections of the magnetic structure of the machine cause deviations of the invariant spin axis  with respect to the ideal vertical direction, even in the absence of an EDM. Such imperfection-induced tilts of the local invariant spin axis, however, leave a trace by also affecting the spin tune of the particles orbiting in the ring. Using two solenoids installed in COSY as makeshift imperfection fields, a dedicated experimental study of these effects has been conducted at COSY\,\cite{Saleev:2017ecu}. Ultimately,  to extract an EDM limit from the data obtained in the precursor experiment, a good understanding of the effects that lead to a tilt of the invariant spin axis is mandatory.

As another example, we comment briefly on the spin-coherence time. \Figure[b]~\ref{fig:sct} shows the normalized polarization in the horizontal plane as a function of time (or turn number). With careful adjustment of the sextupole magnet settings in the machine, after a time interval of $\SI{1000}{s}$, approximately 50\% of the initial polarization survived\,\cite{Guidoboni:2016bdn}. Each particle undergoes betatron and synchrotron oscillations with different amplitudes. This leads to slightly different spin tunes of each particle; as a consequence, over time periods lasting many turns, the individual particle spins precess by different amounts. As shown in \Fref{fig:sct}, the degree of spin coherence in the particle ensemble thus decreases as a function of time. For further details, see Appendix\,\ref{appa:longsct} .

\begin{figure} [hb!]
    \centering
    \includegraphics[width=0.8\textwidth]{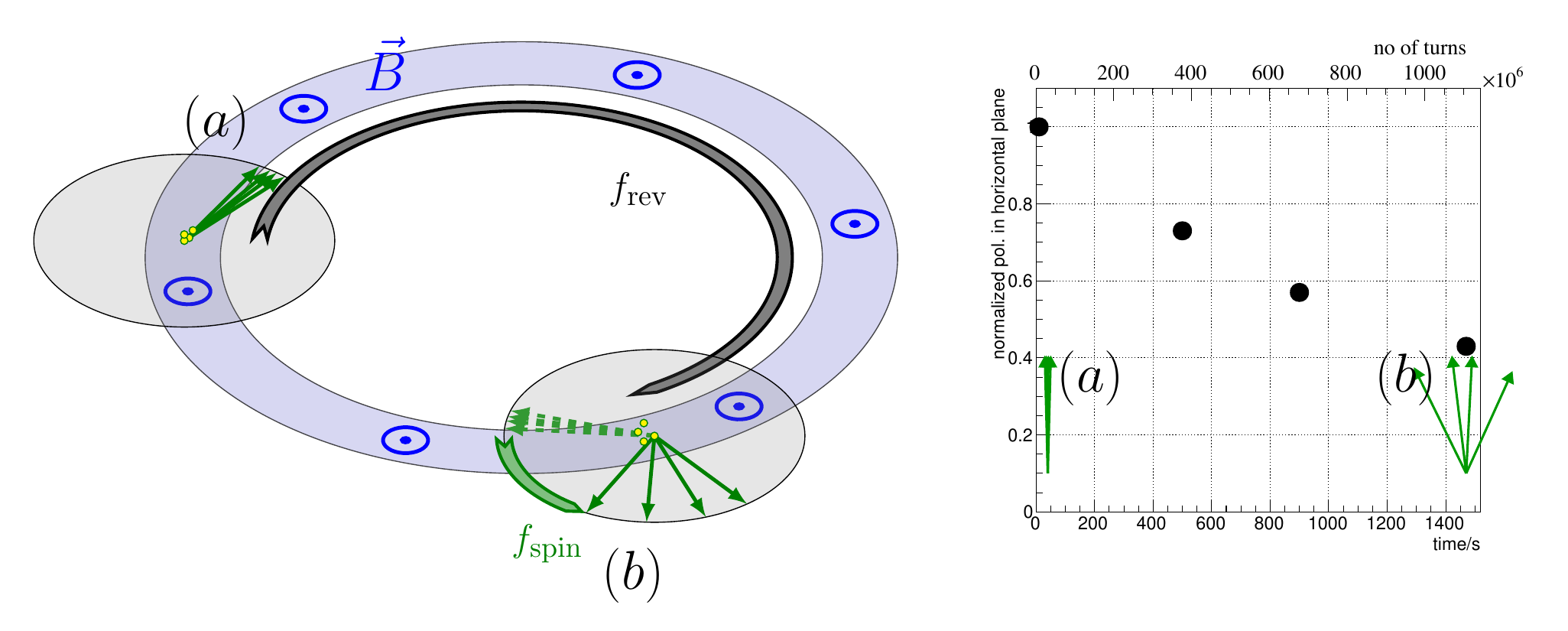}
    \caption{Left: Initially, shortly after beam injection, all spins point in the same direction (a). The differences of the individual spin precession frequencies $f_\text{s}$ lead to decoherence (b). Right: After optimization using various sextupole magnets in COSY, a spin-coherence time in excess of \SI{1000}{s} is routinely achieved\,\cite{Guidoboni:2016bdn}.}
    \label{fig:sct}
\end{figure}

\section{Current status}
With all the aforementioned tools available, a first precursor test run was performed in the autumn of 2018. The main operating parameters of COSY for the precursor experiment are listed in \Tref{tab:cosypara}. The COSY ring is shown in \Fref{fig:cosy_setup}, with the main components used in the precursor experiment.

\begin{table} [h]
\caption{\label{tab:cosypara} COSY operating parameters for the deuteron precursor EDM experiment}
\centering
\begin{tabular}{l l}
\hline \hline
COSY circumference                       &  \SI{183}{m} \\
Deuteron momentum                        &  \SI{0.970}{GeV/$c$} \\
$\beta (\gamma)$                         &    0.459 (1.126) \\
Magnetic anomaly $G$                     & $\approx$\,$-0.143$ \\
Revolution frequency $f_{\mathrm{rev}}$  & \SI{750.6}{kHz}\\
Cycle length                             &  $100\textrm{--}$\SI{1500}{s}  \\
Number of stored particles/cycle         & $\approx$\,$\SI{e9}{s}$ \\
\hline \hline
\end{tabular}
\end{table}

\begin{figure}
\centering
  \includegraphics[width=0.9\textwidth]{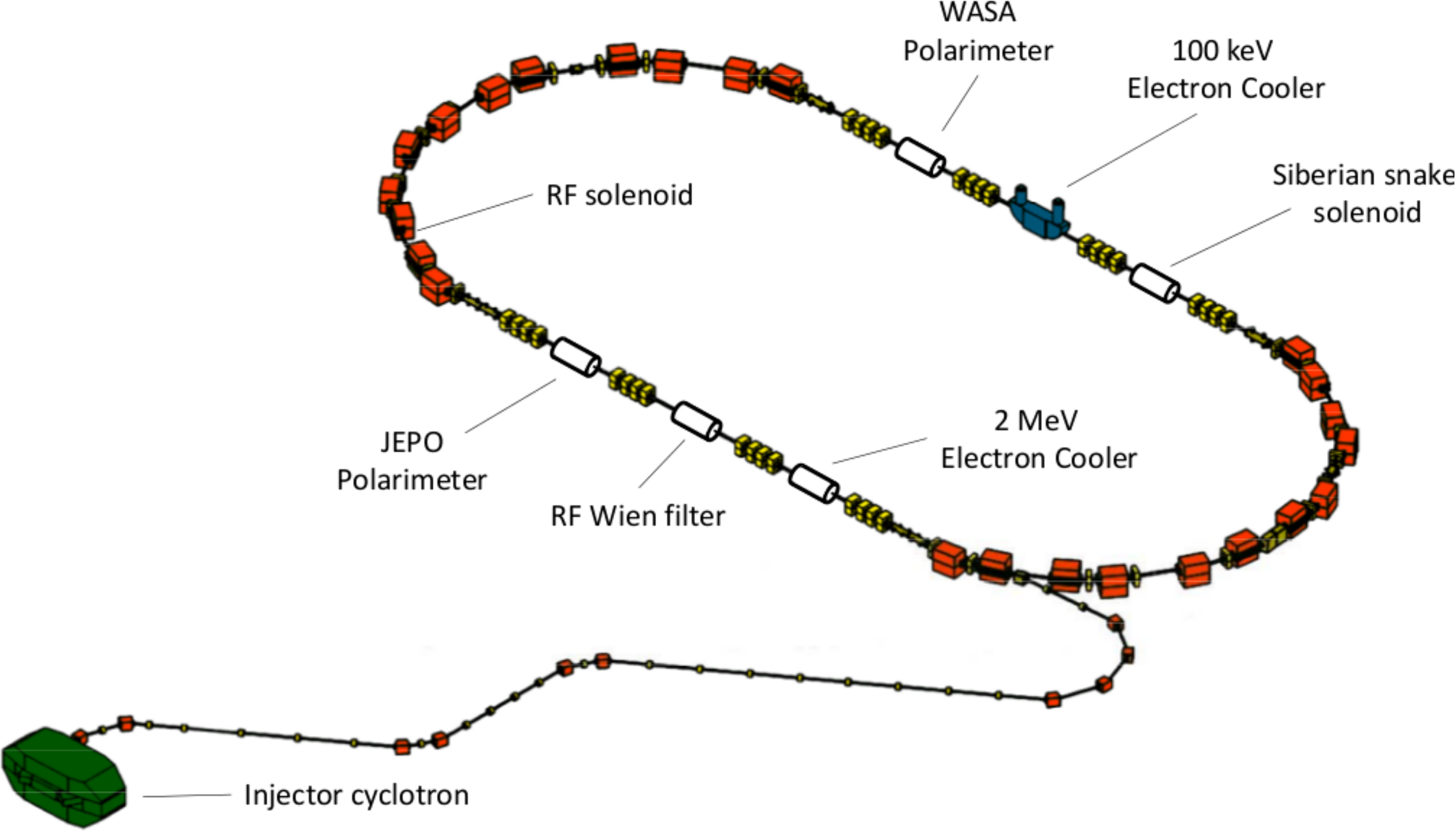}
  \caption{COSY ring with the main components used in the precursor experiment. The new JEDI Polarimeter (JEPO)\cite{Keshelashvili:2019ggh} replaces the decommissioned EDDA detector\cite{Altmeier:2004qz}, as well as the WASA forward detector\cite{Adam:2004ch,Calen:1996ft}, which served as a polarimeter in an intermediate stage.
    \label{fig:cosy_setup}}
  \end{figure}

The spin resonance tune $\varepsilon$ depends on the orientation of the invariant spin axis $\vec n$ and the magnetic axis of the RF Wien filter $\vec n_\text{WF}$ (see \Eref{eq:spin-resonance-strength}). Intentional variations of $\vec n$ and $\vec n_\text{WF}$ allow for  investigations to provide a better understanding of systematic effects. In an ideal ring with no EDM, $\vec n = (n_r, n_y, n_z)$ points in the vertical $y$-direction\footnote{As introduced in \Sref{sec:spin-evolution-electric-magnetic-fields}, the unit vector $\vec u_r$ points radially outwards, $\vec u_z$ is co-linear with the particle velocity, and $\vec u_y$ is defined as $\vec u_y = \vec u_z \times \vec u_r$.}. An EDM adds a radial component, such that $\angle(\vec n,\vec u_y) = \xi = \arctan[\beta \eta/(2G)]$. The net effect of the rotation of the RF Wien filter around the longitudinal beam direction is an additional contribution, $\phi^\text{WF} = \angle(\vec n_\text{WF}, \vec u_y)$, to the EDM and magnetic imperfection terms in $n_r$. The Siberian snake solenoid, located in the straight section opposite the RF Wien filter (see \Fref{fig:cosy_setup}), contributes only to $n_z$, with $\angle(\vec n, \vec e_z) = \chi^\text{sol}/(2\sin \pi \nu_\text{s})$, where $\chi^\text{sol}$ depends on the snake current (see Eq.\,(101) in Ref.~\cite{PhysRevAccelBeams.23.024601}).

A feedback system that locks the relative phase $\Phi$ between the RF field in the Wien filter and the horizontal spin precession, based on a continuous measurement of the in-plane polarization component, will no longer work
when the spins are rotated into the direction perpendicular to the ring plane. Therefore, one may confine the measurement  to the determination of the initial slope of the vertical polarization
\begin{equation}
 \dot \alpha = \left. \frac{\dot P_\text{vertical}}{P_\text{horizontal}} \right|_{t = 0}\,.
 \label{eq:alpha-dot-definition}
\end{equation}

\Figure[b]~\ref{fig:build-up} shows the build-up rate $\dot \alpha$ of the vertical polarization component as a function of the relative phase $\Phi$. For every single data point, the relative phase $\Phi$ was set using the feedback system. The expected sinusoidal shape is observed. To provide an idea of the statistical sensitivity, the hypothetical signal of an EDM of $d = \SI{e-18}{\text{$e$}.cm}$ is indicated by the grey line. The statistical error of the measurement is of the order of the symbol size, indicating that
the measurement is \textit{statistically}  sensitive to EDMs well below the $\SI{e-18}{\text{$e$}.cm}$ level.

\begin{figure} [hbt!]
\centering
    \includegraphics[width=0.65\textwidth]{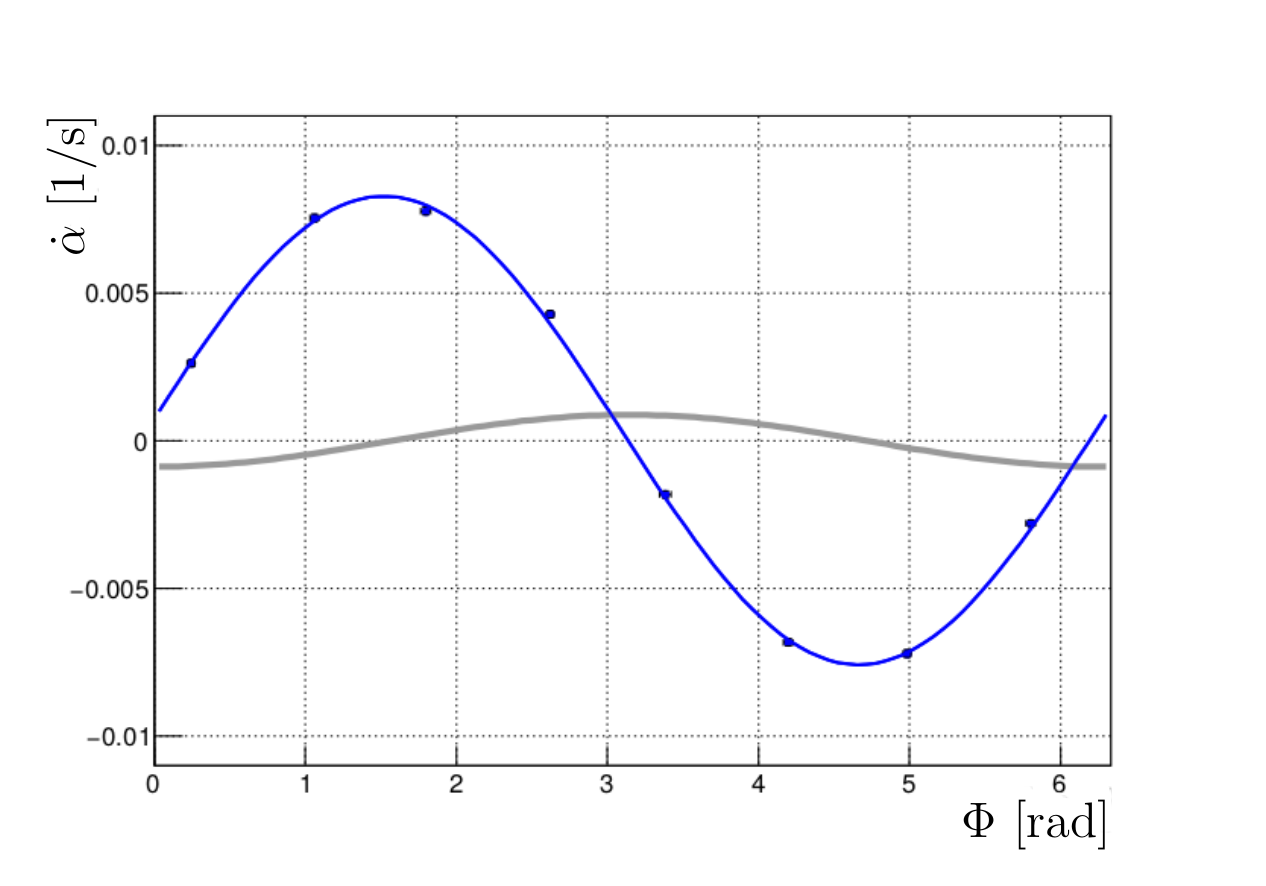}
\caption{Build-up rate of the vertical polarization $\dot \alpha $ (see \Eref{eq:alpha-dot-definition}) plotted as function of the relative phase $\Phi$ between the RF Wien filter fields and the horizontal polarization. The RF Wien filter was set to provide $\phi^{\mathrm{WF}}=0$ and the Siberian snake solenoid in the opposite straight section was switched off ($\chi^\text{sol}=0$). The approximate effect of a hypothetical EDM of  $d = \SI{e-18}{\text{$e$}.cm}$ is indicated by the grey line. The statistical errors of the measurement are of the order of the symbol size. The data shown were obtained in about 4\,h of beam time.
    \label{fig:build-up}}
\end{figure}

Yet higher accuracy for the EDM determination calls for a continuous feedback system that allows one to fully exploit the long spin-coherence time of $\approx$\,$\SI{1000}{s}$,  not just the very beginning of the polarization evolution using $\dot \alpha$ at $t=0$. An option that is currently being investigated involves using a
number of bunches in the machine and exclusively reserving the spin precessions detected in one of the bunches for feedback purposes, \ie gating out the Wien filter RF field so that this particular bunch  is never actually
exposed to these fields. In this way, the gated-out bunch  effectively serves as a co-magnetometer for the experiment. As a function of $\phi^\text{WF}$ and $\chi^\text{sol}$, the square of the resonance tune takes the form of a sum of two independent quadratic functions (see the appendix of Ref.~\cite{PhysRevAccelBeams.23.024601}). As an illustration, for an ideal ring, \Eref{eq:spin-resonance-strength} yields
\begin{equation}
 \varepsilon^2 = \frac{1}{16 \pi^2} \psi_\text{WF}^2 \left[ \left(\xi - \phi^\text{WF} \right)^2  + \left( \frac{\chi^\text{sol}}{2\sin \pi \nu_\text{s}}  \right) ^2\right]\,.
 \label{eq:analytic-epxression-eps-squared}
\end{equation}
The resulting funnel-shaped resonance strength $\varepsilon (\phi^\text{WF}, \chi^\text{sol})^2$ has a tip at $\chi^\text{sol} = 0$ and $\phi^\text{WF} = \xi$.

An example of the experimentally measured dataset of the initial slope $\dot \alpha_{|t=0}$ for one specific setting, $\phi^\text{WF} = \chi^\text{sol} = 0 $, as a function of the relative phase $\Phi$ is shown in \Fref{fig:build-up}. These data were recorded during the test measurement in 2018 with a feedback system that keeps the slope of the growth angle constant, $\dot \alpha = \dot \alpha|_{t=0}$\,\cite{Hempelmann:2017zgg}. Such a feedback system imposes on $\Phi$ a time-dependence, such that it provides a full rotation of the spin from the horizontal direction to the vertical one, but at the expense of the oscillation frequency being proportional to the initial slope $\dot \alpha$. This situation is not perfectly identical to the case of constant phase $\Phi$, when $\dot \alpha = a \omega \cos \Phi$, with $a$ being the amplitude, $\omega$ the frequency of the vertical oscillations in $P_y$, and $\Phi$ the relative phase between the RF Wien filter  and the  in-plane polarization vector (see Eq.\,(78) of Ref.~\cite{PhysRevAccelBeams.23.024601}). 

The experimental data of $\dot \alpha$, extracted from measurements like that shown in \Fref{fig:build-up}, for values of $\phi^\text{WF}$ and $\chi^\text{sol}$ in the range  $\SI{-1.5}{\degree}$ to $\SI{+1.5}{\degree}$ are depicted in \Fref{fig:build-up_3D}. These data exhibit a similar funnel-shape surface, as a function of $\phi^\text{WF}$ and $\chi^\text{sol}$, to $\varepsilon$ (\Eref{eq:analytic-epxression-eps-squared}), and the tip gives the orientation of the invariant spin axis. Strictly speaking, the minimum corresponds to a situation where the invariant spin axis $\vec n$ and the Wien filter axis $\vec n_\text{WF}$ coincide. In general, both of these axes are tilted as
a result of various effects, such as intentional spin and device rotations, an EDM, and machine imperfections. In an ideal ring, the orientation of the invariant spin axis is given in the radial direction by $n_r = \xi = \arctan(\eta\beta/(2G))$ and $n_z = 0$. The surface is a fit to the data using the analytical expression of \Eref{eq:analytic-epxression-eps-squared} for the build-up, but with the weights of the two terms in the square brackets taken as free parameters.

\begin{figure}
\centering
\includegraphics[width=0.65\textwidth]{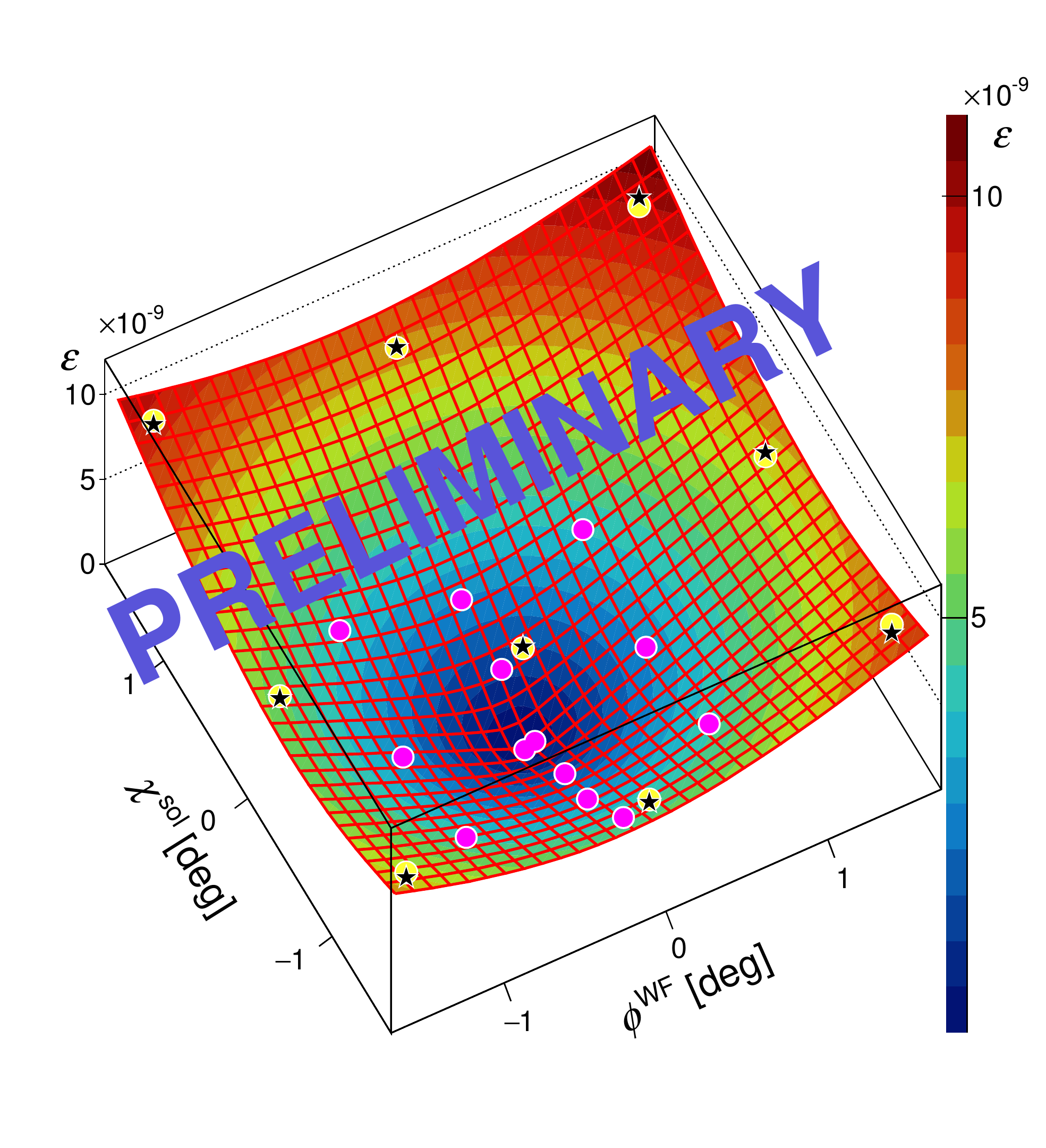}
\caption{Experimentally observed parametric resonance strength $\varepsilon \simeq \dot \alpha/\omega_\text{rev}$, plotted in terms of the initial slope $\dot \alpha_{|t=0}$, for various values of $\phi^\text{WF}$ and $\chi^\text{sol}$. The surface is a fit to the data using the analytical expression of \Eref{eq:analytic-epxression-eps-squared}, modified as described in the text. The minimum of this graph gives the orientation of the invariant spin axis. The three recorded datasets are indicated by the different symbols (yellow circles, black stars, and magenta bullets).}
\label{fig:build-up_3D}
\end{figure}

It should be emphasized that three datasets were recorded in total, which are indicated by the three different types of symbol in \Fref{fig:build-up_3D}. The fact that these three datasets yield \textit{consistent} results, although they were taken several days apart, indicates that the stability of COSY is sufficient to perform this kind of precision study, down to sensitivities corresponding to EDM values well below $\SI{e-18}{\text{$e$}.cm}$. Of course, at this stage, the deviation of the minimum of \Fref{fig:build-up_3D} from the ideal orientation $(n_r,n_z) = (\xi,0)$ is mostly attributed to contributions from magnetic imperfections to $n_r$ and $n_z$, misalignment of magnets, and unknown beam position monitor offsets, causing deviations from the ideal orbit. Work is presently ongoing to try to minimize these effects using beam-based alignment and to quantify systematic contributions with the help of simulations.

\section{Outlook}
A second precursor run on the deuteron EDM is planned for the first half of 2021, to be conducted by  the JEDI collaboration. During 2019, beam-based alignment procedures were developed and  implemented, in order to calibrate
the offsets of the beam position monitors. This shall lead to an improvement in the COSY orbit, and will probably reduce systematic effects on the orientation of the invariant spin axis.

At the same time, simulation tools are being developed   to estimate the contribution of systematic effects on the invariant spin axis (see Chapter\,\ref{Chap:SpinTracking}). The goal is to make, with COSY, a first EDM measurement with a precision similar to that of the muon, \ie $\SI{e-19}{\text{$e$}.cm}$.

It should also be clear that a further increase in the precision of storage ring EDM experiments, by orders of magnitude, will only be possible if a dedicated storage ring with counter-rotating beams is available, in which many of the aforementioned systematic effects cancel each other out.

\begin{flushleft}

\end{flushleft}
\end{cbunit}

\begin{cbunit}

\csname @openrighttrue\endcsname 
\chapter{EDM prototype ring (PTR)}
\label{chap:ptr}

\section{Introduction}
\subsection{Need for a prototype ring (PTR)}
Intense discussions within the CPEDM collaboration have concluded that the final ring cannot be designed and built in one step (see Chapter\,\ref{Chap:Strategy}). Instead, a smaller-scale prototype ring (PTR) must be constructed and operated
as the next step.

In the following chapter, the state of preparedness for a full-scale all-electric proton EDM of approximately \SI{500}{m} circumference is discussed. Ideally containing only electric fields and no magnetic fields, this ring needs to be capable of storing \SI{232.8}{MeV} frozen spin protons circulating in clockwise (CW) and counterclockwise (CCW) directions. For the best possible suppression of systematic EDM errors, the beams have to circulate concurrently.

As part of the preparation of this report, the level of preparedness for constructing was studied in considerable detail, with the results distilled in \Tref{tbl:Preparedness-Chap5}, in which the `lacks of preparedness' are sorted by perceived `degrees of severity'.

The present understanding of possible limitations of the proposed proton EDM measurement does not allow confirmation that the target sensitivity ($\SI{e-29}{\text{$e$}.cm}$) can be reached. Thus, the community agrees on the need for a prototype ring to study key features of the proposal, such as the  operation with counter-rotating beams, and the determination of the average radial magnetic field from the orbit separation of the two beams. Furthermore, the proposed prototype ring comprises the possibility to add, in a second stage, magnetic fields for a first `frozen spin' EDM measurement with electric and magnetic fields

\subsection{Considerations leading to two PTR stages}
The goals of stage\,1, after reconfirming beam control procedures that have already been developed at COSY, will be to turn all `red' flags in Table\,\ref{tbl:Preparedness-Chap5} to at least `yellow\,($+$)' or even `green\,($-$)'. The goals of stage~2 will be more diverse, but their common
thread will be to gain the experience needed to complete the design of the full-scale ring. This has to include acquiring information needed to predict the potential precision with which the proton EDM can be measured.

Certainly, as a prototype, the ring should be small and simple, and as inexpensive as possible.  Yet the ring must
be designed to be capable of achieving its claimed goals. The primary goal of stage\,1 is to demonstrate that performance
routinely obtained in magnetic rings can be replicated in an all-electric ring. The goals for stage\,2 mainly require
frozen-spin protons, which are obtained by adding a vertical magnetic field $B_y$ to the radial electric field $E_r$.

Several considerations went into the determination of kinetic energies for stages\,1 and 2. To limit building costs, the
ring circumference was constrained to not exceed \SI{100}{m}. After allowing for adequate drift space for needed equipment, this
led to a bending radius of less than \SI{9}{m}.
A consequence of these requirements is that the proton kinetic energy is limited to \SI{30}{MeV} for operation with only electric fields. The proton polarimeter figure of merit is satisfactory at \SI{30}{MeV}, but decreases with decreasing energy. As a result of these considerations, a nominal proton beam energy of \SI{30}{MeV}  was adopted for the all-electric stage\,1. (Note that this ring could be used with lower electric fields to circulate `frozen spin' electrons.
Except for the quite low efficiency of  currently available electron polarimetry, this means that, in principle, the electron EDM can also be measured in the PTR.) 

From the \SI{30}{MeV} proton energy for stage\,1, the choice of a maximum energy of \SI{45}{MeV} for stage\,2, the frozen-spin operation, followed almost automatically.
To achieve the frozen-spin condition for protons near this energy, approximately one-third of the bending shall be provided by magnetic fields, which increases the beam energy for similar electric fields. Since the magnet needs to be iron-free (to avoid hysteresis and obtain the required reproducibility), air-core magnets must be used. The required magnetic field is sufficiently low that this is not a serious constraint.

Up to this point in PTR design studies, there has been no differentiation between all-electric, \SI{30}{MeV}, stage\,1 optics and
\SI{45}{MeV}, stage\,2 frozen-spin optics. The basic design has sufficient flexibility to meet both goals. In detail, of
course, the working points and other details will  essentially be different. Detailed lattice design and performance is described
 in \Sref{sect:RingDesign}.

\subsection{Basic beam parameters and layout}
This report describes the adopted `square ring', with a bending radius of \SI{8.86}{m}. The basic proton kinematic data and field strengths are given in \Tref{tbl:BasicBeamParams}, and the ring layout is shown in  \Fref{fig:BasicLayout}. The first column shows  parameters for the nominal \SI{30}{MeV}, all-electric operation (appropriate for simultaneously cir\-culating, but not frozen-spin, beams). The second column shows parameters for electric and magnetic fields for frozen spin, single-beam only, operation at the same  kinetic energy, \SI{30}{MeV}. The third column shows  values needed for the nominal maximum operation,
at \SI{45}{MeV}.

\begin{table} [h]
\centering
\caption{Basic beam parameters for the PTR}
\label{tbl:BasicBeamParams}
\begin{tabular}{l llll}
\hline \hline
                                  &    $E$ only  & \multicolumn{2}{l}{$E$ \& $B$,} &   Unit \\
                                  &                    & \multicolumn{2}{l}{frozen spin} & \\ \hline
Bending radius          &   8.86          & 8.86 & {8.86} & m \\
Kinetic energy                    &      30      &   30    &     45      &   MeV        \\ 
$\beta =v/c$                      &    0.247     &  0.247  &    0.299    &              \\
$\gamma$ (kinetic)                &    1.032     &  1.032  &    1.048    &              \\
Momentum                          &     239      &   239   &      294    &   MeV/$c$      \\ 
Electric field $E$      &    6.67      &   4.56  &      7.00   &   MV/m        \\
Magnetic field $B$   &              &  0.0285 &    0.0327   &    T          \\
r.m.s. emittances $\epsilon_x = \epsilon_y $ &  1   & 1 & {1} &  $\si{\pi . \milli \meter . \milli \radian}$ \\
Transverse acceptance $a_x = a_y$ & >10 & >10 & {>10} &  $\si{\pi . \milli \meter . \milli \radian}$ \\
\hline \hline
\end{tabular}
\end{table}

\begin{figure}
\centering
\includegraphics[scale=0.52,viewport=100 350 600 775,clip]{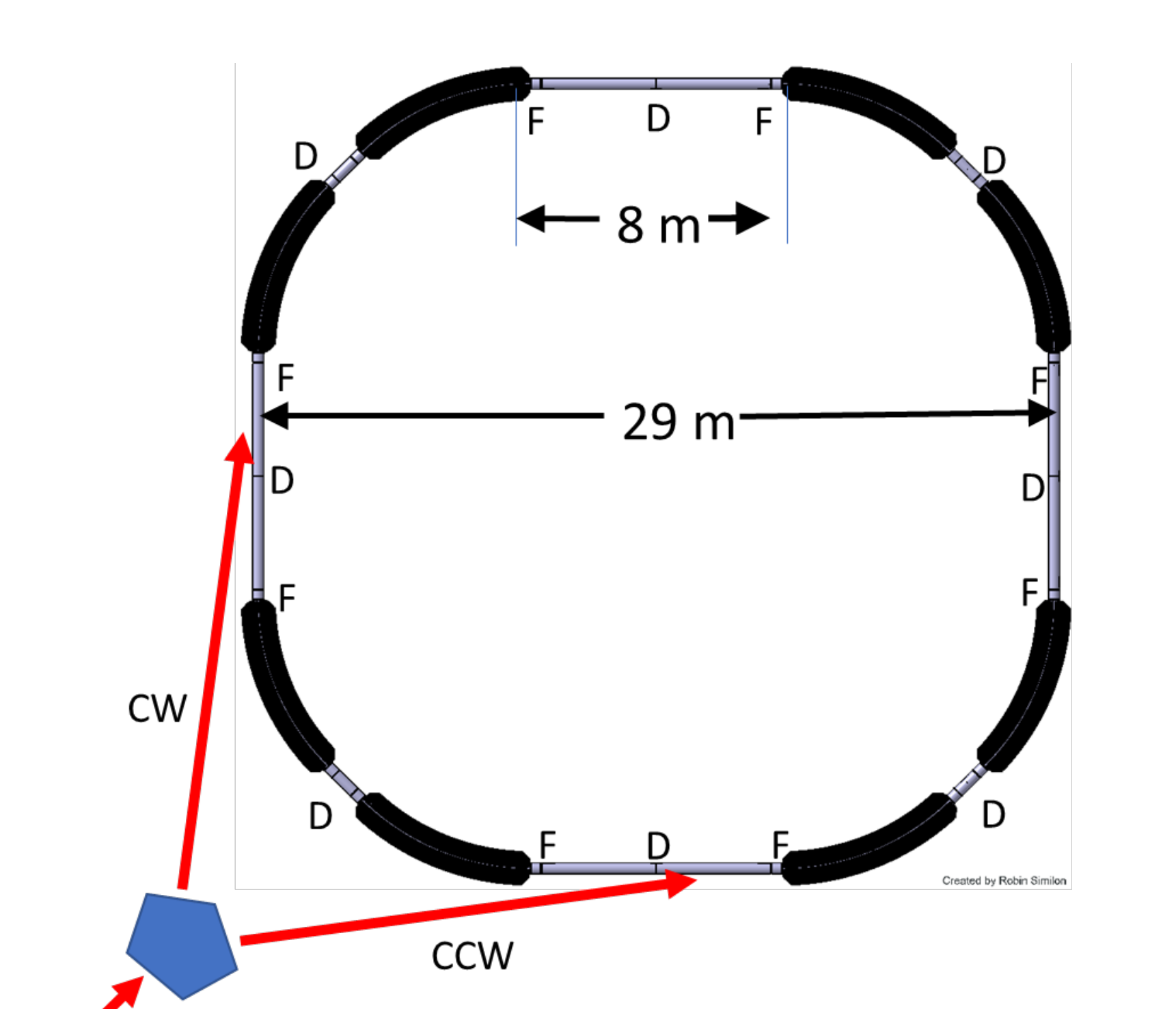}
\caption{\label{fig:BasicLayout} Basic layout of the prototype ring, consisting
of eight dual superimposed electric and magnetic bends and two families of quadrupoles (F, focusing; D, defocusing), with an optional skew quadrupole family at the midpoints of the four \SI{8}{m} long straight sections. The total circumference is about \SI{100}{m}.The separate family of quadrupoles, labelled D at long straight 
sections centers in this figure, are not powered.}
\end{figure}

\section{Goals for the \SI{30}{MeV} all-electric PTR}
\label{ptr:30MeVgoals}
The four \emph{primary}, quantitative, goals for the \SI{30}{MeV} stage are:
\begin{itemize}
   \item to gain experience in operating a large-scale high electric field electrostatic storage ring and, in particular, to push the field to the maximum possible with acceptable reliability (breakdown rate);
   \item to demonstrate the ability to store the $\num{e9}$ polarized protons thought to be the minimum number needed for making proton EDM measurements in a predominantly electric storage ring;
   \item to demonstrate, as necessary to reduce systematic error, the ability to produce and manipulate two polarized beams, each with the same $\num{e9}$ proton intensity, simultaneously countercirculating in the same ring;
   \item to demonstrate that the average magnetic field can be determined from the difference in the vertical position of two counter-rotating beams measured with high-sensitivity pick-ups and very low vertical tune, as described in  \Sref{ssc:BPMs}.
\end{itemize}
%

The proton intensity goal has been set conservatively low to avoid distractions associated with preserving polarization through the injection process---this can be perfected later, using well-understood experimental techniques.

The polarimetry already demonstrated in COSY will be sufficient to complete these goals. As in COSY, the spins will not be frozen;  nevertheless, the spin coherence time (SCT) can be determined. In addition, phase-locked spin control\,\cite{Guidoboni:2016bdn,Hempelmann:2017zgg} can be reconfirmed.

\emph{Secondary}, qualitative goals for stage\,1, therefore, include the
replication of spin-control abilities in an all-electric ring, such as phase-locked loop stabilization of the beam polarization. This capability is required to provide input signals to the external correction circuits needed to manipulate the beam polarization.

Certain \emph{tertiary} goals for stage~1 will also  need to be met, to steer the upgrading of the PTR for a more advanced second stage. However, any such upgrades need to preserve the gross geometry of the ring.  (Mainly to reduce cost, and speed progress) it seems prudent initially, to economise, with flexibility for later upgrades. Investigations in the first stage can shed
light on PTR modification possibilities needed to produce a more productive second stage.  Some examples follow.

It is currently not clear whether a completely cryogenic vacuum will be necessary. Related to this issue is the question of whether or not the beam emittance can be adequately controlled by stochastic cooling, and whether stochastic cooling adversely affects EDM experiments. Also connected with vacuum uncertainty is the possibility of a regenerative breakdown mechanism that could limit the proton beam current. Such a breakdown could commence with  a temporarily free electron being accelerated towards the positive electrode. Secondary electrons, created on impact, would be immediately recaptured, but photons produced could strike the other electrode, producing secondary electron emission  that could lead to regenerative failure. No such phenomenon has ever been observed in  magnetic rings---but this is irrelevant, because the corresponding electric machines do not  exist yet. Some proton intensity limitations in non-relativistic rings seem consistent with such an interpretation. However, no such limitation has been observed in electrostatic separators in either electron or proton high-energy storage rings. Any such breakdown mechanism would presumably tend to be moderated by a superimposed magnetic field.  But weak magnetic fields could be ineffective.

Magnetic shielding is another uncertain issue. Well-understood (but expensive) passive magnetic shielding methods are known, which improve the shielding by several orders of magnitude. However, they require detailed understanding of the apparatus, which can, realistically, only be studied experimentally in situ. Certainly, magnetic shielding could be upgraded in the interval between stages. No active field control based on magnetic measurement is planned for stage\,1, but this could, optionally, be developed for stage\,2.

The possibility of significant upgrading of positioning and alignment is also anticipated between stages\,1 and 2. Ferrite kickers, assumed for stage 1, may need to be replaced by air-core or electrostatic kickers for stage\,2.

Greatly improved critical analysis of beam position monitor (BPM) performance is  expected at stage\,1, possibly informing improvements at stage\,2.
Similar investigations of the stability of basic mechanical and electrical  parameters will  also be performed.

\section{\texorpdfstring{Goals for the \SI{45}{MeV} combined $E$--$B$ PTR}{Goals for the 45 MeV combines E--B PTR}}\label{Goals for E/B PTR}
The following goals are essential.
\begin{enumerate}
   \item To lend confidence to an eventual full-scale EDM ring proposal, experimental methods are to be developed and demonstrated for measuring the proton EDM in a ring with superimposed electric and magnetic bending. Cost-saving measures in the prototype, such as room-temperature operation, minimal magnetic shielding, and the avoidance of obsessively tight manufacturing and field-shape matching tolerances, are expected to limit the precision of the prototype ring EDM measurement. However, data needed for extrapolation to the full scale ring must be obtained from the PTR.
   \item Frequency domain control, for example, a phase-locked spin wheel frozen spin beam control (see Section\,\ref{sect:PTRPhysics}), and measurement capability
are to be demonstrated.
   \item Finally, a first precise storage ring proton EDM measurement can be made. For various reasons, which are mainly due to cost-saving measures in the PTR design, the achieved precision cannot, however, be expected to provide a significant test of the standard model. But information gained from this prototype measurement can be expected to produce  specifications that the nominal all-electric ring needs to meet to reach that goal.
\end{enumerate}

\section{Relation between the PTR and the nominal all-electric ring}
This section provides fine-grained technical details concerning the relation of the proposed prototype to the full scale ring.

The details describe a four parameter lattice design for a complete family of stable all-electric storage rings, ranging from the PTR, at the small-radius, low-energy end, to a full-scale, large-radius, high-energy end. Especially for measuring the EDMs of particles other than protons, there are valid reasons for considering electric rings everywhere in this range. For the proton rings emphasized in this report, when comparing the results of different particle-tracking programmes, it is important for all assumed lattice parameters to be identical, even down to the fine-grained detail given here.

The structure of the PTR was  obtained from the full-scale ring by downscaling from the full-scale   design of Anastassopoulos \textit{et al.}\ \cite{Anas} to 30 or $\SI{45}{MeV}$, trying to keep the two designs as close as possible. After the downscaling, mainly to make element lengths sensible for a low-energy ring, small changes were made to the PTR design before scaling back up to the full-scale ring. In this way, the physical properties of the  scaled-back-up full-scale ring and the ring described in Ref.~\cite{Anas} can be compared, as in Table\,\ref{tbl:HolyGrail-PT-params}. The agreement is quite good for all parameters, well within the ranges of parameter values of the various 2016 ring designs. The skeletal PTR prototype lattice design is shown in Fig.\,\ref{fig:pEDM-square-proto}.

\begin{figure}
\centering
\includegraphics[scale=0.25]{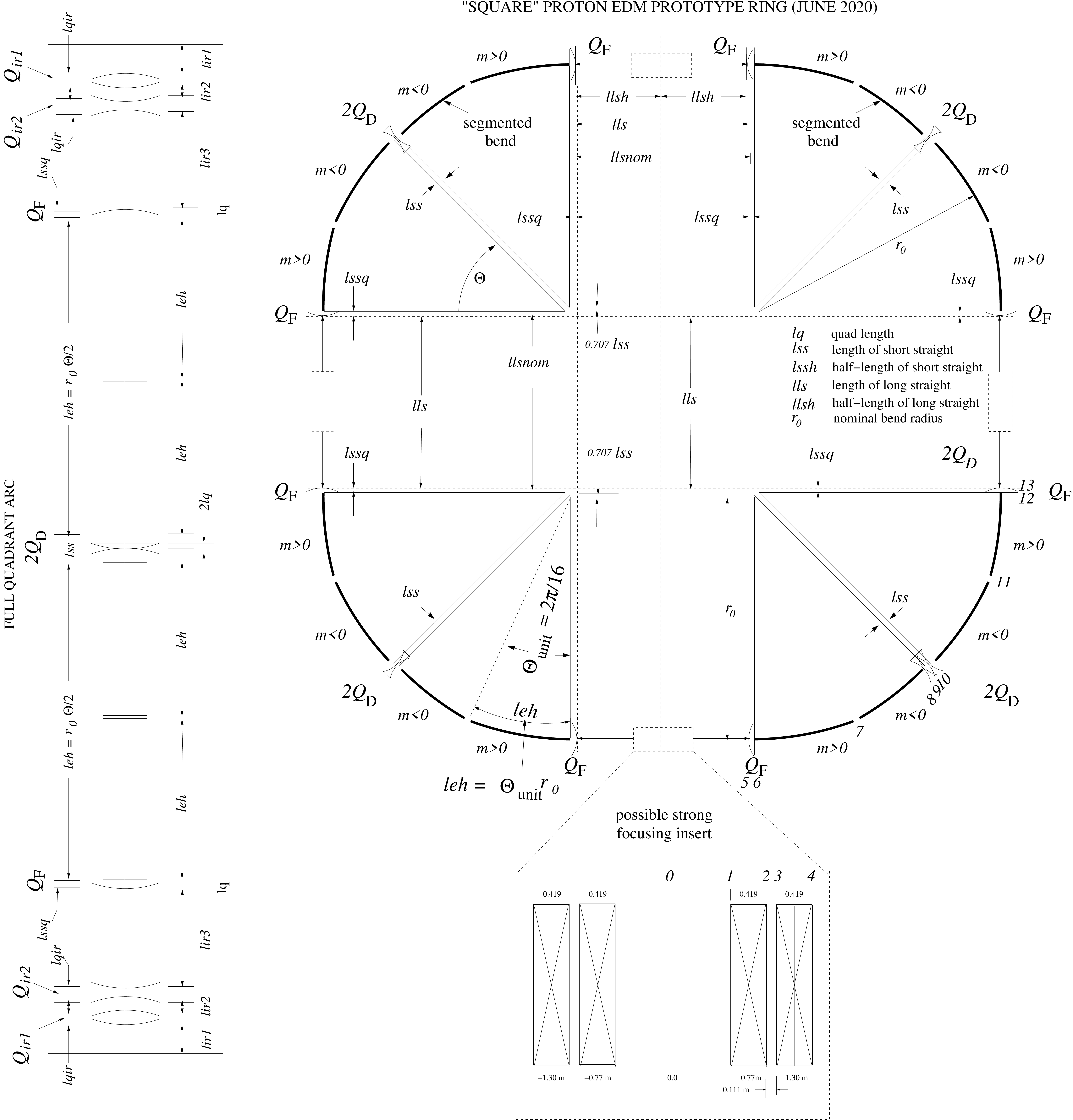}
\caption{\label{fig:pEDM-square-proto} Lattice layouts for (left)\ proposed lattice half-cell  and (right) full ring. The accumulated drift length is not enough for the ring to operate `below transition'.  When scaling up to the final, full-energy, all-electric ring, from four-fold to sixteen-fold symmetry, with drift lengths and bend lengths preserved (but bend angles four times smaller) the total circumference is to be about \SI{500}{m} and operation will be well below transition.}
\end{figure}

In both ring designs, for flexibility, focusing is provided by separated-function electric quadrupoles; additional (very weak)  alternating-gradient, combined-function, electrode shape focusing is  under discussion. (Current designs have favoured focusing  by electric quadrupoles only).

It was decided that the scaling  between the prototype and the full-scale ring would be achieved by relating the ring superperiodicities in the ratio of 4 to 16, while leaving all lengths (except for straight section lengthening, to be explained) within each superperiod constant. This scaling gives the prototype ring the appearance
of a square with rounded corners (see \Fref{fig:pEDM-square-proto}), while the full ring appears very nearly circular
(see \Fref{fig:HolyGrailQuadrant}). In this process, the bend per superperiod was reduced by an integer factor of four. The values of the four main scaling parameters are shown in \Tref{tbl:MagicScaling}.

\begin{table}
 \caption{\label{tbl:MagicScaling}
The four major parameters for scaling between the prototype ring and a version scaled up for operation at the magic energy.}
\centering
\begin{tabular}{lll}  \hline \hline
  Parameter &  Prototype  &  Scaled up \\
 \hline
     Bending radius (m)            & 8.9 & 40 \\
     L\_LONG\_STRAIGHT (m)  & 6.0 & 14.8 \\
     N\_SUPER                         &  4     & 16 \\  
     M\_NOMINAL                     & 0.1 & 0.1 \\
 \hline \hline
\end{tabular}
\end{table}

The adopted scaling relations follow: the field index scales inversely with superperiodicity \linebreak $\text{N\_SUPER}$, with $m = \pm \text{M\_NOMINAL}$ being the field indices of the prototype ring; the scaling relation is $m = \pm \text{M\_NOMINAL} \cdot 4/\text{N\_SUPER}$. Minor scalings are indicated in \Tref{tbl:MagicScaling.2}, with lattice names given in the column headings. 

\begin{table}
 \caption{\label{tbl:MagicScaling.2}
 Minor geometric parameters: $\Theta$, $r_0$, $l_\mathrm{eh}$, $l_\mathrm{ss} = \SI{0.8}{m}$, and $l_\mathrm{lsh}$ are, respectively, bend or half-period, bend radius, bend half-length, short
straight length, and long straight half-length; $K_0$ is the proton kinetic energy; $\pm m_\text{in}$ are alternating field index values. Minor kinetic parameters: $l_q$ is quad length, $q_\text{F}$ and $q_\text{D}$ are quad strengths, and  $Q_x$ and $Q_y$ are tunes.}
\centering
\resizebox{\columnwidth}{!}{%
\begin{tabular}{llllllllllll}  \hline \hline
Lattice name &    $K_0$    &  $m_\text{in}$  &  $\Theta$ &  $r_0$  & $l_\mathrm{eh}$  & $l_\mathrm{lsh}$ &   $l_q$    &   $q_\text{F}/q_\text{D}$     &  Circumference &  Half-gap
 width   &   $Q_x/Q_y$ \\
             &  (MeV)   &         &    (rad) & (m)  & (m)  &  (m) &   (m)   &   (1/m)     &   (m)  &   (m)   &              \\ \hline
E\_30MeV     &  0.0300  &  0.100  &  0.785 &  \phantom{4}9   & 3.53 & 2.60 &  0.2000 & $\mp0.01$   & \phantom{5}83.7   &  0.035  &  1.768/0.093 \\  
EM\_45MeV    &  0.0450  &  0.100  &  0.785 &  \phantom{4}9   & 3.53 & 2.60 &  0.2000 & $\mp0.01$   &  \phantom{5}83.7  &  0.035  &  1.750/0.093 \\ 
E\_233MeV    &  0.2328  &  0.025  &  0.196 & 40   & 3.93 & 7.00 &  0.2000 & $\mp0.0025$ &  501   &  0.015  &  1.815/0.145 \\  \hline \hline
\end{tabular}
}
\end{table}

Detailed lattice descriptions (needed for computer processing) are contained in the following files, available in Ref.~\cite{RTalmanLatts}:
\begin{itemize}
     \item {{\bf\tt EM\_45MeV-con\_xml}:  } `{\tt .xml}' file containing all parameters (both symbols and their values) for a small (\SI{85}{m} circumference) proton EDM prototype ring, including (symbolic) parameters for scaling to the large (\SI{500}{m} circumference) all-electric proton EDM ring;
     \item{{\bf\tt EM\_45MeV-nocon\_xml}:  } symbolic `{\tt .xml}' file describing idealized lattice design;
     \item{{\bf\tt EM\_45MeV.adxf}:  } numerical `{\tt .adxf}' file describing idealized lattice design;
      \item{{\bf\tt EM\_45MeV.sxf}:  } numerical `{\tt .sxf}' file describing fully instantiated lattice design (though without differentiated (\ie individualized) parameter values).
\end{itemize}

Initially, for both the prototype and the full-scale ring, the horizontal tune was expected to be just below 2.0 and the vertical tune less than 1.0, and tuneable to a value as low as 0.02. This ultralow vertical tune was needed to reduce the vertical restoring force, to enhance the beam `self-magnetometry' sensitivity to beam displacement caused by the radial magnetic field.

As an aside, it can now be mentioned that the doubly magic EDM measurement method 
avoids the need for ultraweak vertical focusing, allowing the focusing to be much stronger than was initially anticipated. In a very thorough and valuable 2015 study  \cite{Lebedev}, Lebedev analysed two frozen-spin all-electric designs, one very weakly focusing, the other stronger focusing. With an ultralow vertical tune no longer necessary, the scaled-up PTR can be said to  correspond more closely to the stronger-focusing ring favoured there.

For the full-scale ring, the correspondingly smaller tune advance per superperiod causes the focusing to be weaker. This is what
permits the long straight sections of the full-scale ring to be more than doubled, compared with the prototype (from \SI{6}{m} to \SI{14.8}{m}). This has the beneficial (perhaps even obligatory) effect, for the full-scale ring, of operating `below transition'. This ameliorates intrabeam scattering, as can be explained in connection with stochastic cooling. (Conversely, this is one aspect in which the prototype ring optics is a not-quite-faithful prototype.) This choice was made to reduce the prototype
size. Also, with the COSY ring as a candidate low-energy injector ring, for reasons of beam bunch-to-bunch separation, the EDM prototype ring circumference of \SI{91}{m}, exactly one-half  the COSY circumference, would be a natural choice.

\section{Ring design \label{sect:RingDesign}}
The basic PTR geometric ring parameters are given  in Table\,\ref{tbl:BasicBeamParams}.
The  acceptance of the ring is to be $\SI{10}{\pi.mm.mrad}$ for $\num{e9}$ particles. The lattice is based on fourfold symmetry,
as shown in \Fref{fig:BasicLayout}.
%

%
The bend elements utilize electric and magnetic bending. Pure electric bending can be used for \SI{30}{MeV} protons but, for a (nominal maximum) proton energy of \SI{45}{MeV}, superimposed magnetic and electric bending will be applied. The magnetic part of the bending must be provided by a pure air-coil magnet, to avoid hysteresis effects caused by using  iron for the magnet. It will be possible to store both CW and CCW beams consecutively, but not concurrently.

The present design for the prototype is a `square' ring with four \SI{8}{m} long straight sections. This design was arrived at after lattice studies using different shapes, such as round or race-track shaped. The ring is shown in \Fref{fig:BasicLayout}. It consists of four unit cells, each of them bending $\SI{90}{\degree}$. Each unit cell consists of a focusing structure F-B-D-B-F, where F is a focusing quadrupole, D is a defocusing quadrupole, and B is an electric or magnetic bending unit. The lattice is designed to allow a variable tune of between 1.0 and 2.0 in the radial plane and between 1.6 and 0.1 in the vertical plane, as shown in \Fref{fig:Figure2a}.

\begin{figure} [hb!]
\centering
\includegraphics[height=0.175\textheight]{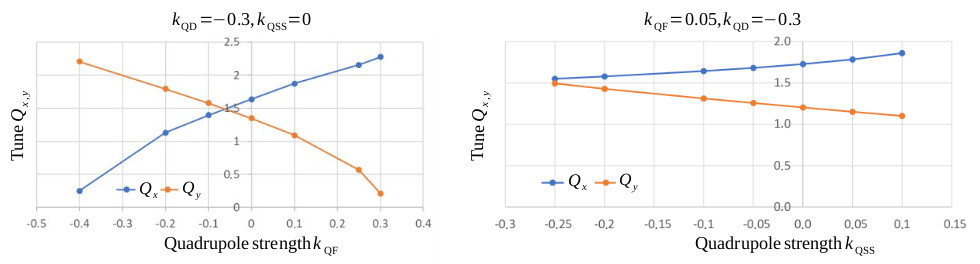}
\caption{\label{fig:Figure2a} Left: Horizontal betatron tune $Q_x$ and  vertical betatron tune $Q_y$ as a function of the strength of the QF quadrupole family; the QD and QSS quadrupole families  are constant, while the QD quadrupole family is varied. Right: The marked points are continued.}
\end{figure}

The straight sections must house separate injection regions for clockwise (CW) and counterclockwise (CCW) beam operation. There will also be a quadrupole of type QSS in the centre of each straight, to provide additional tuning possibilities.
The horizontal gap is determined by the horizontal beam size, which is determined by the maximum acceptance and the maximum beta function as
\begin{equation}
   2 x_\text{max} = 2\sqrt{a_{x,\text{max}} \beta_{x, \text{max}}} =2 \sqrt{\SI{10}{\micro m}\times \SI{50}{m}}  \approx \SI{50}{mm} \, .     \nonumber
\end{equation}
With a safety factor of 1.2, the gap between the plates is then \SI{60}{mm}.
The maximum vertical beta function determines the vertical beam size, as
\begin{equation}
  2 y_\text{max} = 2 \sqrt{a_{y,\text{max}} \beta_{y, \text{max}}} =  2 \sqrt{\SI{10}{\micro m} \times \SI{200}{m}}  \approx \SI{90}{mm} \, .   \nonumber
\end{equation}
The field homogeneity requirements still have to be determined, but may be stringent, owing to the weak vertical focusing and the resulting large vertical betatron functions.
Ring element counts, geometry, and other bend parameters are given in Tables\,\ref{tab:ptr:Table2} and \ref{tab:ptr:Table3}.

\begin{table}
\begin{minipage}[t]{.5\textwidth}
\centering
    \caption{\label{tab:ptr:Table2}Geometry}
    \begin{tabular}[t]{l l l}
   \hline \hline
    & & Unit \\
    \hline 
   No $B$--$E$ deflectors                       & 8       &     \\
  No arc D quads                          & 4       &     \\
    No arc F quads                          & 8       &     \\
    Quad length                             & 0.400   &  m  \\
    Straight length                         & 8.000   &  m  \\
    Bending radius                          & 8.861   &  m  \\
    Electric plate length                   & 6.959   &  m  \\
    Arc length ($\SI{45}{\degree}$)         & 15.7    &  m  \\
    Circumference total                     & 102     &  m  \\ 
    r.m.s. emittance $\epsilon_x =\epsilon_y$  & 1.0     & $\si{\pi.mm.mrad}$ \\
    $a_x=a_y$                               & 10.0    & $\si{\pi.mm.mrad}$ \\
    \hline \hline
    \end{tabular}
\end{minipage}
\begin{minipage}[t]{.4\textwidth}
\centering
    \caption{\label{tab:ptr:Table3}Bend elements, 45\,MeV}
    \begin{tabular}[t]{l l l}
  \hline \hline
    & & Unit \\
    \hline
    {Electric}            &              &          \\
    \hline
    Electric field            & 7.00         & MV/m     \\
    Gap between plates        & 60           & mm       \\
    Plate length              & 6.959        & m        \\
    Total bending length        & 55.673       & m        \\
    Total straight length    & 44.800       & m        \\
    Bend angle per unit       & ($\SI{45}{\degree}$)  &  m       \\
    \hline
    {Magnetic}            &              &          \\
    \hline
    Magnetic field            & 0.0327      &  T       \\
    Current density           & 5.000        & $\si{A/mm^2}$ \\
    Windings/element          & 60           &          \\
    \hline \hline
    \end{tabular}
\end{minipage}
\end{table}

Lattice flexibility is a goal for the design. The betatron working points can be varied over a large range, as shown in \Fref{fig:Figure2a}. A typical plot of the beta functions is given in \Fref{fig:Figure3}.

\begin{figure}
\centering
\includegraphics[width = 0.55\textwidth,clip]{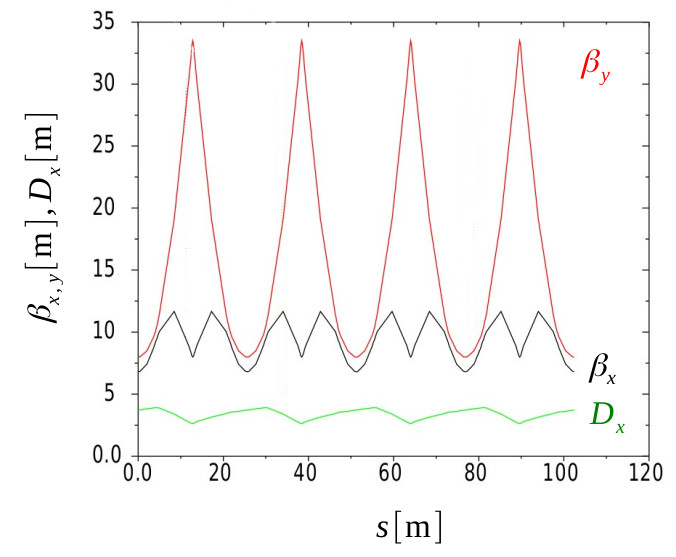}
\caption{\label{fig:Figure3}Beta functions and dispersion for a typical working point: $k_\mathrm{QF}=0.05$, $k_\mathrm{QD}=0.3$, $k_\mathrm{QSS}=0$, $Q_x=1.73$, $Q_y=1.20$.}
\end{figure}

\section{Electric and magnetic bends\label{sect:Soltner}}
\subsection{Electric part}
The simplest electrostatic deflectors consist (ideally) of two cylindrical parallel metal plates with equal potential and opposite sign. Such a design, with sufficiently high-capacitance plates, corresponds to a field index $m \approx 0$. With the zero voltage contour of the electric potential defined to be the centre line of the deflector, the `ideal orbit' of the design particle stays on the centre line. The electric potential is defined to vanish on the centre line of the bends, as well as in drift sections well outside the bends. Thus, the electric potential vanishes everywhere on the ideal particle orbit. With the electric  potential seen by the ideal particle continuous at the entrance and exit of the deflector, its total momentum is constant everywhere (even through the RF cavity).

There are restrictions on the minimum distance between deflectors. Recent candidate ring lattice studies have limited the horizontal good field region for stored particles at \SI{50}{mm}. This requires the minimum distance between electric deflector plates to be about \SI{60}{mm}. The vertical beam size is several times larger than the horizontal one. This imposes restrictions on the vertical dimensions of the flat part of the deflector too. Minimum vertical dimensions of the bending elements will be more than \SI{100}{mm}. To minimize breakdown probability, the shape of the deflectors should follow Rogowski profiles \cite{Rogowski1923, Rogowski1926} at both vertical edges.  The ends of individual deflectors need to be shaped to match  stray fields with subsequent deflectors.

The designed ring lattice requires electric gradients in the range $5$--$\SI{10}{MV/m}$. This is substantially more than the standard values for most accelerator deflectors separated by a few centimetres. Assuming a distance of \SI{60}{mm}  between the plates, to achieve such high electric fields, we have to use high-voltage  (HV) power supplies. At present, two \SI{200}{kV} power converters are available for testing deflector prototypes.  The field emission, field breakdown, dark current, electrode surface,
and conditioning are to be studied using two flat electrostatic deflector plates, mounted on a movable support with the possibility of changing the separation from 20 to \SI{120}{mm}. The residual ripple of power converters is expected to be of the order of $\pm \num{e-5}$ at a maximum \SI{200}{kV}. Such a power converter ripple has been found to excite the beam and to lead to longitudinal emittance blow-up in other electrostatic rings\cite{vonHahn:2016wjl}. Thus, studies on beam blow-up caused
by HV converter ripple are planned and mitigation measures may be required.

\subsubsection{Design of the electric part}
The electric part of the ring can be considered a plate capacitor, whose distance parameter has been determined from beam optics considerations. The 2D cross-section is shown in \Fref{FIG:Figure4}. 
The contours of the upper and lower edges of the plates are rounded according to the Rogowski shape principle. Owing to the finite radius of curvature of the plates, of about \SI{9}{m,} a field gradient is generated. Its magnitude can be estimated for the case of infinitely high capacitance plates because, in this case, the electric potential is purely logarithmic, and its gradient---the electric field---can be obtained analytically. For finitely high capacitance plates, this approach should still provide a good approximation,
\begin{equation}
    U(\rho) = U_\text{i} + (U_\text{o} - U_\text{i})\cdot \frac{\ln(\rho/\rho_\text{i})}{\ln(\rho_\text{o}/\rho_\text{i})}\,,
\end{equation}
where the corresponding electrical field in the radial direction is given by
\begin{equation}
    E_{\rho}(\rho) = - \frac{\partial}{\partial \rho}  U(\rho) = -\frac{U_\text{o} - U_\text{i}}{\rho} \cdot\frac{1}{\ln(\rho_\text{o}/\rho_\text{i})}\,,
\end{equation}
and $U_\text{i}$ and $U_\text{o}$ are the potential values on the inner (i) and  outer (o) capacitor plates, respectively, with  corresponding  radii
of $\rho_\text{i} = \SI{8.831}{m}$ and  $\rho_\text{o} = \SI{8.891}{m}$. Here, $-U_\text{i}$ = $U_\text{o} = \SI{210.2}{kV}$.

\begin{figure} [ht!]
\centering
\includegraphics[scale=0.4]{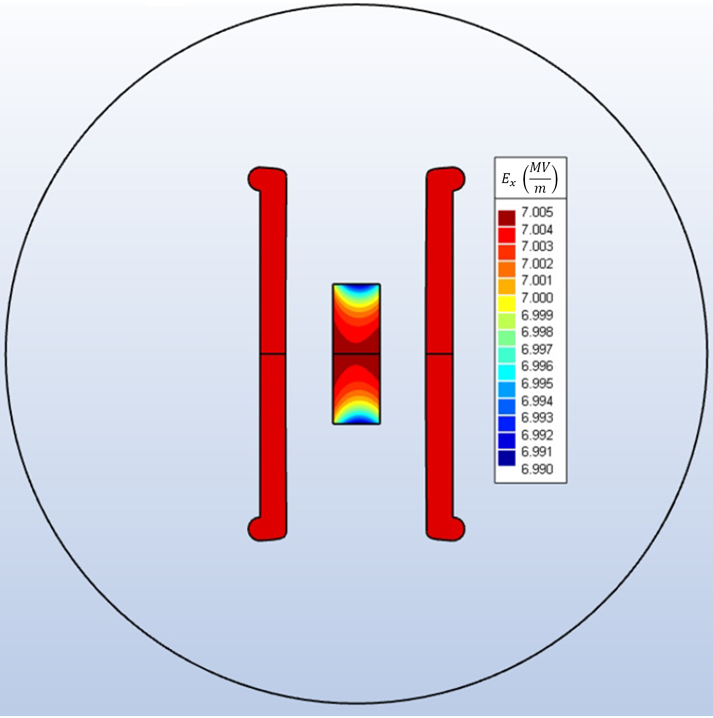}
\caption{\label{FIG:Figure4}Cross-section of the capacitor (in red) inside the beam tube (outer circle). The distance between the plates is \SI{60.7}{mm} and their height (straight part) is \SI{151.5}{mm}. The region of interest  is represented by the two central rectangles.}
\end{figure}

\Figure[b]~\ref{FIG:Figure5} shows the potential  and  electric field strength between the electrode plates calculated using these parameters. It can be noted that the radial electric field is inversely proportional to the distance from the bending centre. This is the behaviour expected for an electric bend with field index $m = 0$. Beam optics studies have not yet concluded whether a small field index $|m| \ll 1$ is preferable. A non-zero field index can, in principle, be implemented by an appropriate shape of the electrodes.

\begin{figure} [ht!]
\centering
\includegraphics[scale=0.28]{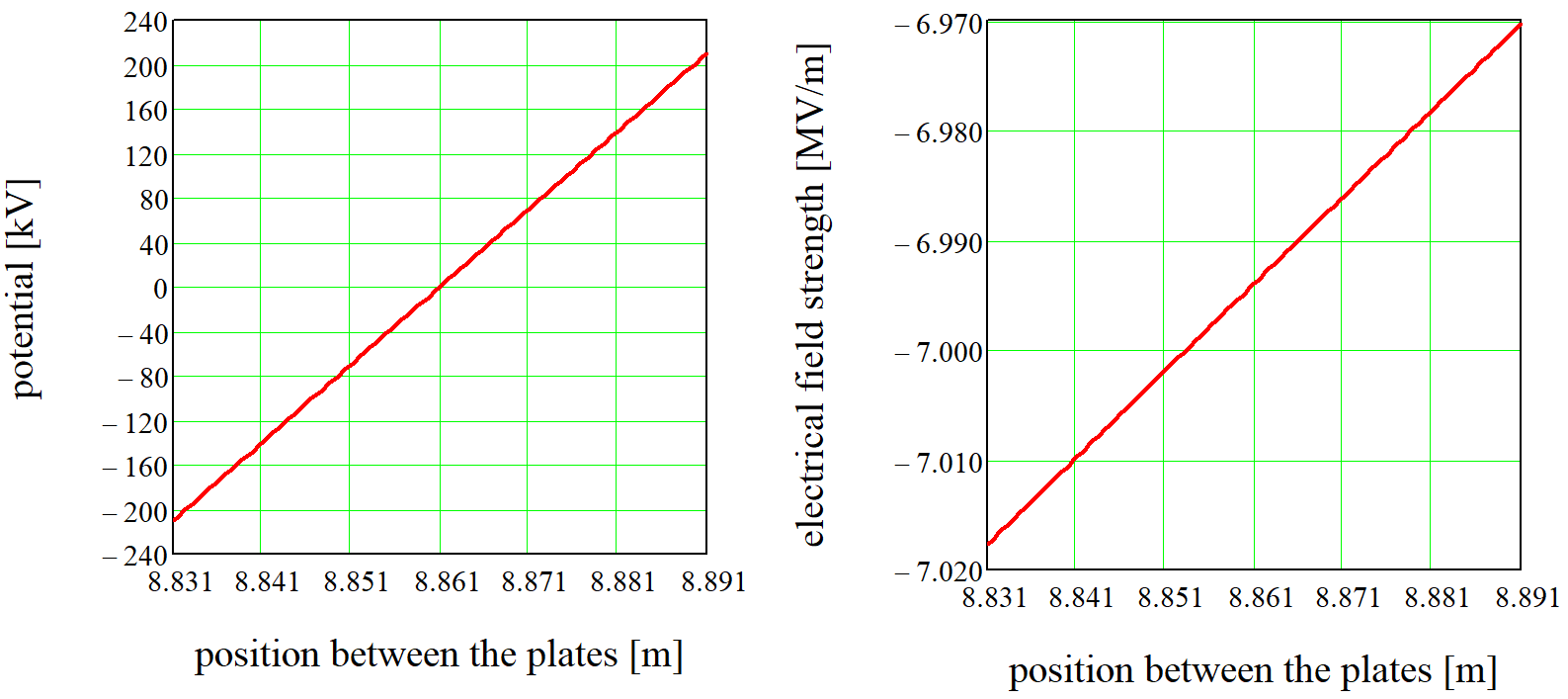}
\caption{\label{FIG:Figure5}Potential  and corresponding electric field strength between the capacitor plates in the case of infinitely high capacitance plates. The average field strength is about \SI{6.998}{MV/m}.}
\end{figure}

The homogeneity profile of the electric field in a region of interest (ROI), which, so far, is smaller than the good field region specified, is shown in \Fref{FIG:Figure7}. The average value is about \SI{7}{MV/m}, the same as predicted by the theoretical considerations leading to the results of \Fref{FIG:Figure5}. The maximum relative difference of the electric field in the ROI is about \num{2.1e-3}. Note that the field homogeneity achieved with this geometry may not yet be sufficient for stable motion of the particles in the ring. \Figure[b]~\ref{FIG:Figure7} shows how the field homogeneity in the ROI depends on the geometry of the deflector electrodes and gives an indication of means to improve the quality. Depending on the requirements still to be defined, further optimizations may be necessary.



\begin{figure}
\centering
\includegraphics[scale=0.5]{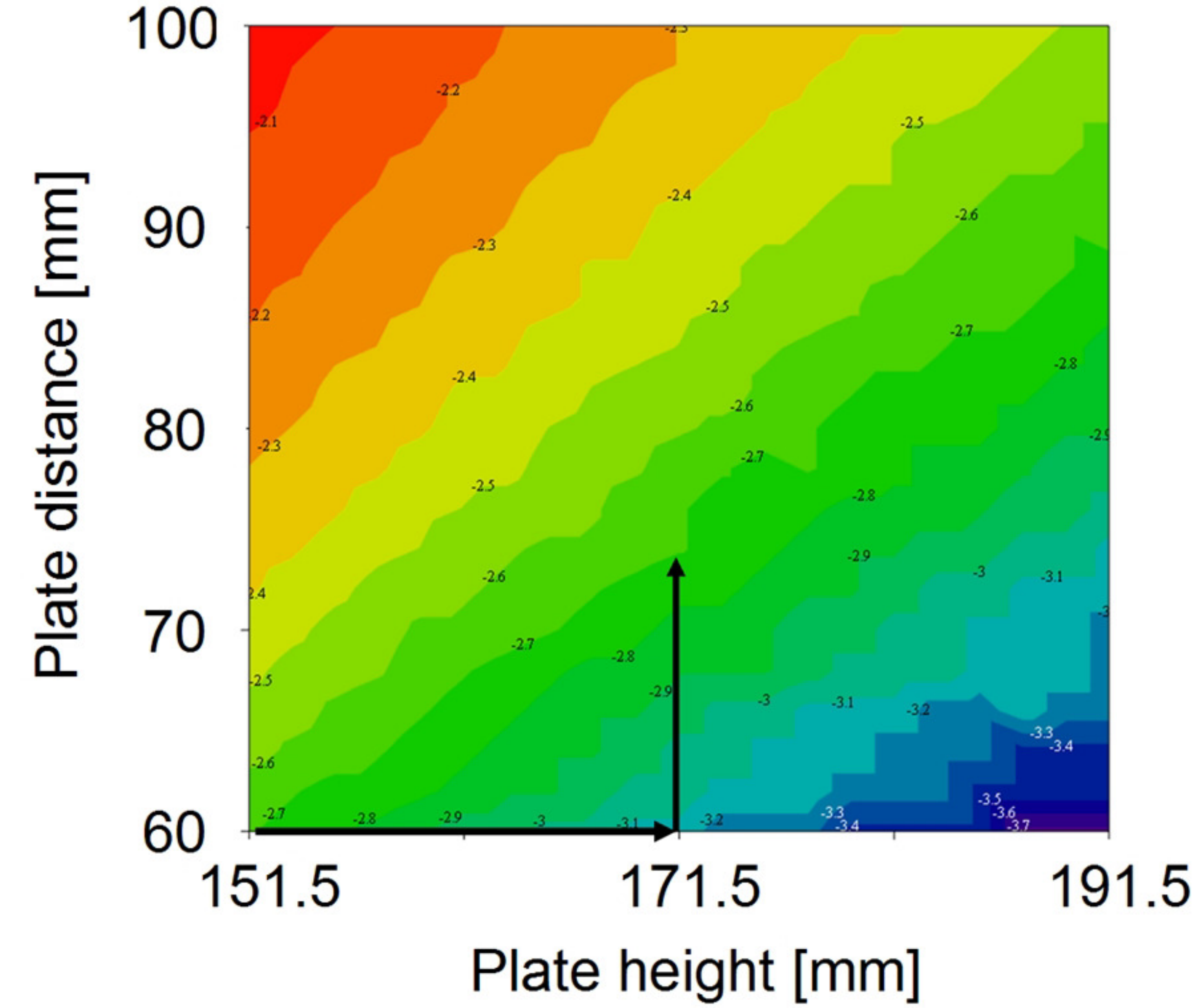}
\caption{\label{FIG:Figure7}
Variation (on a logarithmic scale) of the homogeneity of the electrical field strength as given by the difference of the maximum and minimum values on the circumference of the region of interest for a straight capacitor. For
example, with no change to the geometry, the homogeneity is close to \num{e-2.7} = \num{2.1e-3}. Enlargement of the (nearly) straight section of the plates by about \SI{20}{mm} improves this value to \num{e-3.2} = \num{6.3e-4} (horizontal arrow). A subsequent increase of the plate half distance by \SI{12}{mm} deteriorates this value  to about \num{e-2.7} (vertical arrow).}
\end{figure}

\subsection{Magnetic part}
In this section, we deal with the design of the magnets. The stray field of the magnet is investigated separately, because it determines the shape of the electrode plates for the combined electric and magnetic design. The nominal magnetic bending field is vertical, $B_y=\vec B \cdot \vec e_y$. For the combined $E$--$B$ prototype ring, a first design has been made based on the requirements on integrated electric and magnetic fields. Specifically, the magnetic flux density of the magnet should be $B = \SI{32.65}{mT}$, with
 a corresponding electric field of  $E = \SI{6.998}{MV/m}$. 
The prototype ring comprises electric and magnetic units. The design is shown in \Fref{FIG:Figure1}.

\begin{figure}
\centering
\includegraphics[scale=0.45]{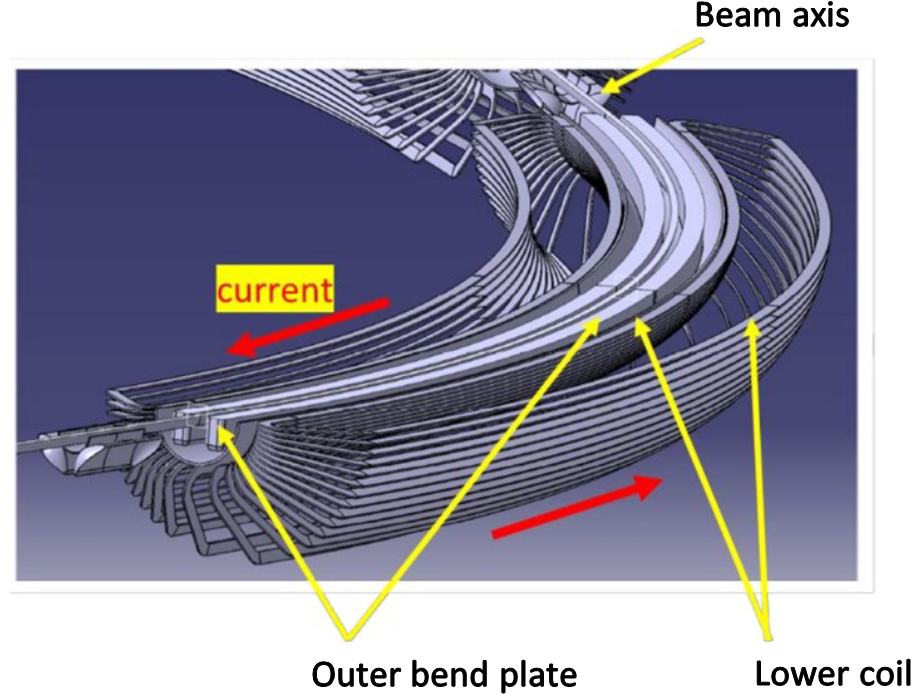}
\caption{\label{FIG:Figure1}One-quarter of the combined electric
and magnetic prototype ring. Two $\cos \theta$ dipoles surround the beam tube, in which the capacitor plates are accommodated.}
\end{figure}

All the magnetic design simulations were carried out using the programs Amperes (3D) and Magneto (2D) by IES\footnote{Integrated Engineering software (IES), \url{https://www.integratedsoft.com}.}. For the electric field simulations, the programs Coulomb (3D) and Electro (2D), by the same company, were used.

\subsubsection{Design of the normal conducting magnets}
The required vertical flux density of $B_y = \SI{32.65}{mT}$ is small enough to envisage a solution with normal conducting, even air-cooled, coils. The magnets are designed according to the $\cos \theta$ scheme to ensure a high level of homogeneity of the magnetic field. To avoid detrimental magnetic fields from the return paths of the cables in the $\cos \theta$ dipole, even these have been distributed in a $\cos \theta$ fashion. This reduces the effective field in the ROI, but is feasible with the modest flux density required. The cross-section of the $\cos \theta$ magnet is depicted in \Fref{FIG:Figure2}.

\begin{figure}
\centering
\includegraphics[scale=0.28]{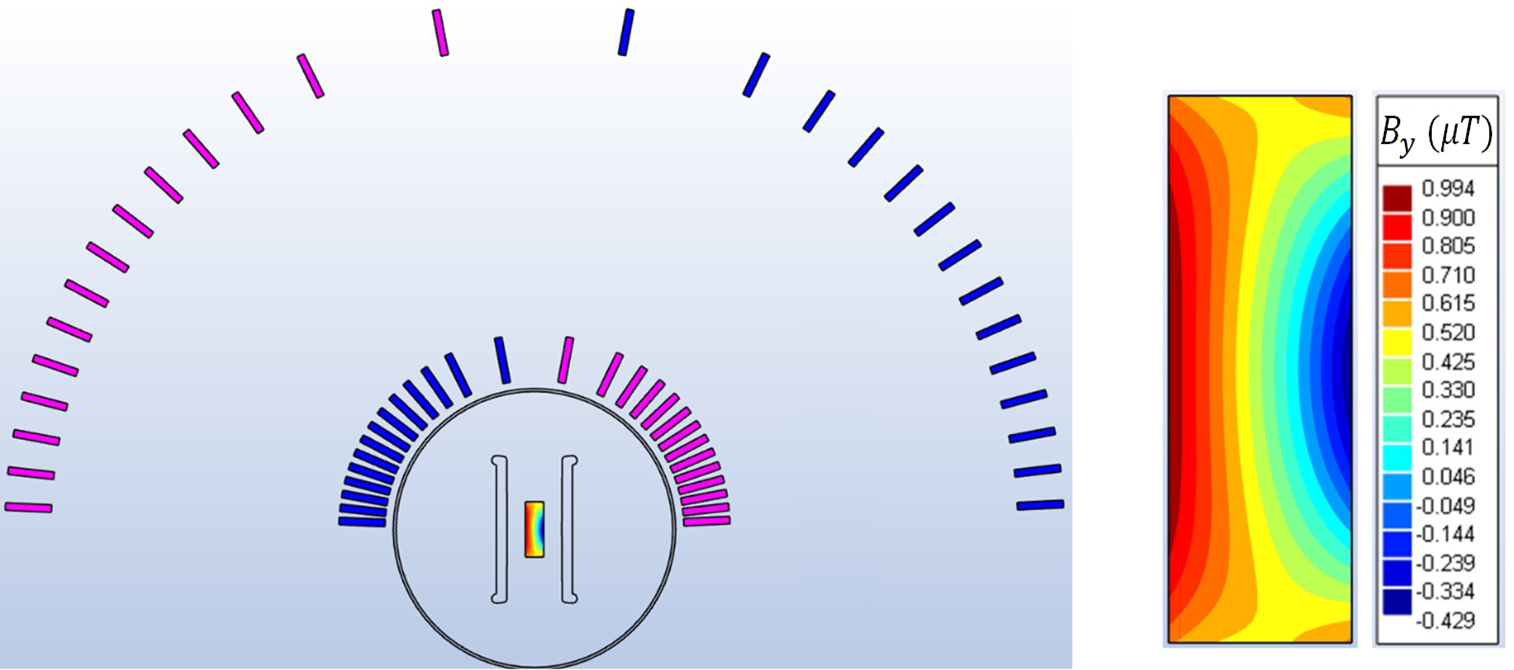}
\caption{\label{FIG:Figure2}Left: Cross-section of $\cos \theta$ dipole (only coils for the upper part are shown) and resulting field homogeneity. The inner conductors dominate the field. The return conductors are placed outside so as not to degrade the field quality, but at the cost of reducing the field strength. The current direction is represented by the colour of the conductors. The beam tube is represented by the two concentric circles, with an inner diameter of \SI{300}{mm}. The outer diameter of the conductor circumference is \SI{1148}{mm}. The ROI can be seen as a rectangle in the centre, surround by two rectangles representing the electrodes. Right: Field homogeneity in the ROI in more detail.}
\end{figure}

In this design, the conductors have a cross-section of $\SI{50}{mm} \times \SI{8.1}{mm}$. The rectangle in the centre of \Fref{FIG:Figure2} represents the ROI, with dimensions $\SI{20}{mm} \times \SI{60}{mm}$. The average flux density in the ROI is $B_y = \SI{32.65}{mT}$.  \Figure[b]~\ref{FIG:Figure2} shows the \emph{deviation} from this value. It is less than \SI{1}{\micro\tesla}, \ie so small that, in reality, the homogeneity will rather be dominated by manufacturing tolerances.
%
The contour plot in the panel on the right in \Fref{FIG:Figure2} is slightly asymmetric, because the magnet is not straight but follows a radius of about $\rho = \SI{8.8}{m}$. This curvature introduces a gradient in the magnetic flux density, leading to a left--right asymmetry. This asymmetry has been reduced by the introduction of a slight rotation of the upper conductors and a reverse rotation of the lower ones by about \SI{0.16}{\degree} around the centre of the arrangement, which cannot be perceived in the figure, because of the smallness of this angle.

The current density in the conductors is about $\SI{2.6}{A/mm^2}$. For this design, the generated power amounts to about \SI{43}{kW} at a current of \SI{1053}{A}, corresponding to a voltage drop of about \SI{41.0}{V}. This may be too large a value to rely on air cooling alone for the removal of the generated heat, but design studies have been conducted that show that the length of the conductors can be enlarged from \SI{8.1}{mm}, thus reducing the current density and the heat load without compromising the field homogeneity. At present, it seems reasonable to assume a water-cooled magnet.  The mass of the copper conductors for a single magnet amounts to about \SI{3000}{kg}. The magnet can be accommodated outside the vacuum tube.

\subsubsection{Matching of magnetic and electric stray fields}
A staged approach was agreed on to match electric and magnetic fields. A global matching of the electric and magnetic fields based on field integrals will suffice in the first stage. This requirement can easily be fulfilled with the designs presented. Nevertheless,  improved matching between electric and magnetic fields may be required to reach the EDM sensitivity level aimed for; this issue is studied here.

Inside the magnet and inside the electrodes, the fields are quite constant in amplitude, and their ratio can be chosen according to the requirements. In the stray field regions, both fields reveal different decay lengths, because the coils generating the magnetic field are much larger in size than the electrodes generating the electric field. For this reason, the magnetic stray field has a much larger decay length, and the geometry of the electrodes can be adapted to the decay of the magnetic field. The decay of the magnetic field can hardly be changed, because the way in which the inner conductors are to be connected to the outer returning counterparts is more or less determined by the cross-section, shown in \Fref{FIG:Figure2}, resulting in the field between adjacent bends plotted in \Fref{fig:BetweenMagnets}.

\begin{figure} [hb!]
\centering
\includegraphics[scale=0.5]{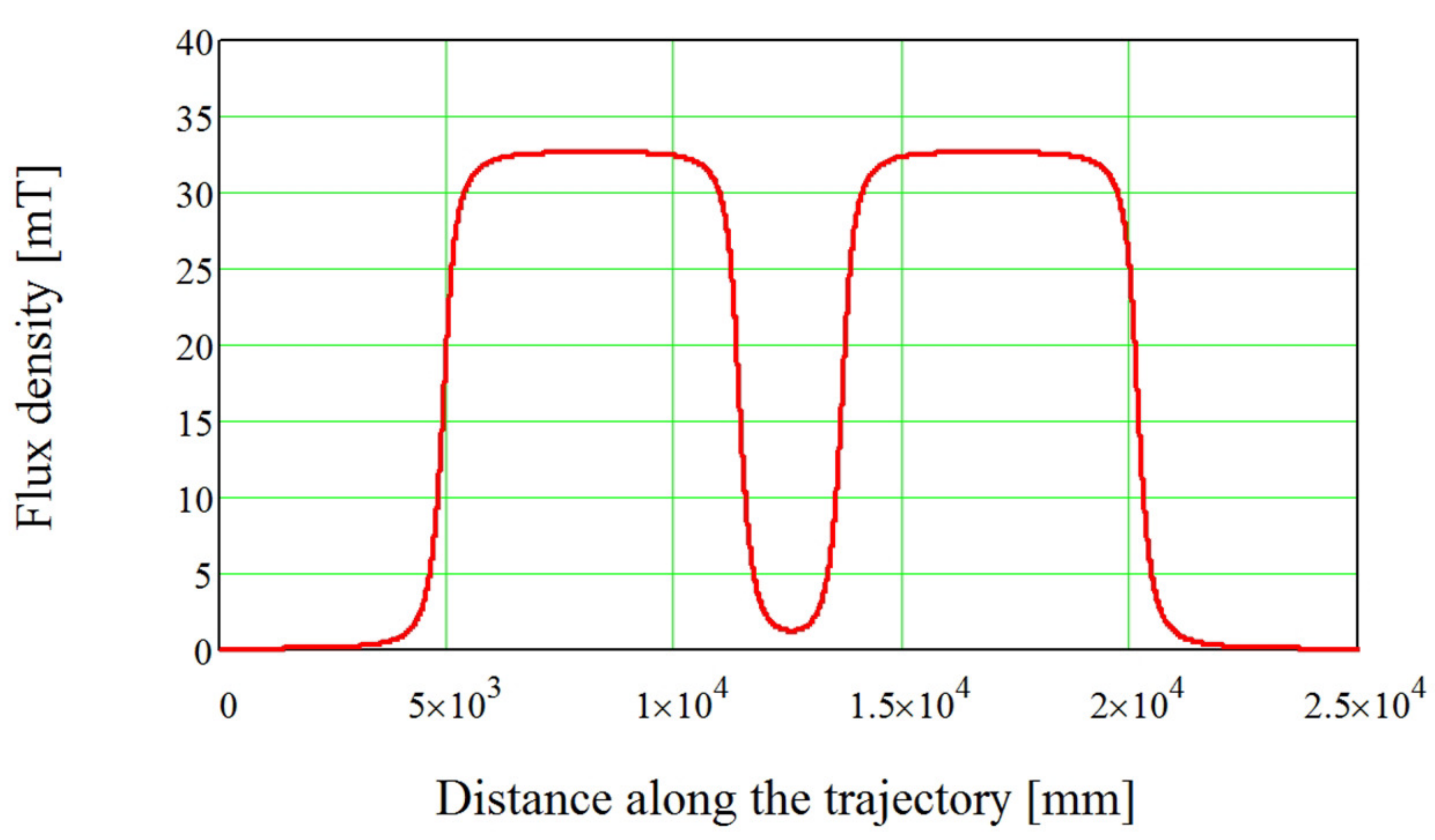}
\caption{\label{fig:BetweenMagnets}Flux density between the two magnets shown in \Fref{FIG:Figure1} along the central trajectory within the ROI. In the centre of the magnets, a flux density of \SI{32.65}{mT} is obtained, whereas midway between adjacent magnets, at $\SI{1.25e4}{mm}$, the flux density drops to  about \SI{1}{mT}. }
\end{figure}

It is well known from electrostatics that the electric field of a plate capacitor is inversely proportional to the distance of the plates for a fixed potential difference. Several simulations for this study have shown that, locally, the electric field also follows this rule. More specifically, as long as the field plates are much higher than the gap distance, the local electric field is inversely proportional to the plate distance at this location. For this reason, a flux density distribution, like the one shown in \Fref{fig:BetweenMagnets}, can be regarded as the inverse gap distance of a capacitor providing the same field behaviour. From this consideration, we can already conclude that it will be difficult to fulfil the requirement of locally matching electric and magnetic fields at all locations on the trajectory, because the magnetic field drops to very low values outside the magnet pairs, which would correspond to a very large gap between electrodes. Obviously, this would require large-aperture vacuum chambers and would not be practical. The situation is even more difficult around the extremities of bends adjacent to straight sections. In this case, the question arises as to  whether several such electrode pairs with stepwise decreasing potentials may be stacked along the trajectory to approximate the magnetic field decay in a stepwise fashion.

\Figure[b]~\ref{FIG:Figure9} shows an example of this stacking principle for the field decay between magnets. This figure shows the normalized electric and magnetic fields obtained with numerical simulations. These normalized field values cannot be distinguished on this scale but the difference values (red curve) show small features in the overlap region where two neighbouring electrode pairs meet. In total, five  capacitors with decreasing potential differences are used, which require the same number of power supplies, unless a solution with voltage dividers is chosen. The number of capacitors is dictated by the maximum expansion factor (the ratio between maximum and minimum opening of the electrode pair) accepted, which, for the example in \Fref{FIG:Figure9}, is about 1.9. This translates into a local distance of the capacitor of $\SI{60}{mm} \times 1.9 = \SI{114}{mm}$.

\begin{figure} [ht!]
\centering
\hspace{9mm} \includegraphics[scale=0.37]{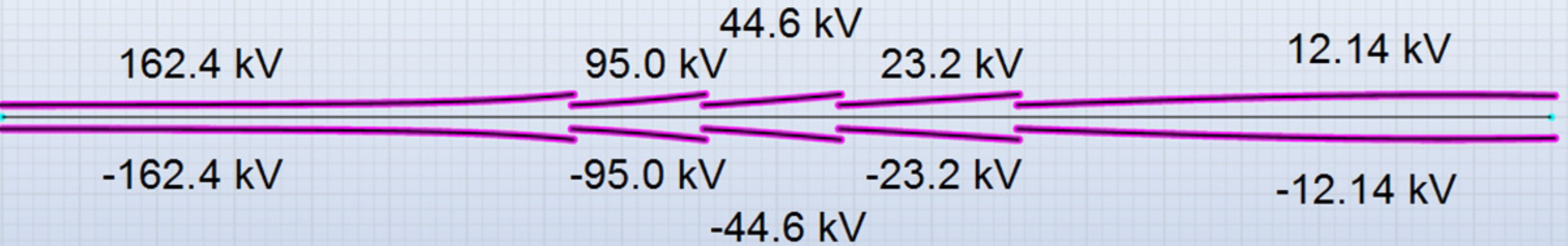}  \vspace{3mm} \\
\includegraphics[scale=0.5]{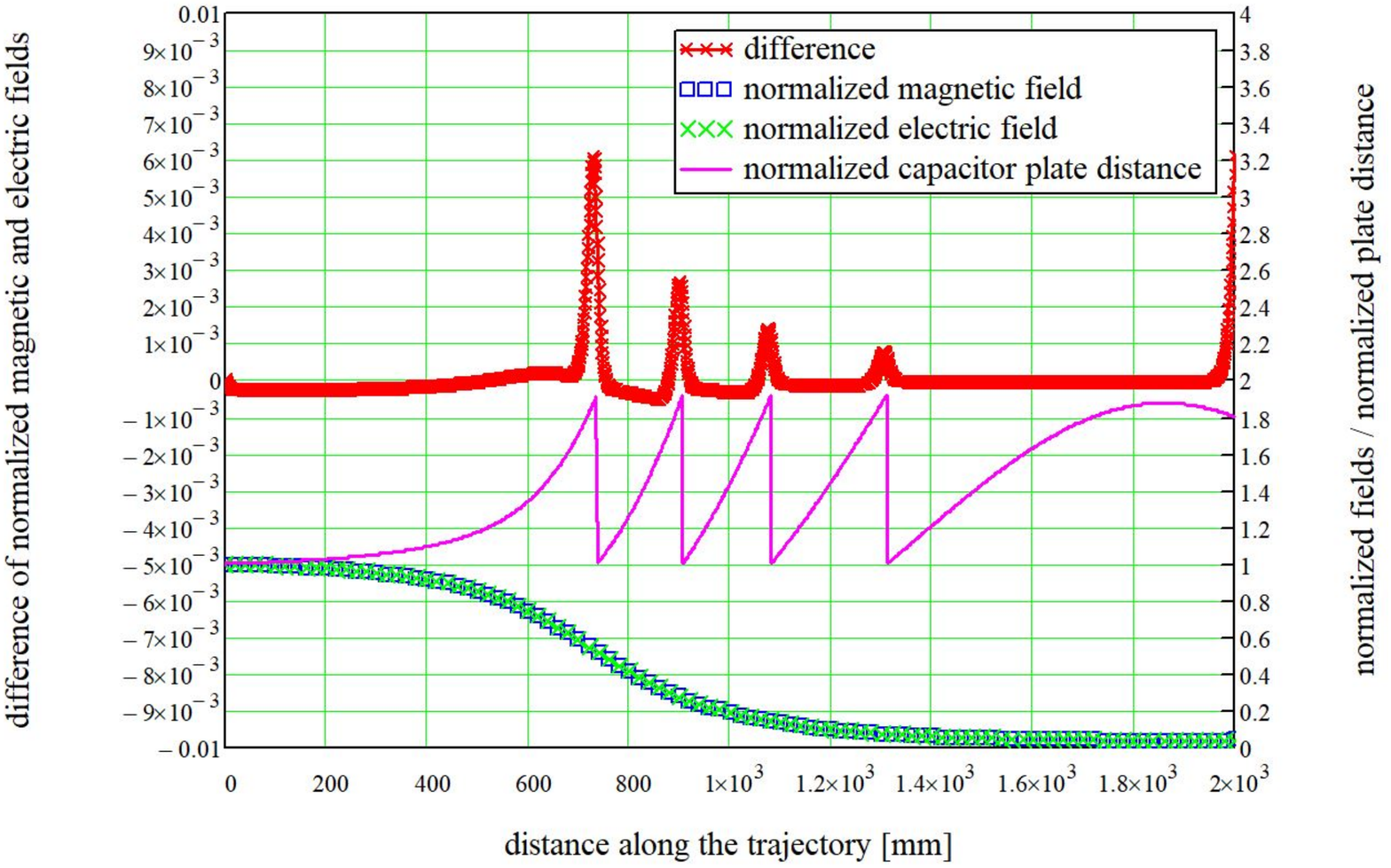}
\caption{\label{FIG:Figure9} Top: Electrode configuration with five pairs of plates with different potentials and openings varying with longitudinal position to improve the matching between electric and magnetic bending; Bottom: results obtained. }
\end{figure}

Preliminary attempts have been made to reduce the amplitude of the mismatch between  electric and magnetic fields (red curve in \Fref{FIG:Figure9}) even further by letting the electrode pairs overlap slightly along the trajectory. This goal seems to be achievable, but should be pursued only after a thorough engineering of the design has been carried out.


If larger expansion factors are acceptable, there may be space in the wide gap between the magnets to accommodate auxiliary devices, such as quadrupoles or beam position monitors. \Figure[b]~\ref{FIG:Figure11} shows such an option with a larger capacitor gap at a coordinate of \SI{1800}{mm}.

\begin{figure}
\centering
\includegraphics[scale=0.45]{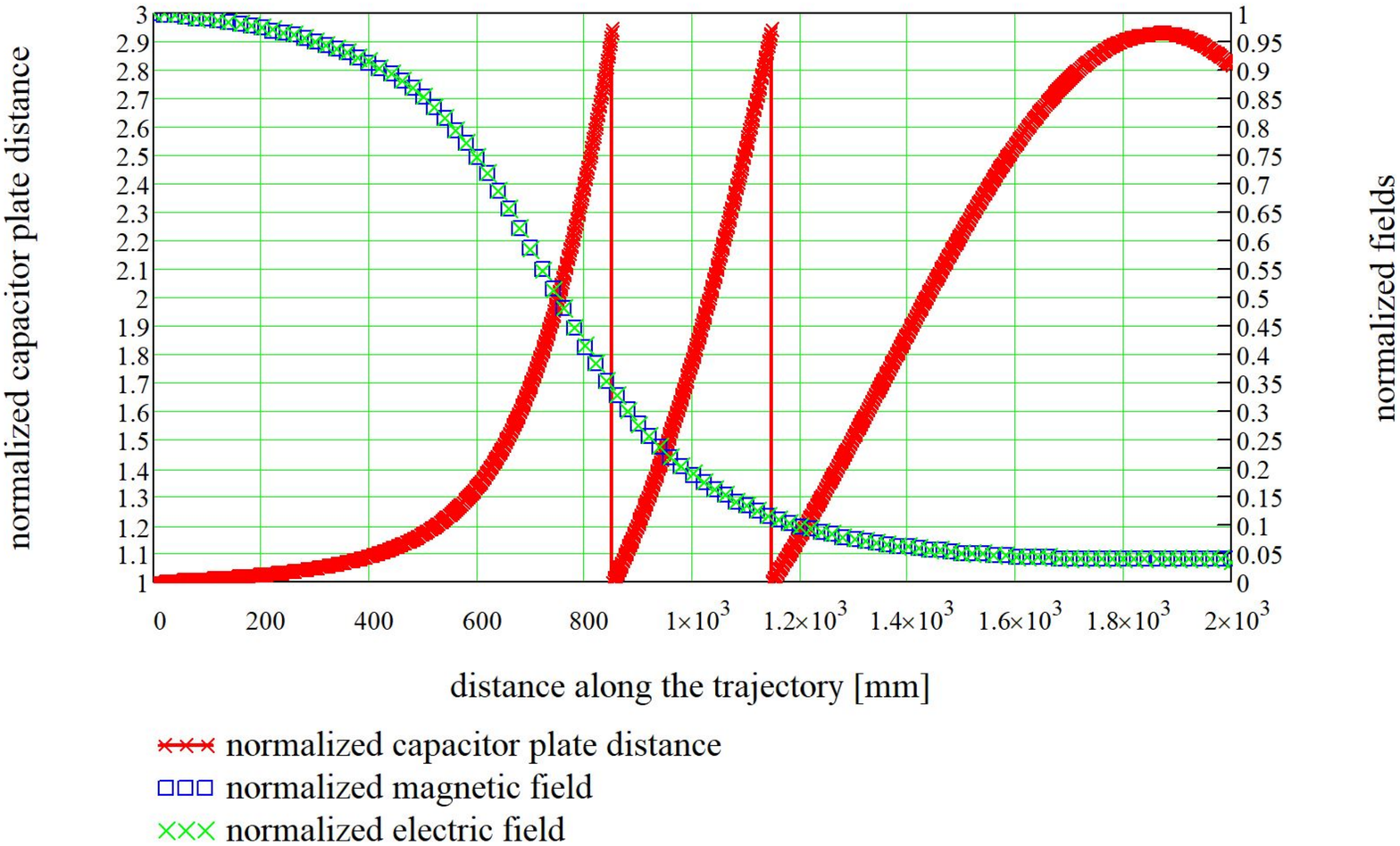}
\caption{\label{FIG:Figure11}As for Fig.~\ref{FIG:Figure9}, except with a larger expansion factor of about 2.95, yielding a larger space at a longitudinal coordinate of \SI{1800}{mm,} with a diameter of $\SI{60}{mm} \times 2.95 = \SI{177}{mm}$. Only three capacitors are required in this case.}
\end{figure}

\section{Components}
\subsection{Beam position monitors } \label{ssc:BPMs}
About 20 beam position monitors (BPMs) are located around the ring, as shown in \Fref{fig:Rogowski}. A BPM is placed at the entrance and the exit of each bending unit. Additionally, one BPM will be placed  close to the quadrupoles in the straight sections. The BPMs must be mounted precisely and rigidly, as close as possible to the quadrupoles, to which they are accurately and rigidly attached.

\begin{figure}
\centering
\includegraphics[scale=0.64]{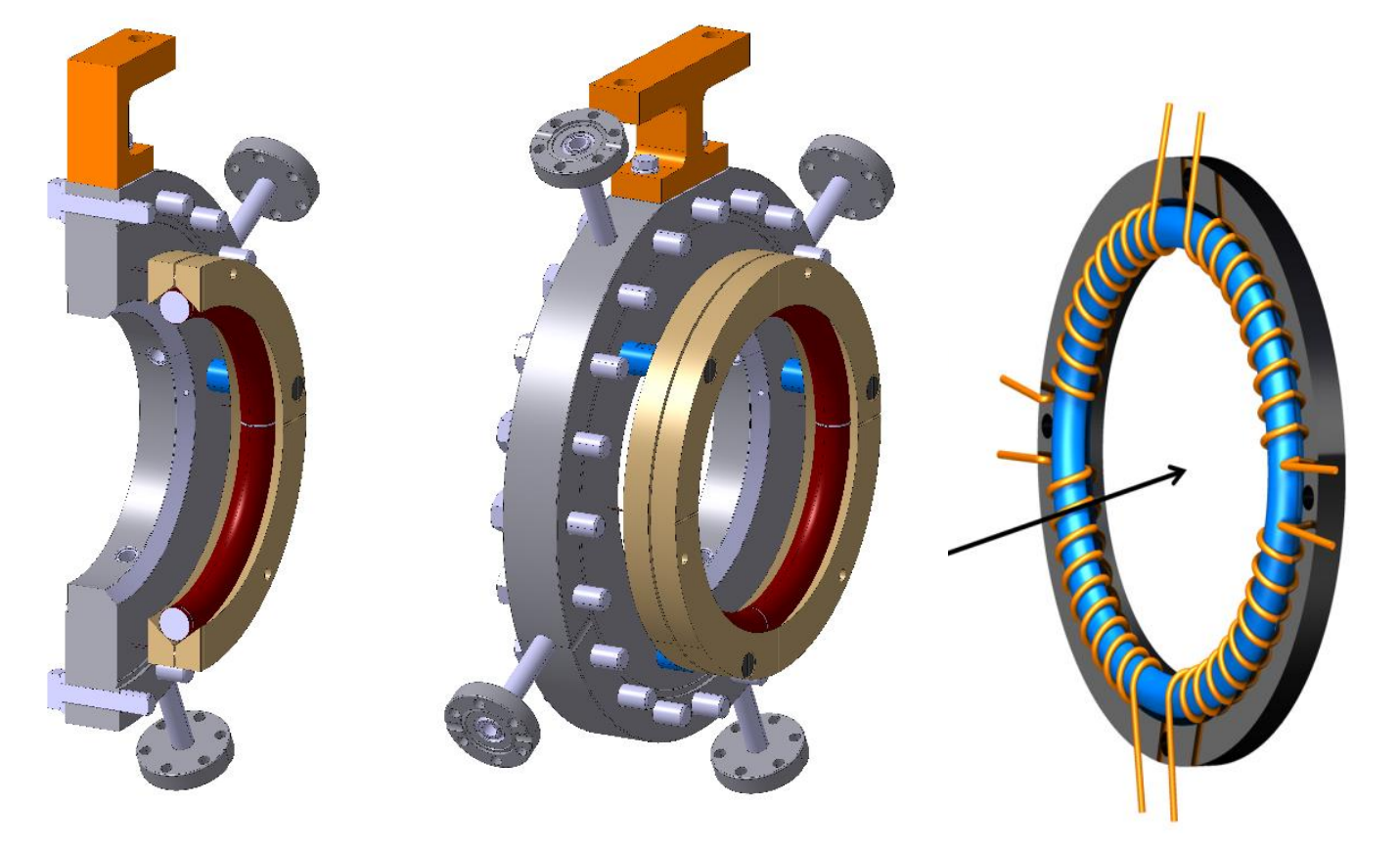}
\includegraphics[scale=0.40]{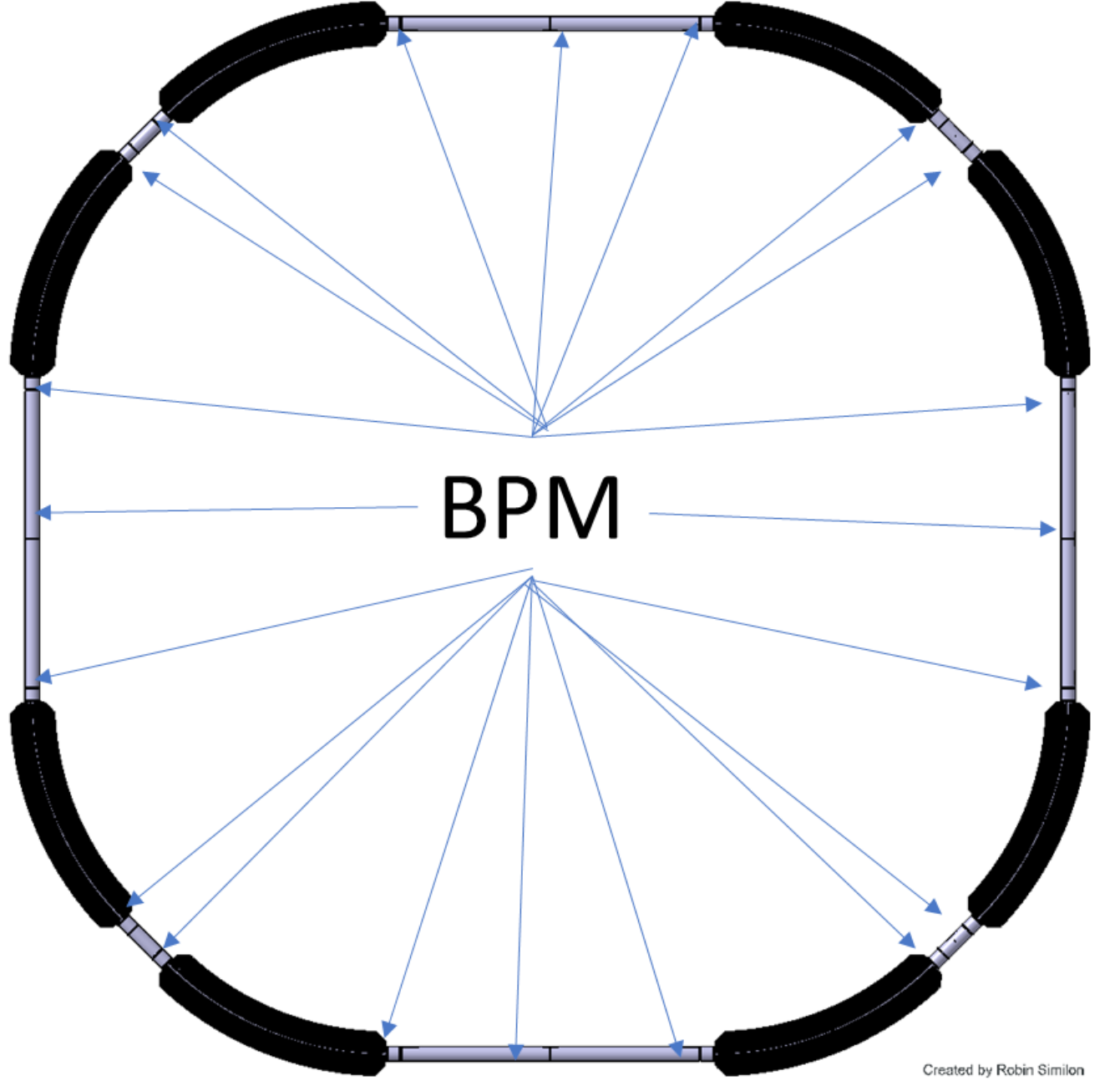}
\caption{\label{fig:Rogowski}Left:  Rogowski pick-up module \cite{Trinkel:2017,Fal-18-10}. The inner diameter is \SI{100}{mm}. Right: Beam position monitor locations around the PTR.}
\end{figure}

A new type of BPM is currently being investigated at the IKP, Forschungszentrum J\"ulich. The position resolution is measured as $\SI{10}{\micro m}$ over an active area of $\SI{100}{mm}\times\SI{100}{mm}$\,\cite{Trinkel:2017,Fal-18-10}. These BPMs, designed on the basis of a segmented toroidal (Rogowski) coil\cite{Rogowski1912}, are very attractive because of their short length, only \SI{60}{mm}, and the expected accurate absence of systematic relative transverse displacement of forward and backward beams.

One of the main purposes of the BPM system in the PTR, and later in the full-scale EDM ring, is the measurement of the relative displacement of CW and CCW orbiting beams, which is proportional to the residual radial magnetic field $B_r$  (see
\Sref{section:dual} and Appendix\,\ref{Chap:MagneticFields} for further details).

\subsection{Electric quadrupoles}
The quadrupoles for the PTR are characterized by an aperture diameter of \SI{80}{mm} powered at $\pm \SI{20}{kV}$. We have simulated a design with a vacuum chamber of  \SI{400}{mm} diameter (see \Fref{fig:ElectricQuad}, left panel). The maximum pole tip potential is \SI{30}{kV}, to allow some margin for  conditioning  the device.  A 3D design has been produced. The calculated sextupole, octupole, and higher harmonics of the integrated field seem very reasonable. The 3D integration model (see  \Fref{fig:ElectricQuad},
right panel) suggests that the device can be built within the allocated \SI{800}{mm} longitudinal length; the radial diameter is \SI{620}{mm}.

\begin{figure}
\centering
\includegraphics[scale=0.32]{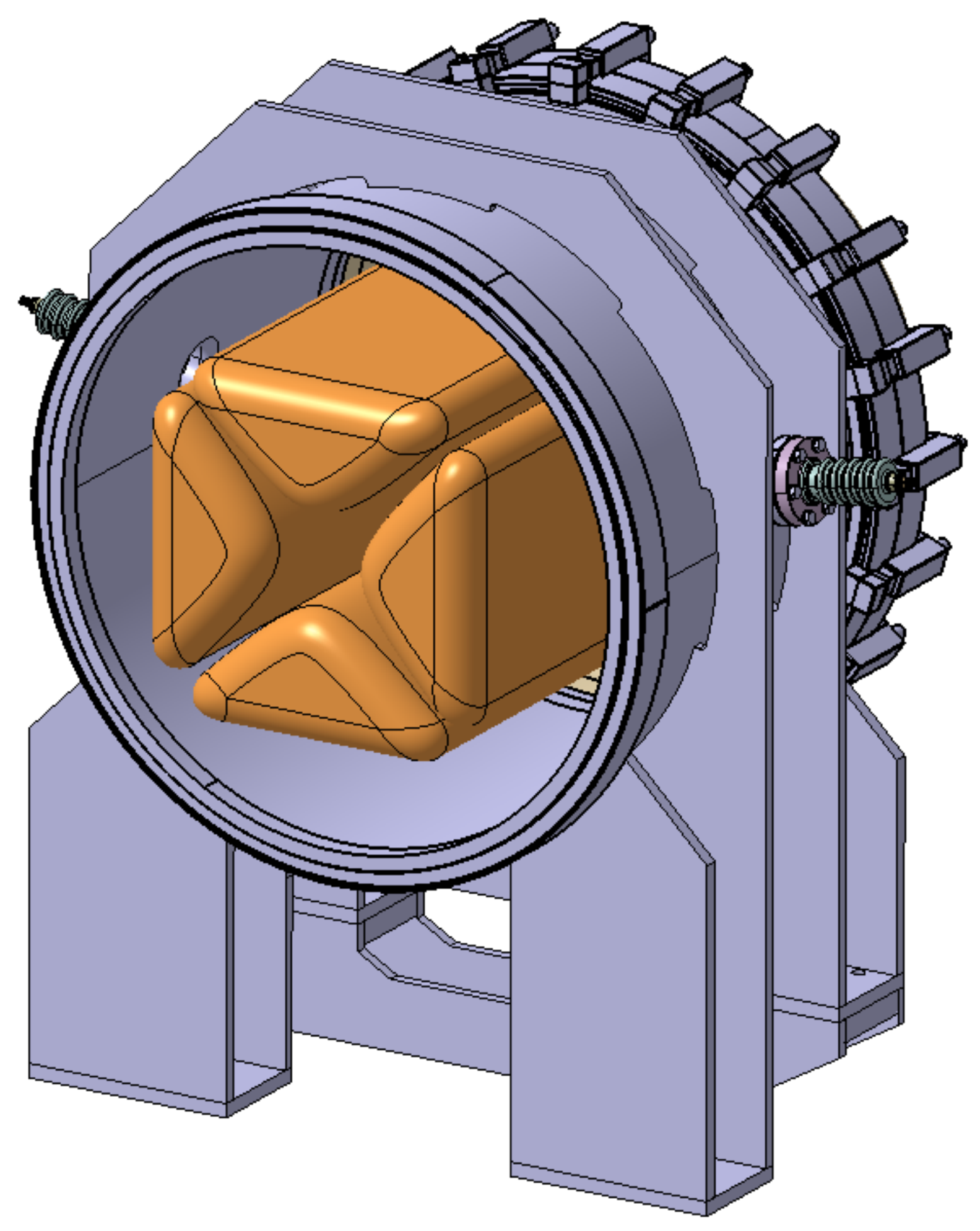}
\includegraphics[scale=0.40]{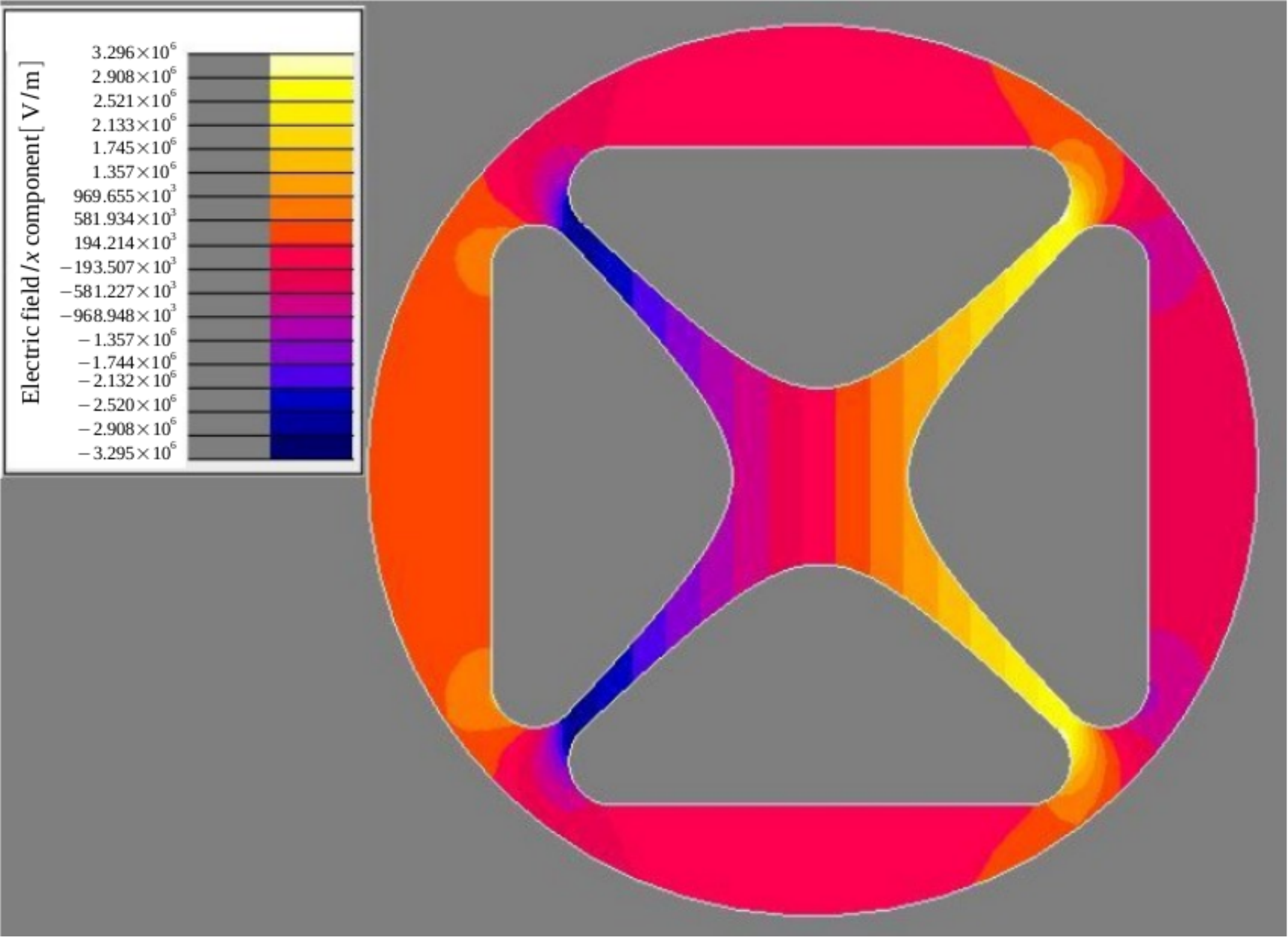}
\caption{\label{fig:ElectricQuad} Design of electrostatic quadrupoles: (left)  hardware; (right), integrated horizontal electric field}
\end{figure}


\subsection{RF solenoids}
The vertical polarization of a stored beam can be rotated into the horizontal plane by the longitudinal field of an RF solenoid. As shown in \Fref{fig:RF-slnd}, the RF solenoid at COSY is a 25-turn air-core water-cooled copper coil with a length of \SI{57.5}{cm} and an average diameter of \SI{21}{cm}. It has an inductance  of about \SI{41}{\micro \henry}, and produces a maximum longitudinal RF magnetic field of about \SI{1.17}{\milli \tesla} (r.m.s.) at its centre. The solenoid is a part of an RLC resonant circuit, which typically operates near \SI{917}{kHz} at an RF voltage of about \SI{5.7}{\kilo \volt} (r.m.s.), producing a longitudinal RF field integral of \SI{0.67}{\tesla\milli\meter}. Typical ramp-up times, from vertical to horizontal polarization, are about \SI{200}{ms}.

\begin{figure}
\centering
\includegraphics[scale=0.7]{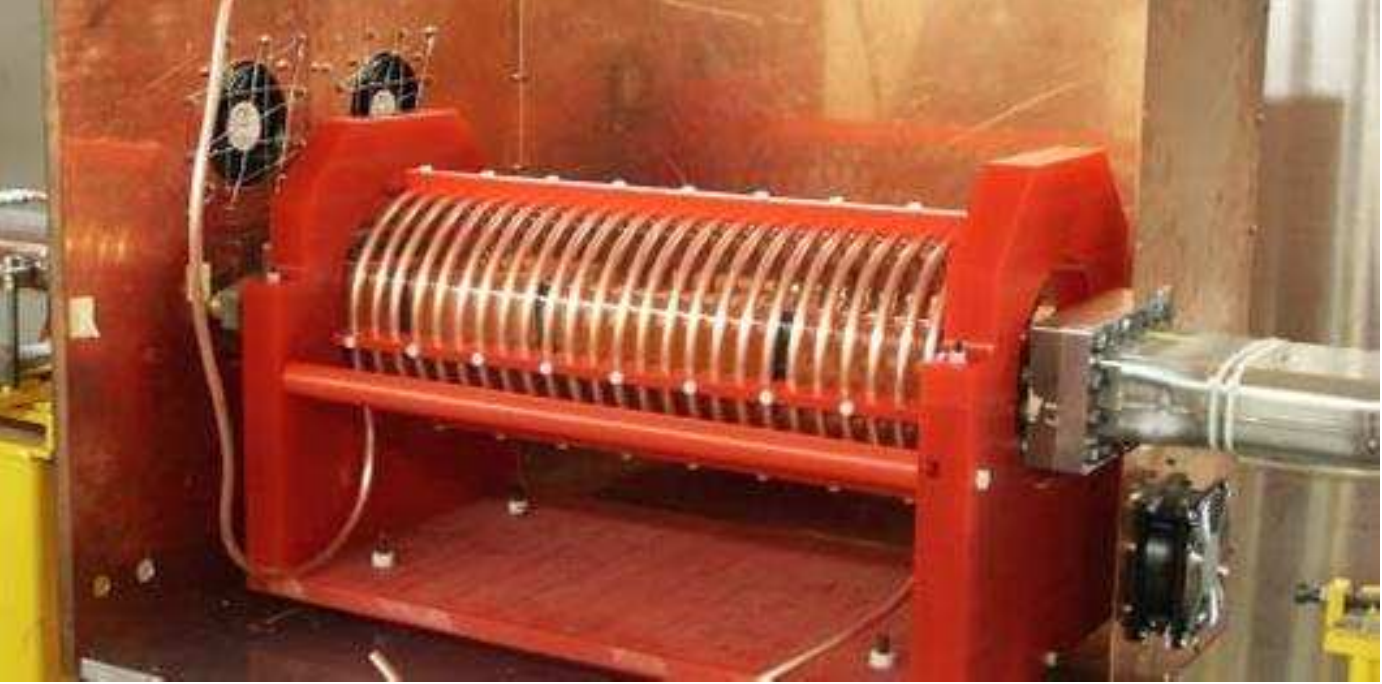}
\caption{\label{fig:RF-slnd}COSY LC-resonant RF solenoid. In COSY, this element precesses the polarization vectors of all particle bunches identically. Its role in the PTR ring will be the same.}
\end{figure}

\subsection{RF Wien filter}
\label{sec:PTR:RF-Wien-filter}
The outer  and inner parts of the COSY waveguide RF Wien filter\,\cite{Slim:2016pim} are shown in \Fref{fig:Figure9_1}.  Beam tuning manoeuvres described previously in this chapter  employed small radial magnetic fields to apply small controlled torque to the beam polarization to control the spin wheel (explained further in \Sref{sect:PTRPhysics}). Such a radial magnetic field also causes an undesirable beam orbit perturbation. In some cases, the applied radial magnetic field causes an acceptably small orbit perturbation. However, when this is not the case, an RF Wien filter must be used instead to minimize the Lorentz force on the stored particles. One way of expressing the Wien filter `strength' is to give the spin wheel angular velocity caused per watt of power applied to the RF Wien filter. In a COSY precursor RF Wien filter experiment, a Wien filter magnetic field integral of $\SI{2e-6}{\tesla\meter}$ caused a spin-wheel (sw) frequency of \SI{0.16}{Hz}, $f^{\rm sw} = \Omega^{\rm sw}/(2\pi) =\SI{0.16}{Hz}$. The power conversion was such that an RF power level of \SI{1}{kW} provided  a magnetic field times length integral equal to $\SI{1.6e-5}{\tesla\meter}$. This calibration factor was deduced from an experiment using \SI{0.97}{GeV/$c$} deuterons stored in COSY.

\begin{figure} [hb!]
\centering
\includegraphics[scale=0.35]{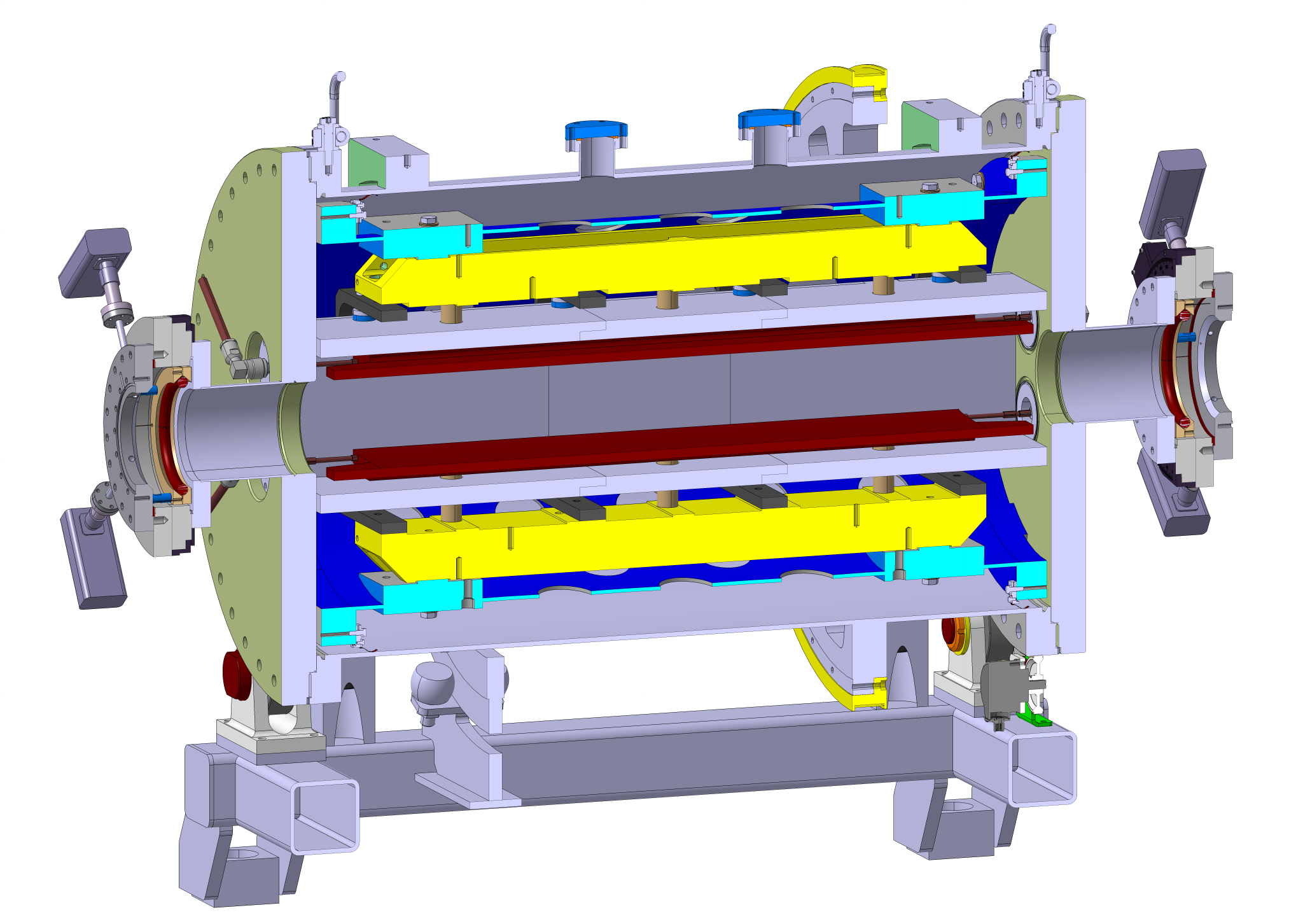}
\includegraphics[scale=0.35]{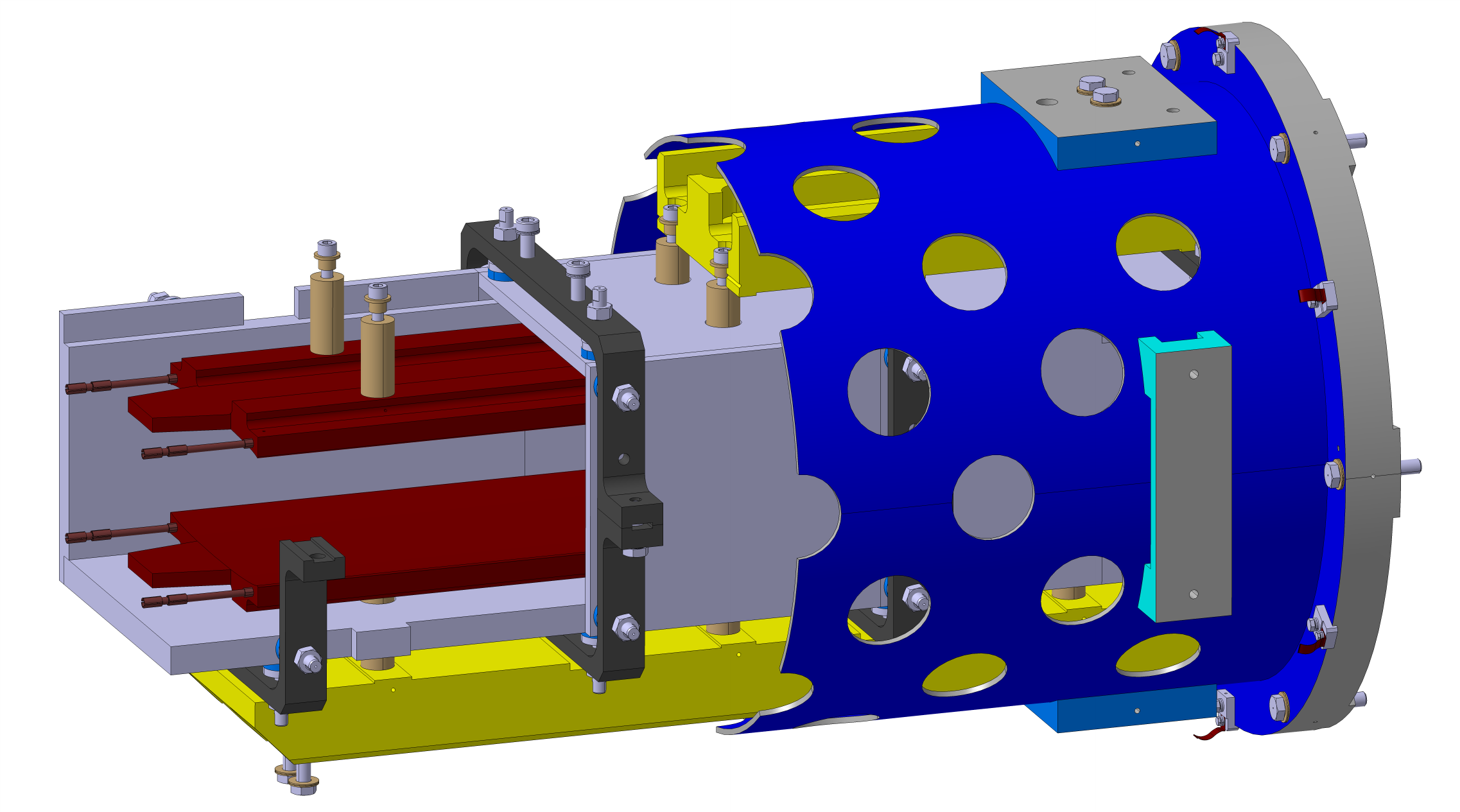}
\caption{\label{fig:Figure9_1}
Outer part (left)  and inner part  of  COSY transmission line RF
Wien filter (right). In COSY, this device allows one to accumulate the spin precession caused by a deuteron EDM (see Chapter\,\ref{Chap:Precursor}). In the PRT ring, it may act identically on all bunches, or to precess individual bunch spins, \emph{without} influencing the other bunches\ \cite{Slim:2016pim}.}
\end{figure}

\subsection{Vacuum}
The requirement for the vacuum is mainly given by the minimum required beam lifetime of about \SI{1000}{s}. The emittance growth in the ring caused by multiple scattering from the residual gas needs to be less than \SI{5e-3}{mm.mrad/s}. With the initial beam emittance assumed to be \SI{1}{mm.mrad}, the beam emittance increases to \SI{5}{mm.mrad} within \SI{1000}{s}. This requires a partial pressures of less than about \SI{e-12}{mbar} for N$_2$ and of about \SI{5e-11}{mbar} for H$_2$. Consequently, the cooling rate for stochastic cooling should be better than \SI{5e-3}{mm.mrad/s}.

For such an ultrahigh vacuum, either cryogenic or NEG pumping systems may be used~\cite{Jagdfeldprivcom}. Bake-out must be foreseen for either cryogenic or NEG systems. The use of NEG systems may introduce some issues.
\begin{itemize}
 \item The NEG material becomes saturated after several pump-downs.
 \item The ageing NEG material may become brittle, leaving dust particles in the vacuum vessel.
 \item The PTR is a prototype ring, and a significant number of pump-downs will be part of the development program.
 \item The high voltage system requires excellent vacuum.
\end{itemize}

A cryogenic vacuum system has also been considered for the PTR ring. The beam pipe would have to composed of a system of three concentric pipes. The inner shell would carry the liquid helium. Next, in the outwards direction, is the 70\,K pipe, while the outer shell would house superinsulation and  heating devices. To avoid these complications and expenses, it might be recommended to use an NEG-based vacuum system. A system based on NEG cartouches, as described in Ref.~\cite{PhysRevSTAB.18.020101}, is currently under discussion.

\section{Beam transfer}

General considerations on beam transfer between an injector and the PTR or the final ring are
given in Appendix\,\ref{app:BeamPrep}. Here, only the resulting PTR injection scheme is described in detail.

\subsection{Beam transfer and injection into the PTR}

The scheme described aims to fill the prototype ring, operated with only electric fields and simultaneously circulating CW and CCW beams. The injection of only CW or CCW beams required for `frozen- spin' operation with superimposed electric and magnetic fields is conceptually simple (standard bunch-to-bucket transfer injection into a synchrotron) and, thus, not described here.

To inject beams in both directions into a machine running with harmonic number $h=6$ without excessive constraints for kicker rise and fall times, only two of the six buckets per direction are filled, with unequal spacing. The first transfer from the injector to the prototype ring, assumed to inject two CW bunches, is conceptually simple and does not lead to particular constraints for the kicker; thus, it is not described here in detail.

The second transfer, assumed to inject CCW circulating bunches, is sketched in Figs.\,\ref{fig:InjectionScheme}, \ref{fig:SpaceTimeDiag}, and \ref{fig:JowettToroidPlot-EDM-protoype}.  \Figure[b]~\ref{fig:InjectionScheme}(a) shows the ring seen from the top with circulating CW  and CCW bunches being injected. The two CW circulating bunches shown in red have already been circulating  for some time, after a previous transfer. The first of two CCW bunches plotted in blue is passing the injection kicker and the second one comes $T_\text{rev}/3$ later. From this plot, at a reference time $t_0$ and the bunch positions, one can deduce the time evolution of the bunches passing the position of the injection kicker, as depicted in Figure\,\ref{fig:InjectionScheme}(b). The injection kicker pulse, which must deflect the two CCW circulating bunches and must not affect the CW bunches, is shown in green.

\begin{figure}
  \centering
  \includegraphics[width=6cm]{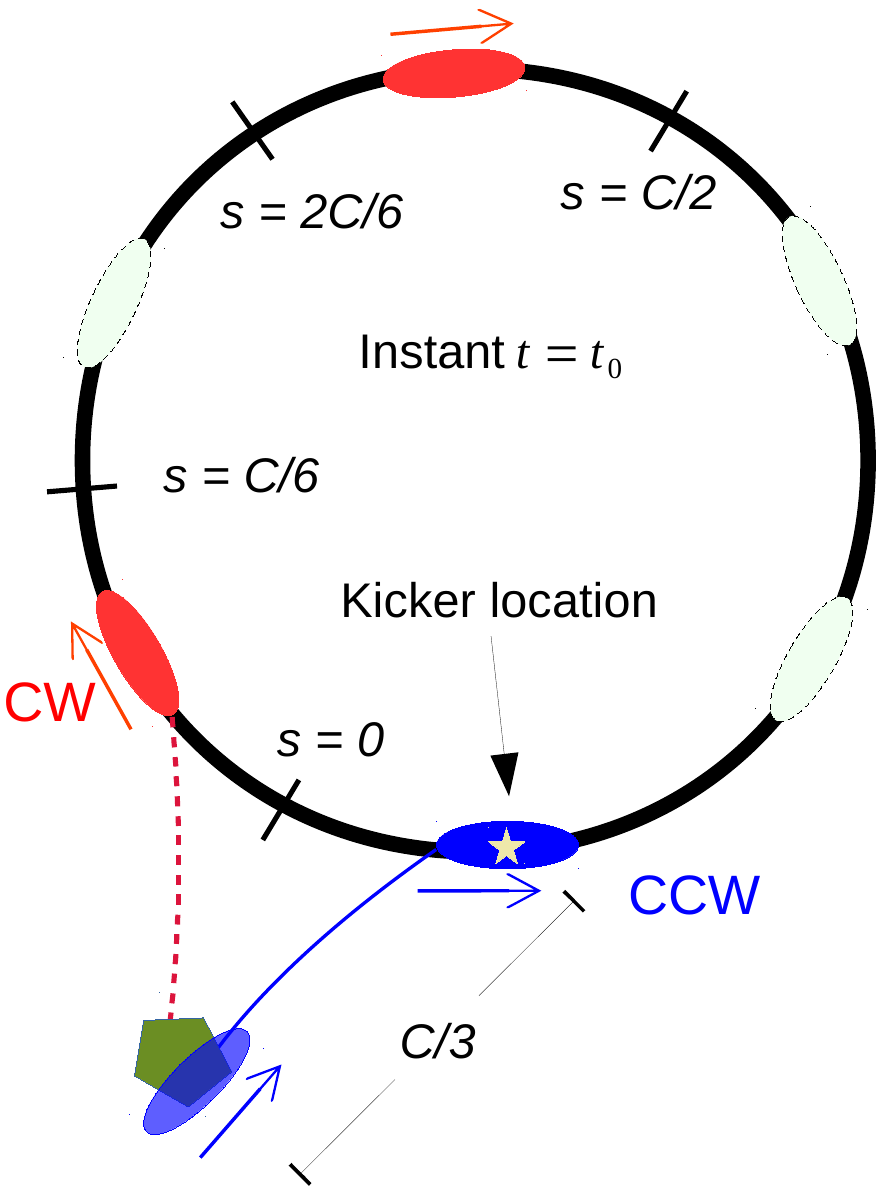}
  \includegraphics[width=9cm]{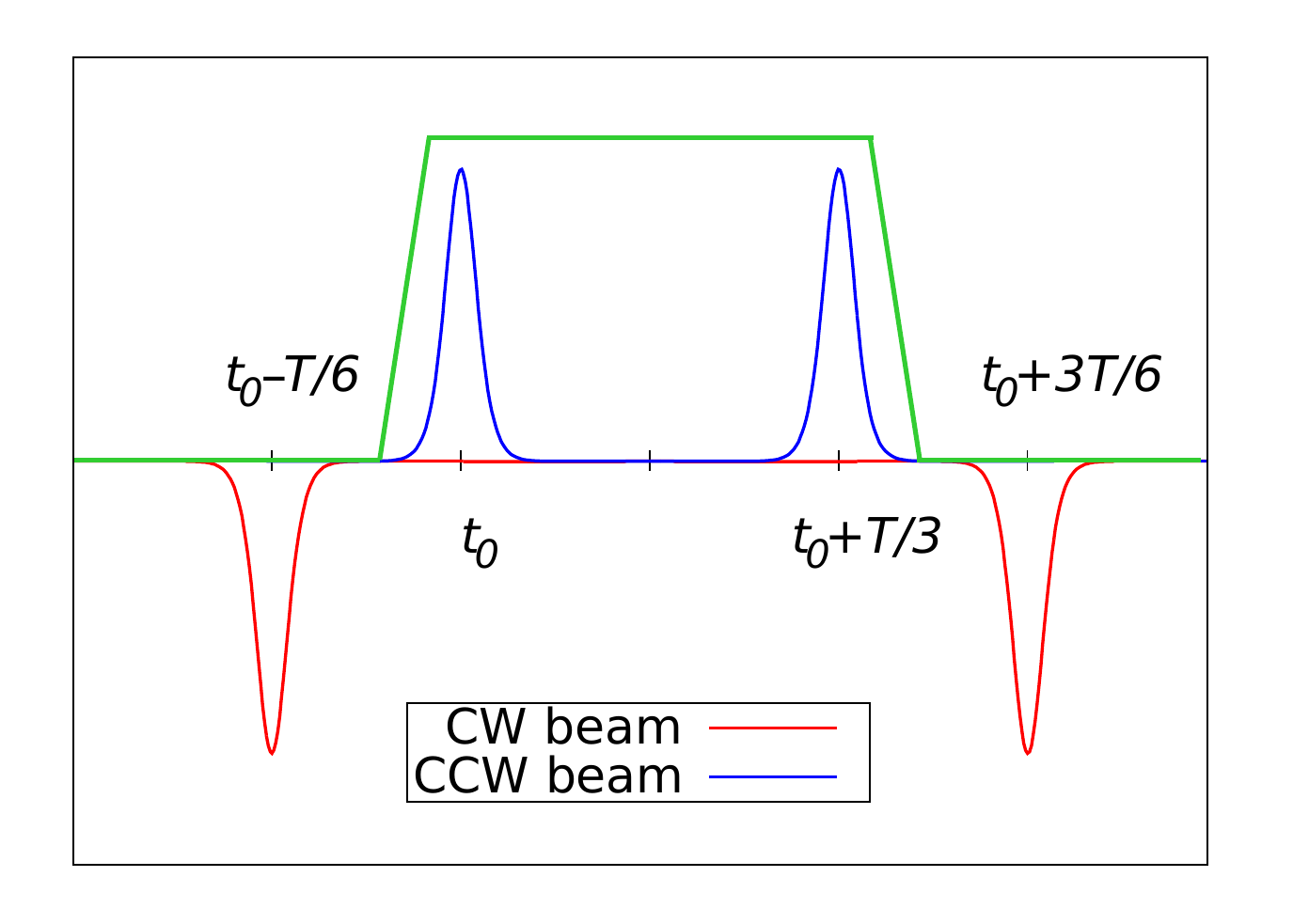} \\
  (a) \hspace{7.3cm} (b) \hspace{1.7cm}
  \caption{(a) PTR at the moment of the second CCW injection; (b) bunches passing at the position of the kicker and kicker pulse as a function of time. Insert (b) shows beam currents versus time in the interval shown as a horizontal 
rectangle outlined at the top of Fig.~\ref{fig:SpaceTimeDiag}.}
  \label{fig:InjectionScheme}
\end{figure}

\begin{figure}
  \centering
  \includegraphics[width=12cm]{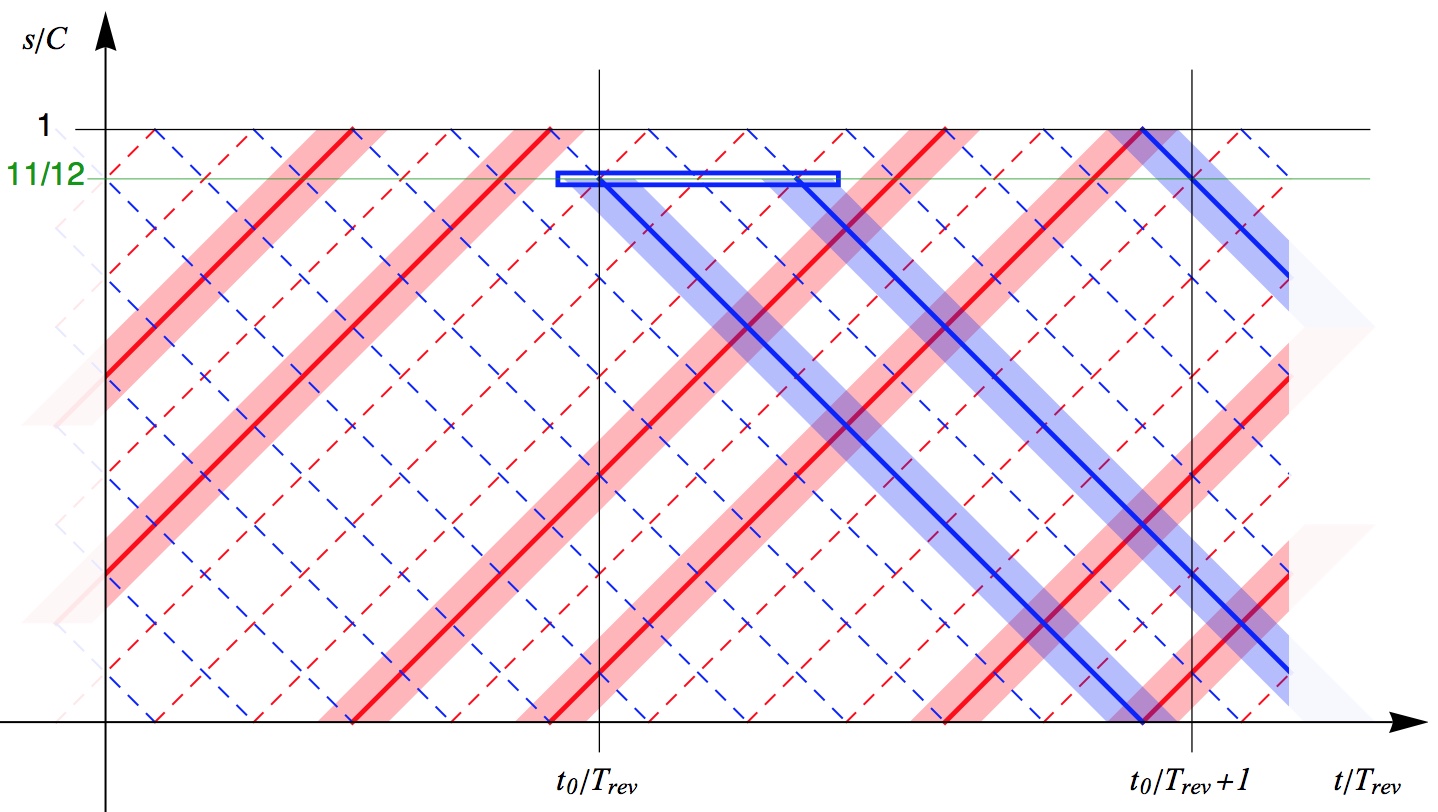}
  \caption{Space--time diagram around the second injection of the CCW rotating bunches. The vertical axis denotes the position $s$ along the circumference, plotted as a function of time. Red and blue lines, and the light red and blue areas, represent CW and CCW rotating bunches, respectively. }
  \label{fig:SpaceTimeDiag}
\end{figure}

\begin{figure}
  \centering
  \includegraphics[width=12cm]{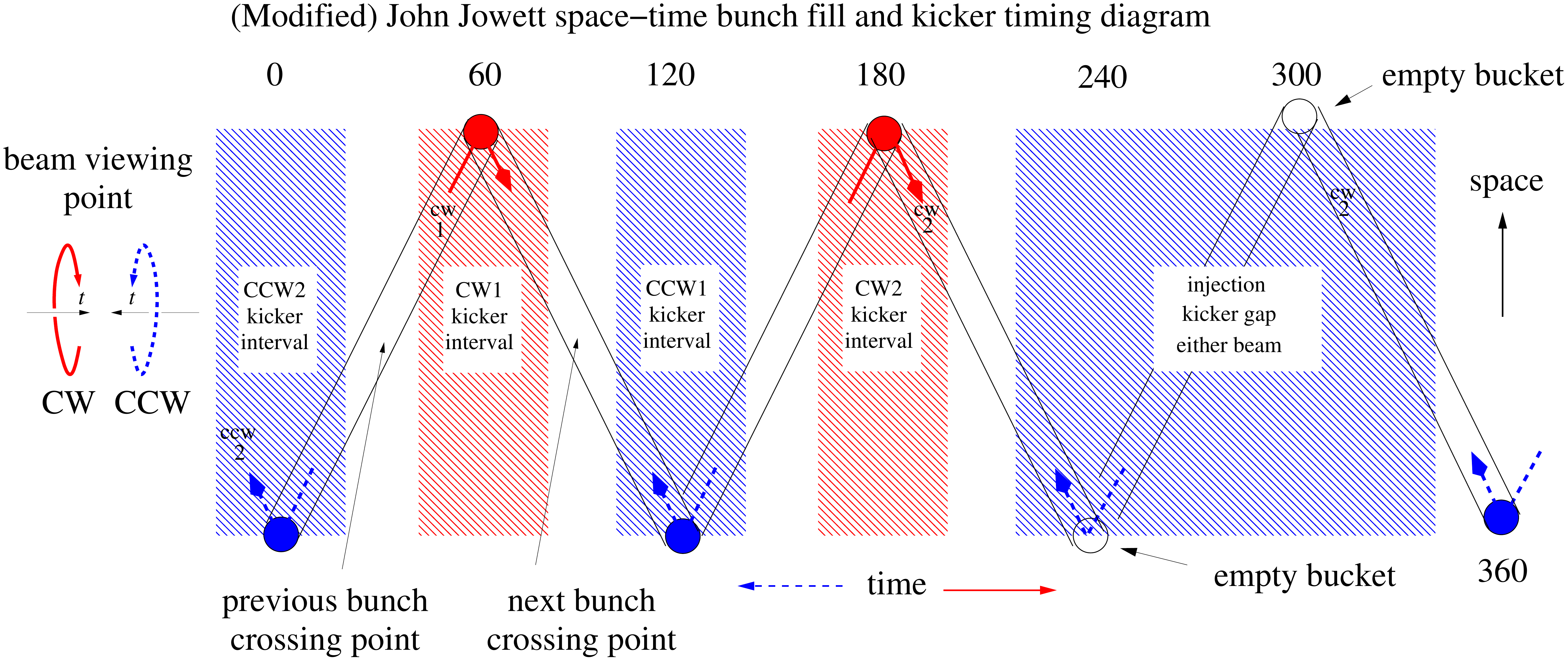}
  \caption{Modified `John Jowett beam bunch space--time plot', illustrating  proposed pattern of countercirculating bunches in the PTR. For this example, there are six stable buckets (for each beam direction) with two CCW blue bunches (separated by an empty bucket), and two CW red bunches (also separated by an empty bucket). There is also a long `gap' on the right, which contains two empty buckets at the instant shown. This gap is required for single-turn injection of a pair of two (either up or down) polarized bunches into one
 beam or the other. This example is directly applicable to the PTR bunch filling scheme explained in the text. As explained in the text, the figure is constructed at exactly the instant when stable RF buckets of the two beams are superimposed. Thus, there are 12 stable RF buckets  in the figure, but every second bucket in each beam is empty and not shown. Also one must count only one of the blue circles in the two lower corners, since they represent the same blue-bunch-filled bucket twice, once entering the figure, and once leaving.}
  \label{fig:JowettToroidPlot-EDM-protoype}
\end{figure}

The time evolution of bunches circulating in the ring is depicted in the `space-time diagram' shown in
\Fref{fig:SpaceTimeDiag} (and in a sightly different manner in \Fref{fig:JowettToroidPlot-EDM-protoype}). The centres of two CW rotating bunches, which have been injected during a previous transfer, are depicted as solid red lines. The distances of the bunches
increase linearly until they reach $s = C$, the circumference, and then appear again at $s=0$. The extensions of the bunches are plotted as red surfaces. The centres of empty buckets are shown as dashed lines. The two CCW bunches plotted in blue are present in the machine from the injection at the kicker position $s = (11/12) C$ and injection times $t_0$ and $t_0 + T_\text{rev}/3$ and their positions decrease linearly. Once the position $s = 0$ is reached, the bunches reappear at $s = C$. The injection kicker pulse is characterized by a duration and length of the device corresponding to the length and width,  respectively,
of the blue rectangle. The vertical line at time $t_0$ corresponds to the snapshot shown in \Fref{fig:InjectionScheme}(a). The horizontal line depicts the time evolution at the injection kicker and corresponds to \Fref{fig:InjectionScheme}(b).  

The proposed injection scheme leads to the following constraints.
\begin{itemize}
  \item From \Fref{fig:InjectionScheme}(b), one deduces that the maximum kicker rise and fall times $T_\text{K}$ and bunch lengths $T_\text{b}$ are given by $T_\text{K} + T_\text{b} < T_\text{rev}/6$. For a $30\UMeV$ beam in a $C = 100 \mbox{ m}$ circumference ring ($T_\text{rev}/6 = \mbox{225 ns}$),  assuming the bunch occupies two-thirds of the buckets, this leads to kicker rise and fall times of $T_\text{K} < 75 \mbox{ ns}$.
  \item The relative phase between CW and CCW bunches (fixed, for example, by the position of buckets assumed
  in \Fref{fig:InjectionScheme}(a)) fixes the position of the RF cavity. The distance between the kicker used for the second transfer (CCW circulating bunches in the description given) and the cavity must be $C/24 + n \cdot C/12$,
  with $n \in \mathbb{Z}$. 
\end{itemize}
 Note that there is no constraint fixing the position of the kicker used for the first transfer between the injector and the prototype ring.

\section{Fundamental physics opportunities for the PTR \label{sect:PTRPhysics} }
To explain the essential differences between stages 1 and 2, it is
useful to expand language that is currently in common use.

Stage 1 discusses spin effects that are understood to imply
in-plane precession, where `in-plane' implies precession in the (horizontal)
plane of the accelerator. In stage 1, with only electric fields, and well below the `magic' energy, the spin cannot be `frozen' (always parallel or antiparallel to the particle trajectory). Thus,  EDM effects do not accumulate monotonically.

Stage 2 concentrates on `out-of-plane' precession, where `out-of-plane' refers to
spin vector precession in the `vertical plane instantaneously tangent to the particle orbit'.
It is precession into this plane, which is driven by a symmetry-violating effect, such as a
proton EDM,  that is the subject of the EDM measurement.
In a paper discussing spin decoherence, Koop\cite{KoopSpinWheel} introduced
the `spin wheel' as a picturesque way of describing precession of the beam polarization vector
in this `out-of-plane' plane. This is very helpful for visualizing the experimental investigations intended for stage~2.
If this `spin wheel' executes a number of revolutions during the observation period, there is a strong
suppression of spin decoherence, with a corresponding increase in the spin coherence
time (SCT).

Regrettably, the magnitudes of the out-of-plane precessions due to the smallest proton EDM one would like to identify are small, of the order of $\upmu \mbox{rad/s}$ for an EDM of \SI{e-26}{\text{$e$}.cm,} 
and the expected size of the PTR.
A run observing a single full turn of the spin wheel for this EDM would take
several days. This means that, from an experimental point of view, the Koop decoherence
argument simply does not apply to any experiment in which the beam polarization is
frozen in all degrees of freedom. However, the Koop wheel picture remains valuable, as the rest of this introduction is intended to explain.

If in-plane precession of the beam polarization vector is
visualized as a propeller blade of a helicopter, and out-of-plane precession as a
blade of a wind turbine propeller, then the remaining possible precession direction
(azimuthal around the beam axis) can be visualized as the propeller of a
propeller-powered aeroplane.

This remaining freedom, precession of the beam polarization around the beam axis,
is driven by a solenoidal (\ie along the orbit) magnetic field acting on the
proton MDM. 
The aeroplane-propeller and helicopter-propeller precessions `do not commute'.
This failure of commutation produces a `wind-turbine propeller-like precession',
which produces a spurious EDM signal; \ie systematic error. Important though this
source of systematic error is, there is currently no plan, other than avoiding
solenoidal fields, to study this effect in stage~2---the importance of this issue must
be addressed theoretically, for example by simulation. This is commonly referred to
as `the geometric phase problem'.

The main thrust of stage~2 is to study `out-of-plane' beam polarization precessions,
which, as just explained, can usefully be visualized as the rolling motion of a `spin wheel'.
After this cartoonish description, a more technically informative discussion can be
based on the matched pair of graphs in \Fref{fig:KoopSlopeGraphs}.
In each graph, the horizontal axis is the magnetic field and the vertical axis describes the Koop wheel response.
Note that the magnetic fields are  several orders of magnitude different  and that an angular frequency is plotted in \Fref{fig:KoopSlopeGraphs}(a), while the angular advance is plotted in \Fref{fig:KoopSlopeGraphs}(b).
The main point of these two graphs is that, in spite their vastly different scales, the
slopes are determined by a single, truly constant, physical constant of nature---the magnetic dipole moment of the proton.

\begin{figure}
\includegraphics[scale=0.40]{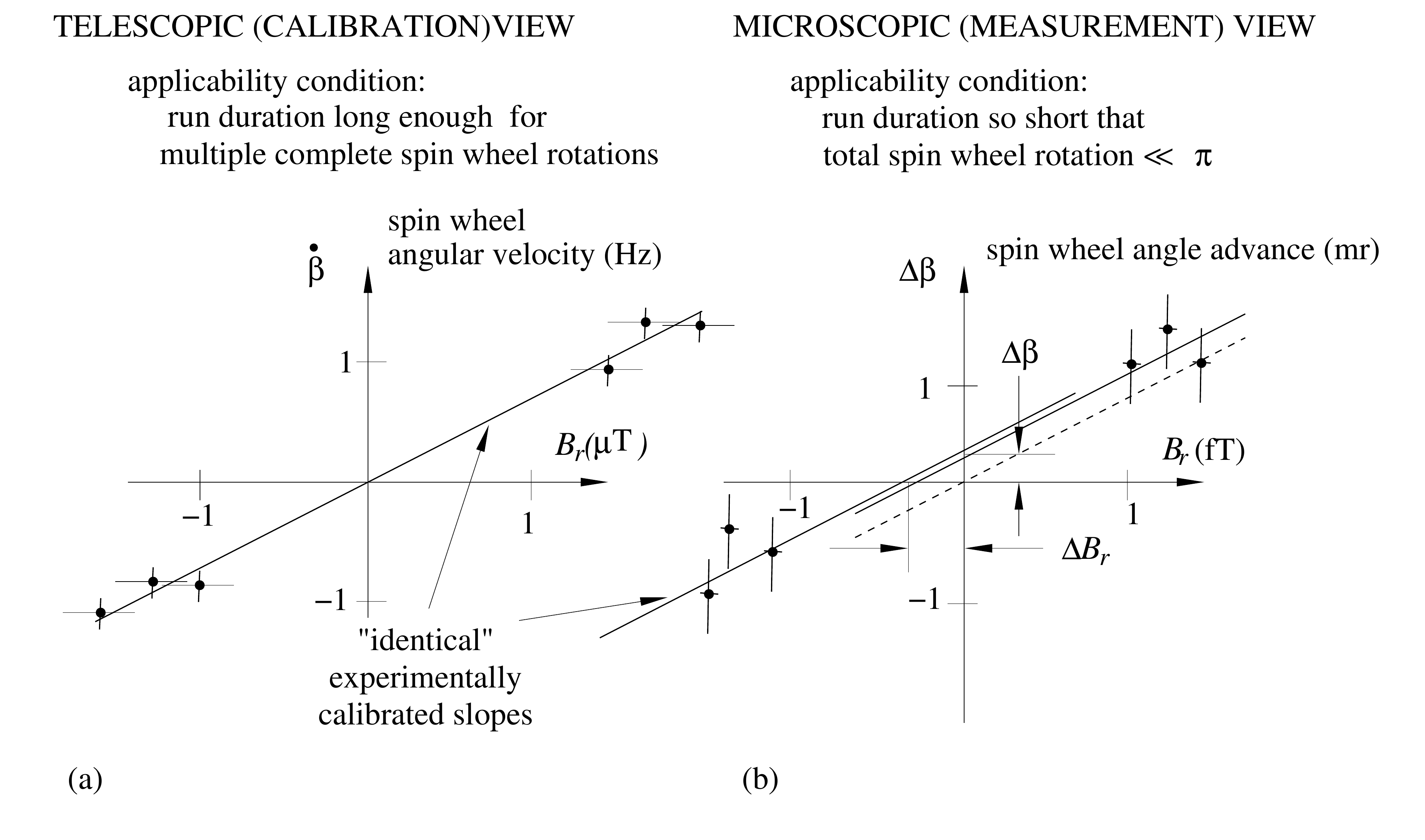}
\caption{\label{fig:KoopSlopeGraphs} Dependence of spin wheel
orientation angle $\beta$ on radial magnetic field $B_r$.  (a) This graph
is especially appropriate for
spin wheel calibration, with large radial torques intentionally applied using
a stripline Wien
filter.  (b) This graph is especially appropriate
for representing the dependence of change $\Delta\beta=\beta_{\rm end}-\beta_{\rm beg}$,
for example, as the result of measuring unknown physically interesting torques during a
long EDM or GR measurement run. For the graphs to be intelligible, the scales must be
unambiguously shown and the error bars need to be shown---here, they are just order-of-magnitude estimates.}
\end{figure}

With the aid of a transmission line Wien filter, such as that shown in \Fref{fig:Figure9_1},
and a frozen spin polarized proton beam, from data implied by  \Fref{fig:KoopSlopeGraphs}(a),
it will be possible to `calibrate' this slope,
as a function of a single, externally imposed current, which is,  itself, experimentally
reproducible to better than parts per million
accuracy\cite{Fernqvist:2003, CERNpreciseCurrent, Fernquist-PowerPoint}\footnote{With the exception of vertical positioning, which needs to be controlled to micrometre accuracy, it is element positioning rigidity, current resettability, and time-independence of all parameters, more than absolute accuracy, that needs to be achieved.}.
The ultimate EDM precision depends on either improving this accuracy, or on scheming to exploit it most effectively.
Appropriately transformed to match
the parameters of the experiment implied in \Fref{fig:KoopSlopeGraphs}(b), this calibration can be applied
to determine the slopes in \Fref{fig:KoopSlopeGraphs}(b), in spite of the nine orders of magnitude difference
in the horizontal scales. The accuracy with which data implied by \Fref{fig:KoopSlopeGraphs}(b) can be used to determine the proton EDM
depends, primarily, on the precision with which the data points are determined, as indicated
by their error bars and point locations---which have been chosen arbitrarily for the figure.
Though just a cartoon, the fact that the two parallel lines in \Fref{fig:KoopSlopeGraphs}(b) do not quite coincide is intended to suggest the presence of errors in the extrapolations.
And there is another significant ambiguity in \Fref{fig:KoopSlopeGraphs}(b); the $\Delta\beta$
ranges may, or may not, include the critical $\beta=0$ point, at which the beams are truly frozen in all degrees of freedom.
Though it is not obvious from the figure, the vast difference in horizontal scales `amplifies' this ambiguity.

One example of the `scheming most effectively' mentioned previously, which can be developed using PTR, would be to exploit the
waveguide Wien filter to isolate just one of the bunches to phase-lock its spin wheel angle $\beta$.
This would, of course, destroy any EDM information contained in this particular bunch. However, the phase-locking would,
to high precision, have no effect on the other bunch polarizations; they would still  respond freely to the EDM torques (including spurious EDM-mimicking torques).

Another possible physics opportunity of the prototype ring is the direct search for an ambient dark matter (DM) field made of axion-like particles, which could generate oscillating EDMs\cite{Chang:2019poy, Abel:2017rtm}. Oscillating EDMs can be searched for with the prototype ring operated such that the spin rotates in the horizontal plane (not in `frozen-spin' mode), with a frequency corresponding to the oscillation frequency of the EDM.


\section{Summary and outlook}
The concept of, and the need for, the prototype storage ring has been outlined in the previous sections to the best of our current understanding.
The next stage is to produce a detailed design report, which should demonstrate the technical feasibility of a prototype storage ring.
The plan of the CPEDM collaboration is to finalize this TDR in 2022 (see Chapter\,\ref{Chap:RoadMap}).

\FloatBarrier




\begin{flushleft}

\end{flushleft}
\end{cbunit}

\begin{cbunit}

\csname @openrighttrue\endcsname 
\chapter{All-electric proton EDM ring}
\label{Chap:allelectricring}

\section{Introduction---BNL design}
It has usually been assumed 
that the most sensitive proton EDM measurement will be made in a dedicated, precision, all-electric storage ring, in which clockwise (CW) and counterclockwise (CCW) beams circulate concurrently at the `magic' kinetic energy of \SI{232.8}{MeV}, for which the proton spins are `frozen', for example, pointing in the forward direction everywhere in the ring. Most recently, a design for this ring has been outlined by  Anastassopoulos \emph{et al.} \cite{doi:10.1063/1.4967465},  evolved from a more detailed earlier proposal \cite{Nominal-BNL}. A similar earlier version of this design has been substantially
analysed by Lebedev\cite{Lebedev}. Parameters for the design described in~\cite{doi:10.1063/1.4967465} are given in \Tref{tbl:HolyGrail-PT-params} (column `full scale'), and one quadrant of the full ring is shown in \Fref{fig:HolyGrailQuadrant}. This report does not attempt to replicate material in that publication in any substantive way. The purpose for any material copied is only for ease of comparison.

\begin{table} [h]
\caption{\label{tbl:HolyGrail-PT-params}Lattice parameter comparison between a lattice upscaled from the prototype PTR lattice, in the last column, and the  same parameters for a full-scale all-electric EDM lattice (in the second-to-last column). Any differences between entries in these two columns lie well within the ranges of values for existing full-scale all-electric proton EDM rings.  }
\centering
\resizebox{\columnwidth}{!}{%
\begin{tabular}{lllll}  \hline \hline
  Parameter                       &  Symbol                  & Unit         & Full scale                 &  PTR scale    \\ \hline
 Bending radius                   &  $r_0$                   & m            & 52.3                       & $47$         \\
Electrode spacing                 &   $g$                    & cm           & 3                          & 3           \\
Electrode height                  &   $d$                    & cm           & 20                         & 20           \\ 
Deflector shape                   &                          &              & cylindrical                & $\approx$\,cylindrical \\
Electrode index                   &   $m$                    &              & 0                          & 0.001         \\
Radial electric field             &   $E_0$                  & MV/m         & 8.0                        & 8.92         \\    
Number long straights             &                          &              & 4                          & 16          \\
Long straight section length      &    $l_\mathrm{ss}$              & m            & 20.8                       & 12.0         \\
Polarimeter sections              &                          &              & 2                          & 2           \\
Injection sections                &                          &              & 2                          & 2           \\
Total circumference               & $\mathcal{C}$            &              & 500.0                      & 500.0         \\ 
Harmonic number                   & $h$                      &              & 100                        & 100           \\
RF frequency                      &                          &              & 35.878                     & 35.878         \\
Number of bunches                 &                          &              & 100                        & 25            \\
Particles per bunch               &                          &              & $\num{2.5e8}$              & $\num{5e8}$            \\
mom.  spread (not/cooled)         &                          &              & $\pm \num{5e-4}/\num{e-4}$ & $\pm \num{5e-4}/\num{e-4}$  \\ 
Max. horizontal\  beta function         & $\beta_{x,\rm max}$      &  m           & 47                         & 48           \\
Max. vertical\ beta function         & $\beta_{y,\rm max}$      &  m           & 216                        & 183           \\
Dispersion                        & $D$                      &  m           & 29.5                       & 46.1          \\
Horizontal\ tune                        & $Q_x$                    &              & 2.42                       & 1.75           \\
Vertical\ tune                       & $Q_y$                    &              & 0.44                       & 0.47           \\
r.m.s. horizontal\ emittance\ (not cooled/cooled)      & $\epsilon_x$             & $\si{\pi.mm.mrad}$ & 3.2/3                      & 3/3           \\
r.m.s. vertical\ emittance\ (not cooled/cooled)     & $\epsilon_y$             & $\si{\pi.mm.mrad}$ & 17/3                       & 17/3          \\
Slip factor                       & $\eta=\alpha-1/\gamma^2$ &              & $\num{-0.192}$             &                 \\
 \hline \hline
\end{tabular}
}
\end{table}

\begin{figure} [hbt!]
\centering
\includegraphics[width=0.6\textwidth]{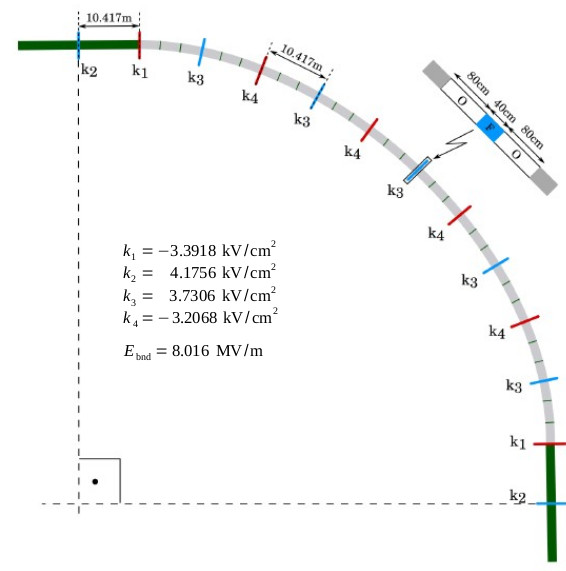}
\caption{\label{fig:HolyGrailQuadrant}One quadrant of a full-scale, all-electric, frozen-spin EDM storage ring. The total circumference is \SI{500}{m}. The deflector radius is \SI{52.3}{m} and the plate spacing is \SI{3}{cm}. The electric field is directed inwards between the plates. The spin and momentum vectors are kept aligned for the duration of storage. A realistic lattice will include 40 bending sections separated by 36 straight sections, each
\SI{2.7}{m} long, with electrostatic quadrupoles in an alternating gradient configuration, and four \SI{20.8}{m} long straight sections for polarimetry and beam injection. It will also include SQUID-based magnetometers, distributed around the ring (see \Sref{sec:tests-with-SQUID}).}
\end{figure}

Planning for the prototype ring (PTR) began by downscaling from the design
of Anastassopoulos~\emph{et al.} by approximate factors of five in both lengths and kinetic energy. The downscaling prescription is described in detail in Chapter\,\ref{chap:ptr}. A result of the downscaling is that, though the full-scale ring quadrant shown in \Fref{fig:HolyGrailQuadrant} looks `round', the PTR ring shown in \Fref{fig:pEDM-square-proto}
looks `square'. This is an artefact resulting from the scaling of lattice functions rather than the scaling of appearances.

After minor changes to match element lengths at the reduced beam energy, the adopted PTR dimensions were upscaled back to the full-scale ring size. Recalculated lattice parameters for the upscaled ring are listed in the last column of \Tref{tbl:HolyGrail-PT-params}. Agreement is quite good for all parameters.
For transverse optical properties, this agreement follows more or less automatically from the scaling. For longitudinal dynamics, the scaling is less transparent, since cavity frequencies and harmonic numbers do not scale automatically. However, the well-established synchrotron oscillation formalism is expected to apply quite directly to both the PTR and the full-scale ring.

The only significant defect of the downscaling has to do with sensitivity to intrabeam scattering.
For the full-scale ring, the correspondingly smaller tune advance per superperiod causes the focusing to be weaker. This is what permits the long straight sections of the full-scale ring to be more than doubled, compared with the prototype (from \SI{6}{m} to \SI{14.8}{m}).
This has the beneficial (perhaps even obligatory) effect, for the full-scale ring, of operating `below transition'. This ameliorates intrabeam scattering, as can be explained in connection with stochastic cooling.

\section{Preparedness for the full-scale ring}
\Tref{tbl:Preparedness}, which is an extension of \Tref{tbl:Preparedness-Chap5}, gives a long (but surely still incomplete) list of requirements that must be satisfied before serious construction of a full-scale EDM ring can begin. Each of these topics has been discussed in preparing this report, at least to the level of formulating criteria for assigning the `preparedness rankings' shown in the table. Though highly abbreviated in this table, as indicated, most of the issues are expanded on elsewhere in the report. The assignment of colour-coded scores is explained in the table caption. These scores are loosely correlated with the PTR prototype ring staging, described in Chapter\,\ref{chap:ptr}.

\begin{table}
   \caption{\label{tbl:Preparedness}Status of preparedness levels for the full-scale all-electric ring: green, `ready to break ground'; yellow, `promising'; red, `critical challenge'.  Plus ($+$) and minus ($-$) signs are to be interpreted as for college course grades. Thus, the ranking, with most prepared first, is \colorbox{green}{$+$G$-$} \colorbox{yellow}{$+$Y$-$} \colorbox{red}{$+$R}. Success in meeting prototype ring goals could amount,  for example, to upgrading all scores to Y($+$) or better. }
\centering
\resizebox{\columnwidth}{!}{%
 \begin{tabular}{llll}   \hline \hline
 Operations                       &  Rank                      & Comment                             & Reference             \\ \hline
 Spin control feedback           & \cellcolor{green}G         & COSY R\&D                           & \Sref{appa:spincontrol} \\
 Spin coherence time              & \cellcolor{green}G($-$)    & COSY R\&D                           & \Sref{appa:longsct} \\
 Polarimetry                      & \cellcolor{yellow}Y        & Polarimetry is destructive          & Chapter\,\ref{Chap:polarimetry} \\
 Beam current limit               &  \cellcolor{red}R        & Enough protons for EDM              & Section\,\ref{ptr:30MeVgoals} \\
 CW or CCW operation                 &   \cellcolor{red}R         & Systematic EDM error reduction      & Ref.\cite{doi:10.1063/1.4967465} \\
 \hline
 Theory                           &                            &                                     &                       \\ \hline
 GR gravity effect                &   \cellcolor{green}G($+$)  & This report, standard candle bonus   & Appendix\,\ref{app:gravity} \\
 Intrabeam scattering             &    \cellcolor{yellow}Y     & May limit run duration              & Ref.\cite{Lebedev} \\
 Geometric or Berry phase theory     &   \cellcolor{yellow}Y     & Needs further study                 & Ref.\cite{Tahar:2019hzv}   \\ \hline 
 Components                       &                            &                                     &                       \\ \hline
 Quads                            &    \cellcolor{green}G      & \eg CSR design                     & Chapter\,\ref{Chap:Efields}  \\ 
 Polarimeter                      &    \cellcolor{green}G      & COSY R\&D                           & Chapter\,\ref{Chap:polarimetry}       \\
 Waveguide Wien filter            &    \cellcolor{green}G      & COSY R\&D  precursor                & \Sref{appa:rfwf}   \\
 Electric bends                   &    \cellcolor{red}R($+$)   & Sparking--cost compromise            & \Sref{appa:deflectors}     \\ \hline 
 Physics and engineering           &                            &                                     &                       \\ \hline
 Cryogenic vacuum                 &    \cellcolor{yellow}Y     & Required?---cost issue only         & Ref.\cite{vonHahn:2016wjl}   \\
 Stochastic cooling               &    \cellcolor{yellow}Y     & Ultraweak focusing issue            & Ref.\cite{Mohl:1993jn}  \\
 Power supply stability           &   \cellcolor{yellow}Y($-$) & May prevent phase lock              & Chapter\,\ref{chap:ptr}  \\
 Regenerative breakdown           &  \cellcolor{red}R($+$)     & Specific to mainly electric;        &  \\
                                  &                            & not seen in $E$ separators            &  \\
 \hline
 EDM systematics                  &                            &                                     &                       \\ \hline
 Polarimetry                      &   \cellcolor{green}G($-$)  & COSY R\&D                           & Chapter\,\ref{Chap:polarimetry}  \\
 CW or CCW beam shape matching       &    \cellcolor{yellow}Y     &                                     & Chapter \ref{Chap:SensSys}  \\
 Beam sample extraction           &    \cellcolor{yellow}Y     & Systematic error?                   & Chapter\,\ref{Chap:polarimetry}, Appendix\,\ref{app:extpol}  \\
 Control current resettability    &    \cellcolor{yellow}Y     &                                     & Ref.\cite{Fernqvist:2003} \\
 BPM precision                    &   \cellcolor{yellow}Y($-$) & Rogowski? SQUIDs?                   & Chapters\,\ref{chap:ptr}, \ref{Chap:SensSys} \\
 Element positioning \& rigidity  &   \cellcolor{yellow}Y($-$) & Must match light source stability   & Ref.\cite{Decker:2005mu}  \\ \hline
 Theoretical analysis             &                            &                                     & Chapter\,\ref{Chap:SensSys} and refs. therein            \\
 Radial $B$ field $B_r$             &                            & Assumed to be dominant              & Ref.\cite{doi:10.1063/1.4967465}       \\
 \hline \hline
 \end{tabular}
 }
 \end{table}

 An inexpensive prototype EDM ring would be needed, in order to investigate, experimentally, issues essential for an eventual  full-scale EDM ring.
  It was 
  decided that the ring designs for the prototype ring and the final ring should be as closely identical as possible.  With the frozen-spin proton kinetic energy being \SI{232.8}{MeV}, prototype proton kinetic energies of 35 and \SI{45}{MeV} were considered.
This scaling is described in full detail in Chapter\,\ref{chap:ptr}.

A detailed list of requirements and goals for the prototype ring is given in Chapter\,\ref{chap:ptr}.
 To check the downscaling by a factor of roughly five in circumference, the entries in the final column of Table\,\ref{tbl:HolyGrail-PT-params} were calculated, for comparison with the values in Ref.~\cite{doi:10.1063/1.4967465}, given in the second-to-last column. Since the focusing
is very weak in both cases, there is little reason to question the reliability of the scaling, as regards transverse optics. Of course, because of the different beam energies, there are substantial differences in the longitudinal dynamics.  However, since
this formalism is very well established in both cases, there is little reason to doubt this aspect of the scaling.

The most obvious need for building a prototype ring is the lack of significant experience with relativistic all-electric accelerators,  especially storage rings. Of course, this is due to the much more powerful bending that is possible with magnetic, rather than electric, fields, and the resulting absence of all-electric examples. To make up for this deficiency, the electric fields have to be increased to a level that is limited by electrical breakdown. Experience in this area is largely based on high-energy particle electrical separators, such as those, \eg used at the Tevatron\cite{prokofiev2005}.  Just one, of many, but perhaps the most important, technical uncertainty has to do with the highest-field, smallest-bending radius, and hence the least expensive, all-electric ring that can be conservatively constructed and guaranteed to store \SI{232.8}{MeV} protons.

The PTR staging can, to some extent, be correlated with the colour-coded entries in \Tref{tbl:Preparedness}.  Red entries in that table represent critical challenges that would necessarily delay commencement of the full-scale ring. They are also referred to as `quantitative goals' for PTR stage\,1. The main goal of stage\,1 is to remove the `R' flags from the table. This includes the `R($+$)' associated with the electric bend--sparking  compromise. This score was increased from `R' to `R($+$)' only to acknowledge that, by increasing the ring radius sufficiently, sparking can be sufficiently suppressed.  However, this could lead to an unacceptable cost increase.

In particular, since the EDM precision is roughly proportional to the proton beam current,  experimental determination of achievable beam intensity is necessary for any future full-scale EDM ring design. Operational experience with electric rings has
been very limited.  There has been a significantly large accumulation of polarized  beam experience, but all in magnetic rings, none in electric rings. In any case, another important goal for stage\,1 is to remove the `R' flag associated with the beam current limit.

Most of the entries in the table with yellow `Y' flags are to be studied in PTR stage\,2. In Chapter\,\ref{chap:ptr}, these are characterized as `qualitative goals', at least partly to acknowledge their indefinite nature. For example, there is a weakness in the PTR downscaling that will limit the extent to which prototype results can be reliably extrapolated to the full-scale ring. Along with well-understood residual vacuum growth, intrabeam scattering is expected to be a significant source of beam emittance growth, which, as well as increasing spin decoherence, can cause beam loss and limit run length. Full 3D equilibration of this growth source is only possible for `below-transition' ring operation. This condition is met in the full-scale design described in Ref.~\cite{doi:10.1063/1.4967465}, but not in the PTR design, as it would have required an approximate doubling of the ring circumference. Investigating this issue will be a goal of the prototype ring.

The bottom (two-line) entry in Table\,\ref{tbl:Preparedness} requires special explanation. This entry is not assigned a colour; it relates to the inevitable residual radial magnetic field average $\langle B_r\rangle$ after all efforts have been exhausted to trim it to its ideally zero value.  This average,  $\langle B_r\rangle$, is expected to dominate  the proton EDM systematic error.  However, because it depends on the uncertain values of all the other entries in the table, the uncertainty in $\langle B_r\rangle$  cannot be compared directly with the other entries in the table---it depends on the accumulated effect of all the other values, and on their theoretical systematic error calculation.

 Note, however, that
one of the main purposes of the BPM system in the PTR and, later in the full-scale EDM ring, is
the measurement of the relative displacement of CW and CCW orbiting beams, which is proportional to
the residual radial magnetic field $B_r$ (see \Sref{section:dual} and Appendix\,\ref{Chap:MagneticFields} for further details).

\section{New ideas}
Of course, one also expects investigations with a prototype ring to give rise to new ideas. In fact, the planning phase itself can motivate the development of new ideas. This study, now well begun, has been no exception. By and large, though, to reduce the proliferation of speculative descriptions, the body of this report concentrates mainly on fleshing out ring design and experimental methods, as established in the first few months of the study. It has also become clear that polarized protons or deuterons stored in a suitable EDM ring may be directly sensitive to an ambient dark matter (DM) field made of axion-like particles, which could produce oscillating EDMs\,\cite{Chang:2017ruk,Abel:2017rtm,Pretz:2019ham}.

Some of the main new ideas that emerged during the preparation of this feasibility study are described in appendices to this report. The titles of these appendices are prefixed with `New ideas' to distinguish them from the preceding, more conventional, appendices. In addition, these appendices are introduced by brief abstracts:
\begin{itemize}
 \item Appendix\,\ref{Chap:hybrid}. New ideas: hybrid scheme;
 \item Appendix\,\ref{Chap:spin-tune-mapping}. New ideas: spin tune mapping for EDM searches;
 \item Appendix\,\ref{Chap:FDM:Full}. New ideas: deuteron EDM frequency domain determination;
 \item Appendix\,\ref{app:fourier}. New ideas: distinguishing EDM effects from magnet misalignment by Fourier analysis;
 \item Appendix\,\ref{app:extpol}. New ideas: sampling polarimeter based on pellet-extracted beam.
\end{itemize}

\begin{flushleft}

\end{flushleft}
\end{cbunit}

\begin{cbunit}

\csname @openrighttrue\endcsname 
\chapter{Electric fields}
\label{Chap:Efields}


\section{Assumptions and boundary conditions}

One proposal for the nominal all-electric EDM ring is a fully electric strong focusing lattice, to obtain a \SI{500}{m} circumference storage ring\cite{doi:10.1063/1.4967465}. It consists of four long straight sections (LSSs), to be used for the injection of each beam (clockwise and counterclockwise) and two polarimeters. The long straight sections are linked with ten cells, each containing three bending sections (bends, see \Fref{MainDipole}), two short straight sections (SSSs) housing magnetometers and separate quadrupoles (quads). This lattice produces the most challenging requirements for the quadrupoles, compared with a `soft
focusing' lattice, where at least some of the focusing is included in the bending elements.

\begin{figure}
    \centering
    \includegraphics[width=\linewidth]{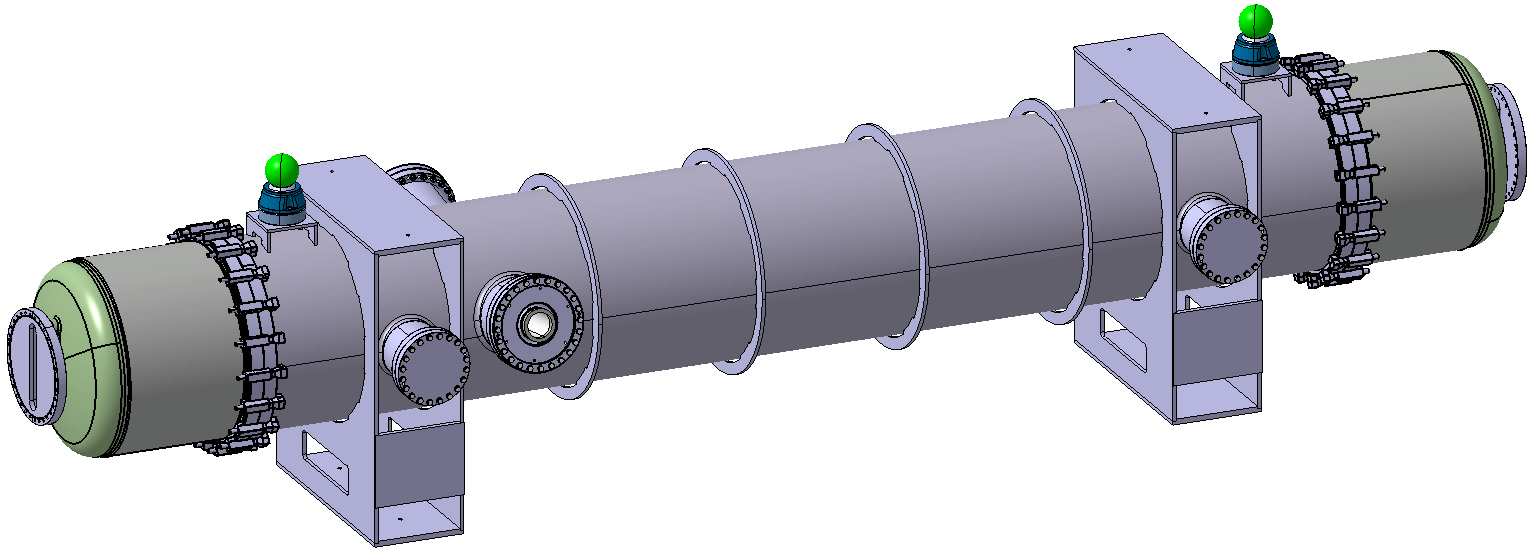}
    \caption{Mechanical dipole concept}
    \label{MainDipole}
\end{figure}

An alternative soft focusing lattice may use bends with soft focusing in the vertical plane\cite{Nominal-BNL}. This will lead to bends with a field index between $m=0.1$ and $m=0.2$, which further increases the challenge for the bend design and manufacturing tolerance, while the quadrupole requirements would be less demanding with respect to the quads of the strong focusing lattice.

The fully electric machine imposes stringent requirements for the background magnetic field in the nanotesla range
\cite{doi:10.1063/1.4967465}. This requirement has an immediate impact on the construction of the ring elements and the materials that can be used. Austenitic stainless steels show a paramagnetic behaviour at room temperature; the relative magnetic permeability is typically of the order of $1.001$-$1.005$ for the fully annealed, fully austenitic grades. One must avoid the use of work hardened or welded components, where magnetic susceptibility could be higher, as a function of the grade used. One could consider fully austenitic grades, such as 316LN, to avoid non-linear behaviour owing to traces of ferromagnetic phases. Alternatively, titanium alloys could be used, at probably higher cost, but the consequences of their relatively poor heat conduction ($17$ compared with \SI{45}{\watt \meter^{-1}\kelvin^{-1}}) are still to be studied  in further detail. Depending on what approach will be retained to achieve the required vacuum level, poor heat conduction may increase the time needed for bake-out or require operation at cryogenic temperatures.

The required vacuum level is in the \SI{e-11}{mbar} range. This implies that the equipment should be compatible with either bake-out at \SI{200}{\degreeCelsius} or \SI{300}{\degreeCelsius}, if the ring is to be operated at room temperature. Alternatively, it should be compatible with cryogenics cool-down, if the ring is to be operated at cryogenic temperatures to avoid too many cold--warm transitions~\cite{vonHahn:2016wjl}. Running the electric devices at low temperatures may lead to a reduced voltage breakdown rate, but this requires a (not-yet-existing design for a) cryogenic $>$$\SI{200}{kV}$ feedthrough (needed to have a margin for conditioning) and possibly a bus bar at cryogenic temperatures. Both are substantial challenges to design and operate reliably.

The aim is to keep the machine's cross-section below $\SI{1}{\metre} \times \SI{1}{\metre}$, including the magnetic shielding that is needed to shield the background magnetic field (Earth's magnetic field and stray fields), and, as such, this has a direct impact on the design of the beam elements. All elements studied have a smaller cross-section but, depending on the space needed for the magnetic shielding or the cryostats, this requirement may have to be revised.

\section{Electrode material}

The electrodes of the bending dipoles, as well as the quadrupoles, are large objects, and need to provide significant fields to achieve the required deflection. The high fields assumed for the ring need to be produced reliably, with large electrode surfaces as well as \SI{30}{mm} gaps. (It should be noted that the designs presented here are for the EDM measurement ring described in Chapter \ref{Chap:allelectricring}, with a nominal horizontal aperture of \SI{30}{mm}; not for the prototype ring described in Chapter \ref{chap:ptr}, with a larger horizontal aperture. The electrodes are already very tall ($200\Umm$
 for the $30\Umm$ wide gap); achieving similar field homogeneities for a $60\Umm$ gap, while respecting the overall cross-section of $\SI{1}{m} \times \SI{1}{m}$ of the system, will become even more challenging.) The high-voltage (HV) breakdown rate is expected to be of the order of once per day for the entire machine, which is very hard to achieve with conventional electrode materials. The choice of electrode material is also strongly influenced by the vacuum requirements and the constraints that these impose on the materials. For example, coated aluminium is commonly used for large septa electrodes at CERN that operate up to  \SI{15}{MV/m}, but is incompatible with bake-out or cryogenics cool-down, owing to crack formation in the oxide coating of the electrode.

Stainless steel and titanium are compatible with the required vacuum conditions.  Older results  demonstrated that titanium has a better voltage holding capacity (VHC) than stainless steel\cite{Manos:2001na, Dunham:2007zz}.  Operational experience with larger electrodes (of about \SI{1}{m} length)  and similar gaps (\SI{30}{mm}) appears, however, limited to around \SI{8}{MV/m} \cite{Borburgh:2004uf, Kramer:2011zd}. Alternative electrode materials may be needed to achieve improved performance for similarly sized electrodes using similarly sized electrode gaps. In this respect, the work done on niobium electrodes\cite{BastaniNejad:2012zz} and TiN-coated aluminium electrodes\cite{Mamun:2015gvm}  using small electrodes, as well as TiN coated stainless-steel electrodes\cite{Mamun2, doi:10.1063/1.5086862} using small electrodes and small gaps, is very encouraging. At CERN, a campaign of breakdown conditioning and breakdown rates for various metals and alloys demonstrates\cite{Descoeudres:2009zz, Descoeudres:2009zza,Descoeudres:1355401} that there is a difference of more than an order of magnitude between the performance achieved in these small-scale laboratory tests and the reported performance of large DC devices. This is a reason to expect that an increase of operational fields in the electric field devices to be used in the EDM ring may be possible, compared with what is used for large DC electric field devices in accelerators so far.

One should not lose sight of the scaling laws for the voltage effect and, more importantly in our application, the area effect\cite{Lorenzi}, where the VHC scales with the surface area as
\begin{equation}
\textrm{VHC} = \sqrt{E\cdot U} = \frac{U}{\sqrt{d}}~\propto~A^{-\frac{1}{\mu}}\,,
\end{equation}
where $E$ is the electric field, $d$ the distance between parallel plates, $U$ the applied voltage, and $A$ the surface of the electrodes, and  $\mu$ can be determined empirically. For objects with an area between \SI{1}{m^2} and \SI{10}{m^2}, typical values of $\mu$ of 5.5 to 8 are reported \cite{Lorenzi}.

\section{Ring elements}
\subsection{Main dipoles}
\subsubsection{Strong focusing lattice main bending dipole}

In the strong focusing lattice, the focusing is entirely left to the quadrupoles, and the dipoles do not focus in the vertical plane. The main dipoles of the strong focusing lattice  use cylindrical electrodes, and an integrated field quality of \SI{1}{ppm} in a central good field region (GFR) of $\diameter = \SI{20}{mm}$ is requested\cite{Borburgh2}. \Tref{Table:Bend} shows the principal requirements assumed for the dipoles.

\begin{table} [h]
\centering
\caption{Main dipole parameters, as assumed for the final strong focusing lattice\cite{doi:10.1063/1.4967465} and  proposed for the strong focusing concept dipole.}
 \label{Table:Bend}
 \begin{tabular}{l ll}
 \hline \hline
  & Strong focusing & Alternative proposed  \\
   & lattice assumption & concept design \\
 \hline
 Physical length (m) &  2.739 & 4.16\\
 Equivalent length (m) &        2.739 & 3.80\\
 Required deflection (mrad) &   52.36 & 78.54\\
 Gap width (mm) & 30 & 30\\
 Electrode height (mm) &        200 &   280\\
 Beam aperture, $a_{x}$ $\times$ $a_{y}$ (mm$^2$) &    30 $\times$ 200 & 30 $\times$ 200\\
 Field homogeneity in GFR of $\diameter= \SI{20}{mm}$ (ppm) &   1 &     700\\
 Main field (MV/m) &     8.00  &        8.67\\
 Voltage per polarity (kV) &    $\pm$120 &      $\pm$130\\
 Electrode radius in H (m) &    52.3 &  48.4\\
 \hline \hline
 \end{tabular}
 \end{table}

 The lattice assumption for the dipoles assumes the equivalent length to be equal to the physical length. Since the 3D design of the dipole concept needs to include space between the electrodes and the beam pipe flanges, the equivalent length of the concept dipole is shorter than its physical length. Therefore, such a lattice assumption of a gap field yields, in reality, a larger gap field when translated into a realistic 3D design.

 Taking the requirements from \Tref{Table:Bend} as a starting point, an alternative dipole design was developed, based on two instead of three bending dipoles per cell (see \Tref{Table:Bend}), to limit the average field increase with respect to the lattice assumption. This design minimizes the required electric field in the dipoles, as well as the impact of the end fields and, as such, improves the integrated field homogeneity. To calculate the integrated field homogeneity, the electric field on straight lines inside the gap is integrated along the longitudinal $z$ axis (parallel to the electrodes) for each position inside the gap. Subsequently, the value of the integrated field in the centre of the gap is taken as a reference, and the deviation of the integrated field in the remainder of the gap is calculated. This can be done by either integrating the absolute field along the lines, or by integrating the vertical or horizontal component of the field, where the latter is more representative of the fields exposed to the beam.

To reduce the impact of end fields (see \Fref{Fig:dipole_fields}, right panel), a first quantification with \SI{20}{cm} tall electrodes and a device with an equivalent length of \SI{3.8}{m} (corresponding to two bends per cell) was simulated \cite{Borburgh2}. These first simulations show an integrated field homogeneity of the central GFR of \num{7e-4}, while, for the full aperture of $\SI{30}{mm} \times \SI{200}{mm}$, the integrated field homogeneity is \num{4e-3}~\cite{Atanasov}. The main electric field is  \SI{8.67}{MV/m}, but the peak fields are around \SI{10}{MV/m}. In \Fref{Fig:3Ddipole_fields}, the fields at the horizontal and vertical midplane of the dipole are shown. Electrodes using Rogowski profile edges\cite{Rogowski1923} should be explored to reduce the peak fields further. These simulation results seem to indicate that, to reach the required homogeneity, the electrodes should be increased further in (vertical) size (see the current design in \Fref{Fig:dipole_fields}, left panel), making the cross-section requirement for the machine (to keep the cross-section below $\SI{1}{\metre} \times \SI{1}{\metre} $) more difficult to fulfil.

\begin{figure}
    \centering
   \includegraphics[scale=0.5]{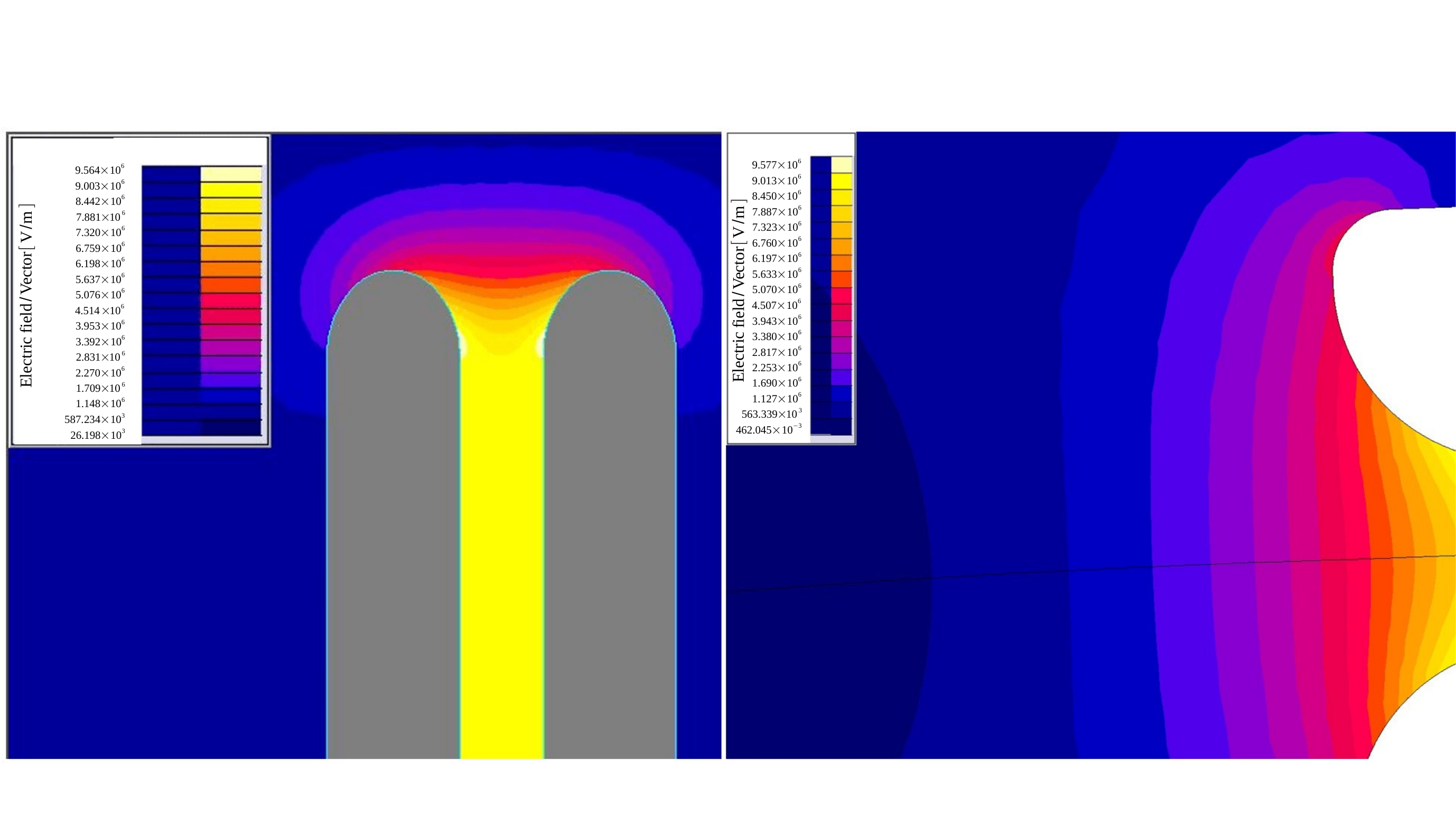}
    \caption{2D field plots of (left) the optimized electrode cross-section
and (right) the top view of the end field}
    \label{Fig:dipole_fields}
\end{figure}

\begin{figure}
    \centering
   \includegraphics[scale=0.5]{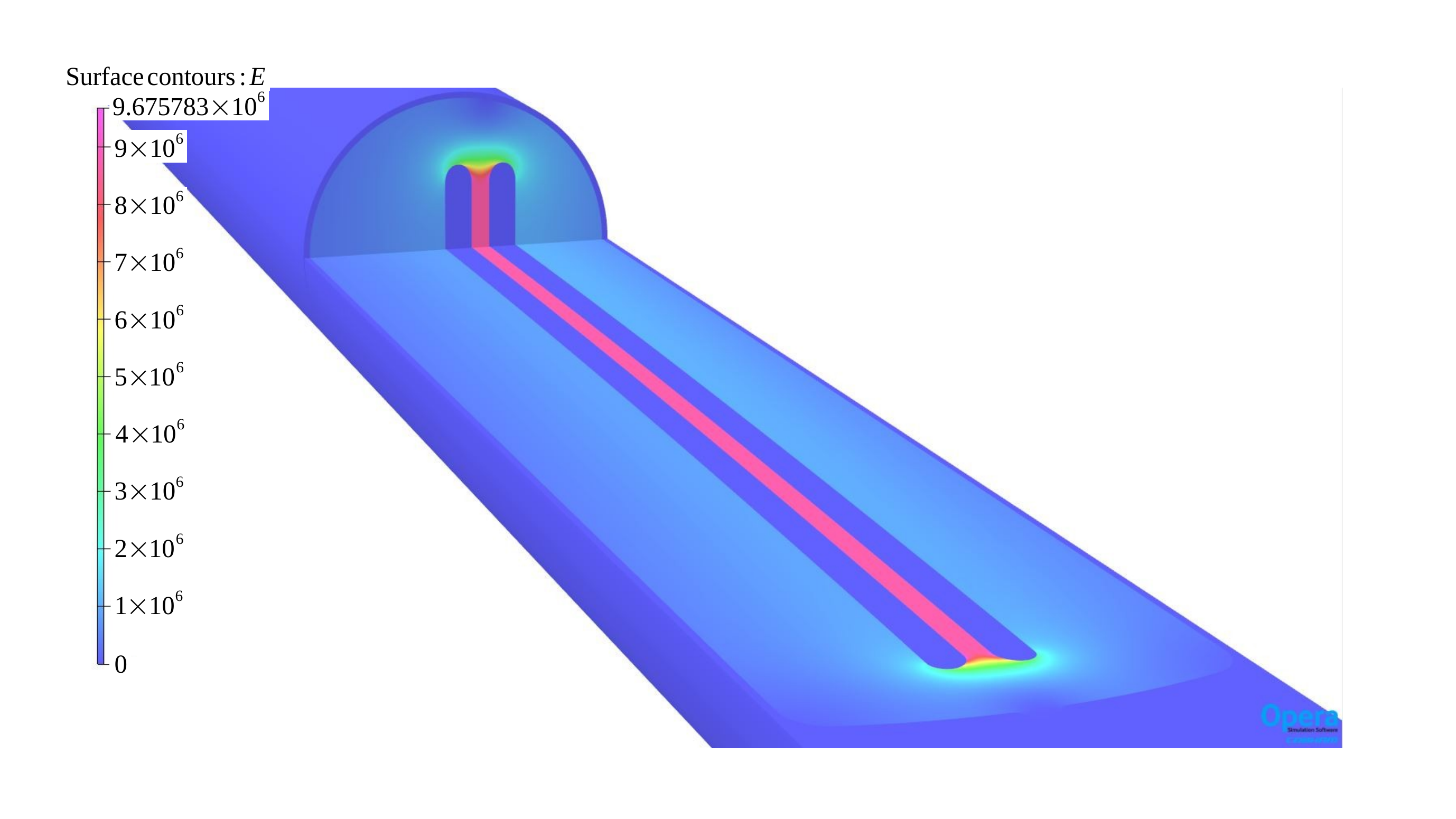}
    \caption{Electric field  computed for the horizontal and vertical midplanes of the main dipole}
    \label{Fig:3Ddipole_fields}
\end{figure}

Based on the vector of phase-space parameters,
\begin{equation}
\label{TMvector}
\begin{pmatrix}X\\X'\\Y\\Y'\end{pmatrix}
=\begin{pmatrix}\text{horizontal displacement} \\ \text{horizontal angular displacement} \\ \text{vertical displacement} \\ \text{vertical angular displacement}
\end{pmatrix}\,,
\end{equation}
the transfer matrix $\mathbf{R}$ of the EDM deflector,
\begin{equation}
\label{k}
   \mathbf{R} =
   \begin{pmatrix}R_{11} & R_{12} & R_{13} & R_{14} \\R_{21} & R_{22} & R_{23} & R_{24}\\R_{31} & R_{32} & R_{33} & R_{34}\\R_{41} & R_{42} & R_{43} & R_{44}\end{pmatrix}\,,
\end{equation}
was computed using the 3D field map\cite{Atanasov} and the beam dynamics code TRACK\cite{TRACK}. The deflector is very weakly focusing in the horizontal plane and essentially behaves as a drift region in the vertical plane. The dynamics are strongly linear inside a region of diameter \SI{20}{mm}. \Figure[b]~\ref{Fig:transferMatrix} shows the dependence of some transfer matrix elements on the initial and final phase-space parameters, as derived from the simulated field map of the proposed concept. The other transfer matrix elements, not plotted in \Fref{Fig:transferMatrix}, are small and are not considered further.

\begin{figure}
    \centering
   \includegraphics[width=0.90\linewidth]{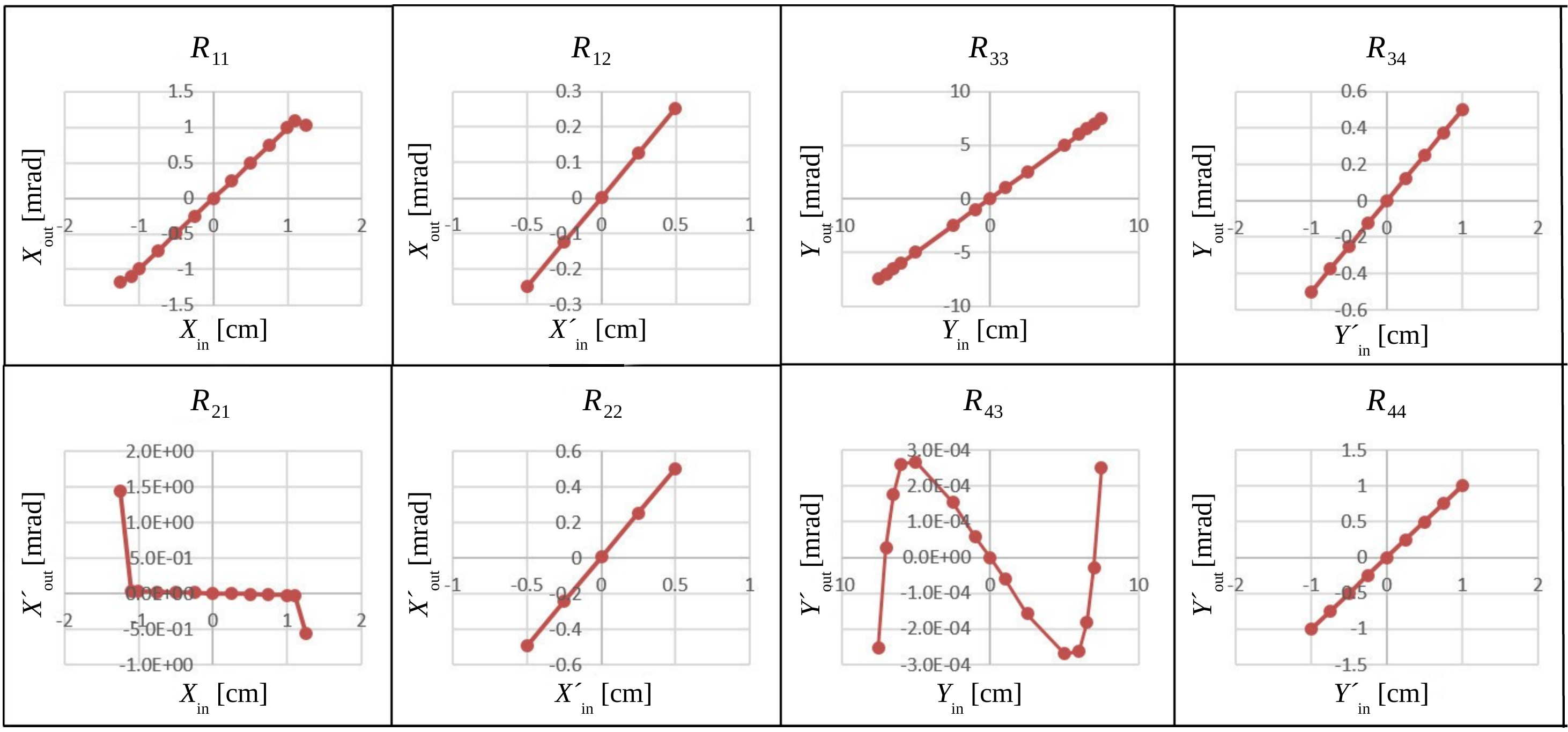}
    \caption{Dependence of the most relevant transfer matrix elements on the initial and final phase-space parameters. The other elements of $\mathbf{R}$ in \Eref{k} are small.}
    \label{Fig:transferMatrix}
\end{figure}

\subsubsection{Soft focusing main dipole}

In addition to the concept without vertical focusing, a concept for a bend with soft focusing in the vertical plane is being studied. Using quasispherical electrodes, this bends with a field index of $m = 0.2$, \ie with  radii of \SI{48.4}{m} in the horizontal plane and \SI{250}{m} in the vertical plane. The electrode curvature amounts to just \SI{24}{\micro \meter} at the top and bottom of the \SI{200}{mm} tall electrodes. This highlights the need for tight manufacturing tolerances, both for the electrodes themselves, and for the electrode fixation inside the vacuum vessel.

\subsubsection{General main dipole considerations}

Electrode manufacturing for both variants will be challenging. To avoid very heavy electrodes (\SI{80}{kg} if made of solid titanium), hollow electrode manufacturing techniques, respecting the required tolerances, are to be developed. The mechanical strength of these electrodes needs to be designed, taking into account the non-negligible force (of the order of \SI{300}{N}) applied on them by the electric field to obtain the required field precision.

A mechanical concept was also developed for the dipole (see \Fref{MainDipole}). The electrode supports are located close to the end of the central tank section. At this location, the vacuum vessel is reinforced with external fins, to optimize its stability and to guarantee that the electrode position is not affected by tank deformation owing to vacuum forces. Three support feet will be mounted on these support fins to allow precise alignment of the tank. To ensure that the requested field quality of \num{e-6} can be reached in the GFR of $\diameter = \SI{20}{mm}$, the electrodes need to be aligned parallel in the vertical plane with a precision of better than \SI{0.3}{\micro \meter}, corresponding to an angular precision of \SI{1.5}{\micro \radian}. Therefore, the electrode supports should be adjustable (in radial position, angle, and height) to facilitate  electrode alignment during assembly, but this will be a very substantial challenge. Upstream, the electrodes are longitudinally fixed to the electrode supports, while at the downstream end the fixation allows for longitudinal movement to limit stress on the ceramic insulators during bake-out or cool-down.

In principle, all dipoles will be powered in parallel to reduce the impact of errors provoked by the instability of the power converter. Conditioning may become challenging if it is to be done with all devices in parallel. This will be even more challenging if the employed electrode materials allow for only a little margin with respect to the required electric field.  Therefore, it is planned to disconnect each device and condition it individually. Since the electrode position is fixed, a two-stage conditioning process could be envisaged. First, each polarity will be separately conditioned, mainly to condition the deflectors on the electrode supports and feedthroughs. This will then be followed by bipolar conditioning, to condition the principal electrode surfaces.

\subsection{Quadrupole}

The principal design assumptions \cite{Carliprivcom} for the strong focusing lattice quadrupoles are summarized in \Tref{Table:Quad}. The present baseline lattice assumes quads of \SI{400}{mm} physical length. Our studies, however, have shown that the required field quality cannot be met with \SI{400}{mm} long quads, partly because of the unrealistically high field gradient and unachievable VHC, but also because of the effect of the end fields on the field quality.  The field requirements can potentially be met with a \SI{1}{m} long device (see \Fref{Fig:Quads}). First, simulations\cite{Borburgh2} have shown integrated field errors of the \SI{1}{m} long (flange-to-flange) device to be of the order of $\num{1E-3}$. The maximum field on the electrodes should still be optimized, but it appears to be very difficult to keep this below \SI{10}{MV/m}.

\begin{table} [h]
\centering
\caption{Principal quadrupole parameters}
 \label{Table:Quad}
 \begin{tabular}{p{5cm} lll}
 \hline \hline
  & Lattice & Simplified & Asymmetric  \\
  & assumption & 3D design & 3D design \\
 \hline
 Physical length (m) &  0.4 &   1.0 & 1.0\\
 Equivalent length (mm) &       400 & 730 & 750\\
 Beam aperture, $a_{x}$ $\times$ $a_{y}$ (mm$^2$) &    30 $\times$ 200 & 30 $\times$ 200 & 30 $\times$ 200\\
 Electrode length (mm) &        n.a. &  700 & 700 (top, bottom)\\
  & & & 834 (left, right)\\
 Field gradient, $g$ (MV/\si{\metre\squared}) & 50 & 27.4 & 26.66\\
 Electrode voltage (kV) &       $\pm$250 &      $\pm$137 & $\pm$133/20\\
 Main field on pole faces (MV/m) &      $\sim$\,2.5 &    $\sim$\,3.5 & $\sim$\,6.6\\
 Quad focal length (m) &        20.97 & 20.97 & 20.97\\
 Field gradient homogeneity in GFR for $\diameter = \SI{20}{\milli\metre}$ & $\num{1e-4}$ & $\num{1.0e-3}$ & $\num{2e-2}$\\
 \hline \hline
 \end{tabular}
 \end{table}

\begin{figure}
    \centering
   \includegraphics[width=\linewidth]{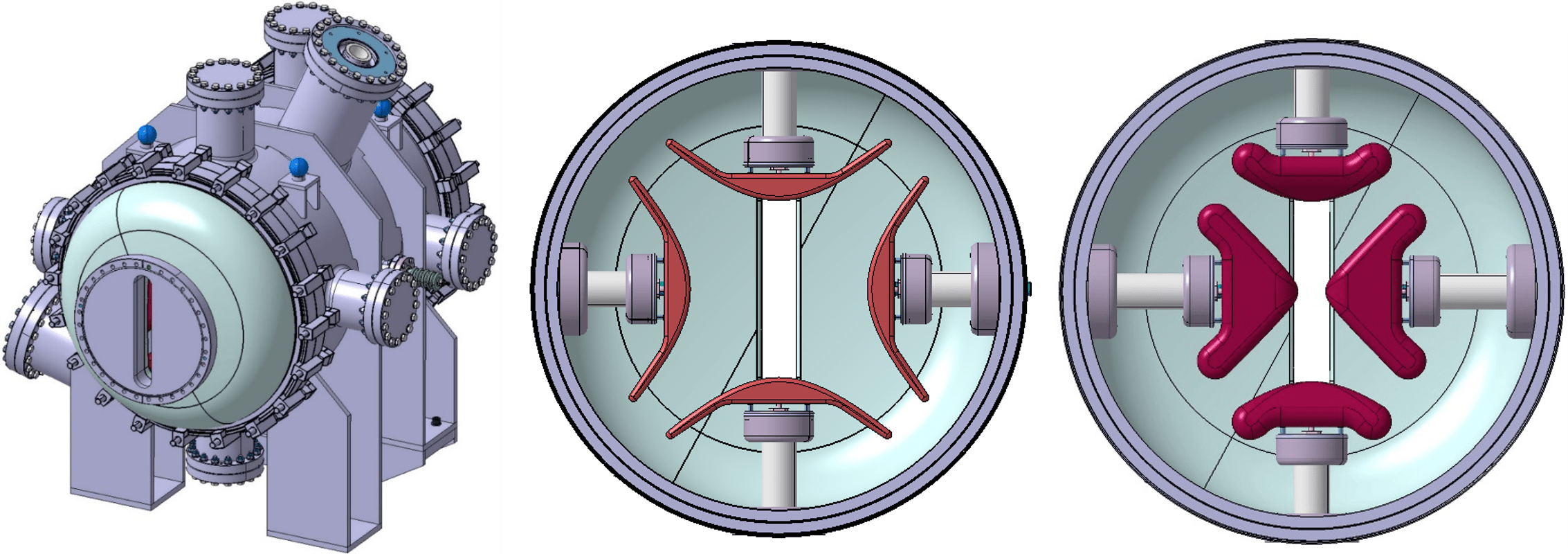}
    \caption{Left: $1\Um$ long quad assembly; Centre: ideal symmetric quad; Right: asymmetric quad}
    \label{Fig:Quads}
\end{figure}

Two quadrupole variants have been studied in further detail. The first is a fully symmetric variant that uses simplified round electrodes to facilitate manufacturing. The integrated field precision required, however, seems  difficult to achieve with cylindrical electrodes\cite{vonHahn:2016wjl}. Therefore,  the second variant uses asymmetric hyperbolic poles: narrow gap poles in the horizontal plane will allow this pair to be powered with a lower voltage, ultimately requiring only one large HV feedthrough. This facilitates integration and reduces  the cost. The principal performance parameters of both variants are also listed in \Tref{Table:Quad}. The 2D field plots\cite{Atanasov} for the quadrupole using the three different pole shapes (hyperbolic, round, and asymmetric) are shown in \Fref{Fig:Quad_fields}. The asymmetric quadrupole's left--right electrode length is longer than the top--bottom electrode length, with the aim of approaching, as closely as possible, the corresponding isopotential surface of an ideal symmetric quadrupole (see \Fref{Fig:AQUAD}). The need for at least two HV feedthroughs makes the requirement to keep the cross-section below $\SI{1}{\meter} \times \SI{1}{\meter}$ challenging, in particular, when a perfectly symmetric (horizontal vs. vertical) electrode design, with two large feedthroughs will be needed.

\begin{figure}
    \centering
   \includegraphics[width=\linewidth]{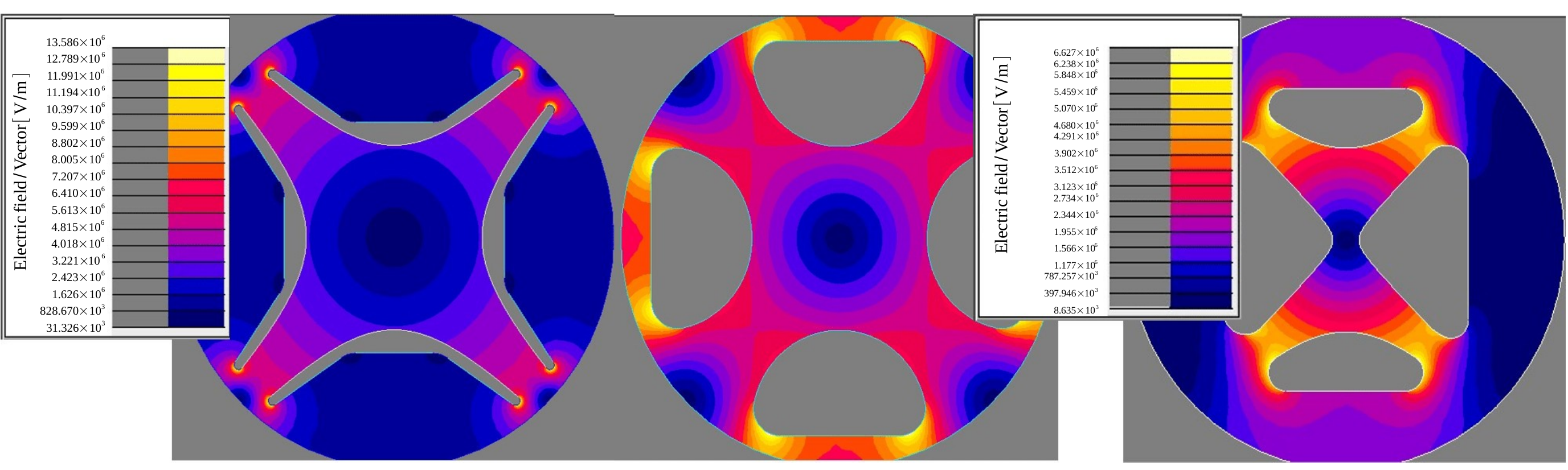}
    \caption{ Electric field of: (left) ideal symmetric quad; (centre) simplified quad; (right) asymmetric quad}
    \label{Fig:Quad_fields}
\end{figure}

\begin{figure}
    \centering
   \includegraphics [scale=0.35]{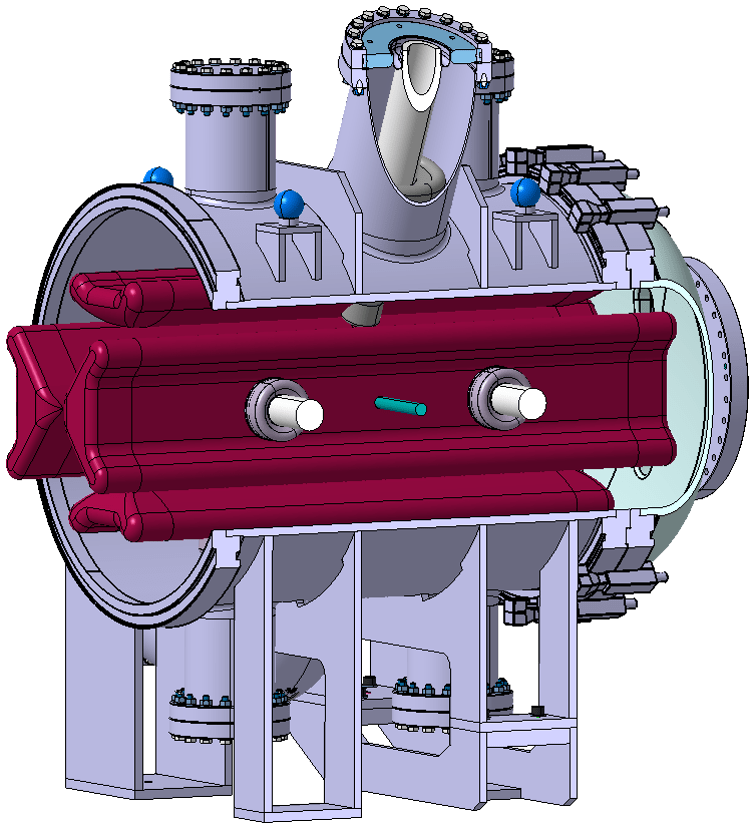}
    \caption{Rendering of the asymmetric quadrupole: the low-voltage electrodes are longer than the high-voltage pair (top and bottom).}
    \label{Fig:AQUAD}
\end{figure}

\subsection{Injection equipment}

Two long straight sections are dedicated to the injection of the two beams. To inject, the beam is deflected by an electrostatic septum followed by a fast-pulsed separator (fast deflector). The principal parameters of these devices are given in \Tref{Table:Injection}~\cite{Carliprivcom}.

\begin{table} [h]
\centering
\caption{Principal injection element parameters}
 \label{Table:Injection}
 \begin{tabular}{l ll}
 \hline \hline
  & Septum & Fast deflector\\
 \hline
 Physical length (m) &  3.5 &   3.0\\
 Equivalent length (m) &        4.0 & 2.5\\
 Deflection angle (mrad) &      57.34 & 10\\
 Gap width, $a_{x}$ $\times$ $a_{y}$ (mm$^2$) &        30 $\times$ 200 & 42.5 $\times$ 200\\
 Field (MV/m) & 8.0 & 1.674\\
 Electrode voltage (kV) &       $-240$ &        $\pm$30.4\\
 Radius of curvature electrodes (m) &   52.32 & $\infty$\\
 $T_\text{rise}$ and $T_\text{fall}$, \SI{0.2}{\percent}$\textrm{--}$\SI{99.8}{\percent} ($\upmu$s) &      $\infty$ & 1.0\\
 Capacitance per electrode (pF) & $\sim$\,$660$ & $<$$500$ \\
  \hline \hline
 \end{tabular}
 \end{table}
 
\subsubsection{Injection septum}

The septum and its (anode) support need to be curved to limit the gap width to \SI{30}{mm}, while displacing the beam by \SI{86}{mm}. The septum can be manufactured from bent or segmented \SI{1}{mm} thick titanium sheets. By limiting the gap  to \SI{30}{mm}, the operational voltage required is approximately \SI{240}{kV}. The vertical acceptance is reduced with respect to the vertical acceptance of the ring to ensure that the septum remains vertically straight when subjected to the mechanical force induced by the electric field. The solid cathode could be made of titanium.

\subsubsection{Fast deflector}

The fast deflector gap width takes into account the beam sagitta using straight electrodes. The external electrode is installed so that the gap at the exit  is \SI{30}{mm} wide and the entrance gap width is \SI{42.5}{mm}. This allows the operating voltage to be limited to \SI{30.4}{kV}. The HV feedthroughs will have to be developed, since these are not commercially available. The feasibility of the fast deflector pulse generator still needs detailed study. In particular, the required rise and fall time feasibility is still to be confirmed. The pulse generator can use semiconductor switch stacks (most
probably MOSFETs), but no commercially available switches have been identified yet.

\section{Required R\&D}

To make sure that all requirements of the electric field elements can be met, the following topics for further research have been identified so far:
\begin{itemize}
\item electrode material performance and their compatibility with bake-out or cryogenic ring operation;
\item feasibility of electrode alignment, providing the required field precision for the ring elements;
\item stable electrode fixation, permitting sufficiently precise adjustment of the electrode position  to obtain the required field quality;
\item feasibility of electrode manufacturing precision;
\item feasibility of the fast deflector pulse generator, in particular with respect to the required rise and fall times.
\end{itemize}

\section{Summary}
The current strong focusing lattice bend concept design achieves \SI{700}{ppm} field homogeneity in the central GFR of $\diameter = \SI{20}{mm}$, which is worse than required. The electric field levels on the bend electrodes might be achievable with titanium, but alternative materials, such as niobium or coated aluminium, may provide a larger margin, and should reduce the spark rate. Further studies of the VHC of large electrode materials are essential to ensure that the proposed elements can be operated at the required fields for extended periods of time, while respecting the desired spark rate.

A design for the quadrupole elements is under development, albeit with a physical length of \SI{1}{m} instead of the \SI{400}{mm} length assumed in the lattice. The asymmetric variant is supplied with \SI{133}{kV} and \SI{20}{kV}, with the maximum voltage close to that used for the dipoles. The achievable integrated field gradient homogeneity in the GFR is, for the time being, insufficient, but, by further optimization of the electrode extremities, it is expected that the required field homogeneity of \num{1e-4} is attainable.  The electric fields on the electrodes are compatible with the choice of titanium as electrode material. The cross-section of this asymmetric quadrupole is somewhat smaller than its symmetric variant, thereby simplifying the integration within the expected cross-section of the ring.

To allow the technical design of the electric field elements and, in particular, to determine the required mechanical tolerances for the dipoles and quadrupoles, such as the electrode profile or alignment, an analysis of these tolerances on the performance reach of the EDM storage ring should be made. Since this is a problem with many input variables, one could use the polynomial chaos expansion (PCE) method, which was already successfully applied to determine the tolerances of an RF Wien filter \cite{Slim:2016dct}.

For the injection of the beams into the storage rings, the feasibility of a curved septum followed by a fast deflector was studied. Both elements operate with conservative fields and voltages, although the feasibility of the fast deflector pulse generator is still to be studied in further  detail with respect to the required rise and fall times.

\begin{flushleft}

\end{flushleft}
\end{cbunit}

\begin{cbunit}

\chapter{Polarimetry}
\label{Chap:polarimetry}

\section{Introduction to polarimetry}

The EDM is a vector-like intrinsic property that measures the asymmetric charge distribution along the spin axis (see footnote 1 on p.\,\pageref{Chap:ExpMethod}). Thus, the experimental connection with the EDM is found through the preparation of beams with spin polarization and the measurement of small spin polarization changes that may be interpreted as evidence for interactions that are signatures of an EDM. Fortunately, polarized beams and the measurement of nuclear spin polarization through strong interaction processes are both mature technologies and capable of high precision. It is also fortunate that polarimeters for spin polarization measurements are at their most sensitive in the range of beam energies where storage ring technology for spin manipulation also works well. This chapter describes the polarization measurements planned for the EDM search in detail.

The chapter begins with a short review of polarization terminology, as codified in the Madison convention, for protons (spin-\sfrac{1}{2}) and deuterons (spin-1)\cite{osti_4726823}. It then moves to a summary of the polarization measurements of the beam as it proceeds through the preparation and acceleration processes. The rest of the chapter deals with the polarimetry planned for the EDM storage ring itself, showing new technology developed for the calorimeter detectors and arrangements for measuring polarization  with high efficiency and precision. Much of this chapter is devoted to ways of handling systematic error problems and limits on the use of counter-rotating beams for identifying the EDM using its  time-reversal violating nature.

\section{Polarimeter spin formalism}

The plan for an EDM-sensitive polarization measurement is to record the horizontal asymmetry in the scattering of protons or deuterons from a carbon target at forward angles. At the energies where the EDM search would be made, the interaction between the polarized particles and the carbon nucleus contains a large spin-orbit term. This gives rise in elastic scattering to an asymmetry between left- and right-going particles when there is a vertical polarization component present.

For spin-\sfrac{1}{2}, the polarization along any given axis is given in terms of the fraction of the particles in the ensemble whose spins, through some experiment, are shown to lie either parallel or antiparallel to that axis. If these fractions are $f_+$ and $f_-$ for the two projections of the proton's spin-\sfrac{1}{2}, the polarization becomes $p = f_+ - f_-$, which ranges between 1 and $-1$, with $f_+ + f_- = 1$. The scattering cross-section $\sigma_\text{POL}$ may be written in terms of the unpolarized cross-section as
\begin{equation}
\sigma_\text{POL}(\theta ) = \sigma_\text{UNP}(\theta) \left[ 1 + pA_y(\theta ) \cos\phi \sin\beta \right] \,,
\label{eq:spin-dependent-cross-section}
\end{equation}
with the vertical polarization component
\begin{equation}
p_y=p\ \cos\phi\sin\beta\,,
\label{eq:orientation-of-polarization}
\end{equation}
where the angles are defined with respect to the coordinate system shown in \Fref{fig:polcoord-2}\footnote{Note that \Fref{fig:polcoord-2} is a duplicate of \Fref{fig:polcoord}, inserted here again for easier reading.}. For polarized protons scattering from an unpolarized target, the complete spin-dependent differential cross-section as a  function of polar and azimuthal angle is given in \Eref{eq:pC-sigma-px-py}.

\begin{figure}
\centering
\includegraphics[scale=0.45]{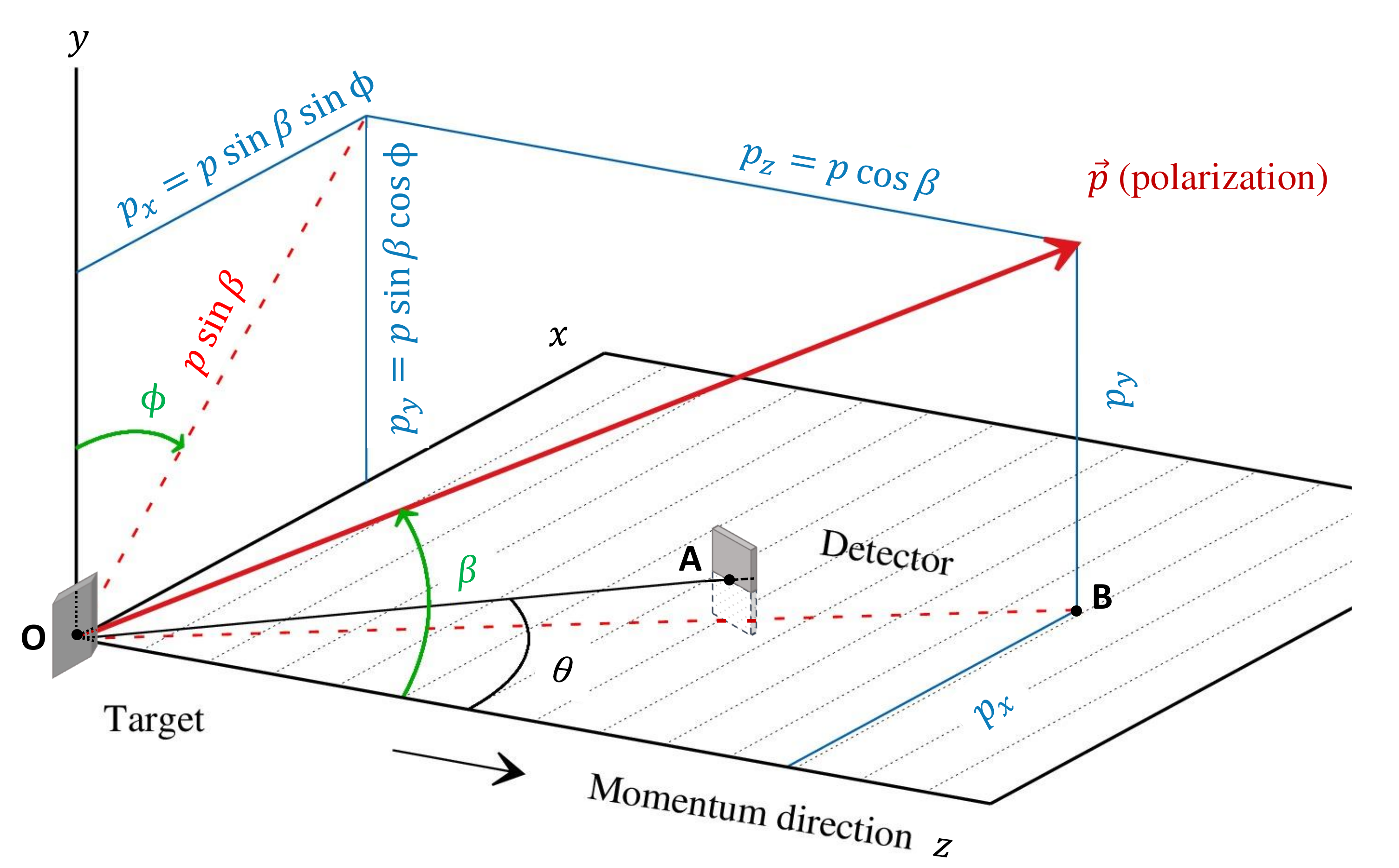}
\caption{\label{fig:polcoord-2}
Coordinate system for polarization experiments, where the incoming beam defines the $z$ axis, and  the particle is scattered in the $xz$ plane (see, \eg Fig.\,6.1 of Ref.~\cite{schieck2014nuclear}). The points $\mathbf O$, $\mathbf A$, and $\mathbf B$ lie in the $xz$ plane, as
does  the detector located at the scattering angle $\theta$, which is used to determine the spin-dependent cross-section (see \Eref{eq:spin-dependent-cross-section}). The angles defining the orientation $\vec p$ are indicated (see \Eref{eq:orientation-of-polarization}). Note that $|\vec p\,| = 1$.}
\end{figure}

The left--right asymmetry measures the vertical polarization component $p_y$. The size of the signal is governed by the strength of the spin-orbit interaction, which gives rise to the asymmetry scaling coefficient $A_y(\theta )$, otherwise known as the analysing power. The left--right asymmetry arises from the $\cos \phi$ dependence of the cross-section on the azimuthal angle of the polarization. If two identical detectors are placed symmetrically about the $z$ axis and their rates are $\text{L}$ and $\text{R}$,  the asymmetry is given by
\begin{equation}
\epsilon =p_y A_y(\theta )=\frac{\text{L} - \text{R}}{\text{L} + \text{R}}\,.
\end{equation}

In the case of the deuteron, which is spin-1, there are three fractions that describe the magnetic substate population,
$f_+$, $f_0$, and $f_-$, where $f_+ + f_0 + f_- = 1$.  The two independent polarizations are vector, $p_V = f_+ - f_-$, and tensor, $p_T = 1 - 3f_0$; the latter can range from 1 to $-2$. If we are interested only in the EDM, then the vector polarization suffices as a marker and the deuteron polarized cross-section (in Cartesian coordinates) becomes
\begin{equation}
\label{eq:Fred}
\sigma_\text{POL}(\theta )=\sigma_\text{UNP}(\theta )\left[ 1+\frac{3}{2}\ p_V A_y(\theta )\ \cos\phi \sin\beta \right] \,.
\end{equation}
For polarized deuterons scattering from an unpolarized target, the complete spin-dependent differential cross-section, as a function of the polar and azimuthal angles, is given in \Eref{eq:dC-sigma-px-py}. Tensor polarization is usually present to a small degree in polarized deuteron beams. There are three independent tensor analysing powers that each add another `$p_TA$' term to \Eref{eq:Fred}
 \cite{Mueller2020}. Their effects may prove useful in polarization monitoring or checking for systematic errors.

\section{Beam preparation}

The essential spin-related parts of the EDM storage ring injector beam line are shown in \Fref{fig:Ed-fig-1_1}. These components are site-independent in the following description. The diagram shows a polarized  proton source, with its associated low-energy polarimeter, spin rotation, and proton acceleration equipment, a trip through the storage ring in order to reduce the phase-space distribution of the beam through electron cooling and bunching, and suitably located polarimeters that confirm that all this works and that calibrate the polarization of the proton  beam. This section summarizes the polarization features.

\begin{figure}
\centering
\includegraphics[scale=0.65]{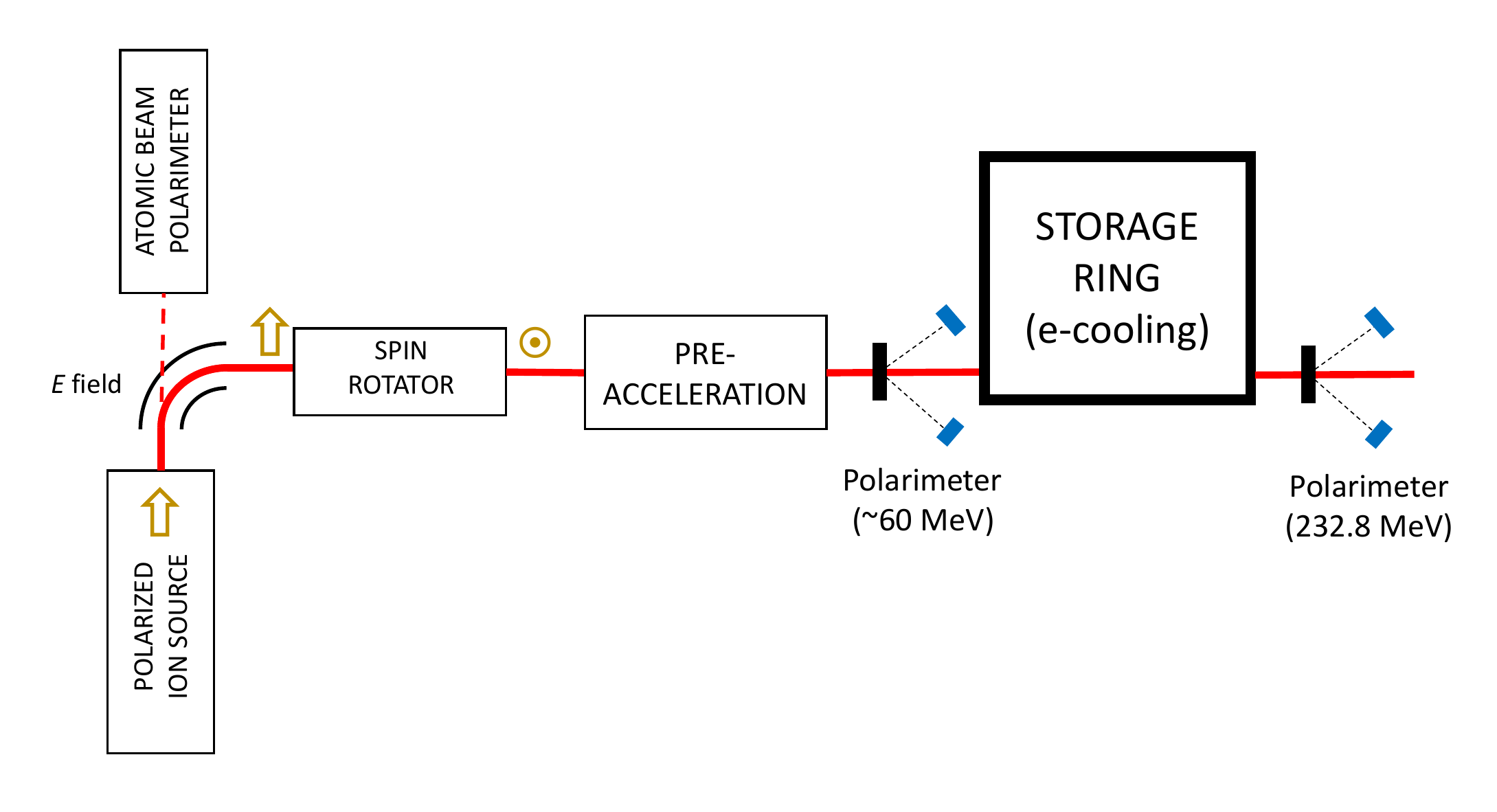}
\caption{\label{fig:Ed-fig-1_1} Main components of the injector beam line that are related to spin manipulation and measurement}
\end{figure}

High-intensity, pulsed polarized proton sources suitable for use at colliders have reached a mature state with adequate beams (\num{e12} per pulse) for a storage ring EDM search\cite{Zelenski:2014vva}. The source creates a high-brightness proton beam in a high-efficiency extraction system before it is neutralized in hydrogen gas to make a well-collimated atomic beam. From there it converges into a pulsed helium ionizer that generates a low-emittance proton beam within a strong axial magnetic field. Inside the high-field region, polarized electrons are added to the protons from an optically pumped rubidium vapour. The neutral atoms proceed to a Sona transition that transfers the electron polarization to the protons\cite{Sona}. The atoms are given an additional negative charge in a sodium vapour and are extracted at \SI{35}{keV} to form a beam for subsequent acceleration. Either state of polarization along the magnetic field axis is possible. The polarization is in excess of 80\%. As an alternative to ionization, an atomic beam polarimeter is present that is capable of measuring the atomic polarization. This allows tuning of the source parameters without requiring acceleration of the beam to higher energy.

For transport through the storage ring, the polarization direction must be perpendicular to the ring plane (aligned with the ring magnetic fields). To accomplish this prior to acceleration, electrostatic plates bend the proton beam without spin precession. This  produces a sideways polarization, and the beam passes through a solenoid where the polarization is rotated into the vertical direction. Initial acceleration is then provided by a linear accelerator, such as an RF quadrupole or a drift tube linac.  Once the protons reach an energy  where nuclear scattering can yield high spin sensitivities, a carbon-target polarimeter becomes feasible. One polarimeter should be installed along the beam line to verify that the ionization, spin rotation, and first acceleration did not alter the polarization. Measurements of the proton--carbon analysing power $A_y$ and figure of merit $\sqrt{\sigma}A_y$\cite{IEIRI1987253}, shown in \Fref{fig:Ed-fig-1_2} over a range of angles and energies between 60 and \SI{70}{MeV}, indicate large values near unity at angles less than \SI{60}{\degree} that are practical for mounting monitor detectors (usually plastic scintillators). This involves the construction of a small scattering chamber. The target may consist of a thin foil of carbon mounted on a movable ladder.

\begin{figure}
\centering
\subfigure[Analysing power, $A_y$]{\includegraphics[height = 0.2\textheight]{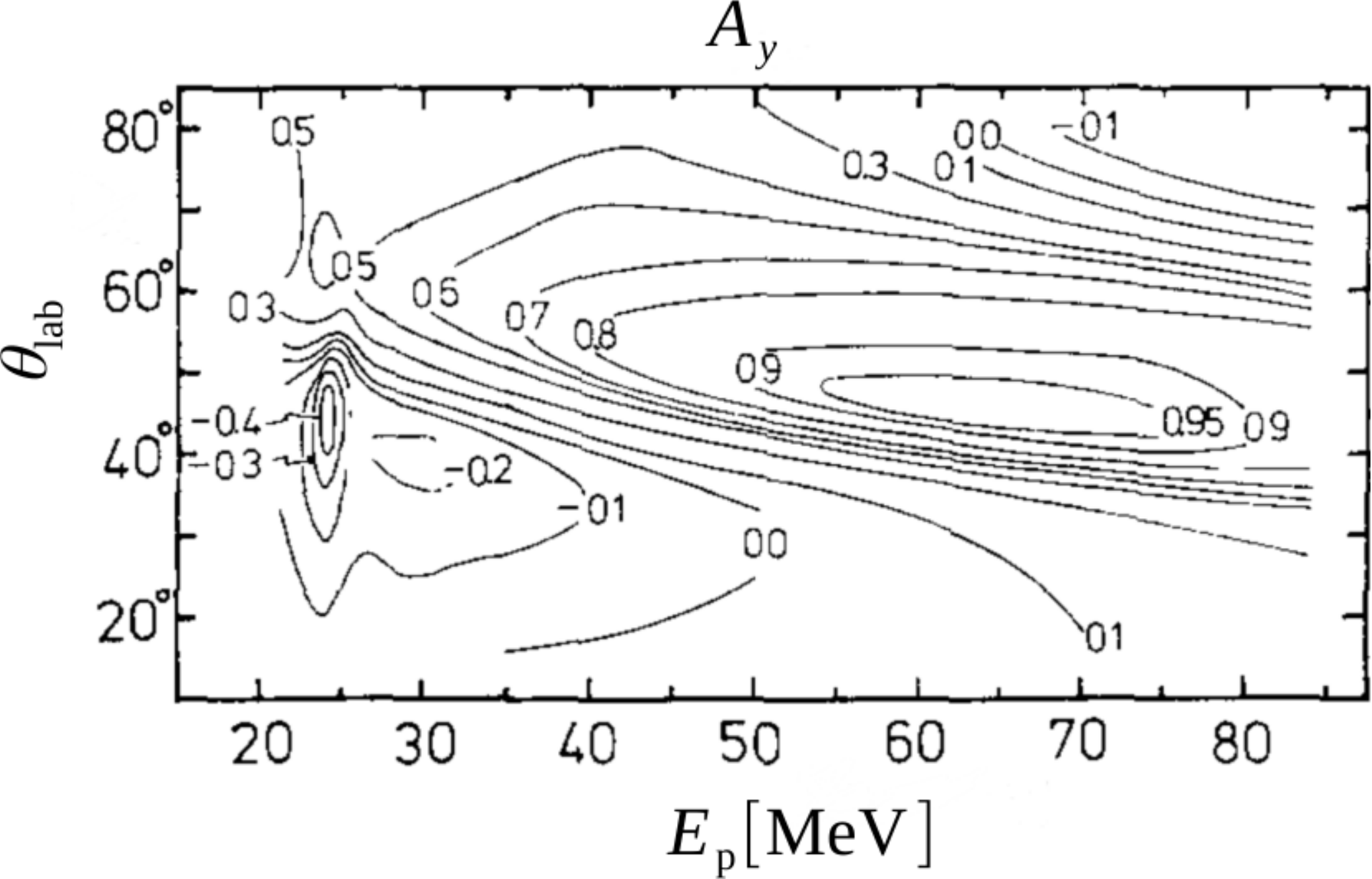}} \hspace{0.2cm}
\subfigure[Figure of merit, $\sqrt{\sigma}A_y$]{\includegraphics[height = 0.2\textheight]{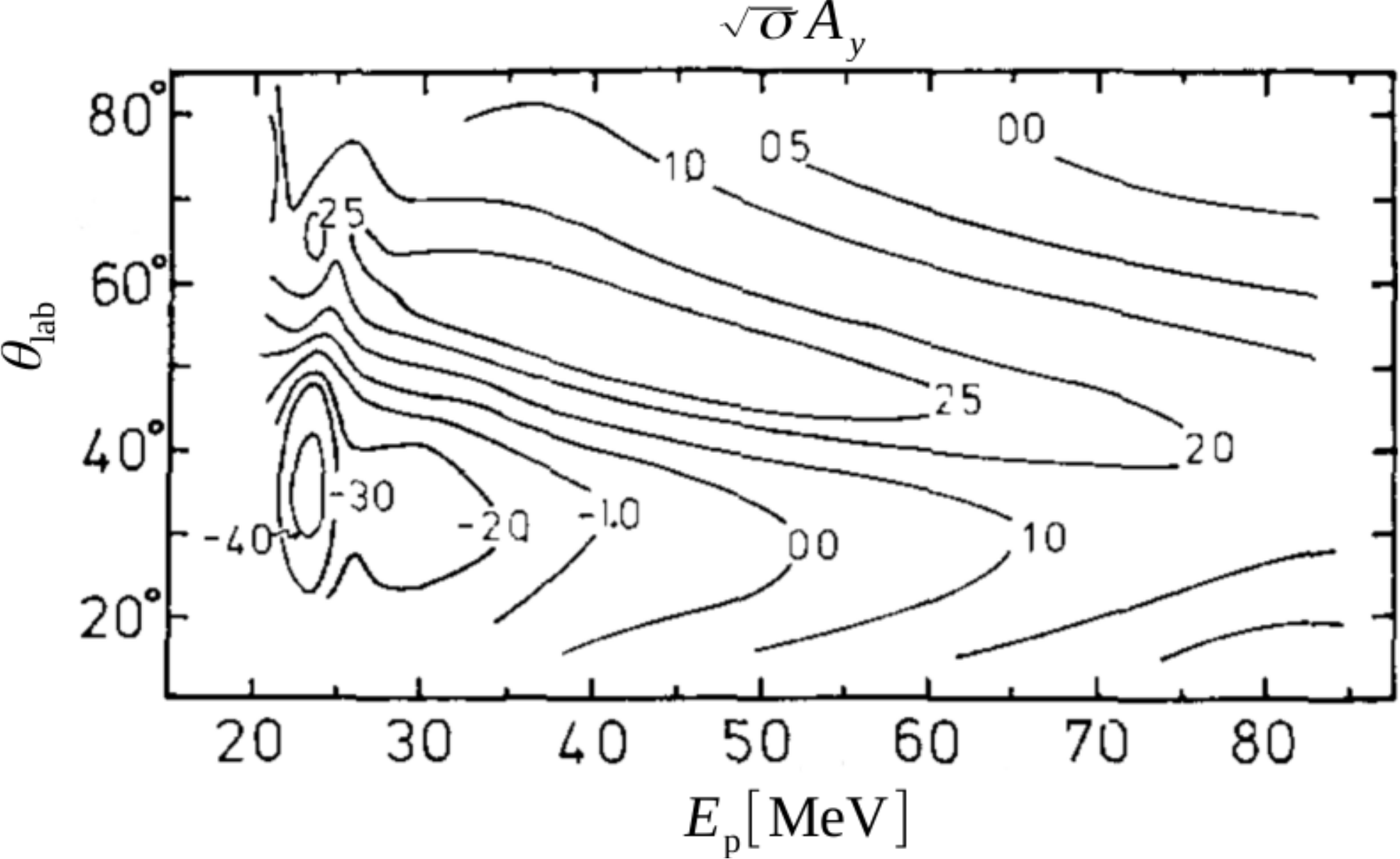}}
\caption{\label{fig:Ed-fig-1_2}Contour maps of (a) analysing power  and (b) figure of merit for proton scattering from carbon in the energy range 20--\SI{84}{MeV}\cite{IEIRI1987253}. The ridge in the \SI{40}{\degree}--\SI{60}{\degree} range is particularly well suited for monitoring beam polarization.}
\end{figure}

The passage of the beam through the storage ring is critical for two reasons. First, the in-plane polarization in the EDM storage ring has a polarization lifetime that is improved if the phase space of the beam is made as small as possible\cite{Guidoboni:2016bdn}. This may be achieved while also using this ring for the second reason, to accelerate the proton beam to an energy of \SI{232.8}{MeV} before injection into the final storage ring. During the acceleration process, the proton energy will pass through the $G\gamma = 2$ imperfection resonance. This resonance, which is driven by magnetic field errors in the ring, is often strong enough to depolarize the beam. Usually, the remedy is to make the resonance even stronger by briefly introducing an imperfection in the form of a vertical steering bump that causes the polarization to completely flip sign as it passes through the resonance\cite{lee1997spin}. A crucial step in the set-up of the beam is to tune the bump so that the maximum polarization survives acceleration. For this, another polarimeter is needed just past extraction from the storage ring. Again, scattering from carbon, shown  in \Fref{Eds-fig_1_3} for a proton beam kinetic energy of \SI{250}{MeV}\cite{Meyer:1988rj}, provides scattering angles with very large analysing powers that may, in fact, be used as the polarization standard for this experiment. A simple foil target and scintillation detectors are again appropriate.

\begin{figure} [hb!]
\centering
\subfigure[Differential cross-section, $({\dd \sigma} / {\dd \Omega})_\text{cm}$]{\includegraphics[height = 0.23\textheight]{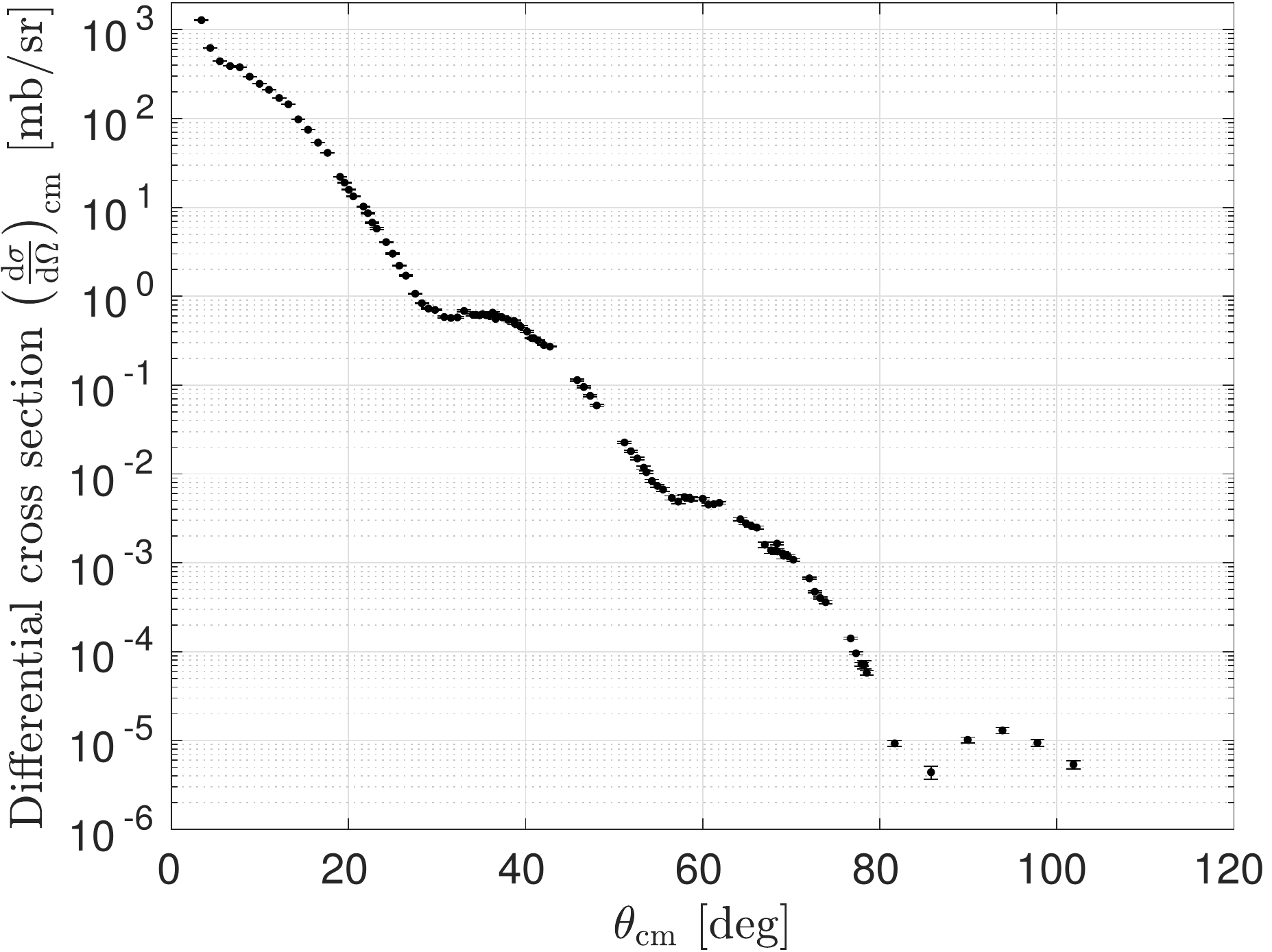}} \hspace{0.2cm}
\subfigure[Analysing power, $A_y$]{\includegraphics[height = 0.23\textheight]{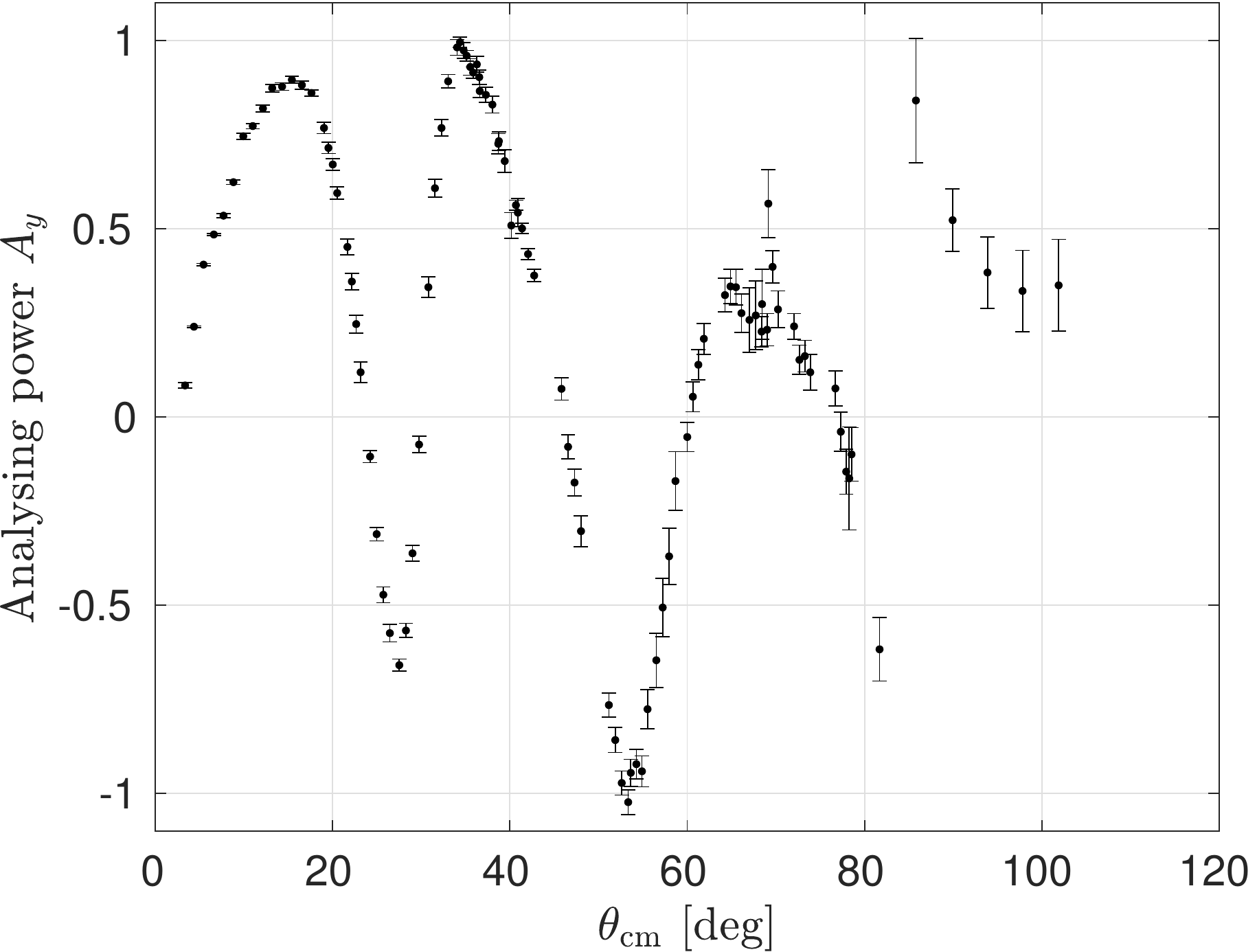}}
\caption{\label{Eds-fig_1_3} Angular distributions of (a) the differential cross-section and (b) the analysing power  for proton scattering from carbon at a beam energy of \SI{250}{MeV}\cite{Meyer:1988rj}. Note that both the first and second interference peaks show very large analysing power values. The second peak is close enough to unity and may serve as a calibration standard for the subsequent use of the beam. (The data \cite{Meyerprivcom} were made available by one of the authors of Ref.~\cite{Meyer:1988rj}.)}
\end{figure}

There need to be two paths by which the injection of the beam is made into the EDM ring,  in order to have both clockwise (CW) and counterclockwise (CCW) beams circulating during the measurement. The switch between these two paths should be relatively rapid so that the requirement that the two beams be as identical as possible is easier to  fulfil. The critical elements are shown in \Fref{fig:Eds-fig-1_4}.

\begin{figure} [hb!]
\centering
\includegraphics[scale=0.55]{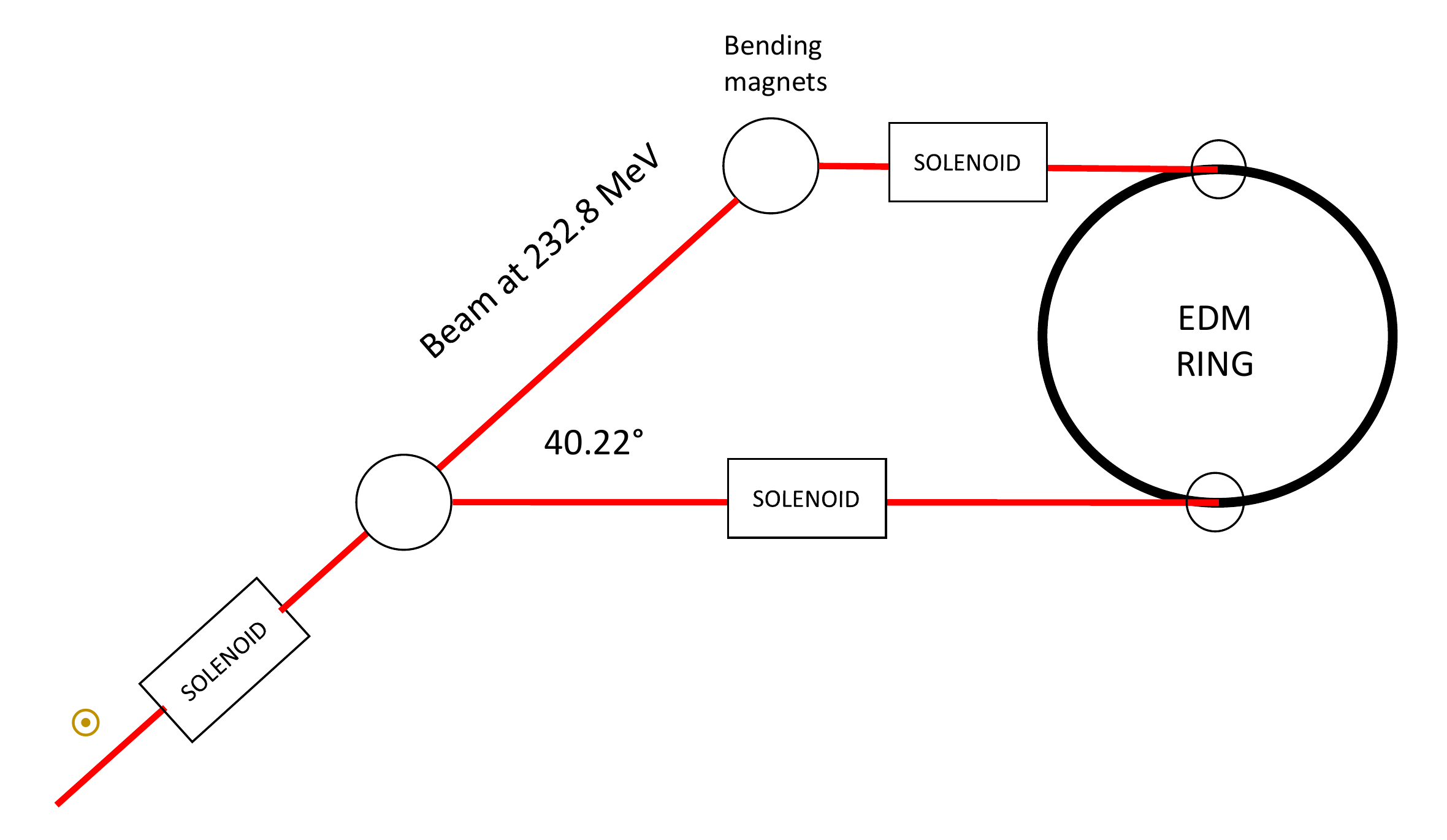}
\caption{\label{fig:Eds-fig-1_4}Elements essential for spin handling during injection. Beams must be injected into the ring in both directions in reasonably rapid succession. It should be possible to have polarization along any direction. The polarization begins perpendicular to the ring plane. A solenoid (up to \SI{2}{T.m} in strength and probably superconducting) is capable of rotating the polarization by at least \SI{90}{\degree} in either direction. This is followed by injecting beamlines in both directions of the EDM ring using a bending magnet of angle \SI{40.22}{\degree}. This aligns the polarization along the beam direction, so that the polarization vector is either parallel or antiparallel to the momentum.  A second solenoid rotates the polarization into the horizontal plane. Some time will be needed for ramping these solenoids if it is desired to have a variety of directions within one beam store. It is assumed that a complete spin flip will be performed at the ion source.}
\end{figure}

\section{Main ring polarimeter design goals}

The goal for the EDM search imposes certain requirements on the polarimeter system, including both target and detector system and the associated data acquisition systems.
\begin{itemize}
   \item The system must make efficient use of the beam particles. A polarization sensitivity at the level of one part per million requires the capture of $10^{12}$ usable polarimeter events, a process that may take many months of data collection. For this, we have explored the use of thick targets located at the edge of the beam at COSY. Particles that enter the front face will be lost from the beam, but the target thickness (\SI{17}{mm} in tests) enhances the probability of scattering into one of the polarimeter detectors. The goal would be to achieve an efficiency near 1\%. Eventually, most of the beam is used up hitting this target.

   \item At the same time, the analysing power $A_y$ should be as large as possible. Values in excess of 0.5 are available for optimal choices of the detector acceptance for either protons or deuterons.

   \item The method of choosing which events to include in the dataset should be relatively insensitive to the choice of cuts, so that small changes have a minimal effect on the measured asymmetry\footnote{In the case of deuterons, it may be important to insert a range absorber ahead of the trigger detector so that most of the break-up proton flux is removed before being processed in the data acquisition system; data acquisition firmware that digitizes the pulse shape and has a high throughput may make this requirement less stringent.}. To ensure a proper early-to-late asymmetry difference, the trigger threshold must be stable over time.

   \item In the EDM search, the left--right asymmetry carries the information on the EDM. At the same time, a monitor is needed for the magnitude of the beam polarization. Such a measurement requires that we rotate the polarization periodically from its frozen spin orientation to the sideways direction. Alternatively, some bunches may be loaded into the EDM ring with a sideways polarization. In this configuration, the polarization is measured through the down--up asymmetry in the scattering. Thus, a full azimuthal acceptance is needed in the polarimeter detectors. If these detectors are segmented, some elements near \SI{45}{\degree} lines may be used for both left--right and down--up measurements; thus, increasing the useful efficiency of the device.

   \item For the counter-rotating beams (CW and CCW, see the following), one block target may serve for the polarization measurement for both beams. This implies that  identical detectors be located up- and downstream of the target. Backscatter from the target is not expected to cause a problem, even with a \num{e-6} sensitivity requirement on the measured asymmetry. Proton--carbon elastic scattering data indicate that such cross-sections are reduced by eight orders of magnitude~\cite{Meyer:1988rj}.

   \item Beam extraction  onto the block target at COSY has usually been achieved by heating the beam to enlarge the phase space in the plane where the target is located. Horizontal and vertical heating may be operated independently, creating the opportunity for two independent polarimeter locations on the EDM ring. More than one polarimeter is useful as a check against systematic errors.

   \item Studies undertaken in 2008 and 2009 demonstrated that the sensitivity of the polarimeter to systematic errors (rate and geometry changes) may be calibrated (for a detailed discussion, see Ref.~\cite{Brantjes:2012zz}). With the use of positive and negative polarization states, such a calibration can be used to remove the effects of these systematic errors. Such a technique thus becomes an important requirement.
\end{itemize}

\section{Implementation of the polarimeter}

While, in principle, the polarization may be deduced from an absolute measurement of the cross-section using a single detector to the left or right of the beam, experience with polarization measurements strongly favours the use of both  left and right detectors simultaneously. In addition, polarized ion sources can provide beams in either positive or  negative polarization states, and the use of both states for the experiment is recommended. In the EDM storage ring, one beam injection scheme envisages filling the ring with both CW and CCW beams, allowing them to come to equilibrium in a coasting state without bunching, and then impose bunching. Another scheme to fill the ring with bunches is described in Appendix\,\ref{app:BeamPrep}. The beam is vertically polarized and both CW and CCW beams are filled using a single polarization state from the ion source. Once bunched into the final pattern, an RF solenoid with multiple harmonics of the bunched beam frequency will be used to precess the bunch polarizations into the ring plane, with alternate bunches polarized in opposite directions. The higher harmonic portions of the RF  solenoidal field will be optimized so that all parts of each bunch are polarized in the same direction following the rotation. Once in plane, the orientation of the polarization in the bunches is maintained using feedback. The feedback system also rotates the polarizations so that the spin alignment axis is parallel to the beam velocity, thus creating the frozen spin condition. The orientation is then maintained by nulling the down--up asymmetry from a continuous polarization measurement. From time to time the polarization is intentionally rotated slightly into the radial direction to monitor the polarization and ensure that the beam is still polarized.

For the rotation of the polarization into the ring plane, the solenoid must carry a complex waveform with several harmonics. The goal is to have the same amount of rotation, given by the solenoid strength,  across the length of each bunch. Outside of the bunch, the solenoid is not constrained. \Figure[b]~\ref{pol:fig6}  shows fits with three and five harmonics that have been adjusted to best reproduce a level equal to one over half of the time ($\sfrac{-\pi}{4}$ to $\sfrac{\pi}{4}$), as indicated by the faint red line. The lower plot is an enlargement of this critical region. Variations are less than 3\%
for three harmonics and less than 0.5\% for five harmonics. The series converges rapidly. Particles within the bunch will sample various areas of this plot as they undergo synchrotron oscillations of different amplitudes, so the size of the differences with respect to unity tells us something about the residual vertical polarization in the beam at the end of this process. The expansion used here contains only odd harmonics and in this way provides two flat-tops. This rotates alternate bunches in opposite directions into the ring plane. Thus, with an even number of bunches, we obtain a condition in which half of the bunches have one polarization and the other half have
the opposite one. Thus, the RF solenoid runs on a harmonic that is half the bunch number. Note that in this scheme each bunch sees the same incremental rotation every time it passes the solenoid because the field is the same. This works because, at the frozen spin condition with a zero spin tune, there is no rotation of the in-plane component of the polarization as the bunch orbits around the ring. It is clearly important to inject the beam at the proper momentum.

\begin{figure}
\centering
\includegraphics[width=275pt]{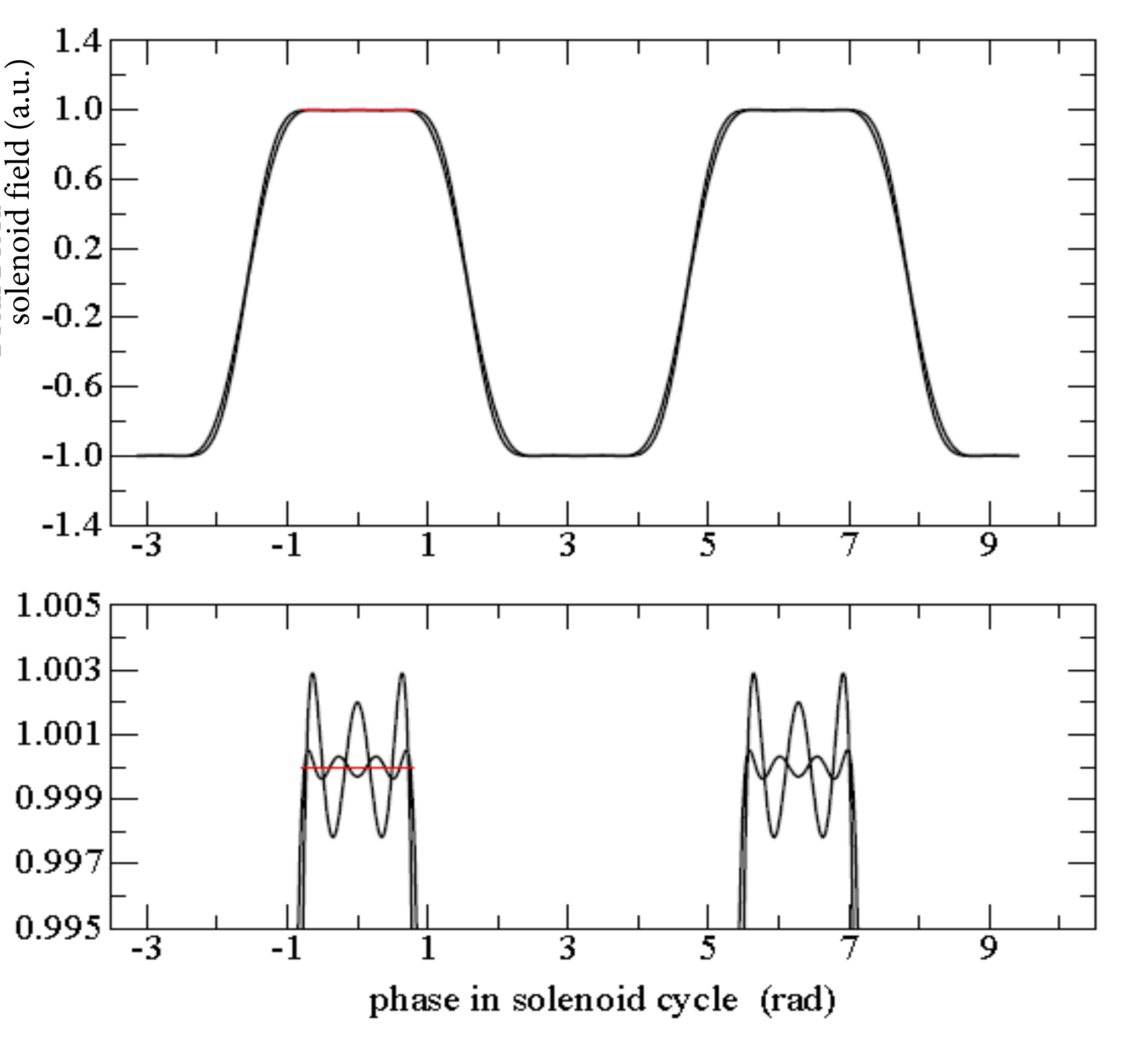}
\caption{\label{pol:fig6}Two curves showing the ability to achieve a flat function over an interval where the beam pulse exists and where its polarization must be rotated uniformly into the ring plane: (top) whole curve; (bottom)  enlargement of the top of the curve. Inclusion up to harmonics 3 and 5 yields larger and smaller variations.}
\label{pol2}
\end{figure}

If the polarization states are $+$ and $-$, and the detectors L and R, then the EDM asymmetry, which is the product of the polarization and the analysing power, may be given by the `cross ratio' formula
\begin{equation}
\epsilon =\frac{r-1}{r+1} ,\ \ \text{where}\ \ r^2=\frac{\sigma_{\text{L}_+} \cdot \sigma_{\text{R}_-}}
{\sigma_{\text{L}_-} \cdot \sigma_{\text{R}_+}}\,.
\label{eq:cross-ratio-definition}
\end{equation}
Here, for example, $\sigma_{\text{L}_+}$ denotes the differential cross-section for scattering into the left (L) detector for polarization $+$. This formula has the advantage that it cancels to first order common errors that depend on differences in the acceptance of the left and right detector systems and differences in the integrated luminosities for the plus and minus polarization states\cite{Hanna:1965}. At the precision required for an EDM search, higher-order errors still affect this formula for the asymmetry and must be removed, as will be discussed later.

Detectors above and below the beam (`up' and `down) are sensitive to the horizontal $x$-component of the polarization. With frozen spin, this should vanish. Thus, any non-zero value implies that the match between the polarization and velocity rotation rates is not perfect. As has been demonstrated at COSY\cite{Hempelmann:2017zgg}, such information may be fed back to a suitably sensitive adjustment, such as the RF cavity frequency, to correct the misalignment\footnote{To correct the velocity rotation rates for the CW and CCW beams, the feedback system should act on two parameters, \eg the RF frequency and an additional tuneable vertical magnetic field.}. This needs to be done continuously during the EDM experiment. In addition, at regular intervals, the polarization should be rotated to the sideways direction (or allowed to precess through a full circle) to provide a monitor of the polarization magnitude. The accumulation of the EDM signal gives the time integral of this quantity.

For tensor polarized deuterons, it is possible to utilize a comparison between the scattering rates in different polar angle ranges as a beam polarization monitor, if the tensor polarization is made intentionally large. This eliminates the necessity for periodic rotations of the polarization away from the direction of the beam. However, a tensor polarization may also generate a left--right (EDM-like) asymmetry if there is a misalignment between the polarization axis and the velocity. This systematic error will be discussed later.

The necessity to monitor both the vertical and horizontal (sideways) polarization components continuously  during the experiment places a premium on polarimeter efficiency, the fraction of particles that scatter usefully into the detectors divided by the number that are removed from the beam. A sensitivity requirement for the EDM asymmetry of \num{e-6} implies \num{e12} recorded and useful events, which may necessitate as much as a year of running time. Polarimeters used in double-scattering experiments \cite{BONIN1990389, LADYGIN1998129} show that for proton and deuteron beams of a few hundred megaelectronvolts, efficiencies of 1\%--3\% have already been achieved using thick  (a few centimetres of solid material) targets. This makes these energies an ideal choice for the EDM ring. Higher energies imply larger storage rings and additional construction costs. Initial ideas for an alternative approach to beam polarimetry for EDM experiments are described in Appendix\,\ref{app:extpol}. If such a system could be realized, it could also be used to determine transverse polarization distributions of the beam.

\section{Choice of analysing reaction}

Work with highly efficient double-scattering polarimeters at these energies has concentrated on the elastic scattering channel at forward angles (\SI{5}{\degree}--\SI{16}{\degree}) as the best choice for polarimetry. Essentially all polarimeters have employed carbon as the target material. The angular distribution represents an interference pattern created by scattering from opposite sides of the nucleus. This angle range typically encompasses one full analysing power oscillation for deuterons and half of an oscillation for protons. At angles less than \SI{5}{\degree}, Coulomb scattering takes over from nuclear scattering and the analysing power quickly goes to zero. This region should be avoided.

\Figure[b]~\ref{pol3} shows the angular distributions of  the cross-section  and the analysing power  for proton scattering from carbon at a beam energy
of $250\UMeV$ \cite{Meyer:1988rj}. The Coulomb-nuclear interference region lies within \SI{5}{\degree}. Beyond this angle, the cross-section arises almost exclusively from nuclear scattering. The elastic scattering channel shown here dominates all other reactions inside about \SI{15}{\degree}. With a positive spin-orbit interaction and an attractive nuclear field at the surface of the nucleus, there is a strong sensitivity to the polarization of the incident proton that results in a positive analysing power. Both the cross-section and the analysing power show an oscillation pattern that reflects interference from opposite sides of the nucleus (see also \Fref{Eds-fig_1_3}). The relative merits of various parts of the angular distribution for polarimetry purposes is usually evaluated through the use of a figure of merit, $\text{FoM} = \sin\theta_{\rm lab} \cdot \left({{\rm d} \sigma} / {{\rm d} \Omega}\right)_{\rm lab} \cdot A_y^2(\theta_{\rm lab})$, where the factor $\sin\theta$ accounts for the reduced solid angle near \SI{0}{\degree} (see \Fref{pol3}(c)). This leads to a clear peak in the FoM. Beyond about \SI{17}{\degree}, the analysing power passes through zero and little additional information is available. A similar peak characterizes deuteron scattering, but it is a few degrees  narrower.

\begin{figure}[tb]
\begin{center}
\subfigure[Differential cross-section, $\left(\frac{\dd \sigma}{\dd \Omega}\right)_\text{lab}$]{\includegraphics[height = 0.17\textheight]{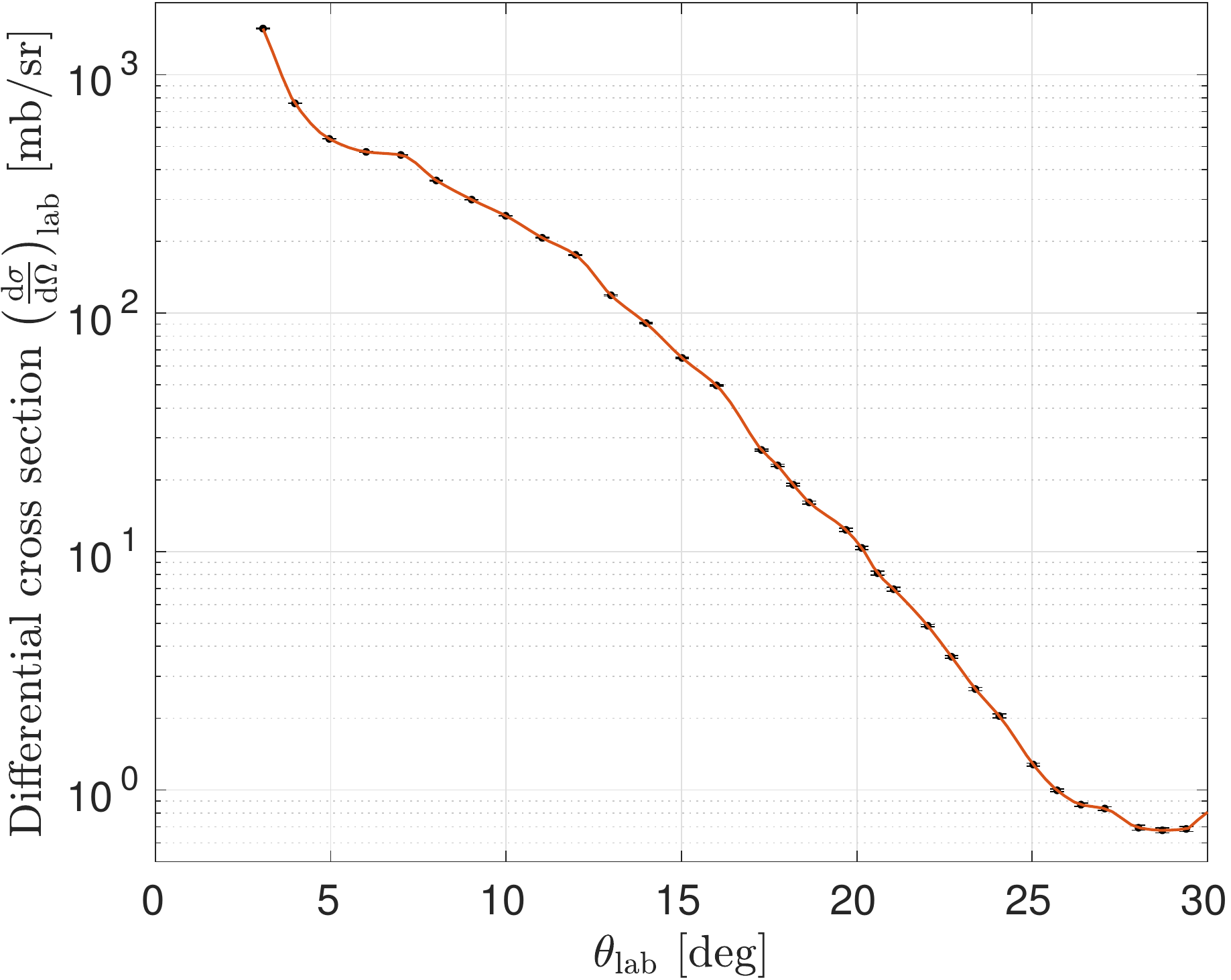}} \hspace{0.1cm}
\subfigure[Analysing power, $A_y$]{\includegraphics[height = 0.17\textheight]{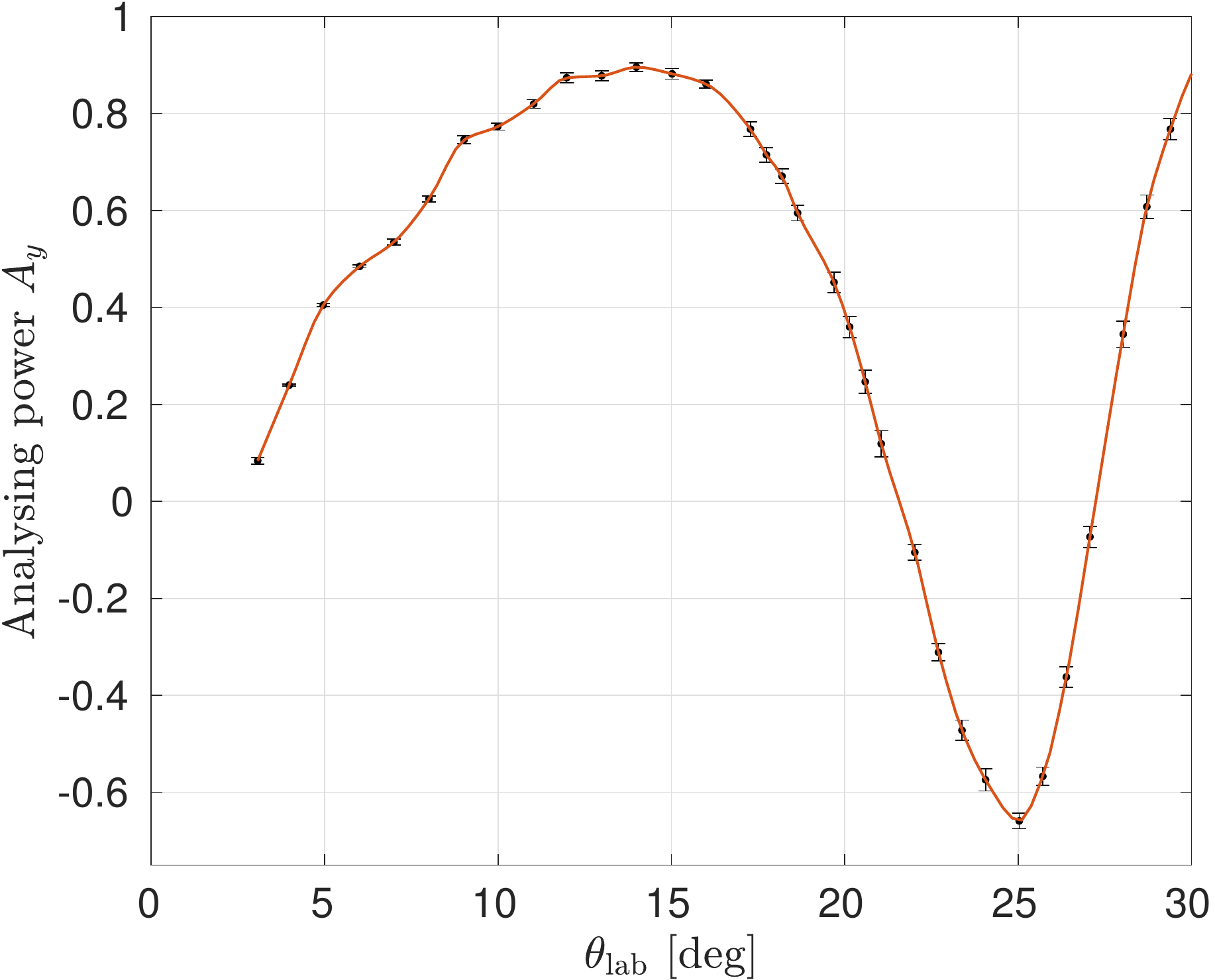}} \hspace{0.1cm}
\subfigure[FoM, $\sin\theta_{\rm lab} \cdot \left(\frac{{\rm d} \sigma}{{\rm d} \Omega}\right)_{\rm lab} \cdot A_y^2(\theta_{\rm lab})$]{\includegraphics[height = 0.17\textheight]{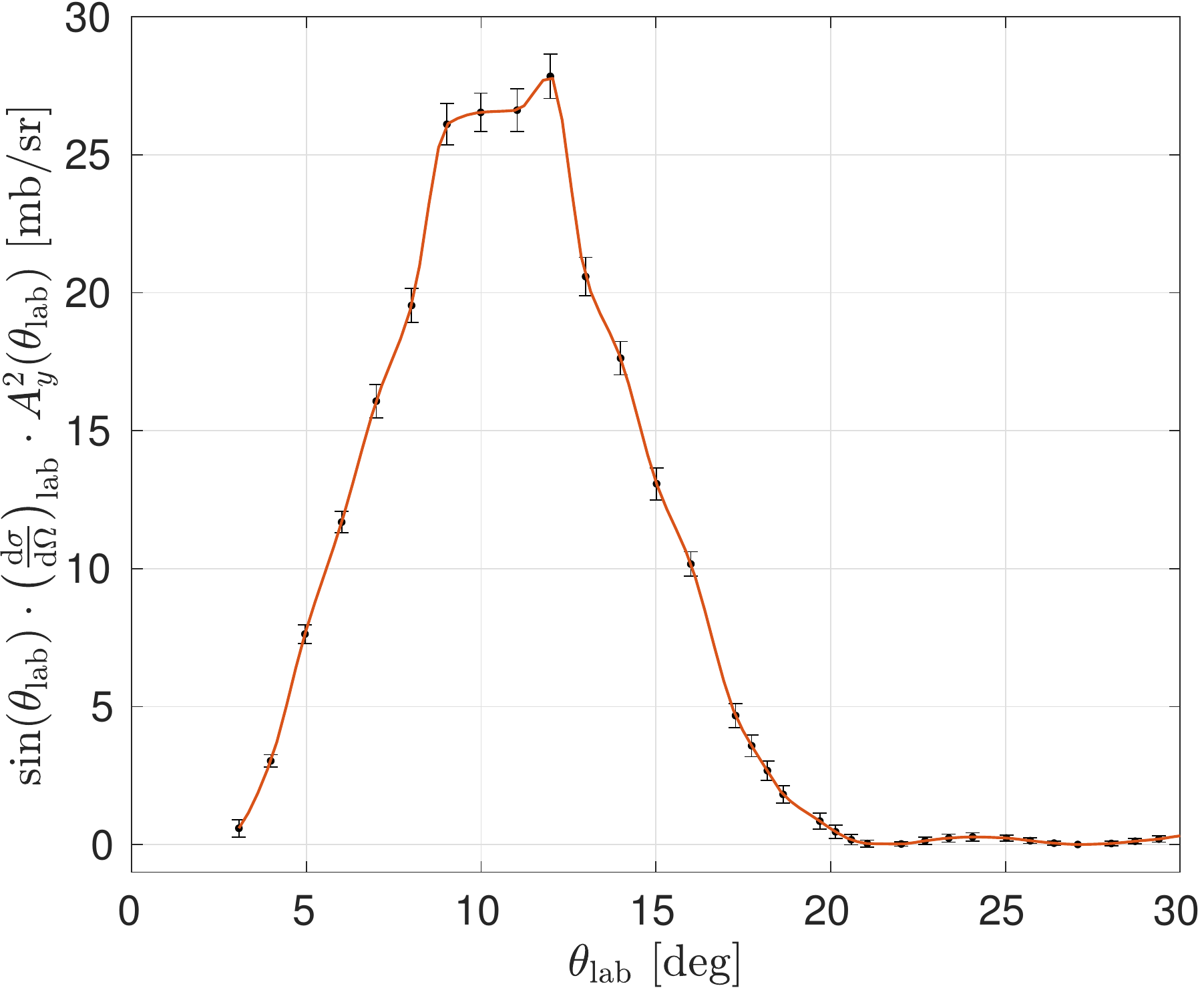}}
\end{center}
\caption{(a) Differential cross-section, (b) analysing power, and (c) figure of merit (FoM) for proton elastic scattering on $^{12}$C at \SI{250}{MeV} bombarding energy\cite{Meyer:1988rj}. The uncertainties of the FoM were obtained through error propagation of the uncertainties of $({{\rm d} \sigma}
/ {{\rm d} \Omega})_{\rm lab}$ and $A_y$. The red line is intended to guide the eye. (The data \cite{Meyerprivcom} were made available by one of the authors of Ref.~\cite{Meyer:1988rj}.)}
\label{pol3}
\end{figure}

Because the forces leading to the large positive analysing power are a property of the nuclear surface and have nothing directly to do with any reactions that might take place (as long as the energy transfer is much smaller than the beam energy), similar features also exist  for a large number of other direct reaction channels. Therefore, there is no particular requirement that the detector be capable of resolving the elastic scattering group exclusively, which requires high resolution in the measurement of the elastic scattering (or other charged particle) energy. This simplifies polarimeter design. The critical feature then becomes the choice of an acceptance that maximizes the FoM, and how stable this acceptance (and trigger threshold) is over time.

Efficient double-scattering polarimeters for protons have been built and used successfully between 100 and \SI{800}{MeV} with a carbon target and simple polarimeter detectors consisting of thin plastic scintillators\cite{Mcnaughton:1986ks}. McNaughton's summary plot, shown in \Fref{pol4}, demonstrates that the largest analysing power appears almost exactly at the magic energy of \SI{232.8}{MeV,} where the proton EDM experiment may run with an all-electric ring. Most proton polarimeters have used carbon as the target because of its ease of handling, wide interference pattern period, and large forward cross-section. All targets in this mass region of the period table tend to give similar results. In \Fref{pol4}, the all-electric frozen-spin energy is marked in red.

\begin{figure} [h!]
\centering
\includegraphics[width=0.5\textwidth]{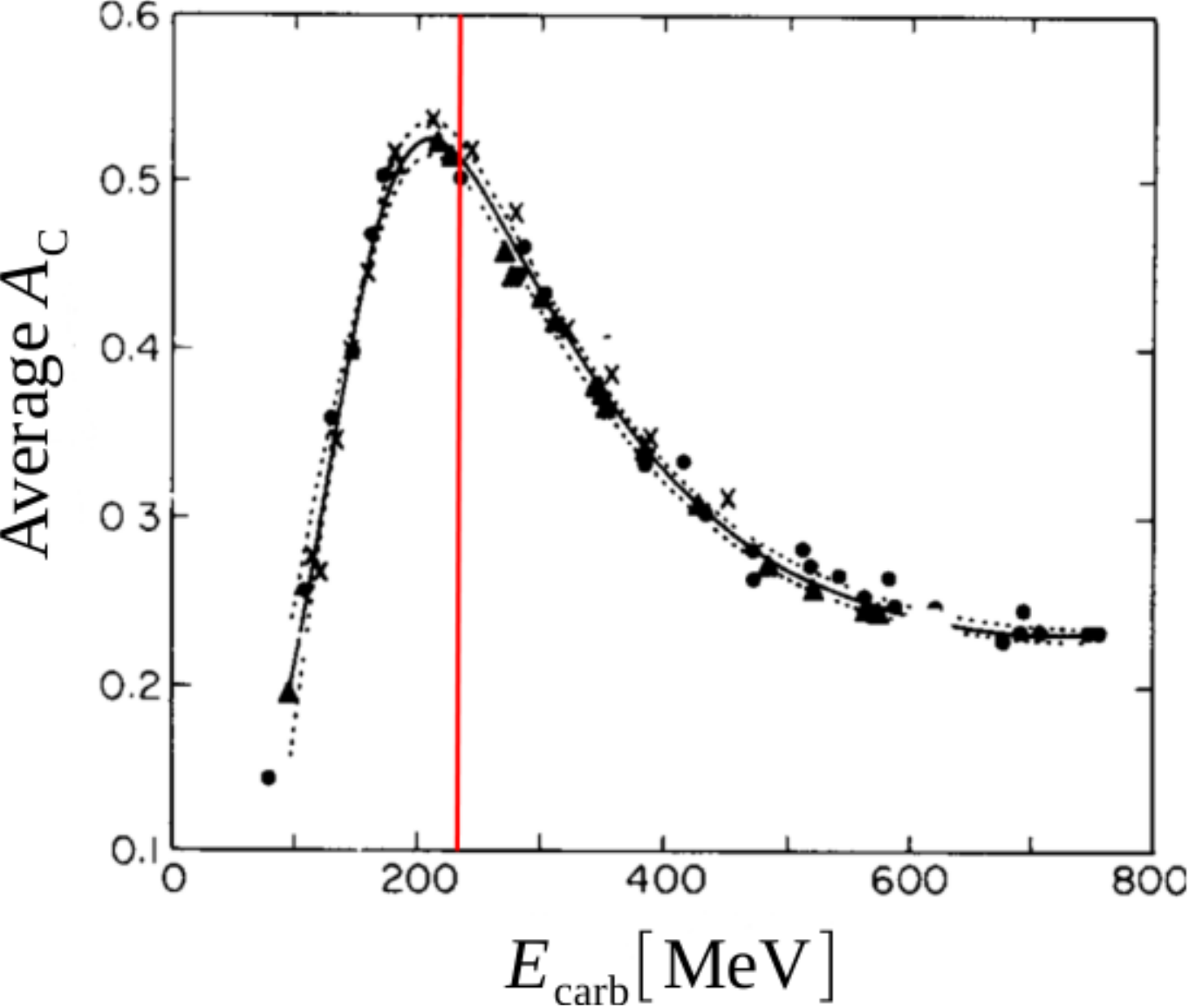}
\caption{Collection of operating point analysing powers for proton--carbon polarimeters at intermediate energies~\cite{Mcnaughton:1986ks}. The curve is a guide to the eye. The red line marks the magic energy of the EDM search. $E_\text{carb}$ is the mean energy at the carbon centre.  Single energy (points) and energy-dependent (solid line) average
$A_y (=A_\text{C})$ for $\theta_\text{lab} = \SI{5}{\degree}$--\SI{20}{\degree} are plotted. Data points are from different experiments, SIN ($\blacktriangle$), TRIUMF ($\times$), and LAMPF ($\bullet$) (see Ref.~\cite{Mcnaughton:1986ks} for details). Dotted lines show the estimated error corridor.
}
\label{pol4}
\end{figure}

These considerations lead to a simple conceptual design for the EDM polarimeter, shown  in \Fref{pol5}. The main detectors consist of an energy loss detector, which identifies the particle, followed by a total energy calorimeter. In many cases, proton polarimeters have used only the $\dd E/\dd x$ detector. For the deuteron, the main background will be protons from deuteron break-up. These protons have almost no sensitivity to beam polarization, and every effort should be made to eliminate them from the trigger rather than relying on post-detector processing. Various groups~\cite{LADYGIN1998129,BONIN1990389} have successfully employed iron absorbers ahead of the scintillator system. If the absorber is appropriately designed, the event trigger from the scintillators may be optimized for large FoM and small sensitivity to scintillator gain drifts.

\begin{figure}
\centering
\includegraphics[width=350pt]{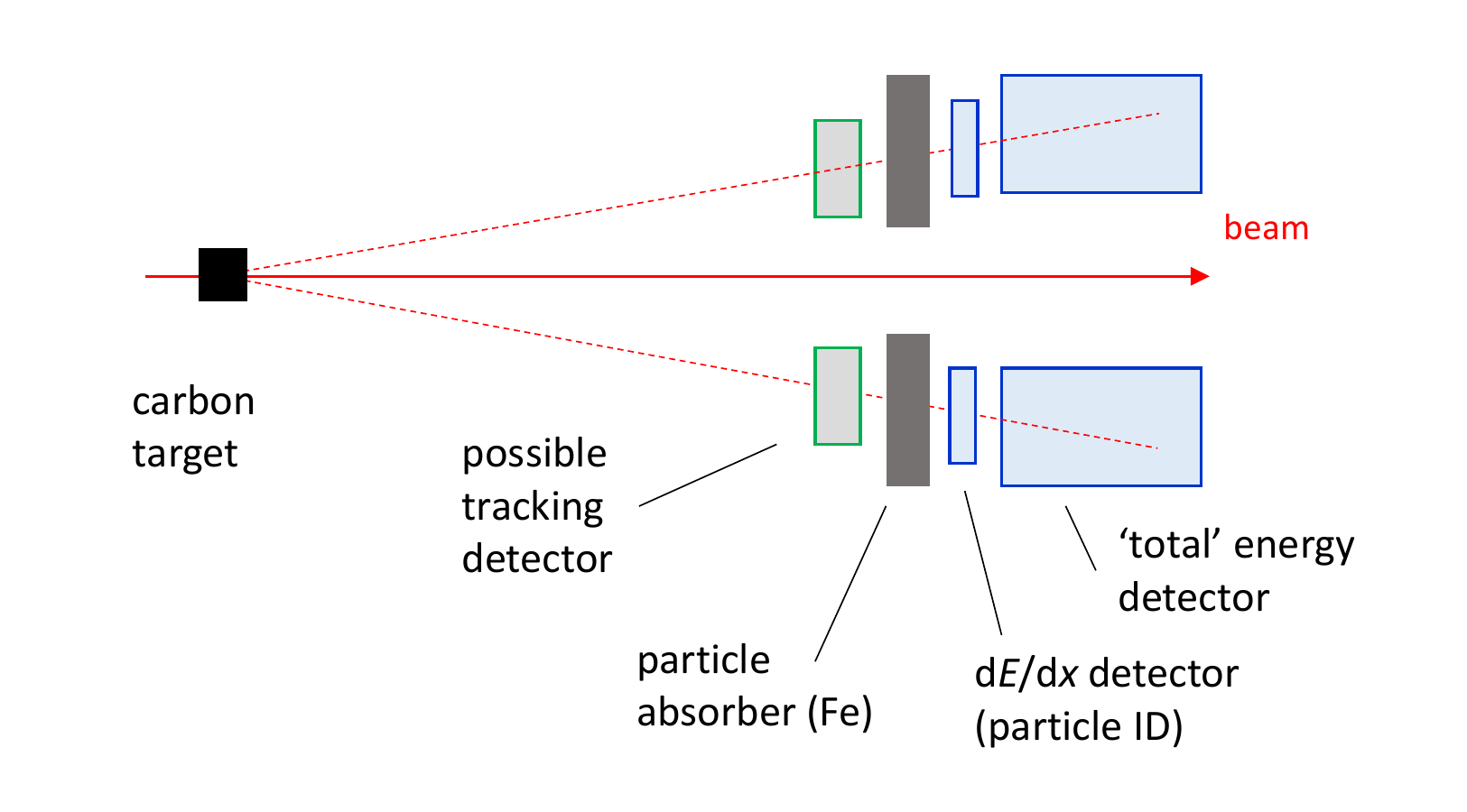}
\caption{Important components of an EDM polarimeter. The beam goes from left to right, and passes through a thick carbon target. Scattered particles first encounter a tracking detector that traces rays back to the target. Next, an absorber removes unwanted events. Lastly, a $\Delta E$ and $E$ detector pair identify the energy of the particles of interest, along with the particle type.}
\label{pol5}
\end{figure}

To make more precise models of any EDM polarimeter, database runs have been conducted at COSY for deuterons at a variety of energies between 170 and \SI{380}{MeV}\cite{Mueller2020}. A similar run for protons was completed in the autumn of 2018. The analysis of these data is in progress. \Figure[b]~\ref{pol6} shows the results of a deuteron database run at the KVI, Groningen, at \SI{110}{MeV}.
The top-left panel shows a 2D representation of the events recorded at \SI{27}{\degree}. Clear bands for protons, deuterons, and tritons appear. The coloured regions indicate regions that should be included in the polarimetry (green) or avoided (purple). The proton band shows large contributions from deuteron break-up with almost no spin dependence. On the right,  panels for deuterons and protons are shown individually. The regions are marked there as well. The proton distribution from break-up is large. These events could be mostly eliminated if absorber material were installed ahead of the detector system.

\begin{figure}
\centering
\includegraphics[width=0.75\textwidth]{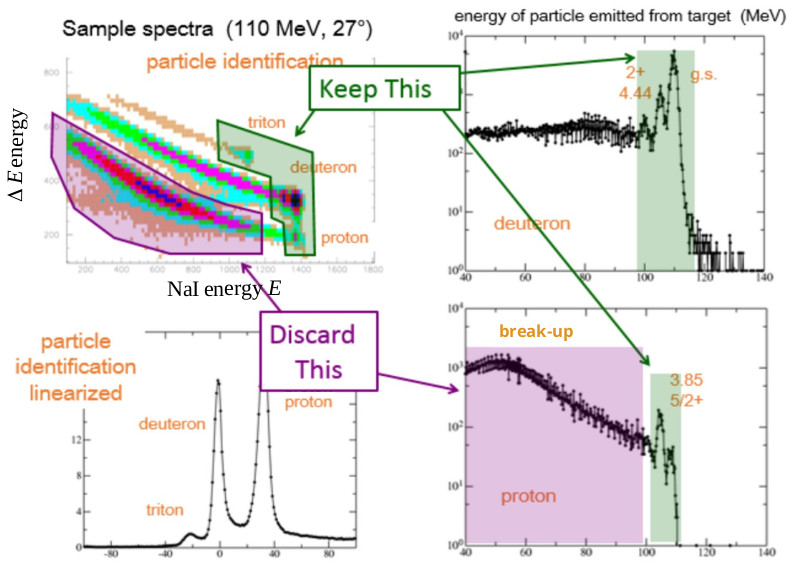}
\caption{Sample from a broad-range deuteron--carbon database run taken at the KVI-Groningen. The deuteron energy was \SI{110}{MeV}. Scattering from a carbon target was observed at \SI{27}{\degree}. Particle type is distinguished in the top panel, which shows $\Delta E$ as a function of $E$. Energy for particles emerging as protons or deuterons are shown in the right-hand panels. Areas outlined in green have significant analysing power and could be used for polarimetry. Areas marked in purple have a low spin sensitivity and should be avoided.}
\label{pol6}
\end{figure}

\section{Target operation in a storage ring}

Prior to the investigations at COSY, no information on highly efficient polarimeter operation in a storage ring  was available. Tests using a thick target were conducted to investigate whether such a target could be operated in this environment, while still allowing the beam to circulate. What worked was placing a square-cornered block about \SI{3}{mm} from the beam centre line. Various schemes were tested to bring the beam to the target slowly, extracting the beam over an extended period of time. It was found that it is better to move the beam than the target, since the beam moves smoothly. A steering bump changes the length of the beam path, creating a problem in maintaining the spin tune (needed for frozen spin). Thus, most of the COSY runs have made use of white noise heating applied through a set of strip-line plates that enlarge the beam through phase space growth. The white noise is applied over a narrow frequency range around a betatron sideband, and this couples well to the beam.

Extraction of the beam using white noise appears to be a two-step process, based on tilted beam studies\cite{Brantjes:2012zz}. In the first step, a beam particle encounters a microscopic ridge on the close face of the target block. If the particle does not undergo  hard scattering, but survives to continue around the storage ring, it will pick up a betatron oscillation. On some subsequent pass by the target, that oscillation takes the particle far enough from the beam centre orbit that it impacts the front face of the block. The tilted beam tests were consistent with the impact point being typically \SI{0.2}{mm} into the target away from the close face. At this distance, the particles would go entirely through the target and therefore have a maximal probability of undergoing a hard scattering event. This picture is confirmed by the observation that the efficiency of the COSY carbon block target is consistent with Monte Carlo calculations that assume that each beam particle has a full interaction with the depth of the target\cite{Brantjes:2012zz}.


The main disadvantage of this target arrangement is that it favours particles that are in the halo of the beam. Below $10^9$~deuterons/fill at COSY, there do not appear to be any issues associated with this; the polarization lifetime measurements show a smooth depolarization curve, as expected. At higher beam currents, structures appear in the time dependence of the polarization that indicate more complex histories in bringing the beam to the target. Modelling of the time dependence confirms this~\cite{Stephensonprivcom}.

Carbon block target thicknesses at COSY were typically \SI{17}{mm,} with a density of approximately \SI{1.7}{\gram \per \cubic \centi \meter}. As the target is made thicker, the energy loss of particles in the target increases. Modelling of the response must therefore involve considering changes in the cross-section and analysing power angular distributions with changing particle energy. These changes, plus considerations of beam alignment, probably restrict carbon block thicknesses to less than \SI{5}{cm}. This thickness, however, is sufficient to achieve efficiencies of the order of 1\%.

\section{Development of calorimeter detectors for an EDM polarimeter}

The EDM polarimeter group at COSY is utilizing dense LYSO crystals as the EDM polarimeter's calorimeter detector\cite{Javakhishvili_2020,Muller_2020}. The $\SI{3}{\centi \meter} \times \SI{3}{\centi \meter}$ modules include a silicon photomultiplier as a readout\cite{JAVAKHISHVILI2020164337} (see \Fref{pol7}). The energy resolution for stopped deuterons at \SI{270}{MeV} is approximately 1\%. These modules will ultimately go into a larger volume array  that surrounds the beam pipe (see \Fref{pol8}).

\begin{figure} [hb!]
\centering
\includegraphics[width=250pt]{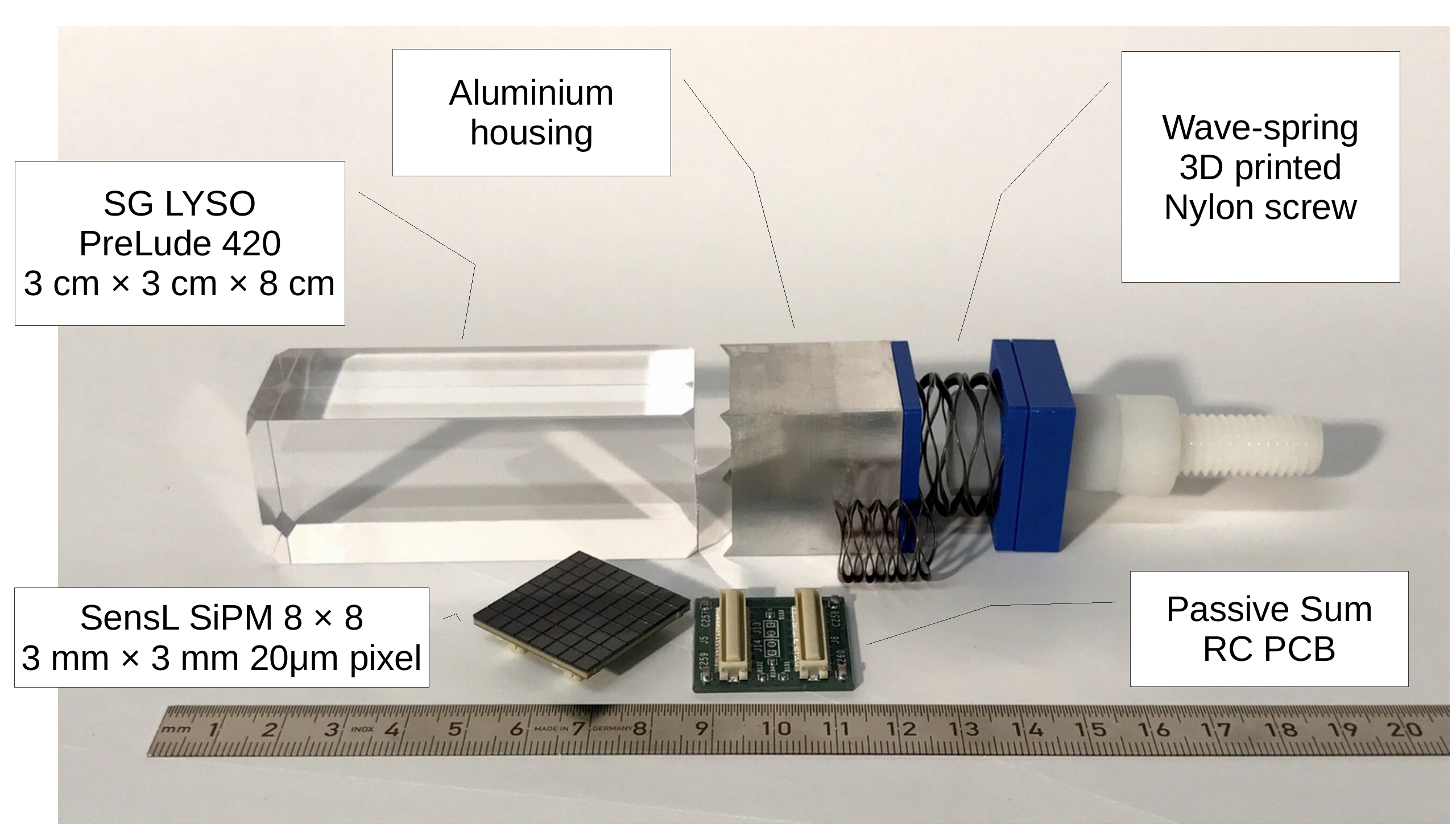}
\caption{Sample LYSO crystal}
\label{pol7}
\end{figure}

\begin{figure} [hb!]
\centering
\includegraphics[width=250pt]{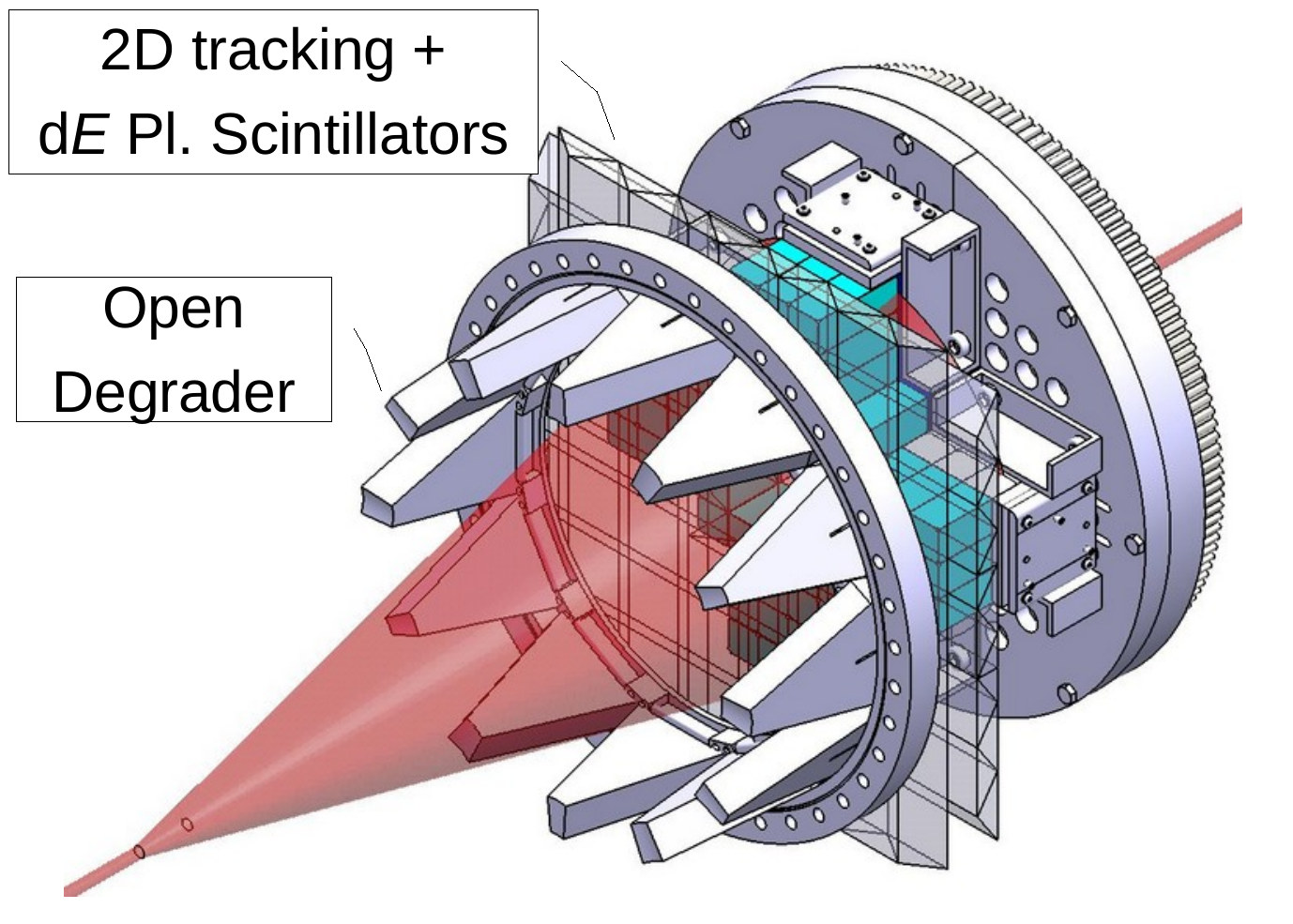}
\caption{Detector system inside the LYSO-based polarimeter. The scattered beam expanding from the target is shown in red. The LYSO-crystal calorimeter detectors are shown in turquoise in the segmented arrangement likely to be used for left--right and up--down asymmetry measurements. Just in front of the calorimeter is an outline sketch of two layers of triangular scintillation detectors. SiPM light collectors located on the ends of the bars are not shown. All particle tracks penetrate both vertical and horizontal layers. Energy sharing between neighbouring scintillators enables more precise position determination. The shutter assembly in front allows for an absorbing layer to be imposed in front of the detectors to remove unpolarized background events. A mechanical system will open and close the degrader or shutter leaves.}
\label{pol8}
\end{figure}

A 48-module mock-up was tested on a beam extracted from COSY. These initial tests of the LYSO modules were made at 93, 196, 231, and \SI{267}{MeV} with a deuteron beam in the external beam of the former Big Karl area at COSY. An overlay of a preliminary series of spectra is shown in \Fref{pol9}. The detector system was moved to the COSY beam line at the beginning of 2019. Various carbon block targets are included in the installation.

\begin{figure}
\centering
\includegraphics[width=350pt]{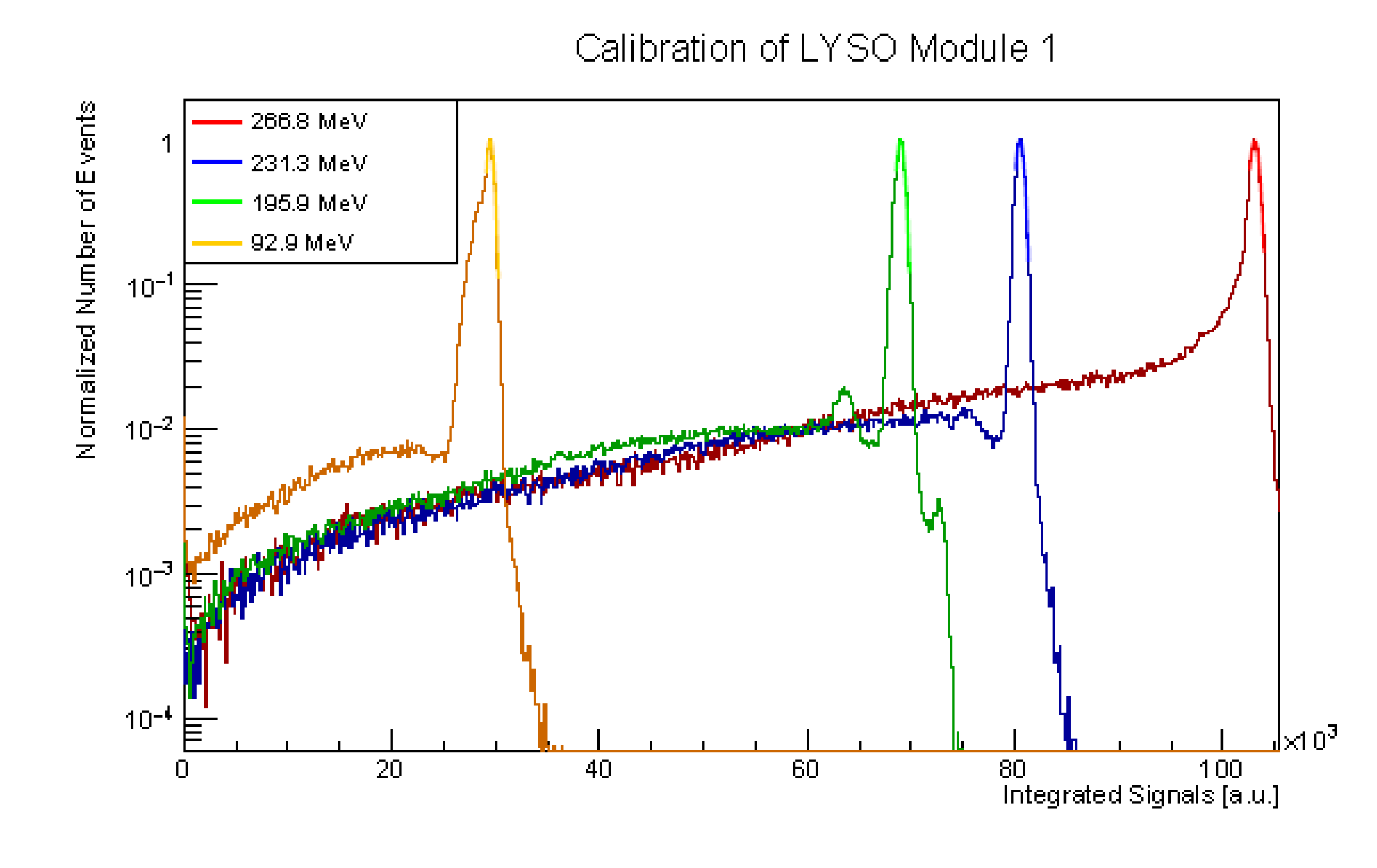}
\caption{Spectra showing the energy deposited in the LYSO crystal for deuteron beams of different energies. The values in the key denote the deuteron energy immediately on entry into the crystal. The energies shown here take into account losses from windows and upstream trigger detectors. The four energies were measured in different set-ups and relative gains have not been reconciled. }
\label{pol9}
\end{figure}

\section{Use of the polarimeter to maintain frozen spin}

For EDM runs of about \SI{1000}{s} in frozen-spin mode, both the beam momentum and the in-plane polarization must rotate about \num{e9} times. The two rotation rates must be sufficiently well matched, so that both the momentum and the polarization undergo exactly the same number of rotations, with a difference at the end of only several degrees. Prior to storing the beam, the value of the spin tune cannot be known to this level of precision, so a feedback mechanism must be used to maintain alignment. One such mechanism was tested at COSY\cite{Hempelmann:2017zgg}, albeit with in-plane polarization, where the RF phase was maintained by a spin-tune feedback\footnote{This situation is similar to keeping the spin aligned along the momentum in frozen-spin mode.}. The analysis of the polarimeter data for in-plane polarization yields a magnitude and phase for each time interval (1 to \SI{4}{s} in duration, for example). A scheme was developed to provide very precise changes to the frequency of the RF cavity controlling the beam, based on a running analysis of the polarimeter data as these were acquired.

\Figure[b]~\ref{pol10} shows an initial situation in which the spin tune is not matched to the rotation of the beam. This result is a slope with time for the phase data. At two times, a signal and its opposite were sent to the RF signal generator requesting a change in frequency. This was immediately reflected in a change in the slope of the phase, which is a measure of the spin tune relative to an assumed value.

\begin{figure}
\centering
\includegraphics[width=230pt]{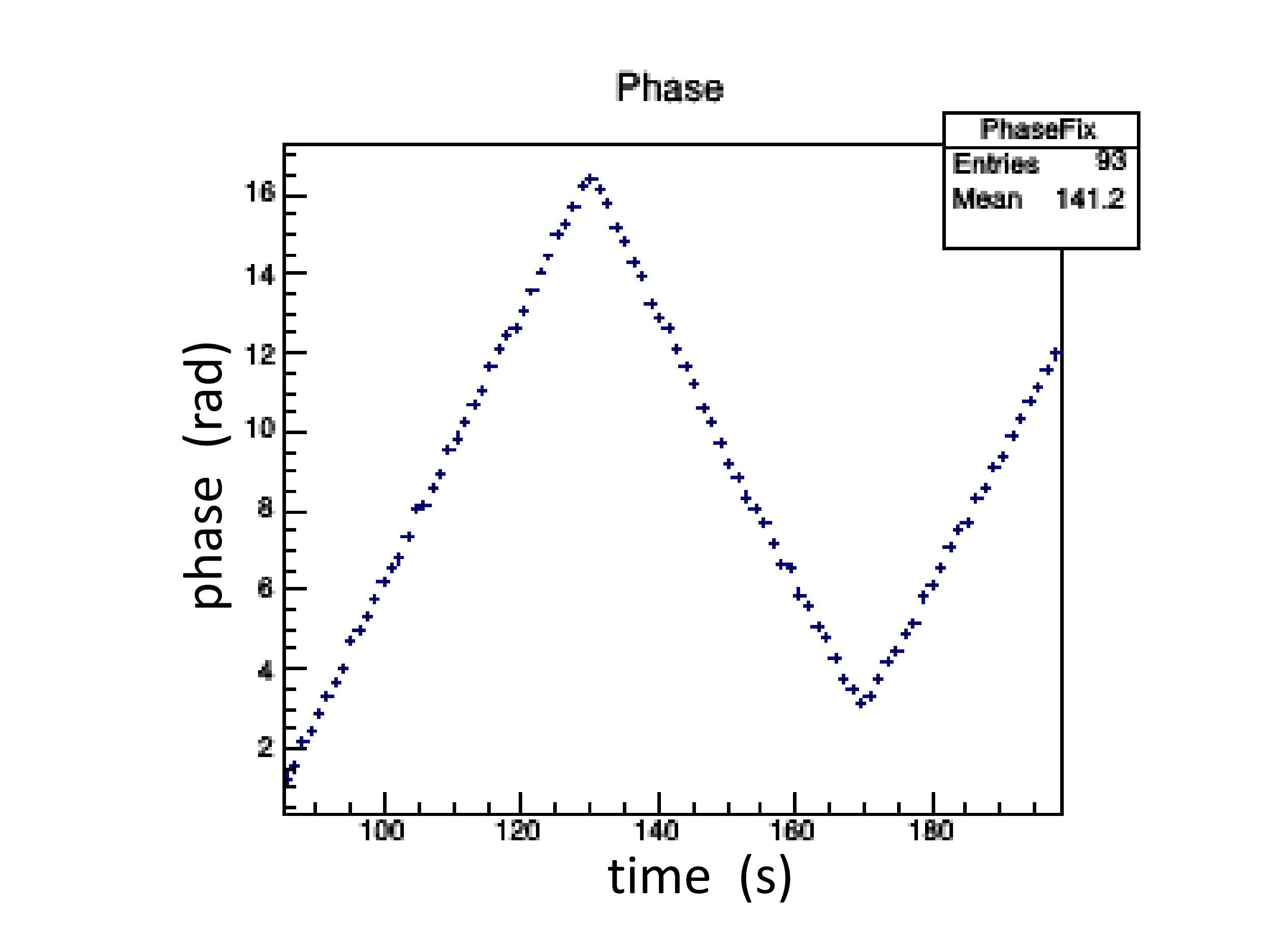}
\caption{Phase of the rotating in-plane polarization \textit{relative} to a standard clock reference as a function of time in the store. Measurements were made by observing the oscillating down--up asymmetry, or sideways polarization, as a function of time. A slope indicates that the spin-tune frequency is not matched to the reference clock. A small change, and then a change back, in the beam revolution frequency changes the spin-tune frequency and hence the slope of the phase as function of time.}
\label{pol10}
\end{figure}

In another test, shown in \Fref{pol11}, a frequency change was sent to the RF generator and then quickly reversed, so that the spin tune afterwards remained the same as before but the pulse caused the RF spin phase itself to shift. With the changes calibrated, the figure shows steps of about \SI{1}{\radian,} resulting from a series of such pulses sent to the RF generator.

\begin{figure}
\centering
\includegraphics[width=200pt]{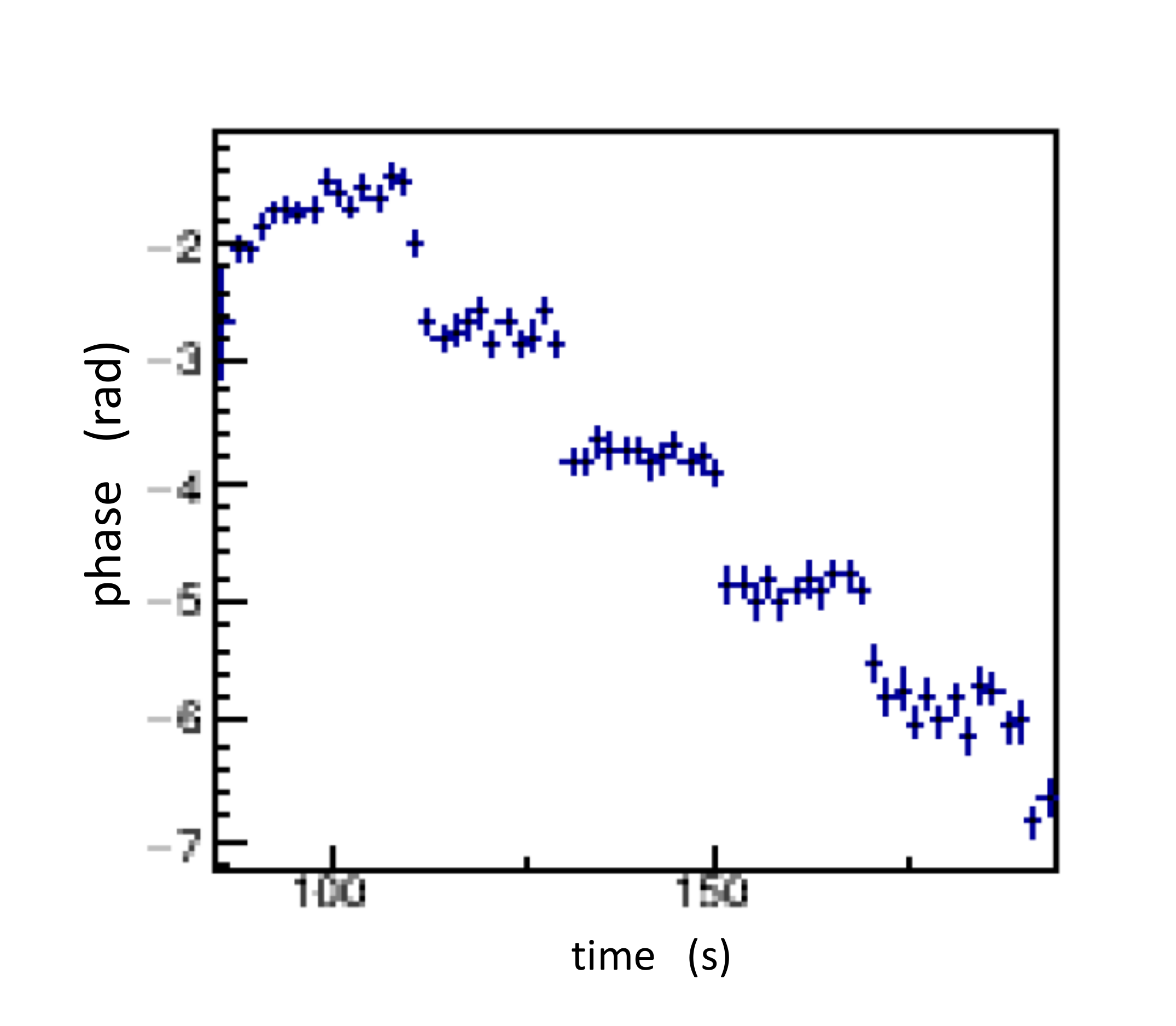}
\caption{Phase of the rotating in-plane polarization \textit{relative} to a standard clock reference as a function of time in the store. Periodically, a pulse is sent to the COSY RF generator that makes a step in the revolution frequency and then quickly reverses it. This pulse makes a step in the phase. Once calibrated, these steps can be tuned to be about \SI{1}{\radian}.}
\label{pol11}
\end{figure}

\Figure[b]~\ref{pol12} shows an example of how this works in a realistic situation, observed during a measurement at COSY (Fig.~1 of Ref.~\cite{Hempelmann:2017zgg}). The top curve is the corrected phase and the bottom curve shows the time sequences of changes made to the RF cavity frequency in order to maintain that level of phase reproducibility. The average deviation, indicated by the grey band, is $\pm \SI{0.21}{\radian}$. The achieved level of control is adequate for the EDM experiment with frozen spin in a dedicated ring\footnote{This technology is essential for maintaining the frozen-spin condition to observe an EDM; it must be in place and operating as soon as the particle spins have been rotated into the plane of the main storage ring.}.

\begin{figure}
\centering
\includegraphics[width=270pt]{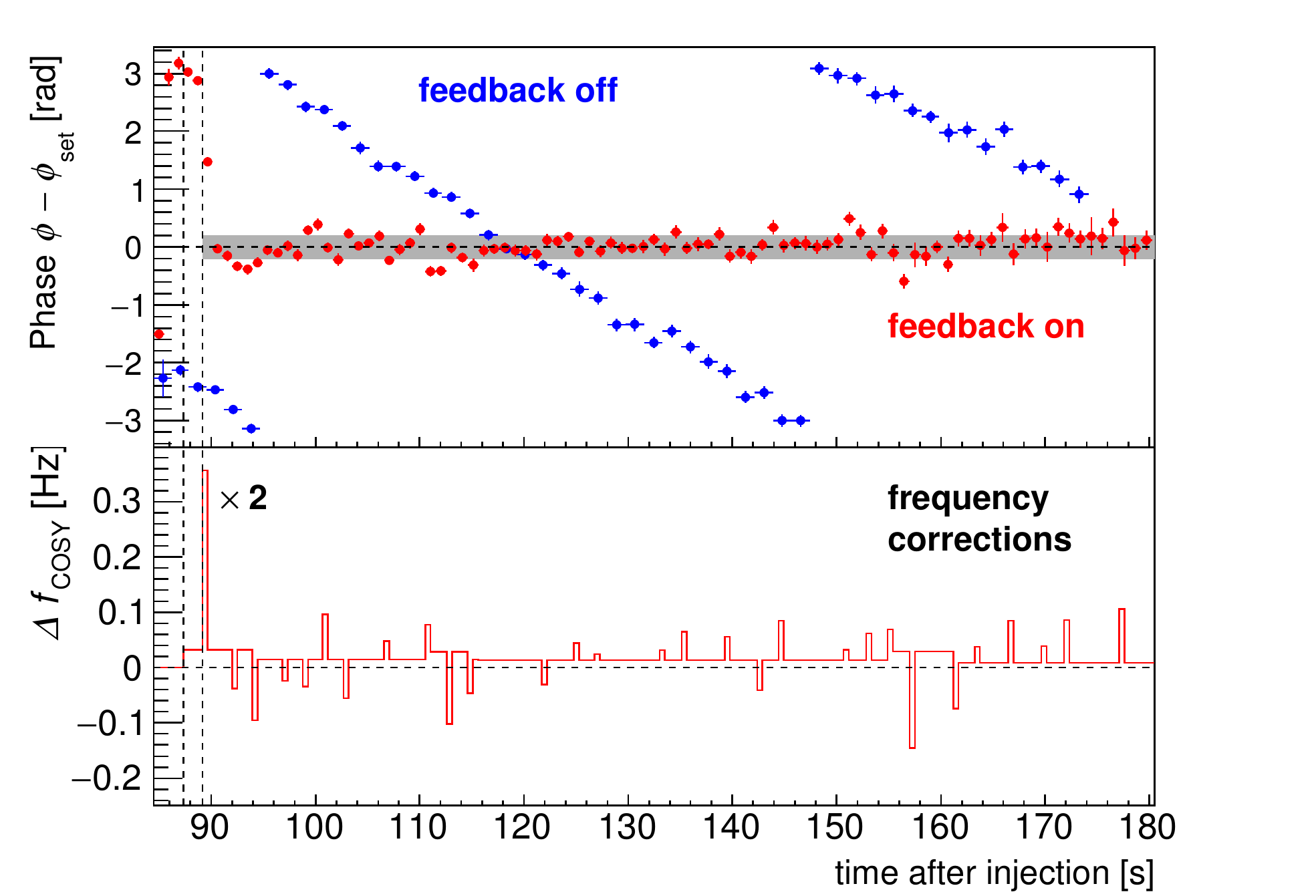}
\caption{Top: Measurements of the phase of the rotating in-plane polarization with reference to an external clock as a function of time, reproduced from
Fig. 1 of Ref.~\cite{Hempelmann:2017zgg}. These measurements are used to correct the COSY revolution frequency in real time so that the phase remains stable at zero (arbitrarily chosen), beginning at \SI{89}{s}. The grey band indicates the r.m.s. deviations of the phase. Bottom: Depiction of the actual changes generated by the feedback system and sent to the RF frequency generator as a function of time.}
\label{pol12}
\end{figure}

\section{Correction of rate and geometry errors in the polarimeter}

A cross-ratio analysis of data from a polarimeter, as described previously, cancels most first-order errors. In a storage ring, the beam itself is continuously changing with time in both intensity and geometric placement, so higher-order effects must be addressed. This is particularly true if sensitivities approaching $\num{e-6}$ need to be probed.

\begin{figure} [hbt!]
\centering
\includegraphics[width=450pt]{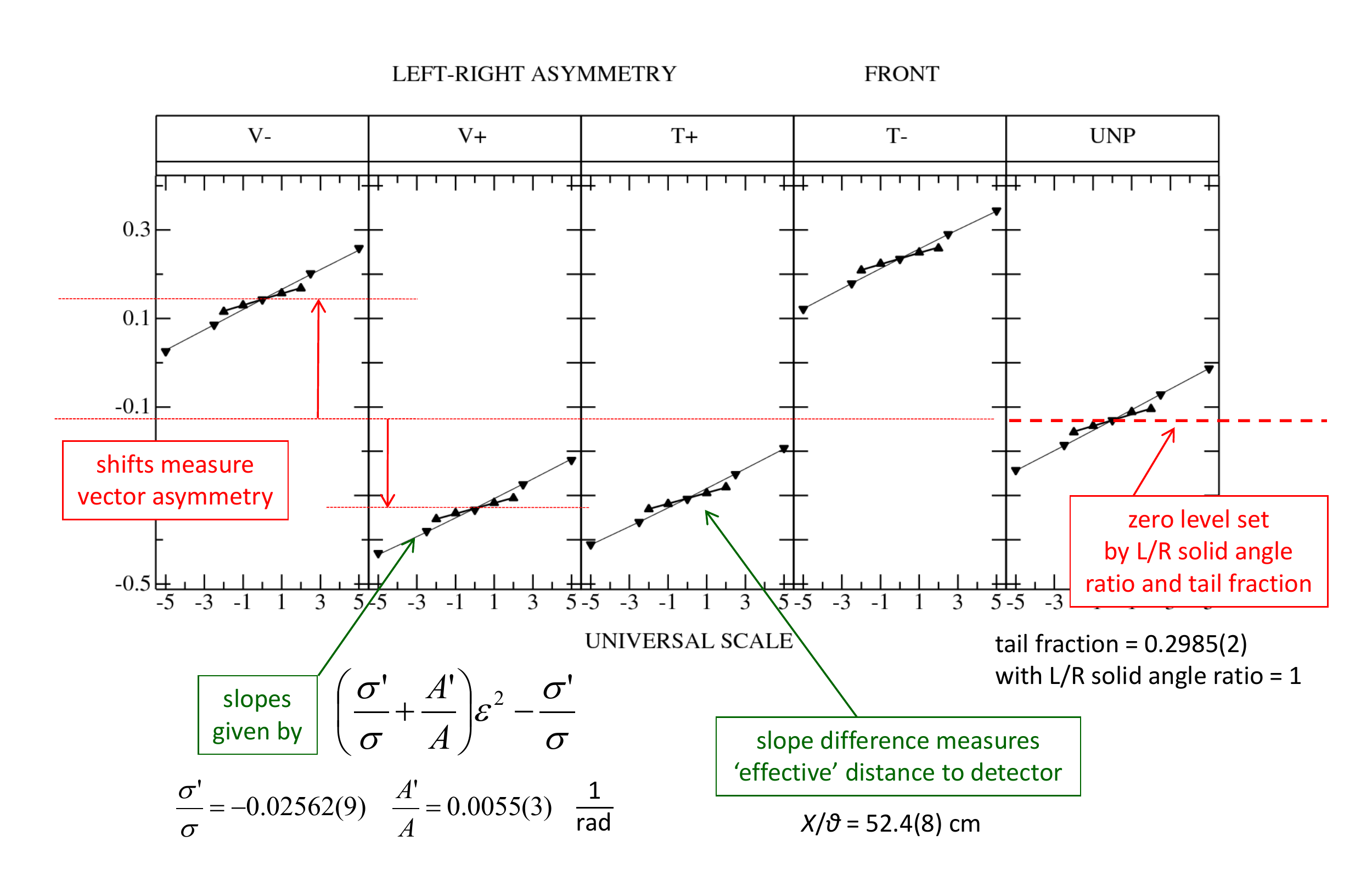}
\caption{Measurements with the EDDA detector of the left--right asymmetry in deuteron carbon scattering as a function of the angle error in milliradians (down triangles) or the position error (up triangles)\cite{Brantjes:2012zz}: V$\pm$ and T$\pm$ denote vector and tensor polarizations; UNP denotes an
unpolarized beam. Various features of the measurements are marked, including some interpretations through model parameters (usually logarithmic derivatives) from the fits through the data. (Figure taken from Ref.~\cite{Brantjes:2012zz}, reused with permission from Elsevier.)}
\label{pol15}
\end{figure}

In 2008 and 2009, the EDDA detector system, then used as a polarimeter for COSY, was calibrated for geometric and rate error sensitivity\cite{Brantjes:2012zz}. The beam was scanned horizontally in both angle and position. The effects of rate were also present in the data, as the rate changed as a function of the time in the store. An example of part of the geometric data for the left--right asymmetry is shown in~\Fref{pol15}. Measurements of a number of polarization observables were made with five different polarization states (V$+$, V$-$, T$+$, T$-$, and unpolarized (UNP) beam). Angular deviations (down triangles, in milliradians) and position variations (up triangles, in millimetres) were recorded. As depicted in the figure, the effects are large and clear. In the same dataset, changes caused by the data acquisition rate were also recorded. A model of all of the error effects was constructed in terms of the logarithmic derivatives of the cross-section and analysing power as geometric parameters; these parameters, as well as other factors, including rate changes, were used to reproduce the data, as shown. The free geometric variable in the model was taken to be the angle deviation from a straight beam. The model was sufficiently robust that it could predict effects for any of the measured polarization observables within the measurement errors.


In the geometric case, \Fref{pol15} shows different effects for angle and position changes. These could be reconciled, provided an effective distance to the detector was assumed, and this became one of the fitting parameters. If this substitution works well,  it can become the basis for reducing the geometry effects to a single parameter. The quality of this result is indicated in \Fref{pol14}. Measurements of the left--right asymmetry correction are overlaid for both angle and position, and are shown to lie along a similar slope.

\begin{figure} [hbt!]
\centering
\includegraphics[width=250pt]{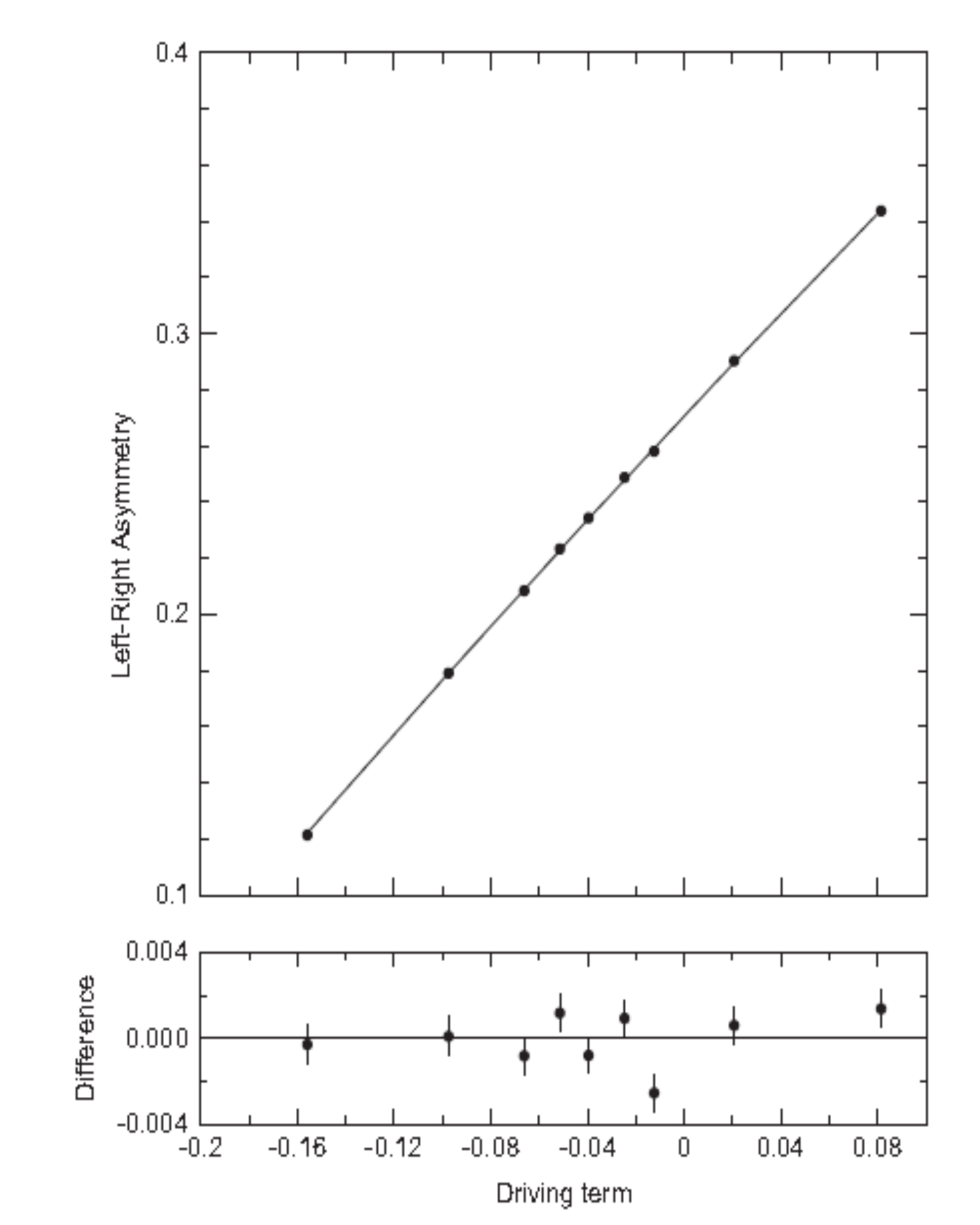}
\caption{Changes to  left--right asymmetry as a function of an index parameter (defined in the text), which is tied to either position or angle variations of the beam on target. The overlap of these two datasets into one universal line indicates that a single index parameter is capable of correcting both types of error. (Figure taken from Ref.\cite{Brantjes:2012zz}, reused with permission from Elsevier.)}
\label{pol14}
\end{figure}

These relationships are evaluated in terms of an index parameter $\phi$, defined as
\begin{equation}
\phi =\frac{s-1}{s+1} \, ,  \qquad \text{ where } \qquad s^2=\frac{\sigma_{\text{L}_+} \cdot \sigma_{\text{L}_-}}
{\sigma_{\text{R}_+} \cdot \sigma_{\text{R}_-}} \,.
\end{equation}
This quantity is available experimentally in real time. Thus, \textit{independently} of the cross ratio (see Eq.~\ref{eq:cross-ratio-definition}), or any other polarization observable, a correction may be applied. The model is used to calculate the correction, such as a change in position along the sloped line shown in \Fref{pol14}. This correction can be applied to any polarization observable.

A similar term, the sum of the four counting rates,
\begin{equation}
 W(t) = \frac{\dd \text{L}}{\dd t} + \frac{\dd \text{R}}{\dd t} + \frac{\dd \text{U}}{\dd t} + \frac{\dd \text{D}}{\dd t}\,,
\end{equation}
tracks the instantaneous flow of data  into the polarimeter, which is largely independent of polarization. As reported in Ref.\cite{Brantjes:2012zz}, the changes in various polarization observables were recorded as a function of the size of $W(t)$ and used as the basis for making a correction to the data. An example is shown in \Fref{pol17}. The measurements of a beam with a constant polarization are given by the red data points. The time dependence is an error that depends on the data rate, as it creates pile-up effects in the detectors. Correction of that error yields the blue data points. But these data are still not fully correct, because of a geometric misalignment. The final correction leads to the black data, which are constant in time to better than one part in $\num{e5}$, which is statistically limited. If the calibrations are known in advance, such corrections may be made in real time during the experiment, a feature that will be essential in maintaining the polarization pointing along the velocity through feedback.

\begin{figure} [hbt!]
\centering
\includegraphics[width=250pt]{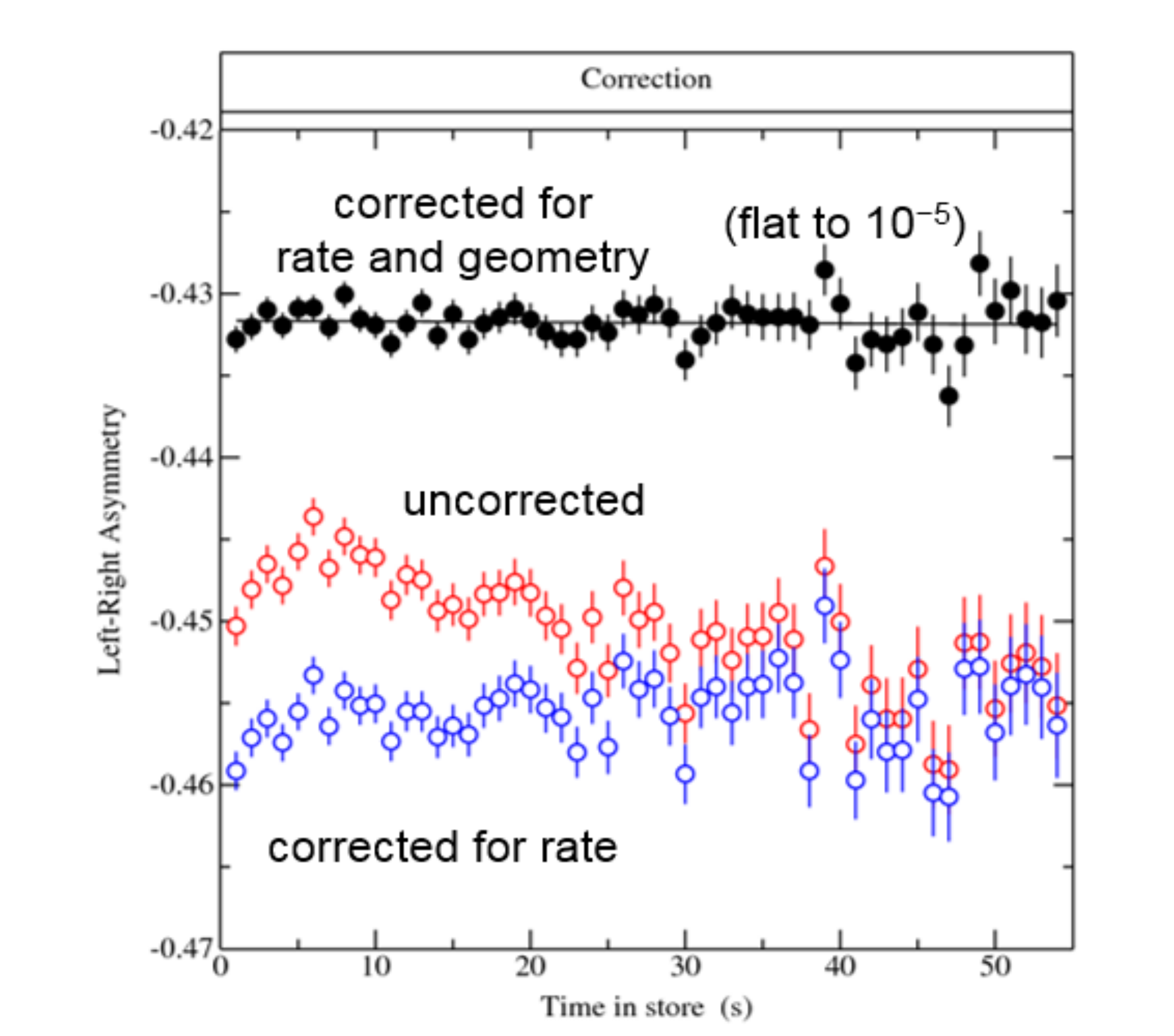}
\caption{Three versions of a set of left--right asymmetry measurements as a function of time during a store in COSY, during which the rate first rose slightly and then started to fall. Red points are uncorrected;  blue points are corrected for rate effects;  black points are corrected for both rate and geometry effects. A line through the black points is indicated and the error in its slope (consistent with zero) is shown. (Figure taken from Ref.~\cite{Brantjes:2012zz}, reused with permission from Elsevier.)}
\label{pol17}
\end{figure}

The example shown in \Fref{pol15} of calibration data for five polarization states is linear in the case of the left--right asymmetry. Higher-order effects appear as curvatures of various ranks, which may also be parametrized using powers of the logarithmic derivatives. The combination of all of these properties of the model-based calibration and driver-term corrections makes it possible to reliably extract a signal as small as $\delta\epsilon = \num{e-6}$  from a series of time-dependent asymmetry measurements.

\section{Polarimeter rotations, energy loss, and deuteron tensor polarization effects}

The polarimeter must be set up so that the coupling between horizontal  and vertical asymmetries, as established by the location of the ring plane, is as small as possible. Such a correlation is measured by stepping the polarization direction (registered as a phase in the feedback circuit) around the in-plane circle and comparing vertical and horizontal asymmetries, as shown in \Fref{pol18} for $t = 0\Us$. Imagined data points for the correlation with an incomplete cancellation are indicated by plus signs. As time proceeds through the beam store, any EDM effect will cause the left--right asymmetry ($\epsilon_{\mathrm{L}-\mathrm{R}}$) to rise as a function of time, taking the correlation with it. This allows, in principle, for a separation of these effects, but it must be remembered that, for a single store, the statistics on each of the data points will be over two orders of magnitude larger than any EDM effect at the expected level of sensitivity. The correlation cancellation is likely to be incomplete because of the long running time needed to establish the correlation.

\begin{figure} [hbt!]
\centering
\includegraphics[width=200pt]{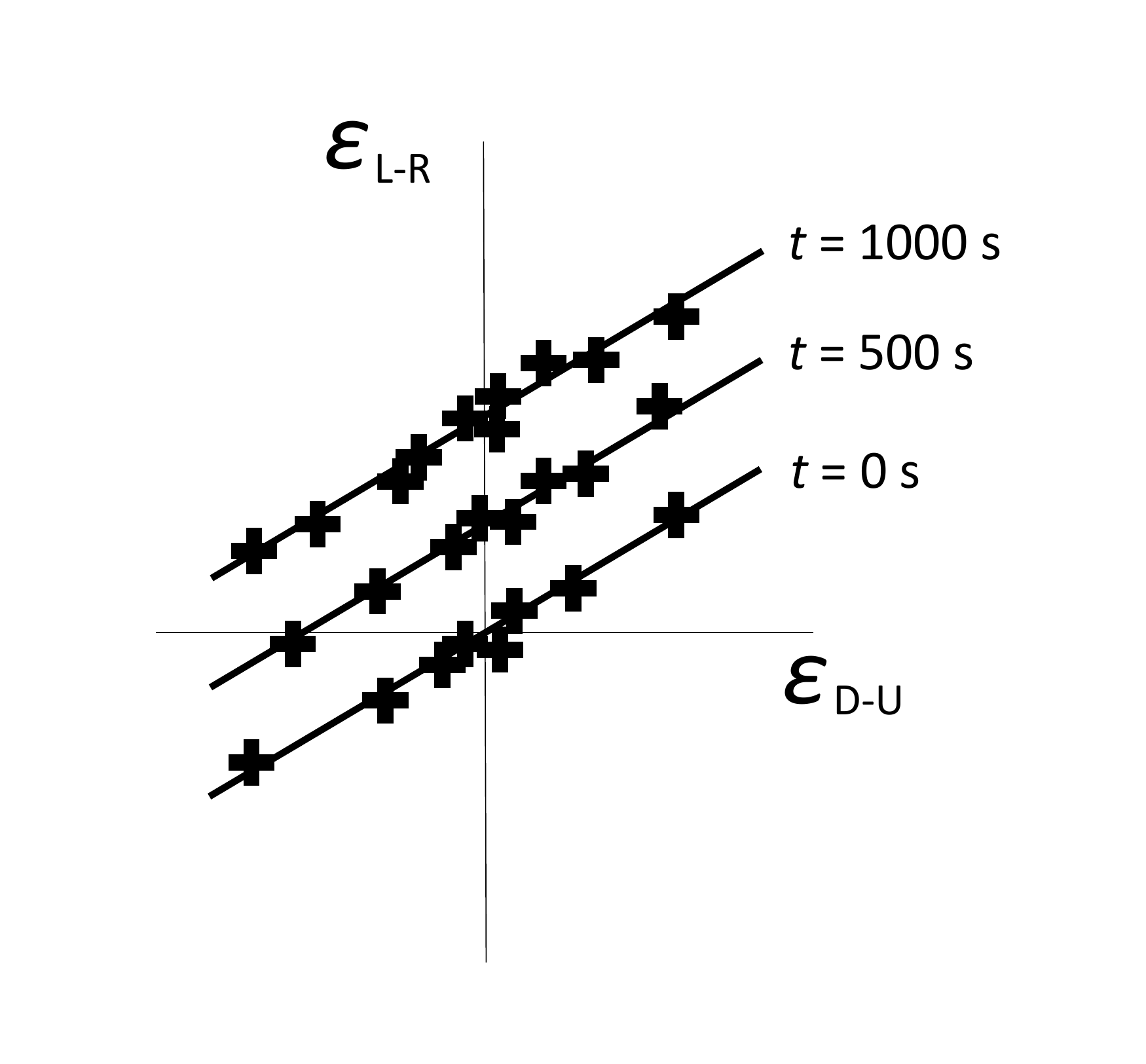}
\caption{Mock data showing a correlation between vertical and horizontal asymmetries as variations in the vertical asymmetry are used (and corrected) in order to maintain frozen spin through feedback. The figure also suggests that this correlation plot rises to include a larger horizontal asymmetry (perhaps from an EDM signal) as time progresses during the store.}
\label{pol18}
\end{figure}

For an off-centre block carbon target, the simple down--up raw asymmetry may be of the order of~0.2. Since left--right sensitivities as small as $\num{e-6}$ may be involved in the EDM signal itself, a cancellation of these polarimeter rotation effects to a similar degree must also be arranged, either by a mechanical adjustment in the polarimeter detector acceptance or by cancellation through terms in the systematic error calibration described earlier. The risk, for example, is that energy loss due to collisions with background gas (or the polarimeter target) will populate lower energy particle orbits, causing, on average, a drift away from the frozen-spin condition that increases with time. While continuous polarization measurements are used to correct the frozen-spin condition, errors that tend towards one direction may produce a bias in the result, and would appear in the figure as data points that
are no longer symmetrically distributed about the vertical axis.

A correlation plot similar to the figure has also been suggested for the elimination of other effects, such as the slow vertical polarization growth associated with a residual sideways magnetic field or vertical electric field in the EDM ring. In that case, the horizontal axis of the plot will be some measurement of the sideways field, such as that indicated by the SQUID readout proposed for the proton ring (see~Appendix~\ref{Chap:MagneticFields}). Because all of these effects operate at the same level as or higher than any EDM signal, a comprehensive analysis must be made at the end for all of the data. Magnetic field errors are apt to appear as changes from store to store, while polarimeter rotation effects always appear within a store and are not as likely to change over time.

A similar effect arises in addition for the deuteron beam, since in most polarized ion sources, it is impossible to eliminate completely a tensor polarization component in the beam. One is often present at the percentage level\footnote{Indeed, there may be arguments for having a large tensor polarization, since it may be monitored as a continuous measure of the beam polarization through the $T_{20}$ analysing power without having to periodically rotate the polarization axis of the beam into the sideways direction.}. A rotation of the tensor polarization axis to either the left or right will directly generate a left--right asymmetry through the $T_{21}$ analysing power. This analysing power is not directly driven by the spin-orbit force, so its values for most forward angle polarimeter geometries are typically less than a few percent.

Systematic errors due to polarimeter rotation and $T_{21}$ may be detected in a running experiment by looking for a correlation between the down--up asymmetry driving the feedback circuit to hold the frozen-spin condition and the left--right asymmetry that, in principle, carries the EDM signal. In the deuteron case, left--right asymmetry sensitivity through either an effective polarimeter rotation or $T_{21}$ sensitivity may be separated in a calibration by using a series of in-plane polarization directions and  looking at the nature of the correlation with rotation angles up to ${\pi}
/ {2}$. A linear relationship indicates a polarimeter rotation effect, while a dependence that follows a sine of twice the in-plane rotation angle indicates a $T_{21}$ sensitivity. These two effects will, in general, have different sizes or slopes for small rotation angles.

\section{Time-reversed experiment}

The EDM violates the symmetries of parity conservation and time reversal. In the case of time reversal, the direction of rotation of the beam around the ring would be changed and all polarizations and magnetic fields would have the opposite sign. Since this is a physically realizable experimental condition, it has been suggested that it be a part of the protocol for the EDM search. In the case of the proton with a positive anomalous magnetic moment, the condition of frozen spin may be realized with only an electric field. This field remains the same under time reversal; thus, it should be possible to operate the storage ring with both CW and CCW beams at the same time. This offers the opportunity to compare beam locations, profiles, intensities, and polarizations in order to verify that they are, in fact, identical. A second polarimeter would need to be installed in the ring in order to capture measurements of the reversed-direction polarization. Some economies of construction and the use of only one extraction mechanism favour a design in which the two polarimeter detector schemes are located on either side of a single block target. Measurements made to large scattering angles of elastic proton scattering from carbon~\cite{Meyer:1988rj} show a reduction of eight orders of magnitude in the cross-section between the forward scattering angles used for polarization measurements and similar backward scattering angles (see \Fref{Eds-fig_1_3}). This should be sufficient to suppress any interference, with small changes being measured through the forward scattering asymmetry.

Several systematic error effects that give rise to an EDM-like signal (changing vertical component of the polarization over time) are time-reversal conserving. This would appear as a rising signal for both CW and CCW beams, while the EDM signal would rise in one instance and fall (become negative) in the other. Therefore, any unsuppressed time-reversal conserving systematic error may be cancelled by subtracting the CW and CCW measurements.

Since the measurement (for small angles of vertical rotation of the polarization) is one of a continuously rising effect, let us denote scattering to the left as
\begin{equation}
\sigma_\text{POL}=\sigma_\text{UNP} \left[ 1 + (S+E) \, pA_y \right]\,,
\end{equation}
where $S$ is the rate of rise due to remaining systematic effects and $E$ is the rate of rise due to the EDM. The simple left--right asymmetries for CW and CCW become
\begin{equation}
\epsilon^\text{CW}=\frac{\text{L} - \text{R}}{\text{L} + \text{R}} = (S+E) \, pA_y\,, \qquad  \text{ and} \qquad
\epsilon^\text{CCW}= (S-E) \,  pA_y\,.
\end{equation}
Therefore, the difference between the two asymmetries yields
\begin{equation}
\epsilon^\text{CW} - \epsilon^\text{CCW}  = 2E \, pA_y\,,
\end{equation}
from which the EDM can be extracted. This subtraction works only to the extent that $pA_y$ values for both CW and CCW are well calibrated. If the average polarization and analysing power values for the CW and CCW beams
 are given by $p =  (p^\text{CW} + p^\text{CCW} ) / 2$
 and $A_y =  (A_y^\text{CW} + A_y^\text{CCW} ) / 2$, and the differences between \textit{calibrated} and \textit{actual} values of polarization and analysing power are given by  $\delta p =   (p^\text{CW} - p^\text{CCW}  ) / 2$ and $\delta A_y =  (A_y^\text{CW} - A_y^\text{CCW}  )
/ 2$, when expanded to first order, one obtains
\begin{equation}
\frac{1}{2}\ \left( \epsilon^\text{CW} - \epsilon^\text{CCW}\right) =
Ep \,A_y + Sp\, A_y \left( \frac{\delta p}{p} + \frac{\delta A_y}{A_y} \right) \,.
\end{equation}
This suggests that the systematic contribution to the EDM signal can be suppressed only to the extent that the unknown fractional errors in the CW and CCW polarization $ ( {\delta p} / {p} )$ and analysing power $ ( {\delta A_y}
/ {A_y}
)$ are small enough to render the systematic error negligible compared with the EDM signal.

In the case of the beam polarization, this introduces the requirement that the CW and CCW beams in the experiment be filled using the same polarization state from the ion source. Likewise, care must be taken in the construction of the polarimeters and the setting up of their detector readout to ensure that the effective analysing powers are also as identical as possible. This puts a premium on other efforts to reduce the systematic error contribution initially.

\begin{flushleft}

\end{flushleft}
\end{cbunit}

\begin{cbunit}



\renewcommand{\DOm}{\Delta\Omega}
\renewcommand{\DOmvec}{\Delta\vec\Omega}

\csname @openrighttrue\endcsname 
\chapter{Sensitivity and systematic effects} \label{Chap:SensSys}

\section{Statistical sensitivity}
\label{app:staterr}

The spin motion, relative to the momentum vector, is governed by the subtracted  Thomas--BMT equation
(see Eqs.\,(\ref{emeq1})--(\ref{emeq3}) and  Chapter~\ref{Chap:ExpMethod}  for  further definitions) in
terms of the angular velocities of the magnetic dipole moment (MDM), $\vec\Omega_{\rm MDM}$,
electric dipole moment (EDM), $\vec\Omega_{\rm EDM}$, and cyclic rotation of the
particle, $\vec \Omega_{\rm cycl}$:
\begin{align}
\frac{{\rm d} \vec{S}}{{\rm d} t} &= \left( \vec \Omega_{\rm MDM} - \vec \Omega_{\rm cycl} +\vec  \Omega_{\rm EDM} \right)\times \vec S \,, \\
\vec\Omega_{\rm MDM} -\vec \Omega_{\rm cycl} &=  \frac{-q}{m}  \left[
  G {\vec B} -  \left( G  -\frac{1}{\gamma^2-1}  \right) \frac{\vec v \times \vec E}{c^2}
  \right]\,,\\
\vec\Omega_{\rm EDM} &= 
  \frac{-\eta q}{2 m c}  \left[   \vec E  +  {\vec v \times \vec B}  \right]\,,
\end{align}
with
$\vec d = {\eta}( {q } / {2 m c} )\vec S$ the EDM vector
and
$\vec \mu = 2(G+1) ( {q } / {2m}) \vec S$ the MDM vector.
For this discussion, $\vec B$ and $\vec E$ denote a vertical magnetic
and a radial electric field, respectively.

For a purely electric ring, the angular precession frequency due to the EDM
is given by
\begin{equation}
\vec \Omega_{\rm EDM} = -\frac{\eta q}{2 m c}  \vec E \, ,
\end{equation}
such that
\begin{equation}
\Omega_{\rm EDM} = \frac{d\, E}{S \hbar}  \label{Omega-vs-EDM}\,.
\end{equation}
Thus, for the case of the proton (\ie spin quantum number $S{=}1/2$), and just  using the value 8 MV/m
of \Tref{tab:para} for the electric field $E$,
one finds \begin{equation}
\Omega_{\rm EDM} = 2.4 \times \SI{e-9}{s^{-1}} \, ,
\end{equation}
if $d = \SI{e-29}{\text{$e$}.cm}$ is assumed as  the EDM.

\begin{table}
\begin{center}
\caption{Parameters used to evaluate the statistical error of the all-electric ring\label{tab:para}}
\resizebox{\columnwidth}{!}{%
\begin{tabular}{l l l}
\hline \hline
$N$  & $4 \times 10^{10} = 2 \times 10^{10} (\text{CW}) + 2 \times 10^{10} (\text{CCW})$  & Particles per fill\\
$E$ & 8 MV/m & Electric field\\
$R_{\text{frac}}$ & 0.65& Fraction of  the ring equipped with $E$ field\\
$T_{\text{mcyc}}$ & 1000\Us &  Period of one measurement cycle  \\
$P$  & 0.8 & Polarization\\
$A$  & 0.6 & Analysing power \\
$f$  & 0.005  & Detection efficiency \\
$S$   & 1/2 (proton), 1 (deuteron)  & Spin quantum number\\
$C_{\rm s}$ & 1 (proton), 3/2 (deuteron) & Weight factor of vertical vector polarization\\
\hline \hline
\end{tabular}
}
\end{center}
\end{table}

\subsection{Three scenarios for the  statistical error evaluation  of EDMs}\label{app:sigd}

\noindent To evaluate the statistical error in the EDM, we discuss three different scenarios.
\begin{enumerate}
\item[A.] There is only a precession due to the EDM, \ie one observes only the initial linear
  rise of the polarization vector because $\Omega_{\rm EDM} \cdot T_{\text{mcyc}} \ll 1$.
  The polarization is continuously measured, as indicated by the points in \Fref{fig:Tom}.
  
\begin{figure} [hb!]
\centering
\includegraphics[width=0.75\textwidth]{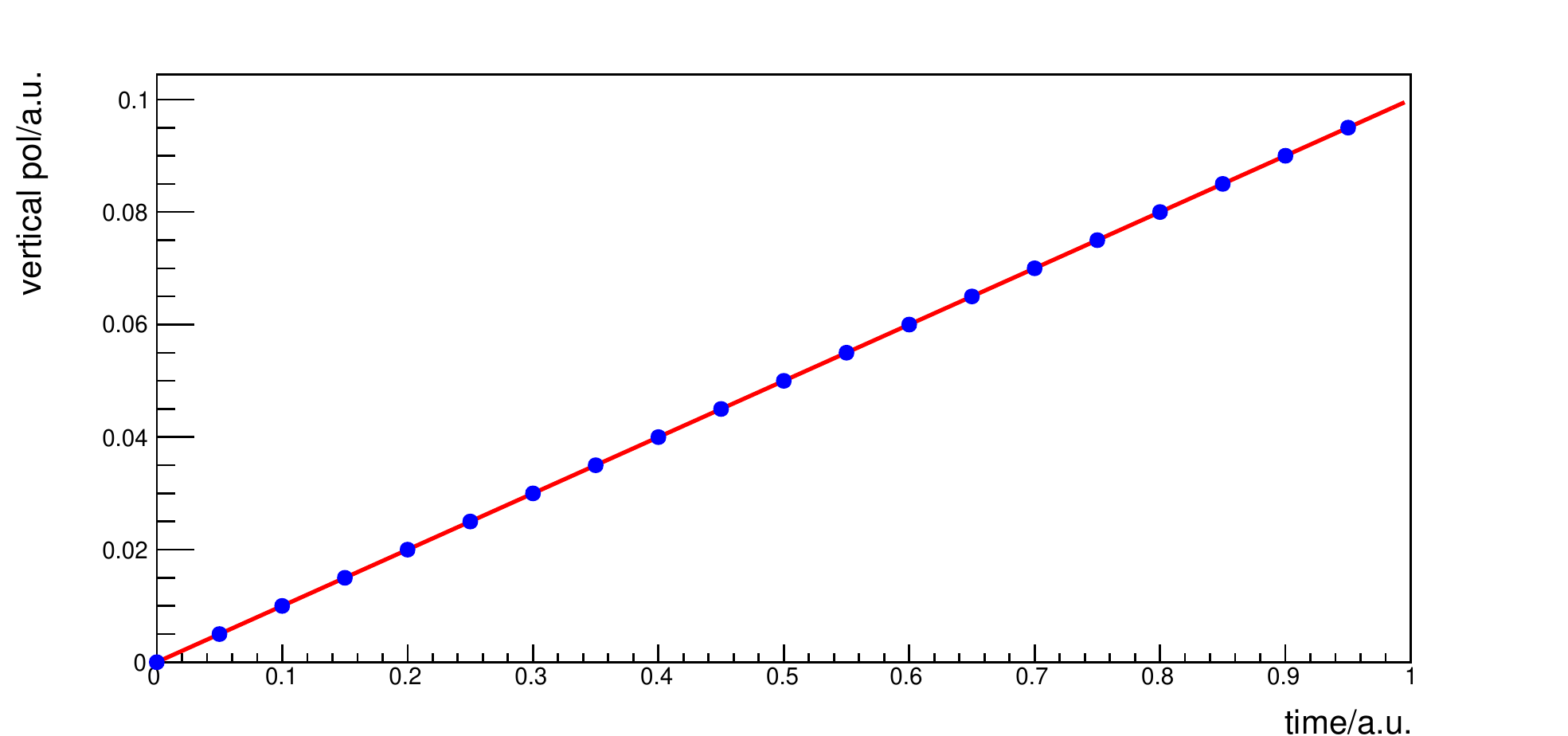}
\caption{Statistical error evaluation of EDMs: scenario A.}
\label{fig:Tom}
\end{figure}

\item [B.] The situation is the same as for scenario A, but half of the beam is extracted at $t=0$, while the other half
  is extracted at $t=T_{\text{mcyc}}$ (\Fref{fig:Dick}).

\begin{figure} [hb!]
\centering
\includegraphics[width=0.75\textwidth]{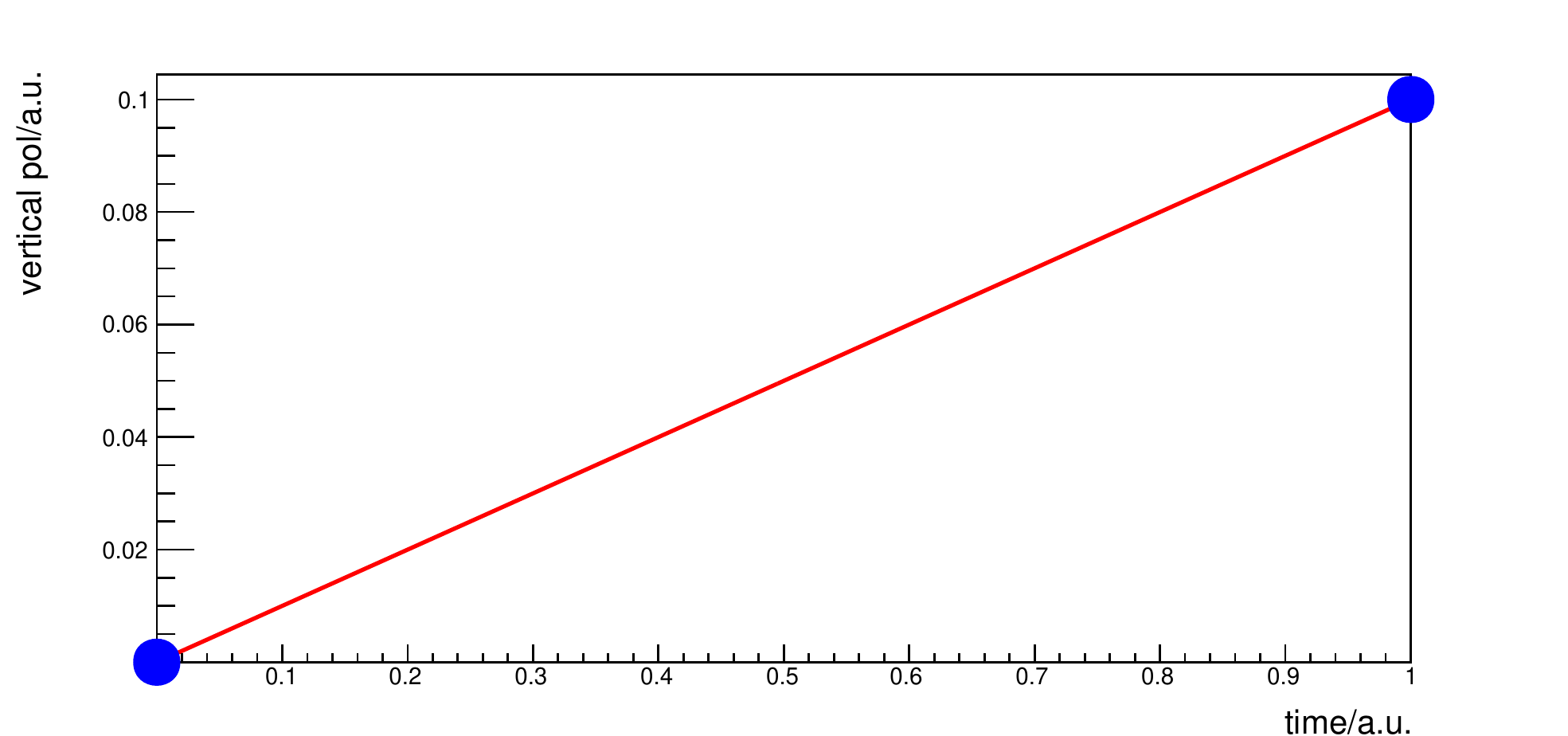}
\caption{Statistical error evaluation of EDMs: scenario B.}
\label{fig:Dick}
\end{figure}

\item[C.] In this scenario, the precession is dominated by systematic effects. One thus observes many oscillations
  during the measurement-cycle  period $T_{\text{mcyc}}$ (\Fref{fig:Harry}).

\begin{figure}
\centering
\includegraphics[width=0.75\textwidth]{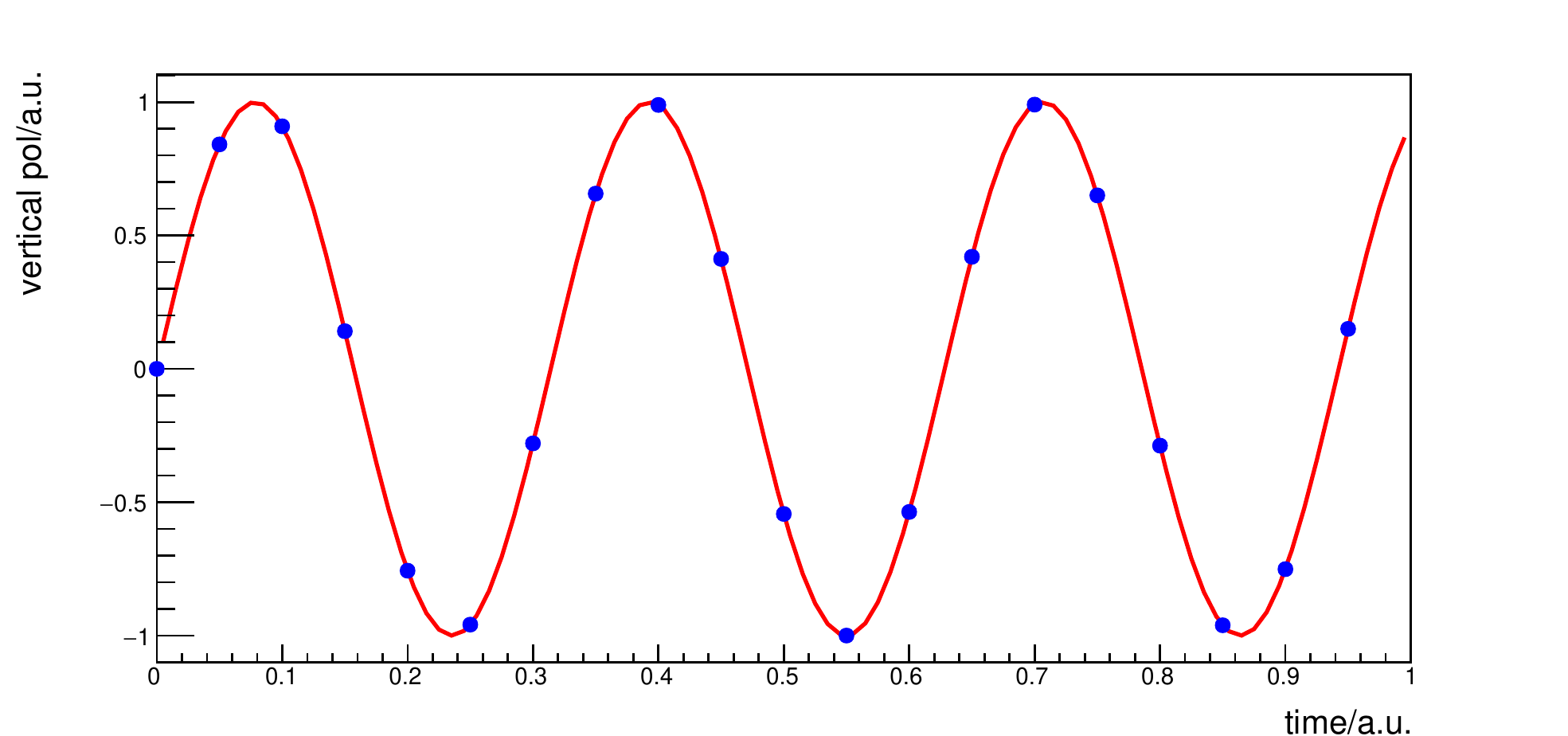}
\caption{Statistical error evaluation of EDMs: scenario C.}
\label{fig:Harry}
\end{figure}
\end{enumerate}

In all three cases, it is assumed that the EDM is extracted from the differences and sums of the polarizations of the CW and CCW measurements and that there is no decoherence during $T_{\text{mcyc}}$.

\begin{enumerate}
\item[A.]
Assuming a polarization vector initially along the momentum vector, we get
\begin{equation} \label{dotPperp}
\dot{P_\perp} = \Omega_{\rm EDM} P =  \frac{ d E}{S \hbar}  \, P \, ,
\end{equation}
and, applying \Eref{Omega-vs-EDM},
\begin{equation}\label{eq:d}
  d = \frac{S \hbar \dot{P_\perp}}{ E P}  \, .
\end{equation}

In general, the variance on the slope parameter $s$
of a straight line is
\[
  V(s) = \frac{\sigma^2}{N_{\text{points}} V(t)}\,,
\]
where $\sigma$ is the error in each individual point,
$N_{\text{points}}$  the number of points entering the fit, and $V(t)$ 
the variance of the points along the horizontal axis (\ie  time in the EDM case).
For evenly distributed values in time, one has
\[
V(t) = T^2_{\text{mcyc}}/12\,.
\]
The slope in the EDM measurement is just $\dot P_\perp$.
The squared error in {\em one} polarization measurement, determined from the {\em azimuthal} distribution of events
(\ie $({1} / {2\pi})\int_0^{2\pi} {\rm d}\Phi\, \cos^2\!\Phi = 1/2$,  cf.
Appendix~\ref{app:stat-err-freq}),  is
\begin{equation}
\sigma_{P_\perp(t)}^2 = \frac{1}{\frac{1}{2}( N f/N_{\text{points}})  C_{\rm s}^2 A^2} = \frac{2}{(Nf/N_{\text{points}}) C_{\rm s}^2 A^2} \, .
\label{sigmaP2}
\end{equation}
The variance on the slope $\dot P_\perp$ is thus
\begin{equation}
\sigma^2_{\dot{P_\perp}} = \,\frac{2}{(Nf/N_{\text{points}})  C_{\rm s}^2  A^2} \cdot \frac{12}{N_{\text{points}} T^2_{\text{mcyc}}}
=\frac{{24}}{(Nf) (C_{\rm s}A T_{\text{mcyc}})^2}\,.
 \label{Pdot-cont}
\end{equation}

From \Eref{eq:d}, we find
\[
  \sigma_{\rm EDM} = \frac{S \hbar}{E P} \sigma_{\dot{P_\perp}} \, ,
\]
which results in
\begin{eqnarray}
\sigma_{\rm EDM} 
&=& 
\sqrt{24} \, \frac{S \hbar}{ \sqrt{N f} C_{\rm s} A P E T_{\text{mcyc}}}
\label{var-EDM-cont}
\end{eqnarray}
as the statistical uncertainty for the EDM measurement $d$.

\item[B.]

Here, we have  $V(t) = T^2_{\text{mcyc}}/4$, which results in
%
  \begin{eqnarray}
  \sigma_{{\text{EDM~}}} 
  &=& 
  \sqrt{8} \frac{ S \hbar}{\sqrt{Nf} C_{\rm s} A P E T_{\text{mcyc}}}
\end{eqnarray}
instead of \Eref{var-EDM-cont} as the statistical uncertainty.

\item[C.]

According to  Appendix~\ref{app:stat-err-freq},   
the error in the frequency is given by
\[
  \sigma_{\Omega}^2 = \frac{24}{\frac{1}{2}(Nf) (C_{\rm s}A P T_{\text{mcyc}})^2}\,.
\]
Using  the relation
\[
 \Omega_{\rm EDM} = \frac{ d E}{S \hbar} \, ,
 \]
 one finds
 \begin{equation}
 \sigma_{\text{EDM}} = \sqrt{48} \, \frac{ S\hbar}{\sqrt{Nf} C_{\rm s}APET_{\text{mcyc}}}
 \end{equation}
as the statistical error in the EDM, $d$.
\end{enumerate}

\subsection{Statistical error evaluation of the EDM in the precursor experiment}

For the precursor experiment,
the build-up of the vertical polarization after $n$ passages of the beam (with $T_{\text{WF}}$ the time of one passage) through the RF Wien filter
of the precursor ring
is given as~\cite{PhysRevSTAB.16.114001,Saleev:2017ecu,Rathmann:2019lwi}
\begin{equation}
   P_\perp  =  n \, \Delta P_\perp = n \, T_{\text{WF}}\, \left(\hat y \cdot \left \langle\vec \Omega_{\text{EDM}}^{\text{prec}}(t) \times \vec P(t)\right\rangle_{\text{turn}}\right) \,,
                  \label{PperpKicks}
   \end{equation}
which is the integrated  version of  $\dot P_\perp  = \Omega_{\text{EDM}} P$,  cf. \Eref{dotPperp}, for the precursor
scenario.
Thus, $n$ is the number of turns of the beam around the precursor ring during the measurement period $T_{\text{mcyc}}$;
$\Delta P_\perp$ stands for a stroboscopic  kick to the polarization build-up
after the  beam passage through the RF Wien filter at the end of  each turn around the precursor ring, where
the vector product of  the polarization, sinusoidally rotating with the frequency of the spin tune,
and the externally driven, sinusoidal RF-dependence of  the Wien filter must  be averaged over one turn (or
over one time-span $T_{\text{WF}}$, since the RF field of the Wien filter vanishes outside this region).

Assuming that the angular frequency for the spin precession of the Wien filter (with $\vec B_{\text{WF}}{\parallel} \hat y$, \ie positioned
in the vertical `EDM position') is in resonance with the frequency
of the precursor ring  in the horizontal plane,  \ie
\begin{equation}
 (\Omega_{\text{MDM}}{-}\Omega_{\text{mcyc}} )^{\text{WF}} =  (\Omega_{\text{MDM}}{-}\Omega_{\text{mcyc}} )^{\text{prec}}
 + K \Omega_{\text{mcyc}}, \qquad \text{where} \qquad K \in \mathbb{Z},
\end{equation}
 cf. \Eref{eq:wien-filter-frequency}, the time-averaging over the sinusoidal terms in \Eref{PperpKicks}
gives \cite{PhysRevSTAB.16.114001}
\begin{equation}
\hat y \cdot \left \langle\vec \Omega_{\text{EDM}}^{\text{prec}}(t) \times \vec P(t)\right\rangle_{\text{turn}}
              =   {\textstyle\frac{1}{2} }\Omega_{\text{EDM}}^{\text{prec}} P \label{Omega-P-averaged} \, ,
\end{equation}
in terms of the pertinent amplitudes.
Moreover,     after each turn around the precursor ring, 
the  amplitude of angular frequency of the spin precession
in the vertical direction  can be expressed as
\begin{equation}
   \Omega_{\text{EDM}}^{\text{prec}} =   \frac{\Omega_{\text{EDM}}^{\text{prec}}}{(\Omega_{\text{MDM}}{-}\Omega_{\text{mcyc}} )^{\text{prec}}  } (\Omega_{\text{MDM}}{-}\Omega_{\text{mcyc}} )^{\text{WF}}
   = \frac{\beta \eta}{2 G} (\Omega_{\text{MDM}}{-}\Omega_{\text{mcyc}} )^{\text{WF}}
   \label{OmegaEDMprec}
\end{equation}
(here for $K=0$  as  harmonic multiple).

In the last relation, we used the fact that the ratio of the EDM to  the (subtracted) MDM angular frequency in the precursor ring  is
$\tan \xi$, with
$\xi = \arctan(\beta \eta/(2G))$  the amplitude of the vertical component of the polarization---see
 \Sref{Precursor-Principle-of-Measurement} and
Ref.~\cite{Rathmann:2019lwi}. Thus,  $\tan\xi$,  which is given
by the ratio of  the right-hand sides of the subtracted Thomas-BMT equations (\Eref{emeq3} and \Eref{emeq2}) for the case $\vec B= \vec B_y$ and $\vec E \equiv 0$, determines the  tilt of the axis around which the spin in the precursor ring is precessing.

Conversely,
for the Wien filter condition,
\[
   \vec E_{\text{WF}} + c\vec\beta \times \vec B_{\text{WF}} = 0\,,
\]
which is consistent with
$\vec \beta \times \vec E_{\text{WF}} = c \beta^2 \vec B_{\text{WF}}$ under the assumption $\vec \beta \cdot \vec B_{\text{WF}} = 0$,
the  very same subtracted Thomas-BMT equations predict 
\begin{align}
\vec\Omega_{\text{EDM}}^{\text{WF}} &= 0\,,  \\
(\vec\Omega_{\text{MDM}}{-}\vec\Omega_{\text{mcyc}} )^{\text{WF}}  &= -
\frac{e}{m}  \,\left( G \vec B_{\text{WF}}  -  \left(G - \frac{1}{\gamma^2 -1} \right) \frac{\vec \beta \times \vec E_{\text{WF}}}{c}  \right)  \nonumber \\
&= - \frac{e}{m}  \, \frac{G+1}{\gamma^2} \vec B_{\text{WF}} =
-\frac{e}{m c} \, \frac{G+1  }{\beta^2  \gamma^2} \,\vec \beta \times \vec E_{\text{WF}}  \,. \label{OmegaMDMWF}
\end{align}

Inserting \Eref{OmegaMDMWF}   in \Eref{OmegaEDMprec} and then in \Eref{PperpKicks}, as modified
by \Eref{Omega-P-averaged},
and using  $T_{\text{WF}} = L_{\text{WF}}/(c \beta)$, with $L_{\text{WF}}$ the spatial length of the Wien filter, we get
\begin{equation}\label{eq:pperp}
 P_{\perp} = -n \, {\frac{1}{2}} \,\frac{\eta \beta}{2 G} \, \frac{e}{m c^2} \,
 \left(\frac{G+1}{\gamma^2 \beta^2} \right) \, E_{\text{WF}} L_{\text{WF}} \, P \,
 = - \frac{T_{\text{mcyc}}}{U} \left(  \frac{\eta e }{2 m c } \right)\,
 \left(\frac{G+1}{2G\gamma^2 } \right) \, E_{\text{WF}} L_{\text{WF}} \, P \, .
\end{equation}
In the last relation, the number of turns $n$
is replaced by the time of the measurement cycle $T_{\text{mcyc}}$
multiplied by the revolution frequency $f_\mathrm{rev}$, $n=T_{\text{mcyc}} f_\mathrm{rev}$, which can in turn  be expressed
as $f_\mathrm{rev}=\beta c/U$, where $U$ is the circumference of the ring. So in total
the replacement $n =   (T_{\text{mcyc}}/U) \beta c$ is inserted.

Applying $d= S\hbar\,  {\eta e}/{(2 mc)}$,  one finds the following expression for the statistical uncertainty of the  EDM
measurement performed in the  precursor ring, according to  scenario A of Section~\ref{app:sigd}:
\begin{equation}\label{eq:lsigd}
 \sigma_{\rm EDM}^{\text{prec}} =\left|  \frac{ 2G \gamma^2 }{G+1}  \right| \times \left|\frac{1}{E_{\text{WF}}L_{\text{WF}} /U} \right|   \,
 \times  \sqrt{24}\frac{ S \hbar}{\sqrt{N f} C_{\rm s} A P T_{\text{mcyc}}}\,.
\end{equation}

Using the parameters of \Tref{table:Fred}  for the case of  horizontally polarized deuterons stored in the precursor ring, one finally arrives at
\[
 \sigma_{\rm EDM}(1 \mbox{ fill}) = 5.7 \times  \SI{e-21}{\text{$e$}.cm} \qquad  \left(8.7 \times \SI{e-21}{\text{$e$}.cm} \right)
\]
per fill of $1000\Us$, assuming no decoherence, or,  in parentheses,  the case of an exponentially decreasing polarization, as discussed in the following subsection.

\begin{table} [h]
\caption{Parameters for the case of  horizontally polarized deuterons stored in the precursor ring}
\label{table:Fred}
\centering
 \begin{tabular}{lll}
\hline \hline
$G$   &  $-0.14$   &Magnetic anomaly of the deuterons \\
$S$  : $C_{\rm s}$  & 1 :  (3/2)& Deuteron spin : weight factor of vertical vector polarization  \\
$\gamma$  & 1.13  & Deuteron beam with $p=1\UGeVc$ \\
$E_{\text{WF}}$ & \SI{2.7}{kV/m} & Averaged $E$ field of Wien filter\\
$L_{\text{WF}}$ & \SI{1}{m} & Length of Wien filter\\
$U$  & \SI{183}{m}    & Circumference of COSY \\
$R_{\text{frac}} $ & $5.4 \times 10^{-3}$&$L_{\text{WF}}/U$ (fraction of the ring equipped with  $E$ field) \\
$T_{\text{mcyc}}$ & \SI{1000}{s} &  Period of one measurement cycle\\
$P$  & 0.8 & Polarization\\
$A$  & 0.6 & Analysing power \\
$N$  & $10^9$  & Particles per fill \\
$f$  & 0.005  & Detection efficiency \\
\hline \hline
\end{tabular}
\end{table}

 \subsection{General form of the statistical uncertainty in EDM measurements in storage rings} \label{Chap:statistics-general}
 The final expression of the statistical error in the EDM is given by
 \begin{equation}
 \sigma_{\rm EDM} =   C_{\text{ring}} \,C_{\text{mod}}  \, \frac{S \hbar}{\sqrt{Nf} \,C_{\rm s} AP R_{\text{frac}} ET_{\text{mcyc}}}\,,
 \label{sigma-EDM-final}
 \end{equation}
 with the coefficient
 \[
 C_{\text{ring}} =
   \begin{cases}
       \left| \frac{2G \gamma^2}{G+1}\right| & \mbox{precursor experiment}, \\
       1 & \mbox{prototype and  final all-electric} 
   \end{cases}
  \]
for the choice of the ring.
The added factor 
$R_{\text{frac}}$, multiplying  $E$,
is the
fraction of the ring equipped with $E$ fields (or Wien filters,
in the case of the precursor experiment), while
 the coefficient  $C_{\text{mod}}$ depends on the way in which the polarization is measured (see scenarios A, B, and C from before)   and on  further
 modifications from the finite spin coherence time $\tau$. Finally, $C_{\rm s} = 1$  (3/2)  is the weight factor of the vertical vector  polarization of the spin 1/2 (spin 1) particles,  cf. \Sref{Chap:polarisation-observables}.

  In  \Sref{app:sigd}, it was assumed that the polarization is constant over $T_{\text{mcyc}}$, \ie $T_{\text{mcyc}} \ll \tau$.
 However, if $T_{\text{mcyc}} \approx \tau$, the average polarization is smaller by the factor
 \[
  \frac{\int_0^{T_{\text{mcyc}}} \mbox{e}^{-t/\tau} {\rm d}t}{T_{\text{mcyc}}} = 1-\mbox{e}^{-1} \approx 0.63  \, ,
  \]
  assuming an exponential decrease. Thus, $C_{\text{mod}}$, and therefore
  the error  in \Eref{sigma-EDM-final}, are increased accordingly. In \Tref{tab:mod},
the factor $C_{\text{mod}}$ is listed  
  for the  scenarios A, B, and C discussed in \Sref{app:sigd}  for the cases where the polarization is constant during
  $T_{\text{mcyc}} \ll \tau$
   or exponentially decreasing over the period \mbox{$T_{\text{mcyc}} \approx \tau$}.

    \begin{table} [h]
  \centering
  \caption{Factor $C_{\text{mod}}$ for  scenarios A, B, and C if either $T_{\text{mcyc}}\ll \tau$ or $T_{\text{mcyc}}\approx \tau$
  \label{tab:mod}}
      \begin{tabular}{llll}
     \hline \hline
 Scenario  & A  &  B   &  C  \\
     & \includegraphics[width=0.22\textwidth]{Figures/Statistics/pol1.pdf} &
     \includegraphics[width=0.22\textwidth]{Figures/Statistics/pol2.pdf}&
     \includegraphics[width=0.22\textwidth]{Figures/Statistics/pol3.pdf}\\
  \hline 
  & & &\\[-3mm]
  Polarization $P$ constant   &  $\sqrt{24}\approx 4.9$  &  $\sqrt{8}\approx 2.8$  & $\sqrt{48}\approx 6.9$ \\
   ($T_{\text{mcyc}} \ll \tau$ ) &  &  & \\ [3 mm]
    $P=P_0 \mbox{e}^{-t/\tau}, T_{\text{mcyc}} \approx \tau$ &   7.8    & 4.5  &  11\\
   \hline \hline
   \end{tabular}
      \end{table}

For the best and worst  cases in \Tref{tab:mod},  the statistical errors in the EDM  $d$ in units of $\si{\text{$e$}.cm}$, with input from
\Tref{tab:para}, are for the  all-electric proton ($S=1/2$) ring, \ie $C_{\text{ring}} =1=C_{\rm s}$ are given in \Tref{table:Tom}.

   \begin{table} [h]
   \caption{Statistical errors in the EDM  $d$ for the best and worst  cases in \Tref{tab:mod}}
\label{table:Tom}
    \centering
      \begin{tabular}{l ll}
        \hline \hline
  Case    &One cycle & One year ($10^4$ cycles) \\
        \hline
   Best:   scenario B, $T_{\text{mcyc}}\ll \tau$               &      $2.6 \times 10^{-27}$       &    $2.6\times 10^{-29}$   \\
   Worst: scenario C,  $T_{\text{mcyc}}\approx \tau$    &     $1.0\times 10^{-26}$      &    $1.0 \times 10^{-28}$  \\
        \hline \hline
        \end{tabular}
      \end{table}


Finally, results of the statistical uncertainty for the different stages proposed  in Chapter~\ref{Chap:Strategy}
for storage ring EDM measurements are given in \Tref{tab:sys:staterr}, where
it is assumed that the beam is constantly extracted on a target in order to measure the polarization as in scenario A and that
the polarization does not decohere during one measurement-cycle time $T_{\text{mcyc}} \ll \tau$.

\begin{table}
\begin{center}
 \caption{Results of the statistical EDM uncertainty  (\Eref{sigma-EDM-final})
  for the three ring stages proposed  in Chapter~\ref{Chap:Strategy}, evaluated according to scenario A  of Section~\ref{app:sigd}
 and  no decoherence of polarization.
  Note that the   factor $R_{\text{frac}}$ was omitted in the executive summary.\label{tab:sys:staterr}}
  \begin{tabular}{l  l  l  l}
\hline \hline
 Stage  proposed in Chapter~\ref{Chap:Strategy} &  Pure magnetic ring \&   & Combined $E$--$B$ ring &  All-electric ring \\
    & Wien filter (precursor) & (prototype ring)               &                     \\
\hline
$C_{\text{ring}}$ \ \  (spin $S$ : weight $C_{\rm s}$)       & $0.42$  \ \ (1 :  3/2)  & 1 \ \ (1/2 :  1) & 1 \ \ (1/2 : 1)\\

Scenario factor $C_{\text{mod}}$        &  {4.9} & {4.9} & {4.9} \\
Fraction of particles detected $f$      &  {0.005}  &  {0.005} &  {0.005} \\
Polarization     $P$       &  {0.8}  &  {0.8} &  {0.8} \\
Measurement cycle $T_{\text{mcyc}}$ & 1000\Us & 1000\Us & 1000\Us \\
 Number of particles stored $N$   &  $10^9$   &  $2 \times 10^9$  &  $4 \times 10^{10}$ \\
Average analysing power $A$   &   0.6     &      0.2         &    0.6        \\
Electric, magnetic field $E,B$  &   $2.7~\si{kV/m},  19~ \si{\micro T} $   &$  7.3~ \si{MV/m}, 0.03~ \si{T}$             &  $8~ \si{MV/m}$, ---         \\
Fraction of ring with fields   $R_{\text{frac}}$      &  $1/183$  &   0.55               &     0.65   \\
$\sigma_{\mathrm{EDM}}(\mathrm{1\,fill})$  &
       $5.7 \times \SI{e-21}{\text{$e$}.cm}$ &  $5.5 \times \SI{e-26}{\text{$e$}.cm}$ & $4.6 \times \SI{e-27}{\text{$e$}.cm}$ \\
$\sigma_{\mathrm{EDM}}(\mathrm{1\,year/ 10000\,fills})$ &
        $5.7 \times \SI{e-23}{\text{$e$}.cm}$ &  $5.5 \times \SI{e-28}{\text{$e$}.cm}$ & $4.6 \times \SI{e-29}{\text{$e$}.cm}$ \\
\hline \hline
  \end{tabular}
%
%
  \end{center}
   \end{table}

\section{Systematic effects}
\label{chap:systematics}

Systematic effects are any phenomena other than an EDM generating a vertical component of the polarization limit sensitivity, \ie the smallest detectable EDM, of the proposed experiment. Such systematic effects may be generated by unwanted electric fields owing to imperfections in the focusing structure,
such as misalignments of components, by magnetic fields penetrating the magnetic shielding or generated inside the shield (for example by the beam itself or the RF cavity), or gravity. A combination of several such phenomena or a combination of an average horizontal polarization and one of these phenomena may also lead to such systematic effects. This chapter describes the current stage of the understanding of systematic effects limiting the sensitivity of the experiment focusing on the measurement of the EDM in an electrostatic `frozen spin' ring \cite{BasicBNL, BasicRevSci}, which is considered in the present baseline proposal. Nevertheless, many of the mechanisms described are relevant for other proposals, such as a hybrid ring with electric bending and magnetic focusing \cite{Hybrid} and the `double magic' ring \cite{DoubleMagic}.

Studies of systematic effects have been conducted and are underway by several teams of the CPEDM collaboration to further improve the understanding of basic phenomena to be taken into account and to estimate the achievable sensitivity. Thus, this report is only a snapshot, aimed at describing the current understanding. The preliminary conclusion is that achieving the sensitivity target of $10^{-29} \, e \, \mbox{cm}$ (smallest identifiable proton EDM) is very challenging and will probably not be possible with the current baseline fully electrostatic `frozen spin' synchrotron.

\subsection{Recap of the proposal}
\label{subsect:Recap}

The basic concept of the proposal to measure the proton EDM in an electrostatic ring \cite{BasicBNL, BasicRevSci} is depicted in \Fref{fig:SketchProposal}. Bunches of protons polarized in the
longitudinal direction, represented by red and blue arrows, circulate in an electrostatic ring. The bending electric field pointing towards the centre is represented by green arrows. Bunches circulating clockwise (CW) are represented by blue arrows and bunches circulating counterclockwise (CCW) by red arrows. The direction of the arrows indicates the orientation of the polarization. For the case sketched in \Fref{fig:SketchProposal}, both the CW and the CCW beams have bunches polarized parallel to the direction of movement and opposite to the direction of movement. Such a bunch structure is advantageous in avoiding some of the systematic effects compromising the sensitivity of the experiment, but  is conversely difficult to generate\footnote{Note that there are proposals that, in addition to the bunches polarized parallel or antiparallel to the direction of movement,  have bunches polarized orthogonal to the direction of movement; such bunches allow some systematic effects to be quantified and, possibly with appropriate feedback systems, to be reduced.}. The signature of an electric dipole moment $\vec{d}$ (aligned with the spin of the particles and, for the rotation indicated in the sketch, parallel to the spin), is a rotation of the spin out of the horizontal plane.

\begin{figure}
   \centering
   \includegraphics[width=10cm]{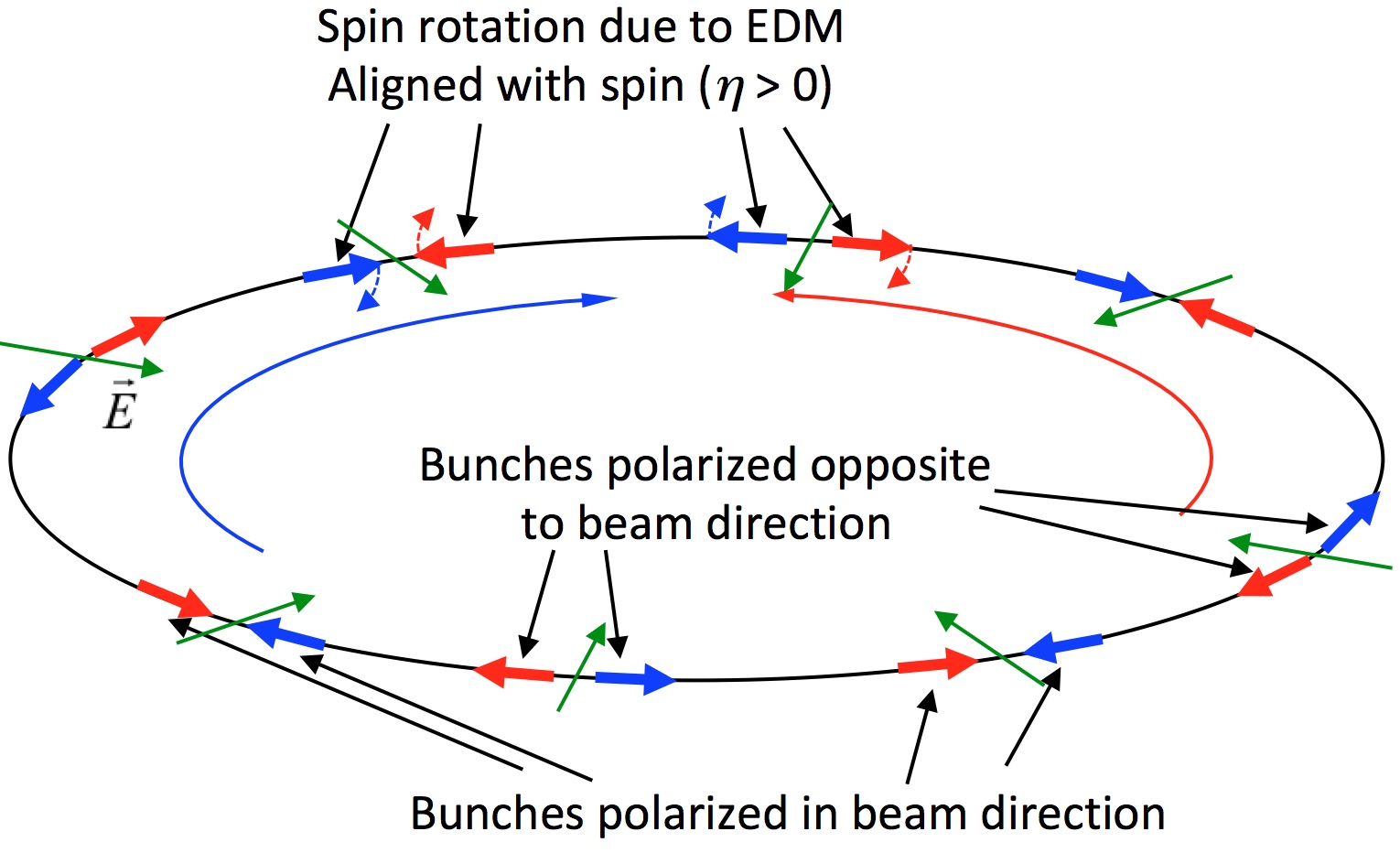}
   \caption{Proposal to measure the proton EDM in a frozen spin `magic energy' electrostatic ring}
   \label{fig:SketchProposal}
\end{figure}

The basic equation, used for most of the considerations presented, is the subtracted form of the Thomas--BMT equation, \ie the difference between the angular frequency $\vec \Omega_{s}$ of the spin rotation and the angular frequency $\vec \Omega_\mathrm{p}$ of the rotation of the direction of movement of the particle. With the choice of a vanishing longitudinal component of the angular frequency describing the rotation of the direction of the particle, this quantity is given by
\small
\begin{equation}
\begin{split}
\DOmvec & = \vec\Omega_{s} - \vec\Omega_\mathrm{p} 
= - \frac{q}{m} \Bigg[ G \vec B_\perp + \left(G + 1 \right) \frac{\vec B_\|}\gamma \\[1.2mm]
& - \left( G - \frac 1{\beta^2 \gamma^2} \right) \frac{\vec \beta \times \vec E}c  +
     \frac \eta 2 \left( \frac{\vec E_\perp} c + \frac 1 \gamma \frac{\vec E_\|} c 
     + \vec \beta \times \vec B \right) \Bigg] \, ,
\end{split}
\label{eq:Dom}
\end{equation}
\normalsize%
where $q$ and $m$ are the charge and mass of the particle, $\vec B$ and $\vec E$ the magnetic and electric field, $\beta$ and $\gamma$ the relativistic factors, and $\vec\beta$ a vector with length $\beta$ and a direction parallel to the velocity;
$\vec B_\parallel = (\vec \beta \cdot \vec B ) \vec \beta / \beta^2$ and $\vec E_\parallel = (\vec \beta \cdot \vec E ) \vec \beta / \beta^2$ denote the longitudinal components in direction of the velocity of the magnetic and electric fields, and $\vec B_\perp = \vec B  - \vec B_\parallel$ and $\vec E_\perp = \vec E  - \vec E_\parallel$ are the components of the magnetic and electric field perpendicular to the direction of movement\footnote{Using these definitions for the longitudinal and perpendicular field components and $\vec \Omega_\mathrm{p} = (q / \gamma m ) ( (\vec \beta \times \vec E /c )/\beta^2-\vec B_\perp )$,
it is simple to show that this equation is consistent with the Thomas--BMT
equation (\Eref{eq2} in the executive summary).}.
The quantities $G$ and $\eta$ describe the magnetic dipole moment, which is, in general, well known, and the electric dipole moment to be measured.
For the case of protons, $G = 1.79285$. Note that for a proton EDM of $d_{\text{s}}$ = \num{e-29}~$e \,$cm, which is often quoted as the expected sensitivity of the proposed facility, $\eta$ is as low as $\eta_{\text{s}}$ = \num{1.9e-15}.

The transverse component of the angular frequency describing the rotation of the (tangential) particle direction  $\vec{t} = \vec{v}/|\vec{v}|$, with $\vec{v}$ the particle velocity, is given by
$\vec\Omega_{\mathrm{p},\perp} = \vec{t} \times ( \mathrm{d} \vec{t}/\mathrm{d}t )$. An arbitrary longitudinal component $\vec\Omega_{\mathrm{p},\parallel} = \kappa \, \vec{t}$ can be added, such that the angular frequency describing the rotation of the particle direction is given by $\vec\Omega_\mathrm{p} = \vec{t} \times  ( \mathrm{d} \vec{t}/\mathrm{d}t ) + \kappa \, \vec{t}$. It is easy to show that $\mathrm{d}\vec{t}/\mathrm{d}t = \vec\Omega_\mathrm{p} \times \vec{t}$ for any value of the free parameter $\kappa$. Here, $\vec\Omega_{\mathrm{p},\parallel} = 0$ is assumed. Nevertheless, the longitudinal components of $\Omega_\mathrm{p}$ and  $\Delta\Omega$ must be interpreted with care;
this is the topic of ongoing discussions.

Considerations presented in this chapter implicitly assume that the particle follows a closed orbit (betatron oscillations are neglected) and that a coordinate system is attached to the closed orbit. The rotation of this coordinate system can be described by a unique angular frequency $\Omega_\mathrm{p}$, with an appropriate choice of  (small) longitudinal component. This is somewhat inconsistent with the choice $\vec\Omega_{\mathrm{p},\parallel} = 0$ made here. Studies are ongoing and results will be published soon.

In a fully electrostatic machine, installed inside a perfect magnetic shielding to reach $\vec B = 0$, and without an EDM, a particle spin aligned with the direction of the movement will rotate together with the particle velocity, \ie fulfil the `frozen spin' condition, if
\begin{equation}
  \beta \gamma = \beta_{\text{m}} \gamma_{\text{m}}  = \frac{1}{\sqrt{G}}
  \, ,
  \label{eq:FrozenSpin}
\end{equation}
where $\beta_{\text{m}}$ and $\gamma_{\text{m}}$ denote the `magic' relativistic factors. For protons,
$\beta_{\text{m}} = 0.598379$ and \mbox{$\gamma_{\text{m}} = 1.24811$}.
With the `magic' relativistic factors, one obtains the magic proton momentum,
\[
p_{\text{m}} = \beta_{\text{m}} c \gamma_{\text{m}} m = 700.74~\text{MeV}/c
\,
\]
 and the `magic' kinetic energy \mbox{$(\gamma_{\text{m}} - 1) m c^2 = 232.792$~MeV}. Note that real `magic' relativistic factors are obtained only for positive values of the quantity $G$, as is the case for protons (or electrons). Thus, a purely electrostatic ring fulfilling the `frozen spin' condition is not possible for particles with negative $G$, such as deuterons or helions.

An electric dipole moment described by a non-zero $\eta$ generates a rotation of the spin from the longitudinal direction to the vertical direction. In an electrostatic ring with circumference  \mbox{$C = 500 \mbox{ m}$}, which is about the minimum required with the given beam energy and in order to keep feasible maximum electric field strength, the angular frequency for
$\eta = \eta_{\text{s}}$ is about \mbox{1.6\,nrad/s}. This small vertical spin rotation must be detected by precise polarimetry in order to identify the particle EDM.

Additional ingredients to the `magic energy' proton EDM measurement concept are (i)  simultaneous circulation of polarized beams in both clockwise (CW) and counterclockwise (CCW) directions, (ii)  operation of the synchrotron with a very weak vertical tune $Q_V$ (proposals vary between $Q_V < 0.1$ and  $Q_V = 0.44$, with some variants even envisaging periodically varying $Q_V$ by, say, about $\pm 10 \mbox{\%}$) and (iii) the use of a measurement of the vertical separation of the two counter-rotating beams to estimate the average radial magnetic field, which causes the most important systematic measurement error. Furthermore, the average horizontal polarization will be continuously monitored using the polarimeters. A feedback loop will be implemented to bring the measured horizontal polarization back to zero. Note that this feedback system must be able to correct horizontal polarization of both the CW and CCW beams. Thus, for simultaneous operation with CW and CCW beams, two parameters, for example the RF frequency acting on the beam energy and a small vertical magnetic field, must be used by this feedback loop.

Note that individual particles have energies that are slightly different from the `magic' energy and consequently their spin rotates in the horizontal plane away from the longitudinal direction. Thus, it is important that an RF system is present and that the beam is bunched so that the particles execute synchrotron oscillations. During about half of the synchrotron oscillation, the particle will have an energy greater than the `magic' energy, so that the spin rotates faster than the direction of movement.  During the other half of the synchrotron oscillation, below the `magic' energy condition, the spin rotation is slowed down. Thus, the synchrotron oscillations lead to periodic rotations of the spin in the horizontal plane, which average to zero over long periods, but introduce horizontal spin components with opposite signs in the head and the tail of a bunch.

\subsection{Sources of systematic errors and general comments}

Effects considered so far and contributing either alone or in combination with other effects to such a rotation are as follows.
\begin{itemize}
  \item \textit{Magnetic fields.} Even small magnetic fields, around $\Delta B = 1 \mbox{\,nT,}$ penetrating state-of-the-art multilayer shielding after degaussing procedures may lead to spin rotations into the vertical plane, which are orders of magnitude larger than those due to the smallest EDM one would like to be able to detect. In particular, an average radial magnetic field as low as $B_{s} = 9.3 \mbox{\,aT}$ for a $C = 500$\,m  machine generates the same vertical spin component as an EDM of \num{e-29}~$e$\,cm.
  \item \textit{Imperfections of the electrostatic machine}. Typical imperfections of electrostatic synchrotrons are misalignments of bends and quadrupoles or mechanical imperfections of components (\eg small errors in the spacing between electrodes of bends alter the electric field and, as a consequence, the deflection), which lead to deformations of the so-called closed orbit, \ie the average transverse offset of the circulating beam and, in consequence, to local shifts of the kinetic energy of the particle directly impacting spin rotations, as described in \Eref{eq:Dom}. A combination of several such imperfections can lead to a rotation of the spin from the longitudinal to the vertical direction.
  \item \textit{Gravity.} Gravity leads to a spin rotation from the longitudinal  to the vertical direction of $\mbox{44\,nrad/s}$ for protons \cite{GravityOrlovAl, GravityObukhovAl, GravitySilenkoAl, GravityLaszloAl}. Nevertheless, the phenomenon does not mimic an EDM in the sense that the spin rotations due to gravity correspond to an EDM of opposite sign for the CW and CCW beams. This effect is unrelated to other sources of systematic effects and will not be treated here any further. More details can be found in Appendix \ref{app:gravity}.
  \item \textit{Small radial polarization.} A small polarization (average over the circumference), which may not be seen by the polarimeter, owing to an asymmetry or even the result of a feedback loop aimed at rotating the spin in the horizontal plane into the longitudinal direction may lead to a generation of vertical spin in combination with vertical closed orbit perturbations.
  \item \textit{Cavity misalignment and closed orbit perturbation (offset of the transverse beam position) at the cavity location.} The azimuthal magnetic field of the cavity is a special case of magnetic field, which (i) creates strong effects along with small offsets between the beam position and the centre of the cavity and (ii) has a strong gradient.
\end{itemize}

The following phenomena  might  generate systematic measurements errors compromising the sensitivity of the experiments but have not yet been studied in detail.\begin{itemize}
  \item \textit{Betatron oscillations and different beam emittances of the two counter-rotating beams.} Studies described here are for particles following the `closed orbit'. Thus, betatron oscillation and possible additional systematic effects caused by them have been neglected at present.
  \item \textit{Inhomogeneous beam distributions.} A small vertical polarization, which is different for particles at the centre of the bunch and particles executing large synchrotron or betatron oscillations, could be generated by the beam preparation process. If particles with large oscillation amplitudes tend to be intercepted by the polarimeter earlier than particles from the centre of the bunches, the average observed vertical spin will change over a measurement cycle, even in the absence of an EDM.
  \item \textit{Electromagnetic fields}. These may be  generated by other particles in the same bunch or by particles of the beam rotating in opposite direction.
They may also be   generated by the interaction of the circulating beams with the surrounding vacuum chambers (image currents, \etc).
\end{itemize}

For numerical evaluations, the $C$ = 500 m  strong focusing lattice \cite{LatticeValeri} will be used. This lattice has been optimized to obtain beam lifetimes close to the project requirements with intrabeam scattering (IBS) and envisaged intensities and, amongst all proposals, is the closest to a ring that could be constructed.

\subsection{Radial magnetic field leading to a systematic error proportional to the perturbation}

Horizontal or `radial' magnetic fields are the only perturbation generating a rotation of the spin from the longitudinal into the vertical direction, which is directly proportional to the perturbation\footnote{Disregarding gravity \cite{GravityOrlovAl, GravityObukhovAl, GravitySilenkoAl, GravityLaszloAl},
which has been mentioned already, is well understood, and is not a concern for the EDM measurement.}. There are two major sources of magnetic fields not generated by the beam itself and acting on the beam, which also have different impacts on the measurement: residual static magnetic fields penetrating the shielding and magnetic fields from the RF cavity.

\subsubsection{Residual magnetic field penetrating the magnetic shielding}
Typical residual magnetic fields inside state-of-the-art multilayer shielding with degaussing procedures are around $\Delta B = 1 \mbox{\,nT}$, which is about eight orders of magnitude larger than horizontal magnetic fields $B_{s}$ = 9.3\,aT generating the same effect as the EDM sensitivity aimed for in typical proposals \cite{BasicBNL, BasicRevSci}.  The average  radial magnetic field around the ring circumference will be somewhat smaller than $\Delta B$, but still orders of magnitude larger than $B_{s}$. The radial magnet field will vary strongly over a distance comparable to the transverse size of the shielding,
which is expected to be around 1\,m.  Assuming, optimistically, that the circumference $C$ = 500 m can be divided into 500 sections with a length of \mbox{1\,m} with about constant field and that there is no correlation of the fields between different sections, one comes to an r.m.s. value of the transverse field of about $\Delta B_{s}/\sqrt{500} \approx 45 \mbox{\,pT}$. Note that static average horizontal fields coupling to the known proton magnetic moment mimic an EDM, in the sense that the contributions from the two counter-rotating beams do not cancel for the final result.

An essential ingredient of the proton EDM measurement proposals in an electrostatic ring is to operate the machine with weak vertical focusing, such that horizontal magnetic fields lead to a vertical separation of the two counter-rotating beams, which is measured with ultrasensitive SQUID-based pick-ups. Note that for the strong focusing lattice proposal \cite{BasicRevSci, LatticeValeri}, with a vertical tune of $Q_V$ = 0.44, an average horizontal magnetic field $B_{s}$ leads to an average vertical separation of the two counter-rotating beams of $\Delta y_{s} \approx 0.26 \mbox{ pm}$. (Other proposals envisage weaker vertical focusing and lower tunes \cite{BasicBNL}.  An average horizontal magnetic field $B_{s}$ with a vertical tune of $Q_V = 0.1$ would give a vertical separation of the beams of $\Delta y_{s} \approx 5 \mbox{\,pm}$. However, with the foreseen intensities, they feature IBS growth rates not compatible with typical assumptions of the machine cycle length of around \mbox{1000\,s}; optimizing a machine with such a low vertical tune to obtain IBS rates compatible with expected cycle lengths and intensities leads to excessive vertical beam sizes.) The measured beam separation will be compensated by additional magnetic fields generated by electrical currents inside the shielding, as much as is possible with the achievable measurement accuracy. The average vertical beam separation after correction over the full duration of the experiment  may still have to be used to reduce the systematic measurement error. Even after averaging over several pick-ups installed around the ring, extensive averaging over durations comparable to the machine cycle, and, further, averaging over many machine cycles, the determination of the remaining average horizontal magnetic field will be most challenging and, probably, prevent the experiment from reaching the sensitivity aim
$d_{\text{s}}$ = \num{e-29}\,$e \,$cm. The following effects may compromise the sensitivity of the experiment.
\begin{itemize} 
  \item \textit{Limited accuracy of orbit difference measurements}, even with averaging over many pick-ups and over long durations.
  \item \textit{Observation of orbit difference only at discrete positions around the circumference}. Even under the assumption that the focusing is perfectly constant around the circumference, the average of the orbit difference measured by a finite number of equally spaced pick-ups is, in general, slightly different from the average \cite{LatticeValeri}.
  A rough estimate for the proposed strong focusing ring  \cite{BasicBNL, BasicRevSci, LatticeValeri}, where the pick-ups are not perfectly spaced,\footnote{This lattice, with four-fold symmetry, and each quarter consisting  of five arc cells and one straight section cell, has 36 orbit difference pick-ups adjacent to quadrupoles in arcs only.}
  leads to the conclusion that this effect limits the uncertainty of the final result to an EDM value about four orders of magnitude greater than $d_{\text{s}} \approx \num{e-29}\,e\,$cm. The effect can be mitigated, in theory, by an optimized spacing of the orbit difference pick-ups and a modulation of the vertical tune \cite{TuneMod, BasicRevSci}. The feasibility of the latter implies that the working point has to regularly cross betatron resonances, which is delicate and may lead to unacceptable beam losses.
  \item \textit{Wanted and unwanted variations of the Twiss betatron functions around the circumference.} In general, the transverse focusing is not homogeneous around the circumference. Even the `smooth focusing' lattices feature field-free straight sections with no focusing at all. In consequence, the so-called Twiss betatron functions vary around the circumference. Thus, the effect of a local horizontal magnetic field on the average orbit separation, which depends on the local betatron function, will depend on the position. A rough estimate based on the strong focusing lattice proposal leads to the conclusion that this effect limits the uncertainty of the final result to an EDM value about five orders of magnitude greater than $d_{\text{s}} \approx \num{e-29}\,e\,$cm. The effects can be mitigated by designing a lattice with small variations in the vertical betatron functions.  Note that these considerations triggered the proposal of a hybrid ring\cite{Hybrid} with electric fields bending the beam and magnetic quadrupoles for focusing. The situation is even more delicate for a realistic ring with `beta beating', \ie unwanted and unknown variations of the betatron function with respect to the lattice design, owing to unknown focusing errors. Careful studies assuming realistic focusing errors and realistic procedures to quantify and correct the resulting betatron beating are required to assess the effect and the implication on the achievable sensitivity.
  \item \textit{Coupling of the betatron motion between the two transverse planes}. Unavoidable skew quadrupolar components resulting from mechanical imperfections (\eg rotation of quadrupoles around the longitudinal axis or
the electrodes of bending units not being perfectly parallel) couple the betatron oscillation in the two transverse planes. A horizontal separation between the CW and  CCW beams caused by residual vertical magnetic fields at the location of such skew quadrupolar components will generate different vertical deflections for the two counter-rotating beams. The resulting vertical separation between the two counter-rotating beams can be misinterpreted as the signature of a horizontal `radial' magnetic field and lead to a systematic measurement error.
\end{itemize}

\subsubsection{Magnetic fields due to the cavity}
Typical azimuthal magnetic fields of RF cavities are orders of magnitude larger than the fields relevant for a study of systematic errors of a proton EDM measurement. Even in the case of a perfectly aligned cavity, individual particles will `see' horizontal magnetic fields and spin rotations into the vertical (and the horizontal) direction. However, the effect on the final result of the EDM measurement will be strongly suppressed, owing to cancellation of the effect for  (i) different particles of a bunch crossing the cavity with different transverse positions or (ii)  one particle crossing the cavity gap in each turn with different betatron phases and transverse positions.

The situation is different for an offset of the electrical centre of the RF cavity with respect to the vertical closed orbit of, say, $\Delta y$ = \SI{100}{\micro\metre}. The integrated horizontal field seen by the CW beam in a  ring operated below transition, owing to a cavity operated with harmonic $h$ and peak RF voltage $V_{\text{RF}}$, is $\Delta B\mathrm{d}l = ( {\pi \, \beta_{\text{m}}}
/ {c \, C}) h \, V_{\text{RF}} \, \Delta y$. Inserting the parameters $V_{\text{RF}} = 6 \mbox{\,kV}$ and $h = 100$ for the strong focusing EDM ring proposal, one obtains $\Delta B \mathrm{d}l = 0.75 \mbox{\,nT\,m}$, which must be compared with the integrated field around the circumference $B_{s} \, C = 4.7 \mbox{\,fT\,m}$ generating the same rotation of the spin into the vertical direction as an EDM of $d_{\text{s}} = \num{e-29}\,e \,$cm.
Thus, an offset of $\Delta y$ = \SI{100}{\micro\metre} between the electrical centre of the cavity and the vertical closed orbit leads to a rotation of the spin into the vertical direction that is a factor  of \num{1.6e5} larger than the effect for a proton EDM $d_{\text{s}}$.
As the direction of the magnetic field is inverse for CW and CCW beams, the effect does not mimic EDM in the sense that contributions from the two counter-rotating beams to the final result cancel each other in a perfect measurement set-up. Nevertheless, this cancellation relies on a measurement of the vertical polarization build-up with high precision for both beams, which may be very challenging.

Another mitigating measure to be discussed in the case of imperfections of the polarity measurement 
is a feedback loop that detects spin rotations of the CW and CCW beams that
are not compatible with the EDM (not `mimicking' the EDM) and corrects them, for example, acting on the vertical closed orbit at the location of the RF cavity. Note that  other effects, described in the next section, also generate  spin rotations of the two counter-rotating beams that are not compatible with an EDM and would be corrected by such a feedback loop.

\subsection{Second-order effects}
\label{chap:sensitivity:second-order-effects}
Several classes of effect in which two different perturbations, such as, \eg residual vertical and longitudinal magnetic fields penetrating the shielding, generate a vertical spin component will be described in this section.
These phenomena are second-order effects in the sense that the resulting vertical spin for small perturbations is proportional to the square of the perturbation (if both the vertical and the longitudinal magnetic fields in the example are increased by a factor $k$, the resulting vertical spin is increased by a factor $k^2$).
All considerations reported apply, strictly speaking, to beam particles following the closed orbit of the ring and not executing any betatron oscillations.

Several, but not all, of the effects described next have been reported and interpreted in terms of geometric phase effects. The list of second-order effects described is not exhaustive. Priority has been given to phenomena not yet described. For example, the well known vertical polarization build-up caused by vertical and longitudinal magnetic fields reported and interpreted as geometric phase effect in Ref.~\cite{GeomPhasesMagFields} is not described here.

\subsubsection{Rotation of the spin from the horizontal to the vertical direction by a vertical slope of the orbit inside bending units}

A geometric interpretation of the effect of rotating the spin from the horizontal to the vertical direction, which has been described previously\cite{VertSlopeSpinRot1}, is sketched in \Fref{fig:VertSlopeSpinRot}.
If the `frozen spin' condition is fulfilled, the rotation of the spin and the direction of the trajectory are described by the same angular frequency vector $\vec\Omega_{s} = \vec\Omega_{p}$, which is pointing downwards with a small longitudinal component $\Omega_{s,s} = {\beta_{\text{m}} c}  y'_{\text{co}} / \rho$ with $y'_{\text{co}} = {\mathrm{d}  y_{\text{co}}}
/ {\mathrm{d}s}$ the slope of the vertical orbit and $\rho$ the curvature radius.
This yields, even if the `frozen spin' condition is fulfilled,
to a build-up of the vertical spin of $ {\mathrm{d}  s_y} / {\mathrm{d}t} = \Omega_{s,s} s_x = {\beta_{\text{m}} c}  y'_{\text{co}} \, s_x
/ \rho$.
The vertical spin generated over one turn is given by
\begin{equation}
   \Delta s_y = \int \limits_{0}^{C} \mathrm{d}s \frac{y'_{\text{co}}(s)}{\rho(s)} s_x(s) \, .
  \label{eq:VertSlopeSpinRot}
\end{equation}

\begin{figure} [hb!]
   \centering
   \includegraphics[width=8cm]{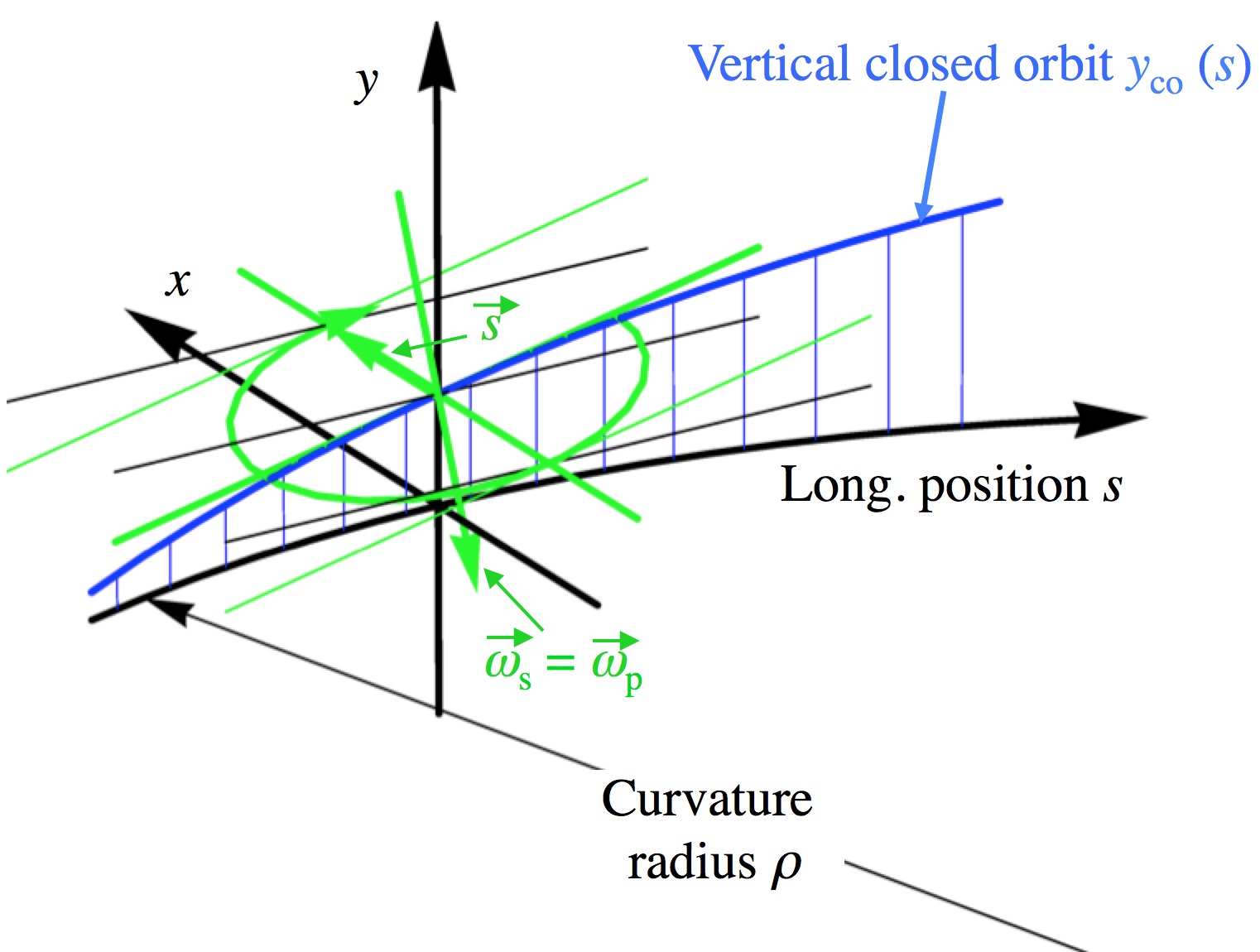}
   \caption{Mechanism rotating the spin from the horizontal to the vertical direction by a slope of the vertical closed orbit inside a bend.}
   \label{fig:VertSlopeSpinRot}
\end{figure}

\subsubsection{Horizontal spin and non-zero average slope of vertical orbit inside bend}
The average horizontal polarization will be monitored continuously using the polarimeter and a feedback loop mentioned in  \Sref{subsect:Recap}.
A small asymmetry of the polarimeter may lead to a non-zero horizontal polarization $\Delta \bar s_x$.
Spin rotations in the horizontal plane, which cancel over one turn, might also lead to non-zero average horizontal polarization, even if the horizontal polarization vanishes at the location of the polarimeter. With the help of \Eref{eq:VertSlopeSpinRot},  and using different nomenclature  from that in Ref.~\cite{VertSlopeSpinRot1}, which gives the same result, the average rate of change of the vertical spin becomes
\begin{equation}
   \frac{\mathrm{d} \, s_y}{\mathrm{d} t} = \frac{2 \pi \, \beta_{\text{m}} c}{C} \, \langle y'_{\text{co}}\rangle s_x \, ,
  \label{eq:SyFromSxandYp}
\end{equation}
with $C$ the circumference (with $N_\mathrm{b}$ bends of length $L_\mathrm{b}$)
 and $\langle y'_{\text{co}}\rangle$
the average slope of the vertical closed orbit inside bends.

An average  radial polarization $\Delta \bar{s}_x = 0.1 \mbox{\,mrad}$ and a vertical misalignment of one vertically defocusing quadrupole at the transition from an arc to a straight section of `strong focusing' lattice ($N_\mathrm{b} = 120$ and $L_\mathrm{b} = 2.7$\,m) by $\Delta y = 0.1 \mbox{\,mm}$ leads to an average slope of $\langle y'_{\text{co}}\rangle = -8.2 \times 10^{-8} \mbox{\,rad}$ and a vertical polarization build-up of $-18 \, \upmu \mbox{rad/s}$. As the average radial polarizations of the two counter-rotating beams might not be correlated (if independent polarimeters are used for CW and CCW beams, with different uncorrelated asymmetries), the systematic EDM measurement error resulting from the effect considered here cannot be reduced by counter-rotating beams. However, even an imperfect polarimeter together with feedback acting on bunches polarized parallel to and opposite from the movement of the same (say CW) beam would generate the same radial residual spin; the effect on the final EDM result from these bunches with opposite polarization cancels.

A thorough investigation of the effect requires a simulation of a machine with realistic imperfections and a correction scheme based on (imperfect) position pick-ups and correctors.

Another mitigation measure proposed for some of the schemes involves using bunches with radial polarization in addition to the bunches with longitudinal polarization to measure and, possibly, to correct a rotation from the horizontal to the vertical direction.


\subsubsection{Vertical spin due to horizontal and vertical quadrupole displacements and orbit distortions in both planes}
\label{MisalignedQuads}

A case considered previously\cite{ElQuadMisalignement} is the simultaneous transverse offset of electrostatic quadrupoles in the vertical and horizontal directions. To better understand the mechanism generating a vertical spin component, a special case with two quadrupoles located at opposite positions in the ring and misaligned with transverse offsets in both transverse planes is considered. The signs of the transverse offsets for the two quadrupoles are opposite. Such transverse offsets, by $\Delta x = \Delta y = \pm 0.1 \mbox{\,mm,}$ of two quadrupoles located in the centres of (opposite) straight sections of the proposed strong focusing ring result in the closed orbits (first-order contributions taken into account only) shown in \Fref{fig:ElQuadMisalignement}.
The energy of a particle following the closed orbit $x_{\text{co}}$ inside a bend is, in general, different from the magic energy, owing to the non-zero electric potential.
This energy offset leads to a rotation of the spin around a vertical axis.
The resulting small horizontal spin component of a proton polarized parallel to its momentum circulating in the CW direction is given by
\begin{equation}
  s_x = \int \limits_{s_0}^s \mathrm{d} s \frac{2}{\gamma_{\text{m}}} \frac{x_{\text{co}}(s)}{\rho^2(s)} \quad .
  \label{eq:SpinOtHor}
\end{equation}
Using \Eref{eq:VertSlopeSpinRot}, the vertical spin obtained over one revolution is given by
\begin{equation}
  \Delta s_y = \int \limits_{0}^{C} \mathrm{d} s \frac{y'_{\text{co}}(s)}{\rho(s)} s_x(s) \, .
  \label{eq:MisalignedQuads}
\end{equation}

\begin{figure}[hbt!]
   \centering
   \includegraphics[width=12cm]{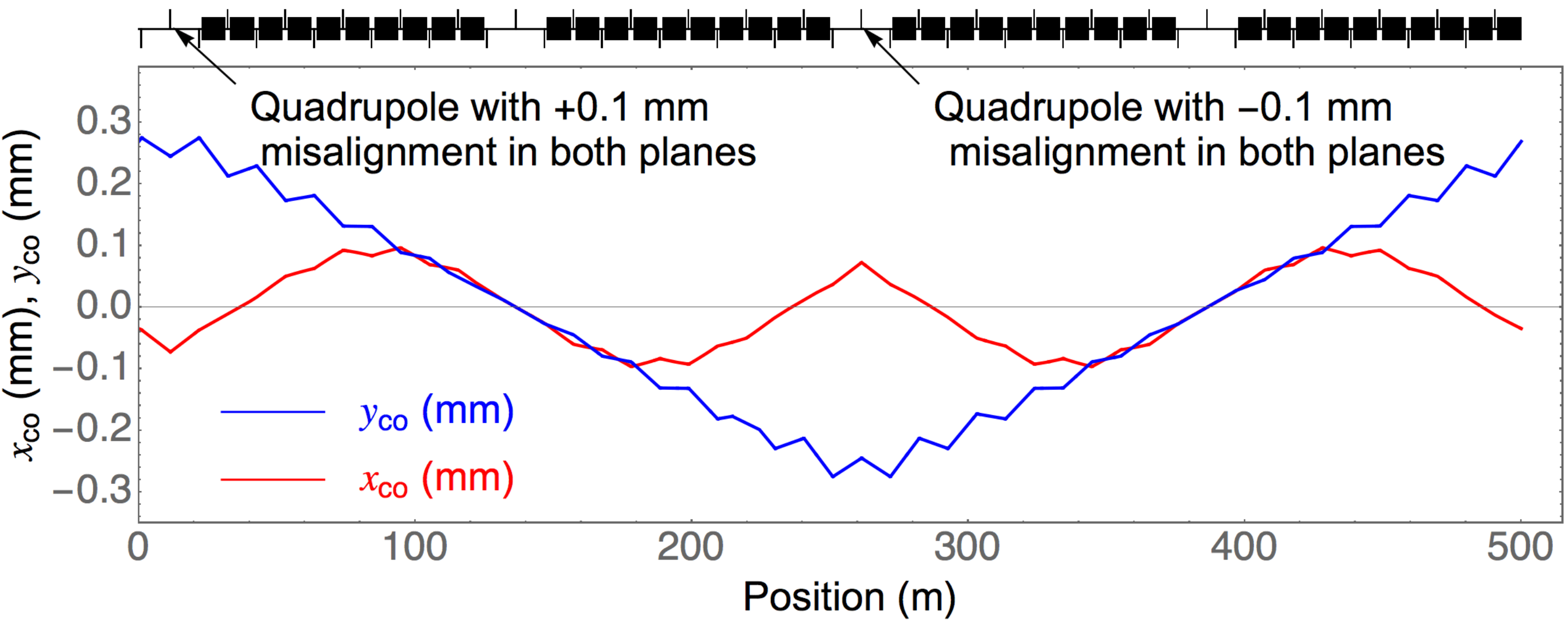} \vspace{3mm}\\
   \includegraphics[width=12cm]{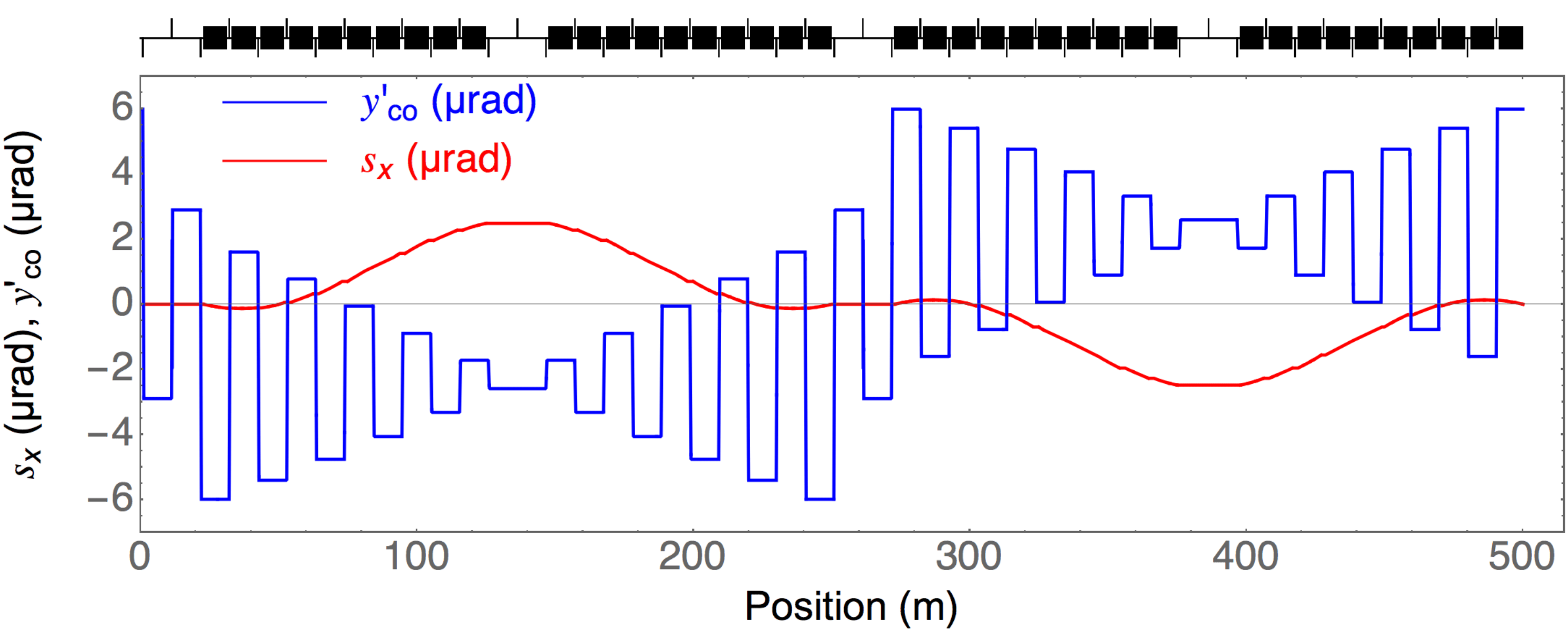}
   \caption{Misalignment in both transverse planes of two quadrupoles at opposite positions in the ring with opposite offsets: (top) closed orbit in radial (red) and vertical (blue) directions; (bottom) radial polarization (red) and the slope of the vertical orbit (blue), \ie the quantities needed to estimate the vertical polarization build-up given in
\Eref{eq:MisalignedQuads}.}
   \label{fig:ElQuadMisalignement}
\end{figure}

The functions required to compute the resulting vertical spin build-up are plotted in the lower panel of \Fref{fig:ElQuadMisalignement}. The result can be interpreted in terms of a geometric phase effect, as it is the result of two rotations, one around a vertical axis and the other around a longitudinal axis, which are out of phase. For the case described, based on the strong focusing lattice, one obtains an average build-up of the vertical spin component of ${\mathrm{d} s_y} / {\mathrm{d} t}$ = \SI{-4.5}{\micro\radian/\second}. The effect to be expected in a realistic machine can only be estimated by thorough simulations with realistic assumptions for misalignments of components and closed orbit correction. Note that this effect does not mimic an EDM, in the sense that the contributions from the CW and  CCW beams to the final result cancel.
Nevertheless, it may be challenging to detect a fast rotation from the longitudinal into the vertical plane in the polarimeter, as the build-up of vertical polarization must be measured with very high relative precision for both beam directions. One might also consider a feedback system to correct  spin rotations (from this and other effects that do not mimic an EDM) into the vertical plane that are not compatible with an EDM, for example, by acting on the vertical position at the location of the RF cavity.

Note that the effect cannot be cured by using a `weak focusing' lattice, as a horizontal offset and rotations around the longitudinal axis result not only in the same phenomena but, in addition, a more direct rotation from the longitudinal to the vertical direction, which will be described in the next section.

\subsubsection{Electric bend with horizontal offset and a rotation around the longitudinal axis}

\label{MisalignedBends}

An electric bend with a horizontal offset $\Delta x$ induces an electric potential at the location of the reference orbit and, in consequence, shifts the kinetic energy of a beam particle away from the `frozen spin' condition. An additional rotation of the same bend around the longitudinal axis through an angle $\alpha$ leads to a vertical electric field component. The consequence is a non-zero radial component of $\DOmvec$ and a rotation of the spin from the longitudinal  to the vertical direction, which differs from the rotation of the direction of movement. In reality, the situation is slightly more complicated, as the misalignment of the bends also affects  the closed orbits in both planes, such that (i) the closed orbit gives an additional contribution to the shift of the kinetic energy from the `frozen spin' condition, and (ii) the effect described in the previous section must also be taken into account. For numerical evaluations, we consider again the strong focusing lattice with (for symmetry reasons) two electric bends on either side of the centre of opposite arcs misaligned by $\Delta x = \pm 0.05 \mbox{ mm}$ and $\alpha = \pm 0.05 \mbox{ mrad}$ (\Fref{fig:ElBendMisalignement}). The net rotation of the spin from the longitudinal to the vertical direction, taking both effects into account, is given by
\begin{align}
 s_x(s) &= \int \limits_{s_0}^{s} \mathrm{d} \hat{s} \frac{2}{\gamma_{\text{m}} } \frac{x_{\text{co}}(\hat{s}) - \Delta x(\hat{s})}{\rho^2(\hat{s})} \nonumber \\
 \Delta s_y &= \int \limits_{0}^{C}  \mathrm{d} s \frac{2}{\gamma_{\text{m}} } \frac{x_{\text{co}}(s) - \Delta x(s)}{\rho^2(s)} \alpha(s) + \int \limits_{0}^{C}
\mathrm{d} s \frac{y'_{\text{co}}(s)}{\rho(s)} s_x(s) \, .
 \label{eq:MisalignedBends}
\end{align}
One finally obtains ${ \mathrm{d} s_y} / { \mathrm{d} t}$ =  \SI{-5.45}{\micro\radian/\second}. Again, this effect does not mimic an EDM, since the contributions to the final result from the CW and  CCW beams  cancel.

\begin{figure} [hb!]
  \centering
     \includegraphics[width=12cm]{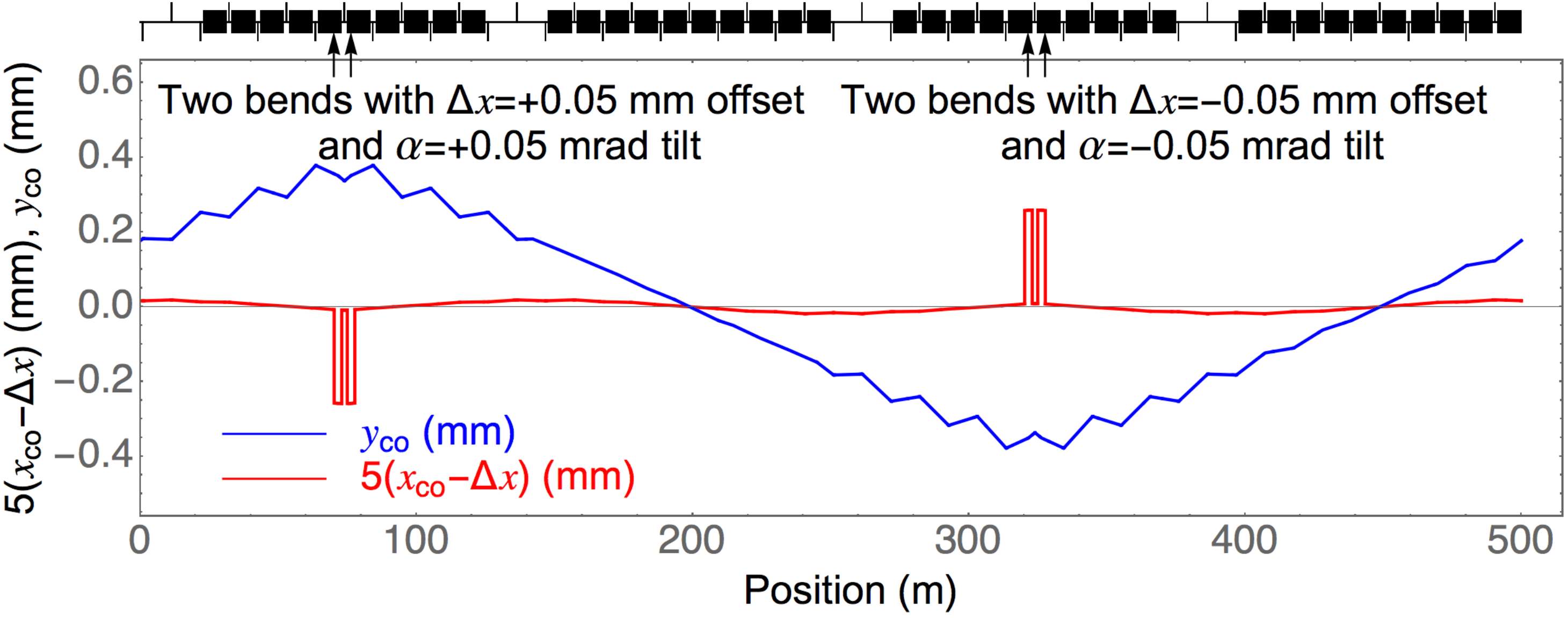}  \vspace{3mm} \\
   \includegraphics[width=12cm]{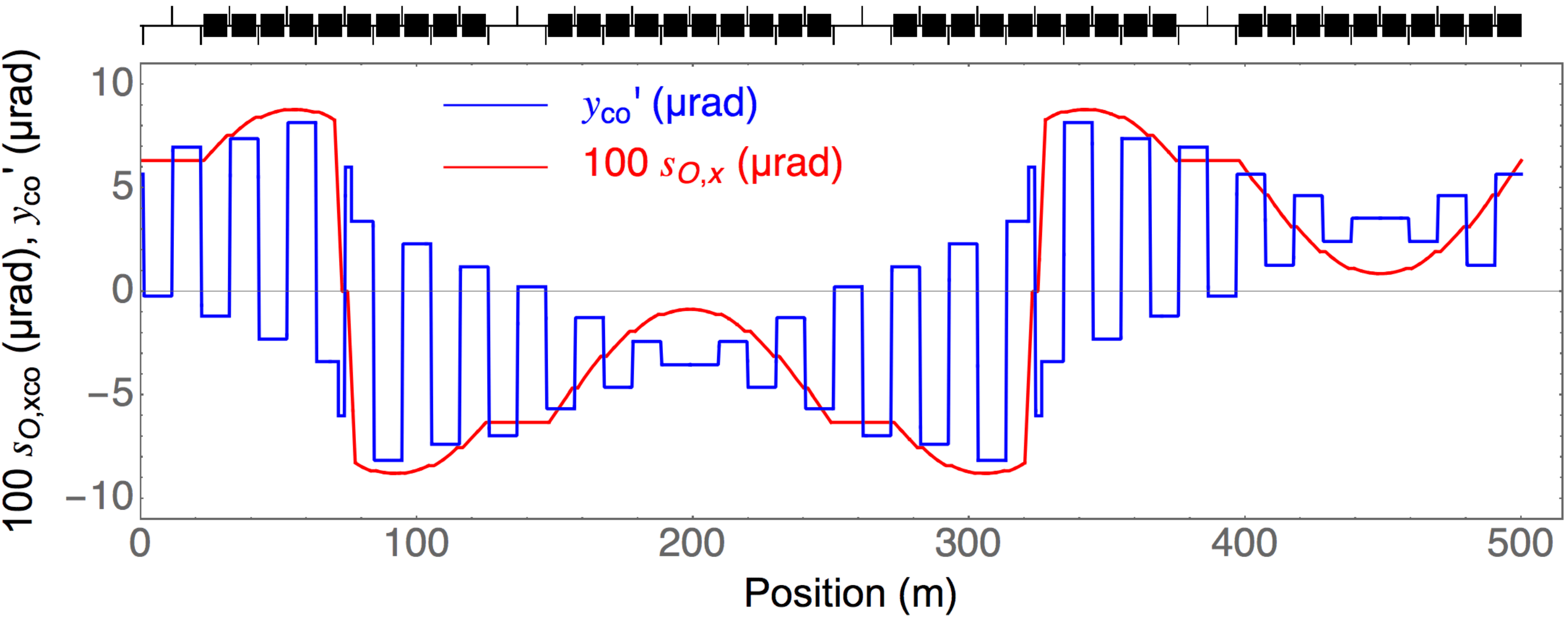}
  \caption{Misalignment of two pairs of electric bends around the centre of opposite arcs: (top)  difference between the horizontal closed orbit and the misalignment of the element (red) and the vertical closed orbit (blue); (bottom) resulting radial polarization (red) and the slope of the vertical orbit (blue), \ie the quantities needed to estimate the vertical polarization build-up given in \Eref{eq:MisalignedBends}.}
  \label{fig:ElBendMisalignement}
\end{figure}

\subsubsection{Vertical polarization build-up from vertical magnetic and electric field components generating closed orbit deformations in both planes}

\label{GeomPhasesMagElFields}

Either an additional (perturbation) vertical magnetic and electric field or an additional horizontal magnetic and electric field leads to orbit deformations in both transverse planes and, in turn, to a build-up of a vertical spin component in a way similar to the mechanism described in  \Sref{MisalignedQuads}. However, these cases, combining perturbations as a result of additional magnetic and electric fields, mimic an EDM in the sense that the contributions from the CW and CCW to the final EDM value do not cancel.

The case treated is a combination of additional vertical electric fields due to misaligned quadrupoles and vertical
magnetic fields, \eg due to imperfect magnetic shielding. The vertical magnetic field also contributes  to the rotation of the spin in the horizontal plane.
An integrated vertical magnetic field of $\pm \Delta B_y = 1 \mbox{\,nTm}$ at the location of quadrupoles located in the centre of opposite straight sections in the strong focusing lattice is assumed. These quadrupoles are vertically misaligned by $\pm 0.1 \mbox{mm}$. The resulting orbit distortions, as well as the spin rotation $s_x$ in the horizontal plane and the derivative of the vertical closed orbit, are plotted in \Fref{fig:VertMagElFields}. The vertical spin component generated over one turn is given by
\Eref{eq:VertSlopeSpinRot} and evaluates to $\Delta s_y = 8.5 \times 10^{-15} \mbox{ rad}$. The resulting vertical spin build-up rate for this probably optimistic case is $3.1 \mbox{\,nrad/s}$, which is almost a factor of two larger than that due to an EDM,
of $d_{\text{s}} = \num{e-29}~e\,$cm.

\begin{figure} [hb!]
  \centering
     \includegraphics[width=12cm]{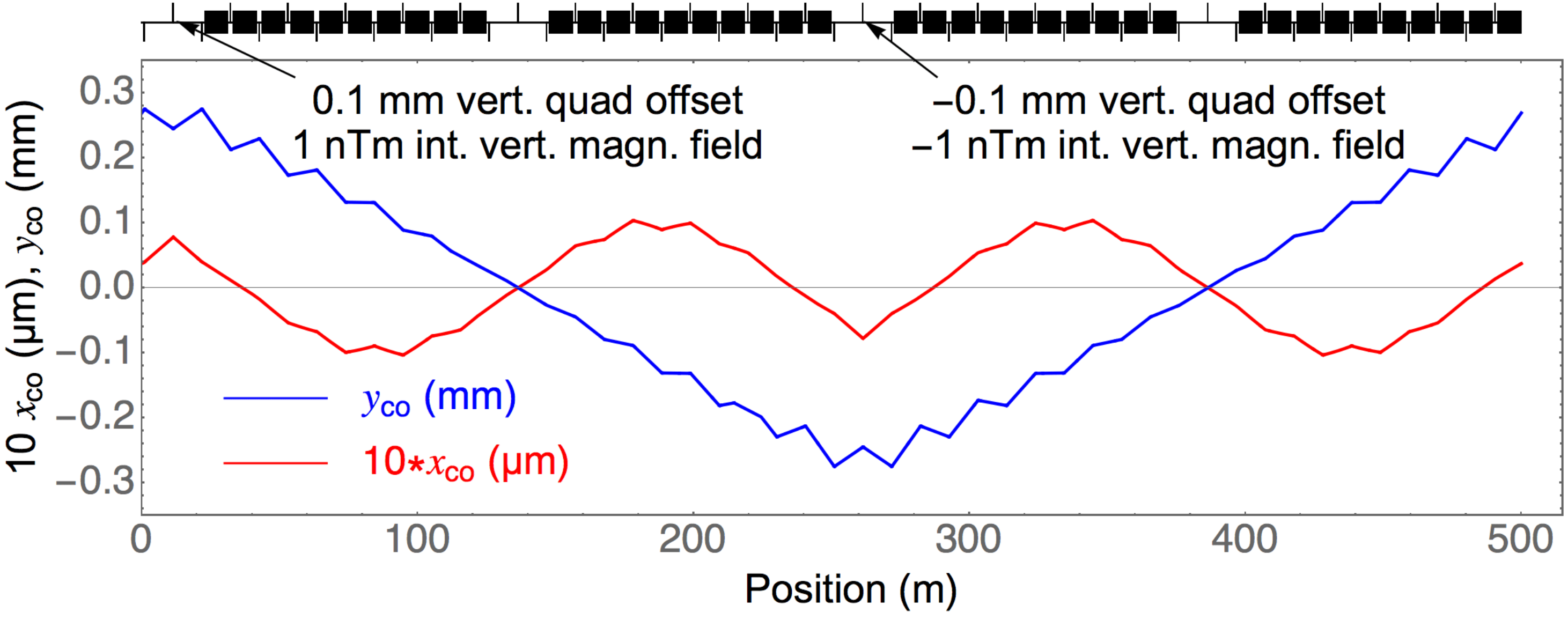} \vspace{3mm} \\
     \includegraphics[width=12cm]{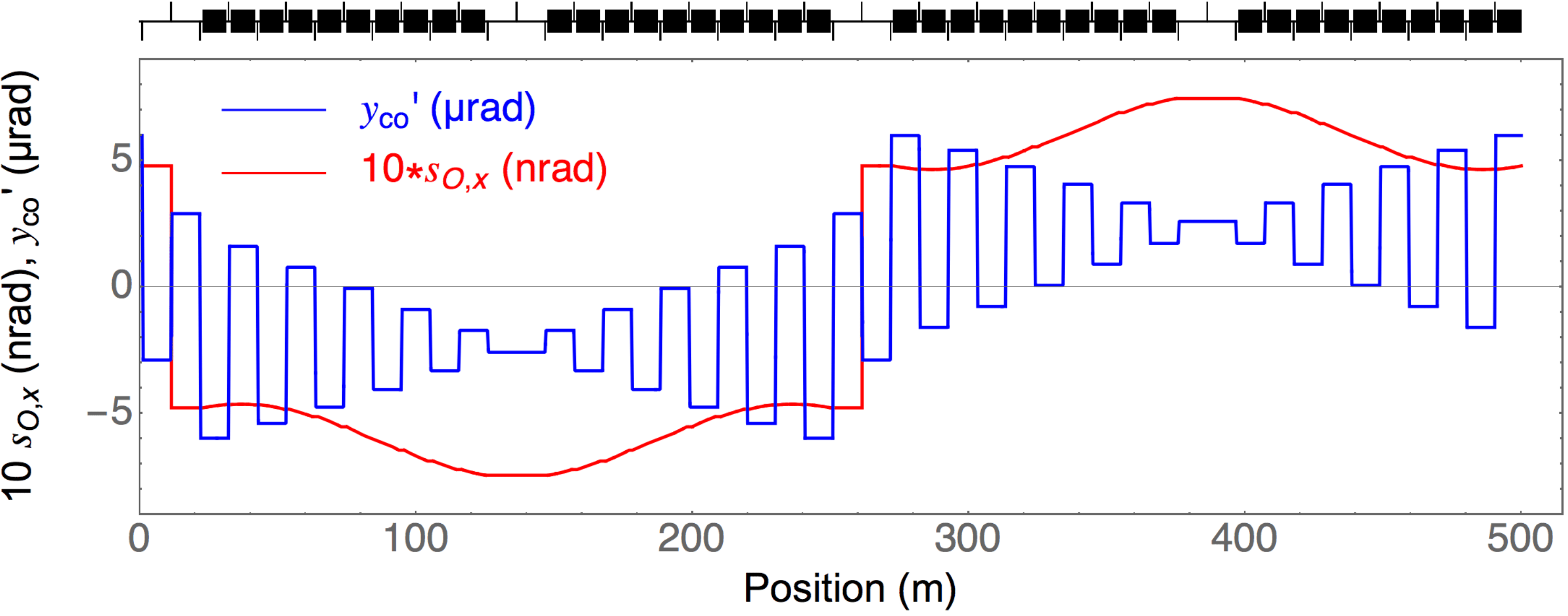}
  \caption{Vertical magnetic and electric fields generating build-up of a vertical spin component: (top)   resulting horizontal (red) and vertical closed (blue) orbits; (bottom) resulting radial spin component (red) and the slope of the vertical orbit (blue), \ie the quantities required to evaluate the vertical polarization build-up using
 \Eref{eq:VertSlopeSpinRot}.}
  \label{fig:VertMagElFields}
\end{figure}

\subsection{Summary}

The phenomena considered as potential limitations to reach the target sensitivity of $d_{\text{s}} =  \num{e-29}~e\,$cm  are summarized in  \Tref{tab:SystEffSummary}. Static (not due to a cavity with a vertical offset) horizontal (radial) magnetic fields are expected to be the main source of systematic errors and to limit the achievable sensitivity to a value larger than this target. In addition, several second-order effects, where two distinct imperfections of the real machine with respect to the perfect design case lead to a spin rotation from the longitudinal to the vertical direction, have been considered. Higher-order effects, as well as betatron and synchrotron oscillations, have not been taken into account and are expected to give smaller contributions to systematic effects.

\begin{table}
  \centering
  \caption{Main systematic effects considered as possible limitations of the achievable sensitivity}
  \label{tab:SystEffSummary}
  \begin{tabular}{ p{5cm}   p{10cm} }
      \hline \hline
      \multicolumn{2}{  l  }{First-order effects} \\
      \hline
         Static horizontal magnetic field & Mimics EDM (no cancellations between contributions from CW and CCW beams on final result);
             effect due to typical magnetic fields inside state-of-the-art shielding about eight orders of magnitude larger than effect due to smallest EDM to be
             detected; expected to be the main limitation to achievable sensitivity, even with orbit separation measurement to estimate (and correct)
             average horizontal magnetic field.  \\
          Horizontal magnetic field due to cavity offset & Does not mimic EDM, but fast rotation of spin, requiring high-precision polarimetry or feedback. \\
          Gravity                             & Effect about a factor of 30 larger than that due to $d_{\text{s}} = 10^{-29}\,e\,$cm. Does not mimic EDM (because of cancellations between contributions from CW and CCW beams); thus, not expected to limit  sensitivity of the experiment.   \\
      \hline 
      \multicolumn{2}{ l }{Second-order effects} \\
      \hline
        Horizontal spin and non-zero average slope of vertical orbit inside bends &
           Depends on polarimeter properties of the two rings, contribution mimicking EDM and incompatibility with sensitivity goal
           $d_{\text{s}} = 10^{-29}\,e$\,cm likely. Mitigation by bunches polarized in forward and backward directions? Mitigation by additional
           bunches with horizontal polarization? \\
            Horizontal and vertical offsets of electric quadrupoles & Does not mimic EDM; large effects expected; high-precision polarimeter or feedback required. \\
             Electric bends with simultaneous horizontal offset and rotation around longitudinal axis &
            Does not mimic EDM; large effects expected; high-precision polarimeter or feedback required. \\
            Static vertical and longitudinal magnetic fields & Does not mimic EDM; moderate effect, probably not limiting sensitivity. \\
            Vertical magnetic field from cavity and static longitudinal magnetic field &
          Mimics EDM; effect small and not expected to limit sensitivity. \\
           Magnetic and electric fields generating orbit deformations in both planes &
           Mimics EDM; worst case with vertical magnetic and vertical electric field probably rules out  achieving sensitivity goal of
           $d_{\text{s}} = 10^{-29}\,e$\,cm. \\
      \hline \hline
  \end{tabular}
\end{table}

For most second-order phenomena, only simple special cases with sometimes optimistic assumptions have been considered, aiming at  an understanding of the underlying mechanisms. This must be followed by more realistic studies with positioning errors of all elements and realistic orbit correction scenarios,
and cross-checked with simulations. Some effects do not mimic the EDM, in the sense that the contributions from the CW and CCW beams to the final results cancel. Nevertheless, there are cases leading to a vertical spin build-up several orders of magnitude faster than the smallest EDM to be measured. This requires either  measurement of the vertical polarization with high accuracy or the implementation of a feedback system to reduce the effect (this could be based on any of the effects generating such a spin rotation, for example, bends with horizontal offsets and rotations around the longitudinal axis).

The optimum operational scenarios depend on the main source of systematic errors in the experiments. In the case where the main contribution comes from the average horizontal magnetic field (after the implementation of mitigation measures), operation with simultaneous CW and CCW beams is important, but the filling patterns of the two rings are not critical. If second-order effects generate a significant contribution to systematic effects, the filling pattern of the two beams becomes important. The optimum operational scenario to control systematic effects would be, in both  CW and  CCW beams, to have some of the bunches polarized in the forward direction, some  polarized in the backward direction, and some  with horizontal polarization.
Schemes to generate such filling patterns may rely either on preparation in the injector, followed by a suitable injection procedure, or spin manipulations inside the ring, using RF solenoids, RF Wien filters, or a combination of both.


\begin{flushleft}

\end{flushleft}

\end{cbunit}

\begin{cbunit}

\csname @openrighttrue\endcsname 
\chapter{Spin tracking}
\label{Chap:SpinTracking}

\section{Introduction}

Spin tracking simulations of the complete EDM experiment are crucial for exploring the sensitivity of the planned storage ring EDM searches and for investigating the systematic limitations. Existing spin-tracking programmes have been extended to properly simulate the spin evolution in the presence of an EDM. The appropriate EDM-induced spin rotations and electric field elements (static and RF) have been implemented and benchmarked. Furthermore, a symplectic description of fringe fields, field errors, and misalignments of magnets has been adapted and verified. For a detailed study during particle storage and spin build-up of an EDM signal, a large sample of particles must be tracked for billions of turns. This constitutes  a challenging task because the anticipated measurement duration of about \SI{1000}{s} requires beam and spin tracking for about $10^9$ turns\footnote{For the COSY precursor experiment with deuterons at \SI{970}{MeV/$c$}.}.

\section{Simulation programs}
\label{sec:spin-tracking:simulation-programs}
Given the complexity of the task, and  to ensure the credibility of the results, various simulation programs using different algorithm have been upgraded and benchmarked with the required accuracy and efficiency.
\begin{itemize}
    \item COSY Infinity\cite{BERZ1990473}, based on map generation using differential algebra and  subsequent calculation of the spin-orbital motion for an arbitrary particle, including fringe fields of elements. COSY Infinity and its updates are used, including higher-order non-linearities, normal form analysis, and symplectic tracking. COSY Infinity contains elements to simulate $\vec E \times \vec B$ elements (static and RF).
    \item COTOBO (COSY Toolbox)\cite{Rosenthal:2015jzr} has been developed to perform simulations,  based on a C\texttt{++} interface for COSY Infinity. The usability of ROOT\cite{Brun:1997pa} enables  fast and easy  analysis
of the simulation results.
    \item MODE (Matrix Integration of Ordinary Differential Equations)\cite{Ivanov:2014aza,Ivanov:2013era} is a software package that provides non-linear matrix map building for spin-orbit beam dynamics simulations. The MODE mathematical model is based on Lorentz and Thomas--MT equations that are expanded as Taylor series up to the necessary order of non-linearity, including fringe fields of elements. The numerical algorithm is based on a matrix representation of the Lie propagator.
    \item Bmad\cite{Sagan:Bmad2006,Sagan:bmad:refmanual} has various tracking algorithms, including Runge--Kutta and symplectic (Lie algebraic) integration. Routines for calculating transfer matrices, emittances, Twiss parameters, dispersion, coupling, and fringe field contributions are also included. Bmad, by interfacing with the PTC tracking code\cite{Schmidt:573082}, can, for example, compute Taylor maps to arbitrary order, as well as normal form analysis.
    \item A customized integration program\cite{Senichev:2012jga}, solving equations of particle and spin motion in electric and magnetic fields using Runge--Kutta integration. The  program  models spin-orbital  motion,  including  fringe   fields  in elements.  The  algorithm  used  in  the  code  is  several  orders  of  magnitude slower than codes based on the generation of maps using differential algebra. Therefore, the program was predominantly used to  investigate short-time phenomena, as well as to benchmark other codes.
    \item The simulation code for numerical integration of beam and spin motion, described in Ref.~\cite{Gaisser:2016ocd}, is a very simple but general approach and integrates the equation of motion, as well as the T--BMT equation, numerically. Standard algorithms, such as the fourth-order Runge--Kutta algorithm, are compared with newer ones and great emphasis is placed on the modular implementation in C\texttt{++}, for maximal flexibility.
\end{itemize}

Particle and spin tracking programs have been benchmarked and simulation results from  different simulation codes were compared with  results from unpolarized and polarized experiments at COSY to ensure the required accuracy of the results\cite{METODIEV2015311,Rosenthal:2015jzr,Maier:2012zzc}.

\section{Status and plans}

Different possible scenarios for EDM measurements have been investigated to explore the sensitivity. In a first step, the resonant method \cite{Rathmann:2011zz,Lehrach:2012eg,Morse:2013hoa} has been developed to enable  EDM measurement at COSY. In parallel, detailed studies have been carried out to  explore the sensitivity of the deuteron precursor experiment at COSY \cite{Rosenthal:2014, Chekmenev:2016cpx, Schmidt:2017wnl,PhysRevAccelBeams.23.024601}. In this context, two different approaches have been investigated to perform deuteron EDM measurements in dedicated storage rings.
\begin{itemize}
   \item \textit{Frozen-spin method} \cite{Farley}. The electrostatic and magnetic bending fields in a storage ring are adjusted according to the particle momentum in such a way that the longitudinally polarized spins of the particle beam are kept aligned (frozen) with their momenta.
   \item  \textit{Quasifrozen-spin method} \cite{Senichev:2016rez, Senichev:2017sry}.
In this case, electric and magnetic field deflectors can be spatially separated. The spins oscillate back and forth  in the horizontal plane every time the particle passes through a magnetic or  electrostatic field. With respect to the momentum vector, the spin oscillations of individual particles in the magnetic and electrostatic sections of the ring compensate each other.
 \end{itemize}
 Various spin-tracking results for deuteron EDM storage rings utilizing different lattice configurations are presented in Refs.~\cite{Senichev:2016rez, Senichev:2015bkf, Senichev:2016zqo, Senichev:2017sry, Skawran_2017}. Results of spin-tracking simulations for the prototype ring (see Chapter\,\ref{chap:ptr} and also Ref.~\cite{Lehrach:2019pgw}) and the final proton EDM ring (see Chapter\,\ref{Chap:allelectricring}) are discussed in Refs.~\cite{selcuk,Michaud:2019,Tahar:2019hzv}.

\section{Spin tracking simulations for deuteron and proton EDM measurements}

As discussed in \Sref{sec:spin-tracking:simulation-programs}, various computer programs are utilized to simulate the generation of a vertical polarization build-up  from  effects other than an EDM. These systematic errors typically stem from field and alignment errors of the electric and magnetic elements employed in the ring. The resulting vertical polarization build-up from unwanted effects must then be compared in detail  with the EDM-induced polarization build-up.

\subsection{Precursor experiment for deuterons at COSY}

To be able to simulate the polarization build-up for the precursor experiment (Chapter\,\ref{Chap:Precursor}), applying the resonant method using an RF Wien filter (see \Sref{sec:PTR:RF-Wien-filter})\cite{Rathmann:2011zz, Lehrach:2012eg, Morse:2013hoa, PhysRevAccelBeams.23.024601}, time-dependent transfer maps have been implemented in COSY Infinity \cite{Rosenthal:2015jzr}. This device provides superimposed electric and magnetic RF fields, applied such that they do not influence the particle trajectories, but lead to an additional rotation of the spin around the magnetic field axis of the device. The polarization build-up is proportional to the small angle between the axis of the RF Wien filter and the invariant spin axis (see \Eref{eq:spin-resonance-strength}). The particle and spin motions in the machine are perturbed by imperfections of the focusing and deflection elements in the ring through shifts, tilts, and rotations\cite{Rosenthal:2015jzr,Chekmenev:2016cpx}. Even though the resulting closed orbits can be corrected by an orbit-correction system to mitigate the effect, resulting spin rotations, through the magnetic moment, limit the sensitivity, and thus have to be well understood.

\subsubsection{EDM build-up with misaligned magnets}

Different magnitudes of the standard deviation of the Gaussian-distributed quadrupole shifts have been simulated; for each of these misalignments, a spin tracking simulation has been performed using different EDM magnitudes. The phase of the RF Wien filter is assumed to be locked to the situation of maximum build-up.  The resulting vertical spin build-up for different r.m.s. values of the vertical orbit displacements of the quadrupoles is shown in \Fref{fig:precursor_spinbuildup}.

\begin{figure} [hb!]
\centering
\includegraphics[width=0.8\textwidth]{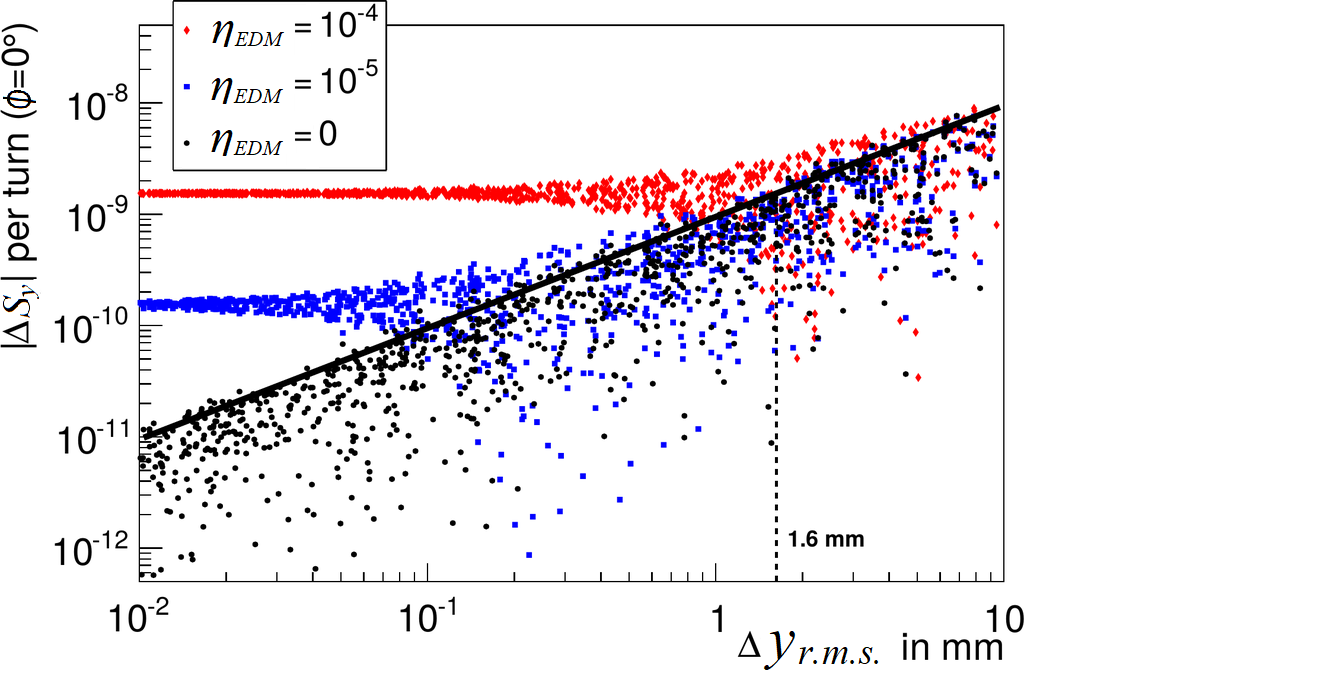}
  \caption{Maximum vertical spin build-up per turn for different EDM magnitudes, parametrized by $\eta_\text{EDM} = \eta$ (see \Eref{emeq5}), as a function of Gaussian-distributed quadrupole shifts with different standard deviations. The r.m.s. value of the vertical orbit displacement is used as a measure for misalignments. (Figure taken from Ref.~\cite{Rosenthal:2016zbf}, reused with permission from the author.)
\label{fig:precursor_spinbuildup}}
\end{figure}

As long as the EDM contribution to the polarization build-up is significantly larger than the spin build-up introduced by misalignments of magnets, a finite EDM may be detected. For a randomized error with a standard deviation of \SI{0.1}{mm}, the r.m.s. value of the displacements is around \SI{1}{mm}. In this case, the contribution to the spin build-up from misalignments of magnets and an EDM are of the same order for $\eta = \num{e-4}$, which corresponds to an EDM of about $\SI{5e-19}{\text{$e$}.cm}$.

Results of benchmarking concerning changes in tune and chromaticity, as well as driven oscillations of the polarization vector, are discussed in Ref.~\cite{Rosenthal:2016zbf}.

\subsubsection{Determination of the invariant spin axis}

To determine the polarization build-up due to the EDM, it is necessary to know the orientation of the invariant spin axis. One current challenge for the precursor experiment is the lack of knowledge about the radial component of the invariant spin axis $\vec{n}$. A possible solution is the determination of $\vec n$ through spin-tracking calculations of the COSY lattice. The EDM, as
well as misalignments of lattice elements, affects the particle trajectories, but also leads to a tilt of the invariant spin axis.

In the case of an ideal ring and a vanishing EDM, the invariant spin axis points in the vertical ($y$) direction, and the spins precess in the horizontal plane. In the presence of an EDM, the invariant spin axis is tilted in the horizontal direction by an angle $\xi_\text{EDM}$, as sketched in \Fref{Fig:EDM_ISA}. This angle is directly proportional to the magnitude of the EDM and can be written as (Eq.~(11) in Ref.~\cite{PhysRevAccelBeams.23.024601})
\begin{equation}
    \tan(\xi_\text{EDM})=\frac{\eta \beta}{2G}\,.
\end{equation}

\begin{figure}
        \centering
        \includegraphics[width=0.6\textwidth]{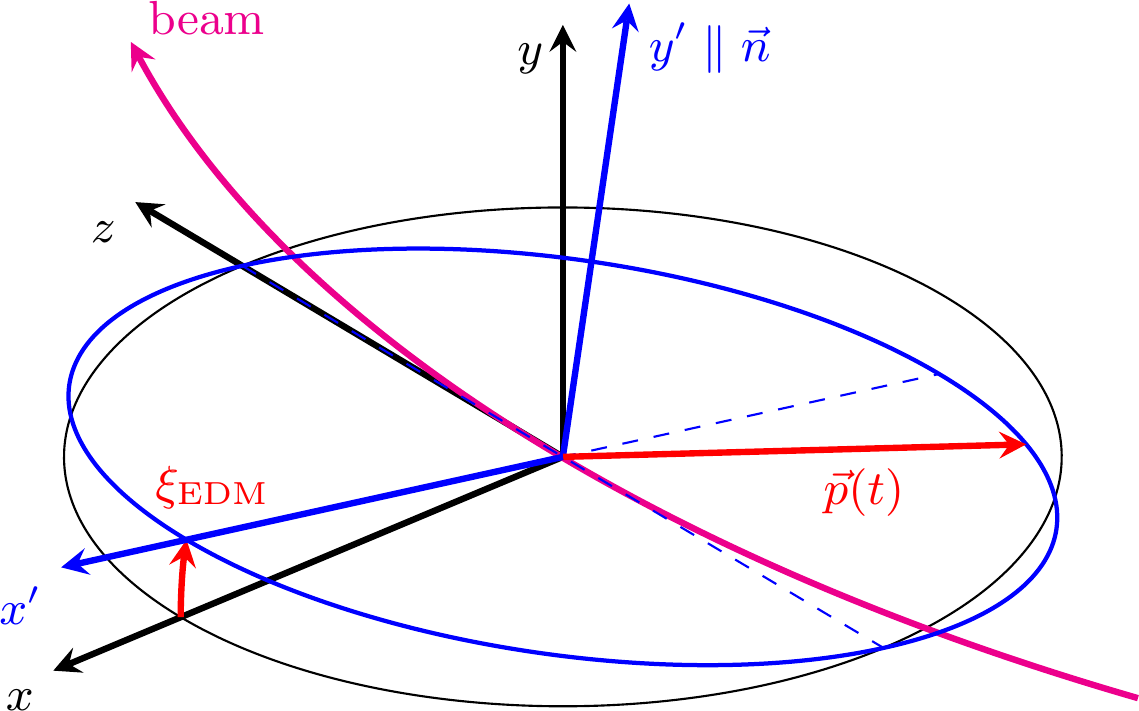}
        \caption{The beam particles move along the $z$ direction. Owing to a permanent EDM, the invariant spin axis $\vec n$ is tilted to the horizontal direction through the angle $\xi_\text{EDM}$. (Figure taken from
Ref.~\cite{PhysRevAccelBeams.23.024601}, licensed under \href{https://creativecommons.org/licenses/by/4.0/}{CC-BY-4.0}.)}
        \label{Fig:EDM_ISA}
\end{figure}

To determine the invariant spin axis, the spin of the reference particle is tracked for $N$ turns, resulting in an ensemble of spin vectors $\vec{s}_j$, where $j \in \mathbb{N}$ and $j \in [ \, 1,N] \,$ (see Sections~III.E.4 and
III.E.5 of  Ref.~\cite{PhysRevAccelBeams.23.024601} and Ref.~\cite{Poncza:2019ith}). For each possible configuration of three spin vectors ($\vec{s}_1$, $\vec{s}_2$, $\vec{s}_3$), an invariant spin axis $\vec{n}_i$ is calculated as
\begin{equation}
\vec{u}_i = \vec{s}_{2,i}-\vec{s}_{1,i}\,, \qquad
\vec{v}_i = \vec{s}_{3,i}-\vec{s}_{1,i}\,, \qquad \text{ and}  \qquad
\vec{n}_i = \frac{\vec{v}_i \times \vec{u}_i}{|\vec{v}_i \times \vec{u}_i|}\,.
\end{equation}
The left panel of \Fref{spins}  illustrates this procedure. The invariant spin axis is  calculated as the average of all $\vec{n}_i$ vectors. Spin tracking is performed using the software library Bmad. The
right panel of \Fref{spins}  shows the spin distribution after tracking through the COSY lattice, including the misalignments of dipoles and quadrupoles, as well as the individual spin vectors $\vec{s}_j$, the invariant spin axes $\vec{n}_i$, and the average invariant spin axis $\langle \vec{n} \rangle$.

\begin{figure} [hb!]
        \centering
        \includegraphics[height=0.25\textheight]{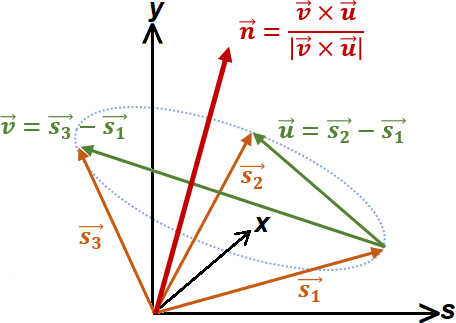}
        \hspace{0.4cm}
        \includegraphics[height=0.25\textheight]{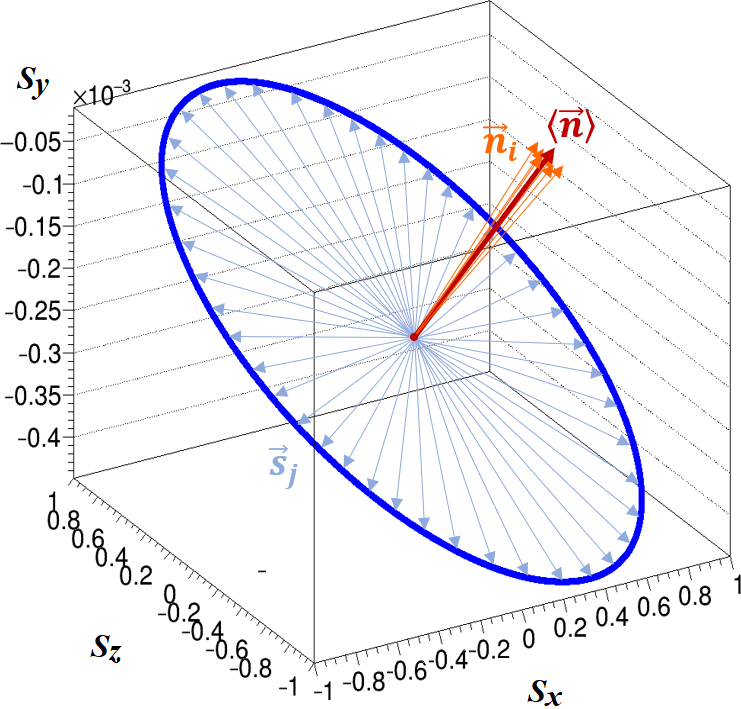}
        \caption{General procedure to calculate the invariant spin axis and spin tracking results. Left: Method for calculating the invariant spin axis from three spin vectors. Right: Spin distribution resulting from misaligned magnets (blue) and average invariant spin axis (red). Note that the magnitude of the spin component $s_y$ has been greatly enhanced, compared with the other two components, to make the tilt of $\langle \vec n \rangle$ better visible.}
        \label{spins}
\end{figure}

\subsection{Prototype proton EDM storage ring}

After design had begun on a prototype EDM storage ring  (see Chapter\,\ref{chap:ptr}  and Ref.~\cite{Lehrach:2019pgw}), operated at proton beam energies between 30 and $50\UMeV$,
spin-tracking simulations were run to study the sensitivity of such a ring for EDM measurements. Spin-tracking simulations have also been run by several groups to investigate the spin motion of $233\UMeV$ frozen-spin protons in dedicated EDM rings\cite{Maier:2012zza,Maier:2012zzc,selcuk,Michaud:2019,Tahar:2019hzv}.

\subsubsection{Simulations using a Runge--Kutta integration method }
 The spin evolution in a storage ring is determined by making use of the Thomas--BMT equation (see \Sref{sec:spin-evolution-electric-magnetic-fields}),
which  describes the spin orientation of relativistic particles in the presence of electromagnetic fields. The latter must be evaluated at each location of the particle following its trajectory. The Thomas--BMT equation is solved together with the equation of motion, for which, in general, a closed-form solution cannot be obtained. Given the high sensitivity aimed at, precise numerical simulations are necessary;  benchmarking these using analytical estimates may help to understand the major systematic effects. An average radial magnetic field of a few attotesla, for instance, yields a vertical spin build-up similar to an EDM  of $\SI{e-29}{\text{$e$}.cm}$. Thus, a precise knowledge of the field at each integration step is mandatory in order to determine its impact on the spin dynamics.

 In the following, several ring lattices are considered, based on the strong focusing, proposed, \eg by V. Lebedev, to achieve the beam requirements suitable for EDM measurements\cite{Lebedev}: the simulated ring consists of four superperiods, each containing five FODO cells\footnote{A structure consisting of focusing and defocusing quadrupole lenses in alternating order with nothing in between is called a FODO lattice, and the elementary cell is called a FODO cell\cite{Holzer:1982419}.}. There are six electric bends per FODO cell, characterized by an \SI{8}{MV/m} radial electric field for \SI{3}{cm} plate separation. In the interfaces between the bending and  straight sections, the  hard  edge  model has been assumed, which means that the electric fields are constant everywhere within each  element and abruptly drop to zero at the edges. Nevertheless, the energy change of the particle was taken into account. This model is particularly useful for simplifying the analysis. Several selected cases of lattice imperfections yielding a vertical spin build-up are discussed next. Further details regarding some of the numerical simulations and their comparison with the analytical estimates can be found in Ref.~\cite{Tahar:2019hzv}, based on the Bogoliubov--Krylov--Mitropolsky method of averages \cite{Bogolyubov}.

\subsubsection{Misalignment of focusing elements}

In the case of misplacement of lattice elements, such as electric quadrupoles, orbit distortions lead to a vertical spin build-up\cite{selcuk}. The latter may exhibit a linear or quadratic increase as a function of time, depending on the amplitude of the perturbation. Several results of spin-tracking simulations for the all-electric ring for the case where one defocusing quadrupole is misaligned by several micrometres are shown in \Fref{vertical_spin_mis2}; a particle with a relative momentum offset of $\Delta p/p = \num{e-5}$ is tracked on the perturbed closed orbit and its spin  recorded after each completed turn. The momentum offset leads to a radial spin component that increases approximately linearly as a function of time. The radial spin component is subsequently rotated into the vertical direction by the constant angular frequency generated by the average slope $y'$ of the vertical orbit inside the electrostatic bending elements. Altogether, this yields a spin build-up that is approximately quadratic as a function of time, and proportional to $s_y \propto y' \Delta p/p$ (see \Fref{vertical_spin_mis2}, and also \Sref{chap:sensitivity:second-order-effects}).

\begin{figure}
\centering
\includegraphics[width=0.9\textwidth]{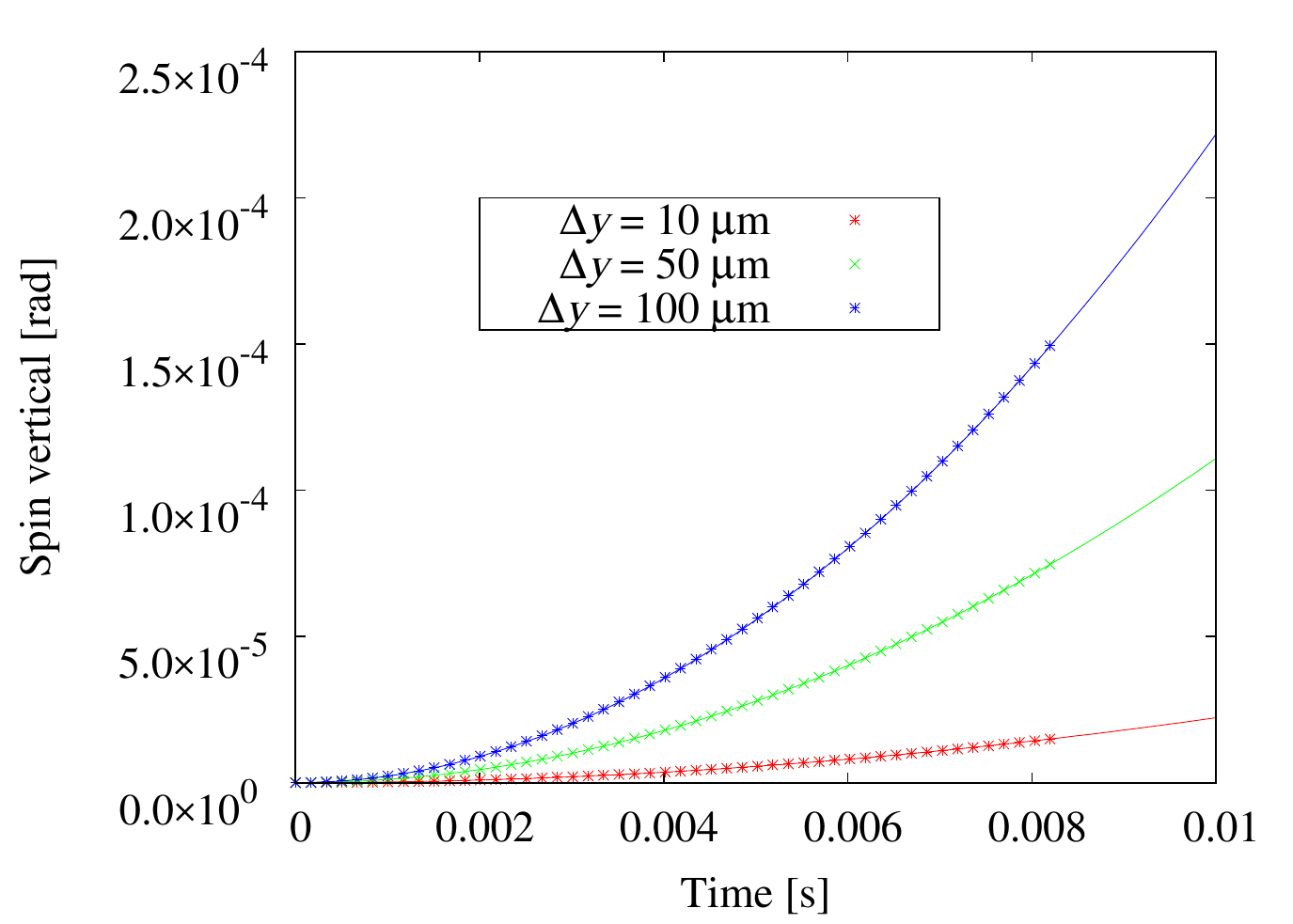}
\caption{Vertical spin as a function of time for a lattice with different misalignment errors and a momentum offset $\Delta p/p = \num{e-5}$. Analytical results are shown as solid lines, while  dotted symbols denote  tracking results. (Figure taken from Ref.~\cite{Tahar:2020gsz}, licensed under \href{https://creativecommons.org/licenses/by/4.0/}{CC-BY-4.0}.)}
\label{vertical_spin_mis2}
\end{figure}

\subsubsection{Geometric phases}
Another simulation example considered here is that of the geometric phases, also referred to as the Berry phases\cite{Berry:1984jv}. Such effects may arise from the non-commutation of spin rotations around different axes, and may dominate a system if the beam energy is very close to the magic one. Let us assume that the beam is injected at the magic energy and that the lattice has alternating magnetic field imperfections, represented by the presence of both vertical and longitudinal magnetic fields that are \SI{90}{\degree} out of phase, as illustrated in \Fref{orb_3d_solen}. In particular, it can be seen that the radial spins are rotated into the vertical plane by means of the longitudinal magnetic fields, such that the leading term of the vertical spin build-up is given by \cite{Tahar:2019hzv}
\begin{align}
\dfrac{\mathrm{d} s_y}{\mathrm{d} t} & \approx \dfrac{1}{c\beta_l C} \left(\dfrac{e}{m} \right)^2 \left(G+\dfrac{1}{\gamma} \right) \dfrac{1+G}{\gamma} B_y L_y B_z L_z \notag \\
                 & \approx \num{5.92e5} B_y L_y B_z L_z \,.
 \label{dsydt1}
\end{align}
Assuming an integrated field perturbation, such that $B_y L_y = B_z L_z = \SI{1}{nT.m}$, yields a spin precession rate of $\approx$\,$\SI{5.92e-4}{nrad/s}$, which is well below the EDM signal level. \Figure[b]~\ref{orb_3d_solen2} shows a comparison of the tracking simulations with the first-order analytical estimate of the spin build-up due to the Berry phases, exhibiting reasonable agreement of both estimates. In addition, it is important to note that such effects may be cancelled using counter-rotating beams.

\begin{figure} [hbt!]
\centering
\includegraphics[width=0.9\textwidth]{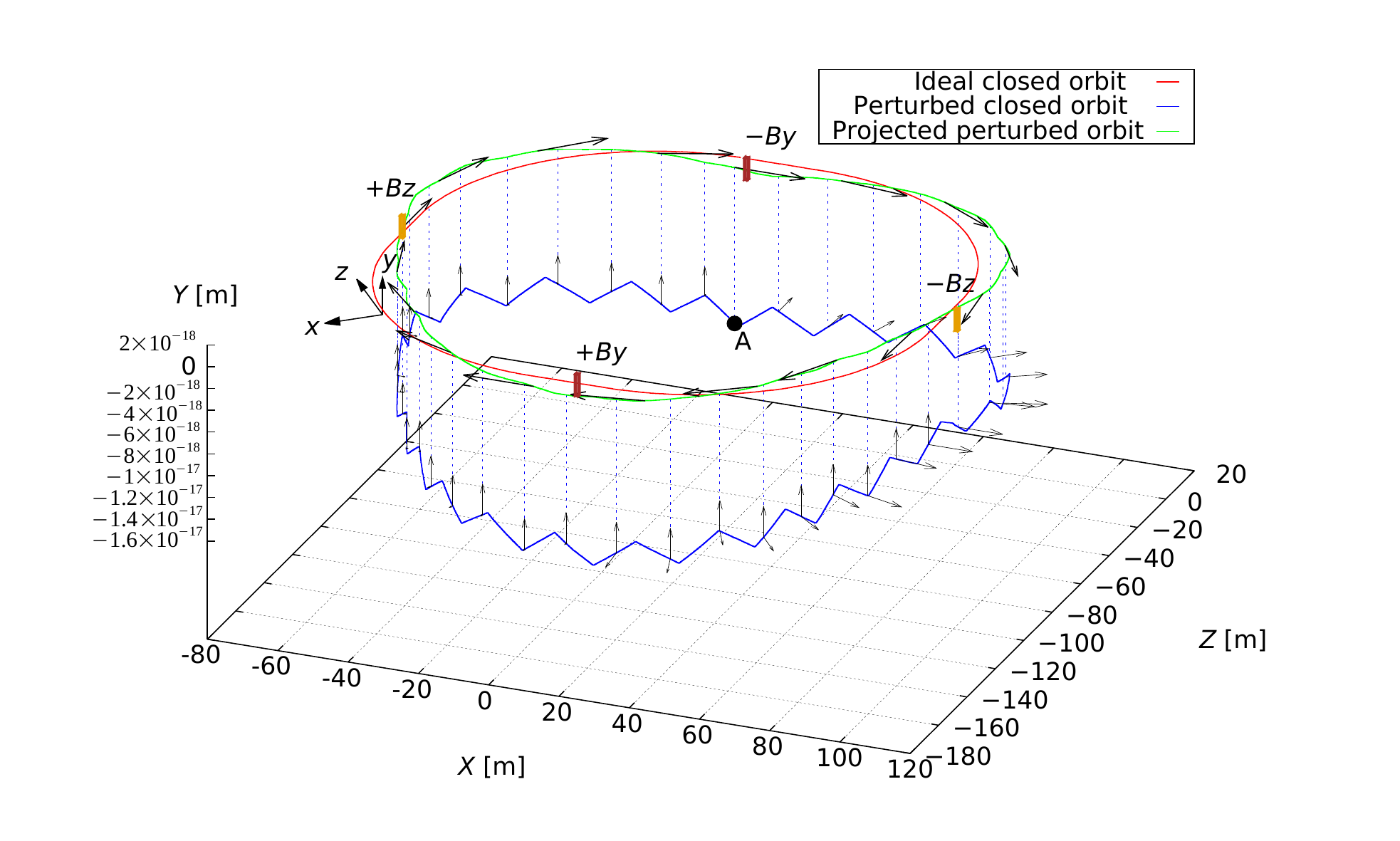}
\caption{Spin and orbit evolution for a lattice with alternating magnetic field imperfections: a vertical magnetic field $B_y$ yields a horizontal spin component, which is rotated into the vertical plane by means of a longitudinal field component $B_z$. The closed orbit of the perturbed motion is shown in blue; the particle motion is clockwise, starting from Point A. (Figure taken from Ref.~\cite{Tahar:2020gsz}, licensed under \href{https://creativecommons.org/licenses/by/4.0/}{CC-BY-4.0}.)}
\label{orb_3d_solen}
\end{figure}

\begin{figure} [hbt!]
\centering
\includegraphics[width=0.7\textwidth]{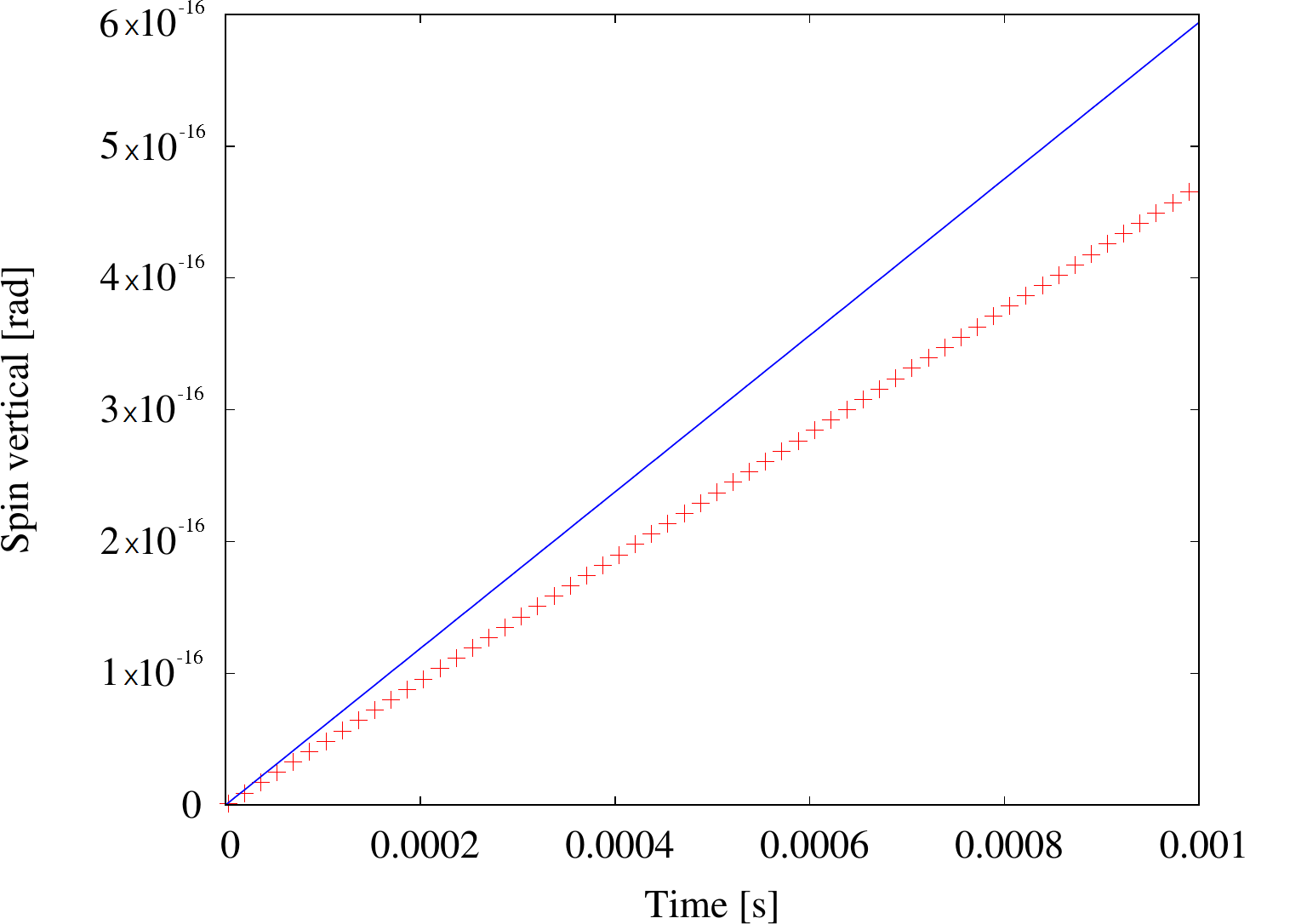}
\caption{Vertical spin build-up from tracking simulations and comparison with the analytical estimate given by \Eref{dsydt1}. (Figure taken from Ref.~\cite{Tahar:2020gsz}, licensed under \href{https://creativecommons.org/licenses/by/4.0/}{CC-BY-4.0}.)}
\label{orb_3d_solen2}
\end{figure}

\clearpage

\begin{flushleft}

\end{flushleft}
\end{cbunit}

\begin{cbunit}

\csname @openrighttrue\endcsname 
\chapter{Roadmap and timeline}
\label{Chap:RoadMap}

\section{CPEDM strategy}

As emphasized in Chapter\,\ref{Chap:Strategy}, this challenging project needs to proceed in stages (outlined in \Fref{fig:figure3}), as follows.

\begin{itemize} 
\item COSY will continue to be used for as long as possible for the continuation of critical R\&D associated with the final experiment design. An important requirement is to test as many of the results as possible with protons, where the larger anomalous magnetic moment leads to more rapid spin manipulation speeds.
\item The precursor experiment carried out at COSY (stage 1 in \Fref{fig:figure3})  will be completed and analysed. Some data will be taken with an improved version of the RF Wien filter with better electric and magnetic field matching.
\item The next stage is to design, fund, and build a prototype storage ring (PTR) to address critical questions concerning the features of the EDM ring design (stage 2 in \Fref{fig:figure3}) .
\begin{itemize}
 \item At \SI{30}{MeV}, the ring, operated with an electric field only, is capable of storing counter-rotating beams, but, in this case, frozen spin will not be possible.
 \item At 30 and \SI{45}{MeV}, with an additional vertical magnetic field, the frozen-spin condition can be met. However, the magnetic fields also prevent the CW and CCW beams from being stored at the same time. Nevertheless, EDM experiments may be performed with these two beams using alternating fills.
\end{itemize}
\item Following stage 2, the focus will be to create the final ring design
(stage 3 in \Fref{fig:figure3}), then fund and construct it.
\item Once the ring is ready, the longer-term aim will be to commission and operate the final ring, improving it with new versions as  understanding and mitigation of systematic errors and other experimental issues are further improved.
\end{itemize}

\begin{figure}
\centering
\includegraphics[width = 1\textwidth]{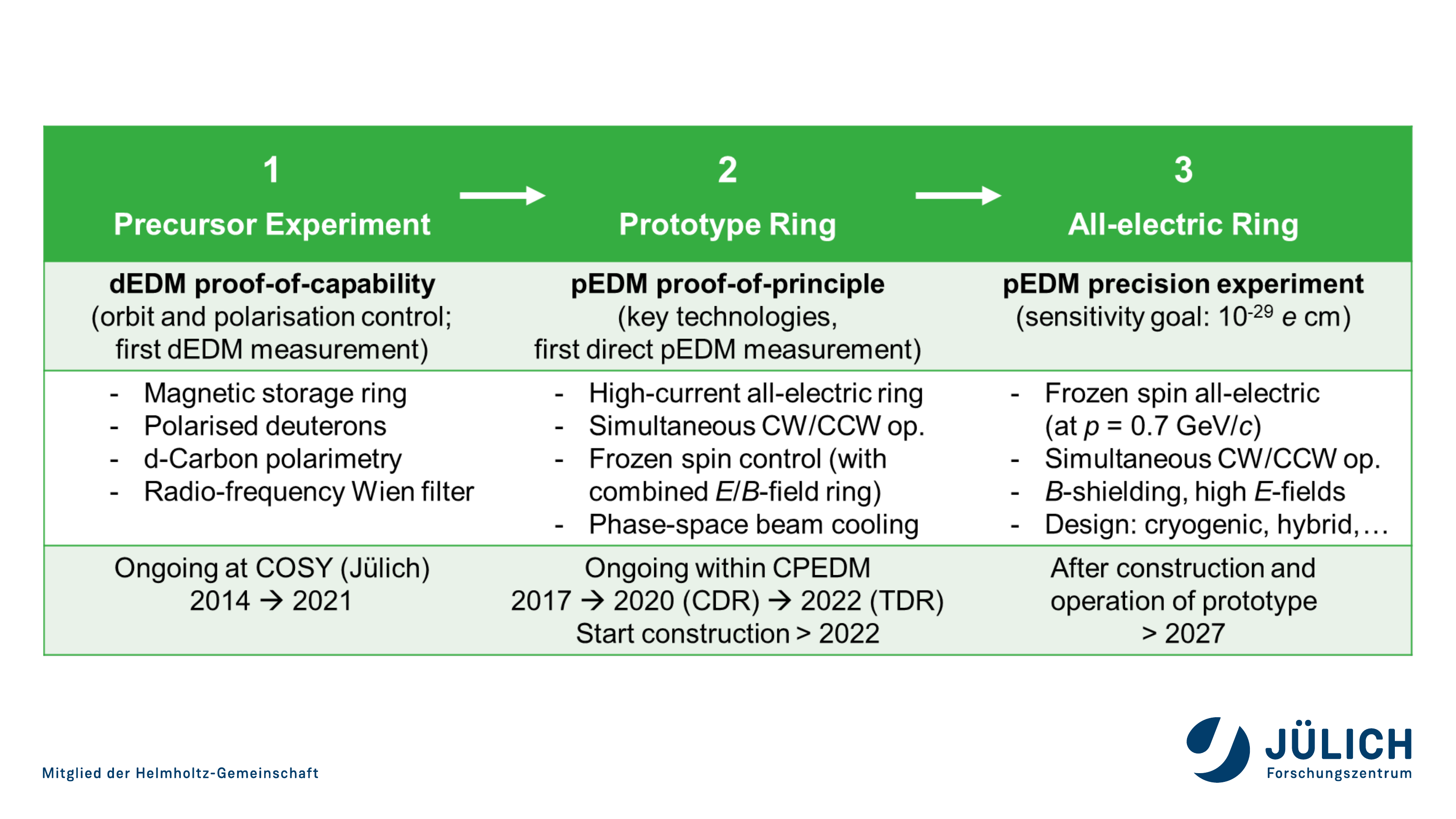}
\caption{\label{fig:figure3}
Important features of the proposed stages of the storage ring EDM strategy (Figure identical to \Fref{fig:strat:figure3}, reproduced here for convenience).}
\end{figure}

Future scientific goals may include conversion of the ring to crossed electric and magnetic field operation so that other particle species besides protons could be examined for the presence of an EDM. An interesting option is the search for \textit{oscillating} EDMs resulting from coupling to axions or axion-like particles (ALPs). To this end, the ring may be operated in a mode where the polarization rotates  with an oscillation frequency of the EDM with respect to individual stored bunches in the ring. Most systematic effects that limit the sensitivity for \textit{static} EDMs are absent  in this mode of operation.

\section{Timeline}

As shown in \Fref{fig:figure3}, a staged approach is pursued, with stage
1 (`precursor experiment') currently ongoing. This investigation is partially funded by an ERC advanced grant, which runs until September 2021. The next stage (stage 2, `prototype ring')  started in 2017; a CPEDM task force is working on a conceptual design report (CDR, due in 2021) and will subsequently finalize a technical design report (TDR, expected in 2022).
Currently, about 5 years are envisaged for building and operating the prototype. Only after that, will it be conceivable to design, build, and operate the final ring (stage 3, all-electric ring).

A more detailed timeline is presented in \Fref{fig:my_label}, which follows the anticipated evolution of the storage ring EDM project through several stages and events.  The main events are as follows.
\begin{itemize}
 \item Experimental work will continue at COSY to study feasibility issues regarding electron cooling.

 \item A first experimental investigation has recently been carried out at COSY towards the development of a search for axions or ALPs.

 \item Refinements are underway for the precursor experiment, as a first measurement of the deuteron EDM.

 \item The next program-oriented funding (PoF) period of the Helmholtz Association (HGF) will start at the beginning of 2021. Work shall begin on a conceptual design report (CDR), followed by a technical design report (TDR), for the creation of the PTR.

 \item Future efforts at COSY shall also be  guided towards enhancing the capabilities with polarized proton beams, to achieve a similar performance
to that already available for deuterons (red band in \Fref{fig:my_label}; the light red band indicates continued optimization work at the PTR to further enhance  polarized proton beam capabilities).

\item Other types of research in related symmetries (axion or ALP search) shall continue at COSY (green band in \Fref{fig:my_label}), and later on
 at the PTR (light green band). Work will also start on the construction of the electric version of the prototype ring (orange).

\item Commission of the PTR  with a first beam at \SI{30}{MeV} may start in 2025 to demonstrate high intensities and counter-rotating two-beam operation.

\item Operation of the PTR  at \SI{45}{MeV} with magnetic bending to allow for frozen-spin operation may begin in 2028.

\item As new feasibility studies with the PTR come to fruition, work will
begin with a CDR and TDR for the proton EDM experiment. This project will be commissioned in the mid-2030s.
\end{itemize}

\begin{figure} [hbt!]
    \centering
    \includegraphics[width = \textwidth]{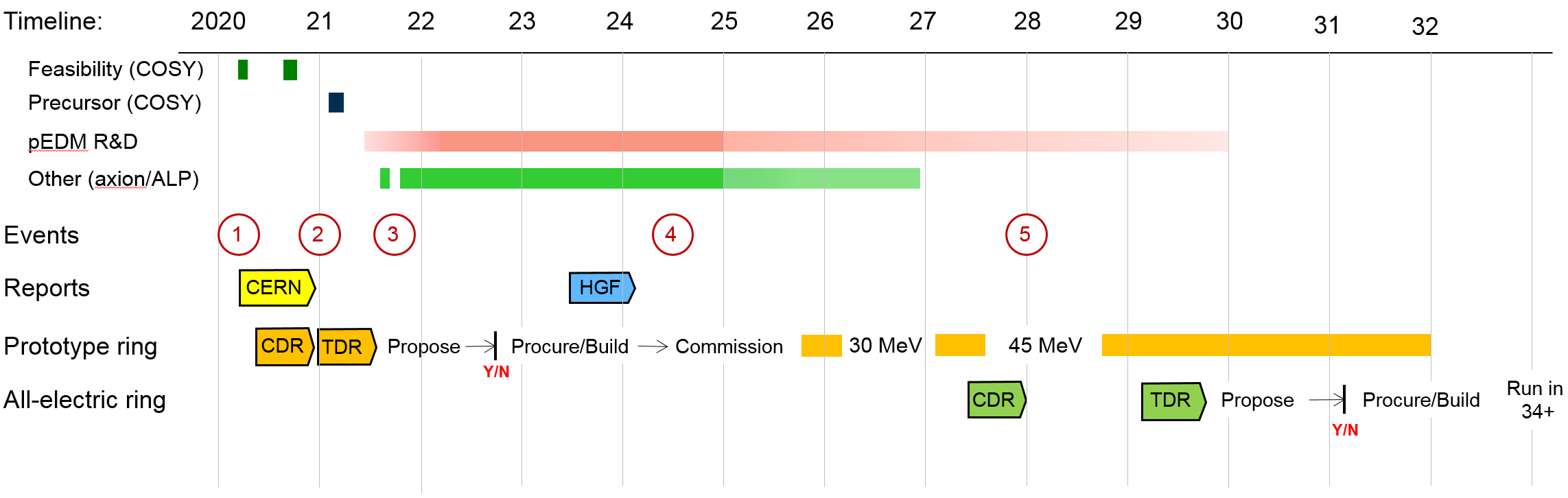}
    \label{fig:my_label}
    \caption{Timeline for  anticipated evolution of the storage ring EDM project: \textcolor{purple}{\protect\circled{1}} strategic programme evaluation
by the Helmholtz Association (HGF); \textcolor{purple}{\protect\circled{2}} start of HGF funding period; \textcolor{purple}{\protect\circled{3}} end of `srEDM' advanced grant from European Research Council (2016--2021)\cite{ERC-694340}; \textcolor{purple}{\protect\circled{4}} HGF mid-term review; \textcolor{purple}{\protect\circled{5}} start of next HGF funding period.
    }
\end{figure}

The prototype ring and the all-electric ring are considered host-independent. If the prototype is built at Jülich, it could take full advantage of the existing facility for the production of polarized proton (and deuteron) beams. The PTR may instead be built at another site, provided that a comparable beam preparation infrastructure is made available. In either case, the lattice design of the PTR will mimic that of the high-precision all-electric final ring, in order to test as many features as possible on a smaller scale.


\renewcommand\bibname{Reference}
\begin{flushleft}

\end{flushleft}
\end{cbunit}

\clearpage


\begin{appendices}

\renewcommand{\appendixname}{Appendix}
\fancyhead[RE]{\textsc{\appendixname\ \thechapter}}

\begin{cbunit}

\chapter{Results and achievements at Forschungszentrum J\"ulich}
\label{app:cosy}

This appendix describes the results and accomplishments achieved so far. It includes results obtained at the Cooler Synchrotron (COSY) at Forschungszentrum J\"ulich, as well as results from the J\"ulich theory group. Some activities and achievements are described in dedicated chapters, see, \eg Chapters\,\ref{Chap:polarimetry}  and~\ref{Chap:SpinTracking}.

\section{Results and achievements at COSY}
For most of the studies,  the parameters listed in \Tref{tab:cosyparam} were used.

\begin{table} [h]
\caption{\label{tab:cosyparam} COSY operating parameters for most of the investigations reported  in this appendix}
\centering
\begin{tabular}{l l}
\hline \hline
COSY circumference                                      &  \SI{183}{m} \\
Deuteron momentum                                       &  \SI{0.970}{GeV/$c$} \\
Lorentz factors $\beta (\gamma)$                        &    0.459 (1.126) \\
Deuteron magnetic anomaly $G$                           &  $\approx$\,$-0.143$ \\
Revolution frequency $f_{\mathrm{rev}}$                 &  \SI{750.6}{kHz}\\
Cycle length                                            &  $100 - \SI{1500}{s}$  \\
Number of stored particles/cycle                        & $\approx$\,$\num{e9}$ \\
\hline \hline
\end{tabular}
\end{table}

\subsection{High-precision spin tune measurements}
Although not directly related to the EDM measurement in a dedicated storage ring using the frozen spin method, the measurement of the fast horizontal \SI{120}{kHz} precession of the polarization vector around the magnetic guiding field in the horizontal plane of the ring constitutes an import step towards understanding and controlling the spin  precession in a storage ring.

In an ideal planar magnetic storage ring, the spin tune---defined as the number of spin precessions per turn---is given by $\nu_s = \gamma G$. For the conditions at COSY, given in \Tref{tab:cosyparam}, $\nu_s \approx -0.16$.  At $p=\SI{970}{MeV}/c$, the deuteron spins coherently precess at a frequency of about \SI{120}{kHz}.  The spin tune was deduced from the up--down asymmetry of deuteron--carbon scattering. In a time interval of \SI{2.6}{s}, the spin tune was determined with a precision of the order of \num{e-8}, and of \num{e-10} for a continuous \SI{100}{s} accelerator cycle\cite{Eversmann:2015jnk}, as shown in \Fref{spintune}.

\begin{figure}
 \centering
 \includegraphics[width=0.65\textwidth]{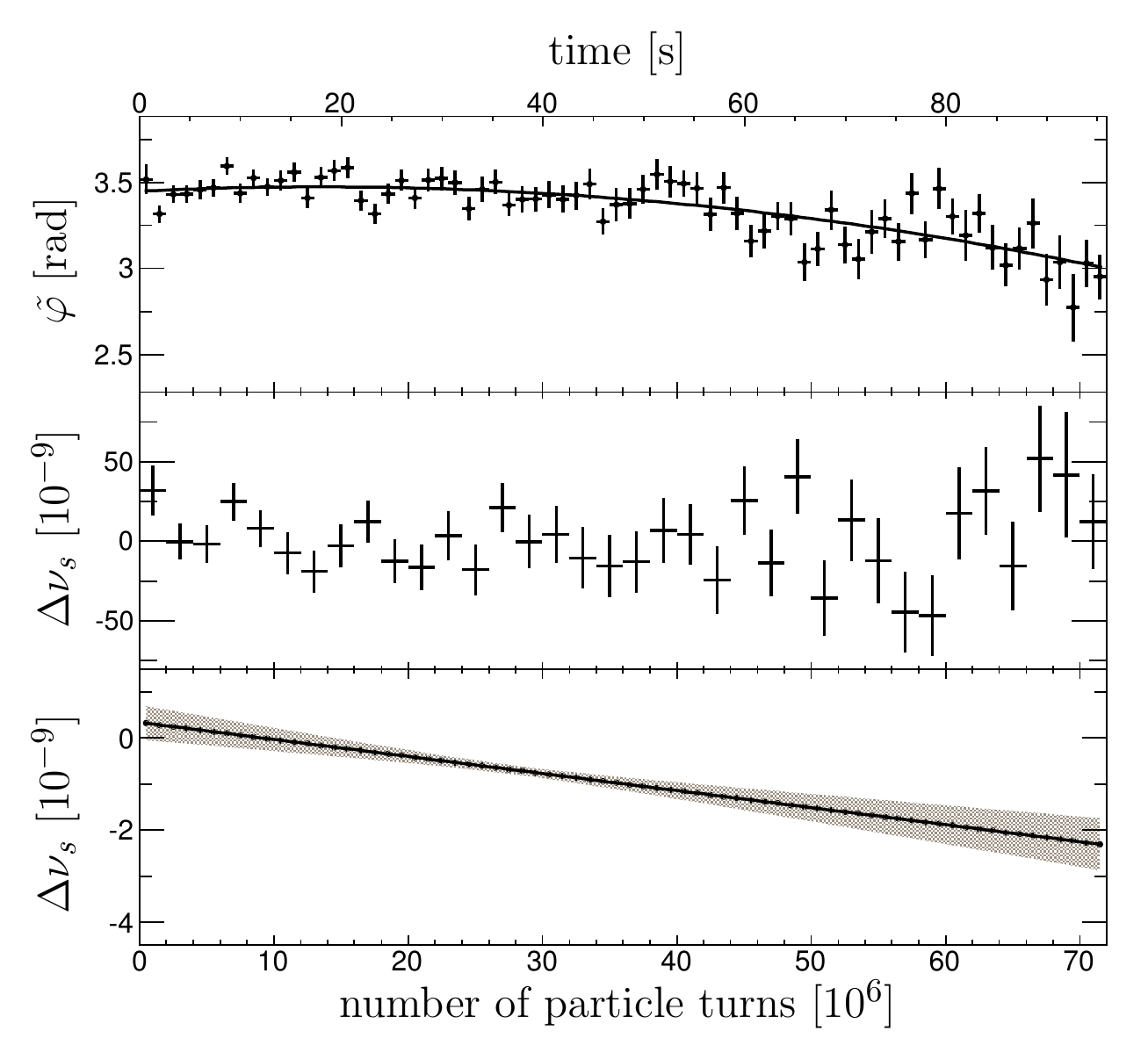}
 \caption{Top: Phase  of the polarization vector in the horizontal plane, evaluated close to the spin revolution frequency of the polarization vector using Fourier analysis over \num{e6} turns.  (centre) Spin tune change  obtained from two consecutive phase measurements. Bottom: Spin tune change obtained from the parabolic fit shown in the upper panel. (See Ref.~\cite{Eversmann:2015jnk} for further details. Figure reused with permission from the American Physical Society.)
 \label{spintune}}
\end{figure}

To appreciate this high relative precision of $\sigma_{\nu_s}/\nu_s \approx \num{6e-10}$ in a \SI{100}{s} cycle, a comparison with the equivalent  quantity obtained in the muon $(g-2)$ experiment\cite{PhysRevD.73.072003} is helpful. Here, the precision reached is about $\sigma_{\nu_s}/\nu_s \approx \num{e-6}$ per year, \ie a ppm measurement of $a=(g-2)/2$ during \SI{1}{year}. The three orders of magnitude higher precision obtained in a much shorter time is mainly explained by the fact that the cycle length of \SI{100}{s} used in Ref.~\cite{Eversmann:2015jnk} is much larger than the \SI{600}{\micro \second} used in Ref.~\cite{PhysRevD.73.072003}.



\subsection{Horizontal polarization lifetime}
\label{appa:longsct}

To achieve a high statistical precision in an EDM experiment in a ring, the spin-coherence time should be as long as possible, of the order of about \SI{1000}{s}. A rough estimate shows that this is not easily accomplished. Without an RF system, the initial momentum spread of a stored beam of typically $\Delta p/p \approx \num{e-5}$, which corresponds to  $\Delta \gamma/\gamma  \approx \num{2e-6}$, leads to a corresponding increase of the spin tune spread $\Delta  \nu_s/\nu_s$. The spin tune is given by $\nu_s = \gamma G$; thus, after $\approx$\,$\num{e6}$ turns, (\ie after $\approx$\,$\SI{1}{s}$), the in-plane polarization is lost. Using a bunched beam, however, first-order effects in $\Delta p/p$ can be cancelled and the spin-coherence time may be a few seconds.

Using a combination of  beam bunching, electron cooling, sextupole field corrections, and the suppression of collective effects through limitation of the beam current, a deuteron beam polarization lifetime of \SI{1000}{s} in the horizontal plane of the magnetic storage ring COSY could be achieved
\cite{Guidoboni:2016bdn}. The results from a recent measurement are shown in \Fref{fig:cosy:sct}.

\begin{figure}
\centering
  \includegraphics[width=0.7\textwidth]{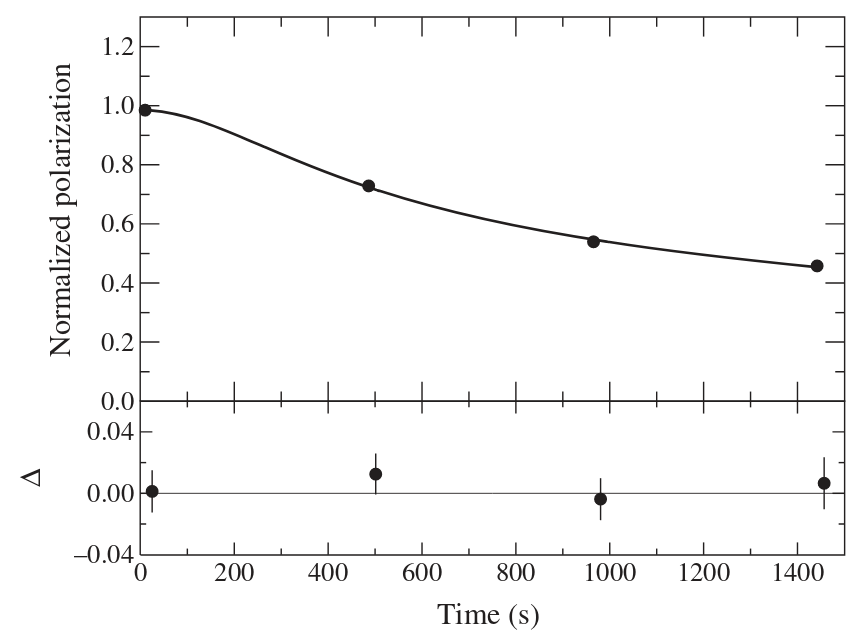}
\caption{Polarization in the horizontal plane as a function of time. The line shows a comparison with a model (see Ref.~\cite{Guidoboni:2016bdn}) and the lower panel shows the deviation with respect to the model. (Figure taken from Ref.~\cite{Guidoboni:2016bdn}, reused with permission from the American Physical Society.) \label{fig:cosy:sct}}
\end{figure}

\subsection{Feedback and control of polarization}
\label{appa:spincontrol}

The precise measurement of the horizontal spin precession, together with long spin-coherence times, allowed us to set up a polarization feedback system. In a dedicated ring, its role would be to ensure that the polarization vector remains always (anti)parallel to the momentum vector of the stored particles,  in order to maximize the statistical sensitivity.

A feedback system making use of the polarization measurement at the revolution frequency of a \SI{0.97}{GeV/$c$} bunched deuteron beam at COSY was constructed in order to control both the frequency of the precession ($\approx$\,$\SI{120}{kHz}$) and the phase of the horizontal polarization component. Real-time synchronization with a radio frequency (RF) solenoid enabled  rotation of the polarization out of the horizontal plane, demonstrating the feedback method to manipulate the polarization, as shown in \Fref{fig:feedback_buildup}. In particular, the rotation rate shows a sinusoidal function of the horizontal polarization phase (relative to the RF solenoid), which was controlled to within one standard deviation range of $\sigma = 0.21$\,rad (see \Fref{fig:feedback_phase}). The minimum possible adjustment was \SI{3.7}{mHz} out of a revolution frequency of \SI{753}{kHz}, which changes the precession rate by \SI{26}{mrad/s} \cite{Hempelmann:2017zgg}. These capabilities fulfil the requirements for employing a dedicated storage ring for EDM measurements.

\begin{figure} [ht!]
 \centering
 \subfigure[\label{fig:feedback_buildup} Left--right asymmetry, proportional to the vertical polarization as a function of time. Initially, the polarization is up (red points) or down (black points), depending on the injected spin state. At $t \approx \SI{88}{s}$, the polarization is flipped into the horizontal plane using the RF solenoid. At $t \approx \SI{116}{s}$, the solenoid is switched on again.]{\includegraphics[width=0.45\textwidth]{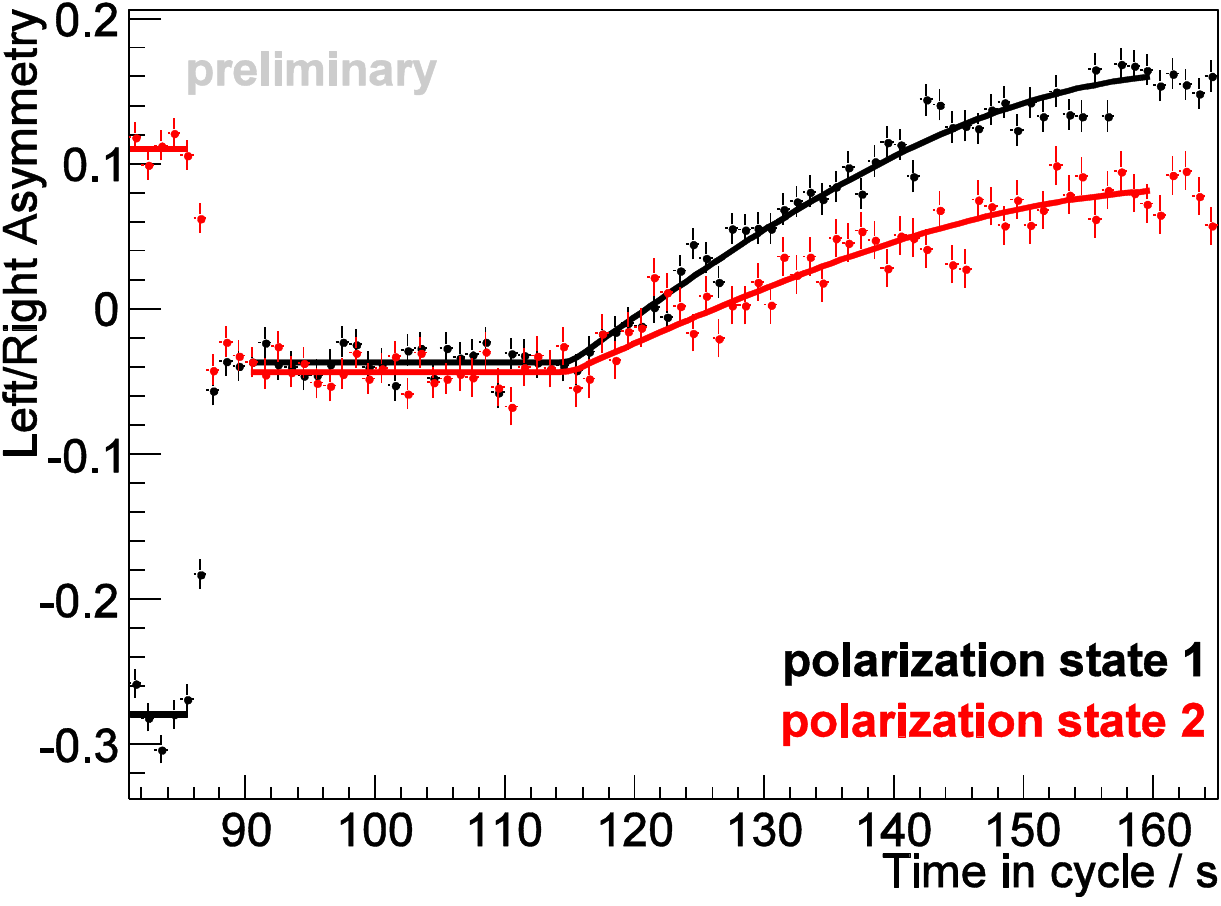}}\hspace{0.2cm}
 \subfigure[\label{fig:feedback_phase} Top: Phase as a function of time with feedback off (blue) and on (red).  The red points stay within an r.m.s. of \SI{0.21}{rad} (grey band). Bottom: Adjustments of the COSY frequency.]{ \includegraphics[width=0.52\textwidth]{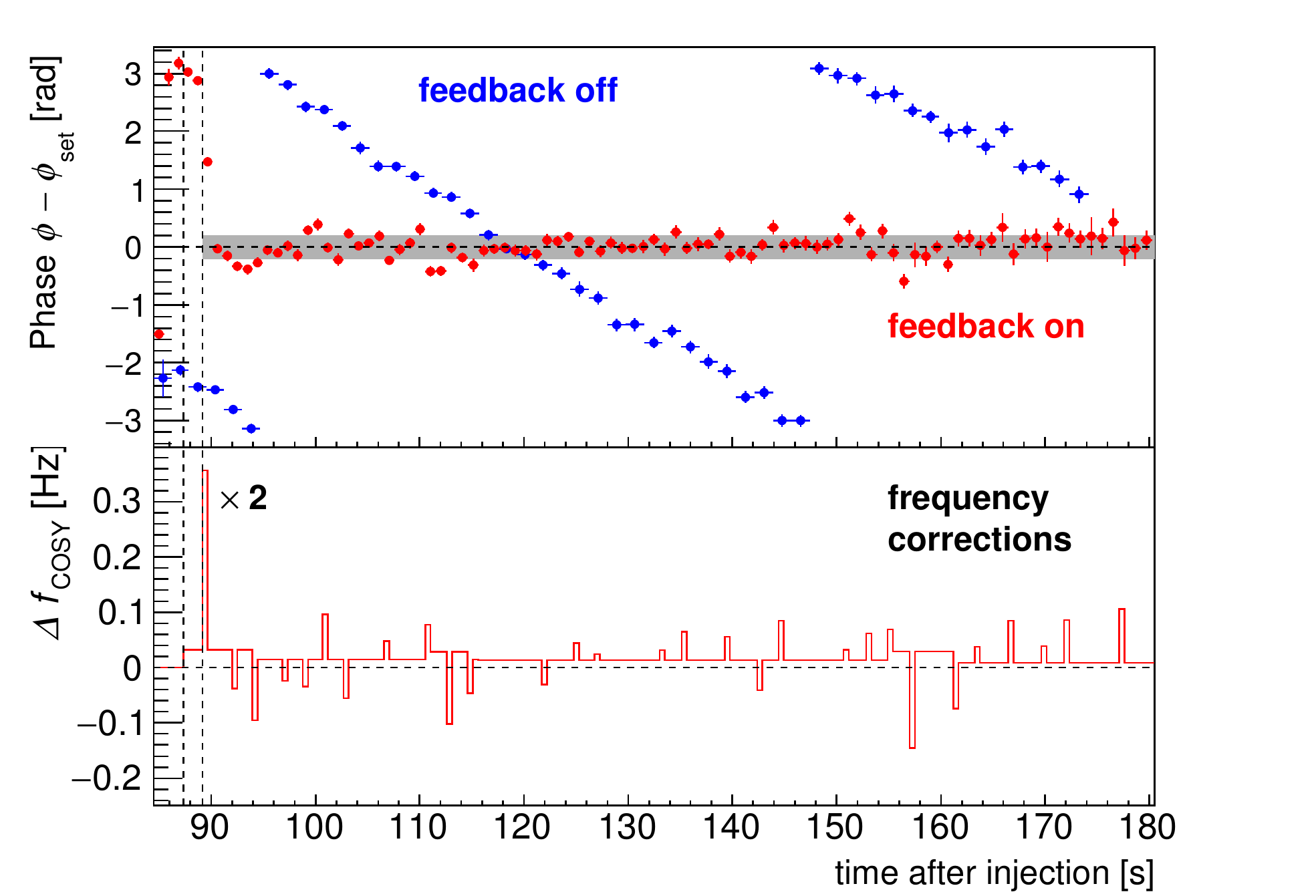}}
\caption{Polarization feedback system and phase control
\label{fig:Fred}}
\end{figure}


\subsection{Invariant spin axis measurements}

Another application of the precise spin tune measurement is the measurement of the invariant spin axis. Reference~\cite{Saleev:2017ecu} describes this
in detail. It is motivated by the fact that precision experiments, such as the search for EDMs of charged particles using storage rings, call for an understanding of the spin dynamics with unprecedented accuracy.  New methods based on the spin tune response of a machine to artificially applied longitudinal magnetic fields, called `spin tune mapping', have been developed. The technique was experimentally tested in 2014 at COSY and, for the first time, the angular orientation of the stable spin axis at two different locations in the ring has been determined to an unprecedented accuracy of better than \SI{2.8}{\micro \radian}\cite{Saleev:2017ecu}. Related ideas are elaborated in some detail in  Appendix\,\ref{Chap:spin-tune-mapping}.


\subsection{Radio-frequency Wien filter for spin manipulation}
\label{appa:rfwf}

In a purely magnetic storage ring like COSY, an EDM will generate an oscillation of the vertical polarization component. For a \SI{970}{MeV/$c$} deuteron beam with a spin precession frequency of about \SI{120}{kHz}, a tiny amplitude of the oscillating polarization is expected, \eg $\num{3e-10}$ for an EDM of $d = \SI{e-24}{\text{$e$}.cm}$. To allow for a build-up of the vertical polarization proportional to the EDM, an RF Wien filter must be operated. Such a device  was developed and constructed\cite{Slim:2016pim,Slim:2016dct}, as shown in \Fref{fig:cosy:wf}. The RF Wien filter was installed in COSY in May 2017. A first commissioning beam time was successfully run in June 2017.  During the 2018 test beam time, the RF Wien filter was operated  with magnetic (electric) field integrals of \SI{0.019}{T mm} (\SI{2.7}{kV}). First results obtained with this device are presented in Chapter\,\ref{Chap:Precursor}.

\begin{figure} [hbt!]
\centering
\includegraphics[height=0.2\textheight]{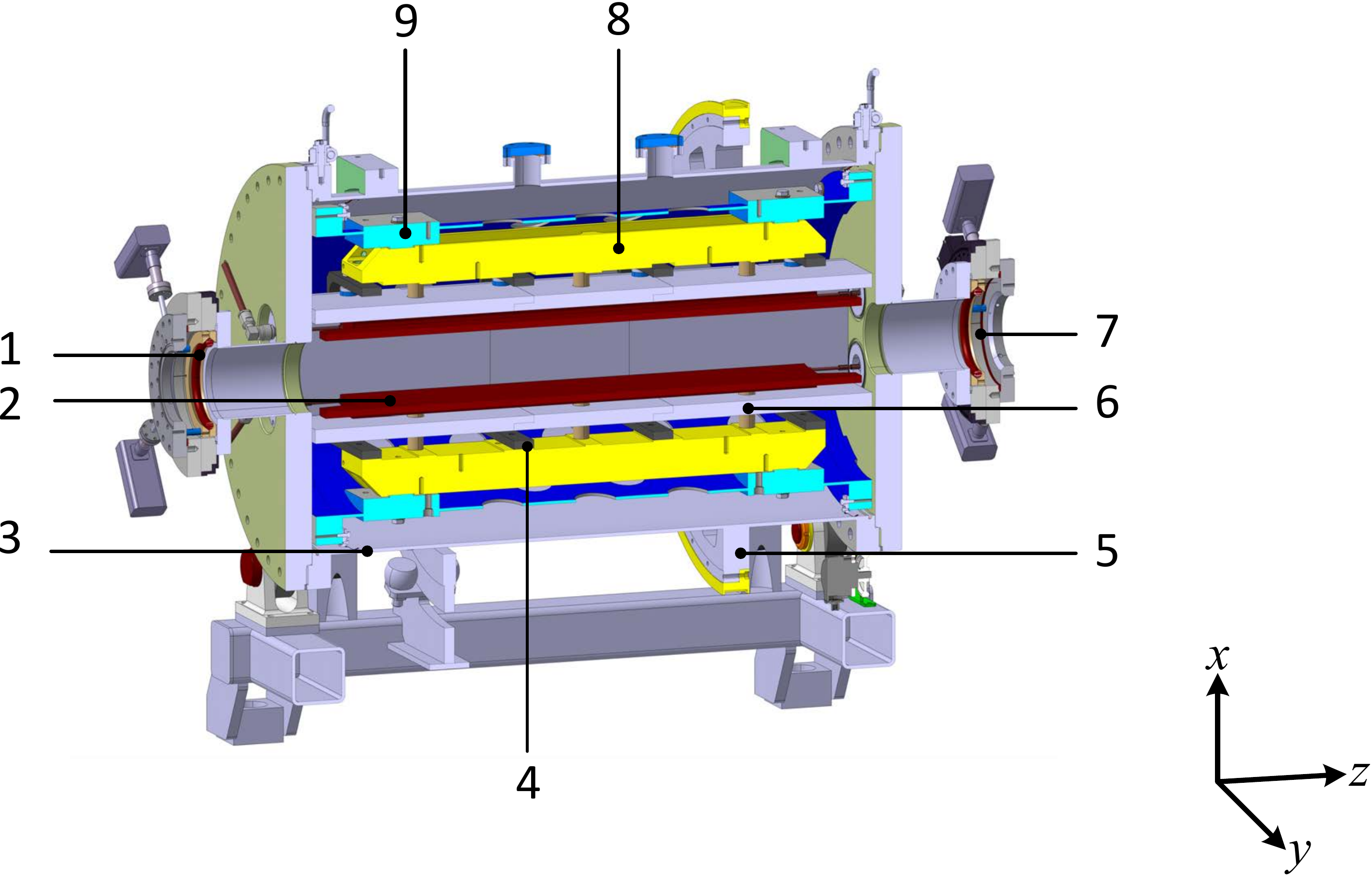}\hspace{0.2cm}
\includegraphics[height=0.2\textheight]{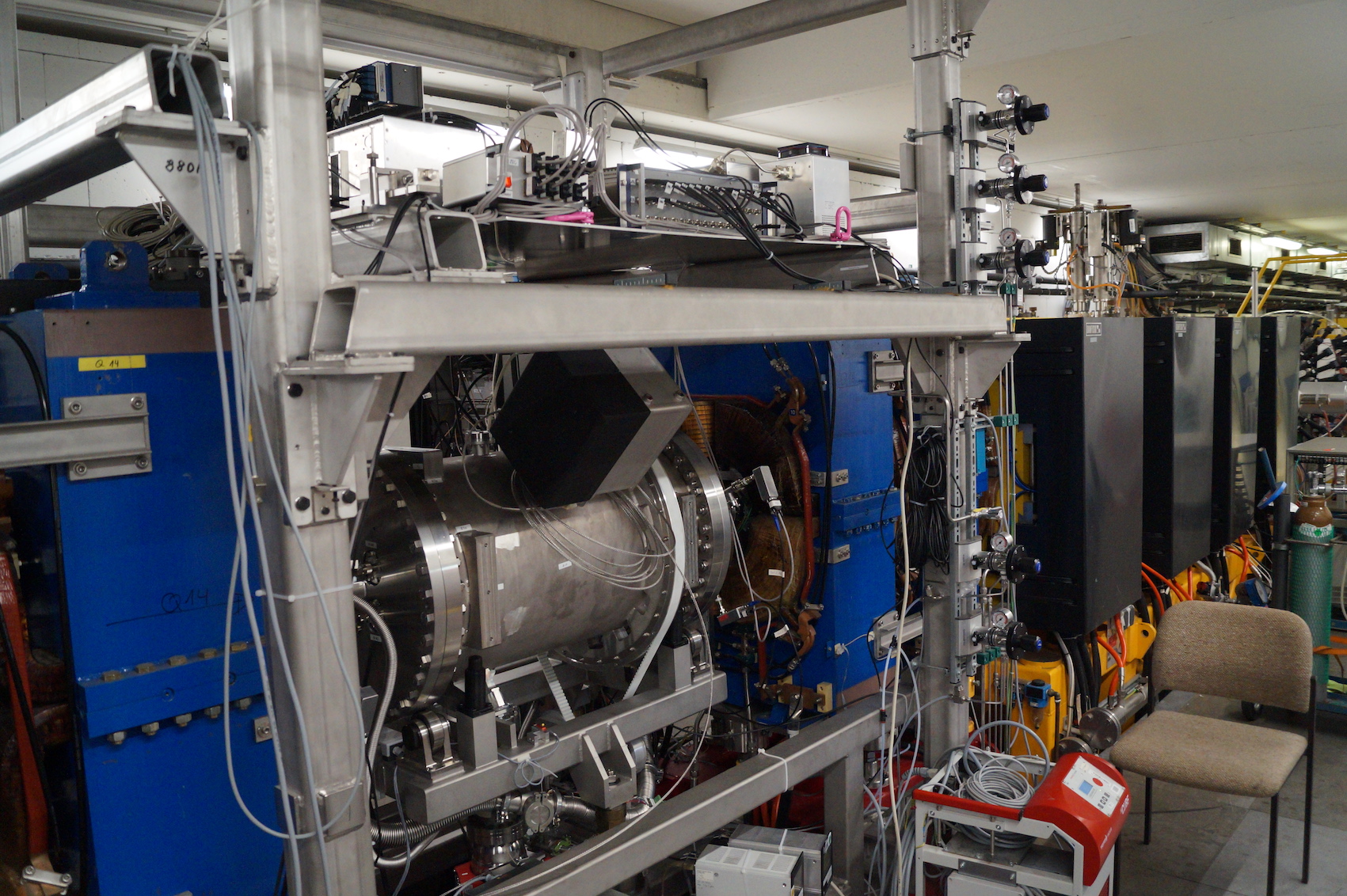}
 \caption{Left: Design model of the RF Wien filter showing the parallel-plate waveguide and the support structure: 1, beam position monitor; 2, copper electrodes; 3, vacuum vessel; 4, clamps to hold the ferrite cage; 5, belt drive for \SI{90}{\degree} rotation, with a precision of \SI{0.01}{\degree} (\SI{0.17}{mrad}); 6, ferrite cage; 7, CF160 rotatable flange; 8, support structure for the electrodes; 9, inner support tube.  The axis of the waveguide points along the $z$-direction, the plates are separated along $x$, and the plate width extends along $y$. During the EDM studies, the main field component $E_x$ points radially outwards and $H_y$  upwards with respect to the stored beam. (Figure taken from Ref.~\cite{Slim:2016pim}, reused with permission from the author.) Right: Photograph of RF Wien filter installed in COSY. \label{fig:cosy:wf}}
\end{figure}







\subsection{Measurement of deuteron carbon and proton carbon analysing powers}
To measure the vertical polarization proportional to the EDM, deuterons or protons
are scattered elastically from a carbon target. To achieve high accuracy, the analysing power should be large and should be known with small uncertainties. A series of measurements were made. \Figure[b]~\ref{fig:deutern_analpow} shows the analysing power of the deuteron--carbon scattering method for various beam energies as a function of the polar angle of the deuteron in the laboratory system\cite{F.Mueller2020}. Data using a polarized proton beam were also taken, which are currently
being analysed.
\begin{figure}[tb]
\centering
\includegraphics[width=0.7\textwidth]{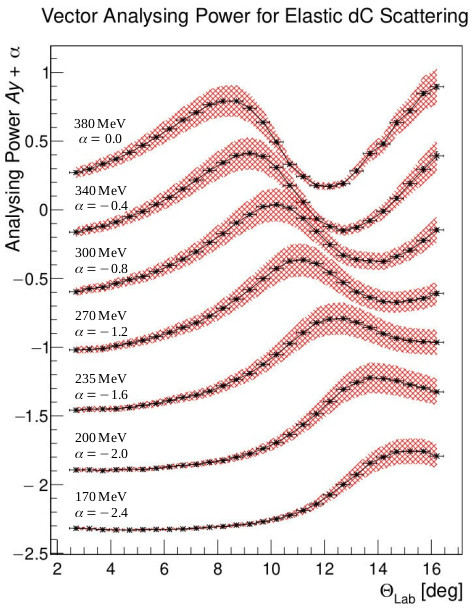}
  \caption{Reconstructed vector analysing power for deuteron beam energies of (from top to bottom) 380, 340, 300, 270, 235, 200, and \SI{170}{MeV}. The curves are sequentially offset by 0.4 for better readability. The statistical errors are indicated by the black error bars on the data points. The systematic errors are shown as red regions. (Figure taken from Ref.~\cite{F.Mueller2020}, licensed under \href{https://creativecommons.org/licenses/by/4.0/}{CC-BY-4.0}.)
\label{fig:deutern_analpow}}
\end{figure}

\subsection{Orbit control}
Systematic errors for EDM measurement occur, for example, as a result of magnet misalignments and orbit offsets. At COSY, many new devices and procedures could be tested and implemented to improve the orbit. First, an automatized  orbit control system was implemented,  allowing the orbit to be corrected in real time. This system reduces the orbit correction procedure from about \SI{10}{\hour} to less than \SI{1}{\hour}. As an example, \Fref{fig:orbit_control} shows the result of an orbit after correction.  The r.m.s. of the horizontal (vertical) orbit is \SI{1.46}{mm} (\SI{0.90}{mm}).

\begin{figure}
    \centering
     \includegraphics[width=\textwidth]{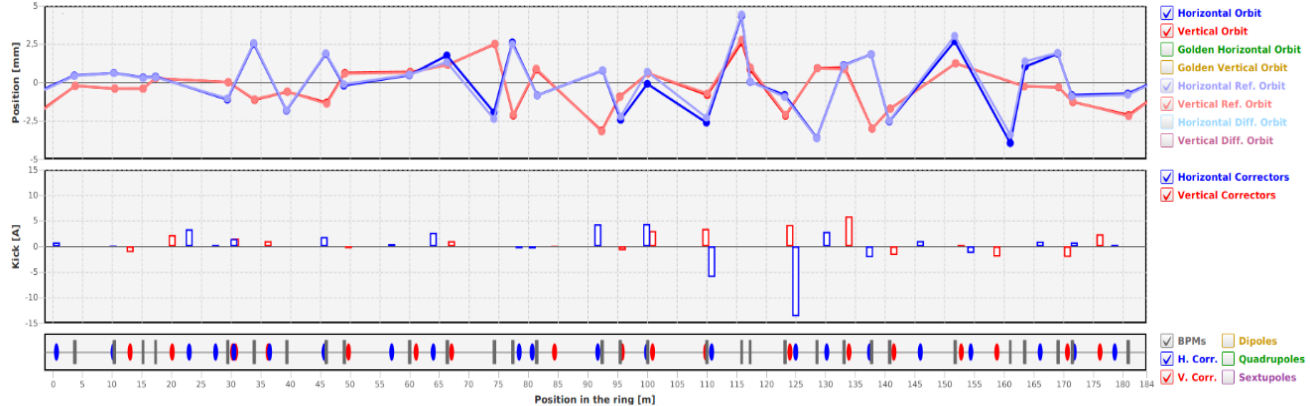}
    \caption{COSY orbit measurement. The upper plot shows the vertical (red) and horizontal (blue) orbit as a function  of the longitudinal position in COSY.  The desired  vertical and horizontal  orbits  coincide with the $x=y=0$ line. The r.m.s. of the horizontal (vertical) orbit is \SI{1.46}{mm} (\SI{0.90}{mm}). The plot in the centre shows the steering magnet currents applied for the correction.\label{fig:orbit_control}}
\end{figure}

\subsection{Beam-based alignment}\label{app:bba}
Beam-based alignment   verifies that the beam passes through the centre of a quadrupole. An off-centre path through a quadrupole results in  deviation of the beam. Modifying the quadrupole strength,
this deviation can be measured. From a survey and alignment campaign of the magnetic elements in COSY, the quadrupole positions are known to approximately \SI{0.2}{mm}. Using the beam-based alignment procedure, the positions of the beam position monitors (BPMs) relative to the quadrupole centres could be determined. \Figure[b]~\ref{fig:bba_results} shows preliminary results. For 12 out of a total of 56 quadrupoles, offsets of the BPMs of a few millimetres were found\,\cite{Wagner:2019xul}. These offsets can now be corrected. This should result in an orbit closer to the design orbit and will reduce the systematic error of the precursor experiment. For further details, see Ref.~\cite{wagner2020beambased}.

\begin{figure}
    \centering
   \includegraphics[width=\textwidth]{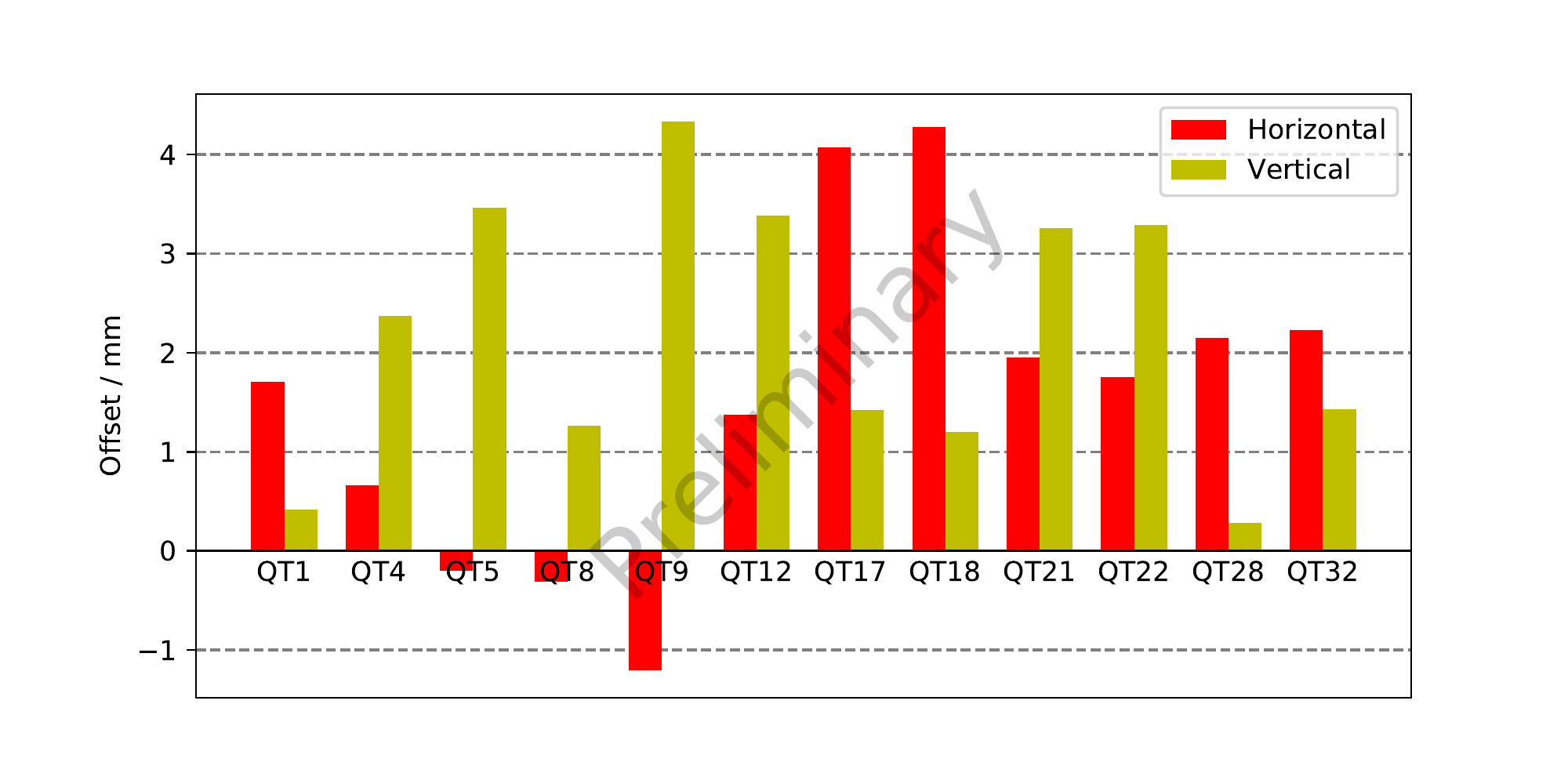}
    \caption{Beam offset  at  various quadrupole positions. Since the quadrupoles are aligned at \SI{0.2}{mm}, these values can be used to calibrate the offsets of the BPMs with respect to the quadrupole centres.
    \label{fig:bba_results}}
\end{figure}

\subsection{Beam position monitors}
New devices, so-called Rogowski coils, were built to determine the beam positions at the entrance and exit of the RF Wien filter. The Rogowski coil BPM consists of four quadrants (up-right, down-right, down-left, down-right). A time-varying beam induces a voltage in the four coils. Combining the four voltages, the beam position can be determined\cite{Fal-18-10}. \Figure[b]~\ref{fig:rogowski} shows a coil installed in COSY and the principle set-up of the coils. First calibration measurements demonstrate that an accuracy of about \SI{100}{\micro \meter} can be reached, as shown in \Fref{fig:rogowski_calibration}.

\begin{figure}
\centering
\includegraphics[height=0.23\textheight]{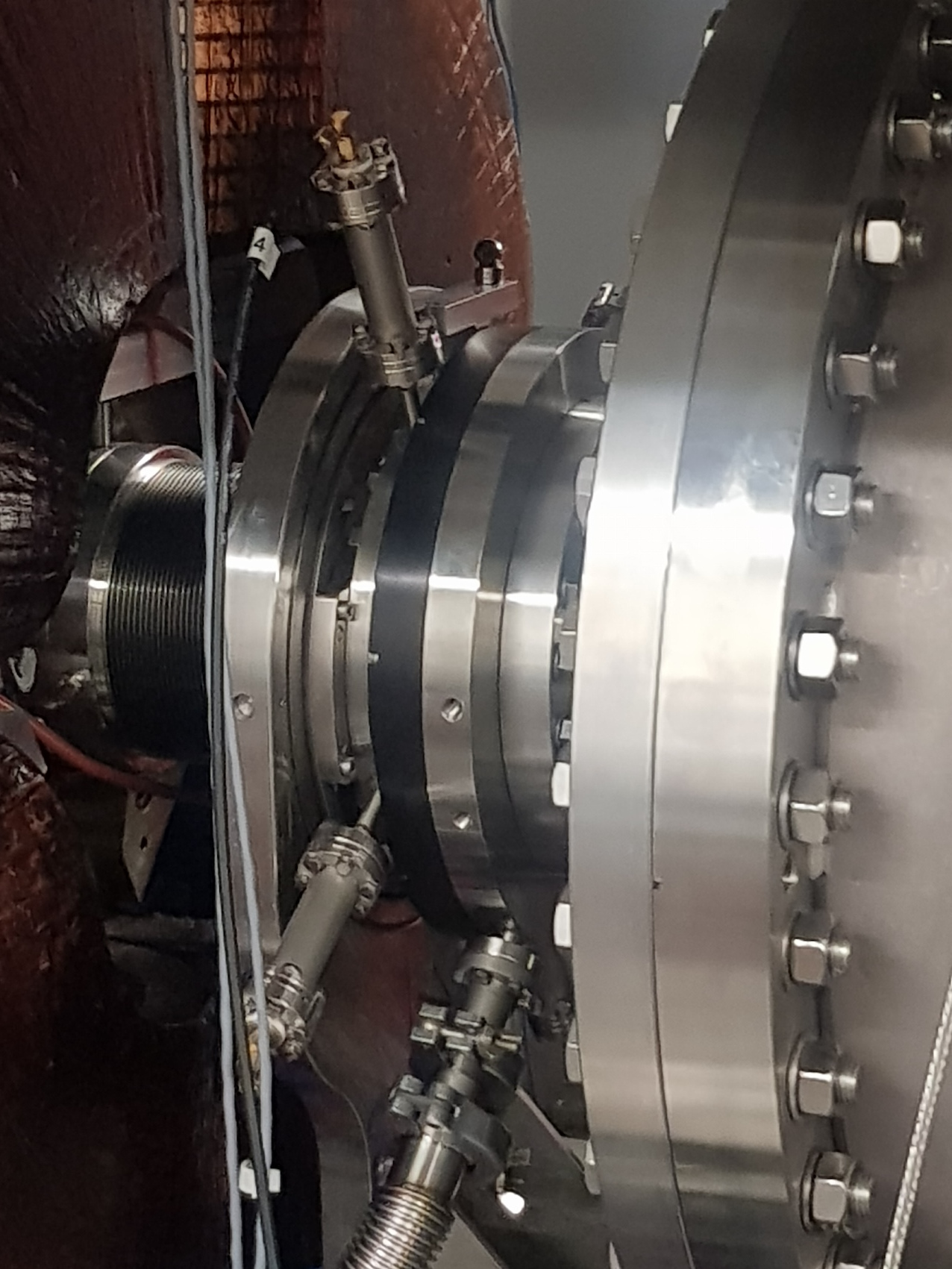}
\includegraphics[height=0.23\textheight,angle=0]{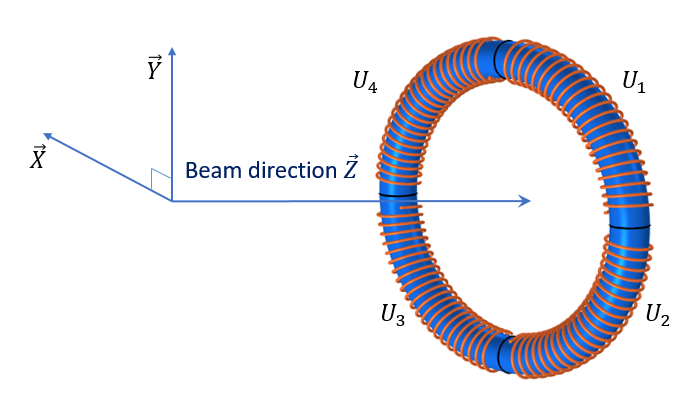}
\caption{\label{fig:rogowski} Left:  Rogowski coil installed in COSY. Right:  Rogowski coil set-up.}
\end{figure}

\begin{figure}
\centering
\includegraphics[height=0.3\textheight]{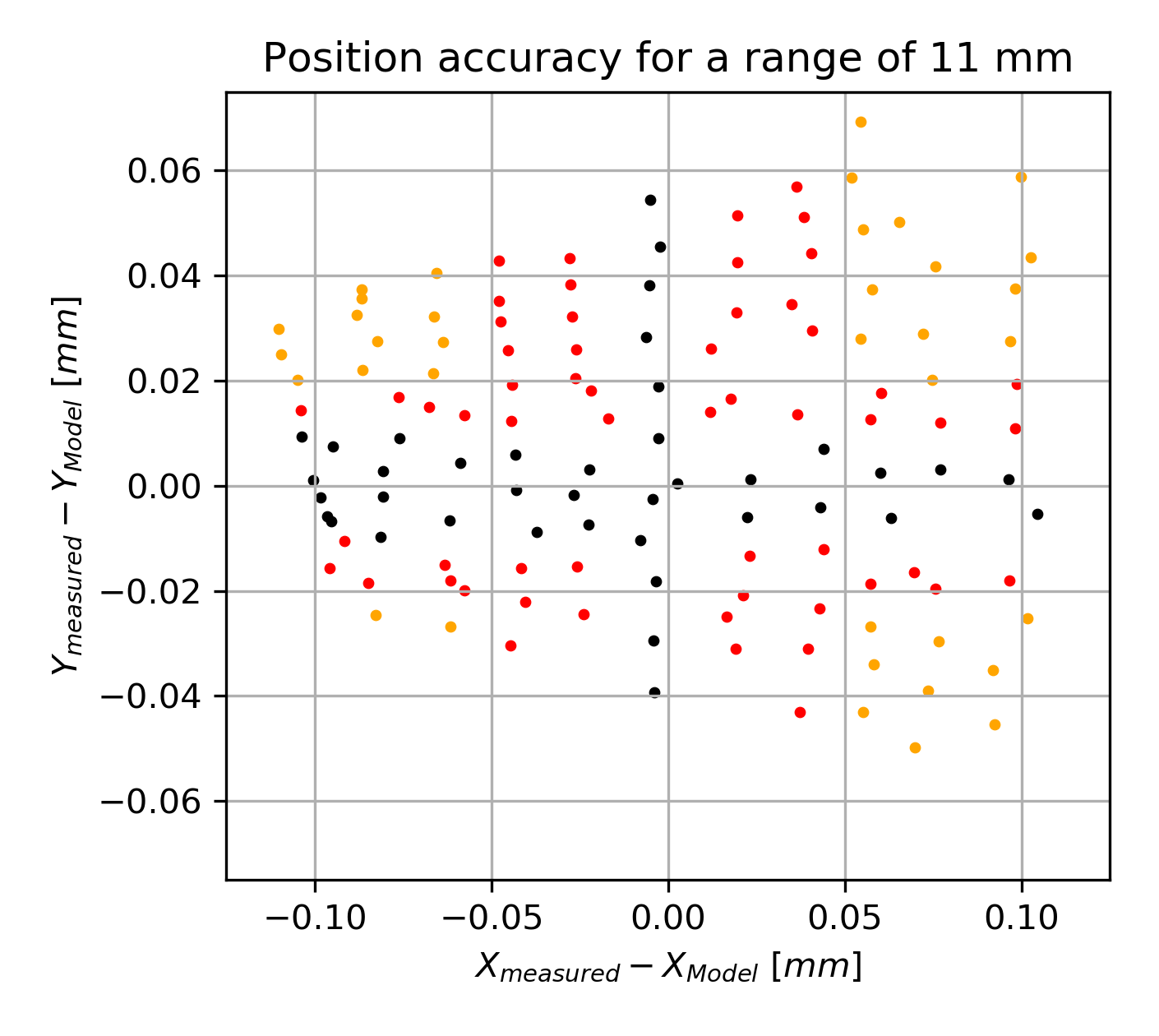}\hspace{0.2cm}
\includegraphics[height=0.3\textheight]{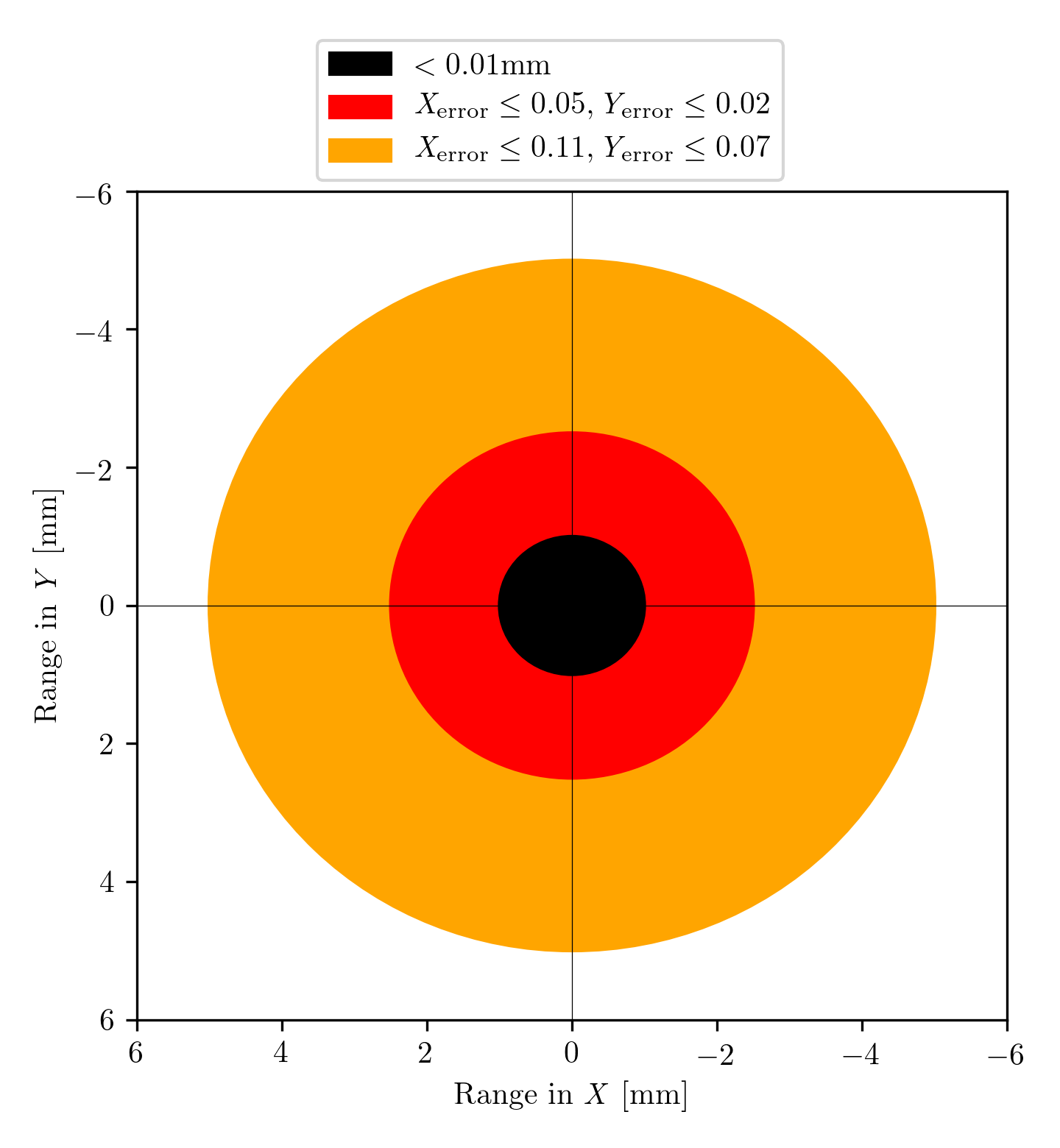}
\caption{ \label{fig:rogowski_calibration} Difference between  measured positions and model prediction  for different regions in the $x$--$y$ plane, as indicated in the right panel.}

\end{figure}


\subsection{Electrostatic and combined deflector development}
\label{appa:deflectors}

Future measurements of the EDM using a primarily electric ring, such as the PTR (see Chapter\,\ref{chap:ptr}), require the development of a prototype of an electrostatic or combined electromagnetic beam-bending element. For a proton beam and a magic momentum of \SI{701}{MeV/$c$}, the bending elements of such a ring can be electric, but for deuterons and, in general, particles for which $G$ is negative, a magnetic field is required to maintain the frozen-spin condition.

The electrostatic deflectors should consist of two parallel metal plates of equal potential but opposite sign. Equal electric potential seen by the particle at the entrance and at the exit of the deflector will not affect the total momentum of the particle. This puts restrictions on the minimum distance between the plates of the deflector. Recent ring lattice studies lead to limitations of the good field region for stored particles of \SI{40}{mm}. This leads to a minimum distance between electric deflector plates of about \SI{60}{mm}. The vertical beam size is several times larger than the horizontal, and this imposes restrictions on the vertical dimension of the flat region of the deflector element as well. Minimum transverse dimensions of the bending elements will be more than \SI{100}{mm}.

To minimize the probability of voltage breakdown between the flat regions of the deflectors and move it to the deflector edge, the shape of the deflector elements should follow a Rogowski profile on both vertical ends. The end caps of individual deflectors should be made to couple the stray fields with subsequent deflector elements. Ring lattices require electric field gradients of the order of 5--$\SI{10}{MV/m}$ (see \Tref{tbl:BasicBeamParams}). This is substantially above the standard values for many electrostatic deflector systems at plate distances of a few centimetres. Assuming \SI{60}{mm} distance between the plates, in order to achieve such high electric fields, we have to use high-voltage power supplies. At present, at COSY, two \SI{200}{kV} power converters are available to test deflector prototypes. The field emission, field breakdown, dark current, electrode surface, and conditioning should be studied using two flat electrostatic deflector plates, mounted on the movable support, allowing for the possibility of changing the plate distance in the range  20--\SI{120}{mm}. The residual ripple of the power converters is of the order of \num{e-5} peak-to-peak at the  maximum voltage of \SI{200}{kV}. This will lead to particle displacements of the order of millimetres.  Achieving a smaller ripple and a better stability control of the high-voltage system will be a dedicated task in the framework of the development of the EDM prototype ring.

\subsection{`Spin-offs'}
This subsection lists a number of publications that were initiated by the studies for a storage ring EDM
measurement but also have applications in other areas.

\begin{enumerate}
\item Polynomial chaos expansion method as a tool to evaluate and quantify the field homogeneity of a novel waveguide RF Wien filter\cite{Slim:2016dct}.
\begin{itemize}
 \item A full-wave calculation demonstrated that the waveguide RF Wien filter is able to generate high-quality RF electric and magnetic fields. In reality, mechanical tolerances and misalignments decrease the simulated field quality, and it is therefore important to consider them in the simulations. In particular, for the electric dipole moment measurement, it is important to quantify the field errors systematically. Since Monte Carlo simulations are computationally very expensive, this paper discusses an efficient surrogate modelling scheme based on the polynomial chaos expansion method to compute the field quality in the presence of tolerances and misalignments, and subsequently provides a sensitivity analysis at zero additional computational cost.
\end{itemize}

\item Computational framework for particle and spin simulations based on the
stochastic Galerkin method\cite{Slim:2017bic}.
\begin{itemize}
 \item An implementation of the polynomial chaos expansion is introduced as a fast solver of the equations of beam and spin motion of charged particles in electromagnetic fields. The investigation shows that, based on the stochastic Galerkin method, the computational framework substantially reduces the required number of tracking calculations, compared with the widely used Monte Carlo method.
\end{itemize}

\item Control of systematic uncertainties in the storage ring search for an EDM by measuring the electric quadrupole moment\cite{Magiera:2017ypv}.
\begin{itemize}
 \item Measurements of the EDM for light hadrons through the use of a storage ring have been proposed. The expected effect is very small; therefore, various subtle effects must be considered. In particular,  interaction of a particle's magnetic dipole moment and electric quadrupole moment with electromagnetic field gradients can produce an effect of a similar order of magnitude as that expected for the EDM. This paper describes a very promising method employing an RF Wien filter, allowing that contribution to be disentangled  from the genuine EDM effect. It is shown that both these effects could be separated by the proper setting of the RF Wien filter frequency and phase. In the EDM measurement, the magnitude of systematic uncertainties plays a key role and they should be kept under strict control. It is shown that particles' interaction with field gradients also offers  the possibility to estimate global systematic uncertainties with the precision necessary for an EDM measurement at the planned accuracy.
\end{itemize}

\item Extraction of azimuthal asymmetries using optimal observables\cite{Pretz:2018bze}.
\begin{itemize}
 \item Azimuthal asymmetries play an important role in scattering processes with polarized particles.  This paper introduces a new procedure that uses event weighting to extract
these asymmetries.  It is shown that the resulting estimator has several advantages in
terms of statistical accuracy, bias, assumptions on acceptance, and luminosities, compared with other estimators discussed in the literature.
\end{itemize}

\item Amplitude estimation of a sine function based on confidence intervals and Bayes' theorem \cite{Eversmann:2015vqn}.
\begin{itemize}
 \item This paper discusses  amplitude estimation using data originating from a sine-like function as probability density function. If a simple least squares fit is used, a significant bias is observed if the amplitude is small compared with its error. It is shown that a proper treatment using the Feldman--Cousins algorithm of likelihood ratios allows  improved confidence intervals to be constructed. Using Bayes' theorem, a probability density function is derived for the amplitude, and is used in an application to show that it leads to better estimates, compared with a simple least squares fit.
\end{itemize}

\item General dynamics of tensor polarization of particles and nuclei in external fields\cite{Silenko:2015qfa}.
\begin{itemize}
 \item The tensor polarization of particles and nuclei becomes constant in a coordinate system rotating with the same angular velocity as the spin, and it rotates in the laboratory frame with this angular velocity. The general equation defining the time dependence of the tensor polarization is derived. An explicit form of the dynamics of this polarization is found for the case when the initial polarization is axially symmetric.

\end{itemize}
\end{enumerate}

\section{Results and achievements from the J\"ulich--Bonn theory group}
\label{Chap:results-theory}
The IKP-3/IAS-4 at the Forschungszentrum J\"ulich, together with the theory group at the Helmholtz-Institut f\"ur Strahlen- und Kernphysik (HISKP) at the University of Bonn,\footnote{Both groups are headed by Ulf Mei{\ss}ner.}  have made a number of benchmark calculations for the EDMs of protons, neutrons, and light nuclei using  chiral effective nuclear field theory (chiral perturbation theory and its extension to few-baryon systems) and lattice QCD simulations.

This project on hadronic electric dipole moments started  in 2009, with the diploma thesis of Konstantin Ottnad (HISKP) on electric dipole form factors of the neutron in chiral perturbation theory~\cite{Ottnad:2009ths}. His work culminated in an analysis\cite{Ottnad:2009jw} of the  QCD $\bar\theta$-angle induced EDMs of the neutron and proton to third order in  ${\rm U(3)}_L \times {\rm U(3)}_R$ baryon chiral perturbation theory, in a covariant and in an
extension by the number of colours ($N_c$).  A new upper bound\footnote{The estimate is modulo the unknown contributions of the contact interactions needed to remove the infinities of the one-loop calculations.} on the vacuum angle, $|\bar\theta | \lesssim   2.5 \times 10^{-10}$, was given and the matching relations for the  three-flavour representation to the SU(2) case were derived. These relations  still comprise today's $\bar\theta$-induced EDM predictions for the neutron and proton in chiral perturbation theory.

In 2012, IAS-4/IKP-3 extended this work to  the QCD $\bar\theta$-term-induced  EDM of the deuteron,  where  the genuine two-nucleon contributions of  the ${P}$- and ${T}$-violating form factor $F_3$ of the deuteron were calculated in the Breit frame of this nucleus using chiral effective field theory  up to and including next-to-next-to-leading order\cite{Bsaisou:2012rg}. In particular, it was found that the difference between the deuteron EDM and the sum of the proton and neutron EDMs  corresponds to a value of $(0.54\pm 0.39)\, \bar\theta \times \num{e-16}\,\si{\text{$e$}.cm}$.   Both the nucleon--nucleon potential and the transition current contributions were calculated, where the ${CP}$- and isospin-violating $\pi N N$ coupling constant $g^\theta_1$ was identified as the source of the dominating contribution to the  uncertainty.  The role that the vacuum alignment plays for the generation of $g^\theta_1$ was outlined and an estimate of the  additional  and previously unknown  contribution to  $g^\theta_1$ was derived from a resonance saturation mechanism involving the odd-parity nucleon resonance ${\rm S}_{11}(1535)$.

In the same year, Guo (HISKP) and Mei{\ss}ner calculated the electric dipole form factors and moments of the ground state baryons in chiral perturbation theory at next-to-leading order\cite{Guo:2012vf}. It was shown that the baryon electric dipole form factors at this order depend only on two combinations of low-energy  constants. This was used to derive various relations for the baryon EDMs that are free of unknown low-energy constants, which can be used to cross-check future lattice QCD results. Thus, for a precision extraction from lattice QCD data, the next-to-leading order terms must be accounted for. In 2014, Akan (HISKP), Guo, and Mei{\ss}ner  revisited  this work by investigating finite volume corrections to the ${CP}$-odd nucleon matrix elements of the electromagnetic current, which can be related to the EDMs originating from strong ${CP}$ violation  in the continuum, in the framework of chiral perturbation theory up to next-to-leading order, taking into account the breaking of Lorentz symmetry\cite{Akan:2014yha}. A chiral extrapolation of the recent lattice results of both the neutron and proton EDMs was also
performed.

In 2014, Jan Bsaisou (IKP-3/IAS-4)  completed his Ph.D. thesis, at the University of Bonn,   on EDMs of light nuclei in chiral effective field theory\cite{Bsaisou:2014goa}\footnote{Part of this work was documented previously\cite{Bsaisou:2012rg}.}. Starting from the QCD $\bar\theta$-term and the set of ${P}$- and ${T}$-violating effective dimension-six operators, he presented a scheme to derive the induced effective Lagrangians at energies below $ \Lambda_{\rm QCD}\sim 200\UMeV$ within the framework of chiral perturbation theory (ChPT) for two quark flavours---applying the formulation of Gasser and Leutwyler.  It was shown that the differences between the sources of ${P}$ and ${T}$ violation  manifest themselves in specific hierarchies of coupling constants of ${P}$- and ${T}$-violating vertices. Bsaisou computed the relevant coupling constants of $P$- and $T$-violating hadronic vertices, which are induced by the QCD $\bar\theta$-term with well-defined uncertainties as functions of the parameter $\bar\theta$. The relevant coupling constants induced by the effective dimension-six operators were given as functions of as-yet unknown low-energy constants (LECs), which cannot be determined by ChPT. Estimates of the coupling constants from naive dimensional analysis (NDA) proved  sufficient to reveal certain hierarchies of coupling constants. The different hierarchies of coupling constants translated into different hierarchies of the nuclear contributions to the EDMs of light nuclei. In this way, Bsaisou  could calculate, within the framework of ChPT, the two-nucleon contributions to the EDM of the deuteron up to and including next-to-next-to leading order and the two-nucleon contributions to the EDMs of the helion ($^3$He nucleus) and the triton ($^3$H nucleus) up to and including next-to-leading order. These computations involved thorough investigations of the uncertainties of the results from both the ${P}$- and ${T}$-violating  and conserving components of the nuclear potential. Quantitative predictions of the nuclear contributions to the EDMs of the deuteron, helion, and triton induced by the QCD $\bar\theta$-term as functions of $\bar\theta$ with well-defined uncertainties were presented, while the EDM predictions for the effective dimension-six sources were given as a function of the unknown LECs with NDA estimates. Several strategies to falsify the QCD $\bar\theta$-term as a relevant source of ${P}$ and ${T}$ violation were presented, whereby a suitable combination of measurements of several light nuclei and, if needed, supplementary lattice QCD input could be used. Bsaisou demonstrated how particular effective dimension-six sources can be tested by EDM measurements of light nuclei---with supplementary lattice QCD input in the future.

While this thesis discussed strategies to separate the various dimension-six EDM operators individually, the IAS4-/IKP-3 publication by Dekens \textit{et al.}
\cite{Dekens:2014jka}, using information from this thesis  and from the paper by Dekens and de Vries~\cite{Dekens:2013zca} on the renormalization group running of dimension-six sources for ${P}$ and ${T}$ violation,  showed that the proposed measurements of the EDMs of light nuclei in storage rings would put strong constraints on models of  flavour-diagonal ${CP}$ violation\cite{Dekens:2014jka}. This analysis was exemplified by a comparison of the Standard Model including the QCD theta term, the minimal left--right symmetric model, a specific version of the so-called aligned two-Higgs doublet model, and, \textit{en passant}, a minimal supersymmetric extension of the Standard Model. Again, by using effective field theory techniques, it was demonstrated to what extent measurements of the EDMs of the nucleons,  the deuteron, and helion could discriminate between these scenarios and how measurements of EDMs of other systems relate to light-nuclear measurements. In particular, the focus was on the most important ${P}$- and ${T}$-violating hadronic interactions that appear in each of the scenarios, especially on the ${P}$- and ${T}$-violating pion--nucleon interactions and the nucleon EDMs.  It was demonstrated that chiral effective  field theory is a powerful tool to study the observables  of light nuclei and that measurements of light-nuclear EDMs can be used to disentangle different underlying scenarios of ${CP}$ violation.

The EDM predictions of IAS-4/IKP-3 up to  2014 are summarized in Ref.~\cite{Wirzba:2014mka}, and   a consistent and complete calculation of the EDMs of the deuteron, helion, and triton  by chiral effective field theory is  given in Ref.~\cite{Bsaisou:2014zwa}.   The ${CP}$-conserving and ${CP}$-violating interactions were treated on equal footing and the ${CP}$-violating one-, two-, and three-nucleon operators  were considered up to next-to-leading-order in the chiral power counting. In particular, for the first time, EDM contributions induced by the ${CP}$-violating three-pion operator were calculated. It was found that effects of ${CP}$-violating nucleon--nucleon contact interactions are larger than those predicted in previous studies involving phenomenological models of the ${CP}$-conserving nucleon--nucleon interactions. The results, which apply to any model of ${CP}$ violation in the hadronic sector, can be used to test various scenarios of ${CP}$ violation. In particular, the implications for the QCD $\bar\theta$-term and the minimal left--right symmetric model were demonstrated. Furthermore, in Ref.~\cite{Bsaisou:2014oka}, the  underlying scheme is presented to derive---within the framework of chiral effective field theory---the list of parity- and time-reversal-symmetry-violating hadronic interactions that are relevant for the computation of nuclear contributions to the EDMs of the \Isotope[2]{H}, \Isotope[3]{He}, and \Isotope[3]{H} nuclei. The scattering and Faddeev equations required to compute electromagnetic form factors in general and EDMs in particular are also documented  in Ref.~\cite{Bsaisou:2014oka}.

In 2015,  Shindler, Luu, and de Vries (IAS-4/IKP-3)  proposed a new method to calculate EDMs induced by the strong QCD $\bar\theta$-term\cite{Shindler:2015aqa}, basing their method on the gradient  flow for gauge fields, which is free from renormalization ambiguities\footnote{In fact, their method was already documented in Ref.\cite{Shindler:2014oha}, in a  broader context.}.  The method was tested by computing the nucleon EDMs in pure Yang--Mills theory at several lattice spacings, enabling a  first-of-its-kind continuum extrapolation, which is theoretically sound.

In the same year,  Guo \textit{et al.} \cite{Guo:2015tla} presented an entirely dynamic calculation of the EDM of the neutron on the lattice.   They computed the EDM $d_n$ of the neutron from a fully dynamic simulation of lattice QCD with 2 + 1 flavours of clover fermions and a non-vanishing $\theta$-term. The latter was rotated into a pseudoscalar density in the fermionic action using the axial anomaly. To make the action real, the vacuum angle $\theta$ was taken to be purely imaginary. The physical value of $d_n$ was obtained by analytical continuation $\left(d_n = -3.9(2)(9) \times \num{e-16} \, \bar{\theta} \,\si{\text{$e$}.cm}\right)$ and an upper bound on the QCD theta angle $\left( |\bar{\theta} | \lesssim 7.4 \times 10^{-11} \right)$ was presented.

In 2016, Mei{\ss}ner and de Vries reviewed the progress  in the theoretical description of the violation of discrete space-time symmetries in hadronic and nuclear systems\cite{deVries:2015gea}. They focused  on parity-violating and time-reversal-conserving interactions, which are induced by the weak interaction of the Standard Model, and on parity- and time-reversal-violating interactions, which can be caused by a non-zero QCD theta term or by BSM physics. In particular, they reviewed  the development of the chiral effective field theory extension that includes discrete symmetry violations and  discussed the construction of symmetry-violating chiral Lagrangians and nucleon--nucleon potentials and their applications in few-body systems. In their review of the parity- and time-reversal violation,  information from the aforementioned HISKP and  IAS-4/IKP3  publications was, of course, used, but  results of three recent publications, coauthored by IAS-4 member de Vries  were also integrated: the first on the constraint of the neutron EDM on the value of the ${CP}$-and isospin-violating pion--nucleon coupling constant $g_1$ in the case of dimension-six interactions\cite{Seng:2014pba}; the second on the extension to SU(3) chiral perturbation theory and the  update on the determination of the ${CP}$-violating isospin-conserving pion--nucleon coupling constant $g_0^\theta$\cite{deVries:2015una}; and the third on direct and indirect constraints on the complete set of anomalous ${CP}$-violating Higgs couplings to quarks and gluons originating from dimension-six  operators\cite{Chien:2015xha}.

In 2017, Wirzba, Bsaisou, and Nogga~\cite{Wirzba:2016saz} gave an update on the predictions of Refs.~\cite{Bsaisou:2014zwa,Bsaisou:2014oka},  especially  by  extending  the  computation of the relevant matrix elements of the nuclear EDM operators  in the deuteron case to the N4LO level of chiral effective field theory. Furthermore, they incorporated a review of the  underlying principle that  the existence of a non-zero EDM of an elementary or composite particle (in fact, of any finite system) necessarily involves the breaking of a symmetry, either by the presence of external fields (\ie  electric fields, leading to the case of induced EDMs) or, explicitly, by the breaking of the discrete parity and time-reflection symmetries, in the case of permanent EDMs.

In a series of publications, a collaboration including a current and two former members  of IAS-4/IKP-3 refined the method of Ref.~\cite{Shindler:2015aqa} by extending it from the calculation of EDMs induced by the strong QCD $\bar\theta$-term\cite{Dragos:2018uzd}  to include the  dimension-six Weinberg term\cite{Dragos:2017wms} and the quark--chromo EDM operator\cite{Kim:2018rce}. This work culminated in Ref.\cite{Dragos:2019oxn};
the EDM of the nucleon induced by the QCD theta term was calculated in the gradient  flow  method with $N_f = 2 + 1$ flavours of dynamic quarks corresponding to pion masses of 700, 570, and \SI{410}{MeV},  which were used  by  performing an extrapolation to the physical point based on chiral perturbation theory. The calculations applied  three different lattice spacings in the range  $\SI{0.07}{fm} < a < \SI{0.11}{ fm}$ at a single value of the pion mass, to enable the control of discretization effects. Finite-size effects were  also investigated using two different volumes.  A novel technique was applied to improve the signal-to-noise ratio in the form factor calculations. The very mild discretization effects observed suggested a continuum-like behaviour of the nucleon EDM towards the chiral limit. Under this assumption, the results read $d_{\sf n}/\bar\theta = -1.52(71) \times \num{e-16}\,\si{\text{$e$}.cm}$ and $d_{\sf p}/\bar\theta = 1.1(1.0) \times \num{e-16}\,\si{\text{$e$}.cm}$. Assuming that the theta term is the only source of ${CP}$ violation, the experimental bound on the neutron EDM yields $|\bar\theta| < 1.98 \times \num{e-10}$ (90\% confidence level) as a limit.

\FloatBarrier

\begin{flushleft}

\end{flushleft}
\end{cbunit}

\begin{cbunit}

\csname @openrighttrue\endcsname 
\chapter[Mitigation of background magnetic fields]{Mitigation of background magnetic fields\footnote{This appendix is based on a text by Y.K. Semertzidis and S. Hac\i{\" o}mero{\u g}lu of the Center for Axion and Precision Physics Research, KAIST, South Korea, which has been published as an extended version \cite{bib:MagneticFields}.
}}
\label{Chap:MagneticFields}

The EDM signal can be mimicked by magnetic fields in several different ways. The most critical effect comes from a uniform radial magnetic field, which is required to be at the attotesla level to reach a sensitivity of \SI{e-29}{\text{$e$}.cm}.  A combination of uniform vertical and longitudinal magnetic fields has a strong effect, but with a few orders of magnitude more flexible restriction. Moreover, several configurations of alternating magnetic fields also result in EDM-like spin precession. We studied each of these scenarios and proposed solutions to cancel the effect. All results presented in this appendix have been obtained for a 300 m circumference ring with smooth focusing, with a small field index for weak vertical focusing. Throughout this appendix, the magnetic field is described in a coordinate system attached to the design trajectory. Uniform fields are constant and, in general, represent the average around the circumference. Alternating magnetic fields have opposite sign at different longitudinal positions (in the examples given, opposite sign at opposite positions in the ring) and vanishing average over the circumference. As an example, the radial and longitudinal components of Earth's magnet field are alternating, whereas the vertical component is uniform.

\section{Uniform magnetic field configurations}
\subsection{Uniform radial magnetic field}

As  \Fref{fig:sy_by_Br_and_Er} shows, the uniform (or average) radial magnetic field should be kept at attotesla level to avoid the systematic error giving an effect similar to an EDM of $d_\mathrm{p} = \SI{e-29}{\text{$e$}.cm}$.  This is obviously not possible with magnetic shielding alone and other means to measure radial magnetic fields and actively compensate them  are required. The scheme proposed is to measure the relative position of the counter-rotating beams, proportional to the average radial magnetic field. For the all-electric baseline ring operated at a low vertical tune of $Q_\mathrm{v} \approx 0.1$, an attotesla level field splits the counter-rotating beams vertically by picometres. The split beams induce a magnetic field in the horizontal direction. The magnitude of this field $B_x$ can be measured using a magnetometer or
gradiometer at a few centimetres horizontal distance (\Fref{fig:cw_ccw_b_field}).

\begin{figure} [hbt!]
\centering
\includegraphics[width=0.7\linewidth]{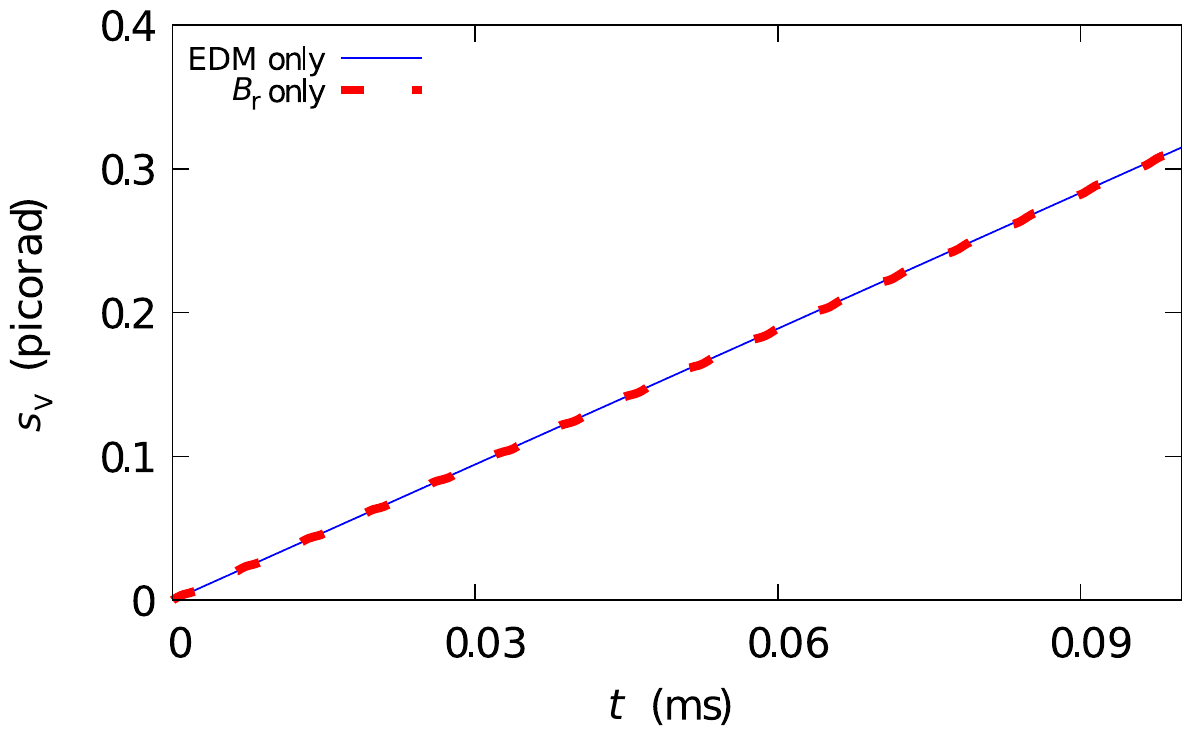}
\caption{Vertical spin component as a function of time for an EDM of \SI{e-29}{\text{$e$}.cm} (blue line) and a radial magnetic field of 17\,aT (red dashed line) \cite{bib:MagneticFields}. Figure reused with permission from the author.}
\label{fig:sy_by_Br_and_Er}
\end{figure}

\begin{figure}
\centering
\includegraphics[width=0.5\linewidth]{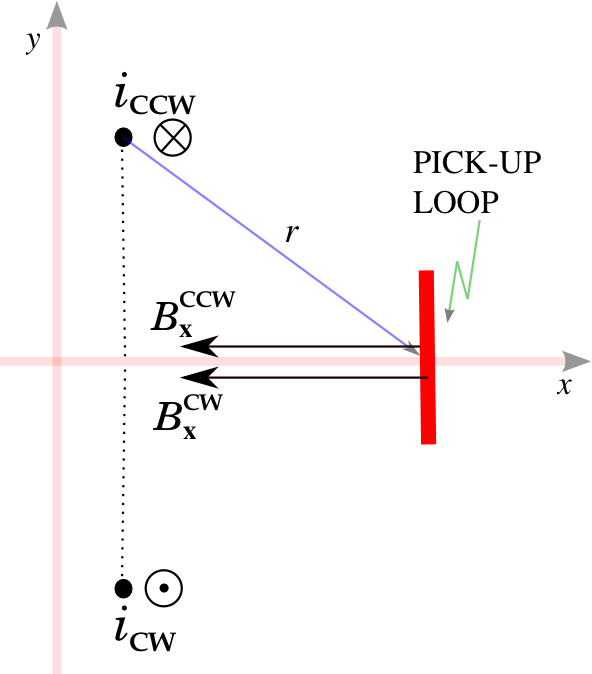}
\caption{A magnetometer can pick up the magnetic field in the horizontal direction that is induced by the vertically split counter-rotating beams \cite{bib:MagneticFields}. Figure reused with permission from the author.}
\label{fig:cw_ccw_b_field}
\end{figure}

We plan to use SQUID-based beam position monitors (BPMs) \cite{bib:SQUIDBPM} to measure $B_x$. To suppress environmental noise, the vertical tune of the beams will be modulated at 1--$10\UkHz$ by means of the quadrupoles\footnote{Another reason to modulate the vertical tune is to alleviate constraints on the number of beam position difference pick-ups and the regularity of their spacing in terms of betatron phase advance.}. The typical white noise of the DC SQUIDs at that range is less than $1 \mbox{ fT}/\sqrt{\mbox{Hz}}$. In such a case, $B_x$ due to the split beams  is given as
\begin {equation} \label{eq_BAtModFreq2Beams}
B_x(t) =\frac{\mu_0 I ~\Delta y}{\pi r^2}2A \cos(\omega_m t) \, ,
\end{equation}
with  beam current $I$, vertical split $\Delta y$, horizontal distance between the pick-up loop and the beams $r$,  modulation amplitude $A$,  and frequency $\omega_m$. Putting $I=10$ mA, $\Delta y=0.5$ pm, $A=0.1$, and $r=2$\,cm into \Eref{eq_BAtModFreq2Beams} gives $B_x \approx 1\,\text{aT} \cos(\omega_m t)$.

As a reference, for an array of eight SQUIDs of $10^{-15}$ T sensitivity at 1\,Hz bandwidth ($1\text{\,fT}/\sqrt{\text{Hz}}$), it requires $2 \times 10^5\Us$  of averaging to achieve SNR $>1$ as $B = \left(1 \mbox { fT}/\sqrt{\mbox{Hz}}\right) / \sqrt{8\times 2 \times 10^5 \mbox{ s}} = 0.8 \mbox{ aT}$.

\subsection{Preliminary tests with SQUID-based BPM}
\label{sec:tests-with-SQUID}
SQUID-based magnetometers can measure magnetic field variations with unprecedented noise levels, less than $1\mbox{\,fT}/\sqrt{\mbox{Hz}}$. This is why they became the best candidates for the beam position monitors in the pEDM experiment. In addition to high resolution, the SQUID-based magnetometers have sufficient bandwidth and compact size that allows the use of multisensor arrays placed along the beam trajectory inside a superconductive shielding structure.

\Figure[b]~\ref{fig:bpm_drawing_inside} shows a BPM. It will operate in vacuum at $4\UK$. The BPMs are positioned in the horizontal plane to measure the vertical split.

\begin{figure} 
        \centering
        \includegraphics[width=0.5\linewidth]{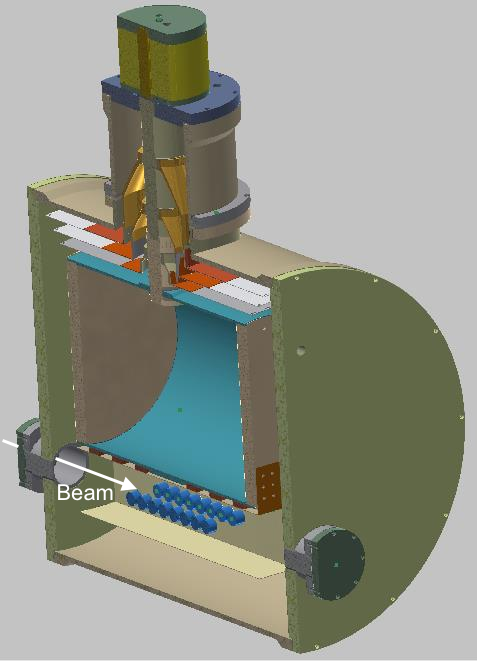}
        \caption{The beam (white arrow) passes between two arrays of SQUID gradiometers \cite{bib:SQUIDBPM}. The SQUID sensor and pick-up loops (dark blue) will be kept at liquid helium temperatures. The liquid helium tank
(turquoise layer) is above the SQUIDs. The whole set-up can fit in a 1\,m$^3$ cube. The figure is licensed under \href{https://creativecommons.org/licenses/by/4.0/}{CC-BY-4.0}.}
        \label{fig:bpm_drawing_inside}
\end{figure}

The BPM works inside a magnetically shielded room (MSR). The data transfer between the SQUIDs and the computer is achieved via fibre lines to minimize electromagnetic noise.  \Figure[b]~\ref{fig:bpm_real} shows  the first prototype.

\begin{figure} [hbt!]
        \centering
        \includegraphics[width=0.5\linewidth]{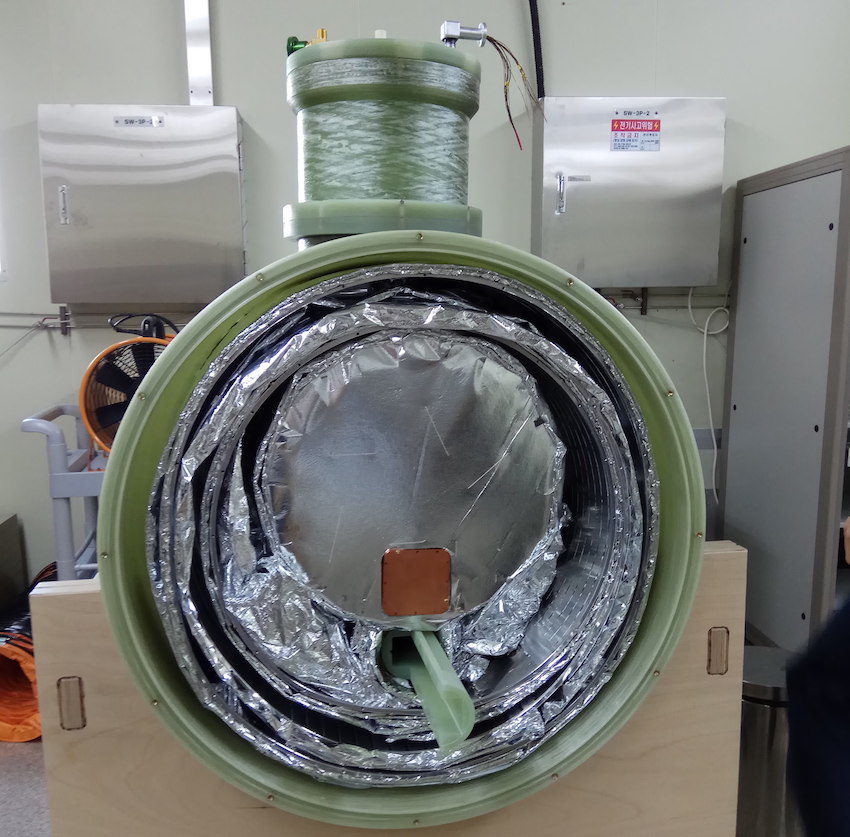}
        \caption{First prototype of the BPM. The three layers of the dewar and the liquid helium tank are covered with aluminized Mylar. The partially inserted half cylinder below the liquid helium tank has casings for the SQUIDs. The other half cylinder has been left out for easier visibility.}
        \label{fig:bpm_real}
\end{figure}

We conducted preliminary tests with a set-up having  similar SQUID electronics but a different design of pick-up loop geometry and dewar. The dewar and the eight SQUID gradiometers are shown in  \Fref{fig:dewar_and_squids}. They were originally designed at KRISS, Korea, for biomagnetic applications. The
set-up has eight axial wire-wound first-order gradiometers positioned along a bottom line inside a fibreglass dewar. Each gradiometer has a 20\,mm diameter and 50\,mm baseline and is bonded to a double relaxation oscillation (DROS) SQUID current sensor. The DROS SQUIDs have a large flux-to-voltage transfer coefficient that minimizes the contribution of  direct read-out electronics noise. The white noise of the gradiometers is about $3\,\text{fT}/\sqrt{\text{Hz}}$ at frequencies above $1\UHz$.

\begin{figure}[hbt!] 
        \centering
        \includegraphics[width=0.7\linewidth]{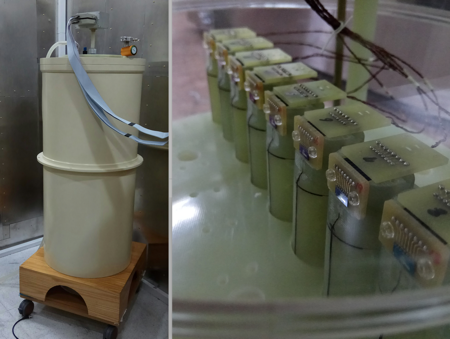}
        \caption{Time-averaging measurements were made with a set-up having the same electronics but different designs of the (left) dewar and (right)  gradiometer  \cite{bib:SQUIDBPM}. The figure is licensed under \href{https://creativecommons.org/licenses/by/4.0/}{CC-BY-4.0}.}
        \label{fig:dewar_and_squids}
\end{figure}

For these measurements with long time averaging, the magnetic field was generated by two parallel traces of $\SI{100}{\micro m}$  separation on a PCB,  carrying opposite currents of $\SI{100}{\micro A}$. The applied current was  $300\UHz$ sinusoidal AC, corresponding to around 200\,fT amplitude field at the pick-up loop location.

The measurements showed more than two orders of magnitude suppression of  white noise from the gradiometers and the SQUID read-out electronics (\Fref{fig:squid_averaging}). This corresponds to $\approx$\,$25\,\text{aT}/\sqrt{\text{Hz}}$ with  5\,h averaging. This indicates very high long-time stability and low intrinsic fluctuation levels in the  instrument, including all cryogenics and semiconductors, and both analogue and digital electronics.

\begin{figure}[hbt!]
        \centering
        \includegraphics[width=1\linewidth]{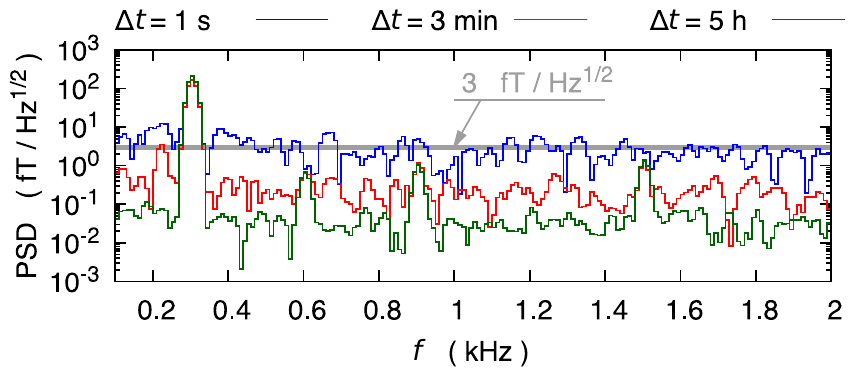}
        \caption{A current of $100\UuA$ was applied through the parallel traces, resulting in 200\,fT on the pick-up \cite{bib:SQUIDBPM}. The noise at $1\Us$ is a few fT$/\sqrt\text{Hz}$, consistent with the 3 fT$/\sqrt\text{Hz}$ sensitivity of the SQUIDs. The noise decreases  to 25\,aT$/\sqrt{\text{Hz}}$ after 5\,h of averaging. The figure is licensed under \href{https://creativecommons.org/licenses/by/4.0/}{CC-BY-4.0}.}
        \label{fig:squid_averaging}
\end{figure}

The real design proposed for the experiment (\Fref{fig:bpm_drawing_inside}) includes 16 magnetometers with two-turn $17\Umm$ diameter pick-up coils bonded to DROS SQUIDs. It allows us to achieve a white noise floor that is more than three times lower, \ie about $8\,\text{aT}/\sqrt{\text{Hz}}$, after $5\Uh$ averaging. For the further noise decrease, we expect to use single-chip integrated magnetometers similar to ML12 reported in Ref.~\cite{bib:sub_micrometer_squid} but with a chip size of $24\,\text{mm} \times 24\,\text{mm}$. Such magnetometers have white noise below $0.2 ~\text{fT}/\sqrt{\text{Hz}}$ at frequencies above $1\UkHz$.

In the hybrid ring design described in Appendix \ref{Chap:hybrid} and Ref.~\cite{HybridRing}, the compensation of the radial magnetic field does not have to be so strict, because the magnetic focusing mechanism leads to a partial cancellation. According to the simulations, the restriction is released by five orders of magnitude.

\subsection{Combination of uniform longitudinal and vertical magnetic fields}
A uniform longitudinal magnetic field can appear in the presence of an electric current passing through the horizontal plane at the inner side of the ring. For instance, a  $12\UmA$ current passing through the centre of a ring with circumference  $\approx$\,$300\Um$ induces $B_\mathrm{L} \approx \mbox{50 pT}$. In addition, an average vertical magnetic field $B_\mathrm{V} \approx \mbox{50 pT}$ is assumed and leads to a rotation of spin around the vertical axis. The resulting radial spin component for a particle with its spin initially
oriented in the longitudinal direction, obtained from a simulation, is plotted in  \Fref{fig:sr_and_sy}. This rotation corresponds to an angular frequency $\omega_\mathrm{a} \approx \mbox{12.5 mrad/s,}$  and the radial spin component can be approximated as $s_\mathrm{R} = \sin \left( \omega_\mathrm{a} \, t \right)$. The radial spin component is rotated around the longitudinal direction and thus leads to the build-up of a vertical spin component. The time derivative of the vertical spin component can be approximated as
\begin{equation}
\frac{\mathrm{d} \, s_\mathrm{V}}{\mathrm{d} t} = \omega_\mathrm{L} \, s_\mathrm{R} = - \frac{e}{m}\frac{g}{2\gamma}B_\mathrm{L} \, s_\mathrm{R} = -\frac{e}{m}\frac{g}{2\gamma}B_\mathrm{L} \, \sin\left( \omega_\mathrm{a} t \right) \, ,
\label{eq:y_due_to_Bl}
\end{equation}
where $g$, $e$, and $m$ are the g factor, electric charge, and mass of the proton, respectively, and $\gamma$ is the relativistic Lorentz factor.
Integrating this differential equation yields the vertical spin component,\begin{equation}
s_\mathrm{V}(t) = \frac{e g B_\mathrm{L}}{2 m \gamma \omega_\mathrm{a}}  \left[ \cos(\omega_\mathrm{a} t) -1 \right] \approx - \frac{e g B_\mathrm{L} \, \omega_\mathrm{a}}{4 m \gamma} \, t^2 \, ,
\label{eq:s_V}
\end{equation}
where the approximation is valid for short durations. The quadratic increase in $s_\mathrm{V}$ over a short duration is clearly visible in  \Fref{fig:sr_and_sy} and quickly exceeds the effect due to an EDM of $10^{-29}e\, \mbox{cm}$.

\begin{figure} 
        \centering
        \includegraphics[width=0.7\linewidth]{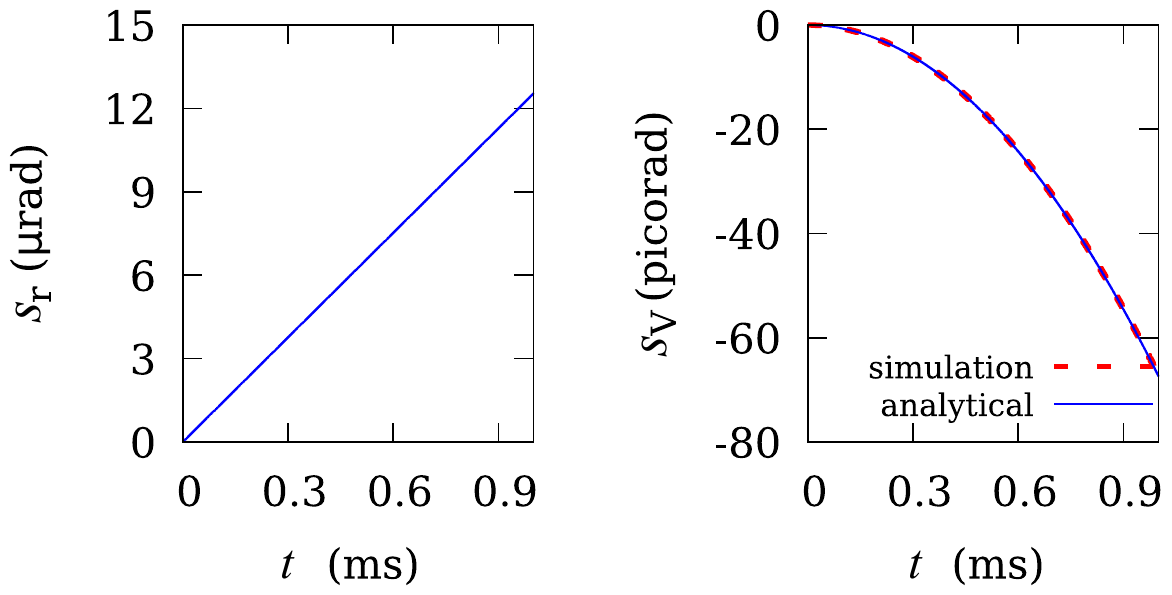}
        \caption{Left: Spin components, simulated for $1\Ums$ storage time with a magic particle in an electric ring \cite{bib:MagneticFields}. Because of the short storage time compared with one cycle of $\omega_\mathrm{a}$, $s_\mathrm{R}$ changes linearly and $s_\mathrm{V}$ approximates to a quadratic function (see \Eref{eq:s_V}). (left) A  vertical magnetic field of 50\,pT causes $\omega_\mathrm{a} \approx 12.5$\,mrad/s on the horizontal plane. Right: Having linear dependence on $\omega_\mathrm{a}$, $s_\mathrm{V}$ has quadratic dependence on time.  A combination of longitudinal and vertical uniform magnetic fields of 50\,pT increases the vertical spin component  to 67\,prad, matching  the analytical estimate well. Figure reused with permission from the author.}
        \label{fig:sr_and_sy}
\end{figure}

Two measures are envisaged to mitigate the effect.\vspace{-3mm}
\begin{itemize}
  \item A feedback system is used to measure the radial polarization (of bunches used for the measurement, which should be polarized only in the longitudinal direction) and rotate the bunches back into the longitudinal direction. Note that this feedback system must act on the rotation of the spin around the vertical axis independently for both beam directions. Thus, the system must act on two different parameters (\eg RF frequency and a small vertical magnetic field added by appropriate windings).
  \item Bunches with polarization in the radial direction, in addition to bunches with polarization in the longitudinal direction used for the measurement, are used to detect and correct a rotation of the spin around the longitudinal axis.
\end{itemize}

\section{Effect of alternating magnetic fields and the geometric phases}
We have studied the major configurations of the magnetic field in a continuous ring. In each case, we have simulated pairs of 1\,nT fields in perpendicular directions with different phases. In some cases, we have seen the spin increase much faster than the EDM signal, as in the case of longitudinal and vertical magnetic fields with  $90^\circ$ phase difference (alternating $B_\mathrm{V}$ and $B_\mathrm{L}$, $90^\circ$ of  \Tref{tbl:summary}). Some configurations are harmless, as they average out themselves. Some of them cancel out, thanks to the counter-rotating beam design.

\begin{table}
        \caption{Summary of the major independent  magnetic field configurations;  $\langle \mathrm{d} \, s_\mathrm{V} / \mathrm{d} t \rangle$ is the average spin precession rate in the vertical plane. Each simulation was run with
a magnetic field strength of 1\,nT.}
        \begin{tabular}{l l p{2.8 cm} p{6.1cm}}
                \hline \hline
                Field & AC phase& $\langle \mathrm{d} \, s_\mathrm{V} / \mathrm{d} t \rangle$ ~~~(rad/s) & Solution \\
                \hline
                Uniform $B_\mathrm{R}$ &n/a& 0.18 & Measurement and active cancellation with BPMs \\
                Uniform $~B_\mathrm{L}$ and $B_\mathrm{V}$ &n/a& $<$$5.5 \times 10^{-6}$, \newline proportional to $\omega_\mathrm{a}$& Current loop around shield to be limited to $<$$1$mA and DC $B_\mathrm{V}$  to be avoided  \\
                Uniform $B_\mathrm{V}$&n/a& 0 & Can be avoided with BPM similar to $B_\mathrm{R}$ case?  \\
                Alternating $~B_\mathrm{V}$ and $B_\mathrm{L}$ \vspace{2mm} &$90^\circ$&$9 \times 10^{-9}$ & CW, CCW average out \\
                Alternating $B_\mathrm{R}$ and $B_\mathrm{V}$ \vspace{2mm} &$0^\circ$ & $3.5 \times 10^{-9}$ & CW, CCW average out  \\
                Alternating $B_\mathrm{R}$ and $B_\mathrm{L}$ \vspace{2mm} &$0, 90^\circ$ & $<$$10^{-10}$ & CW, CCW average out \\
                Alternating $B_\mathrm{R}$ and $B_\mathrm{V}$ \vspace{2mm} &$90^\circ$ & $<$$10^{-10}$ & CW, CCW average out \\
                Alternating $B_\mathrm{V}$ and $B_\mathrm{L}$ &$0^\circ$&Negligible & \\
                \hline \hline
                \label{tbl:summary}
        \end{tabular}
\end{table}

\Table~\ref{tbl:summary} summarizes all of the studied cases, including average and alternating (the average of the magnetic field components around the
circumference vanishes) magnetic field configurations. The following measures have been proposed and are being studied to mitigate systematic effects due to magnetic fields:
\begin{itemize}
        \item SQUID-based BPMs for uniform radial magnetic fields;
        \item less-sensitive BPMs for uniform vertical magnetic fields;
        \item a radially polarized test bunch for the uniform longitudinal magnetic field;
        \item counter-rotating beams.
\end{itemize}
While  coupling between  magnetic fields in the radial and vertical directions is harmless in a continuous ring,  coupling between  beta function and some harmonics of an alternating radial magnetic field splits the beams in the same way as a uniform radial magnetic field\footnote{The origin of the phenomenon is that, even for a vanishing average radial magnetic field, vertical separation of the two beams may occur if, \eg the vertical betatron functions happen to be larger (smaller) at positions with positive (negative) radial field.}. Simulations \cite{bib:MagneticFields} show that the magnetic field must be smoothed down to the 1 pT level to avoid this systematic error. As will be seen in the next section, we have shown that the magnetic field along the shielding prototype is smooth at a level of 10 pT within the storage time. Another one or two orders of magnitude can be gained by flipping the quadrupole signs between runs.

\section{Magnetic shielding}
We are considering magnetic shielding to reduce the magnetic field acting on the beam as much as possible. Magnetic fluctuations inside the shield in the presence of large transient fields must also be kept small. A prototype has been designed in collaboration with P. Fierlinger's group at TUM, Germany (\Fref{fig:prototype_photo}). It contains two layers of Magnifer, a high-permeability material for low-frequency shielding. High-frequency shielding requires a material with high conductivity, such as aluminium. The shielding factor of the system is approximately 500 at low frequencies.

\begin{figure} 
        \centering
        \includegraphics[width=0.7\linewidth]{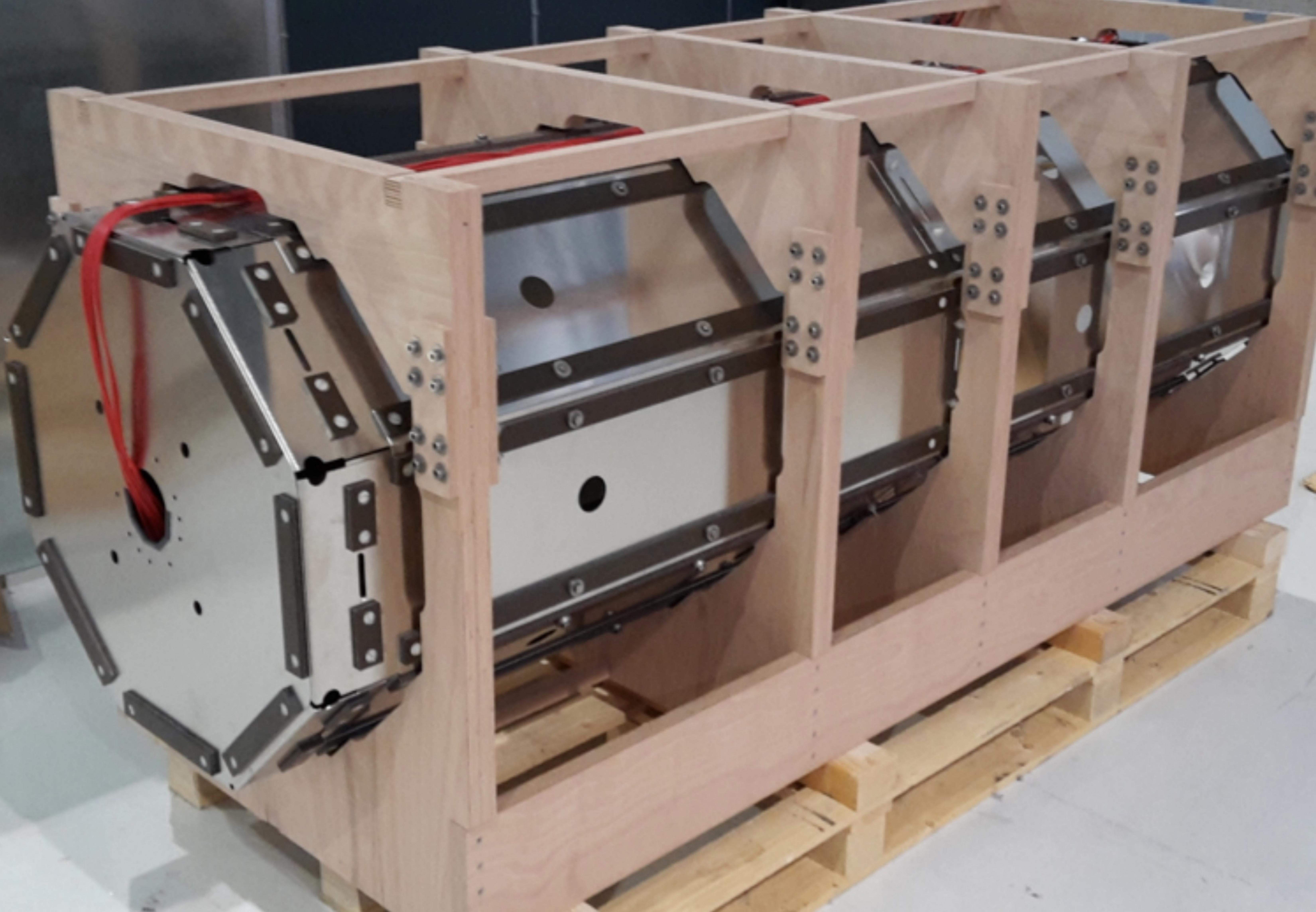}
        \caption{Magnetic shielding prototype, developed in collaboration with Fierlinger Magnetics, Germany. It contains two layers, $1\Umm$ thick, of high-permeability material, separated by $\approx$\,$10\Ucm$ and is approximately $2.5\Um$ long.}
        \label{fig:prototype_photo}
\end{figure}

The working principle of the Magnifer relies on the domain structure inside it. The direction of magnetization is uniform in these small regions, separated by  so-called domain walls. An external magnetic field can move the domain walls, changing the total magnetization of the material. The shielding structure becomes magnetized over time because of this effect. Demagnetization (or degaussing) is commonly used  to avoid it. It is basically achieved by applying an alternating field with a decreasing amplitude. This has an effect similar to shaking,  randomizing the domain magnetization over the material. The red cable shown in  \Fref{fig:prototype_photo} is used to apply a current for degaussing. Our studies showed that the uniformity of the cables along the material matters for the degaussing performance at the inner layer, but not at the outer. Therefore, unlike the outer layer, the inner Magnifer layer has uniformly distributed degaussing cables.

\subsection{Residual field}

There are several key factors in degaussing. First, the amplitude of the applied magnetic field at the beginning of the process should be large enough to saturate the material. The cycles should be slow enough to leave sufficient time for the domains to move ($\approx$\,$10\UHz$ for this prototype). The last steps of degaussing should be smooth enough to obtain an evenly distributed domain configuration. At the end, the material would still have a non-zero magnetization, which results in the so-called `residual field' inside the shielded volume.  \Figure[b]~\ref{fig:residual_field_munich} shows the residual field measurement inside the prototype after degaussing. As seen, a field of 1\,nT  can easily be achieved with two-layer shielding after degaussing.

\begin{figure}
        \centering
        \includegraphics[width=0.7\linewidth]{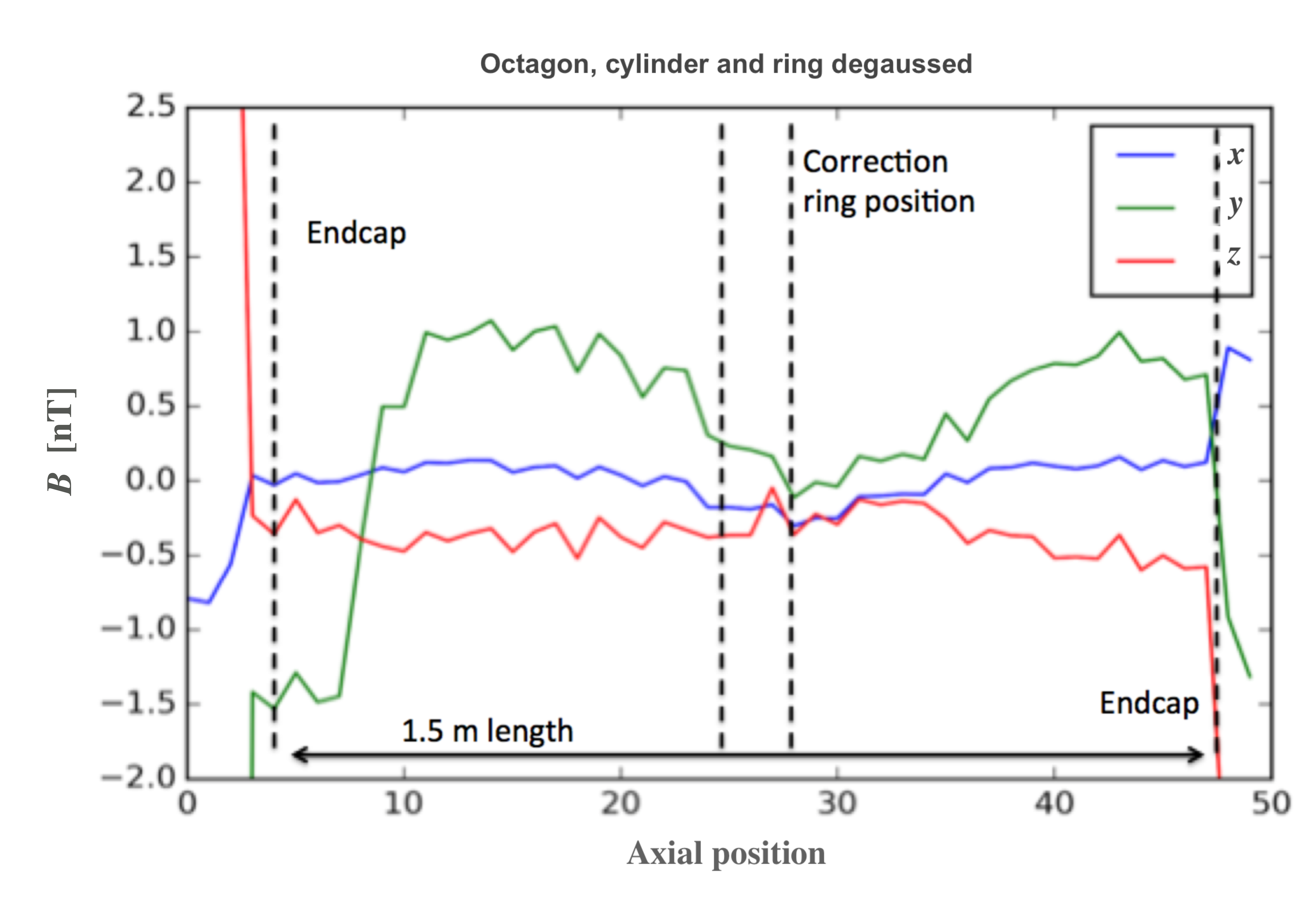}
        \caption{Residual field measurement inside the prototype. The $x$
        axis is the longitudinal position of the fluxgate sensors. The field is larger at the edges, owing to the caps of the prototype, which will not be used when installed at the ring.}
        \label{fig:residual_field_munich}
\end{figure}

\subsection{Time stability of the residual field}

Time stability of the residual field becomes critical, especially when the beta function of the beam is not uniform. Coupling between the varying beta function and the magnetic field moves the beam vertically, mimicking an average radial magnetic field. Runs with alternating quadrupole polarities are proposed to change the polarity of the quadrupoles to cancel this effect. According to simulations, this requires a stable residual field along the ring to $<$$100$\,pT level.

We tested the prototype inside our magnetically shielded room (MSR), as shown in  \Fref{fig:shielding_inside_msr}. In the tests, we used only the outer layer of the prototype. Then, after degaussing it, we measured the magnetic field inside the prototype. \Figure[b]~\ref{fig:Bx} shows the field at three locations along the axis, separated by $70\Ucm$. The measurement lasted almost $25\Uh$. The variation of the field is mainly related to the temperature. It decreases overnight and increases after  sunrise. Of course, the stability during the whole day is irrelevant in the pEDM experiment. Rather, we are interested in the stability within one or two storage runs.

\begin{figure} 
        \centering
        \includegraphics[width=0.45\linewidth]{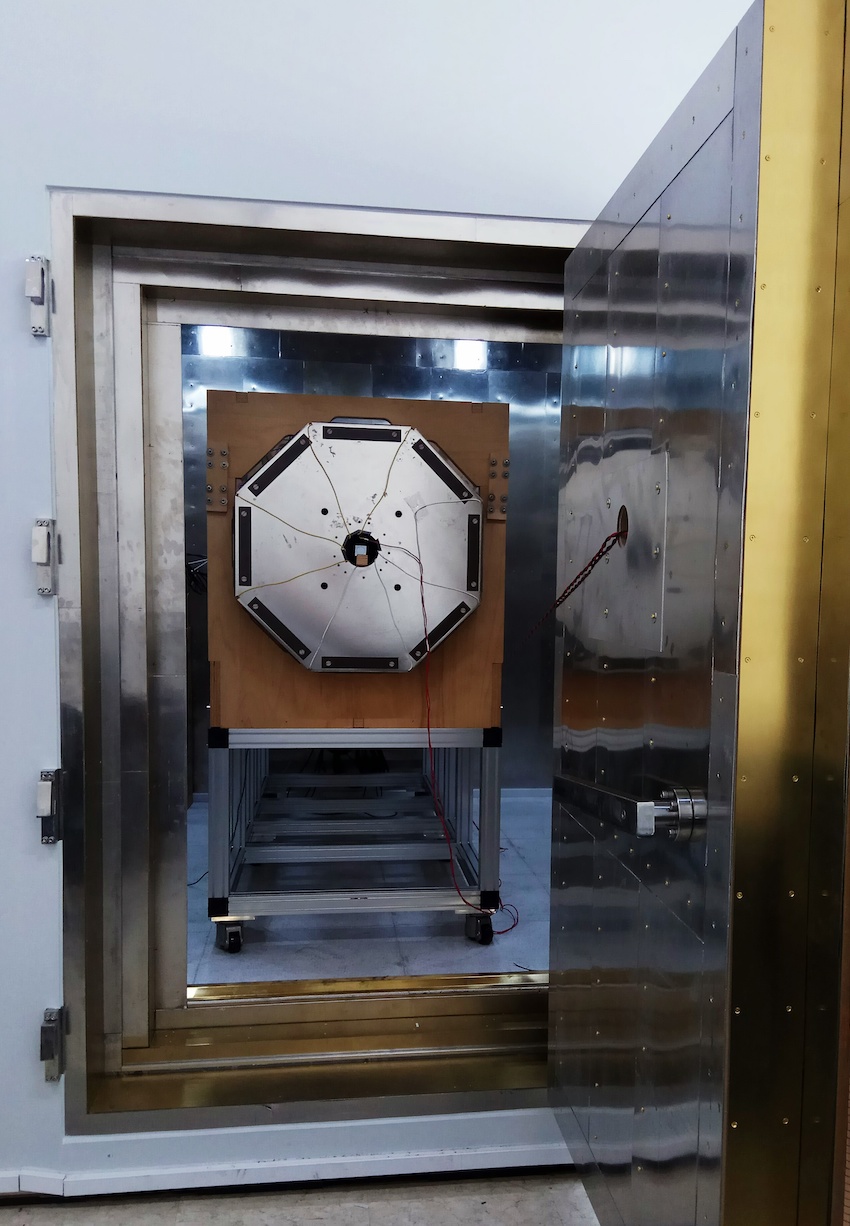}
        \caption{Time stability measurements were made inside the magnetically shielded room.}
        \label{fig:shielding_inside_msr}
\end{figure}

\begin{figure}
        \centering
        \includegraphics[width=0.6\linewidth]{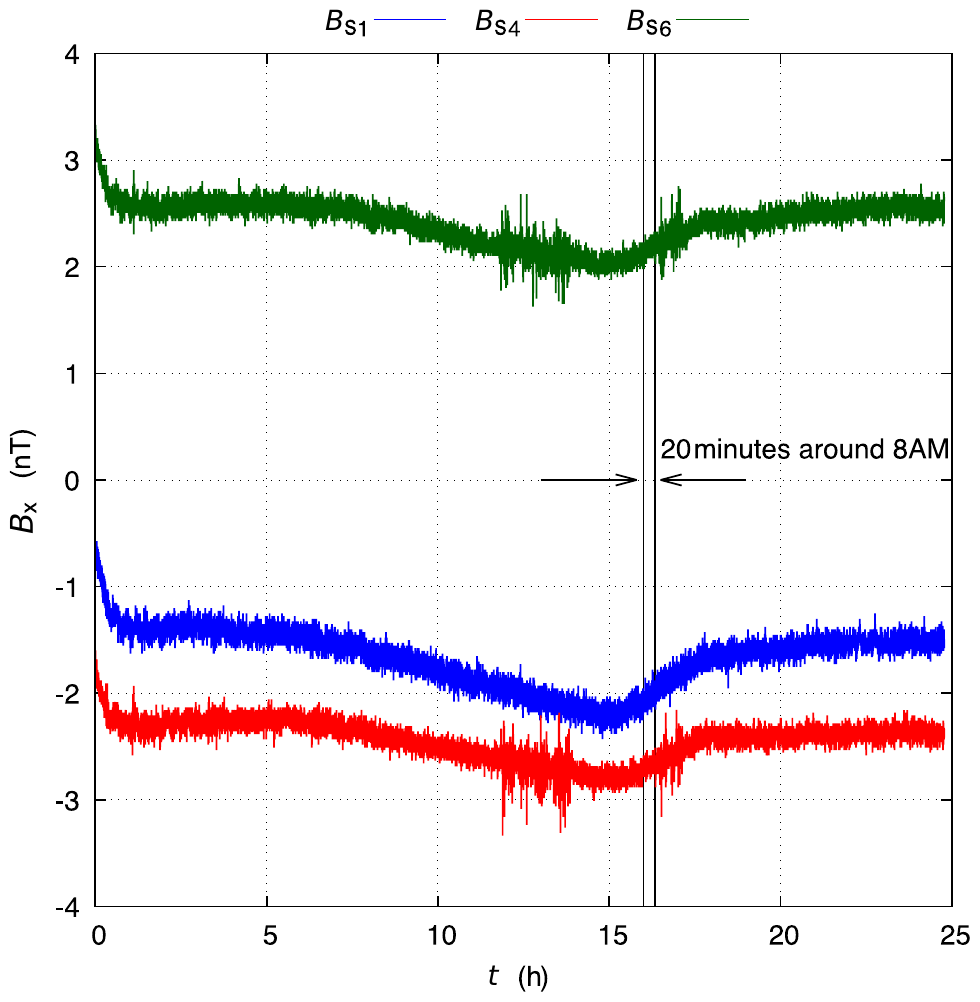}
        \caption{Magnetic field, measured over the course of $25\Uh$. The sensors were located at several locations along the prototype. The dominant reason for the change in the field is the temperature.}
        \label{fig:Bx}
\end{figure}

\Figure[b]~\ref{fig:Bx1_zoom} is an enlargement of  the morning period of \Fref{fig:Bx}, where the temperature changes most rapidly. According to the plot, the change in 20\,min is around 100\,pT. For the effect mentioned previously, the beta function varies at different locations in the ring. Therefore, one needs to look at the correlation between different points.

\begin{figure} 
        \centering
        \includegraphics[width=1\linewidth]{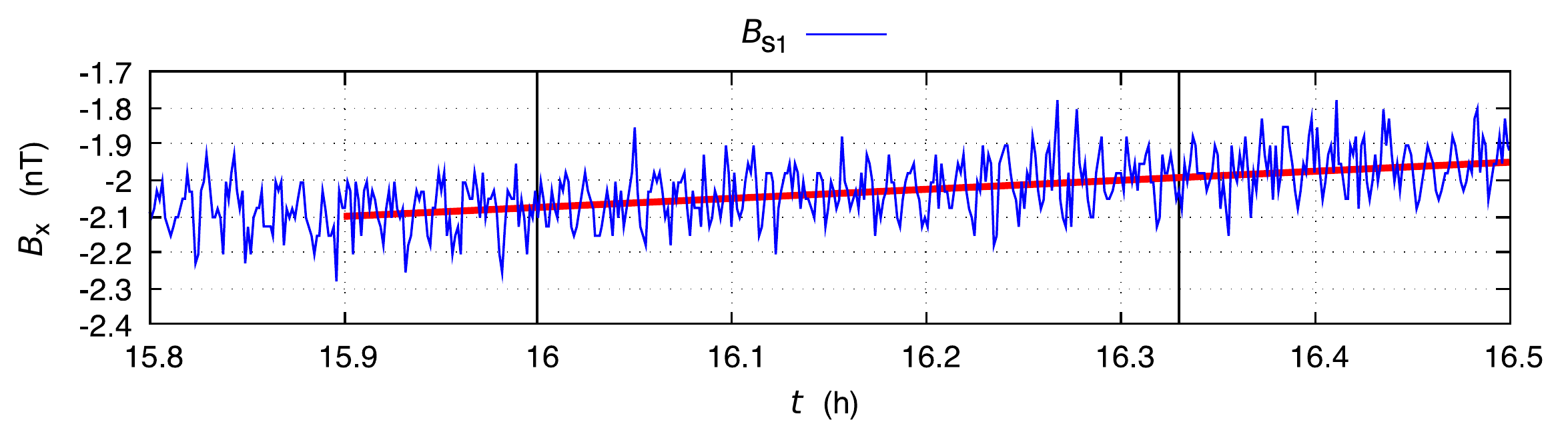}
        \caption{Protons will be stored for 20\,min in the ring. Therefore, the stability of the field at that period is important. Enlarging the marked 20\,min period of  \Fref{fig:Bx}, one sees that the field change is $\approx$\,$100$\,pT. Note that this is the worst period of the measurement, where the temperature changes rapidly.}
        \label{fig:Bx1_zoom}
\end{figure}

\Figure[b]~\ref{fig:Bs1_Bs6_zoom} shows the difference between two sensors in  the same 20\,min period as  \Fref{fig:Bx1_zoom}. The distance between the two sensors is 1.4\,m. The residual field changes by 10\,pT over the  20\,min period.

\begin{figure} 
        \centering
        \includegraphics[width=\linewidth]{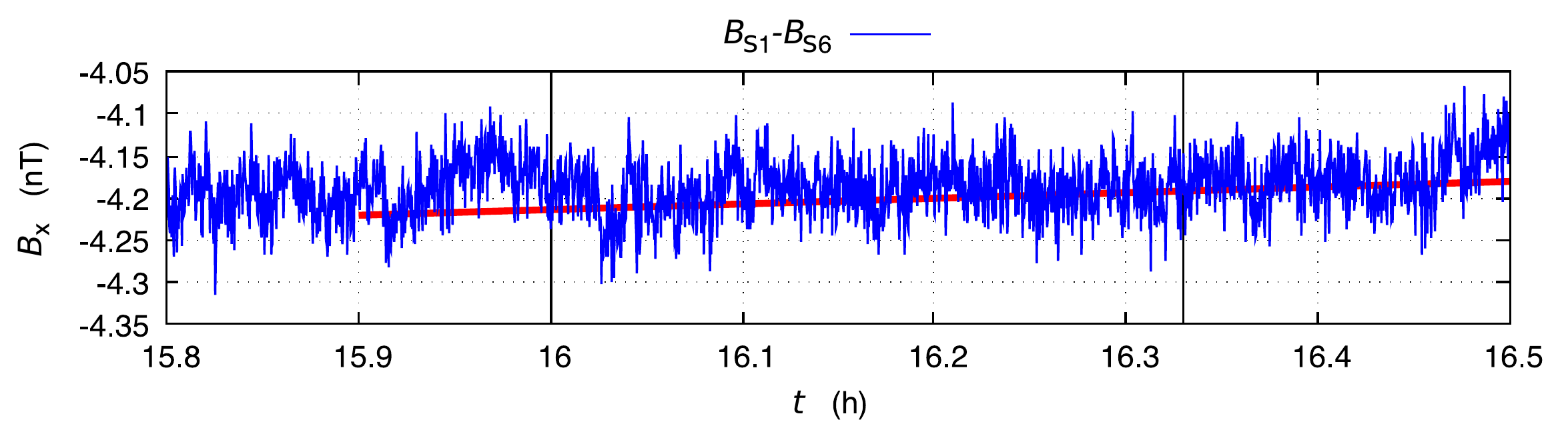}
        \caption{The field at different locations around the ring is quite correlated. The difference between the field at two points 140\,cm apart changes together with the measured field. The difference is $\approx 10$\,pT.}
        \label{fig:Bs1_Bs6_zoom}
\end{figure}

To sum up, we have a prototype to prevent the effects of transient magnetic fields in the pEDM experiment. Its residual field is as low as 1\,nT, with good temporal stability and field uniformity along the cylindrical axis. The temporal stability within 1.4\,m is measured as $\approx$\,$10$\,pT within 20\,min.

\section{Summary}
Our studies show that we can keep the uniform magnetic field under control in an alternating-gradient, all-electric ring. Active and passive cancellation of the magnetic field requires several components, including counter-rotating beams, beam position monitors, a test bunch with horizontal polarization, and magnetic shielding. The tests that we have conducted using SQUID-based BPMs and the magnetic shielding prototype yielded promising results.

Alternating magnetic fields are harmless in a continuous all-electric ring,
but the coupling between the beta function and the alternating radial magnetic field causes a vertical split similar to the uniform radial magnetic field. This can be suppressed by flipping the quadrupoles at every run and keeping the residual field uniformity at the 10\,pT level. However, the hybrid ring is a more efficient solution to this problem. Overall, according to our simulations, the hybrid ring has more flexible requirements for  field cancellation in all scenarios.

\begin{flushleft}

\end{flushleft}
\end{cbunit}

\begin{cbunit}

\csname @openrighttrue\endcsname 
\chapter{Statistical error in a frequency measurement} \label{app:stat-err-freq}
\newcommand{\avrg}[1]{\left<  #1 \right>}

In this appendix,  the statistical error in a frequency measurement,  in particular, for the case of the
vertical polarization  $p_y(t)  = P \sin(\omega t+ \varphi )$ in terms of the total polarization $P$,
is derived
 from the time dependence of counting rates 
\begin{equation}
  n(t) =  n_0 \bigl(1+ C_{\rm s} A \,p_y(t)\cos\Phi\bigr) \,,
\end{equation}
 cf. \Sref{Chap:polarisation-observables}.
The factor $n_0$ includes the unpolarized cross-section, acceptance, and efficiencies, while $A$ is the analysing power and
$\Phi$ is the azimuthal angle of the scattered proton  (weight factor $C_{\rm s}=1$,  cf. \Eref{eq:pC-sigma-px-py}) or
deuteron (weight factor $C_{\rm s} = 3/2$,  cf. \Eref{eq:dC-sigma-px-py}).
The goal  is to determine the frequency $\omega$ and the phase $\varphi$.
In the case of a build-up, we have $|\omega t| \ll 1$  and $\varphi\approx 0$, such that  $\sin(\omega t + \varphi) \approx \omega t + \varphi$.

To determine $\omega$ and $\varphi$, the maximum likelihood method can be used.
The log-likelihood function reads
\begin{equation}
  \ell = \log {\mathcal L} = \sum_i \log \bigl(n_0 ( 1 + C_{\text{s}} A P \sin(\omega t_i+ \varphi) \cos\Phi_i  \bigr) \,,
\end{equation}
where the sum runs over all detected events at times $t_i$.
The likelihood estimators are given by setting the first derivatives of $\ell$ to zero:
\[
\frac{\partial \ell}{\partial \omega} = \sum_i \frac{ t_i C_{\text{s}}AP  \cos (\omega t_i +
  \varphi) \cos\Phi_i}{1+C_{\text{s}}AP \sin(\omega t_i + \varphi) \cos\Phi_i} \stackrel{!}{=} 0 \, , \qquad
\frac{\partial \ell}{\partial \varphi} = \sum_i \frac{C_{\text{s}}AP\cos (\omega t_i +
  \varphi) \cos\Phi_i}{1+C_{\text{s}}AP \sin(\omega t_i + \varphi)\cos\Phi_i} \stackrel{!}{=} 0 \, .
\]

Here, we are more interested in the error of the estimator, which is given by
the expectation values of the second derivatives of the log-likelihood function $\ell$:
\begin{align}
     \frac{\partial^2 \ell}{\partial \omega^2}
      &= -\sum_i \frac{t_i^2 C_{\text{s}}AP\cos(\Phi_i)  \{C_{\text{s}}AP\cos(\Phi_i)+\sin(\omega t_i+ \varphi) \}}
               {\{1+C_{\text{s}}AP\cos(\Phi_i) \sin(\omega t_i+ \varphi)\}^2}
               \label{dloo} \, ,\\
      \frac{\partial^2 \ell}{\partial \varphi^2}
      &= -\sum_i \frac{C_{\text{s}}AP\cos(\Phi_i)\{C_{\text{s}}AP\cos(\Phi_i)+\sin(\omega t_i+ \varphi)\}}{\{1+C_{\text{s}}AP\cos(\Phi_i)
               \sin(\omega t_i+ \varphi)\}^2}
               \label{dlphiphi} \, ,\\
      \frac{\partial^2 \ell}{\partial \omega \partial \varphi}
      &= -\sum_i \frac{t_i C_{\text{s}}AP\cos(\Phi_i)\{C_{\text{s}}AP \cos(\Phi_i)  +  \sin(\omega t_i+ \varphi)\}}
      {\{1+ C_{\text{s}}AP
               \cos(\Phi_i) \sin(\omega t_i+ \varphi)\}^2}
               \label{dlophi} \, .
\end{align}
To evaluate the error, we replace the sums in Eqs.~(\ref{dloo}) to (\ref{dlophi}) with the corresponding
expectation values, \ie
\begin{equation}
  \avrg{\frac{\partial^2 \ell}{\partial \omega^2}}  = - \int_0^T\!\! {\rm d}t \,n(t) \,\, \frac{1}{2\pi}\! \int_0^{2\pi} \!\! {\rm d}\Phi
  \,\frac{t^2 C_{\text{s}}AP\cos(\Phi)\{C_{\text{s}}AP\cos(\Phi)+\sin(\omega t+\varphi)\}}{\{1+C_{\text{s}}AP\cos(\Phi) \sin(\omega t+\varphi)\}^2} \,,   \label{dloo-ev}
\end{equation}
etc. To solve the integral analytically, we assume $C_{\text{s}} A P\ll 1$.
In this approximation, the expectation values can easily be performed:
 using $\omega T \gg 1$, where $T$ is the period of one measurement cycle, the $\sin, \cos$, and $\sin \cdot \cos$ terms average to 0,
and $\sin^2$ and $\cos^2$ give a factor of {\textonehalf}.
Applying, finally, $N=n_0 \cdot T$ as the total number of observed events in the measurement-cycle period $T$, one gets
\begin{align*}
  \avrg{\frac{\partial^2 \ell}{\partial \omega^2}}
  &\approx
   -\frac{n_0 C_{\text{s}} AP }{2\pi} \int_0^T {\rm d}t \int_0^{2\pi} {\rm d}\Phi \ t^2 \cos\Phi \,
      \{C_{\text{s}}AP\cos\Phi+\sin(\omega t +\varphi)\} \\
 &\qquad\qquad\qquad\qquad\qquad\qquad\quad \ \mbox{}\times \{1-C_{\text{s}}AP\cos\Phi \sin(\omega t+\varphi)\} \nonumber \\
&\approx -\frac{n_0 (C_{\text{s}}AP)^2}{4}\int_0^T{\rm d}t\, t^2  = -\frac{1}{12} n_0  (C_{\text{s}}AP)^2 T^3 = -\frac{1}{12} N (C_{\text{s}}AP T )^2 \, ,\\
\avrg{\frac{\partial^2 \ell}{\partial \varphi^2}} &= -\frac{1}{4} N (C_{\text{s}}AP)^2\,, \\
\avrg{\frac{\partial^2 \ell}{\partial \omega \partial \varphi}} & =  -\frac{1}{8} N (C_{\text{s}}AP)^2T \, .
\end{align*}
This leads to the following inverse covariance matrix:
\begin{equation}\label{covinv}
\mbox{cov}^{-1}(\omega, \varphi) = -\avrg{\frac{\partial^2 \ell}{\partial a_i \partial a_j}} =
N (C_{\text{s}}AP)^2 
\begin{pmatrix}
{T^2}/{12}  &   {T}/{8}    \\
{T}/{8} & {1}/{4}  \\
\end{pmatrix}
 \, , 
\end{equation}
with $(a_1,a_2) \equiv (\omega, \varphi)$ and
\begin{equation}\label{cov1}
\mbox{cov}(\omega, \varphi) = \frac{1}{N (C_{\text{s}}AP)^2}
\begin{pmatrix}
{\phantom{-}48}/{T^2}  &   {-24}/{T}  \\
{-24}/{T} & {\phantom{-}16} \\
\end{pmatrix}
 \, .
\end{equation}
The statistical squared errors in $\omega$ and $\varphi$ are given by the diagonal
elements of the matrix in \Eref{cov1}:
\[
 \sigma_{\omega}^2 =  \frac{48}{N (C_{\text{s}}AP)^2T^2}  \, , \qquad
 \sigma_{\varphi}^2 =  \frac{16}{N (C_{\text{s}}AP)^2}  \, .
\]
If the phase is known, the squared error in the frequency is $\sigma_{\omega}^2 = {12}/\bigl(N (C_{\text{s}}APT)^2\bigr)$\footnote{The right-hand side follows from
the element $(1,1)$ of the matrix in \Eref{covinv}.}.

As expected, the  errors scale with ${1} / {\sqrt{N}}$.
The error in $\omega$ scales in addition with $(C_{\text{s}}APT)^{-1}$, \ie to measure a frequency, it is better
to have  a longer measurement time and a higher polarization; this also makes perfect sense.


Up to now we assumed that $\omega T \gg 1$.
In a next step, we consider the case where $\omega T \ll 1$.
This applies, for example, to the case of a spin movement only due to an EDM.
We discuss the case for $\varphi\approx 0$, \ie a build-up of a vertical polarization.
The detector rate is given by
\begin{equation}
n(t) =   n_0 \bigl(1+C_{\text{s}}AP\sin(\omega t + \varphi)\cos(\Phi) \bigr) \approx n_0 \bigl(1+C_{\text{s}}AP\cdot (\omega t + \varphi)\cos(\Phi)\bigr) \,  .
\end{equation}
In this case, one finds, for the inverse of the covariance matrix,
\begin{equation}\label{covinv1}
\mbox{cov}^{-1}(\omega, \varphi) = -\avrg{\frac{\partial^2 \ell}{\partial a_i \partial a_j}} = N (C_{\text{s}}AP)^2
\begin{pmatrix}
{T^2}/{6}  &   {T}/{4}    \\
{T}/{4} & {1}/{2}  \\
\end{pmatrix}
 \, ,
\end{equation}
leading to
\begin{equation}
\mbox{cov}(\omega, \varphi) = \frac{1}{N (C_{\text{s}}AP)^2}
\begin{pmatrix}
{\phantom{-}24}/{T^2}  &   {-12}/{T}  \\
{-12}/{T} & {\phantom{-1}8} \\
\end{pmatrix}
 \, ,
\end{equation}
such that the  squared statistical error in $\omega$  is given by
$ \sigma^2_{\omega} = {24}/ \bigl({N (C_{\text{s}}APT)^2}\bigr)$.
This is an alternative derivation for the error of scenario A of \Sref{app:sigd}.

In addition, under the aforementioned conditions, $\omega T \ll 1$ and $|\varphi| \ll 1$,
there are two alternative ways to rederive  the error of scenario B of \Sref{app:sigd};
either by a direct application of   Eqs.\,(\ref{dloo}) to (\ref{dlophi}) for  $i=1,2$,  with, say, $t_1=0$ and $t_2=T$,
or by simply inserting the ansatz
\begin{equation}
  n(t) =   {\textstyle\frac{1}{2}} N\big( 1+ C_{\text{s}}AP\cdot( \omega t + \varphi) \bigr) \cdot \bigl\{ \delta(t) + \delta(t-T) \bigr\}
\end{equation}
into \Eref{dloo-ev} and the other two analogous equations. The  pertinent inverse covariance matrix is then given by
\begin{equation}\label{covinv1b}
\mbox{cov}^{-1}(\omega, \varphi) = -\avrg{\frac{\partial^2 \ell}{\partial a_i \partial a_j}} = N (C_{\text{s}}AP)^2
  \begin{pmatrix}
            {T^2}/{4}  &   {T}/{4}    \\
             {T}/{4} & {1}/{2}  \\
    \end{pmatrix}
   \, ,
\end{equation}
leading to
\begin{equation}
     \mbox{cov}(\omega, \varphi) = \frac{1}{N (C_{\text{s}}AP)^2}
            \begin{pmatrix}
                    {\phantom{-}8}/{T^2}  &   {-4}/{T}  \\
                    {-4}/{T} & {\phantom{-1}4} \\
      \end{pmatrix}
     \, ,
\end{equation}
as the covariance matrix and  $\sigma^2_\omega = 8/\bigl(N (C_{\text{s}}APT)^2\bigr)$
as the squared statistical error in $\omega$ for scenario B.

The resulting figures of merit on $\omega$ for  the period $T$ of one measurement cycle are given in \Tref{tab:2freq}.
In scenario C (frequency measurement) of \Sref{app:sigd}, the squared error is twice as large as
in scenario A (polarization build-up, continuously measured).
This makes sense, because, in the latter case, usually only those events are considered where the slope is greatest and therefore contains the most information about the frequency\footnote{However, in a frequency measurement with a {\em fixed} phase, the squared error is half as small as in scenario A, while the slope measurement  via two endpoints and
with  vanishing offset (phase) has the smallest statistical error.}.

\begin{table} [h]
\begin{center}
\caption{Figures of merit (FoMs) on  $\omega$ with $\varphi$  the pertinent phase:
$N$ is the total number of observed events in the period $T$ of one measurement cycle;
 $\Phi$ is the azimuthal angle of the scattered  protons ($C_{\text{s}}=1$) or deuterons ($C_{\text{s}}=3/2)$;
$P$  is the polarization of the beam;
 $A$ is the  analysing power.
 The  listed scenarios refer to \Sref{app:sigd}.
 \label{tab:2freq}}
\begin{tabular}{ l   l  l l  l}
\hline \hline
Measurement & Underlying counting rate  $n(t)$   &   FoM$_\omega$  &   Constraints    &  Scenario 
 \\
\hline  
Frequency   & $\frac{N}{T}\bigl(1\,{+}\,C_{\text{s}}AP\sin(\omega t\,{+}\,\varphi) \cos\Phi\bigr)$ &   $\frac{1}{48}{N (C_{\text{s}}AP T)^2}$      &
$\omega T \gg1$ &  C      \\[1mm]
                    &        &     $\frac{1}{12}{N (C_{\text{s}}APT)^2}$     & $\omega T \gg1$,   fixed $\varphi$  & ---  \\[1mm] 
Slope,   &$\frac{N}{T}(1 +C_{\text{s}}AP \cdot (\omega t +\varphi)\cos\Phi)$             & $\frac{1}{24}{N (C_{\text{s}}APT)^2}$   &
     $\omega T \ll 1$,  $|\varphi| \ll 1$ &  A \\[1mm]
continuous     &  & $\frac{1}{6}{N (C_{\text{s}}APT)^2}$ & $\omega T \ll 1$,  $\phantom{|}\varphi\phantom{|} = 0 $ &
---  \\[1mm] 
Slope via  &$\frac{N}{2}(1+C_{\text{s}}AP\cdot (\omega t +\varphi)\cos\Phi)$             & $\frac{1}{8}{N (C_{\text{s}}AP T)^2}$   &
     $\omega T \ll 1$,  $|\varphi| \ll 1$ & B \\[1mm]
endpoints      & $ \mbox{}\ \times \bigl\{ \delta(t) + \delta(t-T) \bigr\}$
& $\frac{1}{4}{N (C_{\text{s}}A P T)^2}$ & $\omega T \ll 1$, $\phantom{|}\varphi\phantom{|} = 0 $  & --- \\

\hline \hline
 \end{tabular}
\end{center}
\end{table}

In all the derivations, we assumed
that  the  product $C_{\rm s} A P$  is known from other measurements,
although for scenario C even this is  not needed. Moreover, the results were obtained under the premise
that the period of the measurement cycle $T$  is small compared with the spin coherence time $\tau$, \ie $T\ll \tau$.
See  \Sref{Chap:statistics-general}  for the case where this constraint is relaxed.

\end{cbunit}

\begin{cbunit}

\csname @openrighttrue\endcsname 
\chapter{Gravity and general relativity as a `standard candle'}
\label{app:gravity}


\vspace{-0.5mm}
 The  subtracted Thomas--BMT equation including corrections for general relativity  is~~\cite{Obukhov:2016vvk,Silenko:2004ad,Silenko:2006er}
\begin{equation}
 \frac{{\rm d}\vec S}{\rm d t} = \left(  \vec \Omega_{\text{MDM-cycl} }+  \vec \Omega_{\rm EDM} +  \vec \Omega_{\rm GRgeo} \right) \times \vec S \,,
\end{equation}
where,
in  the  Frenet--Serret coordinate system, whose axis orientation is   determined
from the local particle motion~\cite{Silenko:2015jqa}  
$\vec\Omega_{\text{MDM-cycl}}$ refers to the angular velocity
from the magnetic dipole moment {\em minus} the cyclotron angular velocity,
$\vec\Omega_{\rm EDM}$ to the angular velocity from the electric dipole moment, and
$\vec\Omega_{\rm GRgeo}$ to the angular velocity of  the geodetic (de Sitter)  minus
the corresponding angular velocity for the particle revolution: 
\begin{eqnarray}
  \vec \Omega_{\text{MDM-cycl}}
   &=& - \frac{q}{m} \left[  G \vec B  - \frac{\gamma G}{\gamma+1}\vec \beta \left(\vec \beta\cdot \vec B\right)
                                  -\left(G - \frac{1}{\gamma^2-1} \right) \frac{\vec \beta\times \vec E}{c}\right]
                                 , \label{MDM}\\[-0.5mm]
 \vec \Omega_{\rm EDM}
   &=& - \frac{\eta q}{2 m c} \left[  \vec E  - \frac{\gamma}{\gamma+1}\vec \beta \left(\vec \beta\cdot \vec E\right)
                                  +c \vec \beta\times \vec B \right]  , \label{EDM}\\[-0.5mm]
      \vec \Omega_{\rm GRgeo}        &=&     -\frac{\gamma}{\gamma^2-1}\frac{\vec \beta \times \vec g}{c}\,.
\end{eqnarray}

Deviating from the local coordinate system used in
Chapter~\ref{Chap:ExpMethod}, here
the right-handed,  beam-comoving coordinate
system $(x,y,z)$ is defined by the unit vectors
$\hat z = \vec \beta /|\vec \beta| \equiv \hat \beta$, $\hat y = -\vec g/|\vec g| \equiv -\hat g$, and
$\hat x = -\hat z \times \hat y =\hat \beta \times \hat g$, \ie the unit vector $\hat y$ is always pointing opposite to the gravitational acceleration $\vec g$. Thus,
for
a clockwise beam, we have $\hat x = \hat r$, while for a counterclockwise beam $\hat x =  -\hat r$, where $\hat r$ is the
outside-pointing radial unit vector inside the storage ring plane.

Note that $\vec\Omega_{\rm GRgeo}$ is calculated from the
difference between the gravity-induced `spin-orbit' precession around
a radial axis  in the Earth's gravitational field, $\vec \Omega_\mathrm{LS}$ (the de Sitter precession, also known as  the geodetic effect)~\cite{Khriplovich:1997ni,Pomeransky:2000pb,Nikolaev_Spin2018},
and the particle revolution around the same axis in the Earth's gravitational field, $\vec\Omega_{\rm rev}$,
 cf. \cite{Obukhov:2016vvk}:
\[
 \vec \Omega_\mathrm{LS} = \frac{2\gamma+1}{\gamma+1} \frac{\vec \beta\times\vec g}{c}\,,
 \qquad \vec\Omega_{\rm rev} =  \frac{1+\beta^2}{\beta^2} \frac{\vec \beta \times \vec g}{c}
                                         = \frac{2\gamma^2-1}{\gamma^2-1} \frac{\vec \beta \times \vec g}{c}\,,\qquad
    \vec\Omega_{\rm GRgeo} =     \vec \Omega_\mathrm{LS} -     \vec\Omega_{\rm rev}   \,.
 \]

 Here, $\vec g$ is the gravitational acceleration at the Earth's surface---for further definitions see Ref.~\cite{Obukhov:2016vvk} and Chapter~\ref{Chap:ExpMethod}.
Furthermore, according to Ref.~\cite{Obukhov:2016vvk}, $\vec E$ and $\vec B$ in Eqs.~(\ref{MDM}) and (\ref{EDM}) must
be replaced by $\vec E  + \vec E_{\vec g}$ or $\vec B +\vec B_{\vec g}$, respectively, where
$\vec E_{\vec g}$ and $\vec B_{\vec g}$  are focusing fields compensating the gravitational downwards  pull on  beam particles of  mass $m$ and velocity $c\vec\beta$,
\begin{equation}
   \vec F_{\vec g} = \gamma\left(1 + |\vec \beta|^2 \right) m \vec g = \frac{2\gamma^2-1}{\gamma}m\vec g \,.
\end{equation}
This follows from the storage ring lattice condition for the closed orbit,
\begin{equation}
 \frac{2\gamma^2-1}{\gamma}m\vec g
 + q \left( \vec E_{\vec g}
 + c \vec \beta \times \vec B_{\vec g}\right) \equiv 0\,,
  \label{lattice-cond}
\end{equation}
\eg the upwards pointing vertical electric field  for a pure electric ring reads
\begin{equation}
\vec E_{\vec g}=  (\vec E_{\vec g} \cdot \hat y) \hat y  = \frac{2\gamma^2-1}{\gamma} \frac{m}{q}(-\vec g)\,,
\end{equation}
while the gravity-compensating radially inwards or outwards pointing magnetic field for a
counterclockwise or clockwise beam would be
\begin{equation}
\vec B_{\vec g} = (  \vec B_{\vec g} \cdot \hat x ) \hat x = (2\gamma^2-1)\frac{\gamma }{\gamma^2-1}
\frac{m}{q}\,\frac{ \vec \beta \times
           \vec g}{c} = \frac{2\gamma^2-1}{\sqrt{\gamma^2-1}} \frac{m |\vec g|}{c q}\, \hat \beta \times \hat g \, ,
           \label{B_g}
\end{equation}
with\footnote{Assuming here, and in the following, that the storage ring plane is normal to $\vec g$.}
\begin{equation}
 c \vec\beta \times \vec B_{\vec g} =  (2\gamma^2-1){\frac{ \gamma}{\gamma^2-1}}\frac{m}{q}\left(\vec \beta (\vec \beta\cdot \vec g) -\vec g (\vec \beta\cdot \vec \beta) \right) =
  \frac{2\gamma^2-1}{\gamma}\frac{m}{q} (-\vec g)\,.
           \end{equation}

\begin{enumerate}
\item
Note that  for the case of the frozen-spin (fs) condition,
$1/(\gamma^2-1)=G$,  in an all-electric ring
we have~\cite{Nikolaev_Spin2018}
\begin{equation}
\vec \Omega_{\rm GR}^{B=0, {\rm fs}}=-\left.\frac{\gamma}{\gamma^2-1}
\frac {\vec \beta  \times  \vec g}{c}\right |_{\rm fs}
  =  -\frac{\hat \beta \times \vec g}{c} \,  G \sqrt{\frac{1+ G}{G}} \frac{1}{\sqrt{1+G}}  =  -\frac{ |\vec g| \sqrt{G}}{c}
  \hat\beta\times\hat g\,,
 \label{eqn:Omega_frozen}
\end{equation}
which agrees with the earlier result of Orlov, Flanagan,
and Semertzidis~\cite{Orlov:2019gtt}.
Thus, by equating $\frac{1}{2} \Omega_{\rm GR}^{B=0, {\rm fs}}/E_{\rm r}$ with
 $d_{\sf p}^{\rm GR}  =
 {\textstyle\frac{1}{2}} \eta_{\sf p}^{\rm GR}    {e\hbar} / {2mc}$ (cf. \Eref{emeq5}),  where $E_{\rm r}$ is the mean radial component
 of the electric field and $m$ and $e$ denote here the proton mass and charge,  respectively.
 One can map the geodetic effect of general relativity
 to a `fake'  proton EDM of, \eg  modulus
 \[
\begin{array}{l     }
 d_{\sf p}^{\rm GR} \approx 1.44 \times 10^{-28} e \,{\rm cm} \ \, (\mbox{\ie}\
 \eta^{\rm GR}_{\sf p}\approx 2.75 \times 10^{-14})    \ \,  \mbox{corresponding to}\  E_{\rm r} = 10 \,{\rm MV/m},\\[1mm]
 d_{\sf p}^{\rm GR} \approx 2.75 \times 10^{-28} e \,{\rm cm} \ \, (\mbox{\ie}\
 \eta^{\rm GR}_{\sf p}\approx 5.22 \times 10^{-14})    \ \,  \mbox{corresponding to}\  E_{\rm r} = 5.27 \,{\rm MV/m},\\[1mm]
  d_{\sf p}^{\rm GR} \approx 1.44 \times 10^{-27} e\, {\rm cm} \ \, (\mbox{\ie}\
  \eta^{\rm GR}_{\sf p} \approx 2.75 \times 10^{-13})
  \ \, \mbox{corresponding to}\ E_{\rm r} = 1\,{\rm MV/m}.
  \end{array}
\]
In this way, the geodetic effect could
serve as a standard source or `standard candle'  for EDM measurements  in frozen-spin all-electric  storage rings,
while the gravity-compensating fields  just correspond to $E_{\vec g} \approx 0.173\, \upmu {\rm V/m}$ or $B_{\vec g} \approx 0.967\,{\rm fT}$.
Such  tiny focusing fields are automatically generated by a
minuscule orbit displacement by the Earth's gravitational pull.

\item
If the radial  component $B_x =\hat x \cdot \vec B$ of the magnetic field
is identical to zero, the $\vec F_{\vec g}$  compensating
field only arises from the vertical electric field, $E_y =\hat y \cdot \vec E$; therefore, we would have,
as the gravity-induced contribution to the angular velocity~\cite{Obukhov:2016vvk,Nikolaev_Spin2018},
\begin{equation}
    \vec\Omega_{\rm GR}^{B_x=0} = \vec\Omega_{\rm GRgeo} - \frac{q}{m} \left(G-\frac{1}{\gamma^2-1}\right) \frac{\vec\beta\times \vec E_{\vec g}}{c} = \frac{1-G(2\gamma^2-1)}{\gamma}
    \frac{\vec\beta\times\vec g}{c}.
    \label{Omega_elec}
    \end{equation}
Obviously, in the frozen-spin scenario of an all-electric ring,  $1/(\gamma^2-1)=G$,
the result  of \Eref{eqn:Omega_frozen} and thus of Ref.~\cite{Orlov:2019gtt} is recovered.

\item
If the vertical electric field $E_y = \hat y \cdot \vec E$ is identical to zero,  the $\vec F_{\vec g}$ compensating field
only arises from the radial magnetic field $B_x =\hat x\cdot \vec B $; therefore, we would find, as the gravity-induced
contribution to the angular velocity~\cite{Silenko:2015jqa,Obukhov:2016vvk,Nikolaev_Spin2018},
\begin{equation}
    \vec\Omega_{\rm GR}^{E_y=0} = \vec\Omega_{\rm GRgeo} - \frac{q}{m} G \vec B_{\vec g} = - \frac{\gamma}{\gamma^2-1}\Bigl (1+ G(2\gamma^2-1)\Bigr)
                                         \frac{\vec \beta \times \vec g}{c} .
   \label{Omega_magn}
\end{equation}
If the frozen-spin condition, $1/(\gamma^2-1)=G$, of the all-electric ring is inserted, the
result of
\Eref{Omega_magn}  is enhanced
by a factor $(3+G)$ in comparison with \Eref{eqn:Omega_frozen}, \ie
\begin{equation}
   \vec\Omega_{\rm GR}^{E_y=0, {\rm fs}} =
   -\frac{ |\vec g|}{c} (3+G)\sqrt{G}\,\hat \beta \times \hat g\,.
   \label{eqn:Omega_frozen_Bx}
\end{equation}

\item
In a mixed ring with $E_y\neq 0\neq B_x$, using
\begin{equation}
  \kappa\equiv   \frac{c\beta B_x}{ E_y} \approx {\rm const.}\,,
\end{equation}
we can derive, from the storage ring lattice condition (\Eref{lattice-cond})
\begin{equation}
 \frac{m |\vec g|}{q} \, \frac{2\gamma^2-1}{\gamma}= E_y +c \beta  B_x = E_y(1+\kappa)\,,
  \label{frozen_cond2}
\end{equation}
the following expression for the gravity-induced angular velocity~\cite{Nikolaev_Spin2018}:
\begin{eqnarray}
 \vec \Omega_{\rm GR}^{E\,B} (\kappa)
  &=&\left\{ -\frac{\gamma}{\gamma^2-1}\frac{\beta|\vec g|}{c}
      -\frac{q}{m}\frac{G}{c\beta}
     \left( c\beta B_x + \left (1-\frac{1}{G\gamma^2-1}\right)  \beta^2 E_y   \right)\right\}
     \hat\beta\times \hat g \ \nonumber\\
     &=& \left\{ -\frac{\gamma}{\gamma^2-1} 
  -\frac{q}{m |\vec g|}\frac{G}{\beta^2}
     \left( c\beta B_x + E_y  - \frac{1}{\gamma^2}\left (1+\frac{1}{G}\right)  E_y   \right)\right\}
    \frac{ \vec \beta\times \vec g}{c}\nonumber\\
     &=&- \frac{\gamma}{\gamma^2-1}\left\{1
     + (2\gamma^2-1) \left(G -\frac{G+1}{\gamma^2(1+\kappa)}\right)\right\} \frac{\vec \beta\times \vec g}{c} \nonumber\\
     &=& \frac{1}{1+\kappa} \left(\vec\Omega_{\rm GR}^{B_x=0} + \kappa \vec\Omega_{\rm GR}^{E_y=0}\right)\,.
  \label{Omega_EB}
   \end{eqnarray}
    Of course, one recovers
   Eqs. (\ref{Omega_elec}) and (\ref{Omega_magn}) from \Eref{Omega_EB}
   if one simply inserts   \mbox{$\kappa\to 0$} or
    $\kappa\to\infty$, respectively, while, by applying
    the `frozen-spin value',
    $1/(\gamma^2-1)=G$,  one would obtain the general
    form $- \hat \beta \times (\vec  g/c)  \sqrt{G}(1+(3+G)\kappa)/(1+\kappa)$.
\end{enumerate}
Note that the contributions  (Eqs. (\ref{eqn:Omega_frozen}) to (\ref{eqn:Omega_frozen_Bx}) and
(\ref{Omega_EB}))  switch sign
if a counterclockwise beam is replaced by a clockwise one.
This clearly separates these contributions from any (MDM-term induced) {\em fake} EDM signal when a {\em radial} magnetic field points, for
both beams,
in the same direction---either in the outward  ($\hat r$) or in the inward radial  ($-\hat r$) direction.
In fact, if  the scenario $E_y=0$ can be
realized  (or the value of $\kappa$ can be determined in the general case of \Eref{Omega_EB} by some means), the lattice orbit condition (\Eref{lattice-cond}) ensures that $B_x$ of each of the
beams is determined, on average, by \Eref{frozen_cond2}. Thus, the
extraction of the gravity-induced spin rotation from the half-sum or half-difference of  counterclockwise and clockwise
beams---assuming that the horizontal spins  of the beams point
in the opposite or same direction\footnote{Here,
the  qualifier `opposite or same spin direction' in
the frozen-spin scenario refers to  the setting
that the {\em components} of the horizontal spin  {\em in the beam directions} of
the clockwise and counterclockwise cases agree or differ in sign.}---would determine the orbit-averaged value of the effective radial magnetic field,  which could
then  be used to correct the EDM signal.

\begin{flushleft}

\end{flushleft}
\end{cbunit}

\begin{cbunit}

\chapter[Beam preparation]{Beam preparation\footnote{This appendix is mostly based on ideas of Richard Talman (Cornell University). }}
\label{app:BeamPrep}

\section{Design principles for bunch polarization patterns}
Before describing proposed injection sequences,
it is useful to establish some principles common to all or most schemes,
whether for a prototype or full-scale ring, or for single- or dual-beam injection.

Assuming that the harmonic number is 80 for the full-scale ring, one can have a
lopsided fill with as many as 60 consecutive stable buckets filled and the other
20 empty. This allows the injection kicker to be pulsed on for half a microsecond
or so, which is comfortably long. To fill the other beam,
the CCW one, the bunch train and
kicker duration can be the same. Similar considerations apply to the prototype ring.

The stored bunches would then be too close to be acted on individually, so maybe one
would prefer to have just 30 filled buckets, alternating with empty buckets. The
spacing between bunches would be too close for single-bunch injection or extraction,
but it could be amply long for `tweaking'  bunches individually.

A useful principle recognizes that the final ring is the `experiment'
and the injection
system is not. Any time spent in the final ring adjusting the bunch structure is time
taken away from the experiment, so time taken to trim the spins after injection should be
minimized.  The
responsibility for best arranging the bunch pattern is therefore delegated to
the injection system. Minimum injection time would be achieved by injecting just
two trains of prepared bunches, which could reduce the set-up time to as little as
$10\Us$  or so.

Most of the following
principles are intended to ensure the uniformity of all
polarized bunch properties, at least to the extent possible, by
assuring that all bunches are subject to identical injection treatment.
\begin{enumerate}
\item
All spin flips should be performed in the low-energy injection ring, where (at COSY)
essentially 100\%\ efficiency has been persuasively
demonstrated\cite{PhysRevSTAB.18.020101}.
\item
During any single data collection sequence, there should be no change in
the low-energy source region (except for test purposes); this includes
maintaining identical bunch polarization.  The reason for this constraint is
to best maintain identical parameters for all bunches. (This constraint
is not actually imposed from the point of view of minimizing the duration of
the entire injection process. In fact, the time needed to change parameters
for a subsequent train is expected to be only
about $5\Us$.)
\item
All injected bunches will have been pre-cooled in the low-energy injection ring. In all cases, only vertically polarized bunches (all up or all down) will be injected into the EDM ring.
\item
Injection as close as possible to the magic frozen spin energy will be desirable, but the injected beam energy will always be off-energy by an amount great enough for the loss of beam polarization (after
betatron and synchrotron equilibration, either by filamentation or by active damping of coherent oscillations) to be negligible.
\item
Finally, and most importantly (not counting special polarimetry
investigations), after all buckets have been populated with vertically
polarized bunches, identical external fields will be applied to every
bunch to bring all polarization orientations into their desired final
injection state---\ie the initial EDM measurement configuration state.
\end{enumerate}

\section{Pattern of bunch polarization in RF buckets}
The polarized bunch filling sequence can be described in general terms without
having frozen the RF frequency or harmonic numbers. The same discussion can also
apply to either a small prototype ring or the eventual full-scale ring.
In both cases, preliminary commissioning will use just a single, say clockwise
(CW), beam. However, since the sequential injection of simultaneously circulating beams
does not greatly complicate the process, only dual-beam injection and bunch polarization
manipulation will be described here. It will be obvious, in the following, which steps are
to be skipped for single-beam injection.

The longitudinal bunch patterns of countercirculating beams in a predominantly
electric EDM ring will be quite similar to the bunching pattern of first generation,
single-ring, electron--positron colliders, such as   the Cornell Electron--Positron Storage Ring
(CESR) or the DESY Doppel-Ring-Speicher (DORIS). In all cases, the RF
timing must be arranged so that all bunches, both CW and CCW, pass through the RF
cavity (or cavities) at stable phases.

Assuming a single RF cavity, there will be a number of stable RF buckets, both CW and
CCW, equal to the harmonic number of the radio frequency. Not all stable buckets will be
filled. Single-turn (or `kick') injection will require the presence of pulsed kickers
in the ring, whose turn-on and turn-off pulse-edge durations will have to be
restricted to time intervals during gaps in the charge distributions of both CW and
CCW beams. The length of each of these gaps has to be at least one RF bucket length (or a
higher integer
multiple of the RF bucket length). We assume that gaps and filled buckets alternate
more or less uniformly around the ring.

Ideally, every bunch will have the same number of particles and be maximally polarized.
But, for reasons of polarimetry, it is optimal for the polarization signs of adjacent
bunches to alternate. When the injection phase has been (almost) completed in each of
the beams, the fill pattern will consist of regular repetitions in a single sequence:
`up-polarized bunch, gap, down-polarized bunch, gap'. For the small PTR, two such
sequences are planned---for a larger ring, probably more.

In a final injection phase, the bunch polarizations will be rotated, but, until this
final injection phase, all bunch polarizations will be up or down, and bunches will
be referred to as `up bunches' in `up buckets' or       `down bunches'
in `down buckets'. One could contemplate an `up bunch' being parked temporarily,
for example, into a `down bucket' but, by an injection principle, this would not be favoured.

\section{Direct beam injection into stable RF buckets}
Injection will proceed in the following steps (for some of which there are
optional procedures).
\begin{enumerate}
\item
At some point, a beam (cooled and at full energy) in the injection ring is selected for one
injection path or the other. It consists of a train of uniformly spaced,
identical, vertically polarized proton bunches---say `up bunches'.
\item
All CW `up buckets' in the EDM ring are then filled by kick injection of
a single train of appropriately spaced, timed, and cooled `up bunches'
from the injection ring. For this injection phase, the EDM ring energy will be slightly different,
say greater, than the magic energy---just enough to prevent decoherence.
\item
Next, with no change in ring energy, all CCW `up' buckets in the EDM ring
are filled by kick injection of a single train of appropriately spaced, timed,
and cooled `up bunches' from the injection ring.
\item
For the next two steps, bunches are identical to the previous train,
except for having been flipped into `down bunches' and, therefore,
having all other properties the same (to the extent possible).
\item
The previous two steps are then repeated, injecting `down bunches' from the injection ring.
\item
After this sequence, all `up buckets' and `down buckets' will have been properly
populated. Up to this point, all bunch polarizations have been vertical, either
up or down.
\item
Then, by ramping the RF frequency down to the magic energy, uniformly and
adiabatically, all bunch energies will have been tuned to the magic energy.
(Though all spins are still vertical, this no longer provides protection against
decoherence).
\item
Then, for a time interval that is an integral number of synchrotron oscillation
periods, by applying an adjustable, uniformly distributed, radial, magnetic,
$B_r$ trim field, all spin orientations will have been rolled through $\pi/2$
around the radial axis, producing longitudinally polarized bunches with
alternating signs. Alternatively, this manoeuvre could be performed using a
waveguide RF Wien filter.
\item
Especially towards the ends of the previous two steps, both horizontal and vertical
polarimetry will probably be needed to control the orientations of all bunches
as intended.
\end{enumerate}


\begin{flushleft}

\end{flushleft}
\end{cbunit}

\begin{cbunit}

\csname @openrighttrue\endcsname 
\chapter{Additional science option: search for axions and ALPs}
\label{Chap:Axions}

\section{Concept of search for axion-like particles}

The theoretical prediction of a neutron electric dipole moment based on QCD
can be estimated as $\vert d^{\bar\theta} _{\rm n} \vert \sim |\bar\theta | \times 10^{-16}e\,$cm,  cf. Chapter \ref{Chap:physics-case},
\Eref{nEDM-theta}. However, the most sensitive experimental result~\cite{Abel:2020gbr}, $|d_{\rm n}| < 1.8\times 10^{-26} e \,$cm (90\% confidence level) for the neutron, sets a very strict upper limit on the  $\bar\theta$
parameter of QCD, which,
by naive dimensional analysis, should have been
of similar size as other dimensionless parameters in the  Standard Model (SM), \eg
the CP-violating angle of the CKM-matrix.    Since there is no natural explanation for the extremely small value of
$\bar\theta$, this is sometimes referred to as  {\em the strong CP problem\/}, see, \eg Ref.~\cite{Peccei:2006as}.

To solve this problem, Roberto Peccei and Helen Quinn~\cite{Peccei:1977hh}
proposed an extension of the SM by
a global chiral  $U(1)$ symmetry, the so-called {\em Peccei--Quinn} symmetry, which dynamically---via spontaneous symmetry breaking---generates a vacuum expectation value (VEV) that renders the \textit{a priori} large value of $\bar\theta$
vanishingly small.
The fluctuating field around this VEV is the axion field, $a(x)$,
and the corresponding  pseudoscalar (pseudo-)Nambu--Goldstone boson is the axion,
as introduced by Weinberg \cite{Weinberg:1977ma} and Wilczek \cite{Wilczek:1977pj}  within 1\,year of the
Peccei--Quinn proposal. The \textit{a priori} massless axion acquires   its  small mass $m_\mathrm{a}$ via the instanton mechanics
at the QCD phase transition, see, \eg Refs. \cite{Peccei:2006as,Tanabashi:2018oca}.
 It interacts  very weakly with itself and other SM particles, where the scale of the interaction strength is  governed,
 as  is the case for any non-linearly realized pseudoscalar particle, by
 the inverse of its decay constant $f_\mathrm{a}$, the axion analogue of the pion decay constant $f_\pi$.
  As $f_\mathrm{a}$ must be
 very large compared with the Higgs VEV $v_{\rm F} = 246 \,{\rm GeV}$  \cite{Peccei:2006as},
 it would be nearly impossible to detect axions or ALPs in conventional experiments, see, \eg  Ref.~\cite{Tanabashi:2018oca}
 for an overview. But it would be an ideal dark matter (DM) candidate, as it interacts gravitationally with the matter around it.

The axion couples weakly to gluons, fermions, nucleons, \etc This coupling induces an oscillating electric dipole moment (EDM) in nucleons, driven by the axion field in our local galaxy~\cite{Graham:2011qk,Graham:2013gfa}.  In the laboratory,
this oscillating EDM may be expressed as
\begin{equation}
    d_\mathrm{n}(t) \approx 1.2 \times 10^{-16}\, \frac{a(t)}{f_\mathrm{a}}\, e \,
    \text{cm} = 1.2\times 10^{-16}\, \frac{a_0\, \cos (m_\mathrm{a} t+\phi_\mathrm{a}
    )}{f_\mathrm{a}}\, {e}\,\text{cm} \, ,
\end{equation}
\noindent
where $a(t)$ is the local axion (dark matter) field in the laboratory and $\phi_\mathrm{a}$ is an unknown local phase that we will need to consider later. In the case of the original axion, its mass and
decay constant are correlated, $m_\mathrm{a}  \approx 0.5 m_\pi f_\pi / f_\mathrm{a}$, in terms of the pion mass and decay constant~\cite{Tanabashi:2018oca}, while for axion-like particles (ALPs) the masses can be (much) smaller  for
a given value  of `coupling' $1/f_\mathrm{a}$.

Three conditions must be met in order to consider using the horizontally polarized deuteron beam at COSY to search for axions or ALPs. First, the density of the axions passing through the laboratory must be large enough to form a classical oscillating field.
This is the case for axions or ALPs of mass $m_\mathrm{a} < 1\,{\rm meV}$, assuming that the dark matter density in our galaxy, which
is $\sim$\,$10^5$ times larger than the average DM density in our Universe~\cite{Tanabashi:2018oca}, is
at least partially saturated by these  pseudoscalar bosons.
Second, the axions or ALPs in the local  field $a(x)$  must act coherently across a large spatial range so that, as the beam circulates, it remains under the influence of a coherent, spatially constant  axion or ALP field. This will also mean that all of the deuterons in the beam will show the oscillating EDM property simultaneously. Thus, an EDM parallel to
the spin orientations of the deuterons in the beam (on average, the  polarization of the beam) may be used to test for the presence of the local axion field $a(t)$. Third, this interaction must remain present in the COSY experimental hall for a time long enough for the beam to respond. Crossing an axion resonance in a scanning search would probably require a few seconds. Any axions or
ALPs in the neighbourhood of Earth are probably bound to the Milky Way galaxy; thus, there is a lower limit on how quickly they will vanish from view, governed by the virial velocity of our solar system with respect to the centre of our galaxy $\sim$\,$10^{-3}c$.
Estimates~\cite{Graham:2011qk,Graham:2013gfa} based on the confinement of the axions or ALPs to our region of the Milky Way galaxy suggest that these coherence requirements are met at the frequency where we would make a feasibility study ($\sim$\,$630$~kHz) with a quality factor $Q$
exceeding $10^6$.
Note that  630\,kHz would correspond to $2.60\,{\rm neV}/c^2$ as axion or
ALP mass,
such that the pertinent coherence length of the axial field would be about 70\,km, while its coherence time is
limited to 1.4\,s.

The experiment to search for the DM axions or ALPs would consist of a series of runs in which the revolution frequency  $f_{\rm rev}$ of the machine is changed continuously in a slow ramp~\cite{Chang:2019poy}. Measurements of the polarization components would be made during the ramp. If the in-plane polarization precession frequency
happens to match the axion or ALP frequency, which just corresponds, with a quality $Q\gtrsim 10^6$,
 to the pertinent axion or ALP  mass,     a resonance between the two will cause the vertical component of the polarization to undergo a jump proportional to the ratio of the size of the oscillating EDM to the square root of the speed of the ramp. A comparison of polarization asymmetries collected during non-ramped times at the beginning and end of the scan would suffice to quantify
any suspected change.

Experimental signals based on a subatomic EDM depend on a torque about an electric field along the radial direction in a storage ring that lifts the polarization direction out of the ring plane, giving it a small and rising vertical component. Despite the large electric field that exists in the beam frame from the magnets that confine the deuteron beam to the COSY ring, the continuous rotation of the in-plane polarization relative to the beam velocity makes it impossible for any static EDM signal to become large enough through a $\vec{d}\times\vec{E}$ torque to observe directly. Progress is cancelled by retreat whenever the projection of the EDM on the tangential direction reverses. But if there is an oscillating EDM that varies at the same frequency as the rotating polarization  (or an odd multiple thereof), then a vertical component of the polarization will start to grow. A proposal was accepted by the COSY Program Advisory Committee to develop and describe techniques that could be used in such an axion or ALP search and to quantify the sensitivity for reasonable operating conditions. The plan is to start with the deuteron momentum of $p = 0.97$~GeV/$c$, where there is COSY experience of the preparation of a horizontally polarized beam with a long polarization lifetime~\cite{Guidoboni:2016bdn}.

\section{Technical considerations for an axion search}

Accumulation of the vertical component polarization signal depends on the alignment of the polarization along the direction of the beam velocity with the maximum of the value of the oscillating EDM. This alignment is controlled by
the unknown phase of
the  axion or ALP field,  $\phi_\mathrm{a}$. If these two oscillations are out of phase by $\pi /2$,  no accumulation will occur. The plan to overcome this difficulty is to operate COSY on the fourth harmonic (for which hardware already exists), producing four circulating bunches in the ring at the same time. If an RF solenoid is used to precess the polarization from the vertical direction (as it is on injection into the ring) into the horizontal plane, then the resulting laboratory-frame polarization pattern of the four beams in the ring is shown in \Fref{fig:AxFig1} for $f_{\rm sol}=f_{\rm rev}(1+G\gamma )$,
where $G$ is the deuteron's magnetic anomaly and $\gamma$ is the usual relativistic parameter.

\begin{figure}
\centering
\includegraphics[scale=0.55]{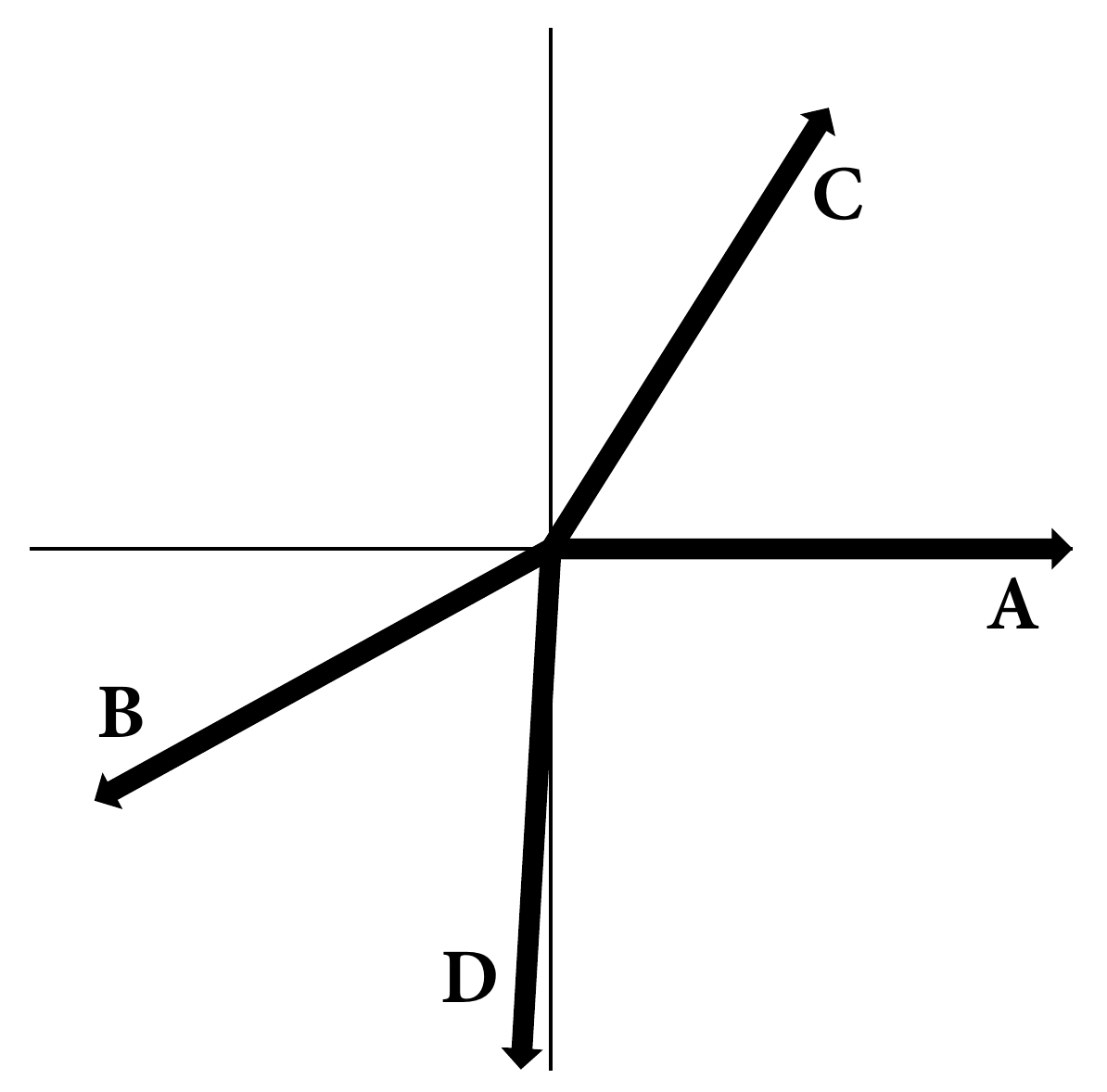}
\caption{Laboratory-frame polarization directions of the four beam bunches as seen from above the plane of the storage ring. The labels show the order (A, B, C, D) in which they were generated by the RF solenoid operating on the $1+G\gamma$ harmonic.}
\label{fig:AxFig1}
\end{figure}

 This pattern features two directions (A and D) that are nearly orthogonal. This means that the experiment carries sensitivity to both components of the phase of the oscillating EDM (sine and cosine). The remaining two polarization directions may be used to verify that the amplitude of any prospective axion signal varies in a sinusoidal pattern around the circle in \Fref{fig:AxFig1} in a manner consistent with the two phase components present in the A and D directions. In addition, there are pairs of polarization directions that are nearly opposite. This provides an opportunity to use them as opposite polarization states in a `cross ratio', which would serve to reduce or eliminate first-order errors in the scanning process resulting from geometric or rate-dependent systematic errors that can develop during the beam store~\cite{Brantjes:2012zz}. Bunch B may be compared with the average of A and C, and bunch C may be compared with the average of B and D.

One way to search for an axion-like particle is to vary the polarization rotation rate continuously while monitoring the vertical polarization. If the frequency of rotation happens to match the axion frequency at some time during the scan, the resonance condition will create a jump in the polarization, as shown in \Fref{fig:AxFig2}.

\begin{figure} [hb!]
\centering
\includegraphics[scale=0.32]{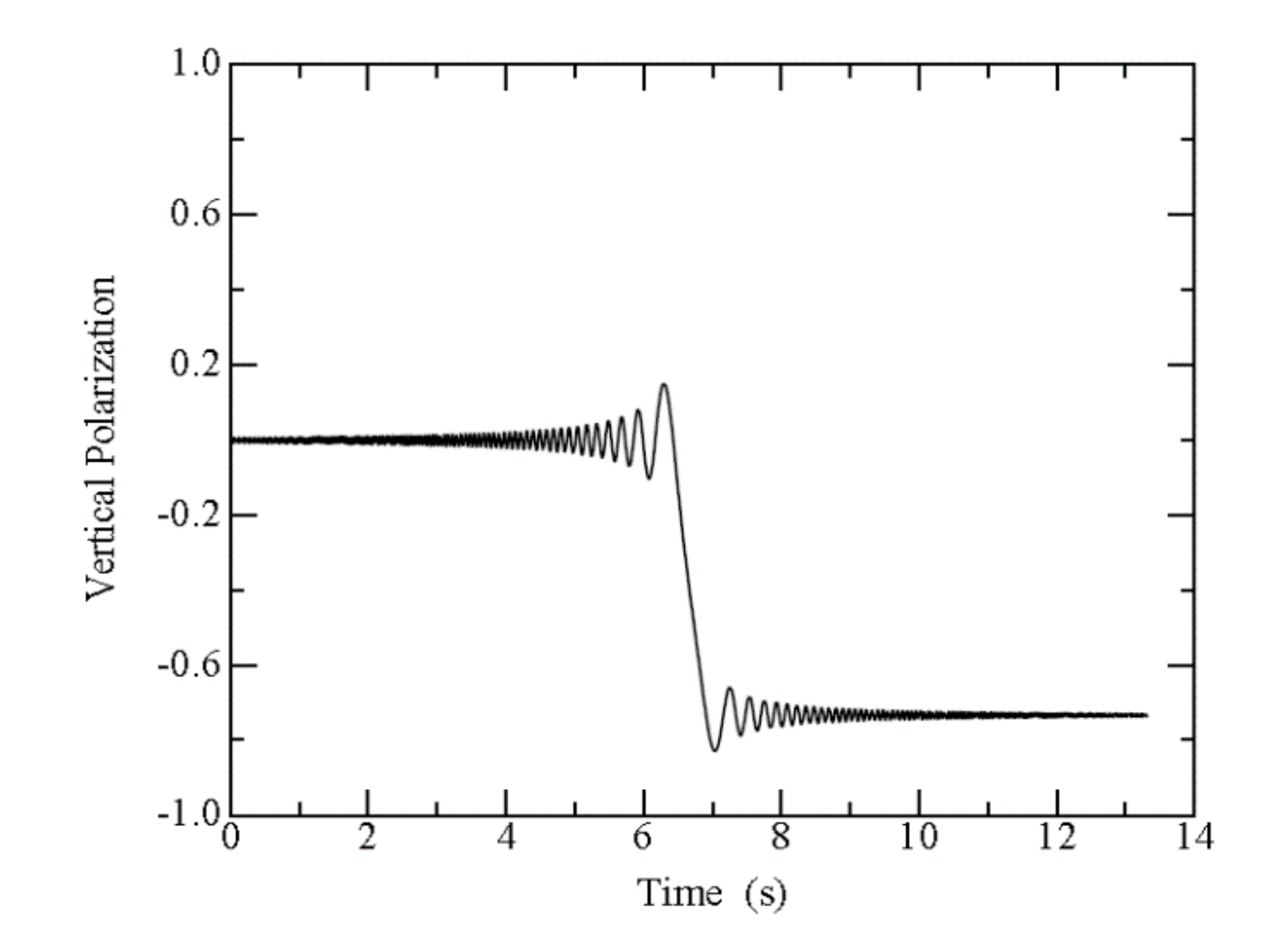}
\caption{Calculation of the resonance crossing with a scan rate of 0.5\,Hz/s.
The strength of the oscillating EDM is assumed to be $1.6 \times 10^{-21} e \,$cm.
Within a span of less than $1\Us$, this causes a jump of $-0.75$ in the
$p_y$ component of the beam polarization (assumed to be initially completely polarized in the ring plane).}
\label{fig:AxFig2}
\end{figure}

A practical scheme for producing such scans would require that the range of the scan is not  so large that it passes outside the acceptance of the storage ring. In addition, the frequency of the RF solenoid that initially precesses the polarization from the vertical to the horizontal direction must be adjusted to match the $1+G\gamma$ spin tune resonance. The easiest way to organize the scan is to vary the revolution frequency of the beam. Critical magnetic ring components, such as the dipole magnets, would be programmed to follow.

\section{Initial tests with beam}
\label{Chap:Axions-inital-test-with-beam}

In December 2018, there was an opportunity to switch COSY to operate on the $h = 4$ harmonic. At that time, the RF solenoid was running on the $1-G\gamma$ harmonic and the sextupole magnets, along with electron cooling, had been set for a long in-plane polarization lifetime.  \Figure[b]~\ref{fig:AxFig3} shows a representation of the count rate in the WASA detector as a function of time in the store (horizontal) and position around the ring (vertical).

\begin{figure} [hb!]
    \centering
    \includegraphics[scale=0.4]{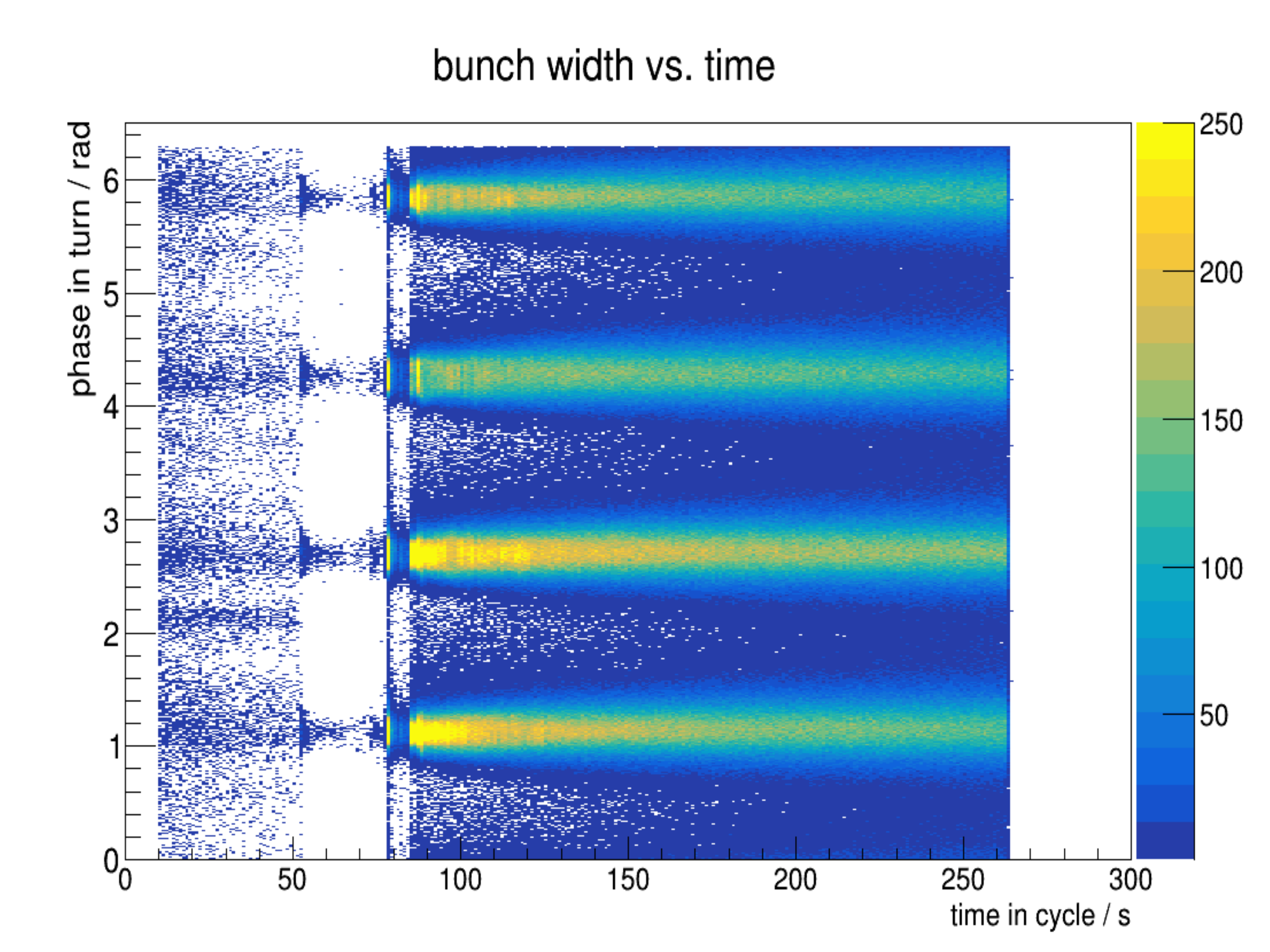}
    \caption{Count rate in the polarimeter as a function of time in the store (horizontal) and position around the circumference of COSY (0 to $2\pi$). Extraction of the beam onto the WASA polarimeter begins at $90\Us$. Prior to $80\Us$, the beam is  electron-cooled. There are four horizontal ridges, corresponding to the four beam bunches.}
    \label{fig:AxFig3}
\end{figure}

The four beam bunches show clearly after $80\Us$, following a period of electron cooling. At this time, the RF solenoid frequency was associated with the $1-G\gamma$ harmonic. This yields a different pattern of polarization directions from that in \Fref{fig:AxFig1}. In the laboratory frame, we have the pattern
shown in \Fref{fig:AxFig4}.

\begin{figure} [hb!]
    \centering
    \includegraphics[scale=0.23]{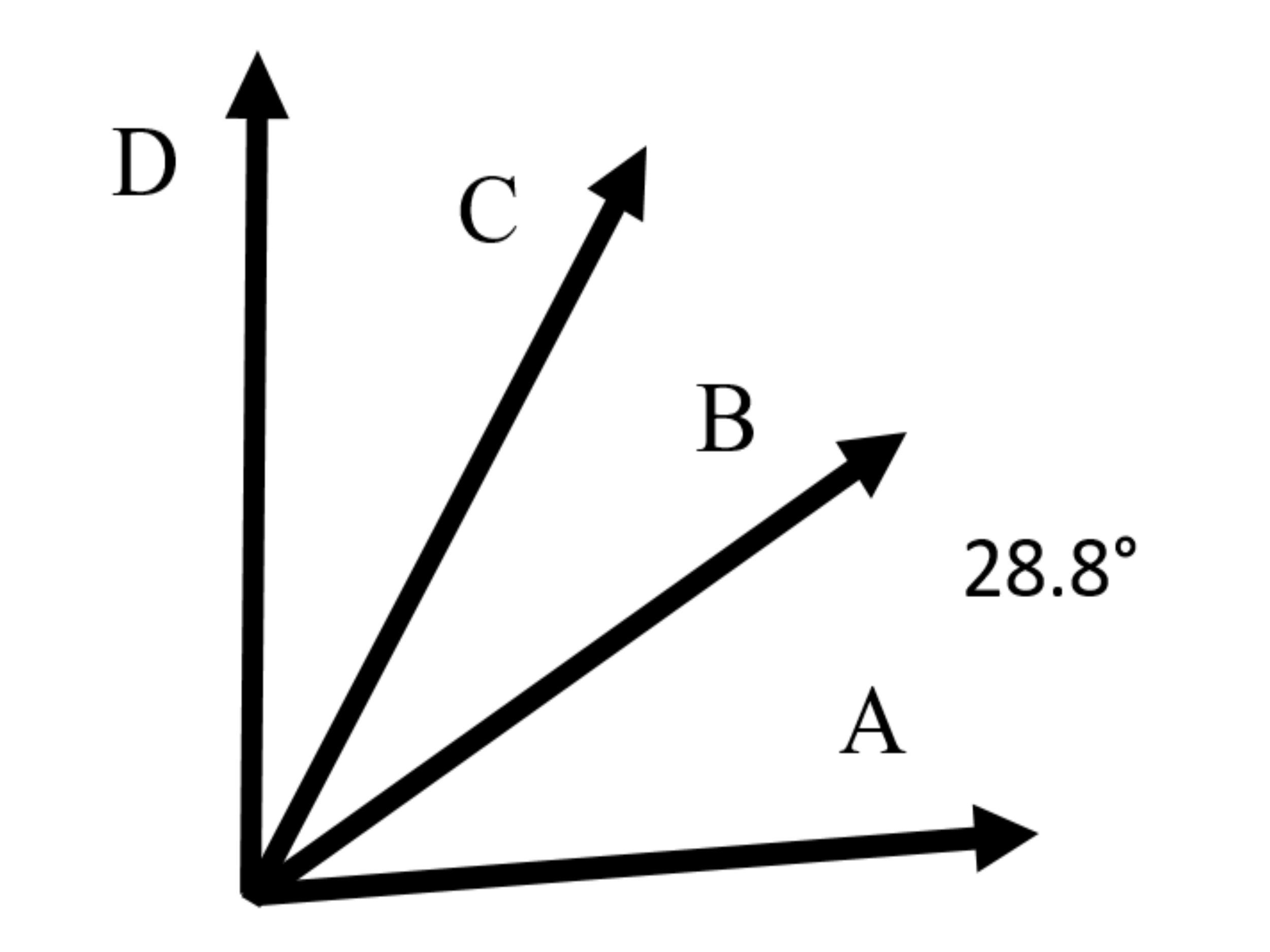}
    \caption{Directions of the in-plane polarizations in the laboratory frame for the case of an RF solenoid operating on the $1-G\gamma$ harmonic. The labels follow the scheme of \Fref{fig:AxFig1}. The opening angle for adjacent pairs is shown as $28.8^\circ$.}
    \label{fig:AxFig4}
\end{figure}

Like the pattern shown in \Fref{fig:AxFig1}, this pattern also presents bunches A and D with polarization directions that are nearly perpendicular. So this pattern also suffices to detect the axion for any value of the axion phase. But the other two polarization directions, B and C, lie in the same quadrant. Their polarization directions are similar, and any axion signal will tend to have a similar signature as A and~D. Thus, we cannot use these signals in a cross-ratio treatment to eliminate systematic errors in the measurements of the asymmetry.

Experimental verification of these polarization directions depends on measurements made with a polarimeter located at one spot on the COSY storage ring. The
polarimeter will see the four bunches sequentially at different times. Given that the polarization continues to rotate in the ring plane, this leads to a different set of directions measured at the polarimeter at different times.
Since the polarization is rotating at about 630\,kHz, it is most useful to consider expressing this polarization as a magnitude and a phase with respect to a starting time that is the beginning of data acquisition. Normally, trimming the fields in the ring, especially the sextupole components, is very useful in maintaining the size of the in-plane polarization (IPP). Then the important task is measurement of the phases for the four beam bunches. An example is shown in \Fref{fig:AxFig5}.

\begin{figure} [hb!]
    \centering
    \includegraphics[scale=0.4]{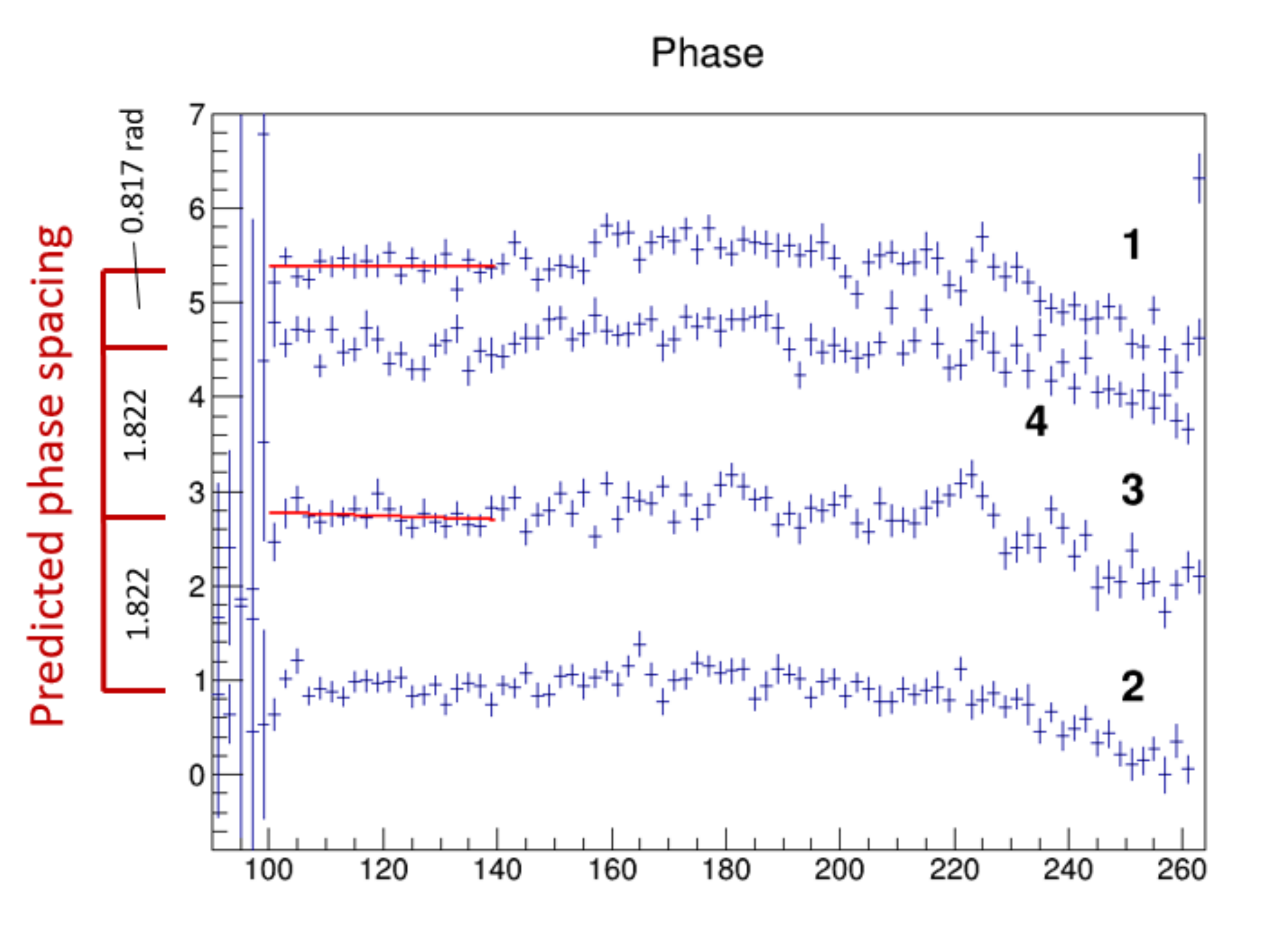}
    \caption{Measurements of the polarization phases for the four beam bunches in a test run made in December 2018. The phase, measured along the horizontal axis, is shown as a function of time in the store, and is relative to a calculation of the polarization direction based on an assumed value of the spin tune frequency ($f_{\rm rev}G\gamma$) that yields a prediction of the phase at any moment in time during the store. A perfect match between the prediction and the measurements yields phase values that remain constant with time. The numbers on the curve correspond to the four bunches (A to D). Along the left-hand axis, a diagram using red lines shows the predicted relative phase separations that corresponds to the polarization pattern shown in \Fref{fig:AxFig4}. Given the value of $G\gamma$, the separation of the phase lines should be either 1.822\,rad (for pairs A--B, B--C, and C--D, including wrapping through $2\pi$) or 0.817\,rad (for pair D--A). This diagram gives a good account of the phase separations as measured.}
    \label{fig:AxFig5}
\end{figure}

The match (red lines) with a prediction consistent with \Fref{fig:AxFig4} is good (see caption). The pattern on phases shows three angular separations of 1.822\,rad and a final separation of 0.817\,rad.  This set of unequal gaps indicates that phase A is uniquely identifiable as the bunch synchronized with the maxima in the 871\,kHz RF solenoid pattern at $t = 0$ (start of solenoid operation). Like the phase pattern in \Fref{fig:AxFig5}, the angular separations in \Fref{fig:AxFig4} are also the same ($28.8^\circ$), except for the separation between bunches D and A, which is much larger.

In the case of the $1+G\gamma$ harmonic recommended for this process, the pattern of angular separations, three wide and one narrow, in the polarimeter measurements changes in the phase pattern, to three narrow and one wide.
The narrow angle is $1.32$\,rad and the wide angle is $2.32$\,rad. This leads to a separation angle for the polarimeter measurement of $201.6^\circ$ between successive beam bunches in \Fref{fig:AxFig1}, rotating counterclockwise.

In the scan for the axion or ALP, different polarization phases are distributed with a sinusoidal dependence on the axion phase, as shown in \Fref{fig:AxFig6}.

\begin{figure} [hb!]
    \centering
    \includegraphics[scale=0.4]{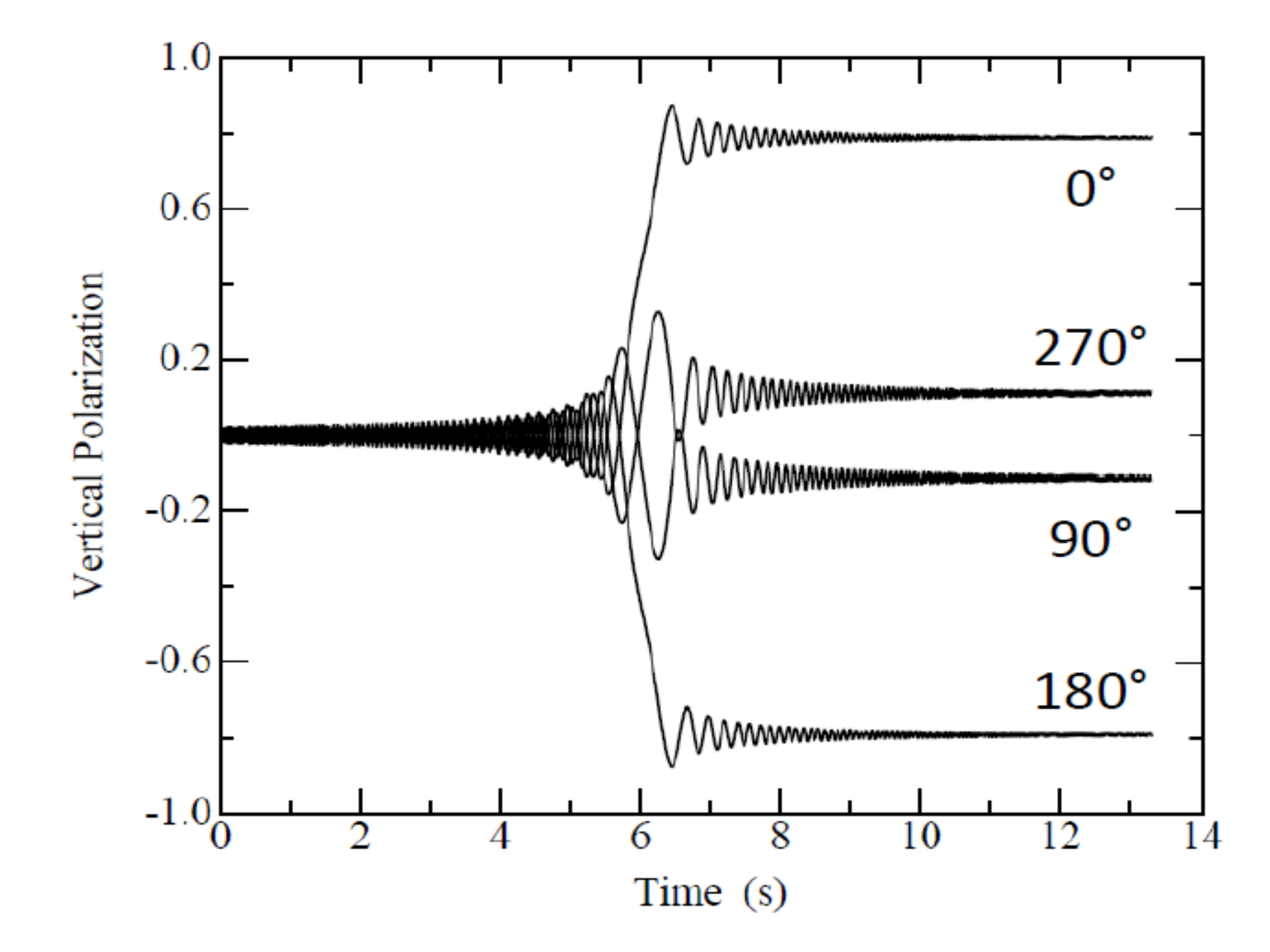}
    \caption{Polarization jump for four different choices of  axion phase; the phase is marked for each curve}
    \label{fig:AxFig6}
\end{figure}

The different polarization jumps form one of the features that distinguishes the detection of an axion from the observation of a machine resonance. In the case of the machine resonance, there is no phase and all four bunches should observe the same polarization jump. The distribution of polarization directions, as shown in \Fref{fig:AxFig1}, ensures that the signals will appear with opposite signs for some pairs of directions.

Machine resonances must also appear at frequencies related to the value of $G\gamma$ through
\begin{equation}
    G\gamma = \ell + m\nu_x + n\nu_y + k\nu_{\rm sync} \, ,
\end{equation}
\noindent where $\ell$, $m$, $n$, and $k$ are integers and the tunes ($\nu$) are connected to horizontal ($x$) and vertical ($y$) betatron oscillations as well as synchrotron oscillations, see p.\,26 in Ref.~\cite{Lee:1997mz}. Smaller integers generally indicate stronger resonances. These checks should allow for the separation of axion signals from other effects.

\section{Immediate plans}

A running period started on April 1st, 2019, with COSY for the purpose of testing the feasibility of creating a four-beam set-up and a frequency ramp with properties appropriate for conducting an axion search.  The set-up includes previously developed conditions for a long IPP lifetime, which involves electron cooling as well as trimming the ring fields with sextupole components such that the $x$ and $y$ chromaticities are simultaneously set to zero~\cite{Guidoboni:2016bdn}.

The new features begin with the four-bunch set-up. The bunches must be well separated spatially so that there is no significant transfer of beam particles from one bunch to the next. This would tend to depolarize the bunches, as the pattern in \Fref{fig:AxFig1} requires nearly opposite polarization directions for neighbouring bunches. Polarization measurements would ensue to check that the understanding of the relative phases between bunches is correct. This would constitute a confirmation of the patterns shown in the previous section.

The next step in the preparation would be the creation of conditions for ramping the machine revolution frequency to make frequency scans possible. Speeds would be slow, perhaps 0.1\,Hz/s. Storage times of $150\Us$ during the ramp means a frequency step of 15\,Hz per scan. Since nearly opposite polarization directions are represented in the laboratory polarization pattern, only one polarization state is needed from the ion source. Note that the data from one scan cannot be directly combined with another, since the relative phase may change, even if the axion or ALP field remains the same.
 The relative phase between scans also depends  on the start time for the RF solenoid, and this cannot be synchronized with the  phase of the axion
or ALP field. Repeated scans of the same frequency range are advisable since, at any given time, the axion field may vanish or not be detected. Ramping the magnetic field of  COSY along with the frequency is necessary, in order to maintain the circumference of the orbit. This allows the spin tune ($G\gamma$) to be known from the revolution frequency. However, the development of the software for dealing with IPP also provides for a direct measurement of the spin tune at any time during the process~\cite{Eversmann:2015jnk}; this will act as a confirmation that the machine conditions are being maintained.

With each scan, a comparison of the vertical polarization component difference between the beginning and the end of the run is needed to determine whether or not there is evidence for a polarization jump during the scan. The statistics of this comparison may be improved if there are times of no ramping before and after the actual ramp. Some threshold (such as two or three standard deviations) must be chosen. If passed, the scan should be repeated to determine whether or not it is an outlier. Once identified, additional scans are needed in order to have the statistics to determine the time location of the polarization jump with precision.

Initial results from this development period are expected to be modest in terms of both the sensitivity and frequency range covered.

\begin{flushleft}

\end{flushleft}
\end{cbunit}

\begin{cbunit}


\csname @openrighttrue\endcsname 
\chapter[New ideas: hybrid scheme]{New ideas: hybrid scheme\footnote{This appendix was written by Y.K. Semertzidis and S. Hac\i{\" o}mero{\u g}lu of the Center for Axion and Precision Physics Research, KAIST, South Korea.}}
\label{Chap:hybrid}

This appendix examines the possible replacement of electrical quadrupoles with magnetic quadrupoles for the focusing in the full-scale ring, which is then referred to as a `hybrid ring'. Because alternating gradient magnetic focusing is used, simultaneous CW and CCW storage continues to be possible, while still allowing for moderately strong vertical focusing, along with the simultaneous CW and CCW storage needed for cancelling important systematic errors. This promises to greatly reduce the contribution of radial magnetic field uncertainty to the EDM systematic error.

\section{Experimental method using a hybrid ring lattice}

Simultaneous storage in clockwise (CW) and counterclockwise (CCW) directions
allows for the cancellation of important systematic errors\cite{Farley,Nominal-BNL}. In combined electric and magnetic fields, \eg the deuteron ring\cite{bnl-sredm-deuteron-proposal}, it is not possible to store the  CW and CCW beams simultaneously and much of the systematic error work was geared towards fixing potential problems arising as a result.
The all-electric ring allows for it; however, the main potential systematic error is large (a consequence of the large sensitivity on the proton EDM), and  required level to determine the radial $B$ field around the ring is at the 10\,aT level.  High-precision SQUID-based BPMs have been developed to be able to detect the required signal caused by the splitting of the counter-rotating beams\cite{Nominal-BNL,Anas}.  For the method to have high sensitivity to the potential systematic error, the vertical focusing strength is kept low, making it rather difficult to handle.
A hybrid ring, in which alternating magnetic focusing is used, allowing simultaneous CW and CCW storage, allows for strong vertical focusing and simultaneous CW and CCW storage for cancelling important systematic errors\cite{Haciomeroglu:2018nre}.

The counter-rotating beams do not actually go through the same places everywhere, because the vertical focusing includes magnetic focusing. Therefore, those beams may not exactly cancel those systematic errors at all places.  However, we have shown that it is possible to use the same magnetic quads with flipped field directions (opposite sign currents) and, on average, the particles do follow the same trajectories.   This idea seems to work very well, eliminating  the radial $B$-field issue completely.  In addition, the vertical dipole $E$-field effect is cancelled completely in CW and CCW injections, as is the effect of gravity.  The suggested working lattice is shown in  \Fref{lattice2}, which is a modification of the lattice shown in Ref.~\cite{Anas}, which describes the all-electric storage ring method, except this time the electric quadrupoles are replaced with corresponding magnetic ones.  \Figure[b]~\ref{ver_beta_function_fin} shows the vertical beta function of the CW and CCW stored beams, and  \Fref{hor_beta_function_fin} the corresponding function for the horizontal.  Flipping the sign of the currents in the magnetic quadrupoles will produce symmetric beta functions for the CW and CCW beams.

\begin{figure}
\centering
\includegraphics[scale= 0.4]{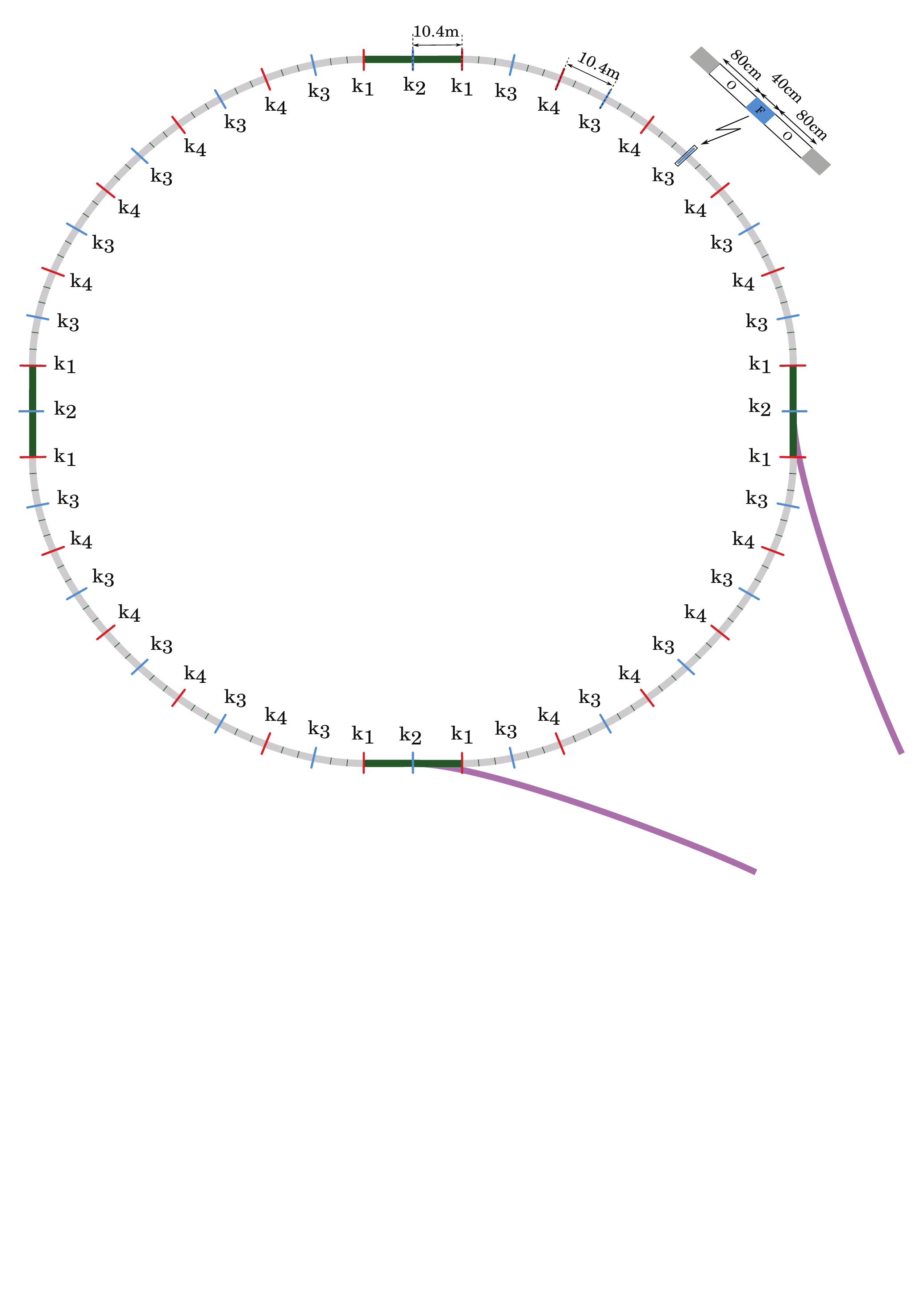}
\vspace*{-4cm}
\caption{\label{lattice2}Detail of the storage ring lattice, with focusing and defocusing quadrupoles ($\rm k_3$ and $\rm k_4$).  The bending sections, including the short straight sections, are 10.417\,m long, with three sections assembled as one unit.  The long straight sections are 20.834\,m long with a quadrupole ($\rm k_2$) in the middle and two half-length quads (k$_1$) at both ends.  The values of the magnetic quadrupole strength are: k$_1 = 0.1$\,T/m; k$_2 = -0.1$\,T/m; k$_3 = -0.1$\,T/m; k$_4 = 0.1$\,T/m.  The vertical tune, when running with these quadrupole strengths, is $Q_y = 0.67$, while the horizontal tune is $Q_x = 1.73$.
}
\end{figure}

\begin{figure}
\centering
\vspace*{0cm}\includegraphics[scale = 1.2]{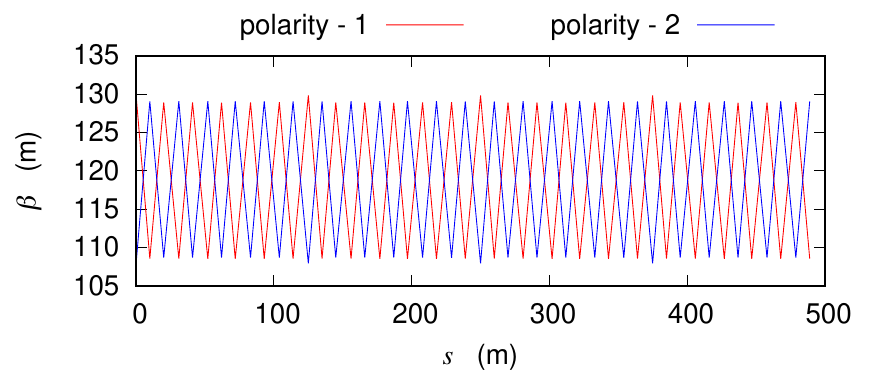}
\vspace*{-0.4cm}\caption{\label{ver_beta_function_fin}Vertical beta function  around the ring for CW and CCW operations.  The sign is  flipped when the magnetic quadrupoles are running with the opposite sign; therefore, the counter-rotating particles, on average, trace the same paths.
}
\end{figure}

\begin{figure}[hbt!]
\centering
\vspace*{0cm}\includegraphics[scale = 1.2]{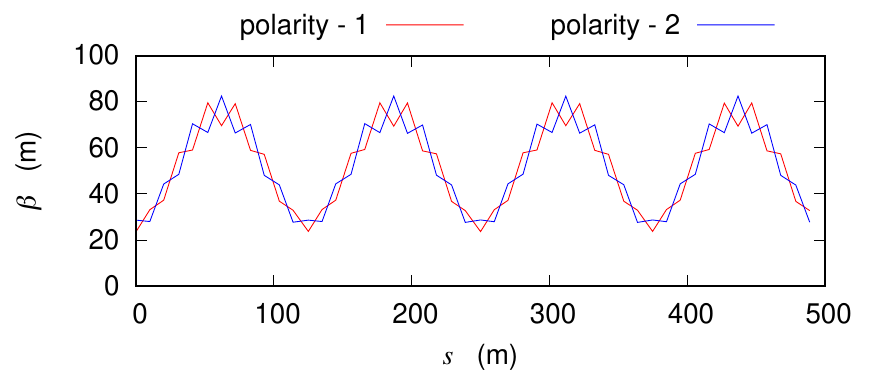}
\vspace*{-0.4cm}\caption{\label{hor_beta_function_fin}Horizontal beta function values around the ring for CW and CCW operations.
}
\end{figure}

However, it is always possible that some electric focusing will be present somewhere in the ring.  This focusing or defocusing could originate from the bending electric field plates, which produce the required radial $E$ field. One or both plates could be misaligned, readily producing a vertical dipole, or a quadrupole or even higher multipole $E$ fields.  There could also be induced charges (image charges) from any horizontally placed metals around the lattice,  tune shift, and tune spread effects due to high beam intensities, \etc  It may be possible to detect some of those systematic errors, \eg by modulating the voltage at the bending $E$ field plates or control them by using beam bunch intensities of various strengths.  At the end of the experiment, however, we need to have high confidence regarding the origin of the effect.  Here we are suggesting the use of a number of runs with different vertical magnetic focusing strengths to differentiate between a systematic error and a genuine EDM signal.

The total effect, \ie the vertical spin precession rate, is going to be in a functional form:
\[R_\mathrm{V} = R_{\rm EDM} + R_{B_r} \times \frac{Q_{\rm Backgr}^2  + \dotsb} {\zeta \times Q_{\rm Magnetic}^2 +Q_{\rm Backgr}^2 + \dotsb} \, ,\]
where $R_\mathrm{V}$ is  the total vertical spin precession rate, $R_{\rm EDM}$ is  the portion due to the particle EDM, $R_{B_{r}}$ is  the vertical spin precession rate due to the radial $B$ field, $Q_{\rm Magnetic}^2$ is the square of the tune due to the magnetic quads,
$Q_{\rm Backgr}^2 =f( Q_{\rm Electric}^2, Q_{\rm ImageCharge}^2, Q_{\rm BeamIntensity}^2,
\dots)$ is the square of the tuning due to non-magnetic effects,  and $Q_{\rm Electric}^2$, $Q_{\rm ImageCharge}^2$, and  $Q_{\rm BeamIntensity}^2$ are the squares of the tunes due to the electric quads, the forces due to induced charges, and the forces due to the beam intensity, respectively.  The point is that a net radial $B$ field can create a vertical spin precession, which can only be cancelled exactly by another $B$ field; in this case, we assume it to be the magnetic focusing.  Magnetic focusing can essentially eliminate this systematic error, provided that it is the only source focusing the beam.    \Figure[b]~\ref{y_vs_n} shows the average vertical offset of the stored beam as a function of a radial $B$ field multipole whose amplitude is always kept at 1\,pT.  \Figure[b]~\ref{wr_vs_n} shows the vertical spin precession rate under the same conditions.  A genuine EDM signal for $ 10^{-29} \, {\it e} \, {\rm cm}$ is larger than 1\,nrad/s, and therefore much larger than this background signal.  However, if, on one of the magnetic quadrupoles, we add an overlapping electrical quadrupole with a strength of 1\,kV/m$^2$, then we get the much larger spin precession rate of 0.4\,nrad/s, for the $N=4$ harmonic case of the radial $B$ field.  This effect will be further and effectively suppressed by applying varying levels of magnetic field focusing, as described in the next section.

\begin{figure} [hbt!]
\centering
\vspace*{0cm}\includegraphics[scale = 1.2]{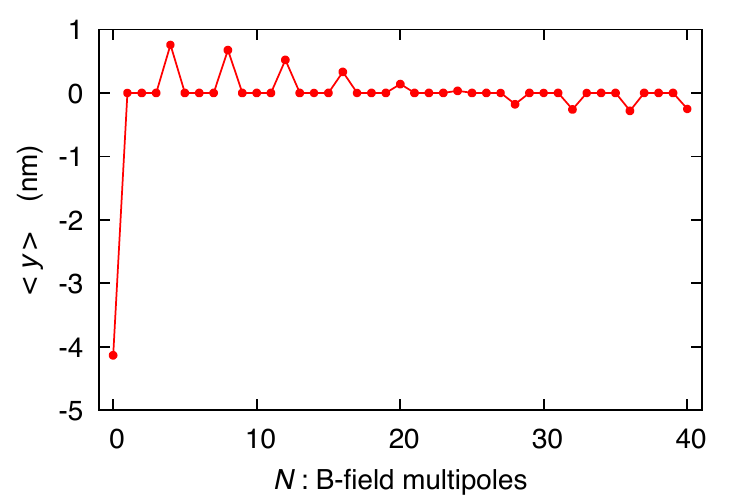}
\vspace*{-0.4cm}\caption{\label{y_vs_n}Average vertical beam offset when only magnetic focusing is used, as a function of the radial $B$ field multipoles ($N$ values).  The amplitude of the background radial $B$ field is always kept at 1\,pT, while the quadrupole strength is kept at $\pm$0.1T/m.}
\end{figure}

\begin{figure}
\centering
\vspace*{0cm}\includegraphics[scale = 1.2]{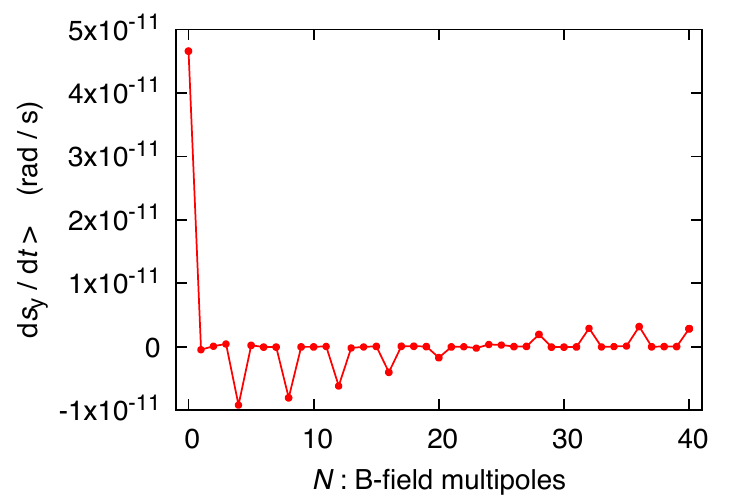}
\vspace*{-0.4cm}\caption{\label{wr_vs_n}Vertical spin precession of the counter-rotating beams when only magnetic focusing is used, for different radial $B$ field multipoles ($N$ values).  The amplitude of the radial $B$ field is always kept at 1\,pT, while the quadrupole strength is kept at $\pm$0.1\,T/m. A genuine EDM signal for $ 10^{-29} \, {\it e} \, {\rm cm}$ is larger than 1\,nrad/s, and therefore much larger than the background signal.}
\end{figure}

\section{Experimental approach}
  We apply a series of $B$ field focusing strengths, from weak to stronger,  to probe the EDM effect.  With magnetic focusing, the main systematic error is the out-of-plane dipole electric field, which is cancelled by CW and CCW beam storage as in the deuteron storage ring EDM experiment~\cite{bnl-sredm-deuteron-proposal}.  Since simultaneous CW and CCW storage is possible in the current configuration,  most of the issues related to $E$ field direction stability vanish.
In addition, any focusing effect from the electric field plates or any other sources is sorted out by running the experiment at different alternating magnetic focusing strengths, as shown in  \Fref{wr_vs_Pm}.  Here, an additional electric focusing exists, together with a DC ($N=0$) radial magnetic field around the ring with strength of 1\,pT.  The electric focusing is originated by shaping all the bending plates, producing a vertical focusing with a field index of $m=0.1$.  The spin precession rate equation, when expanded, can be written as
\[R_\mathrm{V} = R_{\rm EDM} + R_{B_r} Q_{\rm Backgr}^2  P_{m1} - R_{B_r} Q_{\rm Backgr}^4  P_{m1}^2 +  \dotsb \, , \]
with $P_{m1} = 1/(\zeta \times Q_{\rm Magnetic}^2)$, showing clearly that for a large magnetic focusing tune, \ie $P_{m1} \rightarrow 0$, the spin precession rate corresponds to the EDM signal.  Hence, the DC offset in  \Fref{wr_vs_Pm} corresponds to the EDM signal and the obtained value is consistent with the simulations.  In  \Fref{wr_vs_Pm},  the spin precession rate corresponds to the $ 10^{-28} \,  e \, {\rm cm}$ EDM level, to prove the principle of the method.  It will be advantageous to keep the spin precession rate smaller by adding much stronger magnetic focusing cases and keeping the electric focusing below the $m=0.01$ level.  The method will work best, requiring less leverage, when the magnetic focusing dominates all other focusing effects.  In a similar way, we can prove that the sextupole vertical electric field cancels with CW and CCW storage, \etc, provided that the beam emittances are the same, to an adequate level.
From our simulations, we infer that the SQUID-based BPM resolution requirements are relaxed by several orders of magnitude over the lattice where electric focusing is used, which is a major breakthrough.  The new requirements are a well-shaped quadrupole magnetic field in the ring, so that the centres of the CW and CCW beams overlap within 100\,nm at all magnetic quadrupole strengths, using the SQUID-based BPM signals.  In addition, the ring needs to be flat (absence of corrugation) to 100\,nrad, which we achieve by a combination of mechanical alignment, beam-based alignment and the use of bunches polarized in the radial direction.
 A summary of the main systematic errors in the experiment with hybrid fields (electric bending and magnetic focusing) and the current remediation plan is given in \Tref{syserr}.

\begin{figure}[hbt!]
\centering
\vspace*{0cm}\includegraphics[scale = 1]{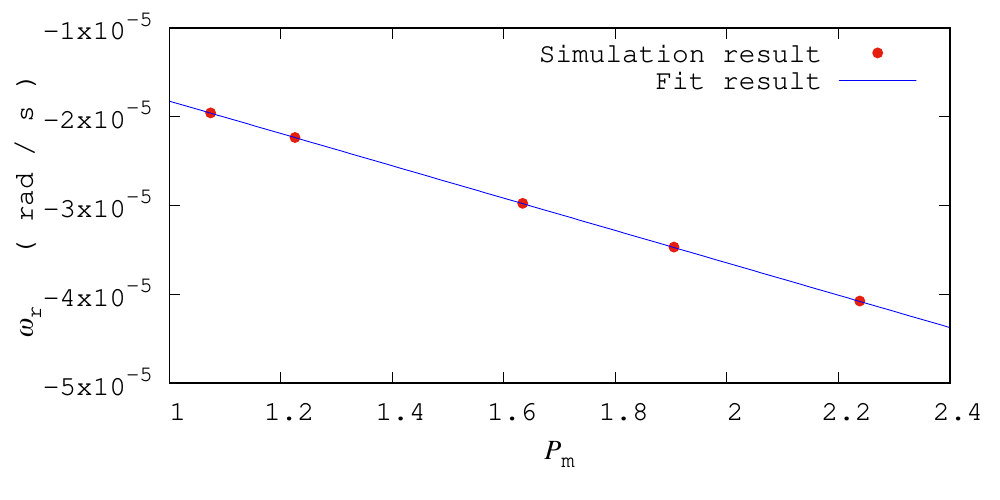}
\vspace*{-0.4cm}
\caption{\label{wr_vs_Pm}Vertical spin precession rate as a function of  $P_m = 1/Q_y^2$ when the background effect is due to a combination of a DC ($N=0$) radial magnetic field around the ring with strength of 1\,pT and a large electric focusing effect of the bending plates. The bending plate focusing corresponds to an (electric) vertical focusing field index of $m=0.1$.   The fit result is from a first-order polynomial.  The DC offset corresponds to the EDM precession rate, which, in this case, is $-1.9 \times 10^{-8}$\,rad/s, consistent within the estimated errors to the input EDM value, corresponding to $-4.1 \times 10^{-8}$\,rad/s.}
\end{figure}

\begin{table}
\caption{\label{syserr}Main systematic errors and their remediation when hybrid fields (electric bending and magnetic focusing) are used.}
\begin{tabular}{ p{5cm}   p{10cm} }
\hline \hline
Effect & {Remediation}  \\ \hline
Radial $B$ field &  Magnetic focusing \\ 
Radial $B$ field when other than magnetic focusing is present &  Varying magnetic focusing and fit for the DC offset in the vertical precession rate\\
Dipole vertical $E$ field &  CW and CCW beam storage \\ 
Corrugated (non-planar) orbit &  Observe CW vs. CCW beam split with magnetometers, \eg SQUID-based BPMs\cite{Nominal-BNL} \newline  Probe with stored beams  with their spins frozen in the radial direction ~\cite{bnl-sredm-deuteron-proposal} \\ 
RF cavity misalignment & Vary longitudinal lattice impedance to probe the effect of the cavity's vertical angular misalignment; CW and CCW beams cancel the effect of a vertically misplaced cavity \cite{Anas} \\
\hline \hline
\end{tabular}
\end{table}

\section{Conclusions}

The hybrid ring, where the radial $E$ field bends the stored beam and an alternating $B$ field provides focusing, allows for simultaneous CW and CCW storage, eliminating the most important systematic error source.  The experiment will also run at various magnetic focusing strengths to eliminate possible electric focusing sources, \etc  In addition, the counter-rotating beams will sense any quad misalignment to better accuracy than needed, as well as the spin precession of a beam with a radial spin direction.  The method needs to be studied by an independent group; this should take less than 6\,months to complete.

\begin{flushleft}

\end{flushleft}
\end{cbunit}

\begin{cbunit}


\csname @openrighttrue\endcsname 
\chapter[New ideas: spin tune mapping for EDM searches]{New ideas: spin tune mapping for EDM searches\footnote{This appendix was written by Artem Saleev (Institut für Kernphysik, Forschungszentrum Jülich).}}

\label{Chap:spin-tune-mapping}

The appendix describes an EDM measurement method that uses a Wien filter to produce a spin phase advance
in the same plane in which the MDM-induced spin precession occurs.
For protons at $30\UMeV$ (non-frozen spin) this plane is the horizontal plane of the ring. For protons
at $45\UMeV$ with `frozen spin', this plane is orthogonal to the radius of the
ring, where the MDM-induced spin precession is produced by horizontal
magnetic and vertical electric fields of the ring lattice imperfections.
The EDM can then be extracted by ultraprecise determination of the shift of
the spin precession frequency when the sign of the MDM component is reversed
between running the beam clockwise and counterclockwise in the ring.  An
important virtue of the method is that, in the case of a pure electric ring
(not necessarily frozen spin), it is free
of the background from imperfect magnetic fields of the ring lattice and also allows protection from the presence of external magnetic fields.  Also, because frozen
spin operation is not required, the method can be used to measure the deuteron EDM.
For searches of proton EDM in the prototype EDM storage ring,  sensitivity
$\approx$\,$2.2 \times 10^{-24} e\, \rm{cm}$ at beam energy $30\UMeV$ can be reached.

\section{Introduction}
\label{intro}
Interaction of the MDM with vertical electric imperfection fields in the pure electric storage ring creates the tilt of the invariant spin axis $\vec{c}=\vec{c}_{y}+\vec{c}^{\rm{MDM}}_{xz}$ away from the vertical direction $\vec{e}_{y}$, where $\vec{c}^{\rm{MDM}}_{xz}\perp\vec{e}_{y}$ and  $\vec{c}_{y}\parallel\vec{e}_{y}$, also $c_y\approx 1$. Projection $\vec{c}^{\rm{MDM}}_{xz}$ is a function of azimuthal angle---it depends on which point in the ring the invariant spin axis is viewed at. The reason for this is the non-commutativity of spin rotations in the imperfection fields.  Reduction of the imperfection fields implies that all elements of the ring are precisely aligned relative to a common vertical axis, which becomes a normal vector to the planar beam orbit. Then $\vec{c}^{\rm{MDM}}_{xz}\to0$ at every point of the ring.

Interaction of the EDM with the electric field in the ring tilts the invariant spin axis $\vec{c}$ towards the $X$-axis (which points against the radius of the ring). This tilt is an indication of an EDM signal. For a purely electrostatic storage ring, the tilt angle $\xi_{\rm{EDM}}$ due to the EDM is defined as
\begin{equation}
\label{eq:xiedm}
\tan\xi_\mathrm{EDM}=\frac{\eta\beta}{2(1-\beta^2(1+G))} \, ,
\end{equation}
where $G$ is the anomalous magnetic moment of the particle and $\eta$ is related to the EDM. For protons with kinetic energy $T = 30\UMeV$, $\xi_{\rm{EDM}}\approx 0.4\eta$.

\section{Mixing of EDM signal with systematic background from MDM}
\label{sec:mixedmmdm}
In a non-ideal storage ring, the tilt due to the EDM adds up to the tilt induced by the MDM (up to a first-order expansion in small $\xi_{\rm{EDM}}$) and the EDM signal mixes with systematic effects of the MDM spin rotation in imperfection fields:
\begin{equation}
\label{eq:isaedmmdm}
\vec{c}=\vec{c}_{y}+\vec{c}^{\rm{MDM}}_{xz}+\xi_{\rm{EDM}}\vec{e}_x.
\end{equation}

The orientation of the invariant spin axis was determined experimentally at COSY\cite{jedi-collaboration}. The method was based on the observation of the most precise quantity measured presently at COSY at the $10^{-10}$ level for $100\Us$ of the beam cycle---a spin tune\cite{Eversmann:2015jnk}. Two static solenoids, one in each straight section of the ring,  acted as artificial imperfections, which induced the change of the spin tune when powered up. The change of the spin tune was predicted by the theoretical model. The unknown parameters were the tilts of the invariant spin axis towards
the $Z$-axis (which points along the momentum) and $c_z$, at the locations
of the   solenoids. A sensitivity to the angular direction of the invariant spin axis of  $2.8 \times 10^{-6}$ rad was achieved.

Determination of $c_x$ projections with this method requires the use of static Wien filters with transverse horizontal magnetic fields ($\vec{B}=\vec{e}_x B$). Such a Wien filter rotates the spin around the $X$-axis through a constant angle at each turn and changes the spin tune. Currently, at COSY there are two Wien filters that can work with a horizontally oriented $B$ field, but both of them are radio-frequency devices. Running such an RF Wien filter at the beam revolution frequency allows it to perform as a static one. The time at which the RF field reaches its maximum should be synchronized with the time that the bunch  passes through the Wien filter. However,  the measurement of $\xi_{\rm{EDM}}\vec{e}_x$ separately from the direction of $\vec{c}^{\rm{MDM}}_{xz}$ would still not be  possible for COSY  \cite{Saleev:2017ecu}. The use of two Wien filters would provide information about the azimuthal dependence of the sum $\vec{c}^{\rm{MDM}}_{xz}+\xi_{\rm{EDM}}\vec{e}_x$. This will give an input to the model of the ring that should be based on  precise knowledge of the fields and beam orbit. Then variations of $\vec{c}^{\rm{MDM}}_{xz}$ from one point to another can be predicted and compared with measured values;
at the same time, $\xi_{\rm{EDM}}$ will be an unknown parameter that must be determined.

\section{Advantage of electrostatic rings}
The advantage of a purely electrostatic machine is that two countercirculating beams can be stored simultaneously. This allows unwanted magnetic fields in the ring to be controlled by observing the relative separation of closed orbits for  clockwise (CW) and counterclockwise (CCW) beams. Then, if unwanted, irreversible magnetic fields are removed, closed orbits become equal and following relations are true:
\begin{align}
\label{eq:ccwvscccwedm1}
\vec{c}^\mathrm{cw}&= \vec{c}_{y}+\vec{c}^{\rm{MDM}}_{xz}+\xi_{\rm{EDM}}\vec{e}_x \\
\label{eq:ccwvscccwedm2}
\vec{c}^\mathrm{ccw}&= -\vec{c}_{y}-\vec{c}^{\rm{MDM}}_{xz}+\xi_{\rm{EDM}}\vec{e}_x
\end{align}
As  already explained in the previous section, $\vec{c}^{\rm{MDM}}_{xz}$ is a function of azimuthal angle;
therefore, this property depends on where in the ring the $\vec{c}^\mathrm{cw}$ and $\vec{c}^\mathrm{ccw}$ are viewed---it should be the same point for both CW and CCW bunches.

Equations~(\ref{eq:ccwvscccwedm1}) and (\ref{eq:ccwvscccwedm2}) are also true for any storage ring operating at a non-frozen spin, be it purely magnetic, purely electric, or a  hybrid electric and magnetic ring, assuming that the correct expression is used for $\xi_{\rm{EDM}}$  in \Eref{eq:xiedm}.

If the condition $\vec{c}^{\rm{cw}}(\xi_{\rm{EDM}}=0)=-\vec{c}^{\rm{ccw}}(\xi_{\rm{EDM}}=0)$ can be guaranteed by making the CW and CCW beams equal, Eqs. (\ref{eq:ccwvscccwedm1})
and (\ref{eq:ccwvscccwedm2}) would enable  extraction of the EDM signal from the sum of the measured $\vec{e}_x$ projections of $\vec{c}^\mathrm{cw}$ and $\vec{c}^\mathrm{ccw}$. In the sum, the systematic effects of MDM spin rotations related to  imperfections of the electrostatic ring lattice are cancelled. Hence, the prototype electrostatic EDM ring offers a unique opportunity to test the principle of separating the EDM signal from the MDM systematic effect using simultaneously countercirculating beams with non-frozen spin.

\section{Effect of the Wien filter on beam and spin}
\label{sec:rfbwien}
Measurement of the $\vec{e}_x$ projection of the invariant spin axis by observation of spin tune perturbations demands the use of a static Wien filter with a horizontal transverse spin rotation axis $\vec{w}=\vec{e}_x$.
But the zero Lorentz force condition for the fields of the Wien filter can only be fulfilled for one direction of the beam:
\begin{equation}
\vec{E}+\vec{\beta}\times\vec{B}=0.
\end{equation}
To fulfil the zero Lorentz force condition for the opposite beam direction, the magnetic field in the Wien filter should change  sign:
\begin{equation}
\vec{E}+(-\vec{\beta})\times(-\vec{B})=0.
\end{equation}
The change of magnetic field direction can be achieved by making it an RF field that oscillates at the beam revolution frequency, in a similar way as proposed in \Sref{sec:mixedmmdm}. The electric field should remain constant for every turn.

There are two points in the ring where CW and CCW bunches are always diametrically opposite to each other and where they intersect, as shown in \Fref{fig:2}. The azimuthal positions of these points are controlled by the RF cavity.
Then the Wien filter should be installed at the point where the CW and CCW bunches are diametrically opposite to each other on every turn, so that, after half of the revolution period, either a CW or a CCW bunch enters the Wien filter (see \Fref{fig:1}).

\begin{figure} [hb!]
\centering
\includegraphics[width=0.35\textwidth]{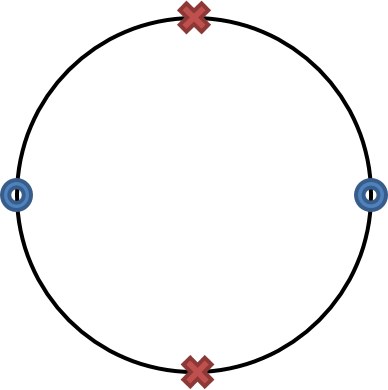}
\caption{The ring. Two crosses define the locations where CW and CCW bunches always intersect. The points where CW and CCW bunches are always located diametrically opposite to each other in the ring are marked as circles. The Wien filter can be installed at one such point.}
\label{fig:2}       
\end{figure} 

\begin{figure}[hb!]
\centering
\includegraphics[width=0.75\textwidth]{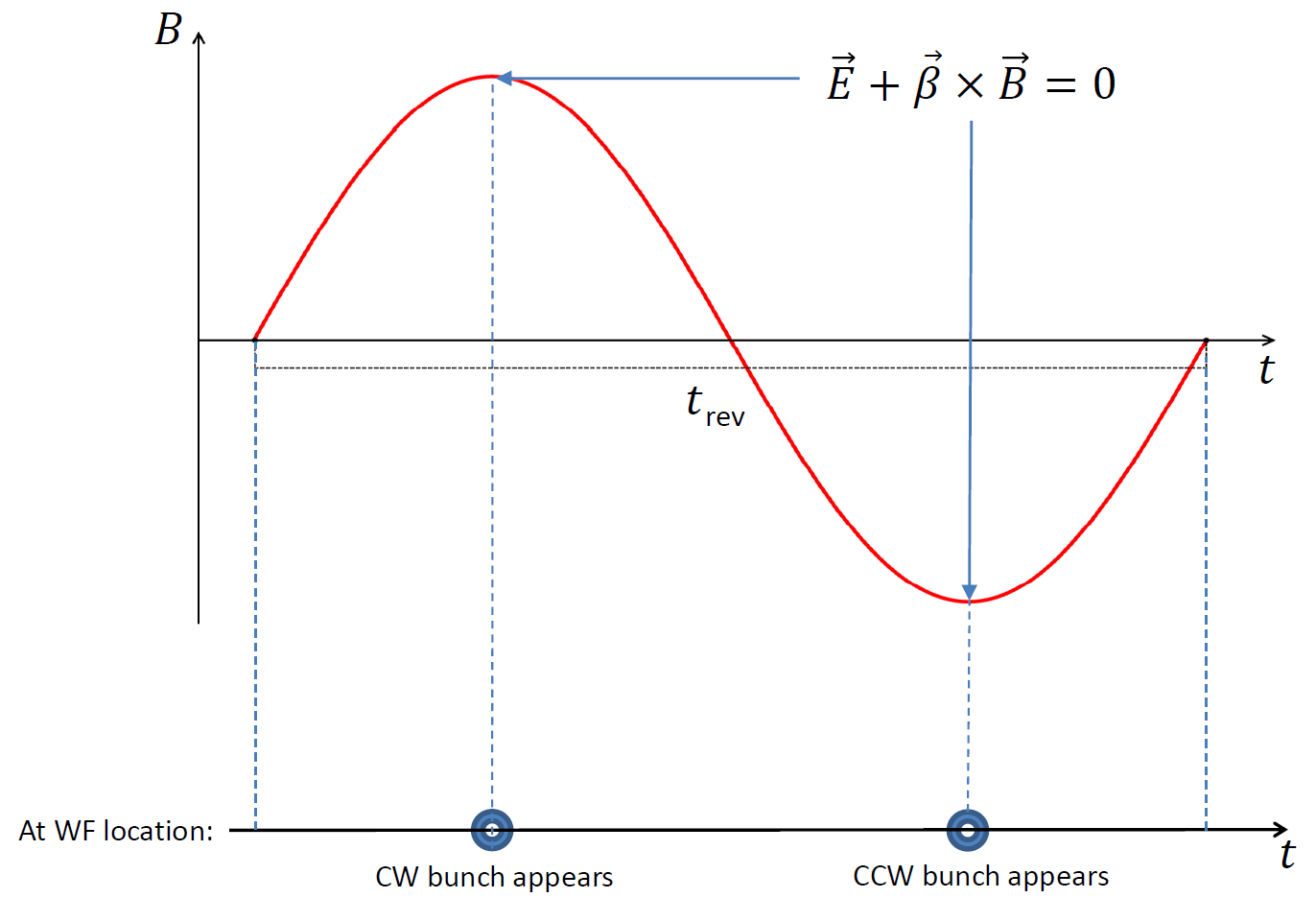}
\caption{Timeline of Wien filter operation}
\label{fig:1}
\end{figure}

The ideal Wien filter has exactly crossed $E$ and $B$ fields matched to a zero Lorentz force and rotates the spin around the $X$-axis. Consider now the case of the imperfect Wien filter, with horizontal magnetic and vertical electric fields that are not strictly orthogonal to each other, where the ratio between $E$ and $B$ does not exactly match the zero Lorentz force condition. Such a Wien filter will steer the beam vertically; it can have other components of spin rotation axis, $\vec{w}=\vec{e}_x w_x +\vec{e}_y w_y+\vec{e}_z w_z$, besides $w_x\approx 1$. The  Wien filter changes the closed orbit, which leads to a change in the direction of the invariant spin axis at the point where the Wien filter is installed. This change adds up to the effect of systematic MDM spin rotations $\vec{c}^{\rm{MDM}}_{xz}$. For both countercirculating beams, these additions are equivalent if the changes in the closed orbits are also equal. This will be the case if the magnetic field precisely reverses direction between the appearances of the CW and CCW beams at the Wien filter. Then Eqs. (\ref{eq:ccwvscccwedm1}) and (\ref{eq:ccwvscccwedm2}) remain valid.

The axis of spin rotation in the Wien filter for the CW beam in comparison with that of the CCW beam  is exactly opposite because of $B$ field reversal:
\begin{equation}
\vec{w}^\mathrm{cw}=-\vec{w}^\mathrm{ccw}.
\end{equation}

\subsection{Spin tune shift by Wien filter and EDM}
The analysis presented here  assumes that the beam in the storage ring has the energy away of the `frozen spin' condition. It is bunched and polarization of the bunch is in horizontal plane. Continuous measurement of the time-dependent horizontal polarization $P_x=\sum_{i=N} S_x^i$  allows  determination of the spin tunes of the CW and CCW bunches ($N$ is the number of particles). The sextupole fields are set up to provide at least $\tau=1000\Us$  of spin coherence time.

The change of the spin tune $\Delta\nu_s$ produced by the spin kick $\psi$ in the Wien filter is given by
\begin{equation}
\label{eq:sptshift}
\cos\pi(\nu_s+\Delta\nu_s)=\cos\pi\nu_s\cos\frac{\psi}{2}-\vec{c}\cdot\vec{w}\sin\pi\nu_s\sin\frac{\psi}{2}
\, .
\end{equation}

The difference between the scalar products $\vec{c}\cdot\vec{w}$ for the CW and CCW beams gives
\begin{equation}
\label{eq:ccww1}
\vec{c}^\mathrm{cw}\cdot\vec{w}^\mathrm{cw}-\vec{c}^\mathrm{ccw}\cdot\vec{w}^\mathrm{ccw}= 2 w_{x}\sin\xi_\mathrm{EDM} \, .
\end{equation}


Then the difference of the spin tunes for the CW ($\nu_s^\mathrm{cw}=\nu_s+\Delta\nu_s^\mathrm{cw}$) and CCW ($\nu_s^\mathrm{ccw}=\nu_s+\Delta\nu_s^\mathrm{ccw}$) bunches is proportional to the EDM tilt angle and the spin kick of the Wien filter, while the effects of the MDM spin rotations cancel:
\begin{equation}
\label{eq:edmeq}
\nu_s^\mathrm{cw} - \nu_s^\mathrm{ccw} = \frac{1}{\pi}\xi_{\rm{EDM}}\psi
\, .
\end{equation}

The time dependence of  the transverse horizontal projection of polarization is measured (see \Fref{fig:3}). The spin tunes of the CW and CCW bunches are determined; each one should depend quadratically on $\psi$. To control the  time-dependent systematic effects within a beam cycle, the phase shift $\Delta Q\cdot t= 2\pi(\nu_s^\mathrm{cw}-\nu_s^\mathrm{ccw})f_\mathrm{rev}t$ between spin oscillations of the CW and CCW beams can be monitored (here $f_\mathrm{rev}=\beta c/U$ is the revolution frequency). The statistical sensitivity to the EDM is given by
\begin{equation}
\label{eq:sigmaedm}
\sigma(|\vec{d}|)=\hbar\gamma^2|\vec{s}|\frac{|1-\beta^2(G+1)|}{G+1}\frac{U\sqrt{12}}{EL\sqrt{Nf}AP\tau}
\, .
\end{equation}
The sensitivity is inversely proportional to the electric field integral $EL$ in the Wien filter.

\begin{figure}
\centering
\includegraphics[width=\textwidth]{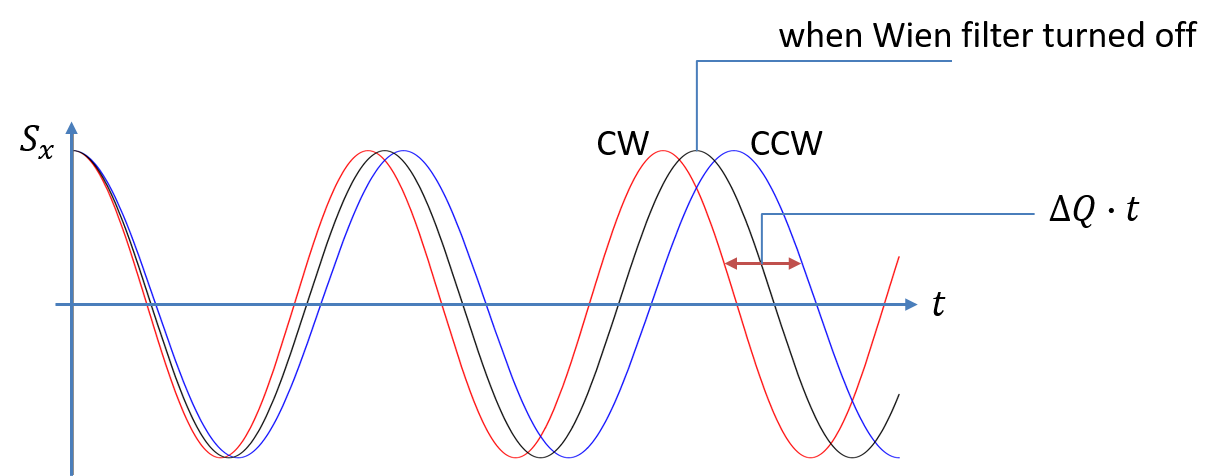}
\caption{Time dependence of horizontal spin projection $S_x$ for CW (red) and CCW (blue) particles. The black curve denotes projection $S_x$ for either CW or CCW particles in the case when the EDM is zero or the Wien filter is switched off.}
\label{fig:3}       
\end{figure} 

Systematic effects that are coming from external magnetic fields (such as the Earth's magnetic field) lead to $\vec{c}^{\rm{cw}}(\xi_{\rm{EDM}}=0)\neq-\vec{c}^{\rm{ccw}}(\xi_{\rm{EDM}}=0)$ for both points in the ring (see \Fref{fig:2}) where the Wien filters could be installed. Moreover, owing to the different direction of the external magnetic field at every element of the ring in the particle rest frame, and because of the non-commutativity of spin rotations, $\vec{c}^{\rm{cw}}_1(\xi_{\rm{EDM}}=0)+\vec{c}^{\rm{ccw}}_1(\xi_{\rm{EDM}}=0) \neq \vec{c}^{\rm{cw}}_2(\xi_{\rm{EDM}}=0)+\vec{c}^{\rm{ccw}}_2(\xi_{\rm{EDM}}=0)$, where indices 1 and 2 distinguish between the two points in the ring for
the location of Wien filters. If the external magnetic field is weak and the orbit separation between the CW and CCW beams is not measurable,  a cross-check of the measurement  $\nu_s^\mathrm{cw} - \nu_s^\mathrm{ccw}$ with a second Wien filter for the same beam can provide another control of such effects. The inequality $\nu_{s1}^\mathrm{cw} - \nu_{s1}^\mathrm{ccw}\neq\nu_{s2}^\mathrm{cw} - \nu_{s2}^\mathrm{ccw}$ indicates that systematic effects are present, while, in the ideal case, the right-hand side of \Eref{eq:edmeq} should be the same for the measurement with every Wien filter and independent of where in the ring it is installed.

The EDM limit of $\sigma_\mathrm{p}\approx 2.2\times10^{-24} e\,\rm{cm}$ can be achieved over the year of measurements, if the new technique is applied at the prototype EDM storage ring for protons at $30\UMeV$.  An integral field of 0.0005\,Tm at the maximum peak of the $B$ field is required in the Wien filter, together with 37.5\,kV of constant integral  electric field. A polarization $P=0.8$ and a beam intensity of $10^9$ particles per fill is assumed. Detection efficiency $f=0.0014$ and analysing power $A=0.47$ can be achieved using a multifoil carbon polarimeter for protons\cite{IEIRI1987253}.

To minimize the effects of synchrotron oscillations, which lead to non-compensated Lorentz forces for head and tail particles in long bunches, it is advisable to have a flat-top pulsed $B$ field.
\section{Spin wheel at the beam energy of frozen spin}
In a special case when the proton energy is such that the `frozen spin' condition is met, the vertical component of invariant spin axis $c_y\vec{e}_y$ vanishes in \Eref{eq:isaedmmdm} and the evolution of the vertical polarization should be measured. If no imperfection fields are present in the ring, the EDM aligns the invariant spin axis with the $X$-axis: $\xi_{\rm{EDM}}=\pi/2$ and $\vec{c}=\vec{e}_x$, while the spin tune becomes $\nu_s=\eta\gamma\beta / 2$.

 When the Wien filter  described in \Sref{sec:rfbwien} works in the ring at `frozen spin', it allows the invariant spin axis to be aligned with the $X$-axis in the presence of imperfection fields. This leads to a `spin wheel' (see Ref.~\cite{KoopSpinWheel} and \Sref{sect:PTRPhysics}) for both CW and CCW bunches that has a frequency proportional to the spin kick of the Wien filter.
Then the difference of the `spin wheel'  tunes is
\begin{equation}
\nu_s^\mathrm{cw} - \nu_s^\mathrm{ccw} = \eta\beta\gamma \, ,
\end{equation}
where $\eta$ is directly proportional to EDM and $\beta$ and $\gamma$ are
Lorentz factors. The wheel frequency is $10\UH$z for the mentioned $E$ and $B$ field integrals at a  proton kinetic energy of frozen spin $T \simeq
232\UMeV$ in the `nominal all-electric storage ring' (see Chapter \ref{Chap:allelectricring}) and $153\UHz$ for protons at $T \simeq 45\UMeV$ in the prototype EDM ring (PTR), where combined $E$ and $B$ fields in the deflectors are used to freeze the spin. In the latter case, either a CW or a CCW beam is stored in consecutive beam cycles.

\section{Other possibilities for EDM measurements with countercirculating beams}
\subsection{An option with two RF Wien filters in the prototype EDM ring}
\label{sec:rfwf}
The development and construction of the Wien filter described in  \Sref{sec:rfbwien} requires time and resources. There is another option to perform the EDM measurement at non-frozen spin energies of the beam in an electrostatic ring. It is based on the method \cite{Rathmann:2014gfa} described in Chapter \ref{Chap:Precursor}. The combined effect of the EDM and MDM on the spin motion in the ring is given by \Eref{eq:isaedmmdm}. Small vertical spin oscillations produced from horizontal components of $\vec{c}$ are resonantly excited by the spin kicks in the RF Wien filter (it has vertical spin rotation axis $\vec{w}=\vec{e}_y$). This leads to much greater amplitude  $S_y$ oscillations, which become accessible for polarimetry. The frequency of the $S_y$ oscillations is proportional to $\|\vec{c}^{\rm{MDM}}_{xz}+\xi_{\rm{EDM}}\vec{e}_x\|$ and the integral electric (or magnetic) field in the Wien filter.

The RF Wien filter is designed such that the $E$ field follows the $B$ field oscillations. This means that the Lorentz force is zero only for one beam direction and that the RF signal should be gated out when the countercirculating beam comes. This can be achieved by installing the RF Wien filters at two points where the countercirculating beams are opposite to each other (see \Fref{fig:2}). Then gating out the RF signal of both Wien filters at the
beam revolution frequency allows the CW and CCW beams to be run simultaneously. The outcome is similar to the one discussed in \Sref{sec:mixedmmdm}: $\|\vec{c}^{\rm{MDM}}_{xz}+\xi_{\rm{EDM}}\vec{e}_x\|$ at two points of the ring is determined, and if direction $\vec{c}^{\rm{MDM}}_{xz}$ can be predicted from the model assumptions, it allows one to find $\xi_{\rm{EDM}}$. However, direct extraction of the EDM signal is also possible when both RF Wien filters are switched to make zero Lorentz force for opposite beam directions. In this case, the directions of $\vec{c}^{\rm{MDM}}_{xz}$ at the Wien filter locations change sign. Spin rotations produced by one RF Wien filter for CW and CCW beams are compared. Additionally, a static solenoid is needed to suppress the $\vec{e}_z$ projection in $\vec{c}^{\rm{MDM}}_{xz}$, otherwise $\|\vec{c}^{\rm{MDM}}_{xz}+\xi_{\rm{EDM}}\vec{e}_x\|\propto\xi^2_\mathrm{EDM}$.

Another advantage of this option is that the RF Wien filter is transparent for the off-momentum particles. The Wien filter RF phase and the amplitude of the field can be adjusted such that only a slow build-up of vertical polarization is observed during the whole beam cycle. This allows  the statistical sensitivity of this method to be increased by 2.5 times in comparison with the method discussed in \Sref{sec:rfbwien}, assuming the same field integrals in the Wien filters. The disadvantage of the method is that direct extraction of the EDM signal from the measured $P_y$ polarization build-ups produced using the same RF Wien filter for the CW and CCW beams depends on the equality of the orbits in the consecutive  CW--CCW beam injections.
\subsection{An option with static Wien filters in the prototype EDM ring}
Instead of RF Wien filters, as described in \Sref{sec:rfwf}, one or more Wien filters with static vertical electric and static horizontal magnetic fields can be used. The placement of Wien filters is not crucial. For a single Wien filter, all conclusions are the same, as stated in \Sref{sec:rfbwien}. The only difference is that \Eref{eq:edmeq} for  $\nu_s^\mathrm{cw} - \nu_s^\mathrm{ccw}$ is calculated for CW and CCW beams that are running consecutively, and the $B$ field of the Wien filters is reversed between the injections in the CW and CCW directions. This can lead to a systematic error if  field reversal is not exact. Assuming that one can achieve two orders of magnitude higher field integrals for static fields, such a method can have the advantage that it allows the statistical error of the EDM measurement to be reduced by two orders of magnitude, compared with that discussed in \Sref{sec:rfbwien}.

\section{Summary and outlook}
Here, we propose a new method for measuring the EDM of charged particles  in electrostatic storage rings. One of the advantages of such rings is that CW and CCW bunches could be stored simultaneously, which allows  the systematic effects of ring lattice imperfections to be cancelled.  The advantage of the method over the BNL proposal (see Chapter \ref{Chap:allelectricring}) is that the ring operation mode is not fixed only to the energy of the `frozen spin', which means that it can be of much smaller size and that different particle species could be studied. The disadvantage is that sensitivity to the EDM signal is suppressed by four orders of magnitude compared with that at `frozen spin', assuming an electric field integral of 37.5\,kV in the Wien filter. Because of this, the method seems to be an intermediate step towards ultimate EDM precision searches, and is applicable at the prototype (PTR) EDM ring. It serves as a complement for the BNL proposal when applied at `frozen spin' for protons. It allows control of the systematic effects of unwanted MDM spin rotations produced by external magnetic fields when two Wien filters are used for spin tune mapping.

\begin{flushleft}

\end{flushleft}
\end{cbunit}

\begin{cbunit}

\csname @openrighttrue\endcsname 
\chapter[New ideas: deuteron EDM frequency domain determination]{New ideas: deuteron EDM frequency domain determination\footnote{This appendix was written by Yuri Senichev (Institute for Nuclear Research, Russian Academy of Sciences).}}
\label{Chap:FDM:Full}

This appendix describes suppression of the systematic geometric phase and machine imperfection errors that are encountered in any frozen spin storage ring EDM measurement method based on observation of a slow, gradual change in the beam polarization vector.
The geometric phase error is caused by non-commutating wobbling precessions of the polarization vector, which are significant only if the polarization vector precession rate is small. The geometric phase can be suppressed by dispensing with operation at the spin resonance (\ie 3D frozen spin) state, in favour of operation at the 2D frozen spin state, represented by a rolling spin wheel.
To eliminate the systematic machine imperfection error, the imperfection fields themselves are utilized as the drivers of the spin wheel.

The method is intended for a combined storage ring; the bend fields are magnetic and the frozen spin condition is met using a number of uniformly distributed, discrete Wien filters. Reversing the bending field (along with the beam direction) reverses the imperfection fields.  The EDM measurement consists of measuring the difference of spin wheel roll rates, which is proportional to the EDM. Though motivated by the need to measure the deuteron EDM, the method can also be applied  to the proton.

\section{Motivation}

Storage ring-based methods of searching for the electric dipole moments (EDMs) of fundamental particles
can be classified into two major categories, which we will call
(1) space-domain and (2) frequency-domain 
methods.

In the space-domain paradigm, one measures a \emph{change in the spatial orientation} of the beam polarization
vector \emph{caused by the EDM}.

The original storage-ring frozen-spin-type method, proposed in Ref.~\cite{bnl-sredm-deuteron-proposal}, is a canonical example of
a methodology in the space domain: an initially longitudinally polarized beam is injected into the storage ring;
the vertical component of its polarization vector is observed. Under ideal conditions, any tilting of
the beam polarization vector from the horizontal plane is attributed to the action of the EDM.

Two technical difficulties are readily apparent with this approach.
\begin{itemize}
\item It poses a challenging task for polarimetry~\cite{Mane:2015jsa}.
\item It puts very stringent constraints on the precision of the accelerator optical element alignment.
\end{itemize}

The former  is due to the requirement of detecting a change of about $5\times 10^{-6}$ to the
cross-section asymmetry $\varepsilon_\mathrm{LR}$ in order to get to the EDM sensitivity level
of $10^{-29}~e\,\Ucm$ (see Ref.~\cite{bnl-sredm-deuteron-proposal}).

The latter  involves minimizing the magnitude of the vertical plane magnetic dipole
moment (MDM) precession frequency (p. 11 of Ref.~\cite{bnl-sredm-deuteron-proposal})
\begin{equation}\label{eq:BNL_syst_err}
\w_\mathrm{syst} \approx \frac{\mu\avg{E_v}}{\beta c\gamma^2},
\end{equation}
induced by machine imperfection fields. According to estimates made by Y. Senichev, if it is to be fulfilled,
the geodetic installation precision of accelerator elements must reach $10^{-14}$ m. Today's technology
allows only for about $10^{-4}$ m.

At the practically achievable level of element alignment uncertainty, $\w_\mathrm{syst} \gg \w_\mathrm{edm}$,
and changes in the orientation of the polarization vector are no longer EDM-driven.

Another crucial problem  faced in the space domain is geometric phase error
(see Ref.~\cite{Anas}).
The problem here lies in the fact that, even if one can somehow make field imperfections (owing to either 
optical element misalignment or spurious electromagnetic fields) zero
\emph{on average}, since spin rotations are non-commutative, the polarization rotation angle due to them
will not be zero.

By contrast, the frequency-domain methodology is founded on measuring the EDM \emph{contribution} to the total
(MDM and EDM together) spin precession \emph{angular velocity}.

The polarization vector is made to roll about a nearly constant, definite direction vector $\nbar$,
with an angular velocity that is large enough for its magnitude to be easily measurable at all times.
Apart from easier polarimetry, the definiteness of the angular velocity vector is a safeguard against geometric
phase error.

This `spin wheel' may be externally applied~\cite{KoopSpinWheel}, or  the machine imperfection fields
may be utilized for the same purpose (wheel roll rate determined by \Eref{eq:BNL_syst_err}).
The latter is made possible by the fact that $\w_\mathrm{syst}$ changes sign when the beam revolution direction
is reversed (see Ref.~\cite{bnl-sredm-deuteron-proposal}).

\section{Universal storage ring EDM measurement problems}

By way of introduction to the proposed measurement methodology, let us briefly summarize some measurement problems
encountered by any EDM experiment performed in a storage ring; they can be grouped into two broad categories:
\begin{itemize}
\item problems solved by a spin wheel:
  \begin{itemize}
  \item spurious electromagnetic fields;
  \item betatron motion.
  \end{itemize}
\item problems having specific solutions:
  \begin{itemize}
  \item spin decoherence;
  \item machine imperfections.
  \end{itemize}
\end{itemize}

\subsection{Spin motion perturbation}
Problems from the first category are those introducing geometric phase error. Indeed, both the spurious
and the focusing fields, when acting on a betatron-oscillating particle, perturb the direction and
magnitude of its spin precession angular velocity vector. The effect is a spin kick in the direction defined
by the perturbation.

Assume that the EDM provides a spin kick about the radial ($\hat x$-) axis. The magnitude of the angular
velocity vector has a general form
\[
\w = \sqrt{\w_x^2 + \w_y^2 + \w_z^2},
\]
where $\w_y$ is minimized by fulfilling the frozen spin condition; $\w_z$ (the constant part of which is
due to machine imperfections) can be minimized via the installation
of a longitudinal solenoid on the optical axis\footnote{Length, $1\Um$; magnetic field approximately $10^{-6}\UT$.}. In the
space domain, one also tries to minimize the $\wimp$ contribution to $\w_x = \w_\mathrm{edm} + \wimp$. Consequently,
spin kicks must be minimized to (significantly) less than $\w_\mathrm{edm}$, so as to reduce the geometric phase to
less than the accumulated EDM phase.

The benefit of having a spin wheel aligned with the EDM angular velocity is that orthogonal MDM contributions
to the total angular velocity vector add up in squares, and hence their effect is greatly diminished:
\begin{align*}
  \w &= \sqrt{(w_\mathrm{edm} + w_\mathrm{SW})^2 + \w_y^2 + \w_z^2} \\
  &\approx(w_\mathrm{edm} + w_\mathrm{SW})\cdot \bkt*{1 + \frac{\w_y^2 + \w_z^2}{w_\mathrm{sw}^2}}^{\sfrac12} \\
  &\approx (w_\mathrm{edm} + w_\mathrm{SW})\cdot \bkt{1 + \frac{\w_y^2 + \w_z^2}{2w_\mathrm{SW}^2}} \\
  &\approx w_\mathrm{SW} + w_\mathrm{edm} + \underbrace{\frac12\frac{\w_y^2 + \w_z^2}{w_\mathrm{SW}}}_{\epsilon}.
\end{align*}

Since our goal is to observe the EDM-related value shift in $\w$, we need to minimize the random variable
$\epsilon$:
\[
\frac12\frac{\w_y^2 + \w_z^2}{w_\mathrm{SW}} < w_\mathrm{edm}.
\]

Let us make some preliminary estimates. Suppose $w_\mathrm{SW} \approx 50$ rad/s (the reason for choosing this
value will be explained shortly), $w_\mathrm{edm}\approx10^{-9}$ rad/s (corresponding to the EDM value
$10^{-29}~e$\,cm). Then $\w_y^2 + \w_z^2$ must be reduced to less than $10^{-7}$ rad/s or, equivalently,
either angular velocity to less than $3 \times 10^{-4}$ rad/s. This is several orders of magnitude greater than
the expected standard error in the angular velocity estimate~\cite{Aksentev:2018krh}, and hence
should not be a problem to achieve.

One case left to be considered is that of MDM spin kicks about the $\hat x$-axis. These are not attenuated, and cause the
most trouble. They come in three varieties:
(a) permanent, not caused by optical element misalignments;
(b) semipermanent, caused by element tilts about the optic axis;
and (c) spurious.

Semipermanent radial spin kicks (be they caused by magnetic or electric fields) change sign when
the beam revolution direction is reversed from clockwise (CW) to counterclockwise (CCW).
Spurious kicks can be dealt with by statistical averaging.
Permanent kicks, insensitive to either the guide field or the beam circulation direction, cannot be controlled.
Looking on the bright side, however, their sources should not be present under normal circumstances.

For more details on spin motion perturbation effects on the measurement of the EDM in the frequency domain,
see Ref.~\cite{Aksentyev:2019jet}.

\subsection{Expected machine imperfection SW roll rate}
In the previous estimates, we used a roll rate $w_\mathrm{sw} \approx 50$ rad/s for the spin wheel. This is
our expected $\w_\mathrm{syst}$ caused by machine imperfections.

Denote the standard deviation of the imperfection radial magnetic field distribution $\sigma[B_x]$.
For the whole ring, MDM precession will be distributed with a standard deviation~\cite{Senichev:2017amn}
\[
\sigma[\w_x^\mathrm{MDM}] = \frac{e}{m\gamma}\frac{G+1}{\gamma}\frac{\sigma[B_x]}{\sqrt{n}},
\]
where $n$ is the number of misaligned elements and  $G=(g-2)/2$ is the anomalous magnetic dipole moment.

For deuterons in lattices~\cite{Senichev:2016rez} of $n$ of the order of 100 elements, rotated about the
optical axis by angles $\Theta_\mathrm{tilt}\sim N(0, 10^{-4})$ rad, Y. Senichev
\cite{Senichev:2017amn} estimates
$\w_x^\mathrm{MDM}$ as between 50 and 100 rad/s.

Our simulations, run in COSY INFINITY, seem to confirm this result. \Figure[b]~\ref{fig:imperfections:SW_roll_rate} shows the results of the simulation
in which we rotated the 32 $E+B$ spin rotator elements used in the frozen spin
(codename BNL) lattice~\cite{Senichev:2016rez} by angles randomly picked from
the distribution $N(\mu_0\cdot(i-5), \sigma_0)$, where $\mu_0 = 10\cdot\sigma_0 = 10^{-4}$ rad,
$i\in\lbrace0,\dots, 10\rbrace$.

\begin{figure}
\centering
\includegraphics[width=0.7\linewidth]{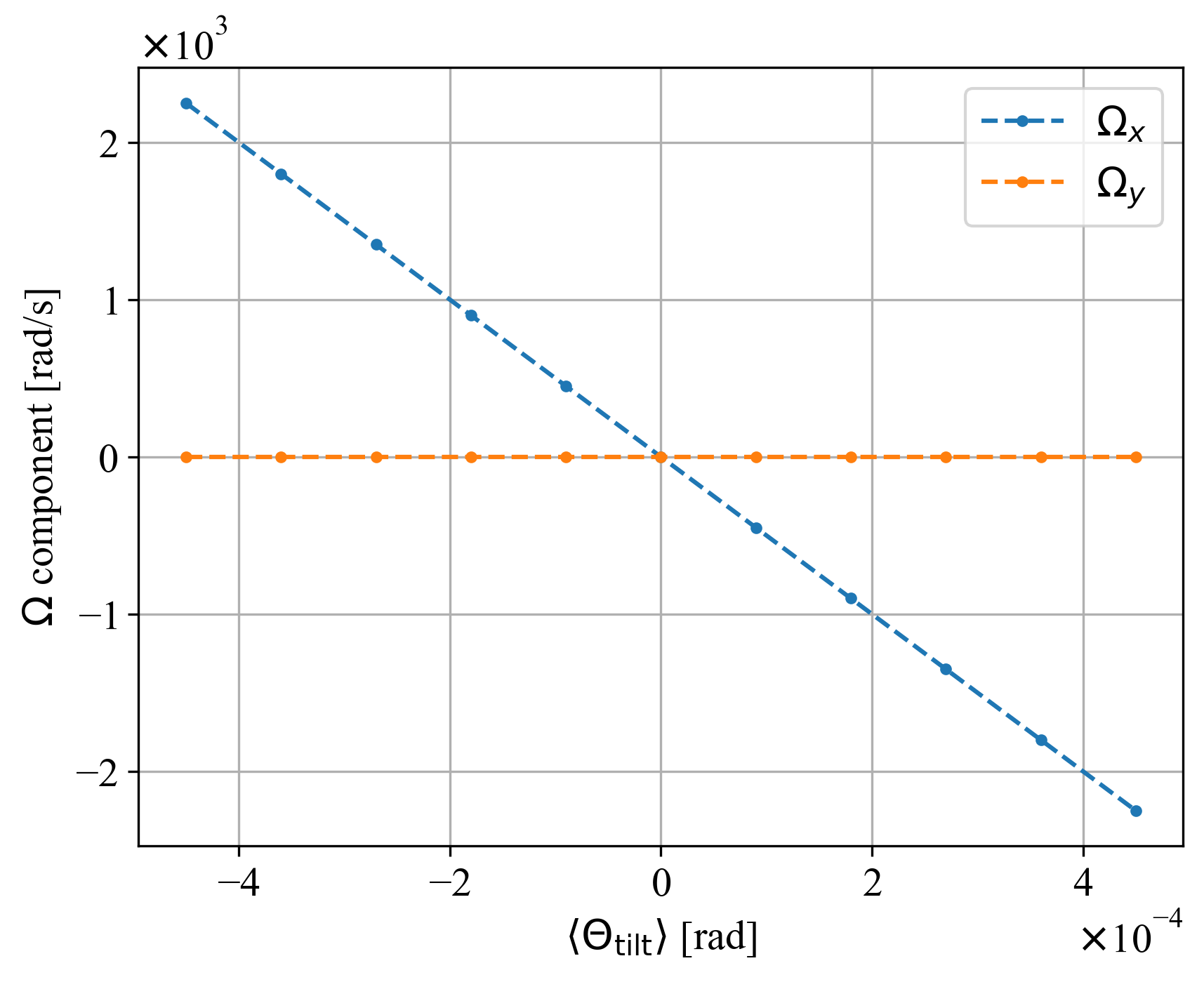}
\caption{Spin precession frequency (radial and vertical components) versus
the mean $E+B$ element tilt angle\label{fig:imperfections:SW_roll_rate}}
\end{figure}

At $\avg{\Theta_\mathrm{tilt}} = 10^{-4}$, we observe a roll rate of 500 rad/s. We should keep in mind,  however, that
Senichev assumes $\sigma_{\Theta_\mathrm{tilt}} = 10^{-4}$ rad, which means, for a lattice with $n=100$ tilted elements,
a standard deviation of the mean $\sigma_{\avg{\Theta_\mathrm{tilt}}} = {\sigma_{\Theta_\mathrm{tilt}}}/{\sqrt{100}} = 10^{-5}$. The dependence of $\w_x^\mathrm{MDM}$ on $\avg{\Theta_\mathrm{tilt}}$
is linear, which means that, in an actual lattice, we would observe  $\w_\mathrm{syst} \le 50$ rad/s with 68\% probability,
and $\w_\mathrm{syst} \le 100$ rad/s with 95\% probability, and, with 27\% probability, $50 \le\w_\mathrm{syst}\le 100$.

\subsection{Spin decoherence}
Spin coherence is a measure or quality of preservation of polarization in an initially fully polarized
beam~\cite{Valetov:2017}. Spin decoherence refers to the depolarization caused by the difference in the
beam particles' spin precession frequencies.

The difference in spin tunes is due to the difference in the particles' orbit lengths, and hence their
equilibrium energy levels, on which spin tune depends. One way in which spin decoherence can be suppressed is by
utilization of sextupole fields. We consider how this can be accomplished in Ref.~\cite{Aksentyev:2019rxq}.

\subsection{Machine imperfections}

As we have seen, the problem with machine imperfections is twofold:
(a) they are practically impossible to remove at the present level of technology; but, even worse,
(b) their removal leaves one in the space domain, and opens the measurement up to geometric phase error.

Fortunately for us, the imperfection spin kicks that they induce change sign when the beam circulation direction
is reversed. Their magnitude is also sufficient for use as a Koop wheel. The one remaining difficulty
is the accuracy of the Koop wheel roll direction flipping. Hopefully, we can make a sufficiently persuasive
 argument how to accomplish this.

\section{Main methodology features}

The method we propose is characterized by two main features.
\begin{itemize}
\item It is a frequency-domain method.
\item The fields induced by machine imperfections, instead of being suppressed,
  are used as a Koop wheel.
  \begin{itemize}
  \item The Koop wheel roll direction is reversed by flipping the direction of the guide field.
  \item Its roll rate is controlled through observation of spin precession in the horizontal plane.
  \end{itemize}
\end{itemize}

The advantages of the frequency domain, such as
(a) ease of polarimetry and
(b) immunity to geometric phase error,
have been discussed in previous sections. Now we will turn to the description of how machine imperfection fields
can be used as a Koop wheel.

\section{EDM estimator statistic}
Since the angular velocity measured using the frequency-domain methodology includes contributions from both the
magnetic and the electric dipole moments, the EDM estimator statistic requires
two cycles to compose:
one in which the Koop wheel rolls forwards, the other backwards.

The change in the Koop wheel roll direction is affected by flipping the direction of the guide field.
When this is done,
$\vec B \mapsto -\vec B$, the beam circulation direction changes from clockwise (CW) to counterclockwise (CCW),
$\vec\beta \mapsto -\vec\beta$, while the electrostatic field remains constant, $\vec E \mapsto \vec E$.
According to the Thomas--BMT equation, spin precession frequency components change as
\begin{subequations}
  \begin{align}
    \w_x^\mathrm{CW} &= \w_x^\mathrm{MDM, CW}   + \w_x^\mathrm{EDM}, \notag\\
    \w_x^\mathrm{CCW} &= \w_x^\mathrm{MDM, CCW} + \w_x^\mathrm{EDM}, \notag\\
    \w_x^\mathrm{MDM, CW} &= -\w_x^\mathrm{MDM, CCW}, \label{eq:CW_CCW_MDM}\\
    \intertext{and the EDM estimator is}
    \hat\w_x^\mathrm{EDM} &:= \frac12\bkt{\w_x^\mathrm{CW} + \w_x^\mathrm{CCW}} \label{eq:FDM_estimator} \\
                  &=  \w_x^\mathrm{EDM} +
          \underbrace{\frac12\bkt{\w_x^\mathrm{MDM, CW} + \w_x^\mathrm{MDM, CCW}}}_{\varepsilon \to 0}.
  \end{align}
\end{subequations}

To keep the systematic error term $\varepsilon$ below required precision, \ie ensure
that \Eref{eq:CW_CCW_MDM} holds with sufficient accuracy, Y. Senichev \cite{Senichev:2017amn}
devised a guide field flipping procedure
based on observation of the beam polarization precession frequency in the horizontal plane. To explain how it works, we need to introduce the concept of the effective Lorentz factor.

\section{Effective Lorentz factor}
Spin dynamics is described by the concepts of \emph{spin tune} $\nu_s$ and \emph{invariant spin axis} $\nbar$.
The spin tune depends on the particle's  equilibrium-level energy, expressed by the Lorentz factor:
\begin{equation}\label{eq:spin_tune_vs_gamma}
  \begin{cases}
    \nu_s^B &= \gamma G, \\
    \nu_s^E &= \beta^2\gamma\bkt{\frac{1}{\gamma^2-1} - G} \\
            &= \frac{G+1}{\gamma} - G\gamma.
  \end{cases}
\end{equation}

Unfortunately, not all beam particles share the same Lorentz factor. A particle involved in betatron
motion will have a longer orbit and, as a direct consequence of the phase stability principle,
in an accelerating structure utilizing an RF cavity, its equilibrium energy level
must increase. Otherwise it cannot remain in the bunch. In this section, we analyse how the particle Lorentz factor
should be modified when betatron motion, as well as non-linearities in the momentum compaction factor, are
accounted for.

The longitudinal dynamics of a particle on the reference orbit of a storage ring are described
by
\begin{equation}
  \begin{cases}
    \ddt{}\D\varphi &= -\w_\mathrm{RF}\eta\delta, \\
    \ddt{}\delta &= \frac{q V_\mathrm{RF}\w_\mathrm{RF}}{2\pi h\beta^2E}\bkt{\sin\varphi - \sin\varphi_0}.
  \end{cases}
\end{equation}
In these equations, $\D\varphi = \varphi - \varphi_0$ and
$\delta = \bkt{p-p_0}/{p_0}$ are, respectively, the deviations of the particle's phase and
the  normalized momentum from those of the reference particle; all other symbols have their usual meanings.

The solutions of this system form a family of ellipses in the $(\varphi, \delta)$-plane, all centred at the
point $(\varphi_0,\delta_0)$. However, if one considers a particle involved in betatron oscillations, and
uses a higher-order Taylor expansion of the momentum compaction factor
$\alpha = \alpha_0 + \alpha_1\delta$, the first equation of the system
transforms into (see Ref.~\cite{Senichev:2013dra})
\begin{align*}
  \ddt{\D\varphi} = -\w_\mathrm{RF} \left[\bkt{\frac{\Delta L}{L}}_\beta + \bkt{\alpha_0 + \gamma^{-2}}\delta  +  \bkt{\alpha_1 - \alpha_0\gamma^{-2} + \gamma^{-4}}\delta^2\right],
\end{align*}
where
\[
 \bkt{\frac{\Delta L}{L}}_\beta = \frac{\pi}{2L}\bkt*{\varepsilon_xQ_x + \varepsilon_yQ_y}
\]
 is
the betatron motion-related orbit lengthening, $\varepsilon_x$ and $\varepsilon_y$ are
the horizontal and vertical beam emittances, and $Q_x$, $Q_y$ are the horizontal and vertical tunes.

The solutions of the transformed system are no longer centred at the same single point. Orbit lengthening
and momentum deviation cause an equilibrium-level momentum shift (see Ref.~\cite{Senichev:2013dra})
\begin{equation}\label{eq:EquLevMom_shift}
\Delta\delta_\mathrm{eq} = \frac{\gamma_0^2}{\gamma_0^2\alpha_0 - 1}\bkt*{\frac{\delta_m^2}{2}\bkt{\alpha_1 - \alpha_0\gamma^{-2} + \gamma_0^{-4}} + \bkt{\frac{\Delta L}{L}}_\beta},
\end{equation}
where $\delta_m$ is the amplitude of synchrotron oscillations.

We call the equilibrium energy level associated with the momentum shift (\Eref{eq:EquLevMom_shift})
the \emph{effective Lorentz factor}:
\begin{equation}\label{eq:EffectiveGamma}
\gamma_\mathrm{eff} = \gamma_0 + \beta_0^2\gamma_0\cdot\Delta\delta_\mathrm{eq},
\end{equation}
where $\gamma_0$ and $\beta_0$ are, respectively, the Lorentz factor and the relative velocity factor of the reference particle.

Observe that the effective Lorentz factor enables us to account for variations in the value of the spin tune
resulting from variations in the particle orbit length. It is crucial in the analysis of
spin decoherence \cite{Aksentyev:2019rxq} and its suppression by means of sextupole fields.

It plays a big role, as well, in the successful reproduction of the MDM component of the total spin precession
angular velocity.

\section{Guide field flipping}
\newcommand{\Traj}{\mathcal T}
\DeclareDocumentCommand{\Stab}{s}{\mathcal{S}\IfBooleanT{#1}{\vert_{\w_y=0}}}
\newcommand{\Fail}{\mathcal F}
\renewcommand{\D}{\mathcal D}

Two aspects of the problem need to be paid attention to:
\begin{itemize}
\item what needs to be kept constant from one measurement cycle to the next;
\item how it can be observed.
\end{itemize}

The goal of flipping the direction of the guide field is to accurately reproduce the radial component
of the MDM spin precession frequency induced by machine imperfection fields. This point should not be overlooked:
a mere reproduction of the \emph{magnetic field strength} would not suffice, since the injection point of the beam's centroid,
and hence its orbit length---and, via Eqs.~\eqref{eq:EffectiveGamma} and~\eqref{eq:spin_tune_vs_gamma}, spin tune---is subject to variation. (Apart from that, the accelerating structure might not be symmetrical, in terms of spin dynamics, with regard to reversal of the beam circulation direction.)

What needs to be reproduced, therefore, is not the field strength, but the effective Lorentz factor of the centroid.

Regarding the second question, we mentioned earlier that the Koop wheel roll rate
is controlled through measurement of the horizontal plane spin precession frequency.
This plane was chosen because the EDM angular velocity vector points
(mainly) in the radial direction; its vertical component is due to machine imperfection fields, and is small compared with
the measured EDM effect. Therefore, to a first approximation, when we manipulate the vertical component of the
combined spin precession angular velocity, we manipulate the vertical component of the MDM angular velocity vector.

Moving on to the effective Lorentz factor calibration procedure, let $\Traj$ denote the set of all trajectories that a particle might follow in the accelerator;
 $\Traj = \Stab \bigcup \Fail$, where $\Stab$ is the set of all stable trajectories
and $\Fail$ are all trajectories
such that if a particle takes one, it will be lost from the bunch.

Calibration is in two phases.
\begin{itemize}
\item In the first phase, the guide field value is set so that the beam particles are injected onto trajectories
  $t\in\Stab$.
\item In the second phase, it is fine-tuned further, so as to fulfil the frozen-spin condition in the horizontal plane.
  By doing this, we physically move the beam trajectories into the subset $\Stab*\subset\Stab$ of trajectories
  for which $\w_y = 0$.
\end{itemize}

The spin tune (and hence the precession frequency) is an injective function of the
effective Lorentz factor $\gamma_\mathrm{eff}$, which means that
$\w_y(\gamma_\mathrm{eff}^1) = \w_y(\gamma_\mathrm{eff}^2) \rightarrow \gamma_\mathrm{eff}^1 = \gamma_\mathrm{eff}^2$. The trajectory space $\Traj$ is partitioned into equivalence
classes according to the value of $\gamma_\mathrm{eff}$: trajectories characterized by the same $\gamma_\mathrm{eff}$ are equivalent
in terms of their spin dynamics (possess the same spin tune and invariant spin axis direction),
and hence belong to the same equivalence class.
Since $\w_y(\gamma_\mathrm{eff})$ is injective, there exists a unique $\gamma_\mathrm{eff}^0$ at which $\w_y(\gamma_\mathrm{eff}^0)=0$,
\[
[\w_y=0] = [\gamma_\mathrm{eff}^0] \equiv \Stab*.
\]

If the lattice did not use sextupole fields for the suppression of decoherence,
$\Stab*$ would be a singleton set. We have shown in Ref.~\cite{Aksentyev:2019rxq} that if sextupoles are
utilized, then $\exists\D\subset\Stab$ such that $\forall t_1,t_2\in\D$:
$\nu_s(t_1) = \nu_s(t_2)$, $\nbar(t_1) = \nbar(t_2)$. By adjusting the guide field strength, we equate
$\D=\Stab*$; hence, $\Stab*$ contains a number of trajectories. Strictly speaking,
  even if sextupoles were used, there would remain some negligible dependence of spin tune
  on the particle orbit length (linear decoherence effects, cf. Ref.~\cite{Aksentyev:2019rxq}).
  Because of this, the equalities for $\nu_s$ and $\nbar$ are approximate, and the set $\Stab*$
  should be viewed as fuzzy:
  we will consider trajectories for which $|\w_y|<\delta$, for some small $\delta$, as belonging to $[\w_y=0]$.

Therefore, once we have ensured that the beam polarization does not precess in the horizontal plane,
all of the beam particles have $\gamma_\mathrm{eff}^0$, equal for the CW and CCW beams.

Guide field flipping procedure simulation results can be found in Ref.~\cite{Aksentyev:2019ajz}.

\section{Statistical precision}

Members of the JEDI collaboration have studied the statistical precision of spin precession angular velocity
estimates from sparse (one detector event per 100 spin revolutions)~\cite{Eversmann:2015jnk} and
dense~\cite{Aksentev:2018krh} polarization data.

According to Ref.~\cite{Eversmann:2015jnk}, the maximum likelihood estimator for the spin precession frequency
estimate has a standard error
\[
\sigma_{\hat\w} = \frac{1}{PT}\sqrt{\frac{24}{N}},
\]
where $N$ is the total number of recorded detector events, $P$ is the beam polarization, and $T$ is the
measurement time.

Assuming $N=7.5\times 10^8$ events, polarization $P=0.4$, and cycle duration
$T=1000\Us$ (the same parameters as in the simulation reported in Ref.~\cite{Aksentev:2018krh}),
we have $\sigma_{\hat\w} \approx 4.5\times 10^{-7}$\,rad/s at the cycle level.
Estimates reported in Ref.~\cite{Aksentev:2018krh} agree with this result.

This precision is sufficient to obtain a mean estimate with statistical uncertainty
 $\sigma_{\avg{\hat\w}} \approx 3\times 10^{-9}$\,rad/s in 1\,year of measurement, with
the accelerator operational 70\% of the time. An EDM of $10^{-29}~e\,$cm should
induce an $\omega_\mathrm{edm}$ of the level of $10^{-9}$\,rad/s in storage rings proposed
in Ref.~\cite{Senichev:2016rez}. Thus, we expect to be able to measure the deuteron EDM
at the $10^{-29}~e\,$cm level in 1\,year of measurement time.

\begin{flushleft}

\end{flushleft}
\end{cbunit}

\begin{cbunit}

\csname @openrighttrue\endcsname 
\chapter[New ideas: distinguishing EDM effects from magnet misalignment by Fourier analysis]{New ideas: distinguishing EDM effects from magnet misalignment by Fourier analysis\footnote{This appendix was written by Andrzej Magiera (Institute of Physics, Jagiellonian University).}}
\label{app:fourier}

This appendix shows that, by measuring vertical
polarization in two properly separated positions in the storage ring,
it is possible to estimate the magnitude of the major systematic
uncertainty induced by ring imperfections. The imprecise positioning
of the magnets causes the creation of a radial field, and the
interaction of the MDM with this field induces the
effect mimicking the EDM signal. The ring imperfections are
distributed rather randomly along the ring, while the dipole magnets
form a very regular pattern. Therefore, the changes in the vertical
polarization induced by magnet misalignment have a non-harmonic
pattern. Conversely, the EDM-induced vertical polarization has
an almost harmonic pattern, since it results from the vertical field
of ring dipoles, which is not strongly affected by their
misalignments. Within a simple model, the vertical polarization induced
by the EDM and ring imperfections is calculated as a function of
time. Then it is shown that Fourier analysis of obtained signals
sampled twice per beam revolution allows  these two
effects to be distinguished. This is achieved through comparison of the Fourier amplitudes for
revolution frequency and for the differences of this frequency and spin
precession frequency. Even for unknown misalignments, it is possible to
predict, with the given likelihood, the magnitude of the systematic
uncertainty induced by ring imperfections.


A reliable limit on the value of the EDM in any experiment can be given
only when systematic uncertainty is under control. In an experiment
measuring the EDM with a storage ring, the most important systematic effect
comes from the radial field arising from magnet misalignment. For
all proposed scenarios for EDM measurements based on the detection of
induced vertical polarization $s_y(t)$, this systematic effect mimics
the expected EDM signal. One might rely on  simulations of
the misalignment effect on $s_y(t)$, but magnet rotations and
displacements are, in fact, unknown. Therefore, direct experimental
estimation of the systematic uncertainty of misalignment effects by means
of Fourier analysis of $s_y(t)$ seems a much better solution.

The presented method of misalignment effect calculations is an
extension of the formalism presented in Ref.~\cite{Magiera:2017ypv}. The
model is limited to particles moving along the central trajectory, but
offers an analytical solution with a detailed insight into the general
features characterizing the time dependence of polarization
$s_y(t)$. In the following example, for simplicity only, misalignments
of COSY dipole magnets resulting from rotation around the beam axis are
considered. With a known placement of magnets and their individual
misalignments, the distributions of all fields are represented by
Fourier series, with $V_0$, $V_j^c$, $V_j^s$ representing the Fourier
coefficients for vertical field $B_\mathrm{V}(t)$ and $R_0$, $R_j^c$, $R_j^s$
for radial field $B_\mathrm{R}(t)$. Then the solution of the BMT equation for
longitudinal spin component $s_z(t)$ is expressed by those
coefficients and two frequencies: orbital $\omega_\mathrm{o}$ and that of spin
precession, $\omega_s$. Finally, $s_y(t)$ is obtained as the integral
over time of the product $s_z(t)B_\mathrm{V}(t)$ for the EDM effect and of
$s_z(t)B_\mathrm{R}(t)$ for the misalignment effect. For more details of
the derivations see Ref.~\cite{Magiera:2017ypv}.

To illustrate the time pattern for $s_y(t)$, the first leading terms
are presented:
\begin{equation}
\begin{split}
s_y(t)=\frac{\omega_X}{2}\left[X_0\frac{\sin(\omega_st)}{\omega_s}+\sum_{j=1}^{\infty} X_j^c\left[\frac{\sin(j\omega_\mathrm{o}t-\omega_st)}{j\omega_\mathrm{o}-\omega_s}+\frac{\sin(j\omega_\mathrm{o}t+\omega_st)}{j\omega_\mathrm{o}+\omega_s}\right]\right.
\\
+ \left.\sum_{j=1}^{\infty} X_j^s\left[\frac{\cos(j\omega_\mathrm{o}t-\omega_st)}{j\omega_\mathrm{o}-\omega_s}+\frac{\cos(j\omega_\mathrm{o}t+\omega_st)}{j\omega_\mathrm{o}+\omega_s}\right]+
\dotsb \right],
\end{split}
\label{eq:sy_t}
\end{equation}

\noindent where for the misalignment effect $\omega_X=\omega_s$,
$X_0=R_0$, $X_j^c=R_J^c$, $X_j^s=R_J^s$ and for the EDM effect
$\omega_X=D\beta cB_0/\hbar$, $X_0=V_0$, $X_j^c=V_J^c$, $X_j^s=V_J^s$,
where $D$ is the EDM value and $B_0$ the dipole magnet field.

Even though the functional time dependence of both effects is the
same, different values of Fourier coefficients lead to different time
histories for the two effects.  In COSY, as in any storage ring, the
magnets form a regular pattern and the vertical field disorders resulting
from magnet misalignment are small, since they scale with the cosine of the
misalignment angle. Therefore, $V_j^c$ for odd $j$ and all $V_j^s$
coefficients are small. Conversely, the radial field scales with
the sine of the misalignment angle; its distribution is quite random
and all radial field Fourier coefficients have arbitrary values.  This
causes some differences in the time dependence of $s_y(t)$ for the EDM and
the misalignment effects seen in \Fref{sy_signal}. The numerical
results presented in this figure and hereafter are obtained for
$D=4.7\times 10^{-21}~e \, \textrm{cm}$ and the measured COSY
dipole misalignment angles.

\begin{figure} [hb!]
\centering
\includegraphics[width=0.84\textwidth]{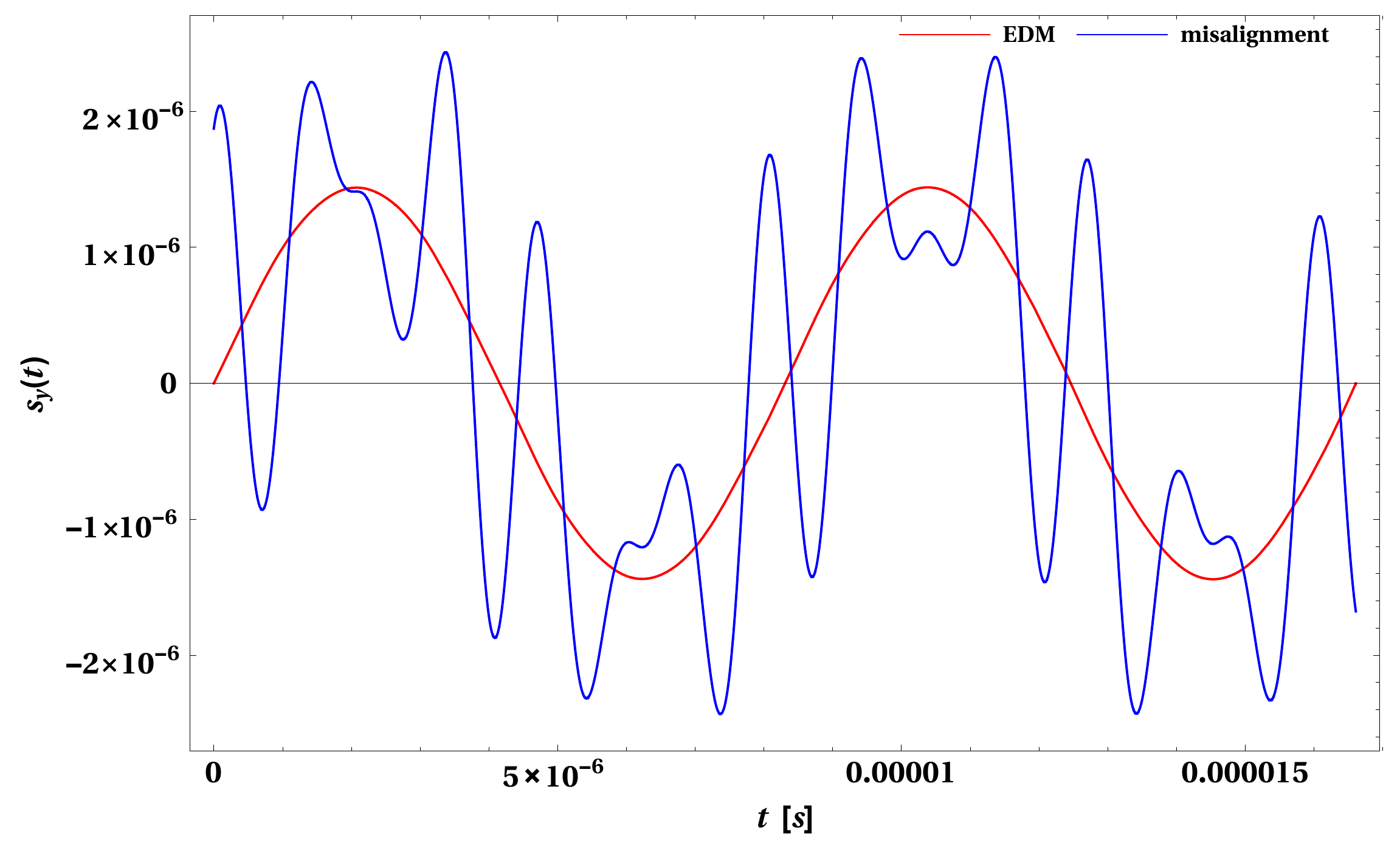}
  \caption{Time dependence of vertical component of polarization
$s_y(t)$ induced by the EDM (red) and magnet misalignment (blue), shown for two periods of spin precession.
\label{sy_signal}}
\end{figure}

The differences in the time dependence of $s_y(t)$ can be quantified
through Fourier analysis of the observed signals. From \Eref{eq:sy_t},
it is seen that Fourier amplitudes for $s_y(t)$ should peak at
frequencies $\omega_s$ and $\omega_\mathrm{o} \pm\omega_s$ (in general,
$j\omega_\mathrm{o}\pm\omega_s$). These maxima can be determined by sampling
(measuring) the vertical polarization with a proper frequency. In
\Fref{fourier_amplitude}, the Fourier amplitudes for $s_y(t)$
sampled with frequencies $\omega_\mathrm{o}$ and $2\omega_\mathrm{o}$ are shown. The first
 corresponds to polarization measurement at one place on the orbit,
while for the second the polarization needs to be measured in two,
reasonably separated, places. It is seen that sampling with $\omega_\mathrm{o}$
is not sufficient to distinguish between the EDM and the misalignment
effects. For the parameters chosen for numerical calculations, the
Fourier amplitudes $F(\omega_s)$ at $\omega_s$ for both effects are
almost the same. Sampling $s_y(t)$ with $2\omega_\mathrm{o}$ frequency,
however, allows observation of a peak in Fourier amplitude
$F(\omega_\mathrm{o}-\omega_s)$ at frequency  $\omega_\mathrm{o}-\omega_s$. In this case,
the amplitude for the EDM effect is  smaller
 by two orders of magnitude than the amplitude for the misalignment effect. Hence, determination of
the $F(\omega_\mathrm{o}-\omega_s)$ amplitude for misalignment effects also
enables 
determination of the magnitude of the amplitude $F(\omega_s)$ for this
effect. Since two polarimeters will be
available for the EDM measurement at COSY, the presented method will allow experimental determination of
the misalignment-related systematic uncertainty in the measured limit
of the EDM value.

\begin{figure}
\centering
\includegraphics[width=0.84\textwidth]{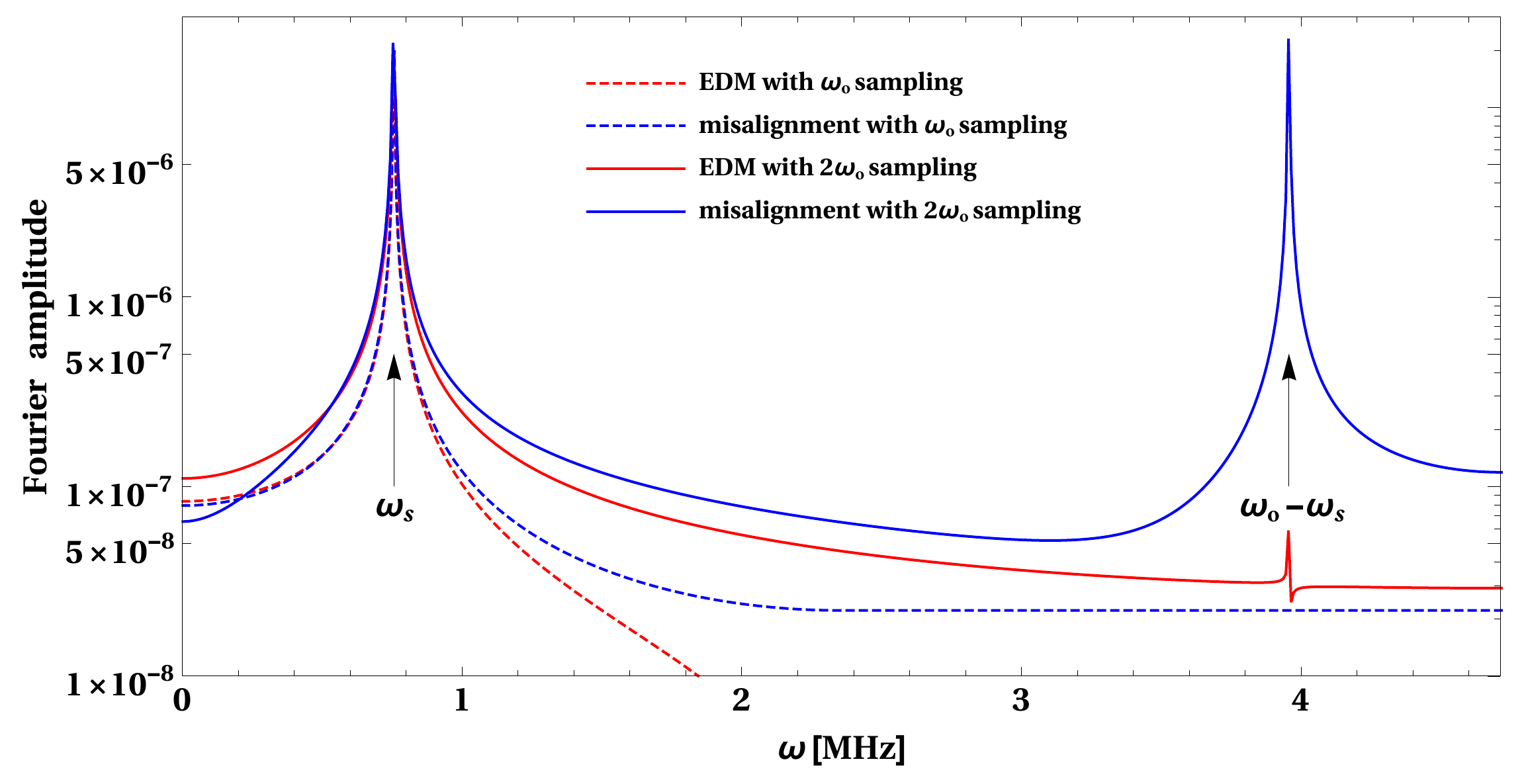}
  \caption{Fourier amplitude of the $s_y(t)$ signal for sampling with frequency
  $\omega_\mathrm{o}$  for EDM (red dashed line) and misalignment effects (blue
dashed line) and for sampling with frequency $2\omega_\mathrm{o}$  for EDM (red solid line) and misalignment effects (blue solid line).
\label{fourier_amplitude}}
\end{figure}

The values of real misalignments of all magnets at COSY are known with rather poor accuracy. In such a case, the presented method allows the probability of occurrence of a certain ratio of Fourier
amplitudes $F(\omega_s)/F(\omega_\mathrm{o}-\omega_s)$ to be calculated. Then, setting a
confidence level, it is possible to determine an upper limit for the
systematic effect contributing to the measured $F(\omega_s)$
amplitude. Since the magnitudes of the Fourier amplitudes for the EDM
effect depend very weakly on magnet misalignments, it is possible to
determine the limit for the EDM value. An example of such an analysis
is shown in \Fref{probability}. The probability distribution of
the ratio $F(\omega_s)/F(\omega_\mathrm{o}-\omega_s)$ was obtained assuming
that the rotation angles of COSY dipoles have a Gaussian distribution with a
standard deviation of $0.01^\circ$.

\begin{figure}[hbt!]
\centering
\includegraphics[width=0.84\textwidth]{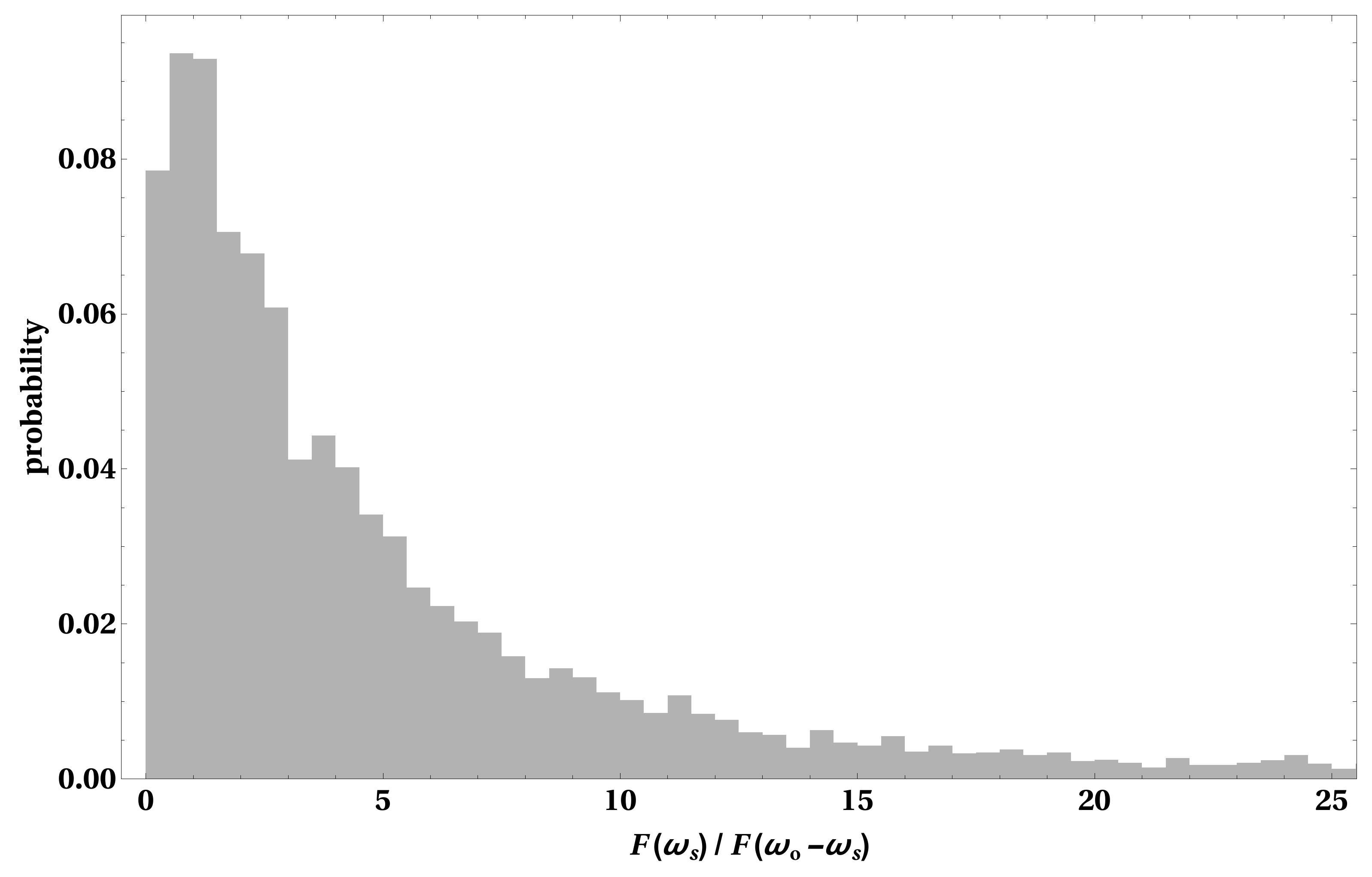}
  \caption{Likelihood of Fourier amplitude ratio
$F(\omega_s)/F(\omega_\mathrm{o}-\omega_s)$, determined for misalignment-induced $s_y(t)$
\label{probability}}
\end{figure}


\renewcommand\bibname{Reference}
\begin{flushleft}

\end{flushleft}
\end{cbunit}

\begin{cbunit}

\chapter[New ideas: sampling polarimeter based on pellet-extracted beam]{New ideas: sampling polarimeter based on pellet-extracted beam\footnote{This appendix was written by Irakli Kesheleashvili (Institut für Kernphysik, Forschungszentrum Jülich) and Richard Talman (Cornell University).
}}
\label{app:extpol}

This appendix describes a pellet extraction
scheme for extracting beam samples from the beam core, rather than from the beam tails
(as had been assumed until now). Though not a new idea itself, pellet beam extraction
has been, until now, very erratic, largely because of the poorly controlled `spray'
of pellet directions. Recent pellet gun developments have made this approach much more
promising. The appendix is largely didactic, collecting formulae needed for the design
of the pellet beam extraction. For EDM, the merit of pellet beam sampling is
the elimination of the need for beam heating to produce the beam tails (with their
dubious lattice function dependence and questionable systematic validity), which enables
internal target polarimetry but cancels stochastic cooling possibilities.
Because the pellets pass approximately through the beam bunch centres, pellet-produced beam
samples will be strongly representative of the true particle distributions (which can be
further monitored by optical tracking of the pellets).

\section{Pellet-extracted beam sampling; qualitative}
\subsection{Successful pellet injector implementation}
Sun \textit{et al.} \cite{8187724} have demonstrated a lithium pellet injector
that can be copied more or less unchanged for the beam sampling requirements
of the EDM experiment. The top part of
\Fref{fig:Improved-LithiumPelletLauncher} (copied
directly from their figure) shows the pellet injector of Sun \textit{et al.}  \cite{8187724}. The bottom
part of the same figure shows the extra focusing (and isolation) stage
needed to send pellets, one by one, through our polarized proton (or other baryon) beam.
This application requires fast lithium pellet microspheres, for
the application of triggering an experimental advanced superconducting tokamak
(EAST). Available pellet speeds range
from 30 to 110\,m/s, ideal for our pellet beam-sample extraction. Our
application requires pellet material with the highest possible charge number $Z$,
for which pellet behaviour is expected to be closely similar.

\begin{figure}
\centering
\includegraphics[scale=0.9]{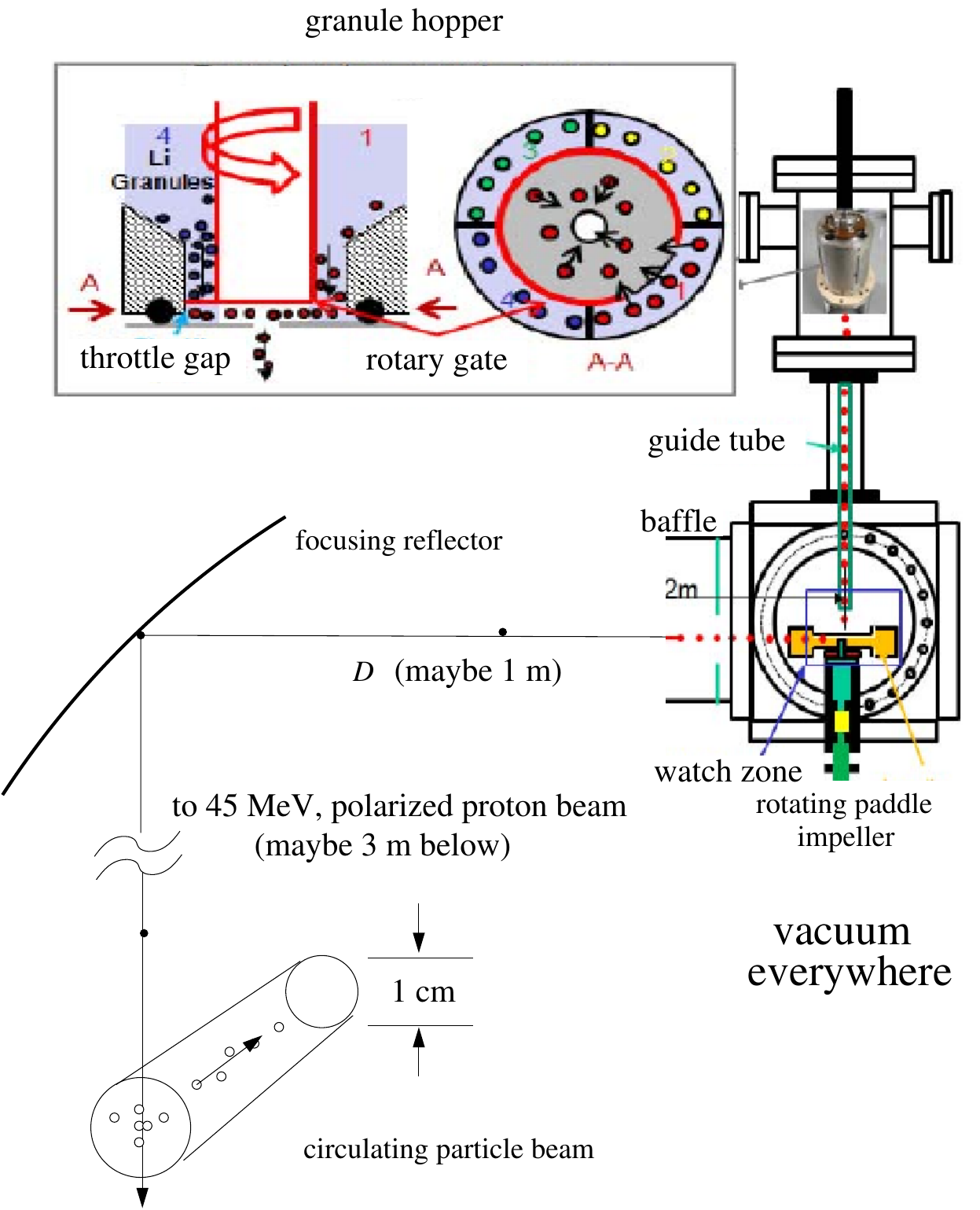}
\caption{\label{fig:Improved-LithiumPelletLauncher}The 
lithium pellet launcher of Sun \textit{et al.} \cite{8187724} adopted for use as the pellet beam sampler
of the EDM prototype storage ring. The ability to switch between four pellet types would
be unnecessary but otherwise the design can just be copied. However, the pellet sizes
needed for the EDM application will be some five times smaller than for the tokamak
triggering application. The apparatus of Sun \textit{et al.}  fed, more than one at a time,
too-small pellets (far smaller than were needed)  but their paper\textit{} \cite{8187724} explains how a single
gap height could be reduced to repair this behaviour. Figure reused with permission from IEEE.
}
\end{figure}

\subsection{General description of pellet-induced beam sample extraction}
The ideal polarimetry for an EDM measurement experiment would be
non-destructive and continuous for hour-long runs, with no beam extraction
sampling required.  However, at present, the only practical form of
polarimetry---left--right asymmetry proton--carbon scattering---consumes
stored particles. One can imagine such scattering polarimetry from an
internal carbon target---for example, from carbon pellets.
It is easy to show that this cannot be practical. A pellet large enough to
have satisfactory polarimetry scattering efficiency will kill the entire
beam within seconds. Beam sample extraction onto a `thick' carbon target
is therefore required---so that the particle can scatter within a thick
external polarimetry target.

It so  happens that the ability just mentioned, of a single pellet to destroy an
entire beam, can actually be exploited to produce very clean and efficient
extraction of controllable samples from the core of a stored proton
(or other baryon) beam. Basically, one person's `suddenly destroyed beam' can be
another person's `efficient slow-extracted beam sampling'. This is illustrated
in \Fref{fig:PelletExtractedPolarimeter}. (The objection to this configuration
for polarimetry, based on the obvious left--right asymmetry of the extraction apparatus,
is to be addressed later.) When a particle in the circulating beam, by chance, passes
through a transitory passing pellet in one straight section, the particle
loses enough momentum that, when it gets to the next straight section, it has
become physically separated from the main beam---\ie it has been `extracted'.

\begin{figure}
\centering
\includegraphics[scale=0.4]{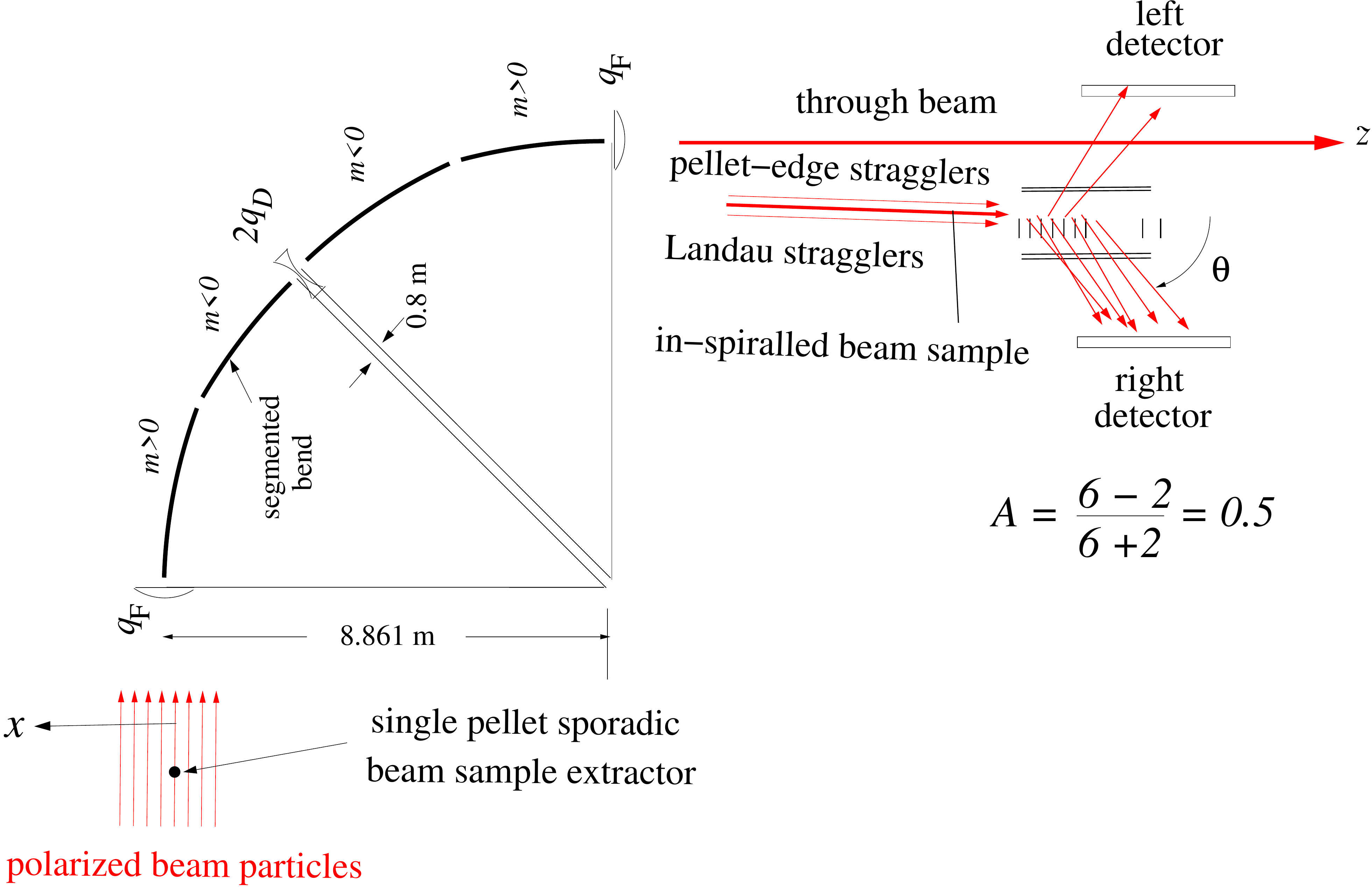}
\caption{\label{fig:PelletExtractedPolarimeter}
Top view of the left--right asymmetry of protons
scattering through angles $\theta$ from the seven graphene foils of
the polarimeter target. (Some polarimeter components are traced from
Fig. 1(a) of Ref.~\cite{IEIRI1987253}.) Short hash marks along the
polarimeter centreline  represent, first, an entry scintillation
counter, followed by seven  polarimeter graphene foil targets
and, finally, two exit scintillation counters. The figure shows how one-quarter
of a ring with horizontal betatron tune $Q_x\approx2$ can act
as a $180^{\circ}$  spectrometer (even though it looks like $90^{\circ}$),
with the point source at the pellet and the polarimeter at the `focus'.
The regular beam focusing also serves to focus the extracted beam. Figure reused with permission from Elsevier.
}
\end{figure}

The most important parameter, for the performance of the sampled beam extraction,
is $\Delta K_\mathrm{p}$ with typical values in the order of $-100 \UkeV$, the kinetic energy change of a
particle (for example, a proton) in its centre passage through a pellet. Approximately
half of the protons hitting the pellet will suffer very nearly this same
energy loss; the rest, because of their more glancing incidence, will suffer
reduced energy loss, from this value all the way down to zero.

For slow protons---for example $45\UMeV$ kinetic energy---the  $\dd E/\dd
x$ stopping
power of protons is large---about seven times minimum ionizing (see
\Fref{fig:DEdx-p-carbon}).
In virtually all cases, the energy loss suffered by a beam particle passing through
any single pellet is far greater than the maximum energy that can be recovered in a
single passage through an RF cavity (should one be encountered along the path).
All such protons will therefore
have been ejected from their stable RF buckets, but their radial positions will not have been
instantaneously  altered in the process. The extracted `beam bunch' duration
will, for example, be about $0.2\Ums$, which is the transit time for velocity $vP$
pellets from entry to exit of the beam bunch. Meanwhile, because of their far greater
velocity $v_\mathrm{p}$, the beam bunches will have made perhaps 100 circulations of the storage ring;
the extracted `bunch' will therefore be made up of 100 `sub-bunches', each of the same
length as the stored bunches, but staggered in time by time intervals equal to the
ring circulation time $T_0\approx 1\Uus$.

\begin{figure}
\centering
\includegraphics[scale=0.5]{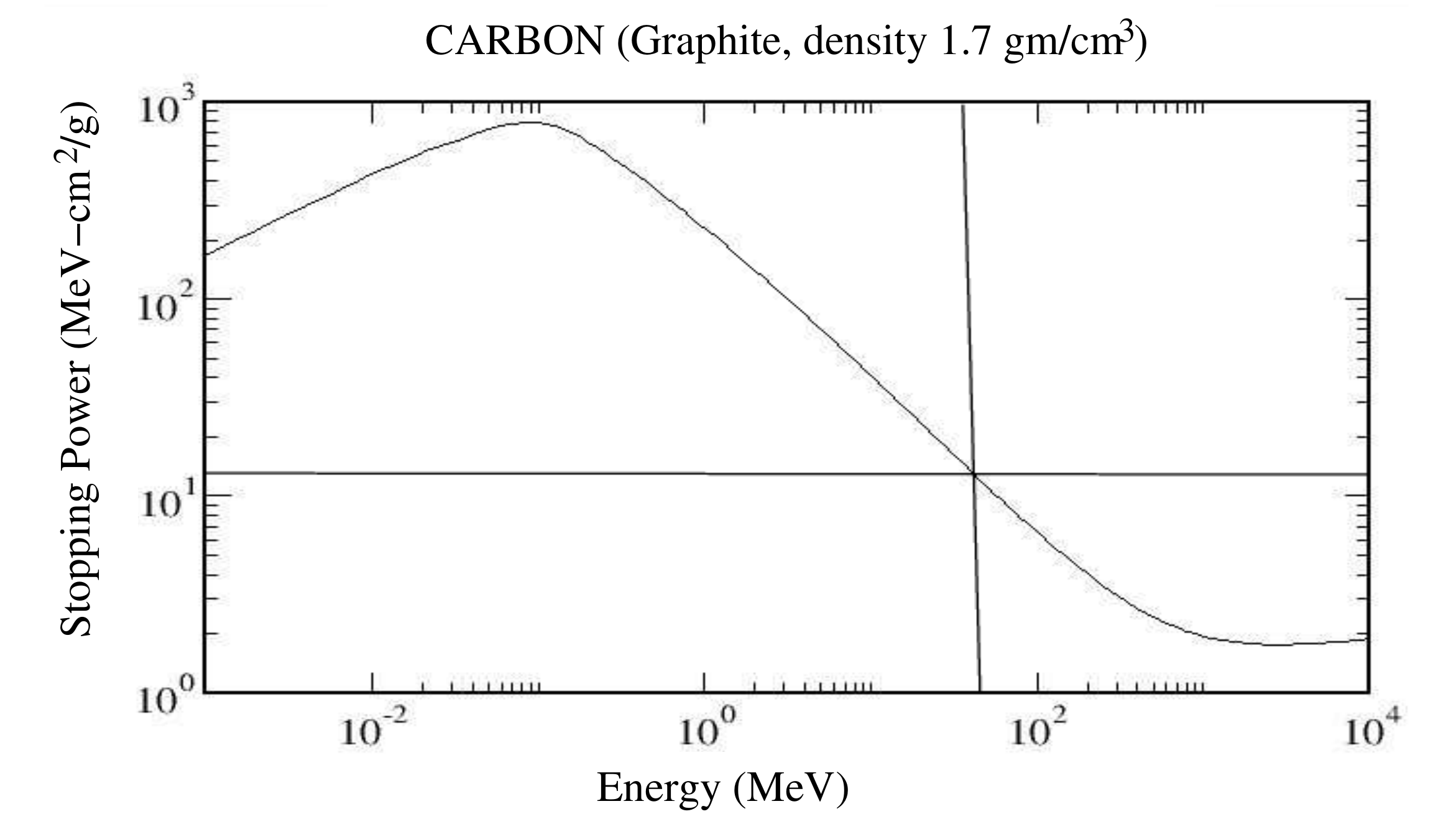}
\caption{\label{fig:DEdx-p-carbon}
NIST Standard Reference Database 124~\cite{NIST124}, $\dd E/\dd x$
for protons incident on carbon, with cross adjusted for $45\UMeV$.
In choosing from different pellet materials, the ratio of $\dd E/\dd x$
at $45\UMeV$ to the ionization minimum will be more or less independent
of the pellet medium, because the beam particle velocity is being held
constant. The figure is NIST open source data.
}
\end{figure}

Apart from this spreading in time, the beam being extracted is still a pencil beam
emerging from a point source. However, most of these protons have off-momentum values near
$\Delta p/p= -0.001$. At a point in the ring with dispersion $D=10\Um$,
these about-to-be extracted protons are initially displaced
from their nominal off-momentum closed orbits by about 1\,cm.
Interpreted as a betatron amplitude, this is almost
twice the nominal beam bunch radius. After a horizontal betatron phase advance of
$\pi$, their radial betatron displacements will be reversed to $-1\Ucm$,
relative to a nominal orbit that is, itself, also displaced by $-1\Ucm$.
As a result, the transverse separation of extracted bunch relative to stored bunch
is about $2\Ucm$.

The extracted beam particles, though all starting from the same point source,
also `remember' their initial betatron slope amplitudes. Downstream, the
extracted sub-bunch transverse particle displacements
(from their appropriately reduced off-momentum closed orbit)
will be approximately the same as those of the co-travelling bunch from which
they were extracted. The separation of stored beam and extracted beam bunches
may be about four times the nominal bunch radius. This is what can pass
as `clean' slow beam sample extraction. (It is not unlike ion-stripping
injection, in which Liouville's law is foiled by a sudden change in
particle rigidity.)

There will also be multiple scattering, suffered by each extracted
proton in its passage through the pellet---for example
$\theta_{\rm r.m.s.}=\pm 2$\,mrad for this angular deflection. Though not a
small angle, at least the core of the extracted bunch remains within
the radial acceptance of the ring, both horizontally and vertically.
The extracted beam will be broadened
somewhat, and acquire transverse tails from this source. It just so happens, though,
that the same horizontal phase advance that doubles the extracted beam separation
also refocuses multiply scattered protons back to a point focus at the
polarimeter scattering target.

In short, when observed at the polarimeter in the next straight section,
most of the protons that have touched the carbon pellet will have been
slowly extracted into a bunch of much the same dimensions as the original bunch,
somewhat broadened, but mainly displaced by $2\Ucm$ from the circulating beam.
A noticeable exception to this analysis concerns protons that have barely
grazed the pellet. Though almost certainly extracted from their stable buckets,
these protons can decohere and form a more-or-less stable coasting beam of
reduced radius, but surely at the percentage level, at most. Though not welcome,
such protons should have an acceptably small effect on the EDM
measurement---to be worried about later.

Suggested starting parameters for an EDM experiment then are: number of stored protons, $10^{10}$;  number of protons
extracted by the first pellet, 25\,million; and total number of pellets, 400
(irrespective of the run length) The pellet
material should have the highest atomic number $Z$ available, with
radius $20\Uum$. However, all parameters mentioned so far apply
only to the starting beam conditions. As the beam intensity falls, say by a factor
of two, to maintain the extracted beam flux it will be necessary for the pellet rate to double.
So the total number of pellets will be larger than has been stated so far. By
controlling the rate at which pellets are launched, the beam attenuation pattern
can be made linear, or whatever is most favourable.

Making, for example, the assumption that the very first pellet is
launched into a beam of $10^{10}$ protons, and the (unduly optimistic) assumption
of 100\% extraction efficiency, the number of extracted beam protons
through the polarimeter from just one pellet will be 25 million. Using
detailed cross-section values copied unchanged from the (invaluable) paper
of M. Ieiri \textit{et al.} \cite{IEIRI1987253}, the polarimeter efficiency is calculated to be $0.00034$,
with analysing power $A_{\rm pol.}=0.78$. From the first pellet, we therefore
anticipate 5500 total polarimeter counts, with $4800\pm70$
scattering to the right (predictably, since we assume the proton beam is
100\% vertically polarized) and $700\pm25$ scattering to the left. This
would produce (statistically) a better than 2\%\ r.m.s. beam
polarization measurement.

\section{Experimental confirmation of wire and pellet beam extraction
 at COSY}
\subsection{Previous moving wire investigations}
Though the pellet extraction of small beam samples from the centre of
a beam bunch has not yet been demonstrated, nor the high quality of the extracted
beam quantitatively confirmed, the concept has, itself, been confirmed
experimentally, as show in \Fref{fig:PelletExtractedPolarimeter2}. In
this test by Keshelashvili \textit{et al.}\cite{Irakli}, a stretched $10\Uum$ carbon fibre
was passed, suddenly and repeatedly, 10 times through a stored COSY beam in
order to show that the basic considerations given here are correct. The
top oscilloscope picture indicates the resulting synchronous counting
rate bursts in counters of the EDDA polarimeter. The bottom figure shows the
beam intensity being reduced in a staircase-like fashion.

\begin{figure} [hb!]
\centering
\includegraphics[scale=0.13]{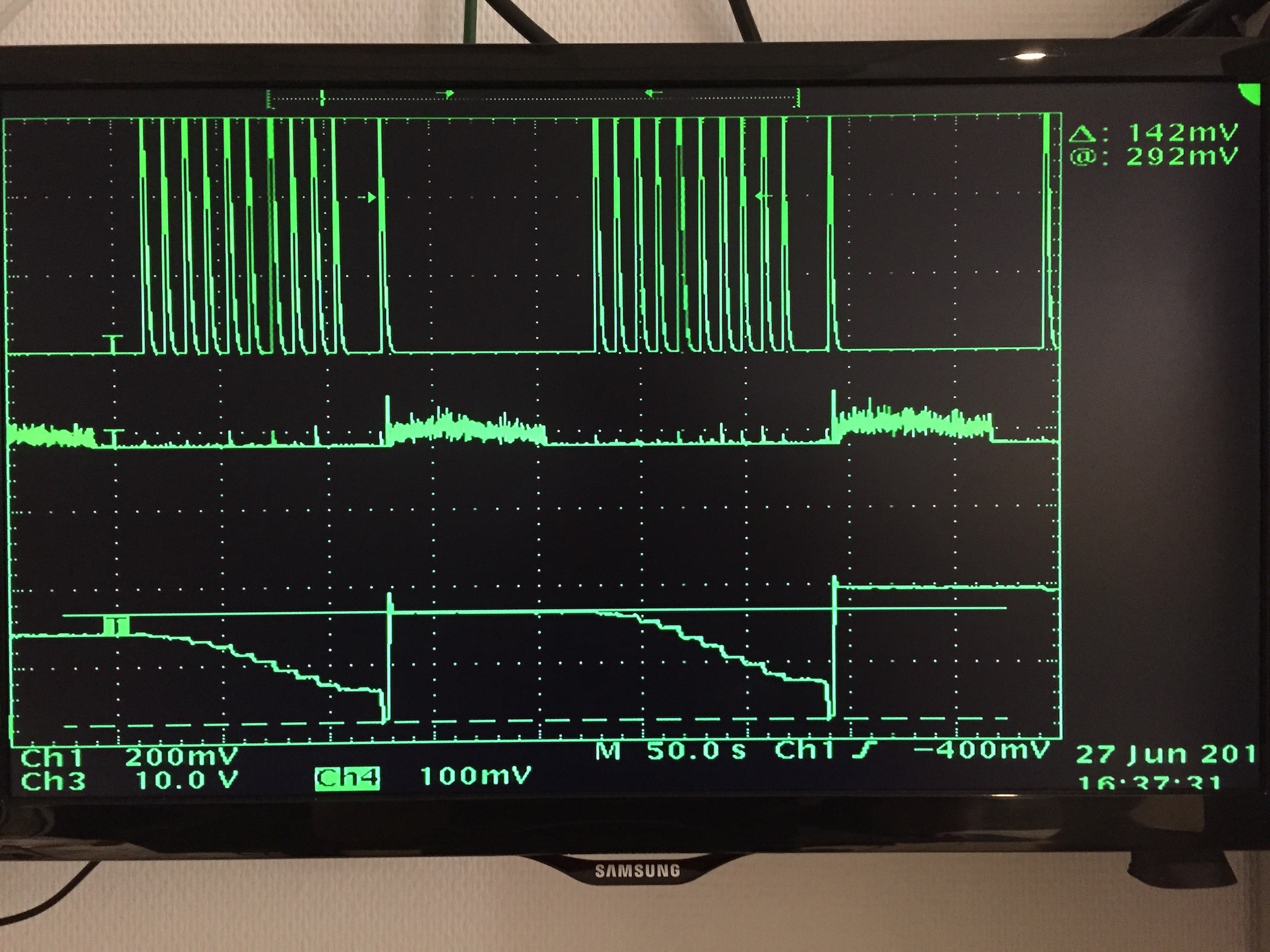}
\caption{\label{fig:PelletExtractedPolarimeter2}Results of an experimental
investigation, by I. Keshelashvili \textit{et al.}\cite{Irakli}
of the interaction of a $10\Uum$ radius carbon wire with the COSY beam for two
consecutive cycles: (top) rate in a detector; (bottom) stepwise reduction of beam intensity for each beam crossing of
the wire.
}
\end{figure}

By reducing the target dimensionality from 2D to 1D, the concept
of beam sampling has been confirmed. But, while a $10\Uum$ carbon fibre
may seem hardly intrusive, the beam attenuation per wire transit is still
three orders of magnitude too great for the intended application. The need
for further dimensionality reduction from 1D to 0D---wire to point---seems
inescapable. The proposed pellets, with radii three orders of magnitude
less than the circulating beam transverse area will provide this needed factor.
Furthermore, the possible performance degradation by electrostatic charging
of an insulator in a beam has been shown to be unimportant, at least for
a wire.

\subsection{Pellet formulation applied to moving wire investigation}
Later in this section, Eq.~(\ref{eq:opacity.1}) is derived, giving the
opacity $O_\mathrm{Bp}$ of a moving pellet.
Here, `opacity' is the fraction of the circulating beam particles that touch a
single pellet (typically over many beam turns) during a single pellet transit.
Here, we simply copy this formula, with minor modification, to give $O^\mathrm{W}_\mathrm{Bp}$,
which is a crude approximation of the opacity of a single transit of a moving wire.
The result is
\begin{equation}
O^\mathrm{W}_\mathrm{Bp} \approx \left(\frac{r_\mathrm{W}}{r_\mathrm{B}^{\perp}}\right)\,\frac{2r_\mathrm{B}^{\perp}/v_\mathrm{W}}{\mathcal{C}/v_\mathrm{p}}
         = \frac{r_\mathrm{B}^{\perp}}{r_\mathrm{W}}\,O_\mathrm{Bp} \, .
\label{eq:opacity-wire.1}
\end{equation}
By the replacement P$\rightarrow$W, pellet radius $r_\mathrm{P}$ becomes wire radius $r_\mathrm{W}$,
pellet transit time $t_\mathrm{P}$ becomes wire transit time $t_\mathrm{W}$, pellet material density
$\rho_\mathrm{P}$ becomes wire material density $\rho_\mathrm{W}$, and pellet velocity $v_\mathrm{P}$ becomes
wire velocity $v_\mathrm{W}$; (of these, only  $r_\mathrm{W}$ and $v_\mathrm{W}$ appear in
\Eref{eq:opacity-wire.1}). Apart from these, purely symbolic,
changes, the only change has been to multiply the pellet opacity by a (large) multiplicative
factor $r_\mathrm{B}^{\perp}/r_\mathrm{W}$. The inclusion of this factor amounts to visualizing the moving
wire as being made up of a (large) number $r_\mathrm{B}^{\perp}/r_\mathrm{W}$, of length $r_\mathrm{W}$, radius $r_\mathrm{W}$
cylindrical pellets stacked end to end. For $r_\mathrm{W} =r_\mathrm{P} =10\Uum$ and $r_\mathrm{B}^{\perp}=1\Ucm$,
the wire opacity is 1000 times greater than the pellet opacity.

As an aside, it can be commented that it is the large factor $r_\mathrm{B}^{\perp}/r_\mathrm{W}$
that makes pellets so much more satisfactory than wires for bunch sample extraction.
But this factor does not impede our purpose here,
of experimentally confirming the moving pellet formalism using moving wire
experimentation.

\section{Reinterpretation and revision of COSY moving wire beam experiments}
The COSY experience with beam sampling by moving an obstacle rapidly through
a circulating beam is summarized in \Fref{fig:PelletExtractedPolarimeter2},
and can be characterized by two qualitative features: the staircase-like reduction
of beam current in equal steps, synchronous with transits of a moving wire, and the
further detection of similarly synchronous bursts of radiation in nearby counters
of the EDDA polarimeter. The constant downward beam current steps prove that
beam particles are hitting the moving carbon wire; the local EDDA counter radiation
bursts suggest that the extracted beam energy is dissipated locally.

The former conclusion is incontrovertible, but the latter is not. It is our understanding
that the EDDA counters are not sensitive to small angle particles less than $10^{\circ}$
 or so. Yet the dominant contribution to the total cross-section for
high-energy charged particles incident on very thin targets is multiple scattering
at angles much less than $10^{\circ}$. This appendix therefore assumes that scattered
beam particles are not contributing significantly to the EDDA signals. This,
and other contentions of this appendix, can be tested experimentally using
existing COSY moving wire apparatus, either with or without new instrumentation.

\subsection{Moving wire investigation without new instrumentation---ready immediately\ }
The simplest suggested experiment is to replace the $10\Uum$ carbon wire with a
$10\Uum$ tungsten wire. According to
\Eref{eq:opacity-wire.1}, the moving wire opacity $O^\mathrm{W}_\mathrm{Bp}$ is independent
of the wire medium density $\rho_\mathrm{W}$. In our model, every particle that touches
a pellet is extracted, irrespective of the wire medium. The switch from carbon to
tungsten wires should therefore not significantly affect the stepwise reduction of
beam current shown in the bottom oscilloscope trace in
\Fref{fig:PelletExtractedPolarimeter2}. Conversely, the local, large-angle radiation should be roughly proportional to the wire medium density. The effect
of switching from carbon to tungsten should therefore increase the ratio of EDDA
counts/pellet to beam current loss/pellet by an order of magnitude.

\subsection{Moving wire investigation with new instrumentation---ready in a few months\ }
The proposed test without
new instrumentation is a significant consistency test, but it does not confirm
our contention that the extracted beam particles can be conveyed with
significantly large efficiency onto a carbon polarimeter scattering target.
What is needed, for example, is a downstream phosphor screen, or other radiation-sensitive imaging device. Judicially placed in the lattice, such an imaging
device can determine, at least approximately, the angular distribution of beam
particles scattered (at small angles) from the moving wire.

The choice of a high-$Z$ medium, such as tungsten, for the moving wire  significantly helps any
such investigation. The sudden betatron amplitude discontinuity,
$\Delta x_{\beta}$, derived later in this appendix, is given by
Eq.~(\ref{eq:SlowingDownKinematics.4-bisp}), which needs only the
symbol conversion $\rho_\mathrm{P}\rightarrow\rho_\mathrm{W}$.

The switch from carbon to tungsten increases $\Delta x_{\beta}$ by an order of
magnitude. Though the dispersion function at the moving wire is, presumably,
more or less fixed, the displacement of the extracted beam is also proportional
to $D_\mathrm{p}$ at the screen location. In the COSY lattice, there are natural high
dispersion points (of order $10\Um$ in the straight sections at arc centres).
It seems natural to consider putting the extracted beam screen at one or other of these points. This is still not enough, though. It is also most
favourable for the horizontal phase advance to be an odd multiple of $\pi$.
To complete even a preliminary design, the true COSY lattice functions must be known, and preferably be tuneable, to optimize the extracted beam
separation.

\subsection{Full demonstration and calibration of pellet extraction---$2+1$\,years}

A pellet extraction test set-up is shown in
\Fref{fig:Yello_Report_Cyclotron}. Using a COSY laboratory cyclotron
(or equivalent spectrometer at any laboratory) a $45\UMeV$
proton beam can be used to confirm, optimize, and calibrate
pellet beam sample extraction. The magnetic spectrometer
mimics one-quarter of the EDM prototype ring. Inset phosphor screen images show
anticipated charge distributions,
with and without the pellet contribution. Charge densities are crudely represented
by greyscale shading. The dark elliptical region is the image of the main beam.
The broken-line rectangle indicates a satisfactory placement
region for an external polarimeter target.  Because the less strongly deflected
intensity overlaps the main beam, the on-target extraction efficiency must be
at least somewhat less than 100\%.

\begin{figure} [hb!]
\centering
\includegraphics[width=0.5\linewidth]{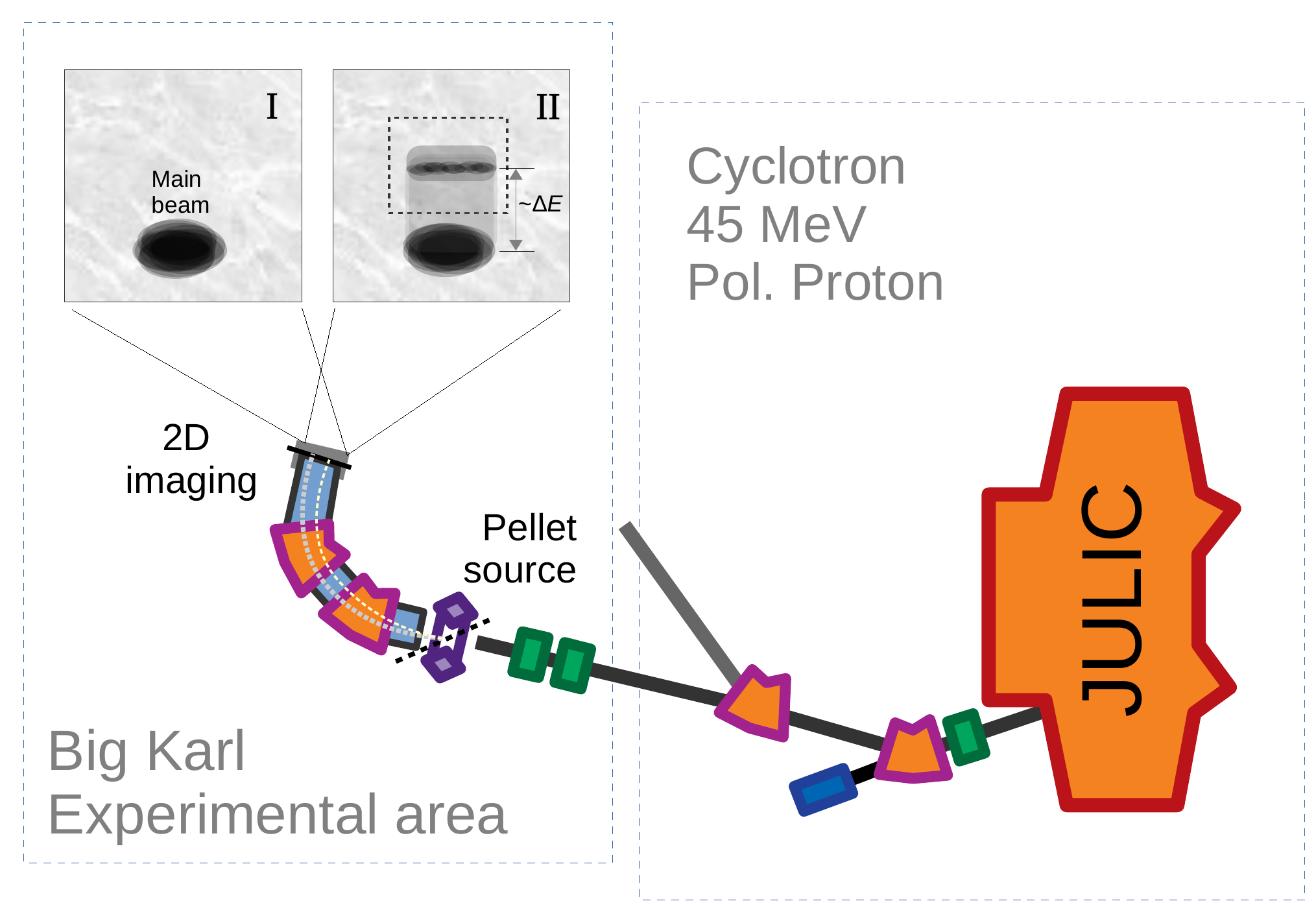}
\caption{\label{fig:Yello_Report_Cyclotron}Practical pellet beam sample
extraction test set-up  }
\end{figure}

\section{Quantitative formulation of pellet beam sampling}
It is necessary to establish many parameters for pellet beam sample extraction.
Symbol definitions for the various parameters and kinematic quantities are given
in \Tref{tbl:Parameter definitions}. Fortunately pellets are `everywhere'
these days, and accurate
microspheres  in a wide range of materials can be acquired. Parameters for materials that seem to be
especially promising are given in \Tref{tbl:PelletMaterialProperties}.
It is not our purpose to determine the parameters with high accuracy. Rather,
the initial purpose is to acquire a sufficiently quantitative understanding
of the relative advantages of low-$Z$ versus high-$Z$ materials. (Surprisingly,
it seems that high-$Z$ pellets are more favourable for our application.)

\begin{table}
\caption{\label{tbl:Parameter definitions} Definitions for  various
parameters and kinematic quantities; $m_\mathrm{p}$ is the proton mass (which is approximately
equal to the a.m.u.).}
\centering
\resizebox{\columnwidth}{!}{%
\begin{tabular}{lll} \hline \hline
 Symbol             &  Definition (MKS unit in formulae, but MeV energies)    & Unit in tables    \\ \hline
     P              & Pellet                                                   &                    \\
 p        & Beam particle (proton, deuteron or helion, not electron) &                    \\
  B        & Beam (of particles)                                      &                    \\ 
$Z_\mathrm{P}$             & Pellet material charge                      &                    \\
$A_\mathrm{P}$             & Pellet mass  number                      &                    \\
                  $\rho_\mathrm{P}$            & Mass density of pellet material                          &  g/cm${}^3$       \\
$n_\mathrm{P}$                & Number density of atoms in pellet material               &  ${\rm cm}{}^{-3}$  \\
$n_\mathrm{e}$               & Electron number density in pellet material               &  ${\rm cm}{}^{-3}$  \\
$X_\mathrm{P}$                & Pellet material radiation `length' (\ie multiplied
by  density) & g/cm${}^2$       \\
$N_\mathrm{P}$                & Number of atoms in a pellet                              &                    \\
$M_\mathrm{P} \approx N_\mathrm{P}\ A_\mathrm{P}\ m_\mathrm{P}$ & Pellet mass                                         & g                 \\
$r_\mathrm{P}$                & Radius of pellet microsphere                             &  \!$\Uum$            \\
$v_\mathrm{P}$                & Speed of pellet                                          &  $\lesssim\,100$\,m/s   \\
$t_\mathrm{P}=2\ r_\mathrm{P}\ \rho_\mathrm{P}$ & Target `thickness' of pellet microsphere    &  g/cm${}^2$       \\ 
$N_\mathrm{p}$               & Total number of stored beam particles                    &  $\lesssim\,10^{10}$    \\
$N_{\rm extr.}$       & Number of beam particles extracted by a single pellet    &   $\sim$\,$10^6$      \\
$r^{\perp}_{\rm B}$   &  Transverse radius of (circular) particle beam           &  cm                \\
$\mathcal{C}$       & Circumference of storage ring                            &  m                 \\
$v_\mathrm{p}$               & Velocity of beam particle                                &  m/s               \\
$\eta_\mathrm{p}(v_\mathrm{p})$ & Slowing-down enhancement factor (relative to minimum ionizing) & $\sim$\,7            \\
$K_\mathrm{p}$               & Kinetic energy of beam particle                          &  m/s                \\
$cp_\mathrm{p}$              & Beam particle momentum (expressed in energy units)       &  MeV                \\
$T_0=\mathcal{C}/v_\mathrm{p}$ & Beam revolution period                                 &  s                  \\
$T_\mathrm{P}=2\ r^{\perp}_{\rm B}/v_\mathrm{P}$ & Pellet transit time through beam                  &  s            \\ 
$O_\mathrm{Bp}$            & `Opacity' of one pellet transit to beam particles       &                     \\
$O^\mathrm{W}_\mathrm{Bp}$          & `Opacity' of one wire transit to beam particles         &                     \\
$\Delta K_\mathrm{p}$        & Ionization energy loss, particle through pellet centre   &  MeV                \\
$\delta_\mathrm{p}=\Delta p_\mathrm{p}/p_\mathrm{p}$ & Corresponding fractional momentum loss of particle  &                     \\
\hline \hline
\end{tabular}
}
\end{table}

\begin{table}
\caption{\label{tbl:PelletMaterialProperties}Material properties of high-quality available
microsphere pellets. They should be hydrogen-free, which rules out plastics. All materials are
crudely treated as single-element metals; quartz treated as silicon, sapphire (aluminium oxide)
as aluminium, stainless steel as iron. Plausible coefficients of restitution ($C_\mathrm{r}$)  are given
in the final column. It must be realized, though, that even though treating, for example,
sapphire as aluminium
may be crudely valid for calculating slowing down and multiple scattering of relativistic particles,
it is not at all a sensible approximation for determining coefficients of restitution\cite{Jackson2010}.
The value given for tungsten, though the result of an actual experiment\cite{Falcon1998}, applies to
bouncing, for which the pellet velocity is much less than we require.}
\centering
\resizebox{\columnwidth}{!}{%
\begin{tabular}{llllllllll} \hline \hline
 Element   & $Z_\mathrm{P}$      &  $A_\mathrm{P}$  & $Z/A$ & $\ln(287/Z)$
 & $Z(Z+1)$ &  $\rho_\mathrm{P}$     &  $X_\mathrm{P}$      & $\dd E/\dd
 x|_{\min}$ & $C_\mathrm{r}$ \\
           &           &        &       &                &        &  g/cm${}^3$  & g/cm${}^2$ & MeV/(g/cm)${}^2$     &         \\ \hline
 Lithium   &   \phantom{1}3       &    \phantom{18}7   & 0.428 &   4.56         &   \phantom{55}12   &    \phantom{1}0.534      & 82.8        &  1.639               & $\sim$\,0.35  \\
 Carbon    &  \phantom{1}6       &   \phantom{1}12   & 0.500 &   3.87         &   \phantom{55}42   &    \phantom{1}2.26       & 42.7        &  1.745               &    \\
 Aluminium  &   13      &   \phantom{1}27   & 0.481 &   3.09         &  \phantom{5}182   &    \phantom{1}2.70       & 24.0       &  1.615               &   \\
 Silicon   &   14      &   \phantom{1}28   & 0.500 &   3.02         &  \phantom{5}210   &    \phantom{1}2.33       & 21.8       &  1.664               & $\sim$\,0.5   \\
 Iron      &   26      &   \phantom{1}58   & 0.448 &   2.40         &  \phantom{5}702   &    \phantom{1}7.87       & 13.8       &  1.451               &   \\
 Tungsten, \etc? & 74  &  184   & 0.402 &   1.36         & 5550   &   19.30       &  \phantom{1}6.76      &  1.145               &  0.97?\cite{Falcon1998} \\
 Gold?    &   79      &  197   & 0.401 &   1.29         & 6320   &   19.32       &  \phantom{1}6.46      &  1.134               &   \\
\hline \hline
\end{tabular}
}
\end{table}

In spite of the ubiquitous availability of high-quality plastic pellets, we
have ruled out all organic materials, because their hydrogen content has
the potential to harm the vacuum. This mainly leaves pure elements, metals, and
ceramics. To simplify the analysis, we pretend that ceramics can be approximated as
pure single-element metals, quartz as silicon, sapphire as aluminium, \etc
\Tref{tbl:PelletMaterialProperties} gives physical properties of an
incomplete list of satisfactory and available pellet materials limited in
this way. There are many possibilities. The main deficiency in the list is
the absence of a really high-$Z$ pellet material, as indicated by question
marks in the table.  If no such pellets exist, it can only be that there
has, as yet, been no commercial application requiring high-$Z$ pellets.

As well as being needed to analyse the kinematics of pellet acceleration, which
is entirely describable by classical and statistical mechanics,
physical properties are also needed  to calculate the slowing down
by ionization loss, as well as the multiple Coulomb scattering of any beam particle that
happens to find itself within the material of a pellet.

We picture our pellet bulk material as being in the condensed liquid state of particles
that would `evaporate' to form an ideal gas if only they could be heated to
a sufficiently high temperature without burning or melting---\emph{which is not even
close to possible.} The requirement to extract one pellet at a time from a fluid
of pellets is the main technical challenge in shooting pellets, one by one,
through our particle beam.  Fortunately, the apparatus of Su \textit{et al.}   \cite{8187724} shown in
\Fref{fig:Improved-LithiumPelletLauncher} shows that it is possible to produce
a reasonably well controllable pellet gun source with the parameters we need.

Ideally, we could dial up our pellet gun, on demand, to deliver exactly one pellet
with an exact speed and direction. In practice, this is unrealistic,
since, once the pellet fluid medium has been shaken enough to make ejecting pellets
one at a time possible, their momentum vectors will have much the same distributions
and uncertainties as given by the Rayleigh--Maxwell distribution of ideal gas molecules.

Fortunately, for our application, the pellet beam requirements are not strict.
The required average pellet rate will be of the order of  $1\UHz$, but the arrival times
can be Poisson-distributed in time. Also, the pellet beam width need only be
comparable with particle beam transverse dimensions of the order of $1\Ucm$.

\section{Derivations of required formulae}
\subsection{Popcorn analogy}
When cooking popcorn on a stove top, the kernels, when they pop, supply enough energy to
stir things up enough to require the sauce pan lid to be kept on. But this also prevents
steam from escaping, which can make the popcorn soggy.  As a compromise, one can leave the top
slightly ajar. As a result, every once in a while, an unpopped kernel comes flying out
through the opening between pan and lid. Voila! A source of fast corn pellets.

In our application, we do not have popping kernels, and it is not even thinkable to supply
enough heat to stir up the pellets thermally---they are far too heavy. We need a moving
`impeller' to `evaporate' the pellets into a `vapour'.  This necessarily makes the
momentum of each particle uncertain, with a Maxwell-like distribution of velocities.  In a
laboratory-scale enclosed vessel with transverse dimensions of order $\ell$, there are
enough micropellets to run all EDM experiments for centuries (if none is wasted).

Say, therefore, that the volume of pellet material is less than the vessel volume by
a factor of 1000, with pellets all sitting, condensed, at the bottom of the vessel.
Some sort of agitator can, however, stir up the pellets enough that any individual pellet
of mass $m$, with gravitational acceleration $mg$, can have acquired a kinetic energy
$mv^2/2$ of order $mg\ell$, enough to have a significant probability of being, for example,
in the top half of the vessel. This establishes a velocity $v\sim\sqrt{g\ell}$,
independent of $m$, which the agitator has to apply randomly to the pellets, in order
for at least some of the pellets to behave like a gas.

This has set a lower limit requirement for the impeller velocity. But this limit is far
lower than the pellet velocity we require. We could, in response, use a much faster impeller.
But this would be a mistake, since this would introduce large and unmanageable transverse
velocities. As always in accelerators, we should start with a low-energy injector, before
applying exclusively longitudinal acceleration. We therefore need two impellers, one to
jiggle pellets free, and another to accelerate individual pellets to `high' speed,
$v_\mathrm{P} \sim 100\,$m/s.

In the apparatus of Sun et al. shown in Figure~\ref{fig:Improved-LithiumPelletLauncher}, the
initial agitation is supplied by the oscillating piezoelectric element, coloured
purple in the figure, and the secondary acceleration is provided by the
rotating `paddle impeller', coloured orange in the figure. (As mentioned earlier, the
capability to switch pellet sizes---indicated by the large red open arrow---is superfluous
for our application.)

\subsection{Pellet acceleration by rotating paddle impeller}
When a micropellet approaches a (not necessarily made of the pellet
material) flat surface at rest at right angles, with momentum $p_{\rm inc.}$,
the pellet bounces with momentum $p_{\rm refl.}$.
The coefficient of restitution\cite{Brach-Dunn} is
defined as the ratio of these momenta:
\begin{equation}
C_\mathrm{r}(v_{\rm paddle}) = \frac{p_{\rm refl.}}{p_{\rm inc.}},
\label{eq:CR.1}
\end{equation}
which is a number in the range 0--1 that depends on the
pellet velocity, and on the pellet and surface materials.
(The notation here is a little garbled; $C_\mathrm{r}(v_{\rm paddle})$ depends on
the paddle speed from which $p_{\rm inc.}$ acquires its value in the
paddle rest frame, and $p_{\rm refl.}$ inherits the same velocity
in its transformation to the laboratory.)
In our case, the flat surface is a paddle, far more massive than
the pellet, and moving in the laboratory with velocity $v_{\rm paddle}$.
In this case, the pellet recoils with velocity
\begin{equation}
v_\mathrm{P} = v_{\rm paddle}(1+C_\mathrm{r}(v_{\rm paddle})),
\label{eq:CR.2}
\end{equation}
which can be as large as $2v_{\rm paddle}$. The pellet will lose
some of its speed in the reflection from the spherical focusing `mirror'.
It will also acquire angular velocity (which will have no significant
effect on the subsequent circulating beam sampling).

\Figure[b]~\ref{fig:SapphireCR} shows the velocity dependence of
sapphire pellets incident on aluminium.  Coefficients of
restitution  for a few possible pellet materials are also
given in \Tref{tbl:PelletMaterialProperties}.

\begin{figure}
\centering
\includegraphics[scale=0.6]{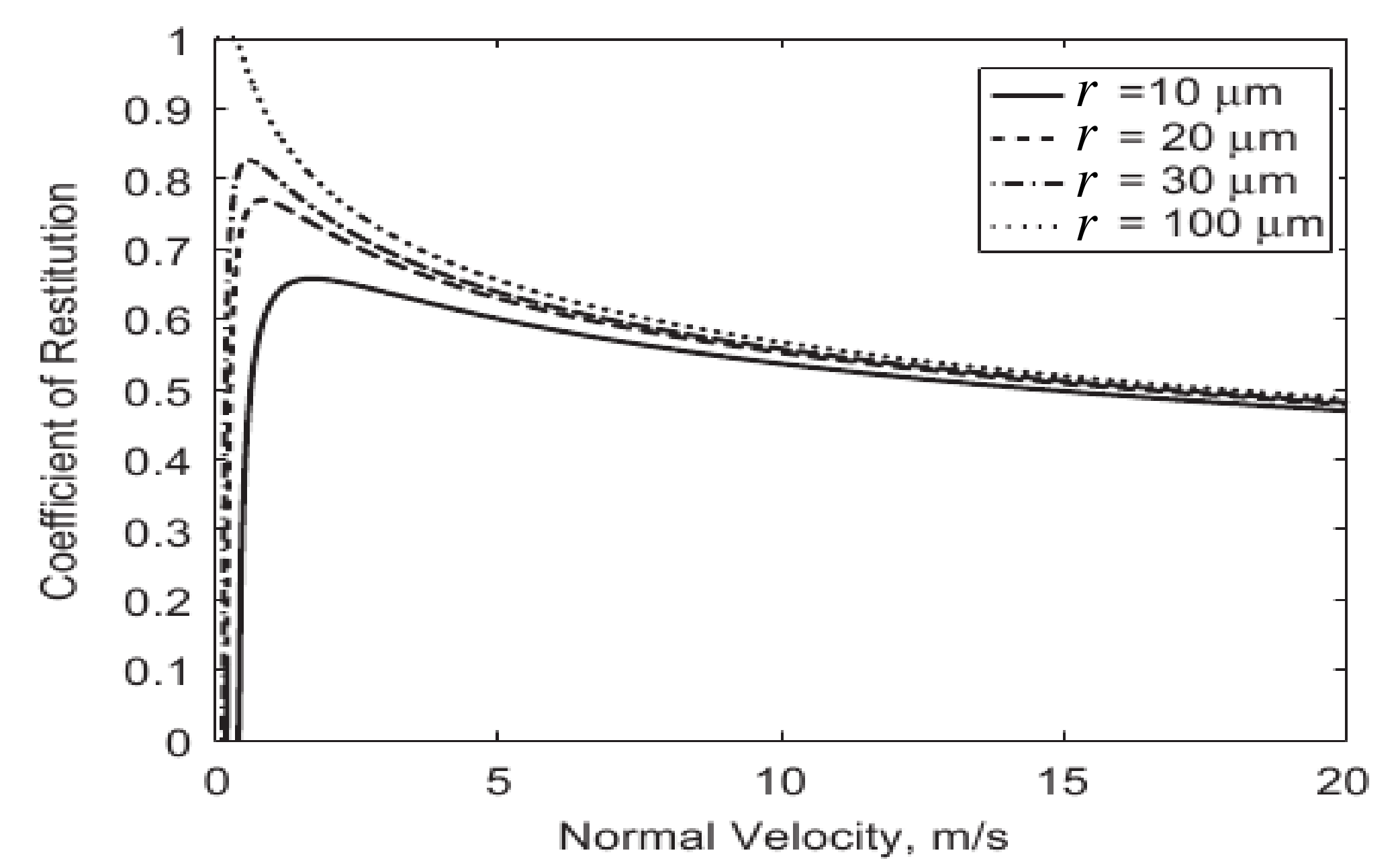}
\caption{\label{fig:SapphireCR}Dependence of coefficient of restitution for
aluminium oxide (sapphire) normally incident on aluminium\cite{Kim-Dunn}. The paddle and pellet media are taken to be identical.
(This is not really legitimate for sapphire on aluminium, since aluminium is
less rigid than sapphire.) Figure reused with permission from Elsevier.}
\end{figure}

\subsection{`Opacity' $O_\mathrm{Bp}$ of a pellet to beam particles}
Our storage ring of circumference $\mathcal C$ has some $N_\mathrm{p}\approx10^{10}$ particles
circulating with period $T_0$ at speed $v_\mathrm{p}$, with very small fractional momentum spread
$\delta_\mathrm{B}\sim 10^{-4}$,
in a beam with circular cross-section of radius $r^{\perp}_\mathrm{B}$. Because a pellet is quite
small, and is moving quickly, it is unlikely for any particular beam particle to come
close enough to a pellet to be affected. In fact, this is a very sharp distinction;
a particle either hits a pellet or it does not. Assuming that the beam is distributed uniformly
in a circle of radius $r_\mathrm{B}^{\perp}$, in a single passage the probability is
$(r_\mathrm{P}/r_\mathrm{B}^{\perp})^2$. However, because the beam particles are relativistic and a pellet
speed is much less, each beam particle has several opportunities, given by the
pellet transit time multiplied by the beam circulation frequency, each time with the same
probability. As a result, the opacity, which is the probability that a proton will encounter
a single pellet, is given by
\begin{equation}
O_\mathrm{Bp} \approx \left(\frac{r_\mathrm{P}}{r_\mathrm{B}^{\perp}}\right)^2\,\frac{2r_\mathrm{B}^{\perp}/v_\mathrm{P}}{\mathcal{C}/v_\mathrm{p}}
      = 2\,\frac{r_\mathrm{P}}{r_\mathrm{B}^{\perp}}\,\frac{r_\mathrm{P}}{v_\mathrm{P}\,T_0},
\label{eq:opacity.1}
\end{equation}
where $T_0$ is the beam revolution period.
Later, we will introduce $\rho_\mathrm{P}$ as the pellet material density, and
$t_\mathrm{P}=2 r_\mathrm{P} \rho_\mathrm{P}$ as the `target thickness' of the pellet, expressed in g/cm${}^2$.
Here, we are anticipating the approximation that the particle path lengths through
the pellet of a substantial fraction of the pellets differ little from the pellet diameter.
In practice, a pellet will be struck by many beam particles, but only a very small number
of beam particles will be aware of the passage of the pellet. Conversely,
because the pellet is so massive, its passage will be unaffected, even though it is hit by
many beam particles. Furthermore, a single beam particle passing through the pellet will,
at first, scarcely notice the interaction. But, because the binding of a particle in a
stable RF bucket is so weak, such a particle is almost certainly doomed or, in less
gloomy terms, `extracted' from its RF bucket.  The number of beam particles extracted by
a single pellet is then given by
\begin{equation}
N_{\rm extr.} = O_\mathrm{Bp} N_\mathrm{p}.
\label{eq:opacity.2}
\end{equation}
Of course, the circulating beam particles will be be reduced by exactly this number,
but the circulating beam will be otherwise unaffected. This has reduced our task to
finding the fate of of the $N_{\rm extr.}$ `extracted' particles. The
quotation marks on `extracted' serve as a reminder that, although the particles
are no longer captured in stable buckets, they have not necessarily been
extracted from the storage ring and delivered to a polarimeter.

\subsection{Expressing pellet mass in terms of pellet `target thickness'}
The role of a pellet is to slow down the beam particles that happen to pass
through it.  This slowing down is caused almost entirely by collisions of the
beam particle with electrons in the pellet. Yet the electrons make only
a negligible contribution to the pellet mass $M_\mathrm{P}$.

The pellet dynamics depends on pellet mass $M_\mathrm{P}$ and the beam particle slowing down
depends on the pellet `target thickness' $t_\mathrm{P}=2\ r_\mathrm{P} {\rho_\mathrm{P}}$.
The number of free parameters can be reduced by relating these two quantities:
\begin{equation}
M_\mathrm{P} = {\rho_\mathrm{P}}\,\frac{4\pi}{3}\,{r_\mathrm{P}}^3 = \frac{2\pi}{3}\,\,{r_\mathrm{P}}^2t_\mathrm{P}.
\label{eq:MP-tP.1}
\end{equation}

\subsection{Slowing down of beam particle passing through pellet}
The slowing down of a weakly relativistic elementary particle passing
through a medium falls inversely with
its squared velocity $v^2_\mathrm{p}$, `bottoming out' at a `minimum ionization' value $\dd E/\dd x$
as the speed approaches $c$. This is illustrated  in
\Fref{fig:DEdx-p-carbon}.
Minimum ionizing values for our promising pellet media are given in
in \Tref{tbl:PelletMaterialProperties}. One sees that these minimum ionization
values are approximately independent of the medium, with approximate value
1.6\,MeV/(g/cm${}^2$). It was commented earlier that, since the beam particle
velocities are significantly less than the speed of light, the slowing down is
enhanced by some voltage-dependent factor $\eta_\mathrm{p}(v_\mathrm{p})\approx7$, where the value
7 is specific to a $45\UMeV$ proton beam energy. This value can be regarded as
constant for current purposes, since we are concentrating only on the determination
of pellet parameters. With longitudinal position variable $z$, we can therefore use
\begin{equation}
\frac{\dd E_\mathrm{p}}{\dd (z\rho)} = -\eta_\mathrm{p}(v_\mathrm{p})\frac{\dd E_\mathrm{p}}{\dd x}\bigg|_{\min}
              \approx -7\times1.6\,\hbox{[MeV/(m/cm$^2$)]}.
\label{eq:dE/dx}
\end{equation}

\subsection{Longitudinal momentum reduction of `extracted' particles}
To track extracted beam particles out of the ring, it is particle
momentum (in the form $\delta=\delta_\mathrm{p}/p$, rather than particle energy)
that is needed. It would not be flagrantly wrong, and it would be consistent with
other relations used in this appendix, to simply use the
non-relativistic relation $K=(1/2)p^2/m$ for this purpose. But, for
greater generality, let us use a formula that is more nearly correct
relativistically, starting with the mass--energy--momentum relationship,
\begin{equation}
\gamma_\mathrm{p}^2m_\mathrm{p}^2c^4 = \mathcal{E}_\mathrm{p}^2 = (m_\mathrm{p}c^2 + K_\mathrm{p})^2 = m_\mathrm{p}^2c^4 + p_\mathrm{p}^2c^2.
\label{eq:SlowingDownKinematics.1}
\end{equation}
Solving for $p_\mathrm{p}$,
\begin{equation}
p_\mathrm{p}^2c^2 = 2m_\mathrm{p}c^2 K_\mathrm{p}\left(1+\frac{K_\mathrm{p}}{2m_\mathrm{p}c^2}\right).
\label{eq:SlowingDownKinematics.2}
\end{equation}
Differentiating this equation, and keeping only the leading
term in $K_\mathrm{p}/(m_\mathrm{p}c^2)$, yields
\begin{equation}
\delta \equiv \frac{\Delta p_\mathrm{p}}{p_\mathrm{p}}
 \approx
\frac{1}{2}\,\left(1 + \frac{K_\mathrm{p}}{2m_\mathrm{p}c^2}\right)\,\frac{\dd
K_\mathrm{p}}{K_\mathrm{p}}.
\label{eq:SlowingDownKinematics.3}
\end{equation}
Substituting from \Eref{eq:dE/dx} produces
\begin{equation}
\delta \approx
 \frac{1}{2}\,\left(1 + \frac{K_\mathrm{p}}{2m_\mathrm{p}c^2}\right)\eta_\mathrm{p}(v_\mathrm{p})\frac{\dd
 E_\mathrm{p}}{\dd
 x}\bigg|_{\rm min}\ t_\mathrm{P},
\label{eq:SlowingDownKinematics.4}
\end{equation}
(which is negative).

\subsection{Transverse displacement of `extracted' beam particles}
Neglecting any pre-existing betatron or synchrotron amplitude of a beam
particle passing through a pellet, let us assume that the beam
particle is in the design orbit as it enters the pellet.
At the location in the ring of the pellet injector, let the particle's
horizontal dispersion function value be $D_\mathrm{p}$, and the dispersion function
slope be zero, meaning that the transverse position of
a particle with fractional momentum offset $\delta$ is given by
\begin{equation}
x_\mathrm{p} = D_\mathrm{p}\,\delta.
\label{eq:MomentumOffset.1}
\end{equation}

On entry, we have assumed $x_\mathrm{p}=x'_\mathrm{p}=0$. Because the pellet is so `short', the
particle will still be in the design orbit (with any non-zero slope having
been caused by multiple scattering, which we are temporarily neglecting)
as it exits the pellet. But, on exit, the pellet's
fractional momentum offset is given by \Eref{eq:SlowingDownKinematics.4};
this means that the pellet is not in its off-momentum closed orbit---the
particle has acquired a (positive) horizontal betatron displacement, given by
\begin{equation}
x_{\beta}^{\rm out} = -D_\mathrm{p}\,\delta^{\rm out},
\label{eq:MomentumOffset.2}
\end{equation}
just right to cancel its sudden, newly established (negative)
off-momentum closed-orbit displacement $D_\mathrm{p}\,\delta^{\rm out}$.
In the absence of any further disturbance, the particle will continue
to oscillate with betatron amplitude given by \Eref{eq:MomentumOffset.1}
about this newly displaced closed orbit. For example, when the
betatron phase has increased by $\pi$, with $D_\mathrm{p}$ assumed constant, the
particle will be displaced from the true, on-momentum design orbit
$2 D_\mathrm{p}\,\delta$, with $\delta$ given by \Eref{eq:SlowingDownKinematics.4}.

In general, the betatron perturbation just calculated will simply be superimposed
on any previously neglected betatron and synchrotron amplitudes.

\subsection{Sudden particle translation expressed in terms of pellet opacity}
The sudden transverse displacement $D_\mathrm{p}\, \delta$
(relative to its off-momentum closed orbit) of a beam particle that has passed
through a pellet causes the particle to be extracted.
For clean sample extraction, we want to maximize this displacement (by increasing
pellet size or atomic number). However, at the same time, we want to minimize the
pellet opacity $O_\mathrm{Bp}$,  in order to minimize beam particle consumption
per pellet---for a bigger sample, one need only send more pellets.

To analyse this compromise, it is useful to express the sudden displacement in terms
of the opacity---that is, to express $D_\mathrm{p}\,\delta$ in terms of $O_\mathrm{Bp}$. Towards this end,
we rearrange \Eref{eq:opacity.1} to give
\begin{equation}
\frac{t_\mathrm{P}}{\rho_\mathrm{P}} = \sqrt{2O_\mathrm{Bp}r_\mathrm{B}^{\perp}v_\mathrm{P}T_0},
\label{eq:opacity.1-bis}
\end{equation}
which, conveniently, depends on pellet parameters only through the opacity.
We also combine Eqs.~(\ref{eq:SlowingDownKinematics.4}) and (\ref{eq:MomentumOffset.2}),
\begin{equation}
\Delta x_{\beta}
 =
-\,\frac{D_\mathrm{p}}{2}\,\left(1 + \frac{K_\mathrm{p}}{2m_\mathrm{p}c^2}\right)\eta_\mathrm{p}(v_\mathrm{p})\frac{\dd
E_\mathrm{p}}{\dd x}\bigg|_{\min}
     \left(\frac{t_\mathrm{P}}{\rho_\mathrm{P}}\right)\,\rho_\mathrm{P},
\label{eq:SlowingDownKinematics.4-bis}
\end{equation}
and substitute from \Eref{eq:opacity.1-bis},
\begin{equation}
\boxed{
\Delta x_{\beta}
 =
-\,\frac{D_\mathrm{p}}{2}\,\left(1 + \frac{K_\mathrm{p}}{2m_\mathrm{p}c^2}\right)\eta_\mathrm{p}(v_\mathrm{p})\frac{\dd
E_\mathrm{p}}{\dd x}\bigg|_{\rm min}
     \sqrt{2O_\mathrm{Bp}r_\mathrm{B}^{\perp}v_\mathrm{P}T_0}\,\,\rho_\mathrm{P},
\label{eq:SlowingDownKinematics.4-bisp}
}
\end{equation}
which is boxed to emphasize its importance. A striking implication of this equation
is that, at fixed opacity,  $\Delta x_{\beta}$ is proportional to the density $\rho_\mathrm{P}$
of the pellet material. The importance of this dependence can be assessed from the
density column of \Tref{tbl:PelletMaterialProperties}.

\subsection{Angular spread caused by multiple scattering}
As well as the loss of momentum just calculated, each extracted beam particle
acquires a multiple scattering angular distribution. The r.m.s. angular
spread can be expressed in terms of the particle momentum $p_\mathrm{p}$, in conjunction
with the radiation length $X_\mathrm{P}$ and target thickness $t_\mathrm{P}$ of the pellet material
The radiation length, expressed in  g/cm$^{2}$, is defined\cite{Gupta:1279627} by
\begin{equation}
X_\mathrm{P} = \frac{716.4}{Z_\mathrm{P}(Z_\mathrm{P}+1)\,{\rm ln}\frac{287}{\sqrt{Z_\mathrm{P}}}}.
\label{eq:MultipleScattering.1}
\end{equation}
Values of $X_\mathrm{P}$ for promising pellet media are given in
\Tref{tbl:PelletMaterialProperties}
The r.m.s. angular spread caused by passage through the pellet
with target thickness $t_\mathrm{P}$ and momentum $p_\mathrm{p}$ is given by
\begin{equation}
\theta_{\rm r.m.s.}
 = \frac{21\,{\UMeV}}{p_\mathrm{p}c\beta_\mathrm{p}}\,\sqrt{\frac{t_\mathrm{P}}{X_\mathrm{P}}}
 = \frac{21\,{\UMeV}}{\sqrt{2m_\mathrm{p}c^2K_\mathrm{p}}\,\big(1+K_\mathrm{p}/(4m_\mathrm{p}c^2)\big)\beta_\mathrm{p}}\,\sqrt{\frac{t_\mathrm{P}}{X_\mathrm{P}}},
\label{eq:MultipleScattering.2}
\end{equation}
where $p_\mathrm{p}c$ has been substituted from \Eref{eq:SlowingDownKinematics.2}.

\subsection{Pellet radius required for efficient bunch sampling extraction}
As explained earlier, with the dispersion function $D_\mathrm{p}$ assumed constant,
when the betatron phase has increased by $\pi$ (or any odd multiple of $\pi$),
a particle passing through a pellet centre will be displaced from its previous
off-momentum closed orbit by an amount $2 D_\mathrm{p}\,\delta$, with $\delta$ given by
\Eref{eq:SlowingDownKinematics.4}. Any polarimeter in the ring is assumed
to be located at such a position.

Even particles touching a pellet will not,
in general, pass through the pellet centre. About one-half of the pellets will be
sufficiently off-centre for their path length through the pellet to be
at least 30\%\ less than the pellet diameter. These pellets we ignore,
under the assumption that their energy loss has been insufficient for them to be
differentiated from the surviving main beam, and therefore unlikely to register
in the polarimetry. The path lengths of the remaining particles will all be
approximately the same. They will be treated as if centred on the pellet.


\begin{thebibliography}{10}

\bibitem{Abusaif:2018oly}
F.~Abusaif {\em et~al.},   \href{https://arxiv.org/abs/1812.08535}{arXiv:1812.08535 [physics.acc-ph]}.

\bibitem{Engel:2013lsa}
J.~Engel \textit{et al.},  {\em Prog.
  Part. Nucl. Phys.} \textbf{71} (2013) 21. \href{https://doi.org/10.1016/j.ppnp.2013.03.003}{doi:10.1016/j.ppnp.2013.03.003}, \href{https://arxiv.org/abs/1303.2371}{arXiv:1303.2371 [nucl-th]}.

\bibitem{Sakharov:1967dj}
A.~Sakharov,  {\em Sov. Phys. Usp.} \textbf{34} (1991) 39. \url{https://doi.org/10.1070/PU1991v034n05ABEH002496}

\bibitem{Afach:2015sja}
J.M. Pendlebury {\em et~al.},  {\em Phys. Rev.} \textbf{D92} (2015) 092003. \url{https://doi.org/10.1103/PhysRevD.92.092003}



\bibitem{Anas}
V.~Anastassopoulos {\em et~al.},  {\em Rev. Sci. Instrum.} \textbf{87}  (2016)
 115116. \url{https://doi.org/10.1063/1.4967465}

\bibitem{Graner:2016ses}
B.~Graner \textit{et al.},  {\em Phys. Rev. Lett.},
  \textbf{116} (2016) 161601.
\newblock [Erratum: \textit{Phys. Rev. Lett.} \textbf{119} (2017) 119901]. \url{https://doi.org/10.1103/PhysRevLett.116.161601}, \url{https://doi.org/10.1103/PhysRevLett.119.119901}

\bibitem{Chang:2017ruk}
S.P. Chang \textit{et al.},
  {\em Proc. Sci.}
  ~\textbf{PSTP2017} (2018) 036.
  \url{https://doi.org/10.22323/1.324.0036}

\bibitem{Abel:2017rtm}
C.~Abel {\em et~al.},  {\em Phys. Rev.} \textbf{X7} (2017)
 041034.
  \url{https://doi.org/10.1103/PhysRevX.7.041034}

\bibitem{Chang:2019poy}
S.P. Chang \textit{et al.}, {\em
  Phys. Rev.} \textbf{D99} (2019) 083002. \url{https://doi.org/10.1103/PhysRevD.99.083002}

\bibitem{Fukuyama:2013ioa}
T.~Fukuyama and A.J. Silenko,  {\em Int. J. Mod. Phys.} \textbf{A28} (2013)
 1350147. \url{https://doi.org/10.1142/S0217751X13501479}

\bibitem{Brantjes:2012zz}
N.~Brantjes {\em et~al.},  {\em Nucl.\ Instrum.\ Methods\ Phys. Res.}
  \textbf{A664} (2012) 49. \url{https://doi.org/10.1016/j.nima.2011.09.055}

\bibitem{Bagdasarian:2014ega}
Z.~Bagdasarian {\em et~al.},  {\em Phys. Rev. ST Accel. Beams} \textbf{17}
(2014)    052803. \url{https://doi.org/10.1103/PhysRevSTAB.17.052803}

\bibitem{Eversmann:2015jnk}
D.~Eversmann {\em et~al.}, 
  {\em Phys. Rev. Lett.} \textbf{115} (2015) 094801. \url{https://doi.org/10.1103/PhysRevLett.115.094801}

\bibitem{Hempelmann:2017zgg}
N.~Hempelmann {\em et~al.},  {\em Phys. Rev. Lett.} \textbf{119} (2017) 014801. \url{https://doi.org/10.1103/PhysRevLett.119.014801}

\bibitem{Guidoboni:2016bdn}
G.~Guidoboni {\em et~al.},  {\em
  Phys. Rev. Lett.} \textbf{117} (2016) 054801. \url{https://doi.org/10.1103/PhysRevLett.117.054801}

\bibitem{Saleev:2017ecu}
A.~Saleev {\em et~al.},  {\em Phys.\ Rev.\ Accel.\ Beams} \textbf{20}
  (2017) 072801. \url{https://doi.org/10.1103/PhysRevAccelBeams.20.072801}

\bibitem{Rathmann:2013rqa}
F.~Rathmann {\em et~al.}, {\em J.\ Phys.\ Conf.\ Ser.}
  \textbf{447} (2013) 012011. \url{https://doi.org/10.1088/1742-6596/447/1/012011}

\bibitem{PhysRevSTAB.16.114001}
W.M. Morse  {\em et~al.},
  {\em Phys. Rev. ST Accel. Beams} \textbf{16} (2013) 114001. \url{https://doi.org/10.1103/PhysRevSTAB.16.114001}

\bibitem{BERZ1990473}
M.~Berz,  {\em Nucl. Instrum. Methods Phys. Res.}, \textbf{A298} (1990) 473. \url{https://doi.org/10.1016/0168-9002(90)90649-Q}

\bibitem{Sagan:Bmad2006}
D.~Sagan, {\em
  Nucl. Instrum. Methods Phys. Res.} \textbf{A558} (2006) 356. \url{https://doi.org/10.1016/j.nima.2005.11.001}


\bibitem{Sagan:bmad:refmanual}
D.~Sagan, Bmad manual.
  \url{https://www.classe.cornell.edu/bmad/manual.html}, last accessed February
  3rd 2021.

\bibitem{vonHahn:2016wjl}
R.~von Hahn {\em et~al.},  {\em Rev. Sci.
  Instrum.} \textbf{87} (2016) 063115. \url{https://doi.org/10.1063/1.4953888}

\bibitem{Trinkel:2017jtf}
F.~Trinkel, \newblock Ph.D. thesis, RWTH Aachen University, 2017.
\url{https://doi.org/10.18154/RWTH-2018-00258}

\bibitem{Haciomeroglu:2018nre}
S.~Haci\"{o}mero\u{g}lu and Y.K. Semertzidis,  {\em Phys. Rev.
  Accel. Beams} \textbf{22} (2019) 034001. \url{https://doi.org/10.1103/PhysRevAccelBeams.22.034001}

\bibitem{Wagner:2020akw}
T.~Wagner {\em et~al.},
{\em J.Instrum.} \textbf{16} (2021) T02001.
\url{https://doi.org/10.1088/1748-0221/16/02/T02001}

\bibitem{Talman:2018dgp}
R.~Talman, \textit{arXiv} (2018) 
  1812.05949.
  \url{https://arxiv.org/abs/1812.05949}

\bibitem{ERC-694340}
Electric Dipole Moments using storage rings, H2020-EU.1.1. grant agreement number 694340).
 \url{http://www.sredm-ercgrant.de}, last accessed February
  3rd 2021.

\end{thebibliography}

\begin{thebibliography}{1}

\bibitem{Guidoboni:2016bdn}
G.~Guidoboni {\em et~al.},  {\em
  Phys. Rev. Lett.} \textbf{117} (2016) 054801. \url{https://doi.org/10.1103/PhysRevLett.117.054801}

\bibitem{Eversmann:2015jnk}
D.~Eversmann {\em et~al.}, 
  {\em Phys. Rev. Lett.} \textbf{115} (2015) 094801. \url{https://doi.org/10.1103/PhysRevLett.115.094801}

\bibitem{Hempelmann:2017zgg}
N.~Hempelmann {\em et~al.},  {\em Phys. Rev. Lett.} \textbf{119} (2017)  014801. \url{https://doi.org/10.1103/PhysRevLett.119.014801}

\bibitem{Brantjes:2012zz}
N.~Brantjes {\em et~al.},  {\em Nucl.\ Instrum.\ Methods Phys. Res.}
  \textbf{A664} (2012) 49. \url{https://doi.org/10.1016/j.nima.2011.09.055}

\bibitem{edms-worldwide}
Searches for permanent electric dipole moments - worldwide.
 \url{https://www.psi.ch/en/nedm/edms-world-wide}, last accessed February
3rd 2021.

\bibitem{Cairncross:2017fip}
W.B. Cairncross \textit{et al.}  {\em Phys. Rev.
  Lett.} \textbf{119} (2017) 153001.
  \url{https://doi.org/10.1103/PhysRevLett.119.153001}

\bibitem{centrex}
CeNTREX: a search for the electric dipole moment (EDM) of the proton.
  \url{https://demillegroup.yale.edu/research/centrex-search-electric-dipole-moment-edm-proton}, last accessed February
3rd 2021.

\end{thebibliography}

\begin{thebibliography}{10}

\bibitem{Sakharov:1967dj}
A.~Sakharov,  {\em Sov. Phys. Usp.} \textbf{34} (1991) 392. \url{https://doi.org/10.1070/PU1991v034n05ABEH002497}

\bibitem{Smith:1957ht}
J.H. Smith \textit{et al.},  {\em Phys. Rev.} \textbf{108}  (1957)   120. \url{https://doi.org/10.1103/PhysRev.108.120}

\bibitem{Baker:2006ts}
C.A. Baker {\em et~al.},  {\em Phys. Rev. Lett.} \textbf{97} (2006) 131801. \url{https://doi.org/10.1103/PhysRevLett.97.131801}

\bibitem{Afach:2015sja}
J.M. Pendlebury {\em et~al.},  {\em Phys. Rev.} \textbf{D92} 
  (2015) 092003. \url{https://doi.org/10.1103/PhysRevD.92.092003}

\bibitem{Abel:2020gbr}
C.~Abel {\em et~al.},  {\em Phys. Rev. Lett.} \textbf{124} (2020) 081803. \url{https://doi.org/10.1103/PhysRevLett.124.081803}

\bibitem{Murthy:1989zz}
S.A. Murthy {\em et~al.}, {\em Phys. Rev. Lett.}
  \textbf{63} (1989) 965. \url{https://doi.org/10.1103/PhysRevLett.63.965}

\bibitem{Regan:2002ta}
B.C. Regan {\em et~al.}, {\em Phys. Rev. Lett.} \textbf{88} (2002)
  071805. \url{https://doi.org/10.1103/PhysRevLett.88.071805}

\bibitem{Chupp:2017rkp}
T.~Chupp {\em et~al.}, {\em Rev. Mod.
  Phys.} \textbf{91} (2019)  015001. \url{https://doi.org/10.1103/RevModPhys.91.015001}

\bibitem{Jungmann:2013sga}
K.~Jungmann,  {\em Ann. Phys.}
  \textbf{525} (2013)  550. \url{https://doi.org/10.1002/andp.201300071}

\bibitem{Hudson:2011zz}
J.J. Hudson {\em et~al.}, {\em Nature}  \textbf{473} (2011) 493. \url{https://doi.org/10.1038/nature10104}

\bibitem{Baron:2013eja}
J.~Baron {\em et~al.},  {\em Science} \textbf{343} (2014) 269. \url{https://doi.org/10.1126/science.1248213}

\bibitem{Baron:2016obh}
J.~Baron {\em et~al.}, 
  {\em New J. Phys.} \textbf{19} (2017) 073029. \url{https://doi.org/10.1088/1367-2630/aa708e}

\bibitem{Andreev:2018ayy}
V.~Andreev {\em et~al.},  {\em Nature} \textbf{562} (2018) 355. \url{https://doi.org/10.1038/s41586-018-0599-8}

\bibitem{Cairncross:2017fip}
W.B. Cairncross {\em et~al.},  {\em Phys. Rev.
  Lett.} \textbf{119} (2017) 153001. \url{https://doi.org/10.1103/PhysRevLett.119.153001}

\bibitem{Pospelov:2013sca}
M.~Pospelov and A.~Ritz,  {\em Phys. Rev.} \textbf{D89} (2014) 056006. \url{https://doi.org/10.1103/PhysRevD.89.056006}

\bibitem{Rosenberry:2001}
M.A. Rosenberry and T.E. Chupp,  {\em Phys. Rev. Lett.} \textbf{86} (2001) 22. \url{https://doi.org/10.1103/PhysRevLett.86.22}

\bibitem{Bishof:2016uqx}
M.~Bishof {\em et~al.},  {\em Phys. Rev.} \textbf{C94} (2016) 025501. \url{https://doi.org/10.1103/PhysRevC.94.025501}

\bibitem{Graner:2016ses}
B.~Graner {\em et~al.}, {\em Phys. Rev. Lett.}
  \textbf{116} (2016) 161601.
\newblock [Erratum: \textit{Phys. Rev. Lett.} \textbf{119} (2017)  119901]. \url{https://doi.org/10.1103/PhysRevLett.116.161601}, \url{https://doi.org/10.1103/PhysRevLett.119.119901}

\bibitem{Dmitriev:2003sc}
V.F. Dmitriev and R.A. Sen'kov,  {\em Phys. Rev. Lett.} \textbf{91} (2003) 212303. \url{https://doi.org/10.1103/PhysRevLett.91.212303}

\bibitem{Sahoo:2016zvr}
B.K. Sahoo,  {\em Phys. Rev.} \textbf{D95} (2017) 013002. \url{https://doi.org/10.1103/PhysRevD.95.013002}

\bibitem{Tanabashi:2018oca}
M.~Tanabashi {\em et~al.},  {\em Phys. Rev.}
\textbf{D98} (2018) 030001. [Updated 2019.] \url{https://doi.org/10.1103/PhysRevD.98.030001}

\bibitem{Bennett:2008dy}
G.W. Bennett {\em et~al.},  {\em Phys. Rev.} \textbf{D80} (2009) 052008. \url{https://doi.org/10.1103/PhysRevD.80.052008}

\bibitem{Khriplovich:1997ga}
I.B. Khriplovich and S.K. Lamoreaux, {\em {CP Violation Without Strangeness:
  Electric Dipole Moments of Particles, Atoms, and Molecules}}
\newblock {(Springer}, Berlin, 1997).
\url{https://doi.org/10.1007/978-3-642-60838-4}

\bibitem{Khriplovich:1985jr}
I.B. Khriplovich,  {\em Phys. Lett.} \textbf{B173} (1986) 193.
\newblock [\textit{Yad. Fiz.} \textbf{44} (1986) 1019]. \url{https://doi.org/10.1016/0370-2693(86)90245-5}

\bibitem{Czarnecki:1997bu}
A.~Czarnecki and B.~Krause,  {\em Phys. Rev. Lett.} \textbf{78} (1997) 4339. \url{https://doi.org/10.1103/PhysRevLett.78.4339}

\bibitem{Gavela:1981sk}
M.B. Gavela \textit{et al.},  {\em Phys. Lett.} \textbf{B109} (1982) 215. \url{https://doi.org/10.1016/0370-2693(82)90756-0}

\bibitem{Khriplovich:1981ca}
I.B. Khriplovich and A.R. Zhitnitsky,  {\em Phys.
  Lett.} \textbf{B109} (1982) 490. \url{https://doi.org/10.1016/0370-2693(82)91121-2}

\bibitem{Seng:2014lea}
C.-Y. Seng,  {\em Phys. Rev.} \textbf{C91} (2015) 025502. \url{https://doi.org/10.1103/PhysRevC.91.025502}

\bibitem{Mannel:2012qk}
T.~Mannel and N.~Uraltsev,  {\em Phys. Rev.} \textbf{D85} (2012) 096002. \url{https://doi.org/10.1103/PhysRevD.85.096002}

\bibitem{Mannel:2012hb}
T.~Mannel and N.~Uraltsev,  {\em J. High Energy Phys.} \textbf{03} (2013) 064. \url{https://doi.org/10.1007/JHEP03(2013)064}

\bibitem{Barr:1990vd}
S.M. Barr and A.~Zee,  {\em Phys. Rev. Lett.} \textbf{65} (1990) 21.
\newblock [Erratum: \textit{Phys. Rev. Lett.} \textbf{65} (1990) 2920].
\url{https://doi.org/10.1103/PhysRevLett.65.21},
\url{https://doi.org/10.1103/PhysRevLett.65.2920}

\bibitem{Peccei:1977hh}
R.D. Peccei and H.R. Quinn,  {\em Phys. Rev. Lett.} \textbf{38} (1977) 1440. \url{https://doi.org/10.1103/PhysRevLett.38.1440}

\bibitem{Baluni:1978rf}
V.~Baluni,  {\em Phys. Rev.} \textbf{D19} (1979)
  2227. \url{https://doi.org/10.1103/PhysRevD.19.2227}

\bibitem{Crewther:1979pi}
R.J. Crewther \textit{et al.},
  {\em Phys. Lett.} \textbf{B88} (1979) 123.
\newblock [Erratum: \textit{Phys. Lett.} \textbf{B91} (1980) 487].
\url{https://doi.org/10.1016/0370-2693(79)90128-X}, \url{https://doi.org/10.1016/0370-2693(80)91025-4}

\bibitem{Bigi:2000yz}
I.I. Bigi and A.I. Sanda, {\em {CP Violation}}
\newblock (Cambridge University Press, Cambridge, 2009). \url{https://doi.org/10.1017/CBO9780511581014}


\bibitem{Abramczyk:2017oxr}
M.~Abramczyk \textit{et al.},
   {\em Phys. Rev.} \textbf{D96} (2017) 014501. \url{https://doi.org/10.1103/PhysRevD.96.014501}

\bibitem{Yoon:2017tag}
B.~Yoon \textit{et al.}, {\em EPJ Web Conf.} \textbf{175} (2018) 01014. \url{https://doi.org/10.1051/epjconf/201817501014}

\bibitem{Guo:2015tla}
F.~K. Guo \textit{et al.}, {\em Phys. Rev. Lett.}
  \textbf{115} (2015) 062001. \url{https://doi.org/10.1103/PhysRevLett.115.062001}

\bibitem{Dragos:2019oxn}
J.~Dragos \textit{et al.}, \textit{Phys. Rev.}  \textbf{C103} (2021) 015202. \url{https://doi.org/10.1103/PhysRevC.103.015202}

\bibitem{Ottnad:2009jw}
K.~Ottnad \textit{et al.}, {\em Phys. Lett.} \textbf{B687} (2010) 42. \url{https://doi.org/10.1016/j.physletb.2010.03.005}

\bibitem{Guo:2012vf}
F.-K. Guo and U.-G. Mei{\ss}ner, {\em J. High Energy Phys.} \textbf{12} (2012) 097. \url{https://doi.org/10.1007/JHEP12(2012)097}

\bibitem{Akan:2014yha}
T.~Akan \textit{et al.}, {\em Phys. Lett.} \textbf{B736} (2014) 163. \url{https://doi.org/10.1016/j.physletb.2014.07.022}

\bibitem{Bsaisou:2014zwa}
J.~Bsaisou \textit{et al.}, {\em J. High Energy Phys.} \textbf{03} (2015) 104.
\newblock [Erratum: \textit{J. High Energy Phys.} \textbf{05} (2015) 083]. \url{https://doi.org/10.1007/JHEP03(2015)104},
\url{https://doi.org/10.1007/JHEP05(2015)083}

\bibitem{Pospelov:2005pr}
M.~Pospelov and A.~Ritz,  {\em Annals Phys.} \textbf{318} (2005)119. \url{https://doi.org/10.1016/j.aop.2005.04.002}

\bibitem{Ng:2011ui}
J.~Ng and S.~Tulin,  {\em Phys. Rev.} \textbf{D85} (2012) 033001. \url{https://doi.org/10.1103/PhysRevD.85.033001}

\bibitem{deVries:2012ab}
J.~de~Vries \textit{et al.},  {\em Ann. Phys.} \textbf{338} (2013) 50. \url{https://doi.org/10.1016/j.aop.2013.05.022}

\bibitem{Weinberg:1989dx}
S.~Weinberg,  {\em Phys. Rev. Lett.} \textbf{63} (1989) 2333. \url{https://doi.org/10.1103/PhysRevLett.63.2333}

\bibitem{Gupta:2019fex}
R.~Gupta,  {\em Proc. Sci.} \textbf{SPIN2018} (2019) 095.
\url{https://doi.org/10.22323/1.346.0095}

\bibitem{Aoki:2019cca}
S.~Aoki {\em et~al.},  {\em Eur. Phys. J.} \textbf{C80} (2020) 113. \url{https://doi.org/10.1140/epjc/s10052-019-7354-7}

\bibitem{Bhattacharya:2015wna}
T.~Bhattacharya \textit{et al.}, {\em Phys. Rev.} \textbf{D92} (2015) 094511. \url{https://doi.org/10.1103/PhysRevD.92.094511}

\bibitem{Alexandrou:2017qyt}
C.~Alexandrou {\em et~al.},  {\em Phys. Rev.}
 \textbf{D95} (2017) 114514.
\newblock [Erratum: \textit{Phys. Rev.} \textbf{D96} (2017)  099906]. \url{https://doi.org/10.1103/PhysRevD.95.114514},
\url{https://doi.org/10.1103/PhysRevD.96.099906}

\bibitem{Gupta:2018lvp}
R.~Gupta \textit{et al.}, {\em Phys. Rev.} \textbf{D98} (2018) 091501. \url{https://doi.org/10.1103/PhysRevD.98.091501}

\bibitem{ArkaniHamed:2004fb}
N.~Arkani-Hamed and S.~Dimopoulos,  {\em
  J. High Energy Phys.} \textbf{06} (2005) 073. \url{https://doi.org/10.1088/1126-6708/2005/06/073}

\bibitem{Giudice:2004tc}
G.F. Giudice and A.~Romanino,  {\em Nucl. Phys.}
 \textbf{B699} (2004) 65.
\newblock [Erratum: \textit{Nucl. Phys.} \textbf{B706} (2005) 487]. \url{https://doi.org/10.1016/j.nuclphysb.2004.08.001}, \url{https://doi.org/10.1016/j.nuclphysb.2004.11.048}

\bibitem{ArkaniHamed:2004yi}
N.~Arkani-Hamed \textit{et al.}, {\em Nucl. Phys.} \textbf{B709} (2005) 3. \url{https://doi.org/10.1016/j.nuclphysb.2004.12.026}

\bibitem{Giudice:2005rz}
G.F. Giudice and A.~Romanino,  {\em Phys. Lett.} \textbf{B634} (2006) 307. \url{https://doi.org/10.1016/j.physletb.2006.01.027}

\bibitem{Wirzba:2016saz}
A.~Wirzba \textit{et al.}, {\em Int. J. Mod. Phys.}
  \textbf{E26} (2017) 1740031. \url{https://doi.org/10.1142/S0218301317400316}

\bibitem{Yamanaka:2016umw}
N.~Yamanaka,  {\em
  Int. J. Mod. Phys.} \textbf{E26} (2017) 1730002. \url{https://doi.org/10.1142/S0218301317300028}

\bibitem{deVries:2015una}
J.~de~Vries \textit{et al.},  {\em Phys. Rev.}
  \textbf{C92} (2015) 045201. \url{https://doi.org/10.1103/PhysRevC.92.045201}

\bibitem{Seng:2016pfd}
C.-Y. Seng and M.~Ramsey-Musolf, {\em Phys. Rev.} \textbf{C96} (2017) 065204. \url{https://doi.org/10.1103/PhysRevC.96.065204}

\bibitem{Mereghetti:2018oxv}
E.~Mereghetti, Electric dipole moments: a theory overview,   13th
  Conf. on the Intersections of Particle and Nuclear Physics (CIPANP 2018),
  Palm Springs,   2018  [\textit{arXiv} (2018) 1810.01320].
  \url{https://arxiv.org/abs/1810.01320}

\bibitem{Chang:2017ruk}
S.P. Chang \textit{et al.}, {\em Proc. Sci.}
\textbf{PSTP2017} (2018) 036.
\url{https://doi.org/10.22323/1.324.0036}

\bibitem{Chang:2019poy}
S.P. Chang \textit{et al.}, {\em
  Phys. Rev.} \textbf{D99} (2019) 083002. \url{https://doi.org/10.1103/PhysRevD.99.083002}

\bibitem{Graham:2011qk}
P.W. Graham and S.~Rajendran,  {\em Phys. Rev.} \textbf{D84} (2011) 055013. \url{https://doi.org/10.1103/PhysRevD.84.055013}

\bibitem{Graham:2013gfa}
P.W. Graham and S.~Rajendran,  {\em Phys. Rev.} \textbf{D88} (2013) 035023. \url{https://doi.org/10.1103/PhysRevD.88.035023}

\bibitem{Graham:2015ouw}
P.W. Graham \textit{et al.},
  {\em Annu. Rev. Nucl. Part. Sci.} \textbf{65} (2015) 485. \url{https://doi.org/10.1146/annurev-nucl-102014-022120}

\bibitem{Abel:2017rtm}
C.~Abel {\em et~al.},  {\em Phys. Rev.} \textbf{X7} (2017)
  041034.
  \url{https://doi.org/10.1103/PhysRevX.7.041034}

\bibitem{Roussy:2020ily}
T.S. Roussy {\em et~al.}, {\em Phys. Rev. Lett.} \textbf{126} (2021) 171301.
\url{https://doi.org/10.1103/PhysRevLett.126.171301}

\bibitem{Budker:2013hfa}
D.~Budker {\em et~al.}, {\em
  Phys. Rev.} \textbf{X4} (2014) 021030.
\url{https://doi.org/10.1103/PhysRevX.4.021030}

\bibitem{JacksonKimball:2017elr}
D.F. Jackson~Kimball {\em et~al.}, \textit{arXiv} (2017) 1711.08999.
\url{https://arxiv.org/abs/1711.08999}

\bibitem{Kahn:2016aff}
Y.~Kahn {\em et~al.}, {\em Phys. Rev. Lett.} \textbf{117} (2016) 141801. \url{https://doi.org/10.1103/PhysRevLett.117.141801}

\bibitem{Ouellet:2018beu}
J.L. Ouellet {\em et~al.},  {\em Phys. Rev. Lett.} \textbf{122} (2019)
  121802. \url{https://doi.org/10.1103/PhysRevLett.122.121802}

\bibitem{Guidoboni:2016bdn}
G.~Guidoboni {\em et~al.},  {\em
  Phys. Rev. Lett.} \textbf{117} (2016) 054801. \url{https://doi.org/10.1103/PhysRevLett.117.054801}

\end{thebibliography}

\begin{thebibliography}{10}

\bibitem{Bennett:2008dy}
G.W. Bennett {\em et~al.},  {\em Phys. Rev.} \textbf{D80} (2009) 052008. \url{https://doi.org/10.1103/PhysRevD.80.052008}

\bibitem{Bailey:1977sw}
J.~Bailey {\em et~al.},  {\em J.\ Phys.} \textbf{G4} (1978) 345. \url{https://doi.org/10.1088/0305-4616/4/3/010}

\bibitem{Farley:2003wt}
F.~Farley \textit{et al.},  {\em Phys.\ Rev.\ Lett.}
  \textbf{93} (2004) 052001. \url{https://doi.org/10.1103/PhysRevLett.93.052001}

\bibitem{Bennett:2006fi}
G.~Bennett {\em et~al.},  {\em Phys.\ Rev.} \textbf{D73} (2006) 072003. \url{https://doi.org/10.1103/PhysRevD.73.072003}

\bibitem{Orlov:2006su}
Y.F. Orlov \textit{et al.}, {\em Phys.\ Rev.\
  Lett.} \textbf{96} (2006) 214802. \url{https://doi.org/10.1103/PhysRevLett.96.214802}

\bibitem{Brantjes:2012zz}
N.~Brantjes {\em et~al.},  {\em Nucl.\ Instrum.\ Methods Phys. Res.}
  \textbf{A664} (2012) 49. \url{https://doi.org/10.1016/j.nima.2011.09.055}

\bibitem{bnl-sredm}
Storage Ring Electric Dipole Moment Collaboration.
  \url{https://www.bnl.gov/edm}, last accessed February 4th 2021.

\bibitem{bnl-sredm-deuteron-proposal}
D.~Anastassopoulos {\em et~al.}, AGS proposal: Search for a permanent electric dipolemoment of the deuteron nucleus at the $\SI{e-29}{\text{$e$ cm}}$ level, (2008).
  \url{https://inspirehep.net/literature/1865067}

\bibitem{bnl-edm-review-2009}
Review of the Storage Ring EDM experiment at BNL, (2009).
 \url{https://www.bnl.gov/edm/review}, last accessed February 4th 2021.

\bibitem{Flambaum:1984fb}
V.~Flambaum \textit{et al.}, {\em
  Sov.\ Phys.\ JETP} \textbf{60} (1984) 873.

\bibitem{bnl-edm-review-2011}
Review of the Storage Ring EDM experiment at BNL, (2011).
 \url{https://www.bnl.gov/edm/review_20110310}, last accessed February 4th 2021.

\bibitem{Maier:1997zj}
R.~Maier,  {\em
  Nucl.\ Instrum.\ Methods Phys. Res.} \textbf{A390} (1997) 1. \url{https://doi.org/10.1016/S0168-9002(97)00324-0}

\bibitem{jedi-collaboration}
JEDI (Jülich Electric Dipole moment Investigations) collaboration.
  \url{http://collaborations.fz-juelich.de/ikp/jedi/},  last accessed February 4th 2021. 

\bibitem{Aksentyev:2017dnk}
A.~Aksentyev \textit{et al.}, {\em Acta Phys.\ Pol.} \textbf{B48} (2017)
  1925. \url{https://doi.org/10.5506/APhysPolB.48.1925}

\bibitem{Eversheim:2017zxl}
D.~Eversheim \textit{et al.}, {\em Proc. Sci.} \textbf{INPC2016} (2017) 177.
\url{https://doi.org/10.22323/1.281.0177}

\bibitem{Keshelashvili:2019ggh}
I.~Keshelashvili \textit{et al.}, {\em J.\
  Phys.\ Conf.\ Ser.} \textbf{1162} (2019) 012029. \url{https://doi.org/10.1088/1742-6596/1162/1/012029}

\bibitem{ERC-694340}
Electric Dipole Moments using storage rings, H2020-EU.1.1. grant agreement number 694340).
\url{http://www.sredm-ercgrant.de}, last accessed February 4th 2021. 

\bibitem{Rathmann:2013rqa}
F.~Rathmann \textit{et al.},   {\em J.\ Phys.\ Conf.\ Ser.}
  \textbf{447} (2013) 012011. \url{https://doi.org/10.1088/1742-6596/447/1/012011}

\bibitem{Morse:2013hoa}
W.M. Morse \textit{et al.},
  {\em Phys.\ Rev.\ ST Accel.\ Beams} \textbf{16} (2013) 114001. \url{https://doi.org/10.1103/PhysRevSTAB.16.114001}

\bibitem{Slim:2016pim}
J.~Slim {\em et~al.},  {\em Nucl.\ Instrum.\ Methods Phys. Res.} \textbf{A828}
(2016) 116. \url{https://doi.org/10.1016/j.nima.2016.05.012}

\bibitem{Slim:2016dct}
J.~Slim \textit{et al.}, {\em
  Nucl.\ Instrum.\ Methods Phys. Res.} \textbf{A859} (2017) 52. \url{https://doi.org/10.1016/j.nima.2017.03.040}

\bibitem{doi:10.1063/1.5086862}
K.~Grigoryev \textit{et al.}, {\em Rev.
  Sci. Instrum.} \textbf{90} (2019) 045124. \url{https://doi.org/10.1063/1.5086862}

\end{thebibliography}

\begin{thebibliography}{10}

\bibitem{BMT}
V. Bargman \textit{et al.},  {\em Phys.
  Rev. Lett.} \textbf{2} (1959) 435. \url{https://doi.org/10.1103/PhysRevLett.2.435}

\bibitem{Courant:1980cf}
E.D. Courant and R.D. Ruth, {The acceleration of polarized protons in
  circular accelerators}, BNL-51270 (Brookhaven National Laboratory, New
York,   1980). \url{https://doi.org/10.2172/7034691}

\bibitem{Silenko:2015jqa}
A.J. Silenko,  {\em Phys. Part. Nucl. Lett.} \textbf{12} (2015) 8. \url{https://doi.org/10.1134/S1547477115010197}

\bibitem{FandS}
{T. Fukuyama and A.J. Silenko},  {\em Int. J. Mod. Phys.} \textbf{A28} (2013) 1350147. \url{https://doi.org/10.1142/S0217751X13501479}

\bibitem{Orlov:2019gtt}
Y.F. Orlov \textit{et al.},  {\em Phys. Lett.} \textbf{A376} (2012)
 2822. \url{https://doi.org/10.1016/j.physleta.2012.08.011}

\bibitem{Obukhov:2016vvk}
Y.N. Obukhov \textit{et al.},  {\em
  Phys. Rev.} \textbf{D94} (2016) 044019. \url{https://doi.org/10.1103/PhysRevD.94.044019}

\bibitem{Silenko:2006er}
A.J. Silenko and O.V. Teryaev,  {\em Phys. Rev.} \textbf{D76} (2007) 061101. \url{https://doi.org/10.1103/PhysRevD.76.061101}

\bibitem{Nikolaev_Spin2018}
N.N. Nikoalev \textit{et al.}, {Spin dynamics in
  storage rings in application to searches for EDM},  {23rd
  Int. Spin Symposium (Spin2018), Ferrara, Italy, 2018}.
\url{https://agenda.infn.it/event/12464/contributions/14394/}

\bibitem{Farley}
F.J.M. Farley \textit{et al.},  {\em Phys. Rev. Lett.}
\textbf{93} (2004) 052001. \url{https://doi.org/10.1103/PhysRevLett.93.052001}

\bibitem{PhysRevAccelBeams.23.024601}
F.~Rathmann \textit{et al.}, {\em Phys. Rev. Accel. Beams}
  \textbf{23} (2020) 024601. \url{https://doi.org/10.1103/PhysRevAccelBeams.23.024601}

\bibitem{Anas}
V.~Anastassopoulos {\em et~al.},  {\em Rev. Sci. Instrum.} \textbf{87} (2016)
 115116. \url{https://doi.org/10.1063/1.4967465}

\bibitem{polarizeref}
M.~Tanifugi, {\em Polarization Phenomena in Physics, Applications to Nuclear
  Reactions}
\newblock  (World Scientific, Singapore, 2018).
\url{https://doi.org/10.1142/10731}

\bibitem{Rathmann:1998zz}
F.~Rathmann {\em et~al.},  {\em Phys. Rev.} \textbf{C58} (1998) 658. \url{https://doi.org/10.1103/PhysRevC.58.658}

\bibitem{schieck2014nuclear}
H.~Schieck, {\em Nuclear Reactions: An Introduction}
\newblock (Springer, Berlin, 2014).
\url{https://doi.org/10.1007/978-3-642-53986-2}

\bibitem{vonPrzewoski:2003ig}
B.~von Przewoski {\em et~al.},  {\em Phys.
  Rev.} \textbf{C74} (2006) 064003. \url{https://doi.org/10.1103/PhysRevC.74.064003}

\end{thebibliography}

\begin{thebibliography}{10}

\bibitem{doi:10.1063/1.4967465}
V.~Anastassopoulos  \textit{et al.},  {\em Rev.
  Sci. Instrum.} \textbf{87} (2016) 115116. \url{https://doi.org/10.1063/1.4967465}

\bibitem{Lebedev:2015}
V.~Lebedev, Limitations on an {EDM} ring design,  EDM collaboration meeting November 10--11, 2014 to March 13, 2015,
  \url{http://collaborations.fz-juelich.de/ikp/jedi/public_files/usual_event/AccPhysLimitationsOnEDMring.pdf},
 last accessed February 11th 2021.

\bibitem{Tahar:2019hzv}
M.H. Tahar and C.~Carli, \textit{arXiv} (2019) 1904.07722.
\url{https://arxiv.org/abs/1904.07722}

\bibitem{vonHahn:2016wjl}
R.~von Hahn {\em et~al.},  {\em Rev. Sci.
  Instrum.} \textbf{87} (2016) 063115. \url{https://doi.org/10.1063/1.4953888}

\bibitem{moehl2013stochastic}
D.~M{\"o}hl, {\em Stochastic Cooling of Particle Beams}
(Springer, Berlin, 2013). \url{https://doi.org/10.1007/978-3-642-34979-9}

\bibitem{Fernqvist:2003}
G.~{Fernqvist} {\em et~al.}, {\em IEEE Trans.
  Instrum. Meas.} \textbf{52} (2003) 440. \url{https://doi.org/10.1109/TIM.2003.809915}

\bibitem{Decker:2005mu}
G.~Decker, {Beam stability in synchrotron light sources},   {7th
  European Workshop on Beam Diagnostics and Instrumentation for Particle
  Accelerators (DIPAC 2005)}, Lyon, 2005, p.~231.
  \url{https://inspirehep.net/literature/704541}

\bibitem{Graham:2011qk}
P.W. Graham and S.~Rajendran,  {\em Phys. Rev.} \textbf{D84} (2011) 055013. \url{https://doi.org/10.1103/PhysRevD.84.055013}

\bibitem{Chang:2019poy}
S.P. Chang {\em et~al.}, {\em
  Phys. Rev.} \textbf{D99} (2019) 083002. \url{https://doi.org/10.1103/PhysRevD.99.083002}

\bibitem{PhysRevD.100.111301}
V.V. Flambaum and H.B.T. Tan,  {\em Phys. Rev. D} \textbf{100} (2019) 111301. \url{https://doi.org/10.1103/PhysRevD.100.111301}

\bibitem{Abel:2017rtm}
C.~Abel {\em et~al.},  {\em Phys. Rev.} \textbf{X7} (2017) 041034.
 \url{https://doi.org/10.1103/PhysRevX.7.041034}

\end{thebibliography}

\begin{thebibliography}{10}

\bibitem{Maier:1997zj}
R.~Maier,  {\em
  Nucl. Instrum. Methods Phys. Res.} \textbf{A390} (1997) 1. \url{https://doi.org/10.1016/S0168-9002(97)00324-0}

\bibitem{PhysRevSTAB.18.020101}
C.~Weidemann {\em et~al.}, {\em Phys. Rev. ST Accel. Beams} \textbf{18} (2015)
020101. \url{https://doi.org/10.1103/PhysRevSTAB.18.020101}

\bibitem{Chislett:2016jau}
R.~Chislett,  {\em EPJ Web
  Conf.} \textbf{118} (2016) 01005. \url{https://doi.org/10.1051/epjconf/201611801005}

\bibitem{PhysRevAccelBeams.23.024601}
F.~Rathmann \textit{et al.}, {\em Phys. Rev. Accel. Beams}
 \textbf{23} (2020) 024601. \url{https://doi.org/10.1103/PhysRevAccelBeams.23.024601}

\bibitem{Rathmann:2013rqa}
F.~Rathmann \textit{et al.}, {\em J.\ Phys.\ Conf.\ Ser.}
  \textbf{447} (2013) 012011. \url{https://doi.org/10.1088/1742-6596/447/1/012011}

\bibitem{PhysRevSTAB.16.114001}
W.M. Morse \textit{et al.},
  {\em Phys. Rev. ST Accel. Beams} \textbf{16} (2013) 114001. \url{https://doi.org/10.1103/PhysRevSTAB.16.114001}

\bibitem{Slim:2016pim}
J.~Slim {\em et~al.},  {\em Nucl. Instrum. Methods Phys. Res.} \textbf{A828}
(2016) 116. \url{https://doi.org/10.1016/j.nima.2016.05.012}

\bibitem{Slim:2016dct}
J.~Slim \textit{et al.}, {\em
  Nucl. Instrum. Methods Phys. Res.} \textbf{A859} (2017) 52. \url{https://doi.org/10.1016/j.nima.2017.03.040}

\bibitem{Saleev:2017ecu}
A.~Saleev {\em et~al.},  {\em Phys.\ Rev.\ Accel.\ Beams} \textbf{20} (2017)
  072801. \url{ttps://doi.org/10.1103/PhysRevAccelBeams.20.072801}

\bibitem{Guidoboni:2016bdn}
G.~Guidoboni {\em et~al.},  {\em
  Phys. Rev. Lett.} \textbf{117} (2016) 054801. \url{https://doi.org/10.1103/PhysRevLett.117.054801}

\bibitem{Eversmann:2015jnk}
D.~Eversmann {\em et~al.}, 
  {\em Phys. Rev. Lett.} \textbf{115} (2015) 094801. \url{https://doi.org/10.1103/PhysRevLett.115.094801}

\bibitem{Hempelmann:2017zgg}
N.~Hempelmann {\em et~al.}, {\em Phys. Rev. Lett.} \textbf{119} (2017) 014801. \url{https://doi.org/10.1103/PhysRevLett.119.014801}

\bibitem{Keshelashvili:2019ggh}
I.~Keshelashvili \textit{et al.}, {\em J.\
  Phys.\ Conf.\ Ser.} \textbf{1162} (2019) 012029. \url{s://doi.org/10.1088/1742-6596/1162/1/012029}

\bibitem{Altmeier:2004qz}
M.~Altmeier {\em et~al.}, {\em Eur. Phys. J.} \textbf{A23} (2005) 351. \url{https://doi.org/10.1140/epja/i2004-10081-1}

\bibitem{Adam:2004ch}
H.-H. Adam {\em et~al.}, \textit{arXiv} (2004) nucl-ex/0411038.
\url{https://arxiv.org/abs/nucl-ex/0411038}

\bibitem{Calen:1996ft}
H.~Cal\'en {\em et~al.},  {\em Nucl. Instrum. Methods Phys. Res.} \textbf{A379}
(1996) 57. \url{https://doi.org/10.1016/0168-9002(96)00468-8}

\end{thebibliography}

\begin{thebibliography}{10}

\bibitem{Guidoboni:2016bdn}
G.~Guidoboni {\em et~al.},  {\em
  Phys. Rev. Lett.} \textbf{117} (2016) 054801. \url{https://doi.org/10.1103/PhysRevLett.117.054801}

\bibitem{Hempelmann:2017zgg}
N.~Hempelmann {\em et~al.},  {\em Phys. Rev. Lett.} \textbf{119} (2017) 014801. \url{https://doi.org/10.1103/PhysRevLett.119.014801}

\bibitem{Anas}
V.~Anastassopoulos {\em et~al.},  {\em Rev. Sci. Instrum.} \textbf{87} (2016)
115116. \url{https://doi.org/10.1063/1.4967465}

\bibitem{RTalmanLatts}
Lattice description files.
  \url{https://seafile.ikp.kfa-juelich.de/d/02fab8becfb14ad2b11a}, last accessed
  February 8th 2021.

\bibitem{Lebedev}
V.~Lebedev, Limitations on an {EDM} ring design,  EDM collaboration meeting November 10--11, 2014 to March 13, 2015,
  \url{http://collaborations.fz-juelich.de/ikp/jedi/public_files/usual_event/AccPhysLimitationsOnEDMring.pdf},
 last accessed February 11th 2021.

\bibitem{Rogowski1923}
W.~Rogowski, 
  {\em Archiv   Elektrotech.} \textbf{12} (1923) 1. \url{https://doi.org/10.1007/BF01656573}

\bibitem{Rogowski1926}
W.~Rogowski and H.~Rengier,  {\em Archiv  Elektrotech.} \textbf{16} (1926) 73. \url{https://doi.org/10.1007/BF01744713}

\bibitem{vonHahn:2016wjl}
R.~von Hahn {\em et~al.},  {\em Rev. Sci.
  Instrum.} \textbf{87} (2016) 063115. \url{https://doi.org/10.1063/1.4953888}

\bibitem{Trinkel:2017}
F.~Trinkel, 
\newblock Ph.D. thesis, RWTH Aachen University, 2017.
\url{https://doi.org/10.18154/RWTH-2018-00258}

\bibitem{Fal-18-10}
F.~Abusaif,  {\em Hyperfine Interact.} \textbf{240} (2019) 4. \url{https://doi.org/10.1007/s10751-018-1543-x}

\bibitem{Rogowski1912}
W.~Rogowski and W.~Steinhaus,  {\em
  Archiv  Elektrotech.} \textbf{1} (1912) 141. \url{https://doi.org/10.1007/BF01656479}

\bibitem{Slim:2016pim}
J.~Slim {\em et~al.},  {\em Nucl.\ Instrum.\ Meth.\ Phys. Res.} \textbf{A828}
(2016) 116. \url{https://doi.org/10.1016/j.nima.2016.05.012}

\bibitem{Jagdfeldprivcom}
H.~Jagdfeld and F.~Klehr, private communication, Forschungszentrum J\"ulich, Germany, August 2018.

\bibitem{PhysRevSTAB.18.020101}
C.~Weidemann {\em et~al.},  {\em Phys. Rev. ST Accel. Beams} \textbf{18}
(2015) 020101. \url{https://doi.org/10.1103/PhysRevSTAB.18.020101}

\bibitem{KoopSpinWheel}
I.~Koop, Asymmetric energy colliding ion beams in the EDM storage ring, 
  Proc. 4th Int. Particle Accelerator Conf. (IPAC
  2013), Shanghai, 2013, Ed. Z.~Dai\textit{ et al.} (JACoW Conferences, Geneva), p.~1961.
  \url{https://inspirehep.net/literature/1338735}

\bibitem{Fernqvist:2003}
G.~{Fernqvist} \textit{et al.},  {\em IEEE Trans. Instrum. Meas.} \textbf{52}
(2003) 440. \url{https://doi.org/10.1109/TIM.2003.809915}

\bibitem{CERNpreciseCurrent}
G.~{Fernqvist} \textit{et al.}, The {CERN} current
  calibrator-a new type of instrument,  Conf. Digest Conf.
  on Precision Electromagnetic Measurements, Ottawa, 2002 (IEEE, Piscataway), p. 410.
  \url{https://doi.org/10.1109/CPEM.2002.1034894}

\bibitem{Fernquist-PowerPoint}
G.~{Fernqvist}, High-precision measurements, CERN Accelerator School : Specialised CAS Course on Power Converters, 2004, pp.299-310.
\url{http://dx.doi.org/10.5170/CERN-2006-010.299},
  \url{https://cas.web.cern.ch/sites/cas.web.cern.ch/files/lectures/warrington-2004/fernqvist.pdf}

\bibitem{Chang:2019poy}
S.P. Chang \textit{et al.},
  {\em
  Phys. Rev.} \textbf{D99} (2019) 083002. \url{https://doi.org/10.1103/PhysRevD.99.083002}

\bibitem{Abel:2017rtm}
C.~Abel {\em et~al.},  {\em Phys. Rev.} \textbf{X7} (2017)  041034.
 \url{https://doi.org/10.1103/PhysRevX.7.041034}
\end{thebibliography}

\begin{thebibliography}{10}

\bibitem{doi:10.1063/1.4967465}
V.~Anastassopoulos \textit{et al.},  {\em Rev.
  Sci. Instrum.} \textbf{87} (2016) 115116. \url{https://doi.org/10.1063/1.4967465}

\bibitem{Nominal-BNL}
Storage Ring Electric Dipole Moment Collaboration.
\url{http://www.bnl.gov/edm/}, last accessed February 8th 2021.

\bibitem{Lebedev}
V.~Lebedev, Limitations on an {EDM} ring design,  EDM collaboration meeting November 10--11, 2014 to March 13, 2015,
  \url{http://collaborations.fz-juelich.de/ikp/jedi/public_files/usual_event/AccPhysLimitationsOnEDMring.pdf},
 last accessed February 11th 2021.

\bibitem{Tahar:2019hzv}
M.H. Tahar and C.~Carli, \textit{arXiv} (2019) 1904.07722.
\url{https://arxiv.org/abs/1904.07722}

\bibitem{vonHahn:2016wjl}
R.~von Hahn {\em et~al.},  {\em Rev. Sci.
  Instrum.} \textbf{87} (2016) 063115. \url{https://doi.org/10.1063/1.4953888}

\bibitem{Mohl:1993jn}
D.~Möhl,  Stochastic cooling, CERN Accelerator School, 5th Advanced Accelerator Physics Course, Rhodes, Greece, p. 587.
\url{http://dx.doi.org/10.5170/CERN-1995-006.587}

\bibitem{Fernqvist:2003}
G.~{Fernqvist} \textit{et al.}, {\em IEEE Trans. Instrum. Meas.} \textbf{52}
(2003) 440. \url{https://doi.org/10.1109/TIM.2003.809915}

\bibitem{Decker:2005mu}
G.~Decker, {Beam stability in synchrotron light sources},  7th
  European Workshop on Beam Diagnostics and Instrumentation for Particle
  Accelerators (DIPAC 2005), Lyon, 2005, p. 231.
   \url{https://inspirehep.net/literature/704541}

\bibitem{prokofiev2005}
O.~Prokofiev, Tevatron Beam Separator R@D, E\&F meeting,  15 Nov. 2005. 
  \url{https://www.bnl.gov/edm/review/files/references/Prokofiev_separator_R&D_dec6_2005.pdf}, last accessed February 8th 2021.

\bibitem{Chang:2017ruk}
S.P. Chang \textit{et al.}, {\em Proc. Sci.}
 \textbf{PSTP2017} (2018) 036.
 \url{https://doi.org/10.22323/1.324.0036}

\bibitem{Abel:2017rtm}
C.~Abel {\em et~al.},  {\em Phys. Rev.} \textbf{X7} (2017) 041034.
 \url{https://doi.org/10.1103/PhysRevX.7.041034}
 
\bibitem{Pretz:2019ham}
J.~Pretz {\em et~al.}, {\em
  Eur. Phys. J.} \textbf{C80} (2020) 107. \url{https://doi.org/10.1140/epjc/s10052-020-7664-9}

\end{thebibliography}

\begin{thebibliography}{10}

\bibitem{doi:10.1063/1.4967465}
V.~Anastassopoulos \textit{et al.},  {\em Rev.
  Sci. Instrum.} \textbf{87} (2016) 115116. \url{https://doi.org/10.1063/1.4967465}

\bibitem{Nominal-BNL}
Storage Ring Electric Dipole Moment Collaboration.
 \url{http://www.bnl.gov/edm/}, last accessed February 9th 2021.

\bibitem{vonHahn:2016wjl}
R.~von Hahn {\em et~al.},  {\em Rev. Sci.
  Instrum.} \textbf{87} (2016)  063115. \url{https://doi.org/10.1063/1.4953888}

\bibitem{Manos:2001na}
C.K. Sinclair \textit{et al.},
  {\em Conf. Proc. C} \textbf{0106181} (2001)
 610--612.

\bibitem{Dunham:2007zz}
B.~Dunham \textit{et al.},
  Performance of a very high voltage photoemission electron gun for a high
  brightness, high average current ERL injector, Proc. Particle Accelerator
Conf., Albuquerque,
2007 (IEEE, Piscataway), p.~1224. \url{https://doi.org/10.1109/PAC.2007.4441037}


\bibitem{Borburgh:2004uf}
L.~Sermeus \textit{et al.}, The design of the special magnets for PIMMS/TERA,   9th European Particle Accelerator Conf., Lucerne, 2004.
  \url{https://accelconf.web.cern.ch/e04/PAPERS/WEPKF020.PDF}

\bibitem{Kramer:2011zd}
J.~Borburgh \textit{et al.},  {\em Conf. Proc.
  C} \textbf{110904} (2011) 3534.
  \url{http://accelconf.web.cern.ch/AccelConf/IPAC2011/papers/THPS048.PDF}


\bibitem{BastaniNejad:2012zz}
M.~BastaniNejad {\em et~al.},  {\em Phys. Rev. ST Accel.
  Beams} \textbf{15} (2012) 083502.
\newblock [Erratum: \textit{Phys. Rev. ST Accel. Beams} \textbf{17} (2014) 029901]. \url{https://doi.org/10.1103/PhysRevSTAB.17.029901}

\bibitem{Mamun:2015gvm}
M.A.A. Mamun \textit{et al.},  {\em J. Vac.
  Sci. Tech. A} \textbf{33} (2015) 031604. \url{https://doi.org/10.1116/1.4916574}

\bibitem{Mamun2}
M.A.A. Mamun, Ph.D. thesis, Old Dominion University, 2016.


\bibitem{doi:10.1063/1.5086862}
K.~Grigoryev \textit{et al.},  {\em Rev.
  Sci. Instrum.} \textbf{90} (2019) 045124. \url{https://doi.org/10.1063/1.5086862}

\bibitem{Descoeudres:2009zz}
A.~Descoeudres \textit{et al.}, {\em Phys. Rev. ST
  Accel. Beams} \textbf{12} (2009) 092001. \url{https://doi.org/10.1103/PhysRevSTAB.12.092001}

\bibitem{Descoeudres:2009zza}
A.~Descoeudres \textit{et al.}, {\em Phys. Rev. ST Accel. Beams} \textbf{12}
(2009) 032001. \url{https://doi.org/10.1103/PhysRevSTAB.12.032001}

\bibitem{Descoeudres:1355401}
A.~Descoeudres \textit{et al.}, {DC breakdown experiments
  with cobalt electrodes}, Tech. Rep. CERN-OPEN-2011-029, CLIC-Note-875
  (CERN, Geneva,  2009).
  \url{https://cds.cern.ch/record/1355401}

\bibitem{Lorenzi}
A.~{De Lorenzi} \textit{et al.}, {HV holding
  in vacuum, a key issue for the ITER neutral beam injector},  
  28th Int. Symp. Discharges and Electrical Insulation in
  Vacuum (ISDEIV), 2018, vol.~2, p.~721. \url{https://doi.org/10.1109/DEIV.2018.8537143}

\bibitem{Borburgh2}
J.~{Borburgh} \textit{et al.},
  {Challenges for the electric field devices for a CERN proton EDM storage
  ring},  28th Int. Symp. Discharges and
  Electrical Insulation in Vacuum (ISDEIV), 2018, vol.~2, p.~765. \url{https://doi.org/10.1109/DEIV.2018.8537088}

\bibitem{Atanasov}
M. Atanasov \textit{et al.}, {EDM dipole and quadrupole field calculations}, Tech.
  Rep. EDMS 2102228 (CERN, Geneva, 2019).
  \url{https://edms.cern.ch/document/2102228/1}

\bibitem{Rogowski1923}
W.~Rogowski, 
  {\em Arch. Elektrotech.} \textbf{12} (1923) 1. \url{https://doi.org/10.1007/BF01656573}

\bibitem{TRACK}
TRACK: The beam dynamics code.
 \url{http://www.phy.anl.gov/atlas/TRACK}, last accessed February 9th 2021.

\bibitem{Carliprivcom} C. Carli, private communication.

\bibitem{Slim:2016dct}
J.~Slim \textit{et al.}, {\em
  Nucl.\ Instrum.\ Methods Phys. Rev.} \textbf{A859} (2017) 52. \url{https://doi.org/10.1016/j.nima.2017.03.040}

\end{thebibliography}

\begin{thebibliography}{10}

\bibitem{osti_4726823}
H.H. Barschall and W.~Haeberli,  Polarization phenomena in nuclear
  reactions. Proc. 3rd Int. Symp., Madison,
  1970.


\bibitem{schieck2014nuclear}
H.~Schieck, {\em Nuclear Reactions: An Introduction}
\newblock (Springer, Berlin, 2014).
\url{https://doi.org/10.1007/978-3-642-53986-2}

\bibitem{Mueller2020}
F.~M{\"u}ller \textit{et al.},  {\em 
  Eur. Phys. J.} \textbf{A56} (2020) 211. \url{https://doi.org/10.1140/epja/s10050-020-00215-8}

\bibitem{Zelenski:2014vva}
A.~Zelenski,  {\em Proc. Sci.} \textbf{PSTP2013} (2013)
  048. 

\bibitem{Sona}
P.~Sona,  {\em Energia Nucl.} \textbf{14} (1967) 
295. \url{https://doi.org/10.1093/nq/14-8-295a}

\bibitem{IEIRI1987253}
M.~Ieiri \textit{et al.}, {\em Nucl.\ Instrum.\
  Methods\ Phys. Res.} \textbf{A257} (1987) 253. \url{https://doi.org/10.1016/0168-9002(87)90744-3}

\bibitem{Guidoboni:2016bdn}
G.~Guidoboni {\em et~al.},  {\em
  Phys. Rev. Lett.} \textbf{117} (2016) 054801. \url{https://doi.org/10.1103/PhysRevLett.117.054801}

\bibitem{lee1997spin}
S.Y. Lee, {\em Spin Dynamics and Snakes in Synchrotrons}
\newblock (World Scientific, Singapore, 1997). \url{https://doi.org/10.1142/3233}

\bibitem{Meyer:1988rj}
H.~Meyer {\em et~al.},  {\em
  Phys. Rev.} \textbf{C37} (1988) 544. \url{https://doi.org/10.1103/PhysRevC.37.544}

\bibitem{Meyerprivcom}
H.O.~Meyer, private communication.

\bibitem{Brantjes:2012zz}
N.~Brantjes {\em et~al.},  {\em Nucl.\ Instrum.\ Methods\ Phys. Res.}
\textbf{A664} (2012) 49. \url{https://doi.org/10.1016/j.nima.2011.09.055}

\bibitem{Hanna:1965}
R.C. Hanna,  Proc. 2nd Int. Sym.
  Polarization Phenomena, Karlsruhe, 1965, Eds P.~Huber and H.~Schopper    (Birkhäuser, Basel, 1966), p. 280.

\bibitem{Hempelmann:2017zgg}
N.~Hempelmann {\em et~al.},  {\em Phys. Rev. Lett.} \textbf{119} (2017) 014801. \url{https://doi.org/10.1103/PhysRevLett.119.014801}

\bibitem{BONIN1990389}
B.~Bonin \textit{et al.}, {\em Nucl.\ Instrum.\
  Methods Phys. Rev.} \textbf{A288} (1990) 389. \url{https://doi.org/10.1016/0168-9002(90)90129-T}

\bibitem{LADYGIN1998129}
V.~Ladygin \textit{et al.},
  {\em Nucl.\ Instrum.\ Methods Phys. Res.} \textbf{A404} (1998)  129. \url{https://doi.org/10.1016/S0168-9002(97)01135-2}

\bibitem{Mcnaughton:1986ks}
M.~Mcnaughton {\em et~al.},  {\em Nucl. Instrum. Methods Phys. Res.} \textbf{A241}
(1985) 435. \url{https://doi.org/10.1016/0168-9002(85)90595-9}

\bibitem{Stephensonprivcom} E.J.~Stephenson, private communication.

\bibitem{Javakhishvili_2020}
O.~Javakhishvili \textit{et al.}, {\em J.
  Phys. Conf. Ser.} \textbf{1561} (2020) 012011. \url{https://doi.org/10.1088/1742-6596/1561/1/012011}

\bibitem{Muller_2020}
F.~Müller \textit{et al.}, {\em J. Instrum.} \textbf{15} (2020) P12005.
\url{https://doi.org/10.1088/1748-0221/15/12/P12005}

\bibitem{JAVAKHISHVILI2020164337}
O.~Javakhishvili \textit{et al.} {\em Nucl.\ Instrum.\ Methods Phys. Res.} \textbf{A977} (2020) 164337. \url{https://doi.org/10.1016/j.nima.2020.164337}

\end{thebibliography}

\begin{thebibliography}{10}

\bibitem{PhysRevSTAB.16.114001}
W.M. Morse \textit{et al.}, 
  {\em Phys. Rev. ST Accel. Beams} \textbf{16} (2013) 114001. \url{https://doi.org/10.1103/PhysRevSTAB.16.114001}

\bibitem{Saleev:2017ecu}
A.~Saleev {\em et~al.},  {\em Phys. Rev. Accel. Beams} \textbf{20} (2017) 072801. \url{https://doi.org/10.1103/PhysRevAccelBeams.20.072801}

\bibitem{Rathmann:2019lwi}
F.~Rathmann \textit{et al.}, {\em Phys. Rev. Accel. Beams} \textbf{23} (2020)
024601. \url{https://doi.org/10.1103/PhysRevAccelBeams.23.024601}

\bibitem{BasicBNL}
Storage Ring EDM Collaboration,
A proposal to measure the proton electric dipole moment with$\SI{e-29}{\text{$e$ cm}}$ sensitivity, (2011).
  \url{https://inspirehep.net/literature/1865072}

\bibitem{BasicRevSci}
V.~Anastassopoulos {\em et~al.},  {\em Rev. Sci. Instrum.} \textbf{87} (2016) 115116. \url{https://doi.org/10.1063/1.4967465}

\bibitem{Hybrid}
S.~Hac\ifmmode \imath \else \i \fi{}\"omero\ifmmode~\breve{g}\else \u{g}\fi{}lu
  and Y.K. Semertzidis,  {\em Phys. Rev. Accel. Beams} \textbf{22} (2019)
034001. \url{https://doi.org/10.1103/PhysRevAccelBeams.22.034001}

\bibitem{DoubleMagic}
R.~Talman, \textit{arXiv} (2018)
  1812.05949.
  \url{https://arxiv.org/abs/1812.05949}

\bibitem{GravityOrlovAl}
Y.~Orlov \textit{et al.}, {\em Phys. Lett.}
 \textbf{A376} (2012) 2822. \url{https://doi.org/10.1016/j.physleta.2012.08.011}

\bibitem{GravityObukhovAl}
Y.N. Obukhov \textit{et al.}, {\em
  Phys. Rev.} \textbf{D94} (2016) 044019. \url{https://doi.org/10.1103/PhysRevD.94.044019}

\bibitem{GravitySilenkoAl}
A.J. Silenko and O.V. Teryaev,  {\em Phys. Rev.} \textbf{D76} (2007) 061101. \url{https://doi.org/10.1103/PhysRevD.76.061101}

\bibitem{GravityLaszloAl}
A.~L{\'{a}}szl{\'{o}} and Z.~Zimbor{\'{a}}s, {\em Classical
   Quantum Gravity} \textbf{35} (2018) 175003. \url{https://doi.org/10.1088/1361-6382/aacfee}

\bibitem{LatticeValeri}
V.~Lebedev, Limitations on an {EDM} ring design,  EDM collaboration meeting November 10--11, 2014 to March 13, 2015,
  \url{http://collaborations.fz-juelich.de/ikp/jedi/public_files/usual_event/AccPhysLimitationsOnEDMring.pdf},
 last accessed February 11th 2021.

\bibitem{TuneMod}
Y.K. {Semertzidis},  \textit{Eur.
  Phys. J. Web Conf.} \textbf{118} (2016) 01032. \url{https://doi.org/10.1051/epjconf/201611801032}

\bibitem{GeomPhasesMagFields}
S.~Haci\"{o}mero\u{g}lu \textit{et al.}, {\em
  Nucl. Instrum. Methods Phys. Res.} \textbf{A927} (2019) 262. \url{https://doi.org/10.1016/j.nima.2019.01.046}, \url{https://arxiv.org/abs/1812.02381}

\bibitem{VertSlopeSpinRot1}
S.~Haci\"{o}mero\u{g}lu, Quadrupole misplacement studies for the pEDM
  rings, CPEDM Meeting, J\"ulich,
  2018.
  \url{https://doi.org/10.5281/zenodo.4812725}

\bibitem{ElQuadMisalignement}
S.~Haci\"{o}mero\u{g}lu and Y.K. Semertzidis, \textit{arXiv} (2017) 1709.01208.
\url{https://arxiv.org/abs/1709.01208}

\end{thebibliography}

\begin{thebibliography}{10}

\bibitem{BERZ1990473}
M.~Berz,  {\em Nucl. Instrum. Methods Phys. Res.} \textbf{A298} (1990) 473. \url{https://doi.org/10.1016/0168-9002(90)90649-Q}

\bibitem{Rosenthal:2015jzr}
M.~Rosenthal and A.~Lehrach, Spin tracking simulations towards electric
  dipole moment measurements at COSY,  6th Int. Particle
  Accelerator Conf., 2015, p.~THPF032.
  \url{https://doi.org/10.18429/JACoW-IPAC2015-THPF032}

\bibitem{Brun:1997pa}
R.~Brun and F.~Rademakers,  {\em Nucl. Instrum. Methods Phys. Res.} \textbf{A389}
(1997) 81. \url{https://doi.org/10.1016/S0168-9002(97)00048-X}

\bibitem{Ivanov:2014aza}
A.~Ivanov and Y.~Senichev, Matrix integration of ODEs for spin-orbit
  dynamics simulation,  5th Int. Particle Accelerator
  Conf., 2014, p.~MOPME011.
  \url{https://doi.org/10.18429/JACoW-IPAC2014-MOPME011}

\bibitem{Ivanov:2013era}
A.~Ivanov \textit{et al.},
  Testing of symplectic integrator of spin-orbit motion based on matrix
  formalism,  4th Int. Particle Accelerator Conf., 2013,
  p.~WEPEA037.
  \url{http://jacow.org/IPAC2013/papers/wepea037.pdf}

\bibitem{Sagan:Bmad2006}
D.~Sagan,  {\em
  Nucl. Instrum. Methods Phys. Res.} \textbf{A558} (2006) 356. \url{https://doi.org/10.1016/j.nima.2005.11.001}

\bibitem{Sagan:bmad:refmanual}
D.~Sagan, Bmad manual. 
  \url{https://www.classe.cornell.edu/bmad/manual.html}, last accessed February
  11th 2021.

\bibitem{Schmidt:573082}
F.~Schmidt \textit{et al.}, Introduction to the polymorphic
  tracking code: fibre bundles, polymorphic Taylor types and `exact
  tracking',  CERN-SL-2002-044-AP. KEK-REPORT-2002-3 (CERN,
  Geneva, 2002).
  \url{https://cds.cern.ch/record/573082}

\bibitem{Senichev:2012jga}
Y.~Senichev \textit{et al.}, Storage ring EDM simulation: methods and
  results,  11th Int. Computational Accelerator Physics
  Conf., 2012, p.~TUADI1.
  \url{http://jacow.org/ICAP2012/papers/tuadi1.pdf}

\bibitem{Gaisser:2016ocd}
M.~Gaisser \textit{et al.},
  Precision spin tracking for electric dipole moment searches, 
  12th Int. Computational Accelerator Physics Conf., Shanghai, 2016,
  p.~THAJI2,
\newblock 
  \url{https://doi.org/10.18429/JACoW-ICAP2015-THAJI2}

\bibitem{METODIEV2015311}
E.~Metodiev \textit{et al.}, {\em Nucl. Instrum. Methods Phys. Res.}
 \textbf{A797} (2015) 311. \url{https://doi.org/10.1016/j.nima.2015.06.032}

\bibitem{Maier:2012zzc}
D.~Zyuzin \textit{et al.}, {\em Conf. Proc. C} \textbf{1205201} (2012) 1335.
\url{http://accelconf.web.cern.ch/AccelConf/IPAC2012/papers/TUPPC071.PDF}

\bibitem{Rathmann:2011zz}
F.~Rathmann and N.~Nikolaev,  {\em
  Proc. Sci.} \textbf{STORI11} (2011) 029.
  \url{https://doi.org/10.22323/1.150.0029}

\bibitem{Lehrach:2012eg}
A.~Lehrach \textit{et al.}, \textit{arXiv} (2012) 1201.5773.
\url{https://arxiv.org/abs/1201.5773}

\bibitem{Morse:2013hoa}
W.M. Morse \textit{et al.},
  {\em Phys.\ Rev.\ ST Accel.\ Beams} \textbf{16} (2013) 114001. \url{https://doi.org/10.1103/PhysRevSTAB.16.114001}

\bibitem{Rosenthal:2014}
M.~Rosenthal, {\em Microscopy Microanal.} \textbf{21S4} (2015) 30,
\newblock 
  \url{https://doi.org/10.1017/S1431927615013094}

\bibitem{Chekmenev:2016cpx}
S.~Chekmenev, {\em Int. J. Mod. Phys. Conf. Ser.} \textbf{40} (2016)  1660099. \url{https://doi.org/10.1142/S2010194516600995}

\bibitem{Schmidt:2017wnl}
V.~Schmidt and A.~Lehrach,  {\em J. Phys. Conf. Ser.} \textbf{874} (2017) 012051. \url{https://doi.org/10.1088/1742-6596/874/1/012051}

\bibitem{PhysRevAccelBeams.23.024601}
F.~Rathmann \textit{et al.},  {\em Phys. Rev. Accel. Beams}
  \textbf{23} (2020) 024601. \url{https://doi.org/10.1103/PhysRevAccelBeams.23.024601}

\bibitem{Farley}
F.J.M. Farley \textit{et al.}, {\em Phys. Rev. Lett.}
 \textbf{93} (2004) 052001. \url{https://doi.org/10.1103/PhysRevLett.93.052001}

\bibitem{Senichev:2016rez}
Y.~Senichev \textit{et al.},
  Investigation of lattice for deuteron EDM ring, 12th
  Int. Computational Accelerator Physics Conf., 2016, p. 17.
  \url{https://doi.org/10.18429/JACoW-ICAP2015-MODBC4}

\bibitem{Senichev:2017sry}
Y.~Senichev \textit{et al.}, Quasi-frozen spin concept of deuteron storage
  ring as an instrument to search for the electric dipole moment, 
  8th Int. Particle Accelerator Conf., 2017, p. 2275.
  \url{https://doi.org/10.18429/JACoW-IPAC2017-TUPVA084}

\bibitem{Senichev:2015bkf}
Y.~Senichev \textit{et al.} Quasi-frozen spin method for EDM
  deuteron search, 6th Int. Particle Accelerator
  Conf., 2015, p. MOPWA044.
  \url{https://doi.org/10.18429/JACoW-IPAC2015-MOPWA044}

\bibitem{Senichev:2016zqo}
Y.~Senichev \textit{et al.} Systematic errors investigation in frozen and
  quasi-frozen spin lattices of deuteron EDM ring, 7th
  Int. Particle Accelerator Conf., Busan, 2016, p. THPMR005.
  \url{https://doi.org/10.18429/JACoW-IPAC2016-THPMR005}

\bibitem{Skawran_2017}
A.~Skawran and A.~Lehrach,  
  {\em J. Phys. Conf. Ser.} \textbf{874} (2017) 012050. \url{https://doi.org/10.1088/1742-6596/874/1/012050}

\bibitem{Lehrach:2019pgw}
A.~Lehrach \textit{et al.},   {\em Proc. Sci.} \textbf{SPIN2018} (2019) 144.
\url{https://doi.org/10.22323/1.346.0144}

\bibitem{selcuk}
S.~Haci\"{o}mero\u{g}lu, Spin tracking studies for pEDM experiment, EDM kickoff meeting, CERN, 13–14 March 2017.  
\url{https://doi.org/10.5281/zenodo.4817374}

\bibitem{Michaud:2019}
J.~Michaud, 
 Ph.D. thesis, Université Grenoble Alpes, 2019.
  \url{https://tel.archives-ouvertes.fr/tel-02481832}

\bibitem{Tahar:2019hzv}
M.H. Tahar and C.~Carli, \textit{arXiv} (2019) 1904.07722.
\url{https://arxiv.org/abs/1904.07722}

\bibitem{Rosenthal:2016zbf}
M.S. Rosenthal,  Ph.D. thesis, RWTH Aachen University, 2016.
  \url{https://publications.rwth-aachen.de/record/671012}

\bibitem{Poncza:2019ith}
V.~Poncza and A.~Lehrach, Search for electric dipole moments at COSY in
  J\"ulich---spin-tracking simulations using Bmad,  10th
  Int. Particle Accelerator Conf., 2019, p. MOPTS028.
  \url{https://doi.org/10.18429/JACoW-IPAC2019-MOPTS028}

\bibitem{Maier:2012zza}
Y.~Senichev \textit{et al.},   {\em Conf. Proc. C} \textbf{1205201} (2012) 1332.
\url{https://accelconf.web.cern.ch/IPAC2012/papers/TUPPC070.PDF}

\bibitem{Lebedev}
V.~Lebedev, Limitations on an {EDM} ring design,  EDM collaboration meeting November 10--11, 2014 to March 13, 2015,
  \url{http://collaborations.fz-juelich.de/ikp/jedi/public_files/usual_event/AccPhysLimitationsOnEDMring.pdf},
 last accessed February 11th 2021.

\bibitem{Holzer:1982419}
B.~Holzer, Lattice design in high-energy particle accelerators, CERN Accelerator School, Advanced Accelerator Physics Course, Trondheim, Norway, August 18--29, 2013, CERN-2014-009, p. 61. 
\url{https://doi.org/10.5170/CERN-2014-009.61}

\bibitem{Bogolyubov}
N.N. Bogoliubov and Y.A. Mitropolsky, {\em Asymptotic methods in the theory
  of non-linear oscillations} (Gordon
  and Breach, New York, 1961).
  \url{https://books.google.ch/books?id=NRwfI8vuoOEC&printsec=frontcover}

\bibitem{Tahar:2020gsz}
M.H. Tahar and C.Carli, \textit{Phys.\,Rev.\,Accel.\,Beams}, \textbf{24} (2021), 034003.
\url{https://doi.org/10.1103/PhysRevAccelBeams.24.034003}

\bibitem{Berry:1984jv}
M.V. Berry,  {\em
  Proc. R. Soc. London} \textbf{A392} (1984) 45. \url{https://doi.org/10.1098/rspa.1984.0023}

\end{thebibliography}

\begin{thebibliography}{1}

\bibitem{ERC-694340}
Electric Dipole Moments using storage rings, H2020-EU.1.1. grant agreement number 694340).
\url{http://www.sredm-ercgrant.de}, last accessed February 11th 2021.

\end{thebibliography}

\begin{thebibliography}{10}

\bibitem{Eversmann:2015jnk}
D.~Eversmann {\em et~al.}, 
  {\em Phys. Rev. Lett.} \textbf{115} (2015) 094801. \url{https://doi.org/10.1103/PhysRevLett.115.094801}

\bibitem{PhysRevD.73.072003}
G.W. Bennett {\em et~al.},  {\em Phys. Rev.} \textbf{D73} (2006) 072003. \url{https://doi.org/10.1103/PhysRevD.73.072003}

\bibitem{Guidoboni:2016bdn}
G.~Guidoboni {\em et~al.},  {\em
  Phys. Rev. Lett.}  \textbf{117} (2016) 054801. \url{https://doi.org/10.1103/PhysRevLett.117.054801}

\bibitem{Hempelmann:2017zgg}
N.~Hempelmann {\em et~al.},  {\em Phys. Rev. Lett.} \textbf{119} (2017) 014801. \url{https://doi.org/10.1103/PhysRevLett.119.014801}

\bibitem{Saleev:2017ecu}
A.~Saleev {\em et~al.},  {\em Phys. Rev. Accel. Beams} \textbf{20} (2017)
072801. \url{https://doi.org/10.1103/PhysRevAccelBeams.20.072801}

\bibitem{Slim:2016pim}
J.~Slim {\em et~al.},  {\em Nucl. Instrum. Methods Phys. Res.} \textbf{A828}
(2016) 116. \url{https://doi.org/10.1016/j.nima.2016.05.012}, \url{https://arxiv.org/abs/1603.01567}

\bibitem{Slim:2016dct}
J.~Slim \textit{et al.},  {\em
  Nucl. Instrum. Methods Phys. Res.} \textbf{A859} (2017) 52. \url{https://doi.org/10.1016/j.nima.2017.03.040}

\bibitem{F.Mueller2020}
F.~M{\"u}ller \textit{et al.}, {\em 
  Eur. Phys. J.} \textbf{A56} (2020) 211. \url{https://doi.org/10.1140/epja/s10050-020-00215-8}

\bibitem{Wagner:2019xul}
T.~Wagner and J.~Pretz, Beam-based alignment at the Cooler Synchrotron
  (COSY), 10th Int. Particle Accelerator Conf.,
 2019,   p.~THPGW024.
 \url{https://doi.org/10.18429/JACoW-IPAC2019-THPGW024}

\bibitem{wagner2020beambased}
T.~Wagner \textit{et al.}, \textit{J. Instrum.} \textbf{16} (2021) T02001.
\url{https://doi.org/10.1088/1748-0221/16/02/T02001}

\bibitem{Fal-18-10}
F.~Abusaif,  {\em Hyperfine Interact.} \textbf{240} (2019) 2019. \url{https://doi.org/10.1007/s10751-018-1543-x}

\bibitem{Slim:2017bic}
J.~Slim \textit{et al.},  {\em Phys.
  Rev.} \textbf{E96} (2017) 063301. \url{https://doi.org/10.1103/PhysRevE.96.063301}

\bibitem{Magiera:2017ypv}
A.~Magiera, 
  {\em Phys. Rev. Accel. Beams} \textbf{20} (2017) 094001. \url{https://doi.org/10.1103/PhysRevAccelBeams.20.094001}

\bibitem{Pretz:2018bze}
J.~Pretz and F.~M{\"u}ller,  {\em Eur. Phys. J.} \textbf{C79} (2019) 47. \url{https://doi.org/10.1140/epjc/s10052-019-6580-3}

\bibitem{Eversmann:2015vqn}
D.~Eversmann \textit{et al.}, {\em J. Instrum.}
 \textbf{11} (2016) P05003. \url{https://doi.org/10.1088/1748-0221/11/05/P05003}

\bibitem{Silenko:2015qfa}
A.J. Silenko,  {\em J. Phys.} \textbf{G42} (2015) 075109. \url{https://doi.org/10.1088/0954-3899/42/7/075109}

\bibitem{Ottnad:2009ths}
K.~Ottnad,  {Diplomarbeit}, Bonn University,  2009. 

\bibitem{Ottnad:2009jw}
K.~Ottnad \textit{et al.}, {\em Phys. Lett.} \textbf{B687} (2010) 42. \url{https://doi.org/10.1016/j.physletb.2010.03.005}

\bibitem{Bsaisou:2012rg}
J.~Bsaisou \textit{et al.},
  {\em Eur. Phys. J.} \textbf{A49} (2013) 31. \url{https://doi.org/10.1140/epja/i2013-13031-x}

\bibitem{Guo:2012vf}
F.-K. Guo and U.-G. Mei{\ss}ner,  {\em J. High Energy Phys.} \textbf{12}
(2012) 097. \url{https://doi.org/10.1007/JHEP12(2012)097}

\bibitem{Akan:2014yha}
T.~Akan \textit{et al.}, {\em Phys. Lett.} \textbf{B736} (2014) 163. \url{https://doi.org/10.1016/j.physletb.2014.07.022}

\bibitem{Bsaisou:2014goa}
J.~Bsaisou, Ph.D. thesis, Bonn University,  2014. 

\bibitem{Dekens:2014jka}
W.~Dekens \textit{et al.},  {\em J. High Energy Phys.} \textbf{07} (2014) 069. \url{https://doi.org/10.1007/JHEP07(2014)069}

\bibitem{Dekens:2013zca}
W.~Dekens and J.~de~Vries,  {\em J. High Energy Phys.} \textbf{05} (2013)
149. \url{https://doi.org/10.1007/JHEP05(2013)149}

\bibitem{Wirzba:2014mka}
A.~Wirzba,  {\em
  Nucl. Phys.} \textbf{A928} (2014) 116. \url{https://doi.org/10.1016/j.nuclphysa.2014.04.003}

\bibitem{Bsaisou:2014zwa}
J.~Bsaisou \textit{et al.}, {\em J. High Energy Phys.} \textbf{03} (2015)
104.
\newblock [Erratum: \textit{J. High Energy Phys.} \textbf{05} (2015) 083]. \url{https://doi.org/10.1007/JHEP03(2015)104},
\url{https://doi.org/10.1007/JHEP05(2015)083}

\bibitem{Bsaisou:2014oka}
J.~Bsaisou \textit{et al.}, {\em Ann. Phys.} \textbf{359} (2015) 317. \url{https://doi.org/10.1016/j.aop.2015.04.031}

\bibitem{Shindler:2015aqa}
A.~Shindler \textit{et al.}, {\em Phys. Rev.} \textbf{D92} (2015) 094518. \url{https://doi.org/10.1103/PhysRevD.92.094518}

\bibitem{Shindler:2014oha}
A.~Shindler \textit{et al.}, {\em Proc. Sci.} \textbf{LATTICE2014} (2014) 251.

\bibitem{Guo:2015tla}
F.-K. Guo \textit{et al.}, {\em Phys. Rev. Lett.}
 \textbf{115} (2015) 062001. \url{https://doi.org/10.1103/PhysRevLett.115.062001}

\bibitem{deVries:2015gea}
J.~de~Vries and U.-G. Mei{\ss}ner,  {\em Int. J. Mod. Phys.} \textbf{E25}
(2016) 1641008. \url{https://doi.org/10.1142/S0218301316410081}

\bibitem{Seng:2014pba}
C.-Y. Seng \textit{et al.}, {\em Phys. Lett.} \textbf{B736} (2014) 147. \url{https://doi.org/10.1016/j.physletb.2014.07.014}

\bibitem{deVries:2015una}
J.~de~Vries \textit{et al.}, {\em Phys. Rev.} 
\textbf{C92} (2015) 045201. \url{https://doi.org/10.1103/PhysRevC.92.045201}

\bibitem{Chien:2015xha}
Y.T. Chien \textit{et al.}, {\em J. High Energy Phys.} \textbf{02} (2016)
011. \url{https://doi.org/10.1007/JHEP02(2016)011}

\bibitem{Wirzba:2016saz}
A.~Wirzba \textit{et al.}, {\em Int. J. Mod. Phys.}
 \textbf{E26} (2017) 1740031. \url{https://doi.org/10.1142/S0218301317400316}

\bibitem{Dragos:2018uzd}
J.~Dragos \textit{et al.},
{\em PoS} \textbf{LATTICE2018} (2019) 259.
   \url{https://doi.org/10.22323/1.334.0259}

\bibitem{Dragos:2017wms}
J.~Dragos \textit{et al.}, {\em EPJ Web Conf.} \textbf{175} (2018) 06018. \url{https://doi.org/10.1051/epjconf/201817506018}

\bibitem{Kim:2018rce}
J.~Kim \textit{et al.},
{\em PoS} \textbf{LATTICE2018} (2019), 260.
\url{https://doi.org/10.22323/1.334.0260}

\bibitem{Dragos:2019oxn}
J.~{Dragos} \textit{et al.},
{\em Phys. Rev.} \textbf{C103} (2021) 015202.
\url{https://doi.org/10.1103/PhysRevC.103.015202}

\end{thebibliography}

\begin{thebibliography}{1}

\bibitem{bib:MagneticFields}
S.~Haci\"{o}mero\u{g}lu \textit{et al.},  {\em Nucl.
  Instrum. Methods Phys. Res.} \textbf{A927} (2019) 262. \url{https://doi.org/10.1016/j.nima.2019.01.046}, \url{https://arxiv.org/abs/1812.02381}

\bibitem{bib:SQUIDBPM}
S.~Haci\"{o}mero\u{g}lu \textit{et al.},  {\em Proc. Sci.}
  \textbf{ICHEP2018} (2019) 279.
  \url{https://doi.org/10.22323/1.340.0279}

\bibitem{bib:sub_micrometer_squid}
M.~Schmelz \textit{et al.}, {\em Supercond. Sci. Technol.} \textbf{24} (2011) 065009. \url{https://doi.org/10.1088/0953-2048/24/6/065009}

\bibitem{HybridRing}
S.~Haci\"{o}mero\u{g}lu   and Y.K. Semertzidis,  {\em Phys. Rev. Accel. Beams} \textbf{22} (2019) 034001. \url{https://doi.org/10.1103/PhysRevAccelBeams.22.034001}

\end{thebibliography}

\begin{thebibliography}{1}

\bibitem{Obukhov:2016vvk}
Y.N. Obukhov \textit{et al.},  {\em
  Phys. Rev.} \textbf{D94} (2016) 044019. \url{https://doi.org/10.1103/PhysRevD.94.044019}

\bibitem{Silenko:2004ad}
A.J. Silenko and O.V. Teryaev,  {\em Phys. Rev.} \textbf{D71} (2005) 064016. \url{https://doi.org/10.1103/PhysRevD.71.064016}

\bibitem{Silenko:2006er}
A.J. Silenko and O.V. Teryaev,  {\em Phys. Rev.} \textbf{D76} (2007) 061101. \url{https://doi.org/10.1103/PhysRevD.76.061101}

\bibitem{Silenko:2015jqa}
A.J. Silenko,  {\em Phys. Part. Nucl. Lett.} \textbf{12} (2015) 8. \url{https://doi.org/10.1134/S1547477115010197}

\bibitem{Khriplovich:1997ni}
I.B. Khriplovich and A.A. Pomeransky,  {\em J. Exp. Theor. Phys.}
  \textbf{86} (1998) 839.
\newblock [\textit{Zh. Eksp. Teor. Fiz.} \textbf{113} (1998) 1537]. \url{https://doi.org/10.1134/1.558554}

\bibitem{Pomeransky:2000pb}
A.A. Pomeransky \textit{et al.}, {\em Phys. Usp.} \textbf{43} (2000) 1055.
\newblock [\textit{Usp. Fiz. Nauk} \textbf{43} (2000) 1129]. \url{https://doi.org/10.1070/PU2000v043n10ABEH000674}

\bibitem{Nikolaev_Spin2018}
N.N. Nikolaev \textit{et al.}, Spin dynamics in
  storage rings in application to searches for EDM, 23rd
  Int. Spin Symp. (Spin2018), Ferrara, 2018.
  \url{https://agenda.infn.it/event/12464/contributions/14394/}

\bibitem{Orlov:2019gtt}
Y.F. Orlov \textit{et al.}, {\em Phys. Lett.} \textbf{A376} (2012)
 2822. \url{https://doi.org/10.1016/j.physleta.2012.08.011}

\end{thebibliography}

\begin{thebibliography}{1}

\bibitem{PhysRevSTAB.18.020101}
C.~Weidemann {\em et~al.},  {\em Phys. Rev. ST Accel. Beams} \textbf{18}
(2015) 020101. \url{https://doi.org/10.1103/PhysRevSTAB.18.020101}

\end{thebibliography}

\begin{thebibliography}{10}

\bibitem{Abel:2020gbr}
C.~Abel {\em et~al.},  {\em Phys. Rev. Lett.} \textbf{124} (2020) 081803. \url{https://doi.org/10.1103/PhysRevLett.124.081803}

\bibitem{Peccei:2006as}
R.~Peccei,  {\em Lect. Notes Phys.}
\textbf{741} (2008) 3. \url{https://doi.org/10.1007/978-3-540-73518-2_1}

\bibitem{Peccei:1977hh}
R.D. Peccei and H.R. Quinn,  {\em Phys. Rev. Lett.} \textbf{38} (1977) 1440. \url{https://doi.org/10.1103/PhysRevLett.38.1440}

\bibitem{Weinberg:1977ma}
S.~Weinberg,  {\em Phys. Rev. Lett.} \textbf{40} (1978)
  223. \url{https://doi.org/10.1103/PhysRevLett.40.223}

\bibitem{Wilczek:1977pj}
F.~Wilczek,  {\em Phys. Rev. Lett.} \textbf{40} (1978) 279. \url{https://doi.org/10.1103/PhysRevLett.40.279}

\bibitem{Tanabashi:2018oca}
M.~Tanabashi {\em et~al.},  {\em Phys. Rev.}
 \textbf{D98} (2018) 030001. [Update 2019]. \url{https://doi.org/10.1103/PhysRevD.98.030001}

\bibitem{Graham:2011qk}
P.W. Graham and S.~Rajendran,  {\em Phys. Rev.} \textbf{D84} (2011) 055013. \url{https://doi.org/10.1103/PhysRevD.84.055013}

\bibitem{Graham:2013gfa}
P.W. Graham and S.~Rajendran,  {\em Phys. Rev.} \textbf{D88} (2013) 035023. \url{https://doi.org/10.1103/PhysRevD.88.035023}

\bibitem{Chang:2019poy}
S.P. Chang \textit{et al.}, {\em
  Phys. Rev.} \textbf{D99} (2019) 083002. \url{https://doi.org/10.1103/PhysRevD.99.083002}

\bibitem{Guidoboni:2016bdn}
G.~Guidoboni {\em et~al.},  {\em
  Phys. Rev. Lett.} \textbf{117} (2016) 054801. \url{https://doi.org/10.1103/PhysRevLett.117.054801}

\bibitem{Brantjes:2012zz}
N.~Brantjes {\em et~al.},  {\em Nucl. Instrum. Methods Phys. Res.}
 \textbf{A664} (2012) 49. \url{https://doi.org/10.1016/j.nima.2011.09.055}

\bibitem{Lee:1997mz}
S.~Lee, {\em {Spin Dynamics and Snakes in Synchrotrons}}
(World Scientific, Singapore, 1997). \url{https://doi.org/10.1142/3233}

\bibitem{Eversmann:2015jnk}
D.~Eversmann {\em et~al.}, 
  {\em Phys. Rev. Lett.} \textbf{115} (2015) 094801. \url{https://doi.org/10.1103/PhysRevLett.115.094801}

\end{thebibliography}

\begin{thebibliography}{1}

\bibitem{Farley}
F.J.M. Farley \textit{et al.},  {\em Phys. Rev. Lett.}
  \textbf{93} (2004) 052001. \url{https://doi.org/10.1103/PhysRevLett.93.052001}

\bibitem{Nominal-BNL}
Storage Ring Electric Dipole Moment Collaboration.
 \url{http://www.bnl.gov/edm/}, last accessed February 15th 2021.

\bibitem{bnl-sredm-deuteron-proposal}
D.~Anastassopoulos {\em et~al.}, AGS proposal: Search for a permanent electric dipolemoment of the deuteron nucleus at the $\SI{e-29}{\text{$e$ cm}}$ level, (2008).
  \url{https://inspirehep.net/literature/1865067},


\bibitem{Anas}
V.~Anastassopoulos {\em et~al.}, {\em Rev. Sci. Instrum.} \textbf{87} (2016), 115116.
\url{https://doi.org/10.1063/1.4967465}

\bibitem{Haciomeroglu:2018nre}
S.~Hacıömeroğlu and Y.K. Semertzidis,  {\em Phys. Rev.
  Accel. Beams} \textbf{22} (2019) 034001. \url{https://doi.org/10.1103/PhysRevAccelBeams.22.034001}

\end{thebibliography}

\begin{thebibliography}{1}

\bibitem{jedi-collaboration}
JEDI (Jülich Electric Dipole moment Investigations) collaboration.
  \url{http://collaborations.fz-juelich.de/ikp/jedi/}, last accessed February
  16th 2021.

\bibitem{Eversmann:2015jnk}
D.~Eversmann {\em et~al.},   {\em Phys. Rev. Lett.} \textbf{115} (2015) 094801. \url{https://doi.org/10.1103/PhysRevLett.115.094801}

\bibitem{Saleev:2017ecu}
A.~Saleev {\em et~al.},  {\em Phys.\ Rev.\ Accel.\ Beams} \textbf{20} (2017)
072801. \url{https://doi.org/10.1103/PhysRevAccelBeams.20.072801}

\bibitem{IEIRI1987253}
M.~Ieiri \textit{et al.}, {\em Nucl.\ Instrum.\
  Methods Phys. Res.} \textbf{A257} (1987) 253. \url{https://doi.org/10.1016/0168-9002(87)90744-3}

\bibitem{KoopSpinWheel}
I.~Koop, Asymmetric energy colliding ion beams in the EDM storage ring, 
Proc. 4th Int. Particle Accelerator Conf. (IPAC
  2013), 2013, Eds. Z.~Dai \textit{et al.},
  (JACoW Conferences, Geneva), p. 1961.
  \url{https://inspirehep.net/literature/1338735}

\bibitem{Rathmann:2014gfa}
F.~Rathmann \textit{et al.} {\em Phys. Part. Nucl.} \textbf{45} (2014) 229. \url{https://doi.org/10.1134/S1063779614010869}

\end{thebibliography}

\begin{thebibliography}{10}

\bibitem{bnl-sredm-deuteron-proposal}
D.~Anastassopoulos {\em et~al.}, AGS proposal: Search for a permanent electric dipolemoment of the deuteron nucleus at the $\SI{e-29}{\text{$e$ cm}}$ level, (2008).
  \url{https://inspirehep.net/literature/1865067}
  
\bibitem{Mane:2015jsa}
S.~Mane, \textit{arXiv} (2015) 1509.01167.
\url{https://arxiv.org/abs/1509.01167}

\bibitem{Anas}
V.~Anastassopoulos {\em et~al.},  {\em Rev. Sci. Instrum.} \textbf{87} (2016)
115116. \url{https://doi.org/10.1063/1.4967465}

\bibitem{KoopSpinWheel}
I.~Koop, Asymmetric energy colliding ion beams in the EDM storage ring, 
Proc. 4th Int. Particle Accelerator Conf. (IPAC
  2013), 2013, Eds. Z.~Dai \textit{et al.},
  (JACoW Conferences, Geneva), p. 1961.
  \url{https://inspirehep.net/literature/1338735}


\bibitem{Aksentev:2018krh}
A.~Aksentev and Y.~Senichev,  {\em J. Phys. Conf. Ser.} \textbf{941} (2017)
012083. \url{https://doi.org/10.1088/1742-6596/941/1/012083}

\bibitem{Aksentyev:2019jet}
A.~Aksentyev and Y.~Senichev, Spin motion perturbation effect on the EDM
  statistic in the frequency domain method, 10th Int.
  Particle Accelerator Conf., 2019, p.~MOPTS011.
  \url{https://doi.org/10.18429/JACoW-IPAC2019-MOPTS011}

\bibitem{Senichev:2017amn}
Y.~Senichev \textit{et al.}, \textit{arXiv} (2017) 1711.06512.
\url{https://arxiv.org/abs/1711.06512}

\bibitem{Senichev:2016rez}
Y.~Senichev \textit{et al.},
 Investigation of lattice for deuteron EDM ring, 12th
  Int. Computational Accelerator Physics Conf., 2016, p 17.
  \url{https://doi.org/10.18429/JACoW-ICAP2015-MODBC4}

\bibitem{Valetov:2017}
E.V. Valetov,  Ph.D. thesis, Michigan State University, 2017.
  \url{https://doi.org/10.2172/1416546}

\bibitem{Aksentyev:2019rxq}
A.~Aksentyev and Y.~Senichev, Spin decoherence in the frozen spin storage
  ring method of search for a particle EDM, 10th Int.
  Particle Accelerator Conf., 2019, p.~MOPTS012.
  \url{https://doi.org/10.18429/JACoW-IPAC2019-MOPTS012}

\bibitem{Senichev:2013dra}
Y.~Senichev \textit{et al.}, Spin tune decoherence
  effects in electro- and magnetostatic structures, 4th
  Int. Particle Accelerator Conf., 2013, p 2579.
  \url{http://jacow.org/IPAC2013/papers/wepea036.pdf}

\bibitem{Aksentyev:2019ajz}
A.~Aksentyev and Y.~Senichev, Simulation of the guide field flipping
  procedure for the frequency domain method, 10th Int.
  Particle Accelerator Conf., 2019, p.~MOPTS010.
  \url{https://doi.org/10.18429/JACoW-IPAC2019-MOPTS010}

\bibitem{Eversmann:2015jnk}
D.~Eversmann {\em et~al.}, 
  {\em Phys. Rev. Lett.} \textbf{115} (2015) 094801. \url{https://doi.org/10.1103/PhysRevLett.115.094801}

\end{thebibliography}

\begin{thebibliography}{1}

\bibitem{Magiera:2017ypv}
A.~Magiera, 
  {\em Phys. Rev. Accel. Beams} \textbf{20} (2017) 094001. \url{https://doi.org/10.1103/PhysRevAccelBeams.20.094001}

\end{thebibliography}

\begin{thebibliography}{1}

\bibitem{8187724}
Z.~{Sun} \textit{et al.},  {\em IEEE Trans. Plasma Sci.} \textbf{46} (2018)
1076. \url{https://doi.org/10.1109/TPS.2017.2773095}

\bibitem{IEIRI1987253}
M.~Ieiri \textit{et al.}, {\em Nucl.\ Instrum.\
  Methods Phys. Res.} \textbf{A257} (1987) 253. \url{https://doi.org/10.1016/0168-9002(87)90744-3}

\bibitem{NIST124}
M.J.~Berger \textit{et al.}, Stopping-power and range tables for electrons, protons, and helium ions, 2017. \url{https://dx.doi.org/10.18434/T4NC7P}

\bibitem{Irakli}
I.~Keshelashvili, J\"ulich ballistic diamond pellet target for storage ring
  EDM measurements,
\newblock ERC Consolidation Grant Research Proposal, 2016. 
  \url{http://collaborations.fz-juelich.de/ikp/jedi/private_files/proposals/Proposal-SEP-210325689.pdf}

\bibitem{Jackson2010}
R.L. Jackson \textit{et al.},  {\em Nonlinear
  Dyn.} \textbf{60} (2010) 217. \url{https://doi.org/10.1007/s11071-009-9591-z}

\bibitem{Falcon1998}
E.~Falcon \textit{et al.},  {\em  Eur. Phys. J.} \textbf{B3} (1998) 45. \url{https://doi.org/10.1007/s100510050283}

\bibitem{Brach-Dunn}
R.M. Brach and P.F. Dunn,  {\em Aerosol Sci. Technol.} \textbf{16} (1992)
51. \url{https://doi.org/10.1080/02786829208959537}

\bibitem{Kim-Dunn}
O.V.~Kim and P.F.~Dunn,  {\em J. Aerosol Sci.} \textbf{38} (2007) 532. \url{https://doi.org/10.1016/j.jaerosci.2007.03.006}

\bibitem{Gupta:1279627}
M.~Gupta, Calculation of radiation length in materials, Technical Report. PH-EP-Tech-Note-2010-013 (CERN, Geneva, 2010).
  \url{https://cds.cern.ch/record/1279627}

\end{thebibliography}
\begin{flushleft}

\end{flushleft}
\end{cbunit}

\end{appendices}

\newpage

\begin{flushleft}
\mbox{}\\[1mm]
\bfseries\LARGE Acronyms and abbreviations\\[1cm]
\end{flushleft}
\addcontentsline{toc}{chapter}{Acronyms and abbreviations}
\markboth{Acronyms and abbreviations}{Acronyms and abbreviations}

\begin{longtable}{@{}lll}

       AC                      &       alternating current  \\
               ALP                     &       axion-like particle \\
               ANKE                    &       name of detector at COSY \\
        
       BMT                     &       Bargmann--Michel--Telegdi (equation) \\
               BNL                     &       Brookhaven National Laboratory \\
              BPM                     &       beam position monitor \\
               BSM                     &       beyond the Standard Model (of elementary particle physics) \\
        
       $C$                     &       charge (symmetry) \\
               CAPP                    &       Center for Axion and Precision Physics (Research) (Daejeon, South Korea) \\
CCW                     &       counterclockwise \\
               CDR                     &       conceptual design report \\
               CeNTREX                 &       name of experiment to search for proton EDM in Tl nuclei \\
               CERN                    &       Conseil Europ\'ean pour la Recherche Nucl\'eaire (European Organization for Nuclear Research)\\
               CESR                    &       Cornell Electron--Positron Storage Ring  \\
              ChPT                    &       chiral perturbation theory \\
               CKM                     &       Cabibbo--Kobayashi--Maskawa (matrix) \\
C.L. & confidence level \\
               COSY                    &       Cooler Synchrotron (storage ring) (Forschungszentrum J\"ulich, Germany) \\
              $CP$                    &       charge-parity (invariance) \\
            CPEDM                   &       Charged Particle Electric Dipole Moment (collaboration) \\
             $CPT$                   &       charge-parity-time reversal (symmetry) \\
              CSR                     &       Cryogenic Storage Ring (Max-Planck Institute, Heidelberg, Germany) \\
             CW                      &       clockwise \\
             $d$                     &       electric dipole moment\\
               DC                      &       direct current \\
              dEDM                    &       deuteron electric dipole moment \\
              DESY                    &       Deutsches Elektronen Synchrotron (Hamburg, Germany) \\
DM & dark matter           \\
              DORIS                   &       name of a detector at DESY \\
             EDM                     &       electric dipole moment \\
              EFT                     &       effective field theory\\
             ELENA                   &       Extra Low Energy Antiproton (ring) (CERN) \\
                       \si{\electronvolt}      &       electronvolt \\
                   
      FNAL                    &       Fermi National Accelerator Laboratory (Chicago, IL, USA) \\
               FRM-II                  &       Forschungsreaktor M\"unchen (Heinz Maier-Leibnitz, M\"unchen, Germany) \\
FZJ & Forschungszentrum J\"ulich \\

      $G$                     &       magnetic anomaly \\
        $g$                     &       g factor \\
   
            GR                      &       general relativity \\

      HGF                     &       Helmholtz-Gemeinschaft Deutscher Forschungszentren \\
HV & high voltage \\
      IBS                     &       Institute for Basic Science (South Korea) \\
             IKP                     &       Institut f\"ur Kernphysik (Institute for Nuclear Physics of FZJ) (J\"ulich, Germany) \\
              ILL                     &       Institut Laue--Langevin (Grenoble, France) \\
              IPP                     &       in-plane polarization\\
              ISOLDE                  &       Isotope Separator On-Line Device (CERN) \\
      JEDI                    &       J\"ulich Electric Dipole Moment Investigations (collaboration) \\
              J-PARC                  &       Japan Proton Accelerator Research Complex (Tokai, Japan) \\
               JULIC                   &       J\"ulich Light Ion Cyclotron (FZJ, Germany) \\
       KAIST                   &       Korea Advanced Institute of Science and Technology (South Korea) \\
             KM                      &       Kobayashi--Maskawa (mixing matrix) \\
            KVI                     &       Center for Advanced Radiation Technology (Groningen, The Netherlands) \\

       LANL                    &       Los Alamos National Laboratory (Los Alamos, NM, USA) \\
              LC                      &       inductance-capacitance \\

       MD                      &       molecular dynamics \\
              MDM                     &       magnetic dipole moment \\
              MeV                     &       megaelectronvolt \\
                    MV                      &       megavolt \\

       NEG                     &       non-evaporable getter (pumps) \\
       $P$ & parity (symmetry) \\
            PAC                     &       Program Advisory Committee \\
            pEDM                    &       proton electric dipole moment \\
           PNPI                    &       Petersburg Nuclear Physics Institute (Gatchina, Russia) \\
                   PSI                     &       Paul Scherrer Institute (Villigen, Switzerland) \\
              PTR                     &       prototype ring \\
       QCD                     &       quantum chromodynamics \\
      R\&D                    &       research and development \\
              RF                      &       radio frequency \\
              RLC                     &       resistance-conductance-capacitance \\
              ROI                     &       region of interest \\
             RWTH                    &       Rheinisch-Westf\"alische Technische Hochschule (RWTH Aachen University, Germany) \\

      SCT                     &       spin coherence time \\
              SM                      &       Standard Model (of elementary particle physics) \\
              SNS                     &       Spallation Neutron Source (Oak Ridge, TN, USA) \\
              SQUID                   &       superconducting quantum interference device \\
            srEDM                   &       storage ring electric dipole moment \\
              SUSY                    &       supersymmetry \\

      T                       &       tesla (unit)\\
             $T$                     &       time-reversal (symmetry) \\
             TDR                     &       technical  design report \\
               TlF                     &       thallium fluoride \\
              TRIUMF                  &       Canada's particle accelerator centre (Vancouver, Canada) \\

     US DOE                  &       United States Department of Energy  \\
       $\beta$                 &       Lorentz factor \\
       $\gamma$                &       Lorentz factor \\
     $\eta$                  &       electric dipole factor \\
       $\mu$                   &       magnetic moment \\
       $\nu_\text{s}$          &       spin tune \\
 
\end{longtable} 
\addtocounter{table}{-1}


\end{document}